\documentclass[final,3p,11pt,times]{elsarticle}
\pdfoutput=1
\usepackage{caption,multirow}
\usepackage[usenames]{color}
\usepackage{pbox}% Include figure files

\usepackage[english]{babel}

\usepackage{booktabs,rotating}
\usepackage{threeparttable}
\usepackage{afterpage}

\usepackage{etoolbox}% http://ctan.org/pkg/{etoolbox,lineno}
\makeatletter
\patchcmd{\@startsection}{\@ifstar}{\nolinenumbers\@ifstar}{}{}
\makeatother

\definecolor{olive}{rgb}{0.0, 0.5, 0.0}

\newcommand{\be}{\begin{equation}}
\newcommand{\ee}{\end{equation}}
\newcommand{\bea}{\begin{eqnarray}}
\newcommand{\eea}{\end{eqnarray}}
\newcommand{\bi}{\begin{itemize}}
\newcommand{\ei}{\end{itemize}}
\newcommand{\ben}{\begin{enumerate}}
\newcommand{\een}{\end{enumerate}}
\newcommand{\la}{\left\langle}
\newcommand{\ra}{\right\rangle}
\newcommand{\lc}{\left[}
\newcommand{\rc}{\right]}
\newcommand{\lp}{\left(}
\newcommand{\rp}{\right)}
\newcommand{\as}{\alpha_s}

\def\frac#1#2{{{#1}\over {#2}}}
\def\gsim{\mathrel{\rlap{\lower4pt\hbox{\hskip1pt$\sim$}}
    \raise1pt\hbox{$>$}}}         %greater than or approx. symbol
\def\lsim{\mathrel{\rlap{\lower4pt\hbox{\hskip1pt$\sim$}}
    \raise1pt\hbox{$<$}}}         %less than or approx. symbol
\newcommand{\mrexp}{\mathrm{exp}}

\newcommand{\art}{\mathrm{art}} 
\newcommand{\rep}{\mathrm{rep}}

\newcommand{\draft}[1]{}

\def\beq{\begin{equation}}  
\def\eeq{\end{equation}}

\usepackage{amssymb}
\usepackage{amsmath}
\usepackage{graphicx}% Include figure files
\usepackage{epsfig}
\usepackage{dcolumn}% Align table columns on decimal point
\usepackage{bm}% bold math
\usepackage{epstopdf}
\usepackage{url}
\usepackage{lineno}
\usepackage{tabu}
\usepackage{longtable}
\usepackage{lscape}

\biboptions{sort,compress}

\usepackage[colorlinks, citecolor=blue,anchorcolor=red,menucolor=red, 
linkcolor=red,filecolor=red,runcolor=red,urlcolor=blue,frenchlinks=red]{hyperref}
\usepackage{bookmark}

\newcommand{\nocontentsline}[3]{}
\newcommand{\tocless}[2]{\bgroup\let\addcontentsline=\nocontentsline#1{#2}\egroup}

\newcommand{\VU}{1}
\newcommand{\VUU}{4}
\newcommand{\UCL}{2}
\newcommand{\INPAC}{3}

\bibstyle{JHEP}

\journal{Physics Reports}

\date{\today}% It is always \today, today,
\begin{document}

\begin{frontmatter}

\title{The Structure of the Proton in the LHC Precision Era}

\author[\INPAC]{Jun Gao}
\ead{jung49@sjtu.edu.cn}
\author[\UCL]{Lucian Harland-Lang}
\ead{l.harland-lang@ucl.ac.uk}
\author[\VU,\VUU]{Juan Rojo}
\ead{j.rojo@vu.nl}

%%%%%%%%%%%%%%%%%%%%%%%%%%%%%%%%%%%%%%
% AFFILIATIONS
\address[\INPAC]{Institute of Nuclear and Particle Physics, \\
Shanghai Key Laboratory
for Particle Physics and Cosmology,\\
School of Physics and Astronomy,
Shanghai Jiao Tong University, Shanghai, China}
\address[\UCL]{Department of Physics and Astronomy, University College London, WC1E 6BT, United Kingdom}
\address[\VU]{Department of Physics and Astronomy, VU University, De Boelelaan 1081, 1081HV Amsterdam, The Netherlands}
\address[\VUU]{Nikhef, Science Park 105, NL-1098 XG Amsterdam, The Netherlands}

\begin{abstract}
  We review recent progress in the determination of the
  parton distribution functions (PDFs) of the proton, with emphasis on the applications
  for precision phenomenology at the Large Hadron Collider (LHC).
  First of all, we introduce the general theoretical
  framework underlying the global QCD analysis
  of the quark and gluon internal structure of protons.
  We then present a detailed overview of the hard-scattering measurements,
  and the corresponding theory predictions, that are used
  in state-of-the-art PDF fits.
  We emphasize here the role that higher-order QCD and electroweak corrections
  play in the description of recent high-precision collider data.
  We present the methodology used to extract PDFs in global
  analyses, including the PDF parametrization strategy and the definition
  and propagation of PDF uncertainties.
  Then we review and compare the most recent releases from
  the various PDF fitting collaborations, highlighting their differences
  and similarities.
  We discuss the role that QED corrections and photon-initiated contributions
  play in modern PDF analysis.
  We provide representative examples of the implications of PDF fits
  for high-precision LHC phenomenological applications, such as Higgs coupling measurements
  and searches for high-mass New Physics resonances.
  We conclude this report by discussing some selected topics relevant
  for the future of PDF determinations, including the treatment of
  theoretical uncertainties, the connection with lattice QCD calculations,
  and the role of PDFs at future high-energy colliders beyond the LHC.
\end{abstract}


\end{frontmatter}

\clearpage

\tableofcontents
% Show running line numbers
%\linenumbers

\clearpage

%\ccode{PACS numbers}
%\pacs{Valid PACS appear here}% PACS, the Physics and Astronomy
                             % Classification Scheme.

% If you want the \paragraph{} sections to be numbered uncomment the following
\setcounter{secnumdepth}{3}

%%%%%%%%%%
\section{Introduction}
\label{sec:introduction}

The determination of the quark and gluon structure of the proton
is a central component
of the precision phenomenology program at the Large Hadron Collider (LHC).
This internal structure is quantified in the collinear QCD factorisation framework by the
Parton Distribution Functions (PDFs), which encode the information related to
the momentum distribution of quarks and gluons within the proton.
Being driven by low--scale non--perturbative dynamics,
PDFs cannot be computed from first principles, at least
with current technology, and therefore they
need to be determined using experimental data from
a variety of hard--scattering cross sections in
lepton--proton and proton--proton collisions.
This global QCD analysis program involves combining the most PDF--sensitive
data and the highest precision QCD and electroweak calculations available within a statistically
robust fitting methodology.
See Refs.~\cite{Rojo:2015acz,Butterworth:2015oua,Ball:2012wy,
Alekhin:2011sk,Forte:2013wc,Forte:2010dt,Perez:2012um,DeRoeck:2011na,
Accardi:2016ndt} for recent reviews on PDF determinations.

One of the main motivations to improve our understanding of the internal
structure of the proton is provided by the fact that
PDFs and their associated uncertainties play a decisive role in
several LHC applications.
To begin with, they represent one of the dominant theoretical uncertainties
for the determination of the Higgs boson couplings~\cite{deFlorian:2016spz}, where any deviation
from the tightly fixed SM predictions would indicate a smoking gun for new physics.
PDF uncertainties also affect the production of new high--mass resonances,
as predicted by many Beyond the Standard Model (BSM) scenarios~\cite{Beenakker:2015rna}, since
they probe PDFs at large values of the momentum fraction $x$, for which
current experimental constraints are scarce.
A third example is provided by the measurement of precision SM parameters at hadron
colliders, such as the $W$ mass~\cite{Bozzi:2015hha,Bozzi:2015zja,Bozzi:2011ww,Aaboud:2017svj}
or the strong coupling constant $\alpha_s(Q)$~\cite{,Khachatryan:2014waa,Aaboud:2017fml,
Chatrchyan:2013txa,Rojo:2014kta,Chatrchyan:2013haa}.
These can be sensitive to BSM effects (for instance via virtual corrections due to new
particles too heavy to be produced directly~\cite{Becciolini:2014lya,Dimopoulos:1981yj})
and in many cases PDF uncertainties
represent one of the limiting factors of the measurements.

Beyond the LHC, there are also several other instances where PDFs play an important
role, for instance in astroparticle physics, such as for the accurate
predictions for signal~\cite{CooperSarkar:2011pa}
and background~\cite{Gauld:2015kvh,Garzelli:2016xmx,Zenaiev:2015rfa,Gauld:2016kpd}
events at ultra--high energy neutrino telescopes.
And needless to say, parton distributions will keep playing an important role
for any future higher--energy collider involving hadrons in the
initial state~\cite{Mangano:2016jyj,AbelleiraFernandez:2012cc},
and therefore improving our understanding of PDFs also
strengthens the physics potential
of such future colliders.

A number of collaborations provide regular updates of their PDF sets, see~\cite{Ball:2014uwa,Dulat:2015mca,
  Harland-Lang:2014zoa,Alekhin:2017kpj,Abramowicz:2015mha,Jimenez-Delgado:2014twa,Accardi:2016qay}
  and references therein.
There are a range of differences between these analyses, arising for example at the level of the selection of the input fitted dataset,
the theoretical calculations of cross sections, methodological choices
for the parametrisation of PDFs, the estimate and propagation of PDF uncertainties, and the treatment
of external parameters.
For instance, while some PDF fits are based on a global dataset, including the widest possible
variety of experimental constraints, some others are based instead
on reduced datasets (for example, without
jet data) or even on a single dataset, as the HERAPDF2.0 set~\cite{Abramowicz:2015mha}
which is based entirely on the HERA inclusive
structure functions.

Despite these differences, it has been shown that, under some well-specified conditions,
PDF sets can be statistically combined among them into a unified set.
The most popular realisation of this paradigm are the PDF4LHC15 sets~\cite{Butterworth:2015oua},
which combine the CT14, MMHT14, and NNPDF3.0 sets using the Monte Carlo (MC) method~\cite{Watt:2012tq},
and are subsequently reduced to small number of Hessian eigenvectors~\cite{Gao:2013bia,Carrazza:2015aoa} or MC
replicas~\cite{Carrazza:2015hva} to facilitate phenomenological applications.

This Report is motivated by the fact that the recent years have seen a number of important breakthroughs
in our understanding of the quark and gluon structure of the proton.
To begin with, the impressive recent progress in NNLO QCD calculations has now made it possible
to include essentially all relevant collider cross sections consistently into
a NNLO global analysis, from top quark differential distributions~\cite{Czakon:2016dgf} to inclusive
jets~\cite{Currie:2016bfm} and dijets~\cite{Currie:2017eqf},
isolated photons~\cite{Campbell:2016lzl}, and the $p_T$ distribution of
$Z$ bosons~\cite{Boughezal:2015dva,Ridder:2016nkl}, among others.
These theoretical developments
have been matched by the availability of high--precision measurements
from ATLAS, CMS, and LHCb at different centre--of--mass energies
$\sqrt{s}$, in several cases with statistical uncertainties
at the per--mile level and systematic errors at the few--percent level.
The combination of these state--of--the art calculations and high--precision
data provides an unprecedented opportunity to constrain
the PDFs, but it also represents a challenge to verify if the global
QCD analysis framework can satisfactorily accommodate them.
Indeed, the validity of the QCD factorisation theorems is being pushed to a level
never reached before.

In addition, there are a number of PDF-related topics that have attracted a lot
of attention recently.
One of these
is the role that QED and electroweak effects, and specifically
the photon PDF $\gamma(x,Q^2)$, play in global fits of parton distributions.
Recent progress has demonstrated that the photon PDF can be computed with
percent level accuracy~\cite{Manohar:2016nzj,Manohar:2017eqh},
improving on existing model~\cite{Martin:2004dh,Schmidt:2015zda,Harland-Lang:2016kog} and data--driven
determinations~\cite{Giuli:2017oii,Ball:2013hta,Bertone:2016ume}, with direct implications
for LHC cross sections.
Another development is related to more sophisticated
theoretical treatments of the charm quark PDF,
moving beyond the assumption that charm is purely
generated by perturbative DGLAP evolution~\cite{Hou:2017khm,Ball:2016neh}.
From the methodological point of view, there have been several improvements in the way that
PDFs are parametrized and the various associated sources of uncertainty are estimated among the
PDF fitting groups.
In addition, there has also been a recent explosion in the number of tools available for PDF studies,
from the open--source fitting framework {\tt xFitter}~\cite{Alekhin:2014irh}, to new
fast (N)NLO interfaces and public codes for the PDF
evolution~\cite{Bertone:2013vaa,Salam:2008qg,Botje:2010ay} and the
efficient calculation of hadronic
cross sections~\cite{Carli:2010rw,Bertone:2016lga,amcfast,Wobisch:2011ij,Czakon:2017dip}.
It is therefore the goal of this Report to present a detailed overview of these
various recent developments, and how they have modified our present understanding
of the quark and gluon structure of the proton, with emphasis
on the resulting phenomenological applications.

This Report focuses on one of the central aspects of the internal structure of
nucleons, namely collinear unpolarized PDFs, which are its
most relevant feature for the exploration of the high--energy frontier at the LHC.
There are however many other aspects of the inner life of
protons that would each deserve their own separate review, and that
due to space limitations cannot be covered here.
These include, among others,
the determination of
its spin structure by means of the polarized
PDFs~\cite{Nocera:2014gqa,deFlorian:2009vb,Sato:2016tuz}, the nuclear modifications
of the free--proton PDFs~\cite{deFlorian:2011fp,Eskola:2016oht,Kusina:2016fxy}, relevant for the understanding of cold nuclear matter effects
at the  RHIC and LHC heavy--ion program, and 
the three--dimensional
imaging of nucleons in terms of transverse--momentum--dependent
PDFs (TMD--PDFs)~\cite{Angeles-Martinez:2015sea}.
We only note here that these fields do not exist in isolation, and progress in these topics
can often directly impact unpolarized PDF fits, and vice versa.

The structure of this Report is as follows.
First, in Sect.~\ref{sec:globalqcd}
we review the theoretical foundations
of the global PDF analysis framework, specifically
the QCD factorization theorems
of lepton--hadron and hadron--hadron collisions, as well
as the scale dependence of the PDFs as encoded in the DGLAP equations.
In Sect.~\ref{sec:datatheory}  we discuss the 
experimental data, as well as the corresponding
state--of--the--art theoretical calculations, that are used
to constrain the PDFs in modern global analyses.
In Sect.~\ref{sec:fitmeth} we present
the methodological framework of PDF fits, including the various
approaches to parametrizing the PDFs and estimating and propagating
the uncertainties from theory and data to physical
cross sections.
In Sect.~\ref{sec:pdfgroups} we summarise the main features
of the different PDF collaborations that provide regular updates
of their PDF fits, and then in Sect.~\ref{sec:structure} we compare
them, assessing their differences and similarities for different
aspects of the proton structure such as the gluon PDF, quark flavour
separation, and the strange and charm content of the proton.

In Sect.~\ref{sec:QED} we then discuss the role
that QED and electroweak corrections play in modern PDF fits,
with emphasis on the photon content of the proton.
In Sect.~\ref{sec:LHCpheno} we highlight a number of
representative examples of the role of PDFs and their
uncertainties for the LHC precision physics program, in particular
the characterization of the Higgs sector, searches for massive New Physics resonances,
and the measurement of precision parameters such as $M_W$.
In the last part of this Report, Sect.~\ref{sec:PDFfuture},
we discuss some of the topics that are likely
to play an important role for the future of PDF determinations,
such as the quantification of theoretical
uncertainties, the interplay with lattice QCD calculations,
and the application of PDFs for future higher energy lepton--proton
and proton--proton colliders.
Finally, we conclude in Sect.~\ref{sec:conclusions}.

%%%%%%%%%%
\vspace{0.6cm}
\section{The global QCD analysis framework}
\label{sec:globalqcd}

In this section
we first present a brief historical account of PDF
determinations.
We then introduce the basic foundations
of the global PDF analysis program, namely the QCD factorization theorems
for lepton-hadron and hadron-hadron collisions.
We also discuss the scale dependence of parton distributions
as encoded in the DGLAP evolution equations.
In the final part of this section
we briefly address the topic of mass effects in
deep-inelastic structure functions.

\subsection{A brief history of PDF fits}\label{sec:pdffitting.history}
The first indications of a non-trivial proton internal
structure were obtained in the the pioneering experiments of Hofstadter {\it et al.}
on elastic electron--nucleon scattering~\cite{Hofstadter:1955ae,Mcallister:1956ng}.
In these experiments,
by examining the deviations from the simple Mott
scattering formulae for point--like particles, the finite extension
of the proton could be resolved.
There, the charge radius of the proton was determined to be $\simeq 0.7$ fm to within a few percent precision.

 Although this result already hinted at an underlying substructure, the serious possibility that the proton is composite originated only later, with the ideas put forward independently by Zweig~\cite{Zweig:1964jf} and Gell--Mann~\cite{GellMann:1964nj} in 1964.
By postulating the existence of three `aces' (Zweig's term) or `quarks' (Gell--Mann's) with fractional electric charge and baryon number, and spin--$1/2$, the complex structure of the hadrons and meson multiplets could be explained in a rather compact way.
However, Zweig and Gell--Mann were understandably cautious about interpreting these objects as physical particles of finite mass, rather than simply convenient mathematical structures, in particular given that the mechanism for binding such quarks together was not understood,
and stable quarks had not been seen experimentally.

This situation changed dramatically in 1967 with the new experimental data on deep inelastic scattering (DIS)  provided by the SLAC 20 GeV linear accelerator.
The physicists of the
SLAC--MIT collaboration were surprised to find that, in contrast to the case of elastic lepton--proton scattering, the two form factors associated with the DIS cross section, the so--called structure functions, were  roughly independent of $Q^2$~\cite{Bloom:1969kc,Breidenbach:1969kd}. Moreover, these appeared to exhibit the scaling behaviour predicted by Bjorken in 1969~\cite{Bjorken:1968dy}, namely that the structure functions should depend only on the ratio of $Q^2$ to the lepton energy loss $\nu$ in the proton rest frame\footnote{Indications of this scaling were also observed at the DESY electron synchrotron in the same year~\cite{Albrecht:1969zy}.}. 
 
 These observations led Feynman to introduce the parton model~\cite{Feynman:1969ej}, in which the incident lepton scatters incoherently and instantaneously with the point--like `partonic' constituents of the proton.
This concept, developed further in~\cite{Bjorken:1969ja}, naturally explains the observed Bjorken scaling behaviour, with the point--like partons in this simple picture providing no additional scale through which Bjorken scaling could be broken. At the same time Callan and Gross~\cite{Callan:1969uq} showed that the DIS structure functions obey a simple relation for the case of spin--$\frac{1}{2}$ quark constituents, a finding that was also supported by the data~\cite{Miller:1971qb}.
These partons were therefore naturally associated with the ``constituent''
quarks of Gell--Mann and Zweig.
The subsequent demonstration of asymptotic freedom in 1973 in strongly interacting non--abelian gauge theories~\cite{Gross:1973id,Politzer:1973fx}  provided a simple explanation for the observed absence of free quarks, through the process of confinement, and the QCD parton model became the established approach to describe scattering processes in the strong interactions.
 
A central ingredient of this QCD parton model are the probability distributions of the momentum
carried by these partons, known as the
parton distribution functions, or
PDFs.
The first studies concentrated on developing simple models for these objects based on the limited experimental input available, for example: in~\cite{Bjorken:1969ja} phase space considerations were used to conclude that the PDFs must also include a contribution from the now well-known sea of quark--antiquark pairs in addition to the valence quarks; in~\cite{Kuti:1971ph} a gluon PDF was introduced to account for the observed quark momentum fractions in a physically reasonable way consistent with energy conservation, and simple $x$ dependencies of the PDFs were predicted according to general Regge theory and phase space expectations. 

The idea of fitting a freely parameterised set of PDF followed soon after these initial studies.
In~\cite{McElhaney:1973nj} the approach of~\cite{Kuti:1971ph} was extended to a more general phenomenological form, and a 4--parameter fit to the quark PDFs was performed to the available data on proton and neutron structure functions.
As the amount and type of data increased, the shape of the PDFs became increasingly general, see for example~\cite{Barger:1973bn,Hinchliffe:1977jy}. Although the momentum fraction carried by the gluon could be determined by the missing contribution to DIS appearing in the momentum sum rule, it was only possible to fit its shape following the observation of scaling violations in the structure functions, first observed at FNAL~\cite{Fox:1974ry} in 1974.
Such $Q^2$ dependent deviations from simple Bjorken scaling occur due to higher--order QCD corrections to DIS and were directly connected through the DGLAP equation~\cite{Altarelli:1977zs,gl,dok,Lipatov:1974qm} to the $Q^2$ evolution of the PDFs. This allowed the first determinations of the shape of the gluon PDF to be made in~\cite{Gluck:1980cp} (see also~\cite{Baulieu:1978ze}).

The subsequent LO fits of~\cite{Eichten:1984eu}  (based on~\cite{Abramowicz:1982re}) to fixed target structure function and neutrino DIS data, and~\cite{Duke:1983gd}, which also included $J/\psi$ meson and muon pair hadroproduction, were widely used for a range of phenomenological applications.
By the late 1980s, PDF fits at NLO in the strong coupling were standard, with the earlier analyses of~\cite{Martin:1987vw,Diemoz:1987xu} fitting to fixed target DIS and the subsequent fits of~\cite{Aurenche:1988vi,Harriman:1990hi,Morfin:1990ck} including prompt photon and Drell--Yan hadroproduction.
By this time, the strategy of the `global QCD analysis'~\cite{Morfin:1990ck}
had emerged,  which emphasized fitting multiple experimental data sets from a diverse range of processes in order to disentangle PDFs of various flavours.
Around the same period,
the `dynamical' PDF set of~\cite{Gluck:1989ze} was produced with the assumption that at low scale the quark sea vanished and the gluon becomes proportional to the valence quark distributions, themselves determined from DIS data.
 
 Up to the early 90s, all DIS measurements was made with fixed target experiments and hence limited to the higher $x\gtrsim 0.01$ region.
This changed in 1992 when the HERA high-energy collider at DESY started taking data.
HERA collided 920 GeV protons with $\approx 27.5$ GeV electrons for most of the run period, allowing the previously unexplored region down to $x\sim 10^{-4}$ region to be probed at high $Q^2$.
By 1994, data from HERA were included for the first in the MRS(A)~\cite{Martin:1994kn} and CTEQ3~\cite{Lai:1994bb} global fits.
These were also the first fits to include data from the Tevatron $p\overline{p}$ collider, with in particular the $W$ asymmetry data providing new information on the quark flavour decomposition, as originally suggested in~\cite{Berger:1988tu}. 
In the years that followed, further public releases within these approaches were produced, including in particular the increasingly precise HERA measurements, as well as Tevatron data on jet production, which placed new and important constraints on the poorly-constrained medium and large-$x$ gluon. 
 
These PDF sets corresponded to the best fit only, that is, no estimate of the uncertainty on the PDFs due to the errors on the data in the fit were included, beyond relatively simple studies where a range of fits under different input assumptions might be performed to give some estimate of the spread.
This was an acceptable situation when the uncertainties on the hadron collider data were sufficiently large, however as the data precision increased, the lack of a reliable
estimate of PDF uncertainties rapidly became an issue. In the 1996 CDF measurement~\cite{Abe:1996wy} of inclusive jet production, for example, there was an apparent excess of events at high jet $E_\perp$ that was interpreted at the time as a possible sign for new physics.
In the subsequent study of~\cite{Huston:1995tw} it was shown that the gluon PDF could be modified in a way that still fit all available data, including the CDF jets. Clearly, a robust evaluation of the PDF uncertainties was needed.

The first attempts to produce such uncertainties, based on linear propagation of the experimental systematic and statistical errors through to the PDFs, considered a restricted set of DIS data~\cite{Alekhin:1996za,Botje:1999dj,Barone:1999yv,Giele:2001mr}.
The extension of these methods to the wider data set included in a global PDF fit was a complicated problem, both from a purely technical point of view, but also more conceptually.
In particular, more conventional statistical approaches to evaluating the uncertainty on the fitted PDF parameters, such as a standard `$\Delta \chi^2 =1$' variation, are only appropriate when fitting perfectly consistent data sets with purely Gaussian errors against a well--defined theory.
For PDF fits based instead on a wide range of experimental measurements, none of these criteria are fulfilled: individual data sets are often found to have a low associated likelihood, with a $\chi^2$ per degree of freedom well above one, the experimental systematic uncertainties will not always be Gaussian in nature, the fixed-order perturbative theory calculation will carry its own (usually omitted) uncertainties, and there will be in general many different
possible choices of the PDF parametrization leading
to comparable values of the $\chi^2$.
These issues were addressed in the CTEQ~\cite{Pumplin:2001ct,Pumplin:2002vw} and MRST~\cite{Martin:2002aw} PDF releases in 2002, with the basic idea being to allow the $\chi^2$ to vary from the minimum by a larger degree, or `tolerance', to properly account for these various effects.

In the mid to late 90s important advances in the theoretical treatment of heavy quark PDFs were also made, via the development~\cite{acot2,Thorne:1997ga} of Variable Flavour Number Schemes (VFNS), which provided a unified treatment of initial--state heavy quarks in the both the $Q^2 \lesssim m_q^2$ and $Q^2 \gg m_q^2$ regimes. The proof of factorization for hard scattering with heavy quarks, which underpins this, was presented in 1998~\cite{Collins:1998rz}. The use of such schemes is widespread in modern PDF sets.

The calculation of the NNLO splitting functions in 2004~\cite{mvvns,Vogt:2004mw} provided the necessary tools to go to NNLO in PDF fits, and by the time of the release of the MSTW08~\cite{Martin:2009iq} and CT10~\cite{Gao:2013xoa} sets (the successors to the MRST and CTEQ sets, respectively) NNLO was the standard for global PDFs. At the same time the ABKM09~\cite{Alekhin:2009ni} NNLO PDFs were released. These were based on the earlier studies of~\cite{Alekhin:1996za,Alekhin:2000ch,Alekhin:2005gq}, and fit to a reduced data set of DIS and fixed target Drell--Yan and dimuon production, with a classical `$\Delta\chi^2=1$' error treatment applied. A further set to consider a reduced data sample to appear at this time was the HERAPDF1.0~\cite{Aaron:2009aa} PDFs. These included only the combined H1 and ZEUS measurements from the HERA Run I phase, with the aim of determining the PDFs from a completely consistent DIS data sample. This allowed the PDF uncertainty to again be described without the introduction of a larger tolerance factor, while the uncertainties due to model assumptions and choice of parameterisation were included in addition. This NLO set was extended to NNLO in the HERAPDF2.0~\cite{Abramowicz:2015mha} PDFs, which used the final combined HERA I + II data sample. The NNLO JR09~\cite{JimenezDelgado:2008hf} set included a range of DIS and fixed target data, applying both a `standard' fitting approach and the `dynamical' approach of~\cite{Gluck:1989ze}. The subsequent JR14~\cite{Jimenez-Delgado:2014twa} set included a range of data updates, including jet production from the Tevatron.

The approaches described above differ greatly in many respects, both in the choice of input data sets, and the treatment of the corresponding theory predictions. However, while there are significant differences in the precise choice of parameterisation, in all cases these rely on parameterising the PDFs in terms of reasonably contained, $O(20-40)$, number of free variables. Moreover, while the precise prescription may vary, these are again all based on the `Hessian' linear error propagation procedure. A different approach, first discussed in~\cite{Forte:2002fg}, has been taken by the NNPDF collaboration.
There, the PDF functional forms are based on artificial neural networks, in particular
feed-forward multi-layer perceptrons, allowing many more, $O(200-300)$, free parameters. In addition, rather than constructing the PDF error from the $\chi^2$ variation about the best fit values, a `Monte Carlo' (MC) approach is taken, with a large enough sample of PDF `replica' sets each fit to randomly distributed pseudo--data generated according to the measured data values and their uncertainties. The first NNPDF1.0 fit was reported in~\cite{Ball:2010de}, at NLO and to a range of DIS and fixed target data. Subsequently, NNPDF2.1~\cite{Ball:2011mu,Ball:2011uy} provided the first NNLO PDF set within this approach, and included Tevatron data for the first time. 

More recently, data from the LHC has played an increasingly important role in PDF determination. The CT14~\cite{Dulat:2015mca}, MMHT14~\cite{Harland-Lang:2014zoa} and NNPDF2.3~\cite{Ball:2012cx} sets included LHC data on jets, $W$ and $Z$ boson production, and top pair production for the first time. In addition, ABM12~\cite{Alekhin:2013nda} was the first set from this group to include constraints from the
LHC, with data on $W$ and $Z$ boson production fit.
As we will discuss in some detail in this review, these
LHC measurements, which are being produced with increasingly high precision,
are now providing some of the most stringent constraints on the PDFs.

In parallel to these developments, there has been increasing focus on the use of PDFs as precise tools for LHC phenomenology, emphasising the need for clear benchmarking exercises between sets and PDF combinations, to provide an overall PDF uncertainty.
The PDF4LHC Working Group, created in 2008, has played a significant role in this, with the benchmarking described in~\cite{Alekhin:2011sk} leading to first so--called PDF4LHC recommendation~\cite{Botje:2011sn} for the use of PDFs and their uncertainties at the LHC.
This has subsequently been updated in~\cite{Butterworth:2015oua}.

\subsection{QCD factorization in deep-inelastic scattering}\label{sec:pdffitting.DIS}
\begin{figure}[t]
\begin{center}
  \includegraphics[scale=0.41]{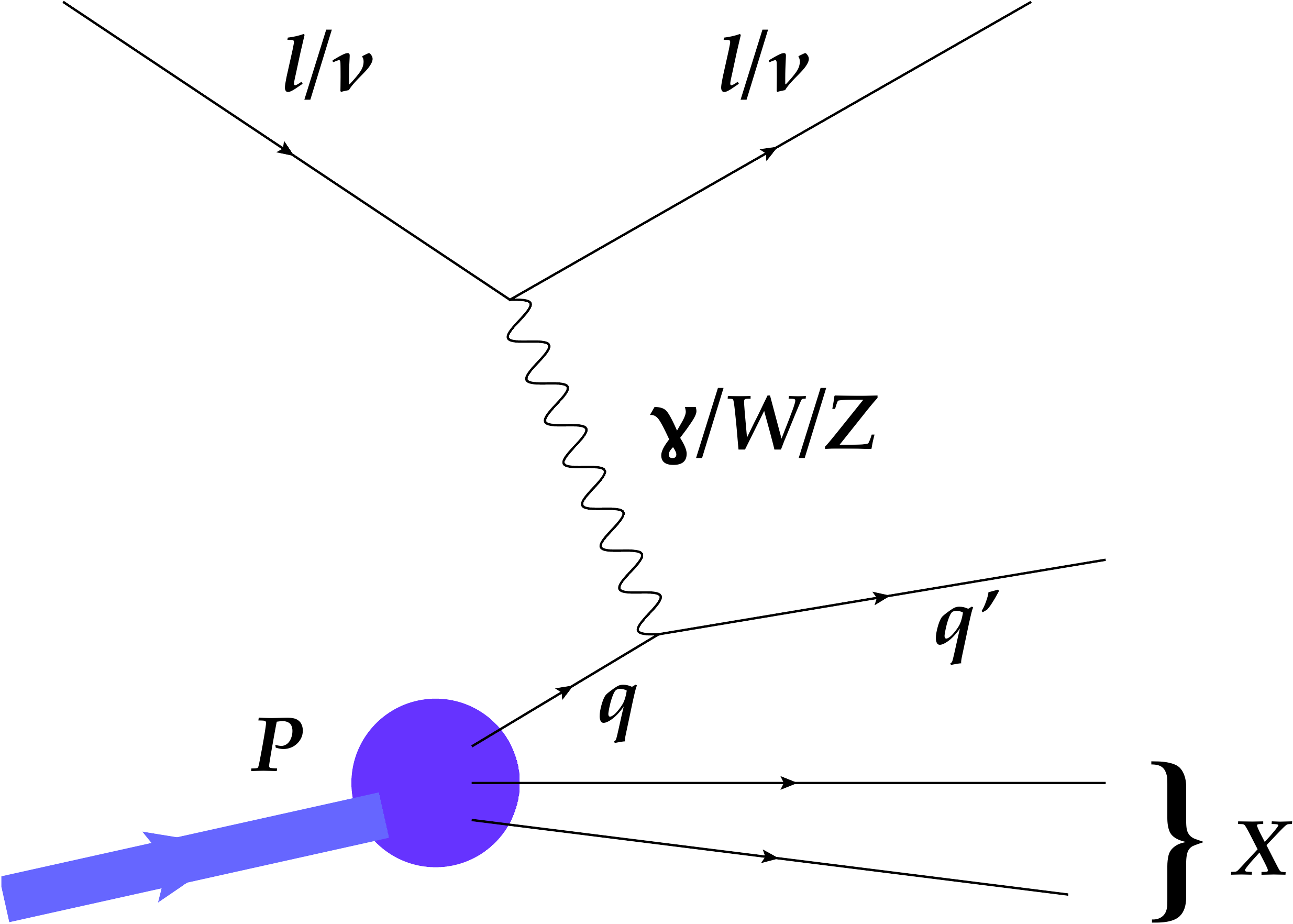}
   \caption{\small Schematic representation of
  the deep--inelastic scattering process.
   An energetic lepton (electron, muon or neutrino) scatters
   off one of the quarks
   in the proton by means of the interchange of an electroweak gauge boson
   ($\gamma, W^{\pm}$ or $Z$).
   The large virtuality carried by the gauge boson, $Q\gg \Lambda_{\rm QCD}$,
   ensures that the process can be described within the perturbative QCD
   factorization framework in terms of coefficient functions
   and PDFs.
    \label{fig:DIS-main}
  }
\end{center}
\end{figure}

The importance of deep--inelastic scattering
for PDF fits cannot be overemphasised.
As discussed in the previous section, this process was instrumental in the discovery of quarks, and has since then represented the backbone of global PDF fits. 
The DIS mechanism is schematically represented
in Fig.~\ref{fig:DIS-main}.
Here, an energetic lepton, which can be either charged
(electron or muon) or neutral (a neutrino) scatters off a proton
or some other hadron by means of the interchange of a virtual photon $\gamma^*$ or
a $W^{\pm}$ or $Z$ boson.
 The large virtuality $Q$ of the gauge boson, $Q\gg \Lambda_{\rm QCD}$,
   ensures that the process can be described within the perturbative
   QCD
   factorization framework in terms of coefficient functions
   and parton distributions, as we show below.

%%%%%%%%%%%%%%%%%%%%%%%%%%%%%%%%%%%%%%%%%%%%%%%%%%%%%%%%%%%%%%%%%%%%%
%%%%%%%%%%%%%%%%%%%%%%%%%%%%%%%%%%%%%%%%%%%%%%%%%%%%%%%%%%%%%%%%%%%%%

The kinematics of the DIS process can be described in terms
of a few invariant quantities, namely
\be
\label{eq:disvars}
x\equiv \frac{Q^2}{2 P\cdot q} \, , \quad Q^2 \equiv -q^2 \, ,
\quad y\equiv \frac{q\cdot P}{k\cdot P} \; ,
\ee
where $k$ and $k'$ are the four--momenta of the incoming and outgoing
leptons, $q$ is the four--momentum of the exchanged gauge boson, and
$P$ is incoming proton's momentum.
Here $x$ is known as the Bjorken variable, and although it is defined purely
in terms of the kinematics of the initial and final--state particles,
it can be shown that in the parton model it corresponds
to the proton momentum fraction carried by the struck parton in the centre--of--mass frame of the high--energy lepton--proton collision. In particular, if we consider
an incoming quark carrying a momentum fraction $\xi$ of the parent proton momentum, then the on--shellness relation for the outgoing quark reads
\be
(\xi P + q)^2 = -Q^2 + 2\xi (q\cdot P)=0\;,
\ee
which immediately implies that $\xi=x$.
Recall that by momentum conservation $q=k'-k$, and thus
all the variables in Eq.~(\ref{eq:disvars}) can be determined
by the knowledge of the incoming momenta of the lepton $k$ and
of the proton $P$ as well as the outgoing momentum of the lepton $k'$
without any reference to the final hadronic state $X$.
In Eq.~(\ref{eq:disvars}), $Q^2=-q^2$ is the virtuality of the electroweak
gauge boson and the variable $y$ is known as the inelasticity.
The centre of mass energy $W$ of the proton--photon collision is given by
\be
W^2 = (P+q)^2 = Q^2 \frac{1-x}{x} + m_p^2\; .
\ee
The value $x=1$ corresponds
to the elastic limit, where the proton remains intact after the collision.

It can be shown that only two of the three variables in Eq.~(\ref{eq:disvars}) are independent, and therefore differential cross sections in DIS are measured for instance as a function of $(x,Q^2)$ or $(x,y)$.
Using Lorentz invariance and kinematic
arguments, it can be shown that the DIS cross sections can be expressed
in terms of a series of independent ``structure functions'' that
describe the dynamics of the interaction between the gauge boson
and the incoming hadron.
In the neutral current (NC) case, that is, where
either a virtual photon $\gamma^*$ or a $Z$ boson
is exchanged, the DIS
differential cross section 
for a charged lepton $\ell^\pm$ scattering off a proton
can be decomposed in terms of structure functions as follows:
\be
\frac{d^2\sigma^{\rm NC,\ell^{\pm}}}{dxdQ^2} (x,y,Q^2)=\frac{2\pi \alpha^2}{ x Q^4}
 \lc
Y_+ F_2^{NC}(x,Q^2) \mp Y_- x F_3^{NC}(x,Q^2)-y^2 F_L^{NC}(x,Q^2)\rc \; ,
\label{eq:ncxsect}
\ee
where we have defined 
\begin{equation}
 Y_{\pm}=1\pm (1-y)^2\; .
\label{eq:ypmdef}
\end{equation}
In most cases, experimental measurements are given
in terms of a reduced cross section, defined as
\begin{equation}
  \label{eq:rednc}
  \widetilde{\sigma}^{\rm NC,\ell^{\pm}}(x,y,Q^2)=\lc
  \frac{2\pi \alpha^2}{ x Q^4} Y_+\rc^{-1}\frac{d^2\sigma^{\rm NC,\ell^{\pm}}}{dxdQ^2}(x,y,Q^2) \; ,
\end{equation}
which is more closely related to the dominant structure function $F_2(x,Q^2)$, and thus to the underlying PDFs of the proton.

In the case of charged current (CC) DIS, when neutrinos
are used as projectiles or when the incoming charged leptons
interact with the proton by means of the exchange of a charged
weak gauge boson $W^{\pm}$,
the differential cross sections are given by:
\begin{eqnarray}
\label{eq:ccxsect}
  \frac{d^2\sigma^{\rm CC,\ell^{\pm}}}{dxdQ^2}(x,y,Q^2) &=&\frac{G_F^2}{4\pi x}
  \lp \frac{M_W^2}{M_W^2+Q^2}\rp^2 \\\nonumber
&\times&\frac{1}{2}\lc
  Y_+ F_2^{CC,\ell^{\pm}}(x,Q^2)\mp Y_- x F_3^{CC,\ell^{\pm}}(x,Q^2)
  -y^2 F_L^{CC,\ell^{\pm}}(x,Q^2)\rc\; .
\end{eqnarray}
which is generally rescaled
to define a reduced cross section 
\begin{eqnarray}
  \label{eq:redcc}
  \widetilde{\sigma}^{\rm CC,\ell^{\pm}}(x,y,Q^2)=\lc
  \frac{G_F^2}{4\pi x}
  \lp \frac{M_W^2}{M_W^2+Q^2}\rp^2 \rc^{-1}
  \frac{d^2\sigma^{\rm CC,\ell^{\pm}}}{dxdQ^2}(x,y,Q^2)\; ,
\end{eqnarray}
similarly to the NC case.
In Eqns.~(\ref{eq:ccxsect}) and~(\ref{eq:redcc}), $\ell^\pm$ labels
either the incoming or outgoing charged lepton.

According to the QCD factorization theorem, the
general expression for the DIS structure functions can be written
schematically as
\begin{equation}
\label{eq:QCDfactorization}
  F (x, Q^2) = x \int_x^1 \frac{dy}{y} \sum_{i}
  C_i \left( \frac{x}{y}, \alpha_s (\mu_R),\mu_F ,Q\right)f_i (y, \mu_F)\; ,
\end{equation}
where the $C_i$ are known as the
coefficient functions, $f_i (y, \mu_F)$ are the PDFs,
and $\mu_F$ ($\mu_R$) are the factorization (renormalization) scales,
typically set to $\mu_R=\mu_F=Q$.
The coefficient functions represent the cross section for the
partonic scattering process $q_i+\gamma^*\to X$, and can be computed in
perturbation theory as a series expansion in the strong coupling
$\alpha_s$, as well as in the electroweak coupling $\alpha_W$ if these
corrections are included.

While the coefficient functions
encode the short distance dynamics of the parton--boson collision,
the PDFs are instead determined 
by long distance non--perturbative QCD dynamics, and 
so cannot be computed using perturbative methods.
On the other hand, the crucial factorization property of Eq.~(\ref{eq:QCDfactorization}) is that while the coefficient
functions (or in general the partonic cross sections) are process
dependent, the PDFs themselves are universal.  This allows us to parameterise and extract the PDFs from a global
analysis of hard scattering measurements. These can then be used to make predictions
 for other PDF--dependent processes.

\subsection{QCD factorization in hadronic collisions}\label{sec:pdffitting.had}
In a similar way to the DIS structure functions for electron--proton collisions, the production cross sections in proton--proton collisions can be factorized in terms of the convolution between
two universal PDFs and a process--dependent partonic cross section.
For example, the Drell--Yan production cross section, $\sigma^{\rm DY}(pp\to l^+l^-+X)$,
can be expressed~\cite{Collins:1987pm,Collins:1989gx} as 
\be
\label{eq:qcdfact1}
\frac{d^2\sigma^{\rm DY}}{dydQ^2}(y,Q^2,\mu^2_R,\mu_F^2)=\sum_{a,b=q,\bar q, g}
\int^1_{\tau_1} dx_1 f_{a}(x_1,\mu^2_F)\int^1_{\tau_2} dx_2 f_{b}(x_2,\mu_F^2
)\,\frac{d^2\hat \sigma^{\rm DY}_{ab}}{dydQ^2}(x_1,x_2,y,Q^2,\mu^2_R,\mu^2_F)\;,
\ee
where $y$ and $Q^2$ are the rapidity and invariant mass squared of the lepton pair,
and $s$ is the centre--of--mass energy of the two
incoming protons, while $\mu_F$ ($\mu_R$) are the factorization (renormalization) scales.
The lower integration limits  are $\tau_{1,2}=\sqrt{Q^2/s}\,e^{\pm y}$.
The partonic cross sections that appear in Eq.~(\ref{eq:qcdfact1}) can be computed as a perturbative expansion in $\alpha_S$:
\begin{equation}
\frac{d^2\hat \sigma^{\rm DY}_{ab}}{dydQ^2}(x_1,x_2,y,Q^2,\mu^2_R,\mu_F^2)=\sum_{n=0}^{\infty}
\left(\frac{\alpha_s(\mu^2_R)}{2\pi}\right)^n
\frac{d^2\hat \sigma^{(n)\,{\rm DY}}_{ab}}{dydQ^2}\;.
\end{equation}
From Eq.~(\ref{eq:qcdfact1}) we observe that the definition of the PDFs, once
perturbative QCD corrections are accounted for, requires the introduction of a factorization scale $\mu_F$,
below which additional collinear emissions are absorbed into a PDF redefinition.
To all orders, the physical
cross section, as a product of the PDFs and partonic cross section,
is independent of the choice of the factorization scale.
However, at any fixed order in the perturbative series, there will be some sensitivity due to the missing
higher orders, which can be minimised by choosing a suitable value of $\mu_F$
so as to maintain a better convergence of the series.
In Drell--Yan production, the conventional scale choice is $\mu^2_F=Q^2$, namely the invariant mass of the
dilepton pair.

In full generality, for
the case of the total inclusive cross section for a narrow resonance
production with mass $M$, the cross section can be factorized as
\begin{equation}
\label{eq:qcdfact2}
\sigma=\sum_{a,b=q,\bar q, g}\int_{M^2}^{s}{d\hat s\over \hat s}
\mathcal{L}_{ab}(\hat s,\mu_F^2)\,
\hat s \hat \sigma_{ab}(\hat s, M^2, \mu^2_R,\mu_F^2)\;,
\end{equation}
where $\hat s$ is the squared centre--of--mass energy of the two incoming partons, and
the parton--parton luminosity can be defined as~\cite{Campbell:2006wx}
\begin{equation}
\mathcal{L}_{ab}(\tau,\mu_F^2)={1\over s}\int_{\tau/s}^1{dx\over x}f_{a}(
\tau/sx,\mu_F^2)\,f_{b}(x,\mu_F^2) \; .
\end{equation}
The partonic cross section depends only on the kinematic variable $z\equiv
M^2/\hat s$ and $\mu_{F,R}$
\begin{equation}
\hat s \hat \sigma_{ab}(\hat s, M^2, \mu^2_R,\mu^2_F)=\sum_{n=0}^{\infty}\left(
\frac{\alpha_s(\mu_R^2)}{2\pi}\right)^nC_{ab}^{(n)}(z,\mu^2_R,\mu_F^2) \; .
\end{equation}
The usefulness of this factorized form Eq.~(\ref{eq:qcdfact2}) is that the complete
PDF dependence of the hadronic cross section is now encoded in the partonic luminosities $\mathcal{L}_{ab}$.
As will be discussed in Sect~\ref{sec:datatheory}, there has been significant
recent progress in the higher--order calculation of partonic cross sections.
In the case of inclusive processes, for instance,
the coefficient functions $C^{(n)}(z,\mu_R^2,\mu_F^2)$ are known to NNLO for Drell--Yan
production~\cite{Hamberg:1990np} and top quark
pair production~\cite{Czakon:2013goa}, and to N$^3$LO for Higgs boson production via
gluon fusion in the limit of infinite top quark mass~\cite{Anastasiou:2015ema}.

\subsection{The DGLAP evolution equations}\label{sec:pdffitting.DGLAP}
\label{subsec:dglap}

As discussed above, the PDFs depend on two variables: the Bjorken
variable $x$, which at leading order can be identified with the momentum fraction carried by the considered parton,
and the scale $Q^2$, which in DIS corresponds to the virtuality of the exchanged gauge boson.
While the dependence of the PDFs on $x$ is determined by non--perturbative dynamics, and therefore
cannot be computed perturbatively, the situation is different for the $Q^2$ variable.
Here, the $Q^2$ dependence of the PDFs is introduced when higher--order initial--state collinear singularities
of the partonic cross section are regularised by means of
a PDF redefinition.
Such singularities arise from universal long--distance QCD dynamics,
and therefore are process--independent.

For this reason,  the $Q^2$ dependence of the PDFs can be computed in QCD perturbation theory up to any given order.
This dependence is determined by a series of integro--differential
equations, known as the Dokshitzer--Gribov--Lipatov--Altarelli--Parisi (DGLAP) evolution equations~\cite{Altarelli:1977zs,gl,dok,Lipatov:1974qm}, which have the generic form
\be
  \label{eq:dglap}
  Q^2\frac{\partial}{\partial Q^2}f_i(x,Q^2) = \sum_j 
                                   P_{ij}(x,\as(Q^2))\otimes f_j(x,Q^2)\; ,
\ee
where $P_{ij}(x,\as(Q^2))$ are the Altarelli--Parisi splitting functions, which can be computed
in perturbation theory
\be
\label{eq:splittingfunctions}
P_{ij}(x,\as(Q^2)) = \sum_{n=0} \lp \frac{\alpha_s(Q^2)}{2\pi}\rp^{n+1}P_{ij}^{(n)}(x) \; ,
\ee
and where  $\otimes$ denotes the convolution
\be
\label{eq:conv}
f(x)\otimes g(x) \equiv \int_x^1 \frac{dy}{y} f(y) g\left(\frac{x}{y}\right)\; ,
\ee
which appears ubiquitously in QCD calculations.
The splitting functions Eq.~(\ref{eq:splittingfunctions}) depend on the type
of initial and final state parton that is  involved in the splitting.
At leading order, the DGLAP splitting functions are given by
\be
P_{qq}=\frac{4}{3}\lc \frac{1+x^2}{(1-x)_+} + \frac{3}{2}\delta (1-x)\rc \; ,
\ee
\be
P_{qg}=\frac{1}{2}\lc x^2+(1-x^2)\rc \; ,
\ee
\be
P_{gq}=\frac{4}{3}\lc \frac{1+(1-x)^2}{x}\rc \; ,
\ee
\be
\label{splittingfunctionsLO}
P_{gg}=6\lc \frac{1-x}{x}+x(x-1)+\frac{x}{(1-x)_+}\rc + \frac{33-2 n_f}{6}  \delta (1-x)\; .
\ee
Note that both $P_{gg}$ and $P_{qg}$ have a singularity at $x=0$: this fact is responsible
for the rapid growth at small $x$ of the gluons and consequently of the sea quarks in this region.
The overall coefficients of the splitting functions are related
to the QCD color factors.
Some splitting functions exhibit an apparent singularity at $x=1$, which cancels against those due to virtual corrections and
is regularized by means of the plus prescriptions, defined as
\be
\int_{0}^1 dx \,f(x)\lc \frac{1}{1-x}\rc_+\equiv \int_{0}^1 dx \,(f(x)-f(1))\lc \frac{1}{1-x}\rc \; .
\ee

The structure of the DGLAP evolution equations is significantly simplified if we use specific
linear combinations of PDFs.
For instance, below the charm threshold, where there are only $n_f=3$ active quarks flavours,
the following combination
\bea
\Sigma(x,Q^2) &\equiv& \sum_{i=1}^{n_f}\lp q_i+\bar{q}_i\rp (x,Q^2) \; , \nonumber \\
T_3(x,Q^2) &\equiv& \lp u+\bar{u} - d-\bar{d}\rp (x,Q^2) \; , \nonumber \\
T_8(x,Q^2) &\equiv& \lp u+\bar{u} + d+\bar{d}- 2(s+\bar{s})\rp (x,Q^2) \; ,  \\
V(x,Q^2) &\equiv& \sum_{i=1}^{n_f}\lp q_i-\bar{q}_i\rp (x,Q^2) \; , \nonumber \\
V_3(x,Q^2) &\equiv& \lp u-\bar{u} - d+\bar{d}\rp (x,Q^2) \; , \nonumber \\
V_8(x,Q^2) &\equiv& \lp u-\bar{u} + d-\bar{d}- 2(s-\bar{s})\rp (x,Q^2) \nonumber \; , 
\eea
has the property that all the quark PDF combinations except for $\Sigma$, known as the total
quark singlet, evolve independently using their own specific splitting functions.
These `non--singlet' flavour combinations therefore obey a particularly simple evolution equation.
As the $g\to q\overline{q}$ splitting can only generate an overall $q+\overline{q}$, only the singlet PDF evolution is explicitly coupled to the gluon.

The splitting functions Eq.~(\ref{eq:splittingfunctions}) are known up to $\mathcal{O}\lp \alpha_s^3\rp$
(NNLO)~\cite{mvvns,Vogt:2004mw},\footnote{Recently, the first
results towards ${\rm N}^3{\rm LO}$ splitting functions have been presented~\cite{Ablinger:2017tan,Moch:2017uml}.} and thus PDF evolution can be performed up to this order.
Several public codes implement the numerical solution of the DGLAP equations, with the {\tt HOPPET}~\cite{Salam:2008qg}, {\tt APFEL}~\cite{Bertone:2013vaa} and {\tt QCDNUM}~\cite{Botje:2010ay} codes using $x$--space methods, while the  {\tt PEGASUS}~\cite{pegasus} code performs the evolution in Mellin (moment) space.
These codes have undergone detailed benchmarking studies, with agreement at the
level of $\mathcal{O}\lp 10^{-5}\rp$ or better being found~\cite{Whalley:2005nh,Dittmar:2009ii}.

\begin{figure}[t]
\begin{center}
\includegraphics[scale=1.1]{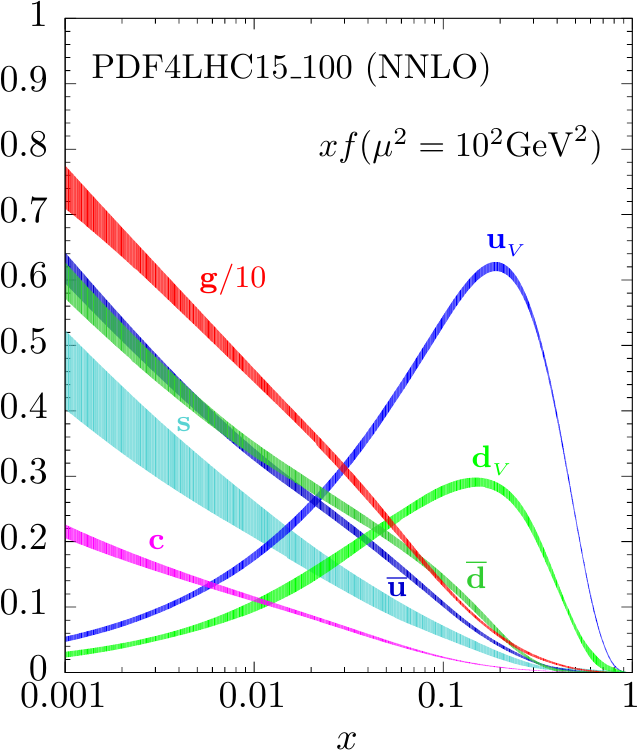}\qquad\quad
\includegraphics[scale=1.1]{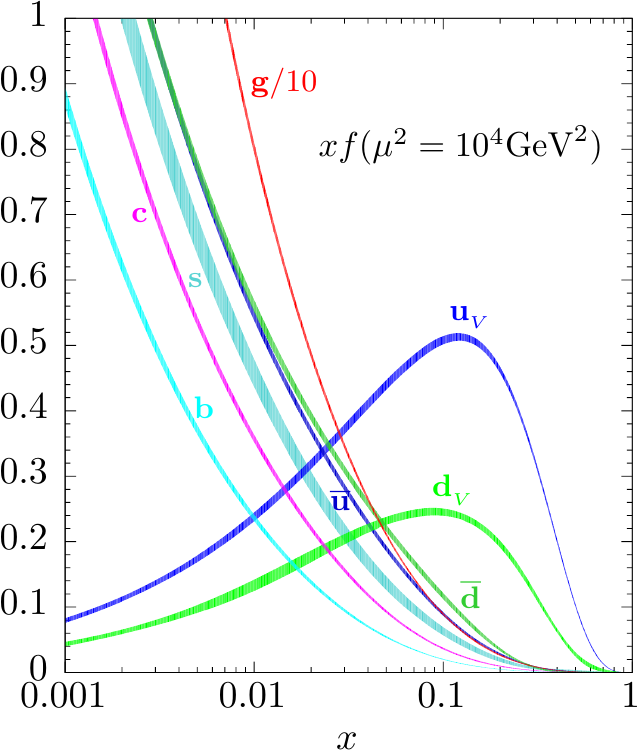}
   \caption{\small The effect of the DGLAP evolution in the PDF4LHC15
   NNLO Hessian set (with 100 eigenvectors).
   We compare
   the PDFs at a low scale of $Q^2=10$ GeV$^2$ (left) with the same PDFs evolved
   up to a typical LHC scale of $Q^2=10^4$ GeV$^2$ (right plot).
   In this plot, the PDFs include the corresponding
   one--sigma uncertainty band.
    \label{fig:pdf4lhcevolution}
  }
\end{center}
\end{figure}

In order to illustrate the impact of the DGLAP evolution on the PDFs,
in Fig.~\ref{fig:pdf4lhcevolution} we show the PDF4LHC15 NNLO Hessian set with 100 eigenvectors,
comparing
   the PDFs at a low scale of $Q^2=10$ GeV$^2$ (left) with the same PDFs evolved
   up to a typical LHC scale of $Q^2=10^4$ GeV$^2$ (right plot).
   In this plot, the PDFs are shown together with the corresponding
   one--sigma PDF uncertainty band.
   From this comparison we see that while the effects of the evolution are relatively mild on the
   non--singlet combinations $u_V=u-\bar{u}$ and $d_V=d-\bar{d}$, they are dramatic on the gluon
   and the sea quarks, where they induce a very steep growth at small $x$.
   This steep growth is driven by the small--$x$ structure of the splitting functions, see
   Eq.~(\ref{splittingfunctionsLO}).
   We also observe from Fig.~\ref{fig:pdf4lhcevolution} is that
   the valence PDFs $xu_V$ and $xd_V$ are integrable, with a similar
   shape but with $u_V \simeq 2 d_V$.
   This behaviour is
a consequence of the valence sum rules, see Sect.~\ref{sec:fitmeth}.

%%%%%%%%%%%%%%%%%%%%%%%%%%%%%%%%%%%%%%%%%%%%%%%%%%%%%%%%%%%%%%%%%%%%%
%%%%%%%%%%%%%%%%%%%%%%%%%%%%%%%%%%%%%%%%%%%%%%%%%%%%%%%%%%%%%%%%%%%%%

\subsection{Heavy quark structure functions}\label{sec:pdffitting.heavyq}
\label{sec:global.heavyq}

The contribution of the charm structure function $F_2^c(x,Q^2)$ to the total inclusive
neutral current structure function
$F_2^p(x,Q^2)$ at HERA can be as high as 25\% at small $x$ and $Q^2$, and therefore
it is crucial to compute it with high accuracy.
In particular, the non--zero value of heavy quark mass must be taken into account.
There are various theoretical schemes that have been proposed for the computation of heavy quark production
in DIS structure functions:
\begin{itemize}

\item The Zero--Mass Variable Flavour Number scheme (ZM--VFNS), where all heavy quark
mass effects are ignored but potentially large logarithms, $\ln Q/m$, are resummed
into the heavy quark parton distribution.
This is also known as the massless scheme.

\item The Fixed--Flavour Number Scheme (FFNS), where the heavy quark is always treated
as a massive particle and never as a massless parton irrespective of the value of the scale $Q$.
In this scheme, the heavy quark PDF does not exist and the number of active flavours is always
kept fixed.
This scheme takes into account heavy quark mass effects in the coefficient functions,
but does not resum logarithmically enhanced terms of the form $\ln Q/m$ that
become numerically relevant at high scales.
This is also known as the massive scheme.

\item The General--Mass Variable Flavour Number scheme (GM--VFNS) combines the advantage
of the massive and massless calculations by means of an interpolated scheme which is valid
for any value of the scale $Q$, and that matches the FFN and ZM--VFN schemes at small and large
values of $Q$, respectively.

\end{itemize}

Here we review the basic steps that enter into the construction of the GM--VFNS calculation
of heavy quark DIS structure functions, using the  FONLL derivation from Ref.~\cite{Forte:2010ta}
for illustration purposes.
Note however that from the phenomenological point of view the
resulting construction turns out to be rather similar
to that of related GM--VFN schemes
such as ACOT~\cite{acot2}, S--ACOT~\cite{Kramer:2000hn} and
TR~\cite{Thorne:1997ga,thornehq}.
Moreover, residual differences can be traced
back to the treatment of subleading corrections, as explicitly
demonstrated in the Les Houches benchmark studies of heavy quark structure
functions~\cite{LHhq,Butterworth:2014efa}.

We start by the expression of a generic DIS structure function $F(x,Q^2)$,
in a kinematic regime where one has $n_l$ light flavours and
 a single heavy
flavour of
mass $m$.
In the massless scheme, accurate when  $W \gg 4 m^2$,
the expression of $F$ in terms of PDFs and coefficient functions is the following
\begin{equation}
  F^{(n_l+1)} (x, Q^2) = x \int_x^1 \frac{dy}{y} \sum_{i = q, \bar{q}, h,
  \bar{h}, g} C_i^{(n_l+1)} \left( \frac{x}{y}, \alpha_s^{(n_l+1)} (Q^2) \right)
  f_i^{(n_l+1)} (y, Q^2), 
\end{equation}
where $q$ are the light quarks and $h$ is the heavy quark.
As indicated from the sum, in this scheme the heavy quark is treated as a massless
parton, with all finite mass effects therefore neglected.

Now, in the massive (or decoupling~\cite{Collins:1978wz}) scheme, which is most suitable
when $W \approx 4 m^2$ and thus where heavy quark mass effects must be accounted for,
this structure function reads
\begin{equation}
  F^{(n_l)} (x, Q^2) = x \int_x^1 \frac{dy}{y} \sum_{i = q, \bar{q}, g}
  C_i^{(n_l)} \left( \frac{x}{y}, \frac{Q^2}{m^2}, \alpha_s^{(n_l)} (Q^2) \right)
  f_i^{(n_l)} (y, Q^2)\; . \label{eq:Fnl}
\end{equation}
where now the massive coefficient functions $C_i^{(n_l)}$ includes the full mass dependence,
and the heavy quark is no longer treated as a massless initial--state parton.
In this scheme, the PDFs and $\alpha_s$ satisfy evolution equations with $n_l$ active quarks.
Massive coefficient functions
are known up to $\mathcal{O}\lp \alpha_s^2\rp$
both for neutral-current~\cite{Laenen:1992zk,Laenen:1992xs} and charged-current~\cite{Berger:2016inr,Gao:2017kkx} scattering, with a number of
partial results  available for the
massive
neutral-current $\mathcal{O}\lp \alpha_s^3\rp$
coefficient functions, see~\cite{Ablinger:2017err}
and references therein. 

The construction of the GM--VFNS structure functions is then based on two steps.
First of all, one needs
to express the PDFs and $\alpha_s$ in the massless scheme by means of the matching conditions
\begin{eqnarray}
  \alpha_s^{(n_l+1)} (Q^2) & = & \alpha_s^{(n_l)} (Q^2) + \sum^{\infty}_{i = 2} c_i (L)
  \times \left( \alpha_s^{(n_l)} (m^2) \right)^i \ ,  \label{eq:matchingalpha}\\
  f_i^{(n_l+1)} (x, Q^2) & = & \int_x^1 \frac{dy}{y} \sum_{j = q, \bar{q}, g} K_{ij} \left( \frac{x}{y}, L,
  \alpha^{(n_l)}_s (Q^2) \right) f_j^{(n_l)} (y, Q^2) \; ,  
  \label{eq:matchingpdf}
\end{eqnarray}
where $L \equiv \log Q^2 / m^2$. Then one uses these transformed expressions to write down $F^{(n_l)}$
in terms of PDFs and $\alpha_s$ in the massless scheme,
\begin{equation}
  F^{(n_l)} (x, Q^2) = x \int_x^1 \frac{dy}{y} \sum_{i = q, \bar{q}, g}
  B_i \left( \frac{x}{y}, \frac{Q^2}{m^2}, \alpha_s^{(n_l+1)} (Q^2) \right)
  f_i^{(n_l+1)} (y, Q^2)\; , \label{eq:Fnlbar}
\end{equation}
Once we have expressed both $F^{(n_l)}$ and $F^{(n_l+1)}$ in terms of PDFs and $\alpha_s$ in the massless scheme,
the second step is to match the two expressions while removing any double counting.
In this way we will maintain the main advantages of the two schemes (heavy quark mass effects in $F^{(n_l)}$,
resummation of large $\ln Q^2/m^2$ logarithms in $F^{(n_l+1)}$) within a single scheme that is valid
for any scale $Q$.
To achieve this, one defines the massless limit of the massive scheme structure function
as follows
\begin{equation}
  F^{(n_l,\,0)} (x, Q^2) = x \int_x^1 \frac{dy}{y} \sum_{i = q, \bar{q}, g}
  B^{(0)}_i \left( \frac{x}{y}, \frac{Q^2}{m^2}, \alpha_s^{(n_l+1)} (Q^2) \right)
  f_i^{(n_l+1)} (y, Q^2)\; , \label{eq:Fnlbarzero}
\end{equation}
where in the coefficient functions $B^{(0)}_i$ all the terms which are power suppressed
of the form $m/Q$ are neglected, and the only dependence on the heavy quark mass $m$
is on logarithms of the form $\ln Q/m$,

The FONLL approximation for $F$ is then given by
\begin{eqnarray}
&&  F^{\rm{FONLL}} (x, Q^2) =  F^{(d)} (x, Q^2)  + F^{(n_l)} (x,
  Q^2)\; , \label{eq:FONLL}
\\
 &&\quad F^{(d)} (x, Q^2)\equiv
\left[ F^{(n_l+1)} (x, Q^2) - {F}^{(n_l,\,0)} (x,
  Q^2) \right]\; ,
\label{eq:fdiff}
\end{eqnarray}
where Eq.~(\ref{eq:fdiff}) is constructed out of the
massless scheme expression $F^{(n_l+1)}$,
and 
the massless limit  $F^{(n_l,\,0)}$ of the massive scheme
expression as in Eq.~(\ref{eq:Fnlbarzero}).
It is thus easy to see that in the limit where $Q \gg m$, the FONLL
structure function reduces to the massless calculation, while for $Q \sim m$
the FONLL result coincides with the massive calculation up to subleading (higher order) terms.
Since close to threshold the difference term $F^{(d)}$ is formally subleading but numerically
non--negligible, it is customary to further suppress it using for instance a damping factor~\cite{Cacciari:1998it}
or $\chi$ rescaling~\cite{Nadolsky:2009ge}, see also
the discussion in~\cite{Chuvakin:1999nx}.

In the specific case of FONLL,
the GM--VFNS formalism can also be generalised to include mass effects
for heavy quark initiated contribution~\cite{Ball:2015tna},
such as those required {\it i.e.} in the
presence of a non--perturbative charm content in the proton.
See~\cite{Hou:2017khm} and references therein
for the corresponding discussion of
massive heavy quark initiated contribution in the ACOT framework,
as well as ~\cite{Ball:2015dpa} for a explicit
comparison between the FONLL and ACOT derivations.

\begin{figure}[t]
\begin{center}
  \includegraphics[scale=0.45]{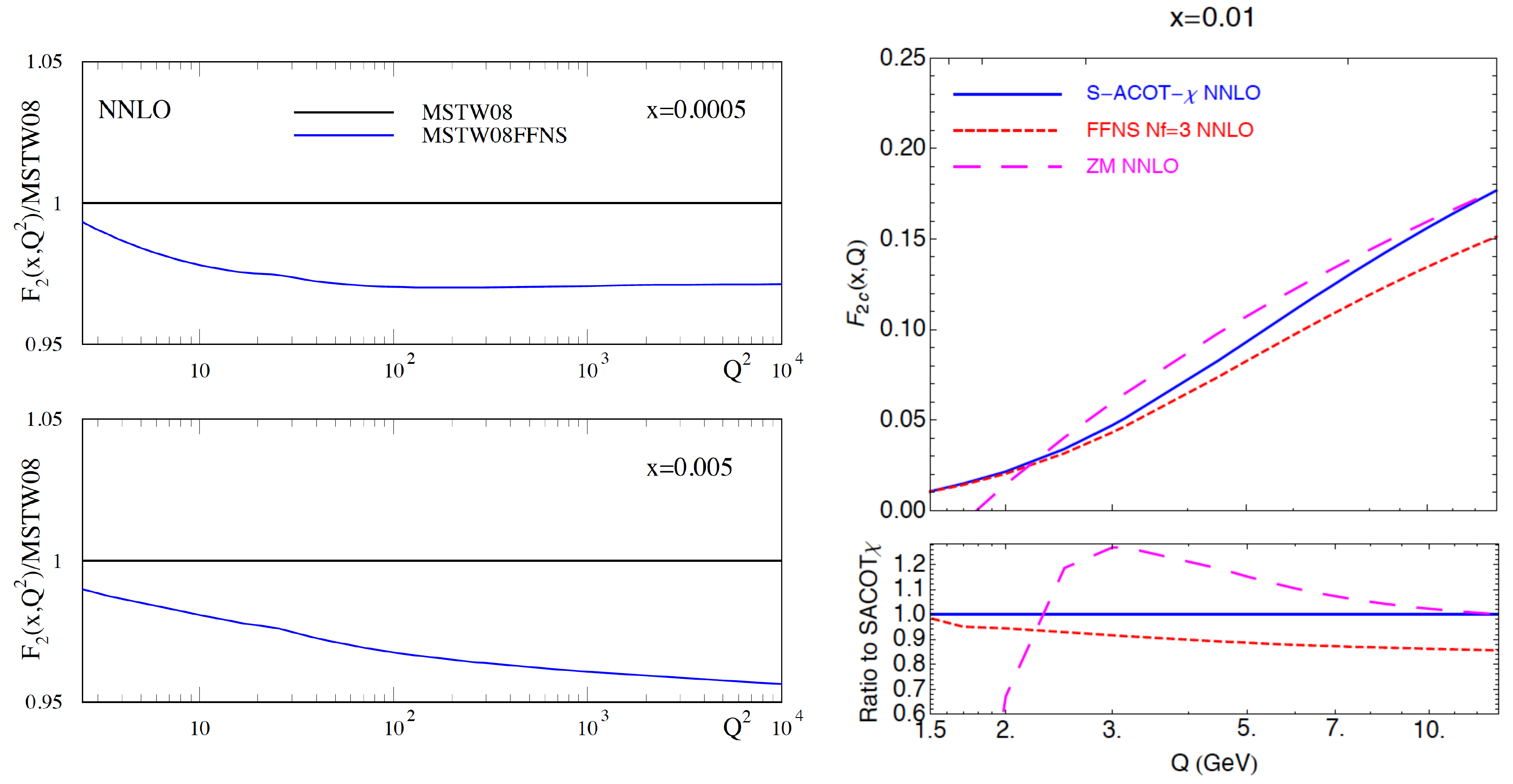}
   \caption{\small Left plot: the inclusive proton structure
   function $F_2(x,Q^2)$ at NNLO as a function of $Q^2$ for two different
   values of $x$ in the TR' GM--VFNS as compared to the FFNS calculation~\cite{Thorne:2014toa}.
   Right plot: the NNLO charm structure function $F_2^c(x,Q)$
   as a function of $Q$ for $x=0.01$ comparing the S--ACOT--$\chi$
   GM--VFNS with the corresponding ZM and FFN scheme calculations~\cite{Guzzi:2011ew}.
    \label{fig:f2c}
  }
\end{center}
\end{figure}

To illustrate the numerical impact that the heavy quark mass effects
have on the DIS functions, in
Fig.~\ref{fig:f2c} we show the
inclusive proton structure
   function $F_2(x,Q^2)$ at NNLO~\cite{Thorne:2014toa}
   as a function of $Q^2$ for two different
   values of $x$ in the TR' GM--VFNS~\cite{Thorne:2006qt}, compared to the FFNS calculation.
   We see that differences can be as large as a few percent, comparable
  to or larger than the uncertainties on the available DIS data.
  In the same figure (right) we show the NNLO charm structure function $F_2^c(x,Q)$
   as a function of $Q$ for $x=0.01$ comparing the S--ACOT--$\chi$~\cite{Guzzi:2011ew}
   GM--VFNS with the corresponding ZM and FFN scheme calculations.
   We can see that the S--ACOT--$\chi$ calculation
   smoothly interpolates between the FFN scheme at low values of
   $Q$ and the massless result at high $Q$.

%%%%%%%%%%%%%%%%%%%%%%%%%%%%%%%%%%%%%%%%%%%%%%%%%%%%%%%%%%%%%%%%%%%%%
%%%%%%%%%%%%%%%%%%%%%%%%%%%%%%%%%%%%%%%%%%%%%%%%%%%%%%%%%%%%%%%%%%%%%

Finally, it is important to note that there are subtleties associated with the definition of $F_2^{c}(x,Q^2)$
starting at $\mathcal{O}\lp \alpha_s^2\rp$~\cite{Guzzi:2011ew,Forte:2010ta}.
From the experimental point of view, all processes leading to a charm quark in the final state
enter the definition of $F_2^{c}$.
On the other hand, from the theory point of view it is more natural 
to define  $F_2^{c}$ as including only
those terms where the charm quark couples to the virtual photon $\gamma^*$, and to include other diagrams with final--state
charm quarks, such as those arising from gluon radiation off light quarks, in the `light' part of the
structure function.
Fortunately, the mismatch between the theoretical and experimental definitions is small and limited
to the large--$x$ and large-$Q^2$ region, where experimental uncertainties on $F_2^c$ are in any case
rather large.
Moreover, this issue does not affect the inclusive structure function measurements and corresponding
theory calculations as it arises only when the charm quark part of $F_2$ is explicitly separated
from the light quark part.

%%%%%%%%%%%%%%%%%%%%%%%%%%%%%%%%%%%%%%%%%%%%%%%%%%%%%%%%%%%%%

%%%%%%%%%%
\vspace{0.6cm}
\section{Experimental data and theoretical calculations}
\label{sec:datatheory}

In this section, we discuss the experimental data that
are used in modern global PDF analyses, as well as the status of the corresponding
theoretical calculations and fast interfaces for their
inclusion in the PDF fits.
For each process, we discuss first the PDF sensitivity, then
the available data and state--of--the--art theory calculations,
and finally illustrate the impact on PDFs.
We start with a general overview of the datasets that are available
for PDF studies and then we move to discuss each process
separately, beginning with DIS and then considering inclusive jet
and weak boson production, the $p_T$ of $Z$ bosons, direct
photon, top-quark pair production, open charm and $W$ boson in association with charm production, and 
central exclusive production.
In the last part of this section we discuss the technical but nonetheless important topic
of fast (N)NLO interfaces, allowing the efficient inclusion of higher-order calculations
in PDF fits.

While the list of processes described in this section focuses
on those that have so far been demonstrated to provide the most
important constraints for PDF determinations, it is certainly not intended to be
exhaustive.
There are other processes that we do not discuss here, such as $W$ and $Z$ boson production in association with jets
and their ratios~\cite{Malik:2013kba}, which can provide PDF constraints.

\subsection{Overview}\label{sec:datatheory.overview}
\begin{figure}[t]
\begin{center}
  \includegraphics[scale=0.60]{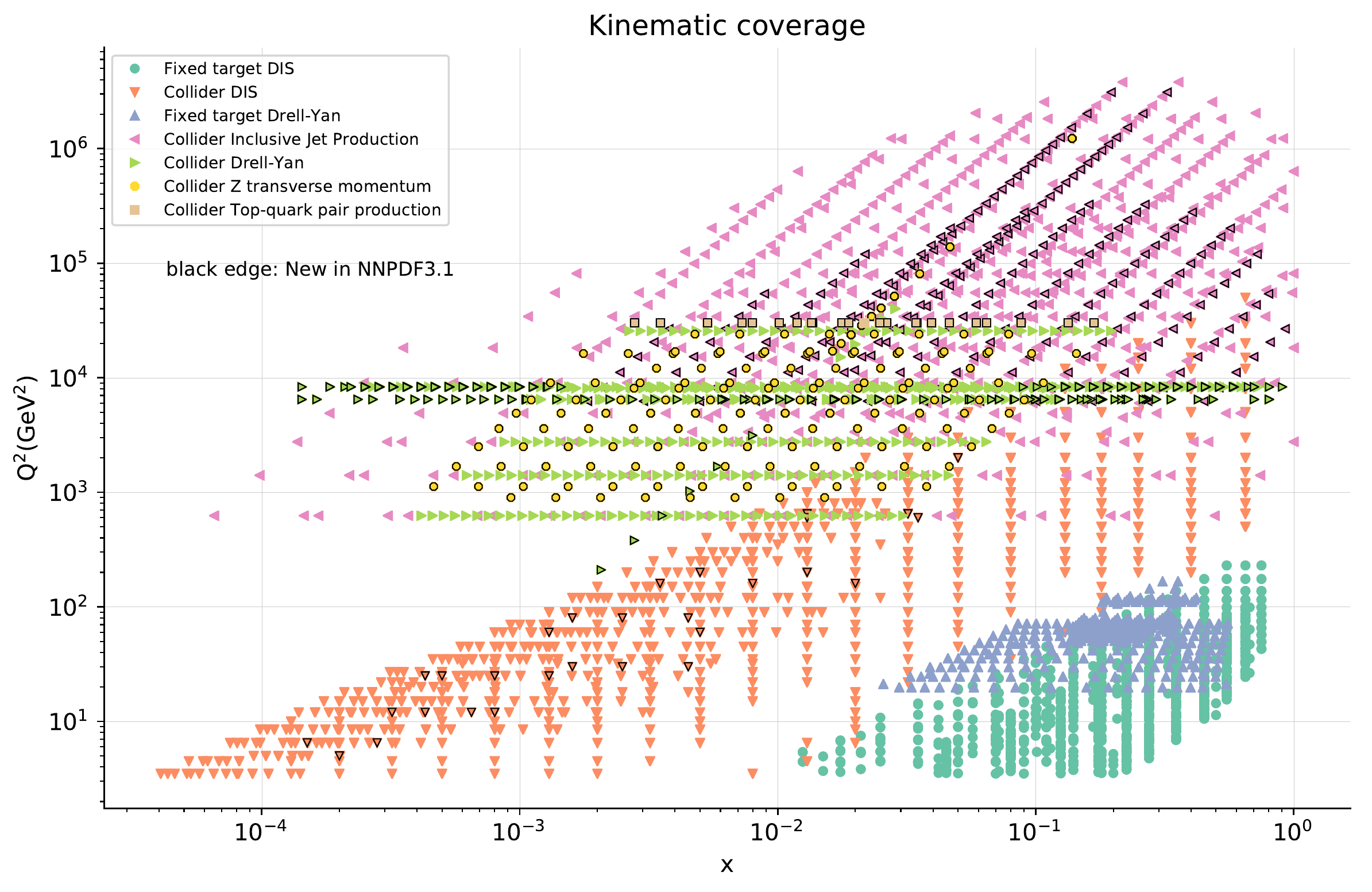}
  \caption{\small Typical kinematical coverage in the
    $(x,Q^2)$ plane for the datasets included in a global
    analysis, in this case NNPDF3.1.
    For hadronic observables, leading order kinematics
    are assumed to map each data bin to a pair of $(x,Q^2)$
    values.
    The various datasets are clustered into families of
    related processes.
    \label{fig:kinplot-report} 
  }
\end{center}
\end{figure}

In Fig.~\ref{fig:kinplot-report} we show the
 kinematic coverage in the
    $(x,Q^2)$ plane of the datasets included in a representative global
    analysis, in this case the recent NNPDF3.1 fit~\cite{Ball:2017nwa}.
 For the hadronic observables, leading order kinematics
    are assumed to map each data bin to a pair of $(x,Q^2)$
    values, while the various datasets are clustered into families of
    related processes.

%%%%%%%%%%%%%%%%%%%%%%%%%%%%%%%%%%%%%%%%%%%%%%%%%%%%%%%%%%%%%%%%%%%%%
%%%%%%%%%%%%%%%%%%%%%%%%%%%%%%%%%%%%%%%%%%%%%%%%%%%%%%%%%%%%%%%%%%%%%

We can see that a global dataset provides a rather wide coverage
in the $(x,Q^2)$ plane.
The low--$x$ and $Q^2$ region is dominated by the inclusive HERA structure function
measurements, which provide information down to $x\sim 3\cdot 10^{-5}$.
The high--$x$ region is covered by various processes, from fixed--target
DIS structure functions at low $Q^2$ to collider jet, Drell--Yan and top quark
pair production at large $Q^2$.
The very high $Q^2$ region, up to a few ${\rm TeV}^2$, is only covered by inclusive jet production
data from ATLAS and CMS.
Until relatively recently, most PDF fits were only based on DIS and
fixed--target data, with some data from the Tevatron included. The breath of experimental
information that is now included in the latest PDF fits is therefore quite impressive, with data from processes such as the $Z$ $p_T$
and the $t\bar{t}$ differential distribution  only recently being considered for the first time.

In Table~\ref{tab:PDFsummarytable} (an extended version of Table~1 from~\cite{Martin:2009iq}) we present another overview of the data entering a modern global PDF analysis. Here, we summarise the various hard scattering processes which are used to constrain PDFs in a global analysis.
 In each case we indicate the hadron--level process,
 the corresponding dominant 
 parton--level process, as well as the partons which
 are constrained in each case and the corresponding range of $x$.
 Note that the latter are necessarily approximate, and only indicate in a qualitative
 way the $x$ region that dominates the PDF sensitivity of each measurement.
 The necessity to include as broad a set of input datasets as possible, in order to constrain all quark flavour combinations and the gluon in the phenomenologically
 relevant region of $x$, is clear.
 We also note that the medium to small--$x$ region, $x\lsim 0.01$, is mostly
 covered by the HERA collider structure functions as well as by some LHC data.
 The very small--$x$ region, below the coverage of the HERA data,
 $x\lesssim 5\cdot 10^{-5}$, is only accessed via $D$ meson production
 and exclusive $J/\psi$ production in the processes listed above.
 
%%%%%%%%%%%%%%%%%%%%%%%%%%%%%%%%%%%%%%%%%%%%%%%%%%%%%%%%%%%%%%%%%%%%%%%%%%%%%
\begin{table}[h!]
\small
  \renewcommand{\arraystretch}{1.15}
  \begin{center}
    \begin{tabular}{lllll}
      \hline
      \hline
 &      Process & Subprocess & Partons & $x$ range \\ \hline
\multirow{7}{*}{Fixed Target}   &   $\ell^\pm\,\{p,n\}\to\ell^\pm+X$ & $\gamma^*q\to q$ & $q,\bar{q},g$ & $x\gtrsim 0.01$ \\
   &   $\ell^\pm\,n/p\to\ell^\pm+X$ & $\gamma^*\,d/u\to d/u$ & $d/u$ & $x\gtrsim 0.01$ \\
   &   $pp\to \mu^+\mu^-+X$ & $u\bar{u},d\bar{d}\to\gamma^*$ & $\bar{q}$ & $0.015\lesssim x\lesssim 0.35$ \\
   &   $pn/pp\to \mu^+\mu^-+X$ & $(u\bar{d})/(u\bar{u})\to \gamma^*$ & $\bar{d}/\bar{u}$ & $0.015\lesssim x\lesssim 0.35$ \\
   &   $\nu (\bar{\nu})\,N \to \mu^-(\mu^+)+X$ & $W^*q\to q^\prime$ & $q,\bar{q}$ & $0.01 \lesssim x \lesssim 0.5$ \\
   &   $\nu\,N \to \mu^-\mu^++X$ & $W^*s\to c$ & $s$ & $0.01\lesssim x\lesssim 0.2$ \\
   &   $\bar{\nu}\,N \to \mu^+\mu^-+X$ & $W^*\bar{s}\to\bar{c}$ & $\bar{s}$ & $0.01\lesssim x\lesssim 0.2$ \\
   \hline
\multirow{5}{*}{Collider DIS}   &   $e^\pm\,p \to e^\pm+X$ & $\gamma^*q\to q$ & $g,q,\bar{q}$ & $0.0001\lesssim x\lesssim 0.1$ \\
   &   $e^+\,p \to \bar{\nu}+X$ & $W^+\,\{d,s\}\to \{u,c\}$ & $d,s$ & $x\gtrsim 0.01$ \\
   &   $e^\pm p\to e^\pm\,c\bar{c}+X$ & $\gamma^*c\to c$, $\gamma^* g\to c\bar{c}$ & $c$, $g$ & $10^{-4}\lesssim x\lesssim 0.01$ \\
   &    $e^\pm p\to e^\pm\,b\bar{b}+X$ & $\gamma^*b\to b$, $\gamma^* g\to b\bar{b}$ & $b$, $g$ & $10^{-4}\lesssim x\lesssim 0.01$ \\	
   &   $e^\pm p\to\text{jet}+X$ & $\gamma^*g\to q\bar{q}$ & $g$ & $0.01\lesssim x\lesssim 0.1$ \\
   \hline
\multirow{4}{*}{Tevatron}    &   $p\bar{p}\to \text{jet}+X$ & $gg,qg,qq\to 2j$ & $g,q$ & $0.01\lesssim x\lesssim 0.5$ \\
   &   $p\bar{p}\to (W^\pm\to\ell^{\pm}\nu)+X$ & $ud\to W^+,\bar{u}\bar{d}\to W^-$ & $u,d,\bar{u},\bar{d}$ & $x\gtrsim 0.05$ \\
   &   $p\bar{p}\to (Z\to\ell^+\ell^-)+X$ & $uu,dd\to Z$ & $u,d$ & $x\gtrsim 0.05$ \\
   &   $p\bar{p} \to t\bar{t}+X$ & $qq\to t\overline{t}$& $q$ & $x \gtrsim 0.1$ \\
       \hline
\multirow{11}{*}{LHC}    &   $pp\to \text{jet}+X$ & $gg,qg,q\bar{q}\to 2j$ & $g,q$ & $0.001\lesssim x\lesssim 0.5$ \\
   &   $pp\to (W^\pm\to\ell^{\pm}\nu)+X$ & $u\bar{d}\to W^+,d\bar{u}\to W^-$ & $u,d,\bar{u},\bar{d},g$ & $x\gtrsim 10^{-3}$ \\
   &   $pp\to (Z\to\ell^+\ell^-)+X$ & $q\bar{q}\to Z$ & $q,\bar{q},g$ & $x\gtrsim 10^{-3}$ \\ 
   &   $pp\to (Z\to\ell^+\ell^-)+X$, $p_\perp$ & $gq(\bar{q})\to Z q(\bar{q})$ & $g,q,\bar{q}$ & $x\gtrsim 0.01$ \\ 
   &   $pp\to (\gamma^*\to\ell^+\ell^-)+X$, Low mass & $q\bar{q}\to \gamma^*$ & $q,\bar{q},g$ & $x\gtrsim 10^{-4}$ \\
   &    $pp\to (\gamma^*\to\ell^+\ell^-)+X$, High mass & $q\bar{q}\to \gamma^*$ & $\bar{q}$ & $x\gtrsim 0.1$ \\	
   &    $pp\to W^+ \bar{c}, W^- c$ & $sg \to W^+ c$, $\bar{s}g\to W^- \bar{c}$ & $s,\bar{s}$ & $x\sim 0.01$\\
   &    $pp \to t\bar{t}+X$ & $gg\to t\overline{t}$& $g$ & $x\gtrsim 0.01$ \\
   &    $pp \to D,B+X$ & $gg\to c\bar{c},$ $b\bar{b}$& $g$ & $x \gtrsim 10^{-6},10^{-5}$ \\
   &    $pp \to J/\psi, \Upsilon+ pp$ & $\gamma^*(gg)\to c\bar{c},$ $b\bar{b}$& $g$ & $x \gtrsim 10^{-6},10^{-5}$ \\
   &    $pp \to \gamma+ X$ & $g q(\bar{q}) \to \gamma q(\bar{q})$& $g$ & $x \gtrsim 0.005$ \\
       \hline
    \end{tabular}
  \end{center}
  \caption{Overview of the various hard--scattering processes which are used to constrain PDFs in a global analysis. In each case we indicate the hadron--level process and the corresponding dominant parton--level process, as well as the partons which are constrained by each specific process in a given range of $x$. This table
  is an extended version of Table 1 of~\cite{Martin:2009iq}.
  The $x$ ranges are merely indicative and based on approximate leading--order kinematics.
  }
  \label{tab:PDFsummarytable}
\end{table}
%%%%%%%%%%%%%%%%%%%%%%%%%%%%%%%%%%%%%%%%%%%%%%%%%%%%%%%%%%%%%%%%%%%%%%%%%%%%%

In the rest of this section, we discuss in turn
the various processes that can be used to constrain the parton distributions
in a global analysis, listed in Table~\ref{tab:PDFsummarytable}. We restrict the discussion to theoretical calculations based on fixed--order perturbative
QCD, see Refs.~\cite{White:2006yh,Bonvini:2017ogt,Ball:2017otu} and~\cite{Bonvini:2015ira} for
studies of the impact of the PDF fit of theory calculations based
on all--order resummations of logarithmically enhanced terms at small-$x$ and large--$x$
respectively.

\subsection{Deep-inelastic scattering}\label{sec:datatheory.DIS}
\subsubsection*{PDF sensitivity}
\label{sec:datatheory.DIS.sensitivity}

Before the establishment of QCD as the renormalizable quantum field theory of the strong interaction, the results of DIS
experiments were interpreted in the context of the so--called quark
parton model.
In this model, the proton is composed of non--interacting, co--moving quarks,
each of them carrying a given fraction $x$ of its total momentum. In this case the
DIS structure functions have particularly simple
expressions in terms of the PDFs.
Moreover, in this model the PDFs have a simple probabilistic interpretation,  with $q_i(x)\Delta x$ 
giving the probability of finding a quark of flavour $i$ inside the proton carrying 
a momentum fraction in the range $\lc x, x+\Delta x\rc$.
The expressions of the DIS structure functions in the quark parton model
therefore provide a useful way to illustrate the PDF sensitivity of this process.

For the NC DIS structure functions $F_2$ and
$F_3$, as defined in (\ref{eq:ncxsect}), the quark parton model expressions
are given by
\be
\label{eq:partonmodel1}
\lc  F_2^{\gamma},F_2^{\gamma Z}, F_2^Z\rc = x\sum_{i=1}^{n_f}\lc e^2_i, 2e_ig_V^i, (g_V^{i})^2+(g_A^{i})^2\rc \lp q_i+\bar{q}_i\rp \, ,
\ee
\be
\label{eq:partonmodel2}
\lc  F_3^{\gamma},F_3^{\gamma Z}, F_3^Z\rc = x\sum_{i=1}^{n_f}\lc 0, 2e_ig_A^i, 2 g_V^i g_A^i\rc \lp q_i-\bar{q}_i\rp \, ,
\ee
while the longitudinal structure function vanishes in this model, $F_L=0$, and the superscripts on the LHS
indicate the gauge boson which is being interchanged, as well as the contribution from the $\gamma Z$
interference term. $e_i$ is the electric charge of the quark of
flavour $i$ and the weak couplings are given by $g_V^i=\pm \frac{1}{2}-2e_i\sin^2\theta^2_W$
and $g_A^i=\pm \frac{1}{2}$, where the $\pm$ corresponds to a $u$ or $d$ type quark.
The sum runs over all the $n_f$ quarks that are active for the specific scale at which the scattering
takes place.
From Eqns.~(\ref{eq:partonmodel1}) and~(\ref{eq:partonmodel2}) we see that the main limitation
of the NC structure functions is that they provide limited access to quark flavour
separation and in particular they cannot separate quarks from antiquarks, unless one goes to very high $Q^2$
values where the suppression induced by the $Z$ boson propagator can be ignored.

In the case of CC DIS, the corresponding expressions for the structure functions
in the parton model, assuming that we are above the charm threshold but below the top quark
threshold, and the CKM suppressed transitions can be neglected, are given by
\bea
F_2^{W^-}&=&2x\lp u+\bar{d}+\bar{s}+c \rp \, , \nonumber
\\
F_3^{W^-}&=&2x\lp u-\bar{d}-\bar{s}+c \rp \, , 
\\
F_2^{W^+}&=&2x\lp d+\bar{u}+\bar{c}+s \rp \, , \nonumber
\\
F_3^{W^+}&=&2x\lp d-\bar{u}-\bar{c}+s \rp \, , \nonumber
\eea
where again the longitudinal structure function $F_L^{W^\pm}=0$ vanishes
in this model.
By comparing the NC and CC expressions, we can see that the main difference between
them is that in the latter case the $F_3^W$ structure function, which provides information
on the difference between quark flavours, is not suppressed with respect to $F_2^W$.
For this reason, CC structure functions, both from HERA and from neutrino
fixed--target experiments,
are generally included in global fits in order to improve the discrimination between quarks and anti--quarks.

These quark parton model expressions are also valid at LO in perturbative QCD, once
the effects of the DGLAP evolution are accounted for as described in Sect.~\ref{subsec:dglap}.
It is only at NLO that the contribution from the gluon PDF must also be included, and therefore the inclusive DIS structure functions will only be  sensitive to the gluon PDF
either through scaling violations (that is, the effect on the quark DGLAP evolution) or via the small $\mathcal{O}(\alpha_s)$ contribution to the coefficient functions.
While scaling violations
represent a small effect at medium and large-$x$, they
are a significantly more important effect  at small-$x$,
in particular in the region covered by HERA.
Therefore, scaling violations from the HERA $F_2^p$ data provide direct
information on the small-$x$ gluon, while at medium to large $x$ the gluon is poorly constrained in a DIS--only fit.

Additional information on the gluon PDF can be obtained
by means of the longitudinal structure function $F_L$, which vanishes at LO, while at
NLO it is non--zero and directly sensitive to the gluon PDF.
Indeed, it can be shown that this structure function is given by
\be
F_L(x,Q^2)=\frac{\alpha_s(Q^2)}{\pi}\lc \frac{4}{3}\int_x^1 \frac{dy}{y}\lp \frac{x}{y}\rp^2F_2(y,Q^2)
+2\sum_i e_i^2 \int_x^1 \frac{dy}{y}\lp \frac{x}{y}\rp^2 \lp 1-\frac{x}{y}\rp g(y,Q^2)\rc \, ,
\ee
which is known as the Altarelli--Martinelli relation~\cite{Altarelli:1978tq}.
For this reason, $F_L$ measurements can, in principle, provide direct constraints
on the gluon in particular at small $x$, provided experimental
uncertainties are competitive.

Finally, in addition to the inclusive structure functions, it is also possible to determine the heavy quark structure functions experimentally, by selecting DIS events with charm or bottom mesons in the final state. The LO process proceeds via the photon--gluon fusion mechanism $\gamma g \to q\overline{q}$, see Fig.~\ref{fig:dis-diagrams} (right), and therefore heavy quark structure functions
offer direct information on the gluon PDF, as well as on the treatment of heavy quark mass effects in the theoretical calculation.
Charm structure functions in addition are an
important ingredient for the determination of the charm
mass $m_c$ together with the PDFs~\cite{Bertone:2016ywq,Gao:2013wwa,Alekhin:2012vu}.
On the other hand, while data on $F_2^b$ are known to have a small impact in the global
 fit, it is relevant for specific applications, for instance
 the determination of the bottom quark mass $m_b$ from
 the PDF fit~\cite{Harland-Lang:2015qea}.

\subsubsection*{Experimental data}\label{sec:datatheory.data}
Since the pioneering DIS experiments at SLAC in the late 60s
and early 70s, there have been many other
measurements of the DIS structure functions.
These have been performed using either electrons,
positrons, muons or neutrinos as projectile, and scattering  off protons, deuterons
and neutrons, either for fixed--target or for collider kinematics.
We now discuss these in turn.

The fixed--target DIS measurements available for PDF fits
can be divided into NC and CC datasets. In the NC case this includes:

\begin{itemize}
\item Proton and deuteron structure function data
by the BCDMS collaboration~\cite{Benvenuti:1989rh,Benvenuti:1989fm}, using muons
as projectiles.
\item Proton and deuteron structure function data
by the NMC collaboration~\cite{Arneodo:1996kd,Arneodo:1996qe},
as well as measurements of the ratio between deuteron
to proton structure functions, $F_2^d/F_2^p$.
\item  SLAC measurements of the proton and deuteron
NC structure functions~\cite{Whitlow:1991uw}.
\item Proton, neutron
and deuteron structure function data at high $x$ and low $Q^2$ performed by 
JLAB experiments such as CLAS~\cite{Tkachenko:2014byy}.
While these are excluded from most PDF fits by the typical DIS cuts
in $x$ and $Q^2$, these are included in the CJ fits~\cite{Accardi:2016qay,Accardi:2011fa,Accardi:2009br}.
\item Older structure function data from the EMC collaboration~\cite{Aubert:1982tt}.
\end{itemize}

In the specific case of the EMC charm
structure functions $F_2^c$, despite their age
this measurement has never
been repeated, and thus it provides unique information on the charm
content of the proton at high $x$.
Several groups have however experienced trouble in fitting this data, which
is known to be affected by a number of problems such as
underestimated systematic errors and an incorrect
charm branching fraction into muons.
The impact of this data on the PDFs, and in particular
on the charm content of the proton, is further discussed in
Sects.~\ref{sec:pdfgroups.NNPDF} and~\ref{sec:structure.charm}.

\begin{figure}[t]
\begin{center}
  \includegraphics[scale=0.42]{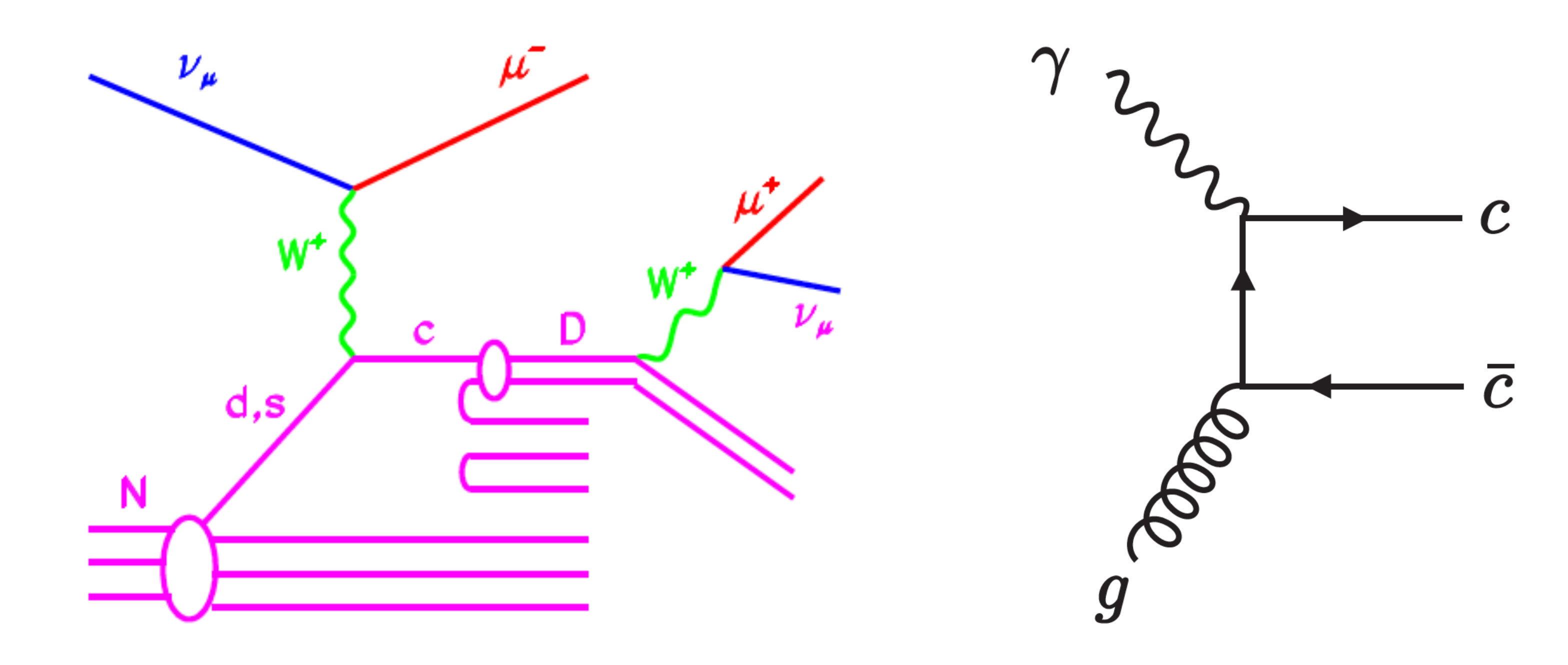}
   \caption{\small Left plot: $D$ meson production
   in CC neutrino-induced DIS.
   This is known as the `dimuon' process, since events are tagged
   when the $D$ meson decays semi--leptonically, with
   the pair of oppositely--charged muons providing a clean signature.
   Right plot:
   charm production in neutral current DIS at leading order proceeds
   via the photon--gluon fusion process, 
   highlighting its sensitivity to the gluon PDF.
    \label{fig:dis-diagrams}
  }
\end{center}
\end{figure}

In the CC case we have:

\begin{itemize}

\item Inclusive structure function measurements due to neutrino beams on nuclear targets, by the CDHSW, CCFR~\cite{Seligman:1997mc,Yang:2000ju} and CHORUS~\cite{Onengut:2005kv} and NuTeV~\cite{Zeller:2001hh} collaborations.

\item Charm production in neutrino--induced DIS.
This process is often referred to as dimuon production, since the charm quark hadronizes into a $D$ meson
which then decays semi--leptonically, see Fig.~\ref{fig:dis-diagrams}.
Data has been taken by the CCFR and NuTeV~\cite{Goncharov:2001qe,MasonPhD} and
and CHORUS~\cite{KayisTopaksu:2008aa} collaborations on the same nuclear
targets as the corresponding inclusive measurements, and also by the NOMAD collaboration~\cite{Samoylov:2013xoa}.

\end{itemize}

%%%%%%%%%%%%%%%%%%%%%%%%%%%%%%%%%%%%%%%%%%%%%%%%%%%%%%%%%%%%%%%%%%%%%
%%%%%%%%%%%%%%%%%%%%%%%%%%%%%%%%%%%%%%%%%%%%%%%%%%%%%%%%%%%%%%%%%%%%%

For the DIS measurements from the HERA lepton--proton collider we have:

\begin{itemize}
\item The final measurements of the NC and CC differential cross sections
using electron and positron projectiles from the combination
of the Run I and Run II data--taking periods~\cite{Abramowicz:2015mha}.

\item The latest heavy flavour measurements of
the combined NC cross sections of charm production
in DIS, $\widetilde{\sigma}_{c}$~\cite{Abramowicz:1900rp}  and
the H1 and ZEUS data on the bottom structure function
 $F_2^b(x,Q^2)$~\cite{Aaron:2009af,Abramowicz:2014zub}.

\end{itemize}

This HERA legacy combination of DIS inclusive structure
functions supersedes all previous
inclusive measurements from H1 and ZEUS, including the Run I
combined dataset~\cite{Aaron:2009aa} as well as the separate
measurements by the two experiments
from Run II~\cite{Aaron:2012qi,Collaboration:2010ry,
Abramowicz:2012bx,Collaboration:2010xc}.
The impact of replacing these individual datasets by the final
HERA combination of inclusive structure functions has been studied
by different
groups~\cite{Rojo:2015nxa,Harland-Lang:2016yfn,Hou:2016nqm},
and is found to be quite moderate in general.
We also note that previous measurements of the longitudinal structure
function $F_L$ by the H1 and ZEUS collaborations~\cite{Collaboration:2010ry,Andreev:2013vha,Abramowicz:2014jak}
are now superseded by the final inclusive HERA combination.

\subsubsection*{Theoretical calculations and tools}\label{sec:datatheory.DIStheory}

The coefficient functions
of the DIS structure functions in the NC case
 are available up to  $\mathcal{O}\lp \alpha_s^3\rp$ in the massless
limit~\cite{Moch:2004xu,Vermaseren:2005qc}, and up to  $\mathcal{O}\lp \alpha_s^2\rp$ taking into account
heavy quark mass effects~\cite{Laenen:1992zk,Laenen:1992xs}, though there has
been considerable recent progress towards the completion of the $\mathcal{O}\lp \alpha_s^3\rp$ calculation
of massive DIS structure functions~\cite{Ablinger:2014nga,Ablinger:2016eyd},
in particular of the terms that dominate in the $Q^2\gg m^2$ limit.
For charged current structure functions, massless
coefficients are available up to $\mathcal{O}\lp \alpha_s^3\rp$ and
massive coefficient functions
up to $\mathcal{O}\lp \alpha_s^2\rp$~\cite{Berger:2016inr}.
For heavy--quark initiated processes, massive coefficient functions are available only
up to $\mathcal{O}\lp \alpha_s\rp$~\cite{Kretzer:1998ju}.

These coefficient functions have been implemented in a number of private and public
codes, which allow the efficient calculation of DIS structure functions using
state--of-the--art theoretical information, such as {\tt QCDNUM}~\cite{Botje:2010ay},
{\tt APFEL}~\cite{Bertone:2013vaa},\footnote{The {\tt APFEL}
program is currently being rewritten into
C++~\cite{Bertone:2017gds}.}
and {\tt OpenQCDrad}~\cite{Alekhin:2010sv}.
The lengthy exact expressions for the NNLO DIS coefficient functions
are also available in the form of more compact interpolated expressions, which
reduce the computational burden of their evaluation and allows for efficient evaluation of DIS cross sections.
Moreover, DIS structure functions can be evaluated either in terms of the heavy quark pole mass
or in terms of the running $\overline{\rm MS}$ mass, as discussed in~\cite{Alekhin:2010sv}.
This statement is valid both in the FFNS as well as in any GM--VFNS, see for instance
the discussion of the FONLL case in~\cite{Bertone:2016ywq}. 

\subsubsection*{Impact on PDFs}\label{sec:datatheory.DISimpact}

From theoretical arguments, we expect that DIS structure functions
in charged-lepton scattering will constrain certain
PDF combinations, in particular the total quark singlet $\Sigma$, the quark
isotriplet $T_3$ from deuteron measurements, and the gluon from scaling
violations at HERA.
The separation between quarks and anti-quarks can then be improved
by using data taken with neutrino beams, which
as discussed above probe different quark combinations
than in charged-lepton DIS.
In particular, the strange PDF can be disentangled
by exploiting the information contained in
the neutrino--induced dimuon cross sections.
The same arguments indicate that other PDF combinations
would be poorly constrained
in a DIS--only fit, such as the gluon
at large $x$.

Given that DIS structure functions
constitute the backbone of any PDF fit, it is not straightforward to gauge their impact by adding
say DIS data on top of a non--DIS dataset, as in the latter case the PDFs will be very poorly constrained to begin with.
Here, in order to illustrate the role that DIS structure functions have in a PDF analysis, we have chosen
instead two other related examples.

\begin{figure}[t]
\begin{center}
  \includegraphics[scale=0.50]{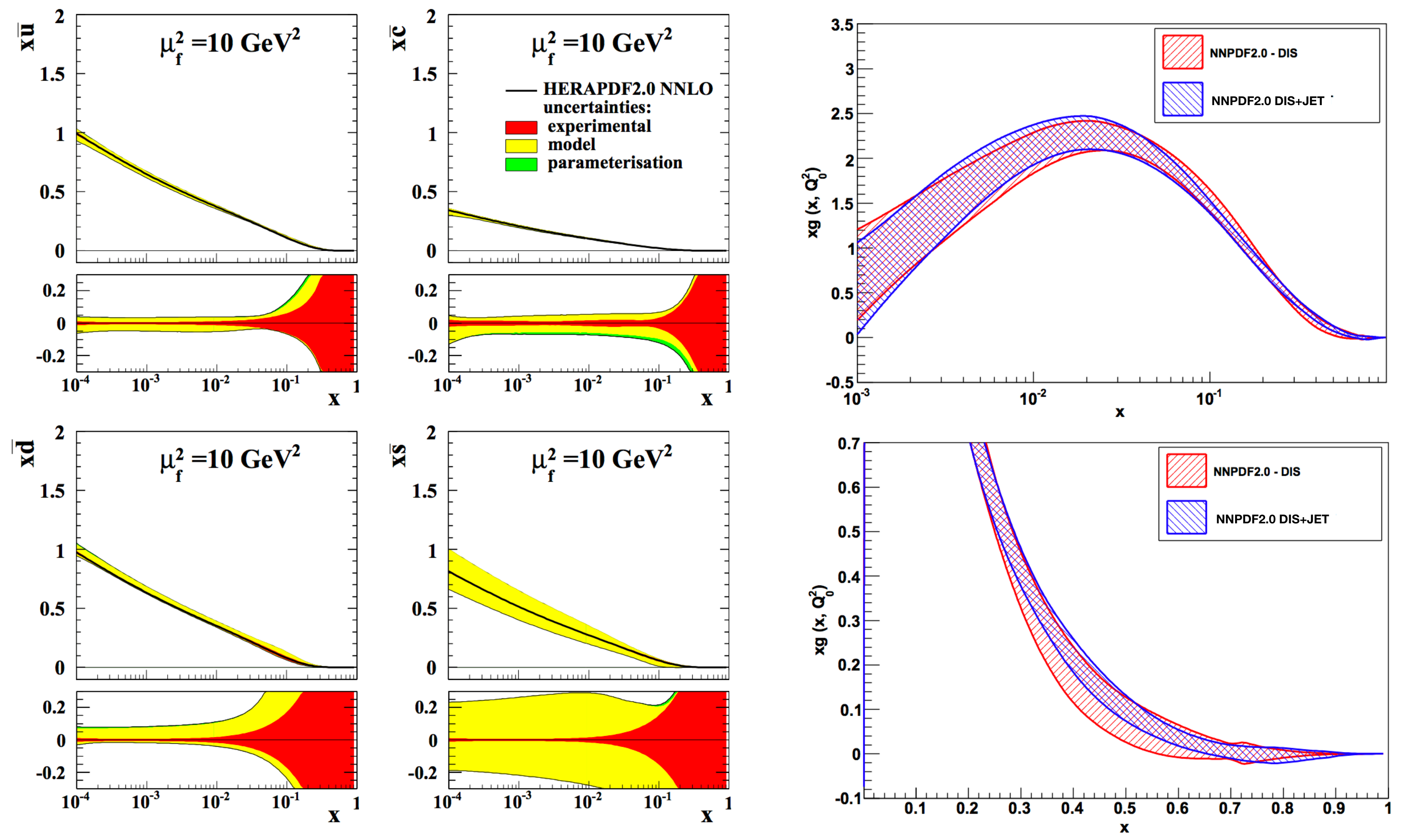}
   \caption{\small
   Left: the relative PDF uncertainties in the HERAPDF2.0NNLO fit, separated
   into experimental, model, and parametrization uncertainties, at
   $Q^2=10$ GeV$^2$, for the various antiquark PDFs.
   Right: the gluon PDF in the NNPDF2.0 analysis at small $x$ (top) and large $x$ (bottom),
   comparing the results of a DIS--only fit with those of a DIS+jet fit.
   See text for more details.
    \label{fig:DISonlyImpact}
  }
\end{center}
\end{figure}

First, in Fig.~\ref{fig:DISonlyImpact} we show
the relative PDF uncertainties in the HERAPDF2.0NNLO analysis, separated
   into experimental, model, and parametrization uncertainties, at
   $Q^2=10$ GeV$^2$, for the various anti--quarks.
   This PDF fit is based only on the inclusive HERA structure function data, and therefore
   it can be used to illustrate the information from quark flavour separation that
   can (and that cannot) be obtained from DIS measurements.
   We observe that while at medium and small $x$ the experimental PDF uncertainties
   are small, the contribution from model choices is larger, adding up
   to between 5\% and 10\%.
   The exception is the strange PDF, which is only poorly constrained by the HERA CC
   cross sections, and where errors are at $\sim 20\%$ or larger.
   Moreover, at large $x$ the PDF uncertainties for all the antiquarks increase rapidly,
   reflecting the fact that the HERA data has no sensitivity to the quark
   flavour separation in this region.

%%%%%%%%%%%%%%%%%%%%%%%%%%%%%%%%%%%%%%%%%%%%%%%%%%%%%%%%%%%%%%%%%%%%%
%%%%%%%%%%%%%%%%%%%%%%%%%%%%%%%%%%%%%%%%%%%%%%%%%%%%%%%%%%%%%%%%%%%%%

While DIS data provide only loose information on quark flavour separation,
unless the information from neutrino-nucleus scattering is used,
they give some of the most constraining information on the gluon PDF
in the medium and small--$x$ region, in particular in the HERA region
from scaling violations.
To quantify this, in the right plot of Fig.~\ref{fig:DISonlyImpact}  we show
 the gluon in the NNPDF2.0 analysis at small $x$ (top) and large $x$ (bottom),
   comparing the results of a DIS--only fit with those of a DIS+jet fit.
   At small and medium $x$, the impact of jet data is minimal, showing that
   in a global fit the DIS data provides the most important constraints
   on the gluon in this kinematic region.
   On the other hand, at large $x$ we observe that the information on the gluon from DIS data is only partial,
   and adding jet production to the fit helps to significantly reduce the PDF uncertainties there.

We also emphasise that neutrino--induced CC structure functions, in
particular the dimuon cross sections, provide valuable constraints on the strange
content of the proton~\cite{Alekhin:2017olj,Alekhin:2014sya,Ball:2009mk,Lai:2007dq,Olness:2003wz}.
The impact of various datasets on the strange PDF will be discussed in
more detail in Sect.~\ref{sec:structure.strange}.

\subsection{Inclusive jets}\label{sec:datatheory.jets}
Since the first run of the Tevatron at Fermilab, inclusive jet production at hadron colliders has provided the dominant 
constraint on the gluon PDF at large $x$.
The definition of jet cross sections starts from a well--defined jet algorithm~\cite{Salam:2009jx}, which
is usually chosen to be infrared and collinear safe so that the corresponding parton--level cross section for hard scattering at high energies can be calculated in
perturbative QCD.
The most commonly used jet algorithm at the LHC is the anti--$k_T$ algorithm~\cite{Cacciari:2008gp}, provided with
the 4--vector recombination scheme.
Other common choices include the $k_T$ algorithm~\cite{Catani:1993hr,
Ellis:1993tq}, the Cambridge--Aachen algorithm~\cite{Dokshitzer:1997in},
the Seedless Infrared-Safe cone algorithm (SIScone)~\cite{Salam:2007xv},
as well as the Midpoint algorithm~\cite{Blazey:2000qt}, which was sometimes used at the Tevatron.
The main parameter entering a jet algorithm is the jet radius $R$, which roughly defines
how large the jet is in the $\lp \eta,\phi\rp$ plane, and whose optimal value
depends on the specific application~\cite{Cacciari:2008gd}.

When comparing the predicted parton--level cross section  to the experimentally measured jet cross section, it is essential to correct these to the hadron level.
That is, additional  non--perturbative corrections due, for example, to the underlying event and
hadronization effects, must be accounted for. These are usually provided by the experimental collaborations as
multiplicative factors derived from leading--order event generators.
The size of such corrections can be significant at low $p_T$, as high as $\sim 20\%$, while at high $p_T$ they are generally small, at the percent level~\cite{Khachatryan:2016mlc}. Variations of these non--perturbative corrections, by considering for example different generator predictions, are then treated as an additional source of correlated systematic
error.
Although PDF fits typically use parton--level predictions, results also exist that include the matching of NLO calculations to parton shower and 
hadronization~\cite{Alioli:2010xa}, and which can then be directly compared with the data at hadron--level.          

\subsubsection*{PDF sensitivity}\label{sec:datatheory.jets.sensitivity}
At leading order,
jet production at hadron colliders includes the following subprocesses
\begin{align}
&gg\rightarrow gg,\, gg\rightarrow q\bar q,\, gq\rightarrow gq,\, q\bar q\rightarrow gg\; ,\nonumber\\
&q\bar q \rightarrow q\bar q,\, q\bar q \rightarrow q'\bar q',\, q\bar q'\rightarrow q\bar q',\, qq\rightarrow qq,\,
qq'\rightarrow qq'\; ,
\end{align}
along with the corresponding
charge conjugate processes.
Therefore, jet production is in general sensitive to both the gluon and quark PDFs.
The kinematics of the two leading jets in the final state
can be characterised by their rapidities $y_{(1,2)}$ and their transverse momenta $p_{T,{(1,2)}}$. At LO we have $p_{T,1}=p_{T,2}=p_T$, and the momentum fractions carried by the
two incoming partons are given by
\begin{equation}
x_1=\frac{p_T}{\sqrt s}(e^{y_1}+e^{y_2}),\, \,\,\,x_2=\frac{p_T}{\sqrt s}(e^{-y_1}+e^{-y_2})\; ,
\end{equation}
where $\sqrt s$ is the centre of mass energy of the two incoming hadrons. If we instead consider the rapidity
of the jet in the centre--of--mass frame of the dijet system, $y^*\equiv (y_1-y_2)/2$, and the boost of the dijet $y_b\equiv(y_1+y_2)/2$,
we have
\begin{equation}
x_1x_2=\frac{4p_T^2\cosh^2 y^*}{s},\, \,\,\,x_1/x_2=e^{2y_b}\; .
\end{equation}
Note that beyond LO there can be multiple jets in the final state from additional QCD radiation,
so that in general the $p_T$ balance of the two leading jets will be lost.

%%%%%%%%%%%%%%%%%%%%%%%%%%%%%%%%%%%%%%%%%%%%%%%%%%%%%%%%%%%%%%%%%%%%%
\begin{figure}[t]
\begin{center}
  \includegraphics[scale=0.19]{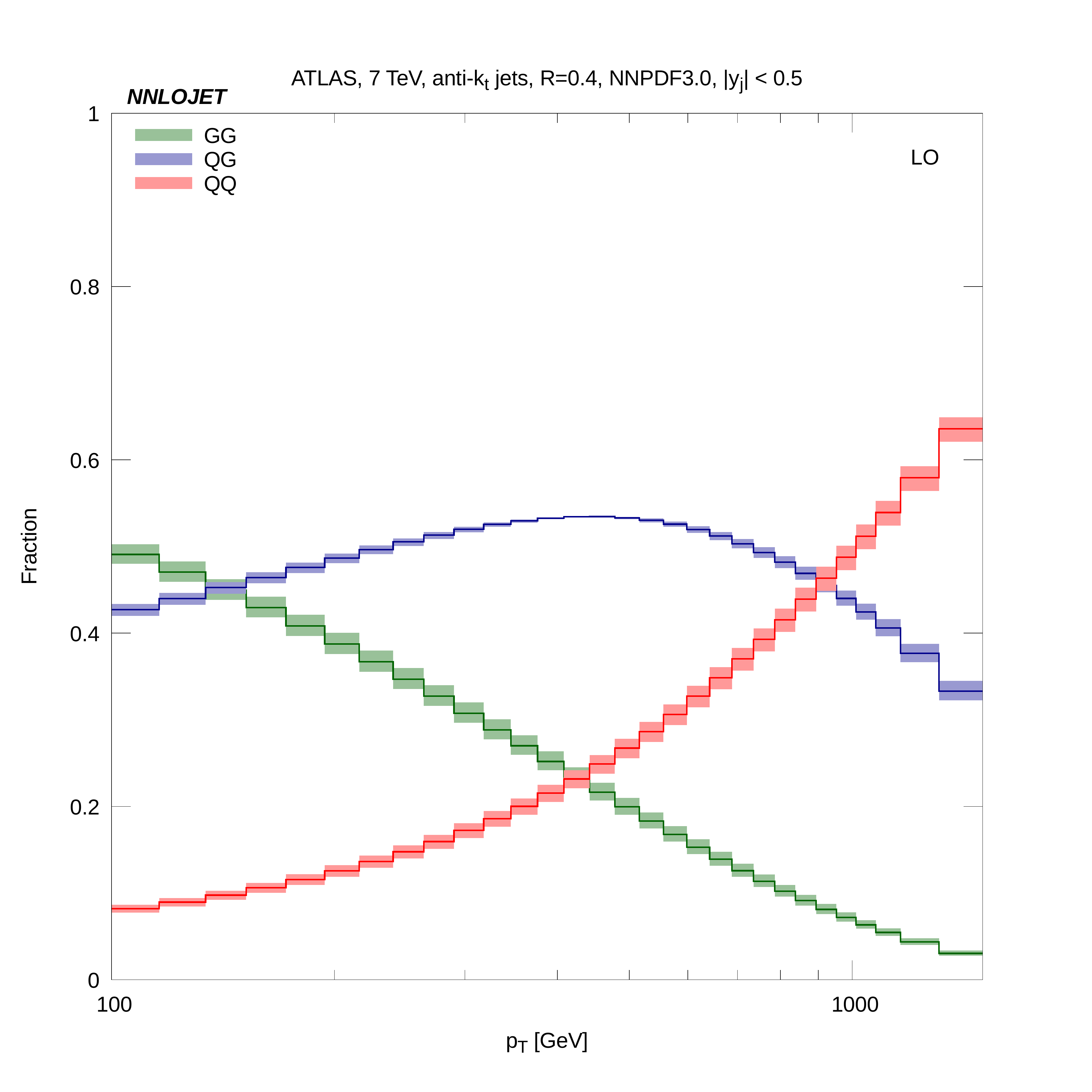}
  \includegraphics[scale=1.35]{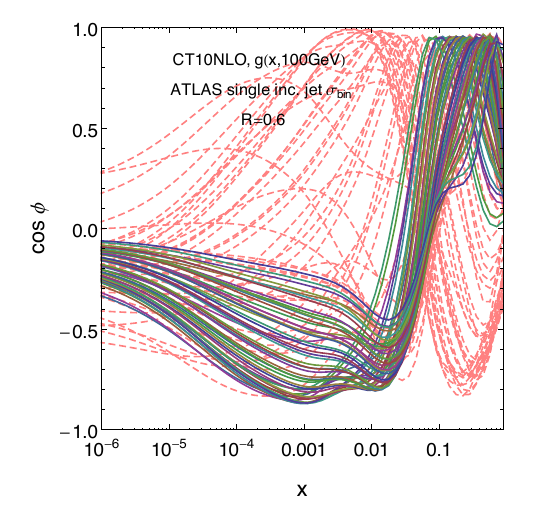}
   \caption{\small
   Left: Fractional contributions from different partonic channels to the single--inclusive jet
   production at the LHC 7 TeV at LO in the central rapidity region, computed
   with NNPDF3.0~\cite{Currie:2017ctp}.
   Right: Correlations between binning cross sections from ATLAS on the single--inclusive jet
   production at the LHC 7 TeV and the gluon PDF; the  dashed curves correspond to
   experimental bins at low $p_T$. 
    \label{fig:jet3}
  }
\end{center}
\end{figure}
%%%%%%%%%%%%%%%%%%%%%%%%%%%%%%%%%%%%%%%%%%%%%%%%%%%%%%%%%%%%%%%%%%%%%

Experimentally, jet production can be measured in various ways. The most common observable for PDF fits is the single--inclusive jet cross section, double--differential in the jet $p_T$ and rapidity $y$.
Here, one counts all jets in a single event and includes them in the same distribution.
Such a double--differential cross section is sensitive to different flavour combinations, depending on the kinematic region considered. 
In Fig.~\ref{fig:jet3} (left) the fractional contributions from the different parton--level subprocesses to the inclusive jet cross section
in the central rapidity region at the LHC is shown, as a function of the jet $p_T$. 
We can see that at low $p_T$ the channels involving initial--state gluons are dominant,
while at higher $p_T$ the $qq$ and $q\overline{q}$ contributions increase with the
former one being dominant,
but nonetheless with a sizeable gluon--induced fraction. 
As the quark PDFs are generally already well constrained by DIS data in
these kinematic regions, jet data is therefore dominantly sensitive to the gluon PDF.
This is illustrated in Fig.~\ref{fig:jet3}, which show the correlation
coefficients (see Sect.~\ref{sec:pdfuncertainties.hess})
between the inclusive jet cross section and the gluon PDF at various $x$ values.
This follows the  ATLAS binning~\cite{Aad:2011fc}, with each curve corresponding to one bin.
From this we can see that the inclusive jet production can potentially
constrain the gluon PDF in a wide range of
$x$ between $x\simeq 10^{-3}$ and $x\simeq 0.7$.

In addition to the single--inclusive case, there are also measurements of the
double--differential cross sections for inclusive dijet production, that is with
respect to $y^*$ and the invariant mass of the two leading jets, or even triple--differential cross
sections, e.g., with respect to $y_b$, $y^*$, and the average $p_T$ of the two leading jets.
Through such refined binning one can probe different initial states more efficiently. The large $y_b$ region usually receives more contributions from gluon initial states, while at large $y^*$ and $p_{T}$  initial states with two valence quarks dominate, allowing the $d$ valence PDF at high $x$ to be further constrained.

Finally, although we do not discuss it specifically here, we note that jet production in DIS at HERA also provides
complementary information about the gluon PDF at large-$x$ see e.g.~\cite{Abramowicz:2015mha} for the most recent analysis.
Significant theoretical progress has also been made quite recently, with the NNLO QCD
corrections to dijet production in DIS presented for the first time in~\cite{Currie:2016ytq}.
 
\subsubsection*{Experimental data}\label{sec:datatheory.jets.data}

The measurements on jet production in hadronic collisions, relevant for constraining the PDFs currently available are:

\begin{itemize}

\item
Double--differential single--inclusive jet production cross section data
from the CDF~\cite{Abulencia:2007ez,Aaltonen:2008eq} and D0~\cite{Abazov:2008ae,Abazov:2011vi} collaborations, at Tevatron Run II (1.96 TeV).

\item
Double--differential single--inclusive jet production cross section data
from the ATLAS~\cite{Aad:2010ad,Aad:2011fc,Aad:2013lpa,Aad:2014vwa} and CMS~\cite{Chatrchyan:2012bja,Chatrchyan:2014gia,Khachatryan:2016mlc} collaborations at LHC Run I
(7 and 8 TeV).

\item
Double--differential inclusive dijet production cross section data
from the ATLAS~\cite{Aad:2010ad,Aad:2011fc,Aad:2013tea} and CMS~\cite{Chatrchyan:2011qta,Chatrchyan:2012bja} collaborations at LHC Run I
(7 and 8 TeV).

\item
Measurements of the ratio of double--differential cross sections in single--inclusive jet
production at different centre--of--mass energies, 2.76, 7 and 8 TeV,
from the ATLAS~\cite{Aad:2013lpa} and CMS~\cite{Khachatryan:2016mlc} collaborations, at LHC Run I.

\item
Double--differential single--inclusive jet production cross section data
from the CMS collaboration~\cite{Khachatryan:2016wdh} at LHC Run II (13 TeV).

\item More recently, measurements of triple differential dijet cross sections are becoming available, see e.g. the recent CMS analysis~\cite{Sirunyan:2017skj} at 8 TeV.

\end{itemize} 

%%%%%%%%%%%%%%%%%%%%%%%%%%%%%%%%%%%%%%%%%%%%%%%%%%%%%%%%%%%%%%%%%%%%%
\begin{figure}[h]
\begin{center}
  \includegraphics[scale=0.3]{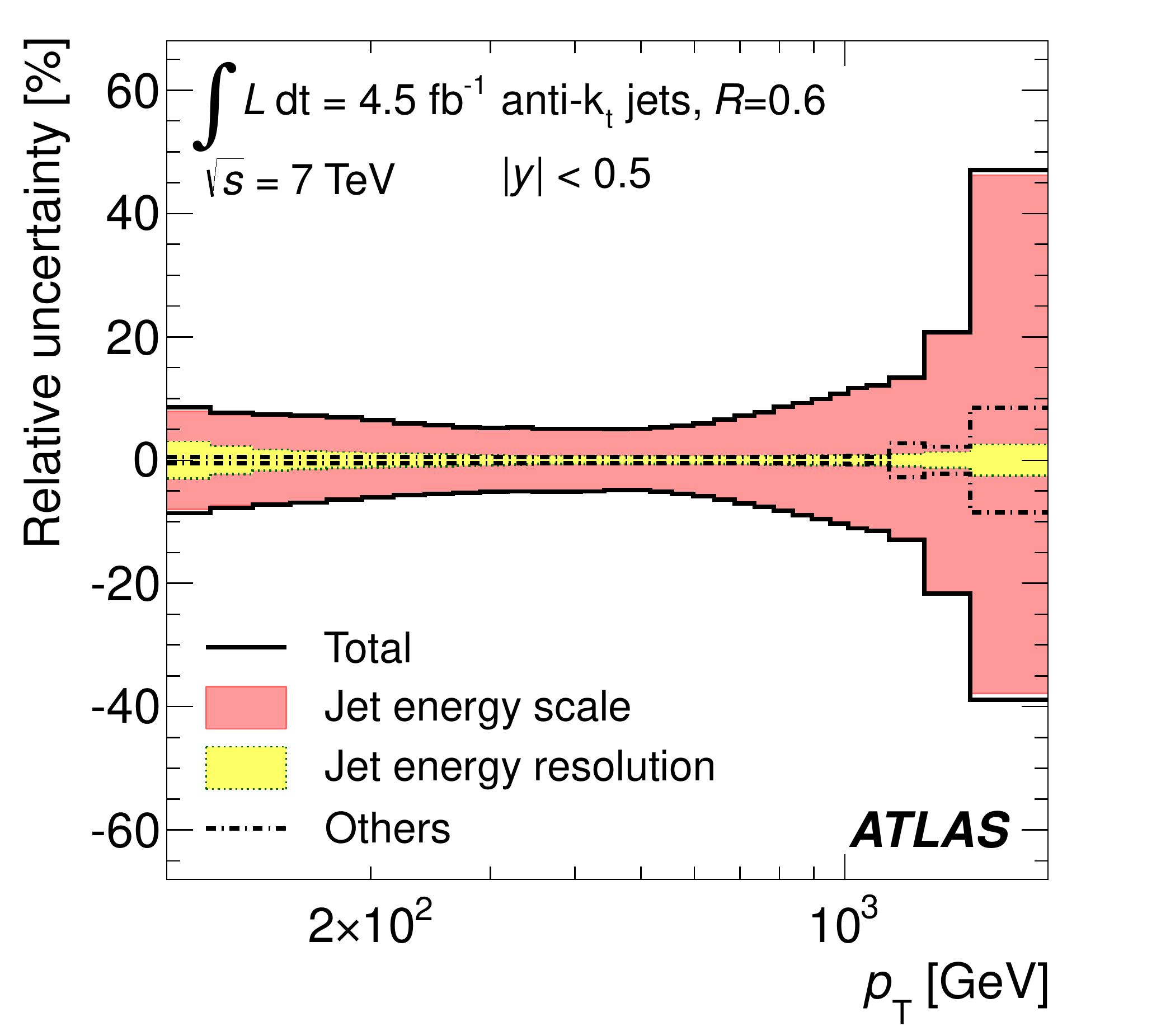}
  \includegraphics[scale=0.45]{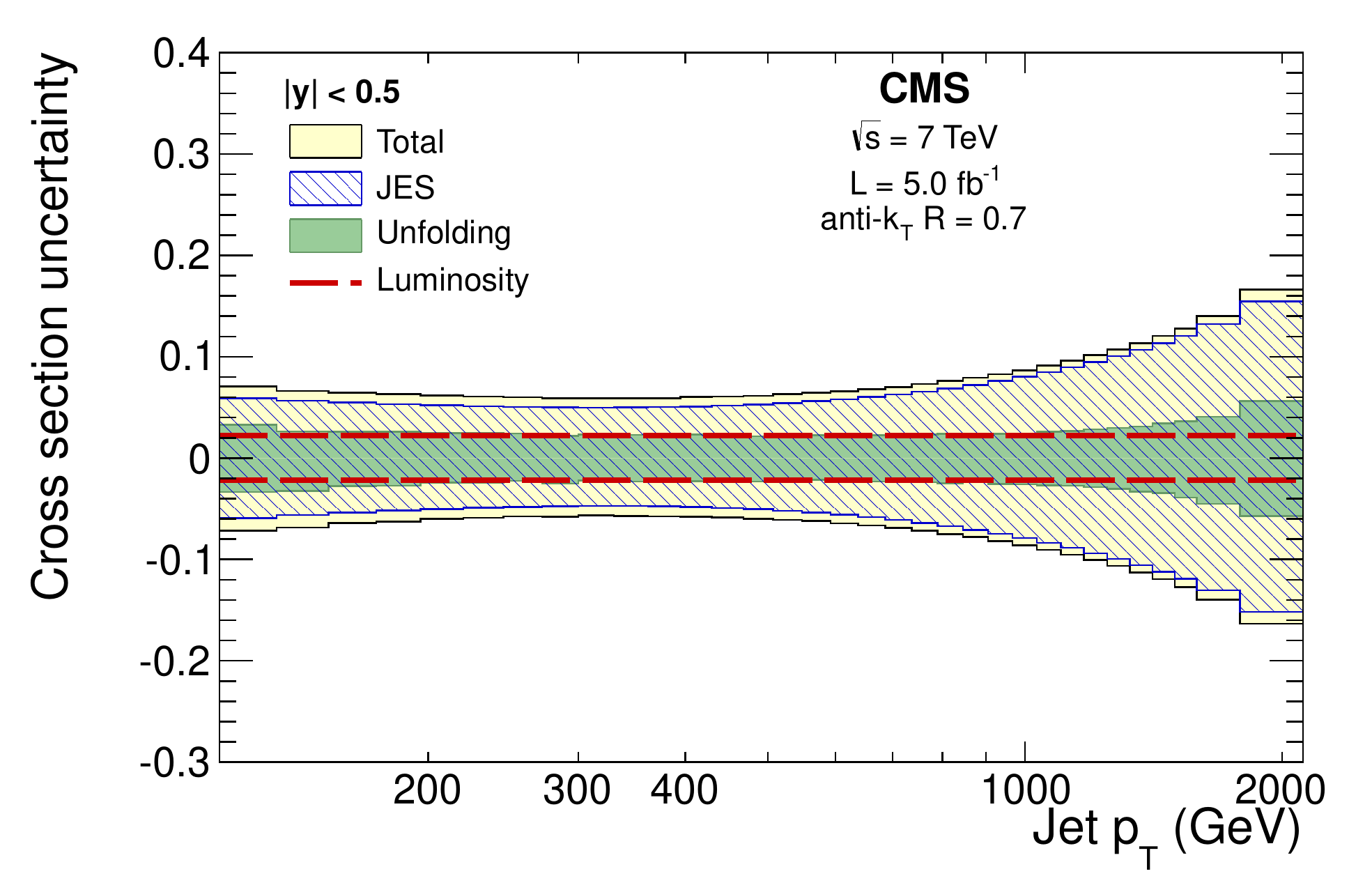}
   \caption{\small
    Representative systematic errors in single--inclusive jet measurement at
    LHC 7 TeV in the central rapidity region, from ATLAS (left)~\cite{Aad:2014vwa}
    and CMS (right)~\cite{Chatrchyan:2014gia}. 
    The luminosity error is not included in the ATLAS plot.
    We note that in both cases the jet energy scale (JES) is the dominant source
    of experimental uncertainties.
   \label{fig:jet4}
  }
\end{center}
\end{figure}
%%%%%%%%%%%%%%%%%%%%%%%%%%%%%%%%%%%%%%%%%%%%%%%%%%%%%%%%%%%%%%%%%%%%%

Due to the complexity of jet reconstruction and calibration there are a large number of experimental systematic uncertainties, with typically $N_{\rm sys}\sim 50 -100$ correlated systematic errors for $N_{\rm dat}\sim 100  - 200$ data points, in the case of both ATLAS and CMS.
In the most recent ATLAS and CMS 7 TeV measurements~\cite{Aad:2014vwa,Chatrchyan:2014gia}, the total correlated experimental uncertainties are at the level of about $5\sim 20$\% in most regions. On the other hand, the uncorrelated systematic errors and statistical errors are at one percent level or less in general, and therefore the uncertainty on such data is generally completely systematics dominated.
The typical experimental systematics from both ATLAS and CMS are shown in Fig.~\ref{fig:jet4}, and are seen to be dominated by the jet energy scale~\cite{Aad:2014vwa,Chatrchyan:2014gia}. The increasing precision of the LHC jet data, and the generally small uncorrelated errors, makes it rather challenging to fit the jet data well across the entire kinematic region in {\it e.g.}
the case of the ATLAS 7 TeV measurement from the 2011 dataset~\cite{Aad:2014vwa} (see~\cite{Harland-Lang:2017ytb} for further discussion).
A full account of these issues will almost certainly require a better understanding of both the experimental systematics and sources of theoretical errors that have not generally been included in PDFs fits previously.

In addition, for measurements of the ratios of the double--differential
inclusive jet production cross sections at different centre--of--mass energies the experimental
systematic errors largely cancel out, although the statistical uncertainties are somewhat larger, see e.g. the CMS 2.76, 7 and 8 TeV~\cite{Khachatryan:2016mlc} and ATLAS 2.76 and 7 TeV~\cite{Aad:2013lpa} measurements.
These are potentially useful for PDF fits~\cite{Mangano:2012mh}, due to the
cancellation not only of experimental systematics but also of theory errors such
as scale variations.

\subsubsection*{Theoretical calculations and tools}\label{sec:datatheory.jets.theory}

The NLO QCD corrections to single--inclusive jet and inclusive dijet production
were first calculated in the early 90's~\cite{Ellis:1992en,Kunszt:1992tn}, and has been implemented
in two numerical programs, {\tt NLOjet++}~\cite{Nagy:2001fj,Nagy:2003tz} and
{\tt MEKS}~\cite{Gao:2012he}. Recently, the
NNLO QCD corrections to the same process have been completed for all partonic
channels~\cite{Currie:2016bfm,Ridder:2013mf,Currie:2014upa,Currie:2017eqf},
with the exception of some sub--leading colour contributions.
The calculation is based on the Antenna subtraction
method~\cite{Ridder:2015dxa,Ridder:2016rzm} for isolating the
infrared singularities in QCD real radiation, and is part of the {\tt NNLOJET} program
for the evaluation of NNLO hadron collider cross sections.

\begin{figure}[t]
\begin{center}
  \includegraphics[scale=0.2]{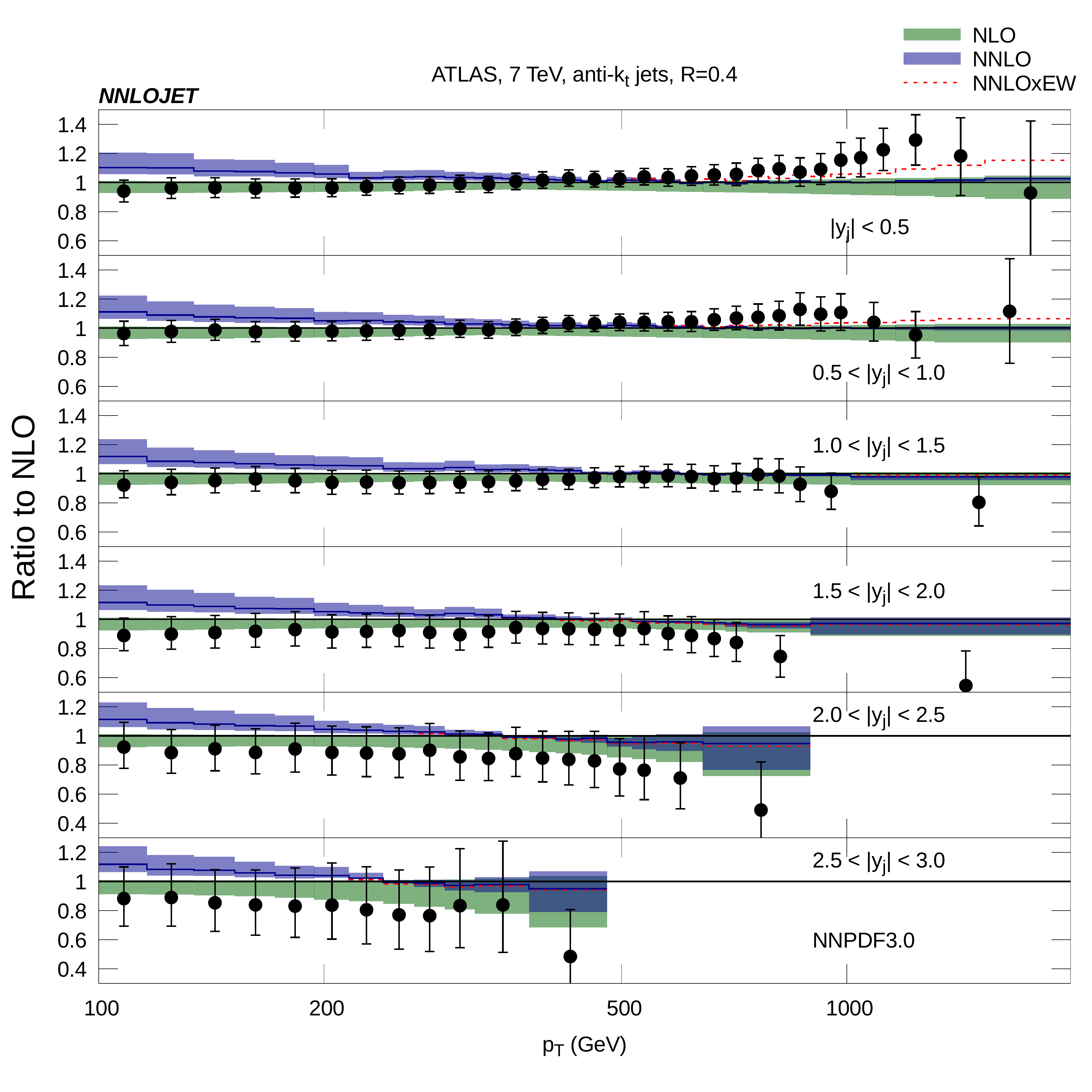}
  \hspace{0.1in}
  \includegraphics[scale=0.2]{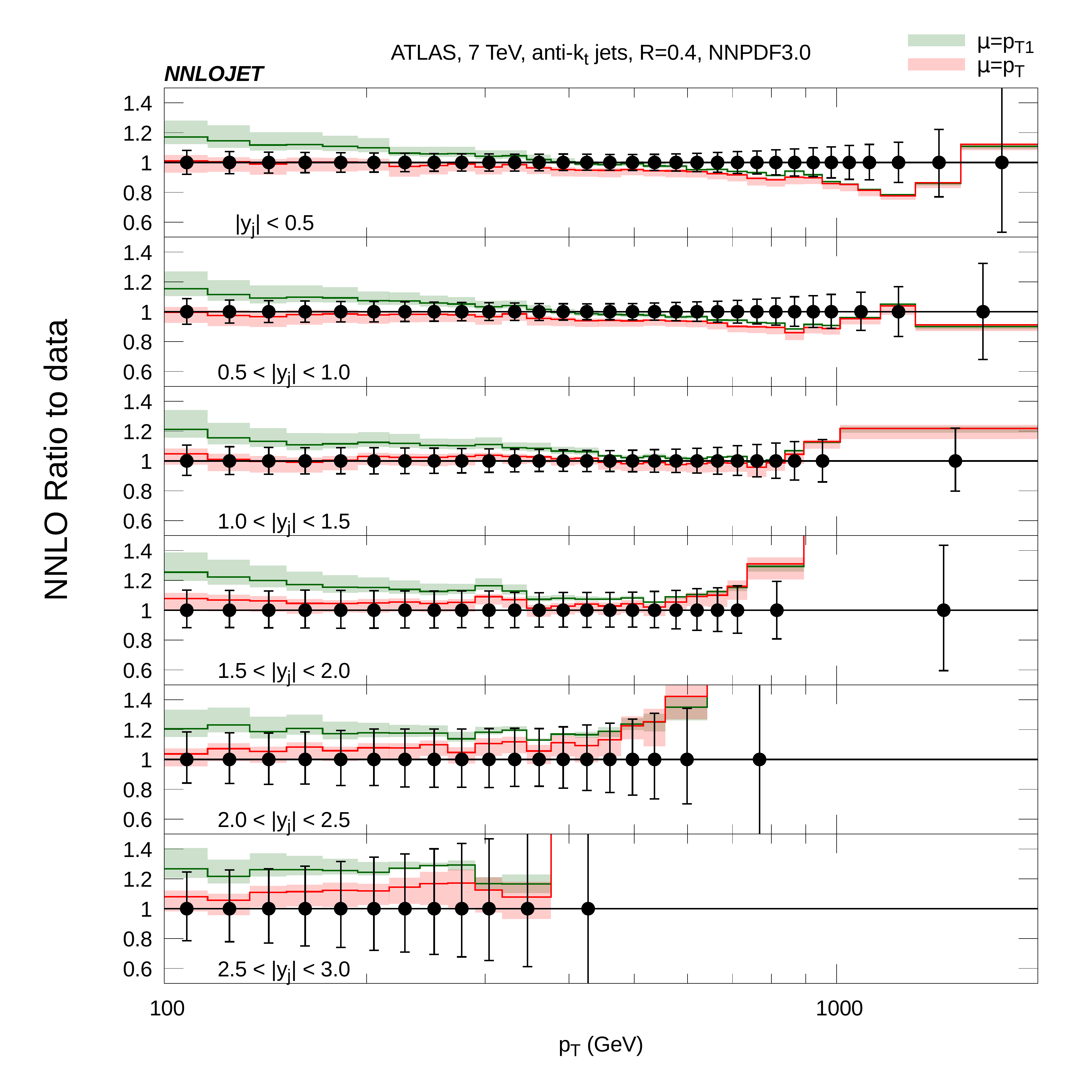}
   \caption{\small
   Left: Predictions for single--inclusive jet production at the 7 TeV LHC
   using ATLAS binning and the anti--$k_T$ algorithm with $R=0.4$. The central scale choice of the leading
   jet $p_T$ is taken and the scale variations, shown by the shaded regions, are found by varying the renormalization and factorization
   scales up and down simultaneously by a factor of 2~\cite{Currie:2017ctp}. 
   Right: As before, but comparing the NNLO predictions using a central
   scale choice of the leading jet $p_T$ (green) and the individual jet $p_T$ (red)~\cite{Currie:2017ctp}.
    \label{fig:jet1}
  }
\end{center}
\end{figure}

Fig.~\ref{fig:jet1} (left) shows
the NNLO QCD corrections to inclusive jet production at the 7 TeV LHC, 
with the anti--$k_T$ algorithm and a central scale choice of the leading jet $p_T$.
The NNLO QCD corrections are seen to be significant at low $p_T$, leading to a
10\% increase with respect to NLO, while at high $p_T$ the NNLO corrections are small.
The NLO scale variations bands are asymmetric at low $p_T$
and, interestingly, largely underestimate the perturbative uncertainties.
Electroweak (EW) corrections
can be significant at high $p_T$ for central rapidities due to the presence of
large EW Sudakov logarithms, but are well below 1\% for a rapidity greater than
unity~\cite{Dittmaier:2012kx}.

There are ambiguities in choosing the appropriate 
QCD scale even in the simplest case of single--inclusive jet production~\cite{Ball:2012wy,Carrazza:2014hra}.
%\footnote{The scale ambiguity was discussed in~\cite{Ball:2012wy} for the NLO predictions.}
%
In particular, one can take either the $p_T$
of the individual jet or the leading jet in the event. While these variables are the same at LO, where the two jets are produced back--to--back, at higher orders there can exist more than two jets which may have large differences in $p_T$. The NNLO predictions using these two choices
for the central scales are studied in~\cite{Currie:2017ctp}, and are found to lead
to non-negligible differences specially at low values of $p_T$.
This is shown in Fig.~\ref{fig:jet1} (right), where at high $p_T$ the two predictions converge, but  at low and intermediate $p_T$, there are
significant differences of the central values in comparison to the size of scale
variations.
Indeed, the two error bands (from scale variations) do not even overlap.
Although it seems that the 
NNLO predictions using the individual jet $p_T$ as the central scale tend to follow the
trend of the ATLAS data better, clearly further investigation is
 needed to resolve the ambiguity of scale choice in the NNLO predictions,
including specifically PDF fits based on exact NNLO theory for the jet data. Interestingly, a first study in this direction, presented in~\cite{Harland-Lang:2017ytb}, indicates that the impact of the scale choice on the extracted PDFs is not significant.

%%%%%%%%%%%%%%%%%%%%%%%%%%%%%%%%%%%%%%%%%%%%%%%%%%%%%%%%%%%%%%%%%%%%%
%%%%%%%%%%%%%%%%%%%%%%%%%%%%%%%%%%%%%%%%%%%%%%%%%%%%%%%%%%%%%%%%%%%%%

As well as fixed--order predictions, there are various theoretical calculations including analytic QCD resummation~\cite{Kidonakis:2000gi,
Kumar:2013hia,Klasen:2013cba,deFlorian:2013qia,Carrazza:2014hra}.
It has been shown in~\cite{deFlorian:2013qia} for the case of inclusive jet production at the LHC, 
that the approximate NNLO predictions from the expansion of threshold resummation
agree well with the exact NNLO predictions for the all--gluon channel at large $p_T$. Over the full rapidity range the threshold
expansion reproduces the fixed--order results down to a $p_T$ of about 400 GeV with
the same value shifted to lower $p_T$ at forward rapidities. Such approximate NNLO
predictions have been used in previous global analysis involving jet data~\cite{Harland-Lang:2014zoa,Ball:2014uwa}.
The jet cross sections are also sensitive to the jet algorithm used, in particular the value
of the jet radius $R$.
The use of a larger jet radius $R$ at the LHC  leads to
a larger inclusive cross section and better convergence in the perturbative expansion.
While this also reduces the non--perturbative corrections from QCD hadronization, it
increases the effects from other non--perturbative phenomena
such as the underlying event (UE) and multiple parton interactions (MPI).
For LHC Run-I, ATLAS used $R=0.4$
and $R=0.6$, while CMS instead used 0.5 and 0.7. In Run-II a uniform of value of $R=0.4$ has been taken.

\subsubsection*{Impact on PDFs}\label{sec:datatheory.jets.impact}

Jet data from the Tevatron and the LHC Run I have already played an important role in global 
analyses~\cite{Watt:2013oha,Ball:2014uwa,Dulat:2015mca,Harland-Lang:2014zoa}, although
in the NNLO fits these have until very recently only applied NLO or approximate NNLO theoretical predictions.
Indeed, it was found that removing all jet data from the global analyses can
lead to an increase of the gluon PDF uncertainties at large $x$ by at least a factor of two~\cite{Ball:2014uwa}, see Ref.~\cite{Rojo:2014kta} for an overview of such studies.

\begin{figure}[t]
\begin{center}
  \includegraphics[scale=0.23]{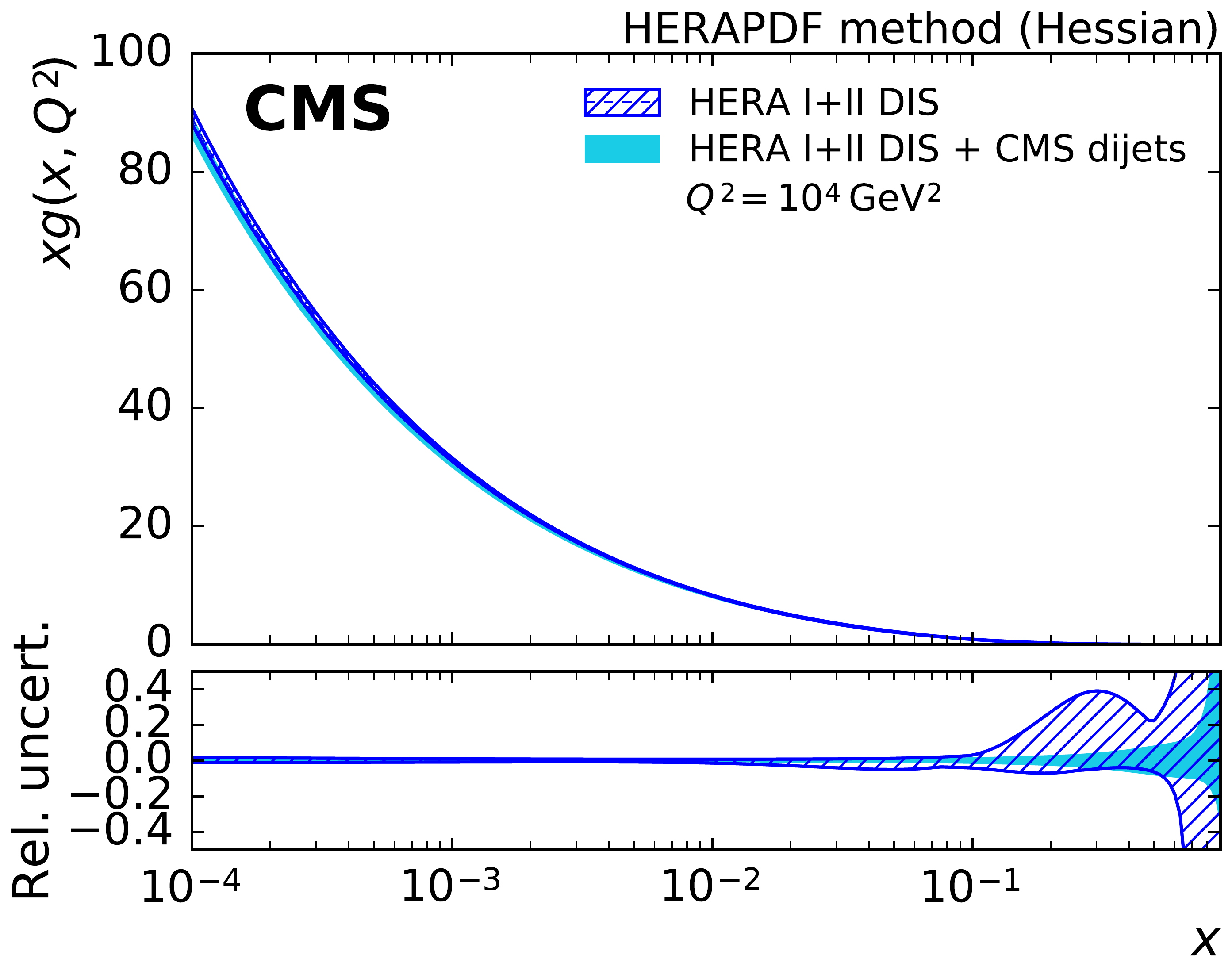}
  \hspace{0.3in}
  \includegraphics[scale=0.23]{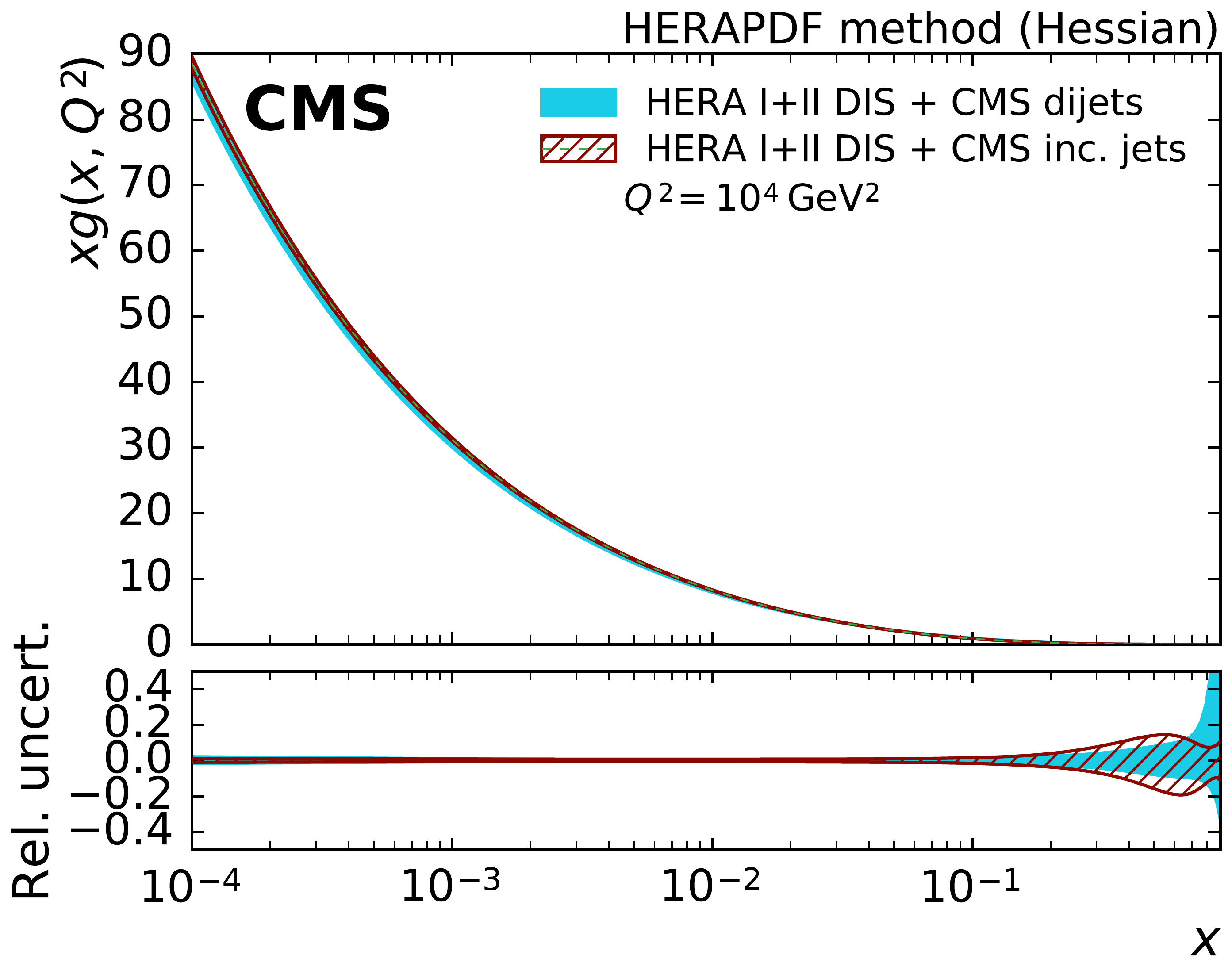}
   \caption{\small
   Effect of including the CMS 8 TeV jet data on the gluon PDF. Studies performed using {\tt xFitter}, with PDF uncertainties obtained
   with the Hessian method. Left: Comparison of HERA DIS fit to that with the CMS dijet data included in addition. Right:  Comparison of HERA DIS and CMS dijet fit  to HERA
   DIS and CMS inclusive jet data fit~\cite{Sirunyan:2017skj}.
   \label{fig:jet5}
  }
\end{center}
\end{figure}

To illustrate the PDF impact of LHC inclusive jet data,
in Fig.~\ref{fig:jet5}, taken from~\cite{Sirunyan:2017skj}, we
show the impact of the CMS 8 TeV jet data
on the gluon PDF, determined by adding the data into a baseline fit with HERA DIS data only~\cite{Abramowicz:2015mha}\footnote{Such comparisons represent a non--trivial and useful indication of the impact of a given data set on the PDFs, and as such we will present examples of these throughout this review. Nonetheless, a more complete analysis should consider the impact in comparison to a truly global PDF baseline, and in general caution is recommended when interpreting the quantitative conclusions of these studies, see Section~\ref{sec:pdfgroups.LHC} for further discussion.}.
The inclusion of both the single--inclusive jet data and the inclusive
dijet data leads to a sizeable reduction in the gluon PDF uncertainty at large $x$.
Meanwhile in the same fit a reduction of the PDF uncertainty in
the valence quark at high $x$ is also observed, providing a complementary constraint
to Drell--Yan
and fixed--target DIS data.
%

%%%%%%%%%%%%%%%%%%%%%%%%%%%%%%%%%%%%%%%%%%%%%%%%%%%%%%%%%%%%%%%%%%%%%
%%%%%%%%%%%%%%%%%%%%%%%%%%%%%%%%%%%%%%%%%%%%%%%%%%%%%%%%%%%%%%%%%%%%%

With the full NNLO predictions on inclusive jet
and dijet production now available, we can foresee significant advances in
 pinning down the gluon PDF at large $x$, in particular using the increasingly precise
inclusive jet data from LHC Run I and Run II.
Moreover, in the latter case, the increased luminosity will allow  jet production to be probed
deep into the TeV region, thus accessing PDFs at larger values of $x$ where
available constraints are rather scarce.

\subsection{Inclusive gauge boson production}\label{sec:datatheory.gauge}
The inclusive production of electroweak
gauge bosons, the so--called Drell--Yan process~\cite{Drell:1970wh}, has been of enormous historical importance, providing the first window on the quark flavour separation in the proton
beyond that contained in DIS structure functions alone.
These now provide the backbone of global
PDF fits together with the fixed--target and HERA structure function data.

\subsubsection*{PDF sensitivity}\label{sec:datatheory.gauge.sensitivity}
The lowest--order contributions to $W$ and $Z/\gamma^*$ production proceed via
the following partonic subprocesses:
\begin{align}
u\overline{d}, \,c\overline{s} \quad (u\overline{s}, \,c\overline{d}) &\to W^+ \;, \\
d\overline{u},\, s\overline{c} \quad (s\overline{u}, \,d\overline{c}) &\to W^- \;, \\
q\overline{q} &\to Z/\gamma^* \;,
\end{align}
where we show the Cabibbo suppressed contributions in brackets and
where $q$ corresponds to all active quark flavours.
These processes can therefore tell us about the flavour decomposition of the proton, given  that each flavour subprocess carries a different weight
in the total cross section.
To examine the dominant PDF sensitivity we can approximate the CKM matrix as diagonal, and thus ignore the bracketed contributions.
In this case it is informative to consider the ratio of $W^+$ to $W^-$ production, differential
in the rapidity $y_W$ of the produced boson~\cite{Berger:1988tu},
\begin{equation}\label{wpwm}
R_\pm = \frac{{\rm d}\sigma(W^+)/{\rm d}y_W}{{\rm d}\sigma(W^-)/{\rm d}y_W}=\frac{u(x_1)\overline{d}(x_2)+c(x_1)\overline{s}(x_2) + 1\leftrightarrow 2}{d(x_1)\overline{u}(x_2)+s(x_1)\overline{c}(x_2) + 1\leftrightarrow 2}\;,
\end{equation}
and the corresponding $W$ asymmetry 
\begin{equation}\label{wasym}
A_W = \frac{{ d}\sigma(W^+)/{ d}y_W-{ d}\sigma(W^-)/{\rm d}y_W}{{ d}\sigma(W^+)/{\rm d}y_W+{ d}\sigma(W^-)/{ d}y_W}=\frac{u(x_1)\overline{d}(x_2)+c(x_1)\overline{s}(x_2)-d(x_1)\overline{u}(x_2)-s(x_1)\overline{c}(x_2) + 1\leftrightarrow 2}{u(x_1)\overline{d}(x_2)+c(x_1)\overline{s}(x_2)+d(x_1)\overline{u}(x_2)+s(x_1)\overline{c}(x_2) + 1\leftrightarrow 2}\;.
\end{equation}
We will for simplicity consider the $W$ rapidity, rather than the experimentally observable rapidity of the charged lepton from the $W$ decay, in what follows. These variables are clearly correlated; we will comment further on this at the end.

Thus these ratios are in general sensitive to a fairly non--trivial combination of quark and antiquark PDFs evaluated at the following values of $x$:
\be
x_{1}=\frac{M_W}{\sqrt{s}}e^{+ y_W}\; ,\qquad
x_{2}=\frac{M_W}{\sqrt{s}}e^{- y_W}\; .
\ee
While these expressions completely define the PDF sensitivity of these observables at LO, it is informative to consider various kinematic limits, where these expressions simplify and more straightforward  approximate dependences become apparent. Including only the (dominant) $u$ and $d$ contributions, we can in particular consider the cases of central and forward $W$ production
\begin{align}\label{wcent}
{\rm Central}&:\qquad y_W\sim 0\qquad x_1\sim x_2 = x_0,\qquad \overline{u}(x_{1,2})\sim \overline{d}(x_{1,2})\;,\\ \label{wfor}
{\rm Forward}&: \qquad y_W \gtrsim 2, \qquad x_1 \gg x_2,  \qquad q(x_1)\sim q_V(x_1),\,\overline{u}(x_2)\sim \overline{d}(x_2)\; ,
\end{align}
where $x_0=M_W/\sqrt{s}$ and $q=u,d$, and we assume that $\overline{u}\sim \overline{d}$ at low $x$ for demonstration here. At the LHC we have $ x_0  = 0.005-0.01$, while in the forward region $x_2 \ll 1$, and therefore the $\overline{d}\sim\overline{u}$ approximation is a very good one. For the case of negative $W$ rapidity we can of course simply interchange $x_1 \leftrightarrow x_2$.

In the central region, applying the simplification of Eq.~(\ref{wcent}) and dropping the $c,s$ contributions we find
\begin{align}
R_{\pm}&\sim \frac{u(x_0)}{d(x_0)}\;,\\
A_W&\sim  \frac{u_V(x_0)-d_V(x_0)}{u(x_0)+d(x_0)}\;.
\end{align}
Thus $A_W$ is sensitive to the valence difference, while $R_{\pm}$ is sensitive to the ratio of $u$ to $d$ at $x_1\sim x_2 \sim x_0$. For these reasonably low $x$ values, the valence $u$ and $d$ quarks are fairly small, and so we roughly expect $R_{\pm}\sim 1$ and $A_W \sim 0$, with the departures from these values being due to the precise flavour content of the proton, in particular the fact that the valence distributions are not completely negligible in this region.

In the forward region, applying the simplification of Eq.~(\ref{wfor}) and again dropping the $c,s$ contributions we find that the following
relations hold
\begin{align}
R_{\pm}&\sim \frac{u_V(x_1)}{d_V(x_1)}\;,\\
A_W&\sim  \frac{u_V(x_1)-d_V(x_1)}{u_V(x_1)+d_V(x_1)}\;.
\end{align}
Thus these provide (equivalent) sensitive constraints on the $u/d$ ratio at high $x$.

Considering now the case of $Z$ production, for forward production we find,
using the same approximations as before,
that the differential ratio of $W$ to $Z$ production is given by
\begin{equation}\label{wzrat}
\frac{{ d}\sigma(W^+)/{ d}y_W+{\rm d}\sigma(W^-)/{\rm d}y_W}{{\rm d}\sigma^Z/{\rm d y_Z}}\approx \frac{u_V(x_1)+d_V(x_1)}{0.29 u_V(x_1)+0.37 d_V(x_1)}\;,
\end{equation}
where the factors in the denominator arise from the electroweak $Z$--quark couplings.
For the central region a similar result, evaluated at $x_0$, is found
up to an overall factor of 2.
Therefore, we can see that the $W^{\pm}$ and $Z$ cross sections provide very similar information about the $u$ and $d$ quarks. 

Up to this point we have omitted the contribution from the strange quarks to $W$ and $Z$ production. Generally speaking this is washed out when considering ratio observables, justifying their omission above, although the $W$ asymmetry displays some sensitivity to the strange difference $s-\overline{s}$.
On the other hand, the contribution to the absolute cross sections is not negligible, in particular at lower $x$. Thus for example the $Z$ cross section at central rapidity becomes, for five active quark flavours,
\begin{equation}
\label{zcross}
\frac{  d\sigma^Z}{{ d y_{ll}}}\sim 0.29\Big(u(x_0)\overline{u}(x_0)+c(x_0)\overline{c}(x_0)\Big)+0.37\Big(d(x_0)\overline{d}(x_0)+s(x_0)\overline{s}(x_0)+b(x_0)\overline{b}(x_0)\Big)\; .
\end{equation}
Therefore, provided the absolute cross section data are sufficiently accurate and the other quark flavours are sufficiently well determined, this may for example be sensitive to the currently less well determined strange quark distribution.
Moreover, this is not a case of a simple overall normalization; as the $Z$ rapidity increases the valence $u,d$ contributions will become increasingly dominant, and the contribution from the strange (and the heavy flavours) will decrease. Thus the shape of the $Z$ rapidity distribution is sensitive to the proton strangeness, as well as to that the contribution from heavy flavour PDFs.
Similar considerations also apply for the absolute cross sections,
where in particular
the ratios of total $W^{\pm}$ and $Z$ cross sections at the LHC are known
to be quite sensitive to the strangeness PDF~\cite{Nadolsky:2008zw}. 

Below the $Z$ peak region, the Drell--Yan process is dominated by an off--shell intermediate photon, with cross section given by
\begin{equation}\label{dy}
\frac{{ d}\sigma^{\rm DY}}{{ d y_{ll}}}\sim \sum_i
e_i^2\Big(q(x_1)\overline{q}(x_2)+q(x_2)\overline{q}(x_1)\Big)\; ,
\end{equation}
where the sum runs over all active quark flavours.
Thus in comparison to Eq.~(\ref{zcross}), a different combination of the quark and anti--quark PDFs is probed, due to the differing electromagnetic couplings.
In particular, the relative $u\overline{u}$ to $d\overline{d}$ contribution is now a factor of $\sim 5$ higher in comparison to the $Z$ cross section.
At the LHC, low mass Drell--Yan production therefore provides complementary flavour information in the low to intermediate $x$ region. In addition, as the cuts on the final--state lepton transverse momenta tend to increase the relative importance of the higher order contributions, for which the $Z$ $p_\perp$ can be non--zero, this can be sensitive to the gluon PDF at lower $x$, which contributes through the NLO $g\to q\overline{q}$ splitting.
In contrast, high mass Drell--Yan production (with $M_{ll} \gg m_Z$) is sensitive to the $q, \overline{q}$ PDFs at high $x$, in particular the antiquarks, which are rather less well determined in this region.

A further constraint is provided by considering the Drell--Yan process on fixed proton and neutron (in practice, deuteron) targets. By using isospin symmetry the PDFs between the proton and the neutron can be related
\be
u^p=d^n \qquad d^p=u^n\;,
\ee
allowing an extra handle on the proton flavour decomposition.
In particular, such fixed target experiments generally have larger acceptance in the $x_1\gg x_2$ region (where $x_1$ is defined with respect to the proton beam) for which the first term in Eq.~(\ref{dy}) is dominant, with $q(x_1)\sim q_V(x_1)$. It is then straightforward to show that
\be
\frac{\sigma^{pn}}{\sigma^{pp}}\sim \frac{\overline{d}(x_2)}{\overline{u}(x_2)}\;.
\ee 
That is, they are sensitive to quark sea decomposition in the intermediate to high $x_2\sim 0.01-0.3$ region probed by these fixed target experiments~\cite{Ellis:1990ti}.
This however comes with the added complication that the cross section must be corrected to account for the fact that the neutron is bound in a deuteron nucleus, and therefore the `free' neutron PDF is not directly probed. 
Fixed--target $pp$ scattering does not suffer from this issue, and is sensitive to the quark sea (dominantly, the $\overline{u}$) in the same $x$ region, but is much less directly sensitive to the $\overline{d}/\overline{u}$ decomposition.

Turning now to the case of $W,Z$ production at the Tevatron, the fact that we have $p\overline{p}$ collisions affects the flavours probed. In particular, we can use charge--conjugation symmetry to write
\begin{equation}\label{ppbar}
q^p=\overline{q}^{\overline{p}}\;.
\end{equation}
In fact, it is straightforward to show that in the region of valence quark dominance, the cross section ratio $R^\pm$ and the asymmetry $A_W$ are again sensitive to the $u/d$ ratio and the valence difference $u_V-d_V$, while the $Z$ cross section again provides similar information to $W^\pm$ cross section sum. 
Nonetheless, these conclusions are only approximately true, and the presence of a $\overline{p}$ beam provides complementary flavour information.

Finally, we emphasise that we have considered here the distributions with respect to the (unobservable) rapidity of the $W$ boson to simplify the discussion.  In general we should correctly account for the kinematics, as well as weight the corresponding $q\overline{q}$ contributions by the appropriate $W$ decay distributions. This is in fact provides a further handle on the flavour sensitivity of this observable, as by changing the $p_\perp$ cut on the charged lepton, different weights of the various quark contributions are achieved, see
{\it e.g.}~\cite{Martin:2009iq,Martin:2012da} for further details.
Nonetheless, the forward and central $W$ rapidity regions are certainly correlated with the equivalent lepton rapidity regions that are measured experimentally, and so the above discussion provides a qualitative guide for the PDF sensitivity of $W$ boson production.
However, as we will discuss below, the current simulation codes for $W$ and $Z$ production include the full kinematics of the leptonic decays, and therefore there is no need to explicitly correct back to the $W$ rapidity.

\subsubsection*{Experimental data}\label{sec:datatheory.gauge.data}

A non--exhaustive list of the available Drell--Yan measurements is as follows:

\begin{itemize}

\item The most precise fixed target Drell--Yan data come from the E866/NuSea~\cite{Towell:2001nh} experiment at Fermilab, while the E906/SeaQuest experiment~\cite{seaquest} will extend out to higher $x$, and is currently taking data. 

\item The Tevatron collider has produced a range of data on $W$ and $Z$ production, including measurements of the $Z$ rapidity distribution~\cite{Abazov:2007jy,Aaltonen:2010zza} and  $W$ production for both the lepton~\cite{Abazov:2013rja,D0:2014kma} and the $W$~\cite{Aaltonen:2009ta,Abazov:2013dsa} asymmetries. 

\item Early LHC measurements of the $Z$ rapidity distribution presented by CMS~\cite{Chatrchyan:2011wt} and ATLAS~\cite{Aad:2011dm}.

\item CMS Drell--Yan data at 7 TeV~\cite{Chatrchyan:2013tia}, for $15 < M_{ll} < 1500$ GeV and at 8 TeV~\cite{CMS:2014jea}, which increased the upper mass limit to 2000 GeV. These are presented double differentially in the rapidity and invariant mass of the lepton pair.

\item ATLAS 7 TeV Drell--Yan invariant mass distribution (integrated over rapidity) at high~\cite{Aad:2013iua} ($116<M_{ll}<1500$ GeV) and low~\cite{Aad:2014qja} ($26<M_{ll}<66$) invariant masses.

\item CMS~\cite{Chatrchyan:2012xt,Chatrchyan:2013mza} 7 TeV $W$ asymmetry, and ATLAS~\cite{Aad:2011dm} $W^+$ and $W^-$ cross section data.

\item ATLAS high precision $W$ and $Z,\gamma^*$ data~\cite{Aaboud:2016btc}, using the full $4.6\,{\rm fb}^{-1}$ data set at 7 TeV. The Drell--Yan rapidity distribution is presented double differentially in three intervals of lepton pair invariant mass, over the $45<M_{ll}<150$ GeV range. In the $Z$ peak and higher mass regions the measurement also extends out to $|y_{ll}|=3.6$.

\item LHCb $Z$ rapidity distributions at 7~\cite{Aaij:2012mda,Aaij:2015gna}, 8~\cite{Aaij:2015vua} and 13~\cite{Aaij:2016mgv} TeV.

\item LHCb lepton rapidity distributions for $W^+$ and $W^-$ production at 7~\cite{Aaij:2014wba}, and 8~\cite{Aaij:2015zlq,Aaij:2016qqz} TeV.

\item In Ref.~\cite{Aaij:2015zlq},
cross section ratios between the 7 and 8 TeV LHCb $W$ and $Z$  measurements are presented, with the cancellation in various systematic uncertainties providing a more precise PDF sensitivity.

\end{itemize}

Thus at the LHC numerous measurements have already been presented, with many others
in preparation.
The $Z/\gamma^*$ data are available over a wide range of invariant masses, providing extensive coverage in $x$.
The $W$ production data are increasingly presented as individual cross sections, including the correlated error information, to provide the maximum possible constraints.
While the majority of cases, such as with the ATLAS and CMS measurements, are limited to the central rapidity region, that is, a lepton pseudorapidity of $|\eta_l|<2.4$, this reach is extended by exploiting the forward acceptance of the LHCb detector, for which $2 <\eta_l < 4.5$ is accessed.
This allows the high and low $x$ region to be probed.

\begin{figure}
  \begin{center}
 \includegraphics[width=0.37\textwidth]{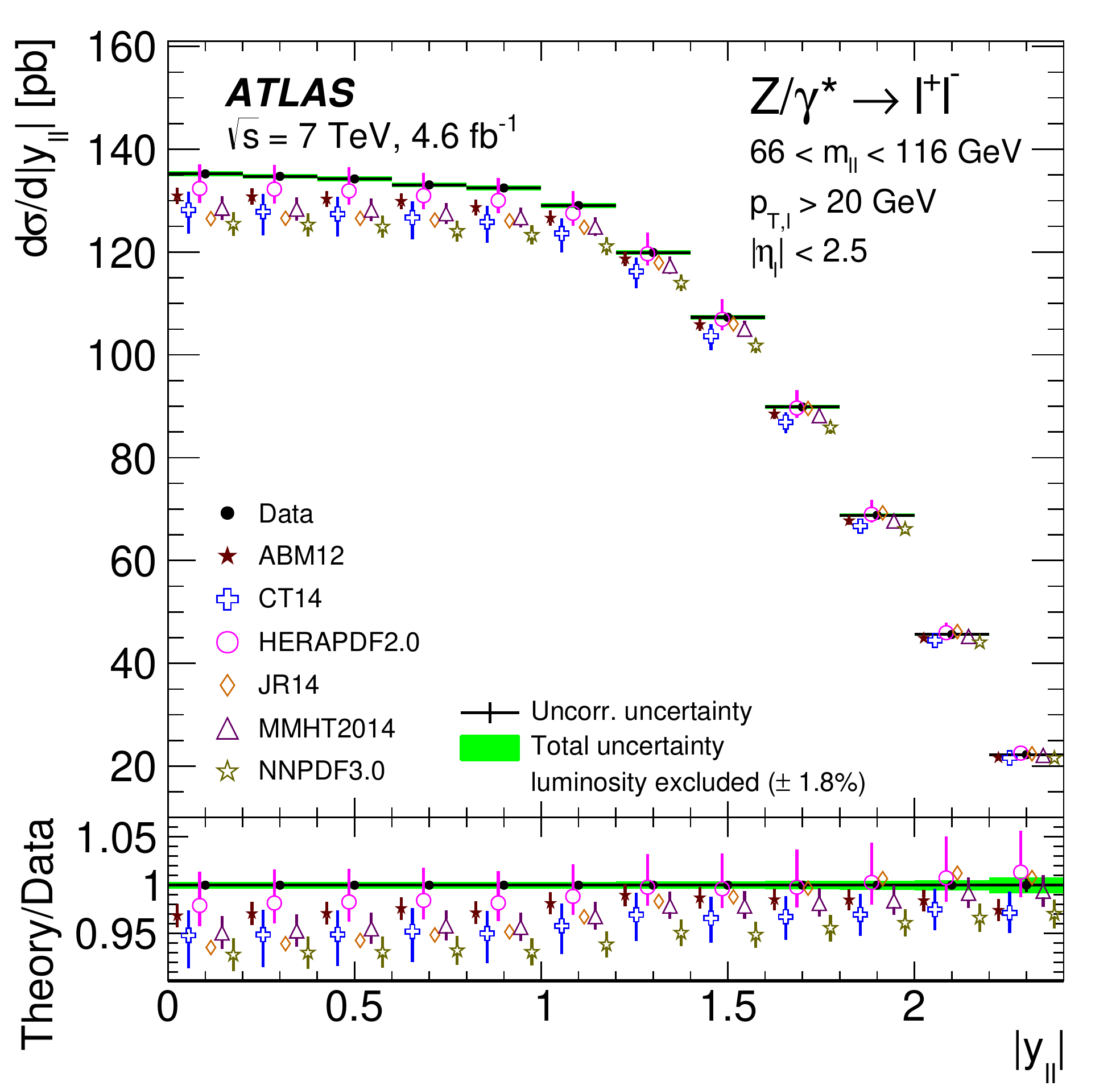}\qquad\qquad
 \includegraphics[width=0.37\textwidth]{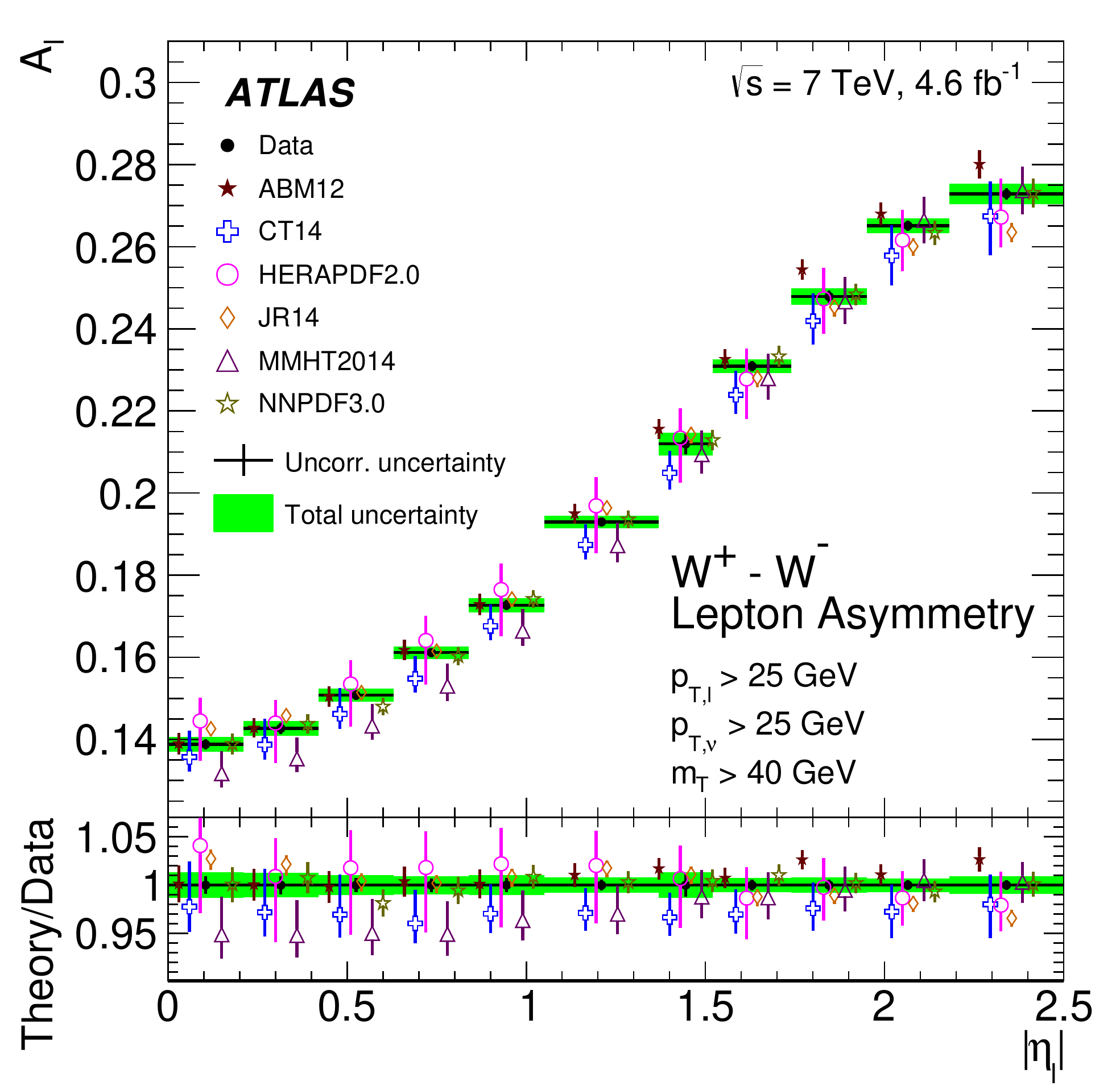}
  \end{center}
\caption{\small The ATLAS 7 TeV measurements of $Z$ rapidity distribution (left) and $W$ asymmetry (right)
based on $\mathcal{L}=4.6\,{\rm fb}^{-1}$, compared
with different NNLO PDF sets. Taken from~\cite{Aaboud:2016btc}.}
\label{fig:ATLASWZdata}
\end{figure}

The most recent ATLAS $W$ and $Z,\gamma^*$ data~\cite{Aaboud:2016btc}, which uses the full $4.6\,{\rm fb}^{-1}$ data set at 7 TeV demonstrates the level of precision that is now being achieved. The $Z$ rapidity distribution and $W$ asymmetry are shown in Fig.~\ref{fig:ATLASWZdata} (in the latter case the individual $W$ measurements are available).
The high experimental precision is clear, in particular in the $Z$ distribution where excluding the luminosity uncertainty it is as low as $\sim 0.3\%$ at central rapidity. The error on the PDF predictions, as well as the spread between sets, is significantly larger. 

%%%%%%%%%%%%%%%%%%%%%%%%%%%%%%%%%%%%%%%%%%%%%%%%%%%%
%%%%%%%%%%%%%%%%%%%%%%%%%%%%%5%%%%%%%%%%%%%%%%%%%%%%%

\subsubsection*{Theoretical calculations and tools}\label{sec:datatheory.gauge.theory}
\label{WZ:theory}

$W$ and $Z$ boson production is arguably the simplest process one can consider at a hadron collider, and indeed it was the first hadroproduction process for which the NNLO calculation became available, with the total cross sections being calculated in the early 90s~\cite{Hamberg:1990np}\footnote{A mistake in the one--loop real emission contribution was reported in~\cite{Harlander:2002wh}.}.
A decade later, in~\cite{Anastasiou:2003ds,Anastasiou:2003yy} the NNLO corrections to the differential $W$ and $Z$ rapidity distributions was presented for the first time.
A more direct comparison with experimental observables was provided in~\cite{Melnikov:2006di,Melnikov:2006kv} which presented the NNLO calculation fully differential in the final--state leptons, including in addition spin correlations, finite width effects and $\gamma - Z$ interference.
This was accompanied by the public release of the \texttt{FEWZ} simulation code, with subsequent improvements reported in~\cite{Gavin:2010az,Gavin:2012sy} and~\cite{Li:2012wna}, where NLO EW corrections (first calculated in~\cite{Berends:1984qa,Berends:1984xv,Baur:1997wa,Baur:2001ze,CarloniCalame:2007cd,Dittmaier:2009cr}) were included.
The \texttt{DYNNLO}~\cite{dynnlo2} parton--level MC provides an alternative tool for generating $W$ and $Z$ production, again including spin correlations, finite width effects, and $\gamma - Z$ interference, but currently without EW corrections.
This code allows for arbitrary user--defined cuts on the final--state partons and leptons to be imposed and histograms to be made, in contrast to \texttt{FEWZ}, where a selection of pre--determined cuts and histograms may be applied. 

These two codes differ in their theoretical treatment of the processes, in particular in the method that is applied to achieve the (non--trivial) cancellation of IR singularities at intermediates steps in the calculation. While \texttt{FEWZ} uses the local `sector decomposition' method~\cite{Melnikov:2006di,Melnikov:2006kv} that provides an automated method for extracting and canceling the IR poles, \texttt{DYNNLO} applies an alternative non--local `$q_T$--subtraction' approach~\cite{dynnlo1} which uses the transverse momentum $q_T$ of the produced $W$ or $Z$ as a cut variable, treating the calculation in a different way above and below some $q_T^{\rm cut}$.

It is important to emphasize that, at the time of writing this Report,
these two codes can still lead to non--negligible differences in their predictions
for the same cross sections even for identical input parameters.
For example, in the recent ATLAS high precision $W$ and $Z/\gamma^*$ analysis~\cite{Aaij:2015zlq} the difference in the fiducial cross section predictions can be as high as $\sim 1\%$, that is larger than the experimental uncertainties. This is due to the differing  subtraction procedures, which affects the predicted boson $p_\perp$ distributions.
The differences between the predicted cross sections are generally more significant when more restrictive cuts on the final--state leptons are imposed; for the total $W$, $Z$ cross sections the codes agree to within $0.2\%$~\cite{Aaij:2015zlq}.
A closer investigation of this issue and its impact on PDF determination will clearly be essential, given the that the experimental uncertainties of existing and
future measurements is at the permille level.

More recently the \texttt{MCFM} event generator~\cite{Campbell:2011bn} has extended the NLO simulation of $W$ and $Z$ production to NNLO~\cite{Boughezal:2016wmq}. This takes a similar non--local approach to \texttt{DYNNLO}, but using the $N$--jettiness variable rather than the $q_T$. Here, it is shown that a careful and process--dependent choice of the cut on the 0--jettiness variable, $\tau_0^{\rm cut}$ (the equivalent of $q_T^{\rm cut}$ above) is required in order to balance the requirements of sufficient statistical precision and control over systematic power corrections that increase in importance as this cut is increased. 

Further progress has subsequently been reported in~\cite{Grazzini:2017mhc}, where the results of the  \texttt{MATRIX} generator, which implements a range of processes including $W$ and $Z$ production at NNLO are presented. This includes an automated extrapolation procedure for implementing $q_T$--subtraction and determining the associated uncertainties.

Finally, public codes including transverse momentum resummation are also available.
The \texttt{DYRes}~\cite{Bozzi:2010xn} program combines NNLO fixed--order with NNLL resummation, while \texttt{ResBos}~\cite{Balazs:1995nz} combines NLO fixed--order with NNLL resummation. However, typical observables that are used in PDF fits are chosen to be largely insensitive to such resummation effects, which are most important as the $W$, $Z$ transverse momentum becomes small, and so these codes are  not generally used in  most PDF fits.
The exception are the CT10 and CT14 global analyses, for which the theoretical predictions
for $W^{\pm}$ and $Z$ production
are based on \texttt{ResBos}, with the motivation that
the NLO+NNLL result is very close to the full NNLO calculation.
Moreover, this choice could be advantageous
since the lepton charge asymmetry may still be sensitive to $q_T$ resummation
effects~\cite{Berge:2004nt} for some specific
choices of the cut $p_t^l$ in the charged lepton transverse momentum.

\subsubsection*{Impact on PDFs}\label{sec:datatheory.impact}

\begin{figure}[t]
 \begin{center}
 \includegraphics[width=0.38\textwidth]{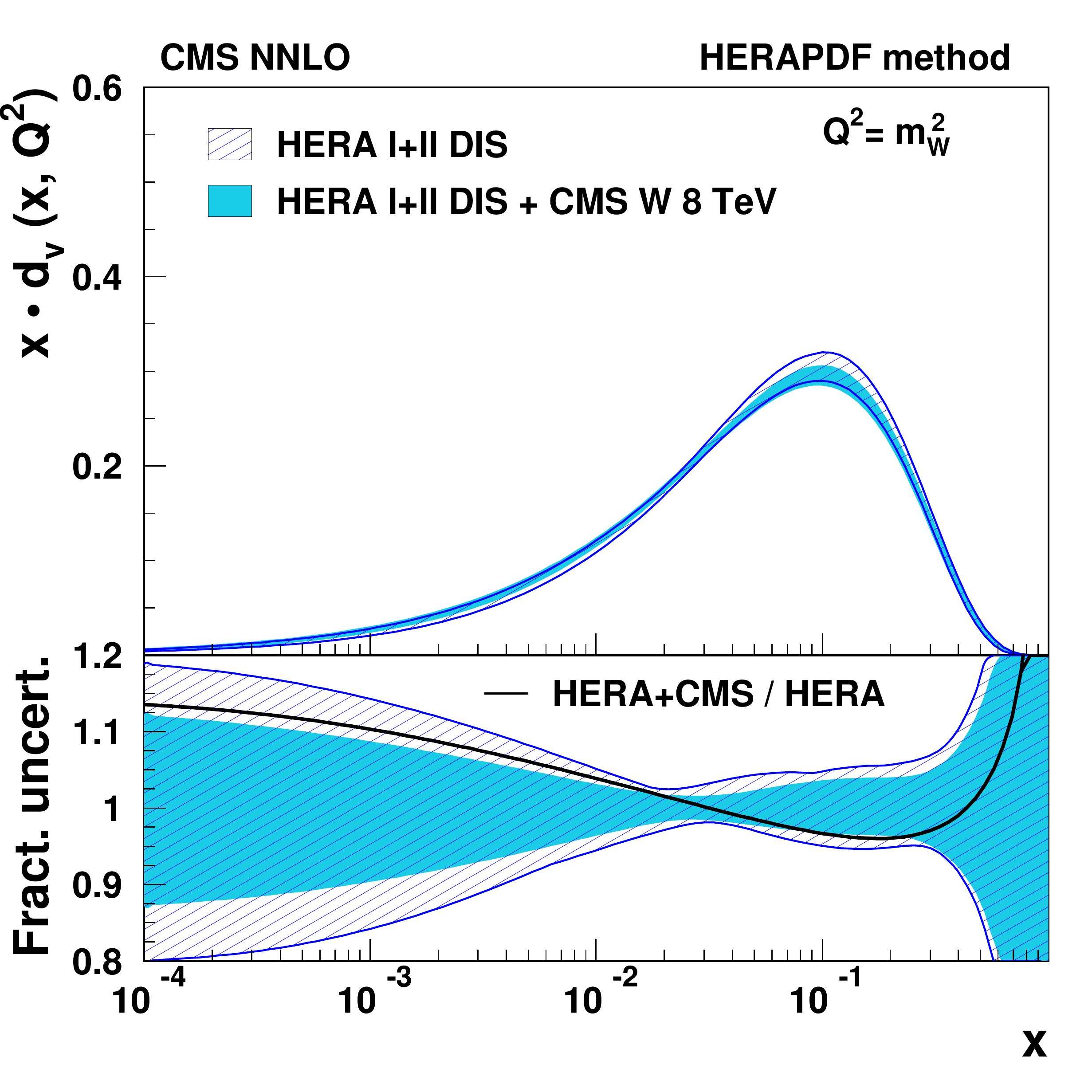}
  \includegraphics[width=0.53\textwidth]{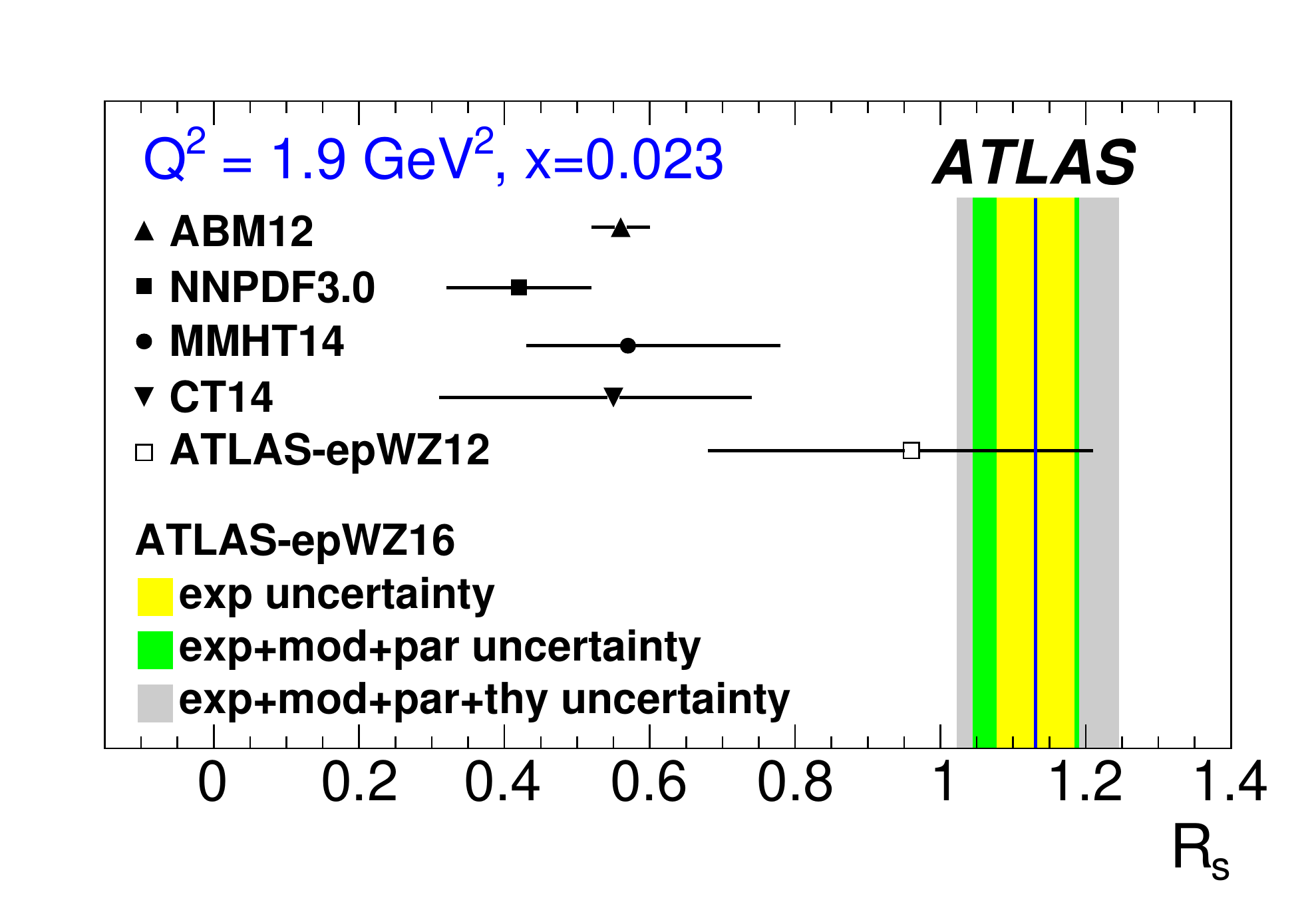}
  \end{center}
\caption{(Left) The $d_V(x,Q^2)$ PDF, with the CMS fit to HERA I+II data only, and including the CMS 8 TeV W production data. Taken from~\cite{Khachatryan:2016pev}. (Right) The ratio of the strange quark to the light quark sea, $R_s$, with different PDF predictions and the result of the ATLAS fit to HERA I+II data and the high precision $W$ and $Z/\gamma^*$ data shown. Taken from~\cite{Aaboud:2016btc}.}\label{fig:WZpdf}
\end{figure}

As described above, inclusive gauge boson production has played a crucial role in determining the quark flavour decomposition of the proton. Indeed, these have been included in all major PDF analyses for some time, from earlier fixed target data through to measurements at the Tevatron and increasingly at the LHC. Two recent LHC results are show in Fig.~\ref{fig:WZpdf}. In the left panel we show the CMS fit~\cite{Khachatryan:2016pev} to the down valence quark distribution. The baseline fit is to the HERA I+II data only, which is compared to the result including the CMS 8 TeV W boson production data. The change in shape and sizeable reduction in the PDF uncertainty over a wide range of $x$ is clear. 

In the right panel we show the impact on the strange quark fraction relative to the light quark sea
\be
R_s=\frac{s+\overline{s}}{\overline{u}+\overline{d}}\;,
\ee
of the ATLAS high precision $W$ and $Z/\gamma^*$ data~\cite{Aaboud:2016btc}. As described in Section~\ref{sec:datatheory.gauge.sensitivity}, provided the light quark flavours are sufficiently well determined, and the data are sufficiently precise, the size and shape of the $W, Z$ rapidity distributions can provide constraints on the strange quark PDFs. This is clear from the figure, where the fit to the ATLAS data predict a significantly higher value of $R_s$ in comparison to previous PDF fits, which do not include the ATLAS data. A hint of this effect is seen in the earlier `ATLAS--epWZ12' result~\cite{Aad:2012sb}, but it is only with the more recent high precision data that a clear effect becomes apparent.
The size and shape of the strange PDF will be further discussed in Sect.~\ref{sec:structure.strange}.

%%%%%%%%%%%%%%%%%%%%%%%%%%%%%%%%%%%%%%%%%%%%%%%%%%%%%%%%%%%%%%
%%%%%%%%%%%%%%%%%%%%%%%%%%%%%%%%%%%%%%%%%%%%%%%%%%%%%%%%%%%%%%

\begin{figure}[t]
 \begin{center}
 \includegraphics[width=0.32\textwidth,angle=-90]{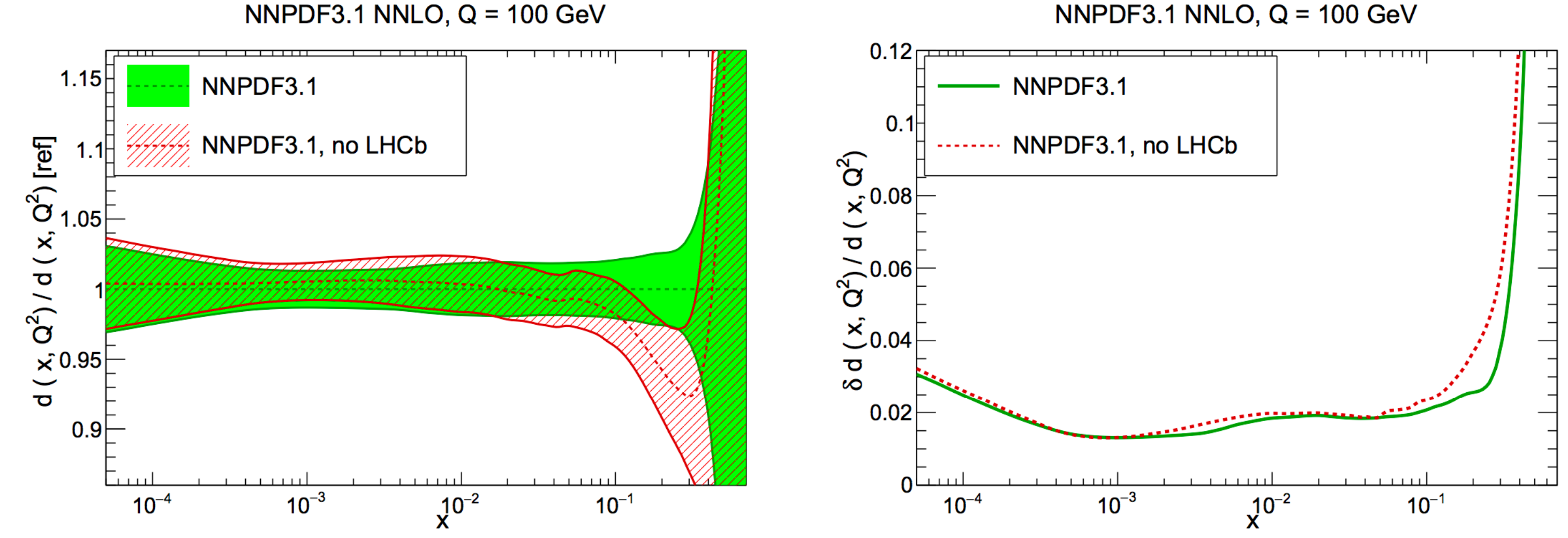}
  \end{center}
\caption{\small The down quark  $d(x,Q^2)$ at $Q = 100$ GeV, comparing the
results of an NNPDF3.1 baseline fit with those of the same
fit without LHCb data~\cite{Rojo:2017xpe}.
We show the PDF
ratio normalized to the central value of NNPDF3.1 (left) and the relative PDF uncertainties (right).
}\label{fig:lhcb-nnpdf31}
\end{figure}

As mentioned above, LHCb data on forward weak boson production provides important information about the quarks and anti--quarks flavour separation in the large--$x$ region, beyond the kinematic coverage of the ATLAS and CMS measurements. To illustrate this,
in Fig.~\ref{fig:lhcb-nnpdf31} we show
the down quark  PDF at $Q = 100$ GeV, comparing the
results of the baseline NNPDF3.1 fit with those of the same fit but
with all LHCb data excluded~\cite{Rojo:2017xpe}.
We find that the LHCb data tends to enhance the value of $d(x,Q^2)$ by almost one--sigma
in the $x\simeq 0.3$ region, with a marked decrease of PDF
uncertainties.
The impact is in fact found to be significant for all quark PDFs in this region, highlighting the important role that such data plays.
%

%%%%%%%%%%%%%%%%%%%%%%%%%%%%%%%%%%%%%%%%%%%%%%%%%%%%%%%%%%%%%%
%%%%%%%%%%%%%%%%%%%%%%%%%%%%%%%%%%%%%%%%%%%%%%%%%%%%%%%%%%%%%%

\subsection{The $p_T$ distribution of $Z$ bosons}\label{sec:datatheory.zpt}
The LHC has provided precision measurements of the inclusive transverse
momentum spectra of the $Z$ boson, which may be exploited for the purposes of PDF fitting,
as well as for other applications such as the tuning of Monte Carlo event generators.
There are three distinct regions of the $p_T$ spectrum.
At small $p_T\ll m_Z$, the fixed--order predictions
diverge due to higher--order logarithms generated by soft gluon radiation.
Here, QCD resummation techniques are needed in order
to achieve reliable predictions, see~\cite{Collins:1984kg,
Davies:1984hs,Ellis:1997ii,Qiu:2000ga,Landry:2002ix,Mantry:2010mk,Becher:2010tm,Catani:2013tia}.
Such predictions require additional non--perturbative input that cannot be calculated from first principles~\cite{Qiu:2000hf,Landry:2002ix,Scimemi:2017etj}, and  therefore the $Z$ $p_T$ distribution cannot be reliably used for the extraction of the collinear PDFs in this region. Moreover, this
distribution is highly correlated with and provides no additional sensitivity relative to
the inclusive Z boson production cross section.

On the other hand, at large $p_T\gg m_Z$, fixed--order predictions
can also receive large logarithmic contributions due to soft gluon radiation
at the partonic threshold of the $Z$ boson and the recoiling jet~\cite{Becher:2013vva,Kidonakis:2014zva}.
It has been shown that those contributions can increase the
integrated cross sections with $p_T>200\,{\rm GeV}$ by $\sim$$5\%$ compared to
the NLO prediction at the LHC~\cite{Becher:2013vva}.
The third region is defined by intermediate $p_T\sim m_Z$ values, where
the standard fixed--order predictions can be better trusted.
Note that resummed/matched calculations can also be reliably used in this intermediate $p_T$ region,
in particular since the impact of the non-perturbative effects
will not be strong enough to affect determination of PDFs, because
the resummed cross section is dominated by the perturbative logs even at $p_T=0$. 
Therefore, is in this region that the $p_T$ distribution of $Z$ bosons
can provide additional constraints on the PDFs, in particular on the gluon.

\subsubsection*{PDF sensitivity}\label{sec:datatheory.zpt.sensitivity}

At leading order,
$Z$ boson production with finite transverse momentum includes the following  partonic
subprocesses
\begin{equation}
q\bar q\rightarrow Zg,\,\, gq\rightarrow Zq,\,\, g\bar q\rightarrow Z\bar q \; .
\end{equation}
In the leptonic channel where experimental measurements are the cleanest,
the kinematics of the $Z$ boson, namely the transverse momentum $p_{T}$ and rapidity
$y_Z$, can be reconstructed from the momenta of the lepton pair produced in the $Z$ decay. The momentum fractions of the initial--state partons are given by
\begin{equation}
x_1=\frac{m_T}{\sqrt s}e^{y_Z}+\frac{p_{T}}{\sqrt s}e^{y_j}\;,\, 
\,\,\,x_2=\frac{m_T}{\sqrt s}e^{-y_Z}+\frac{p_{T}}{\sqrt s}e^{-y_j}\; , 
\end{equation}
where $\sqrt s$ is the centre--of--mass energy of the two incoming hadrons,
$m_T=\sqrt{M_Z^2+p_T^2}$ is the transverse mass of the $Z$ boson and $y_j$ is the
rapidity of the recoiling parton. For inclusive production with respect to the hadronic recoil, that is integrating over $y_j$, these momentum fractions are therefore not uniquely determined, although for LO kinematics lower limits can be derived.
Typically, the LHC experiments
measure double--differential cross sections in $p_T$ and $y_Z$ at the
$Z$ peak, although the
off--shell region, where the contributions from virtual photon may be important,
can also be considered.

%%%%%%%%%%%%%%%%%%%%%%%%%%%%%%%%%%%%%%%%%%%%%%%%%%%%%%%%%%%%%%%%%%%%%
\begin{figure}[t]
\begin{center}
  \includegraphics[scale=0.5]{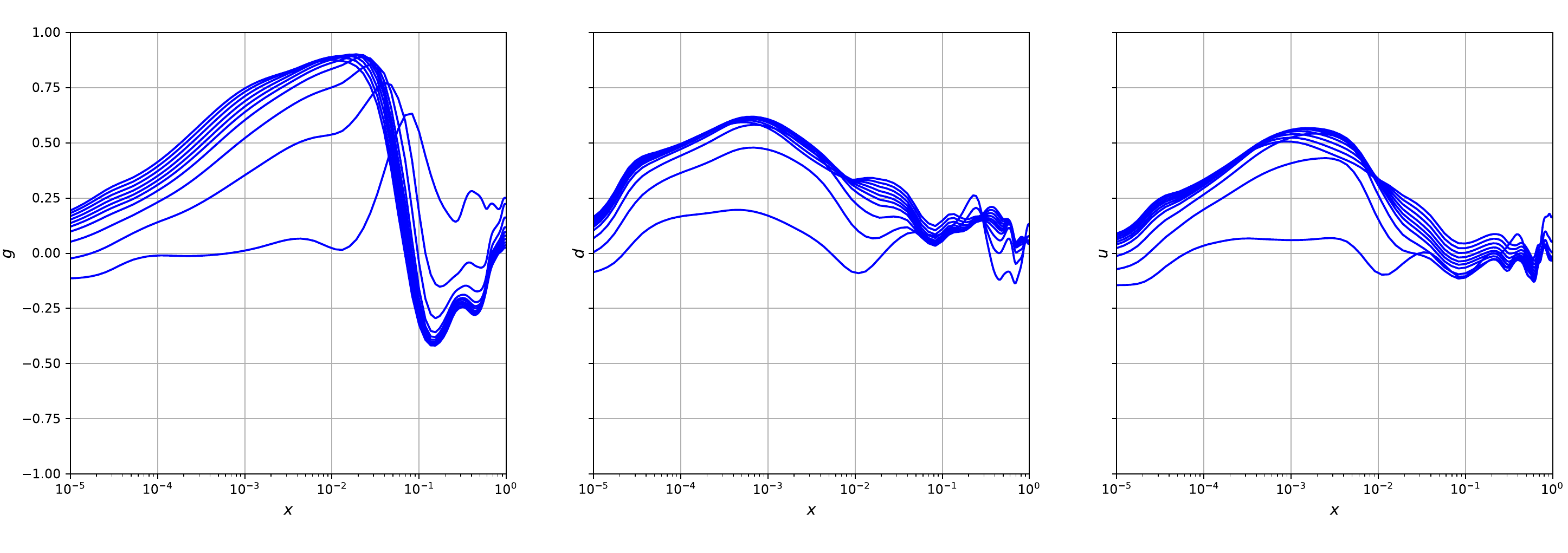}
   \caption{\small 
   Correlations coefficients
   between the cross sections in various $p_{T}$ bins and the gluon, down and
   up quark PDFs as a function of $x$, from Ref.~\cite{Boughezal:2017nla}.
The binning corresponds to the ATLAS 8 TeV
measurement~\cite{Aad:2015auj} within the rapidity interval $0<|y_Z|<0.4$. 
    \label{fig:ptz4}
  }
\end{center}
\end{figure}
%%%%%%%%%%%%%%%%%%%%%%%%%%%%%%%%%%%%%%%%%%%%%%%%%%%%%%%%%%%%%%%%%%%%%

The cross sections at moderate and large transverse momentum are dominated
by contributions from gluon and quark scattering and are strongly
correlated with the gluon PDF in the
region relevant to Higgs boson production at the LHC.
This is 
illustrated in Fig.~\ref{fig:ptz4}, which shows the PDF--induced correlations
between the cross sections in different $p_T$ bins, with $0<|y_Z|<0.4$,
and the gluon, down and up quark
PDFs at various $x$ values~\cite{Boughezal:2017nla}.
We can see that indeed the correlations with the gluon at $x\sim 10^{-2}$
almost reach 0.9.
Moderate correlations with the quark PDFs at $x\sim 10^{-3}$
are also observed.
This highlights the fact that the $Z$ $p_T$ distributions provides
in principle a handle on the gluon in a region of $x$ that lies
between that covered by HERA structure functions, at smaller
values of $x$, and that
covered by inclusive jets and $t\bar{t}$ production, at larger $x$.

\subsubsection*{Experimental data}\label{sec:datatheory.zpt.data}

%%%%%%%%%%%%%%%%%%%%%%%%%%%%%%%%%%%%%%%%%%%%%%%%%%%%%%%%%%%%%%%%%%%%%
\begin{figure}[t]
\begin{center}
  \includegraphics[scale=0.3]{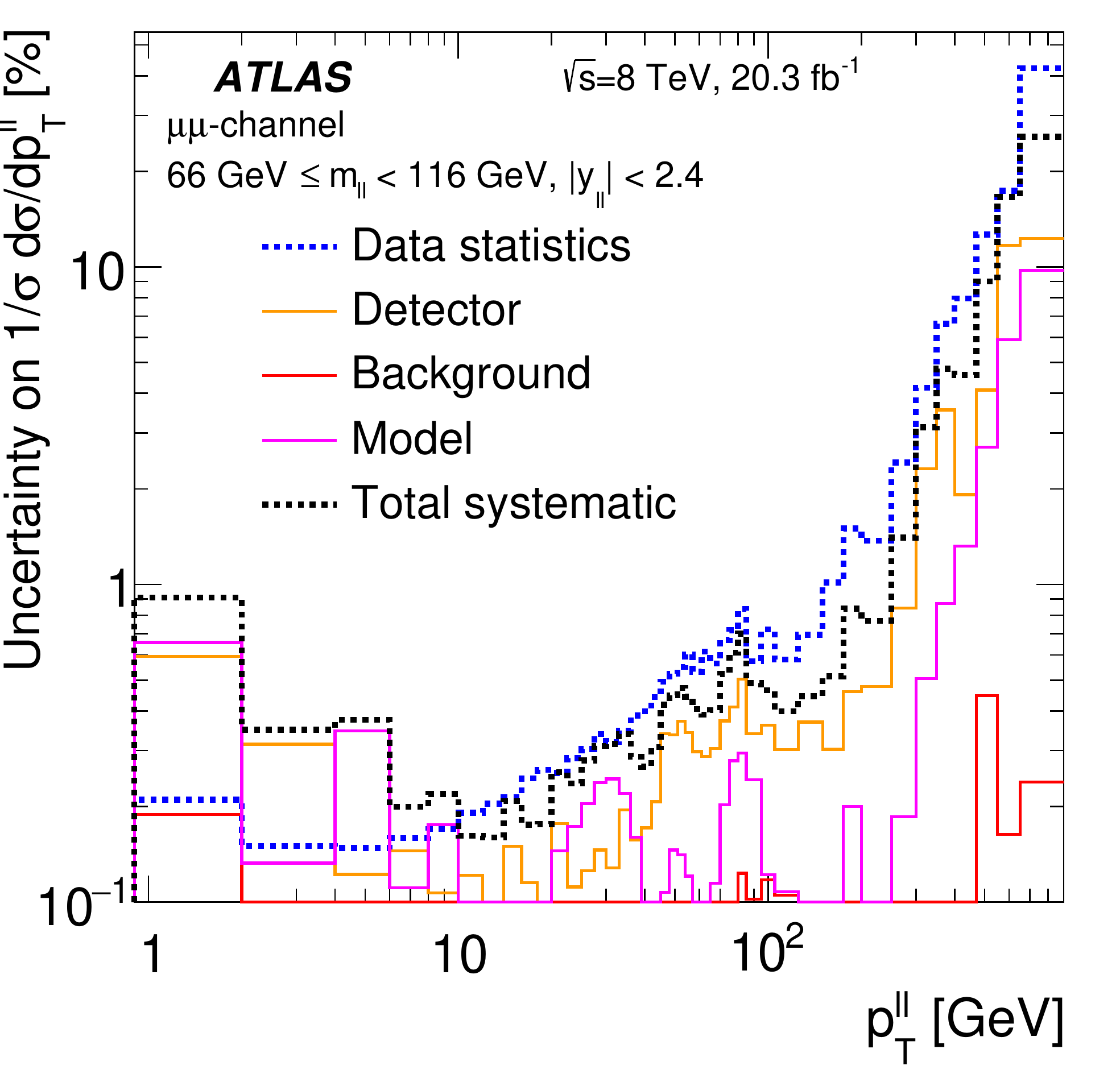}
  \hspace{0.5in}
  \includegraphics[scale=0.3]{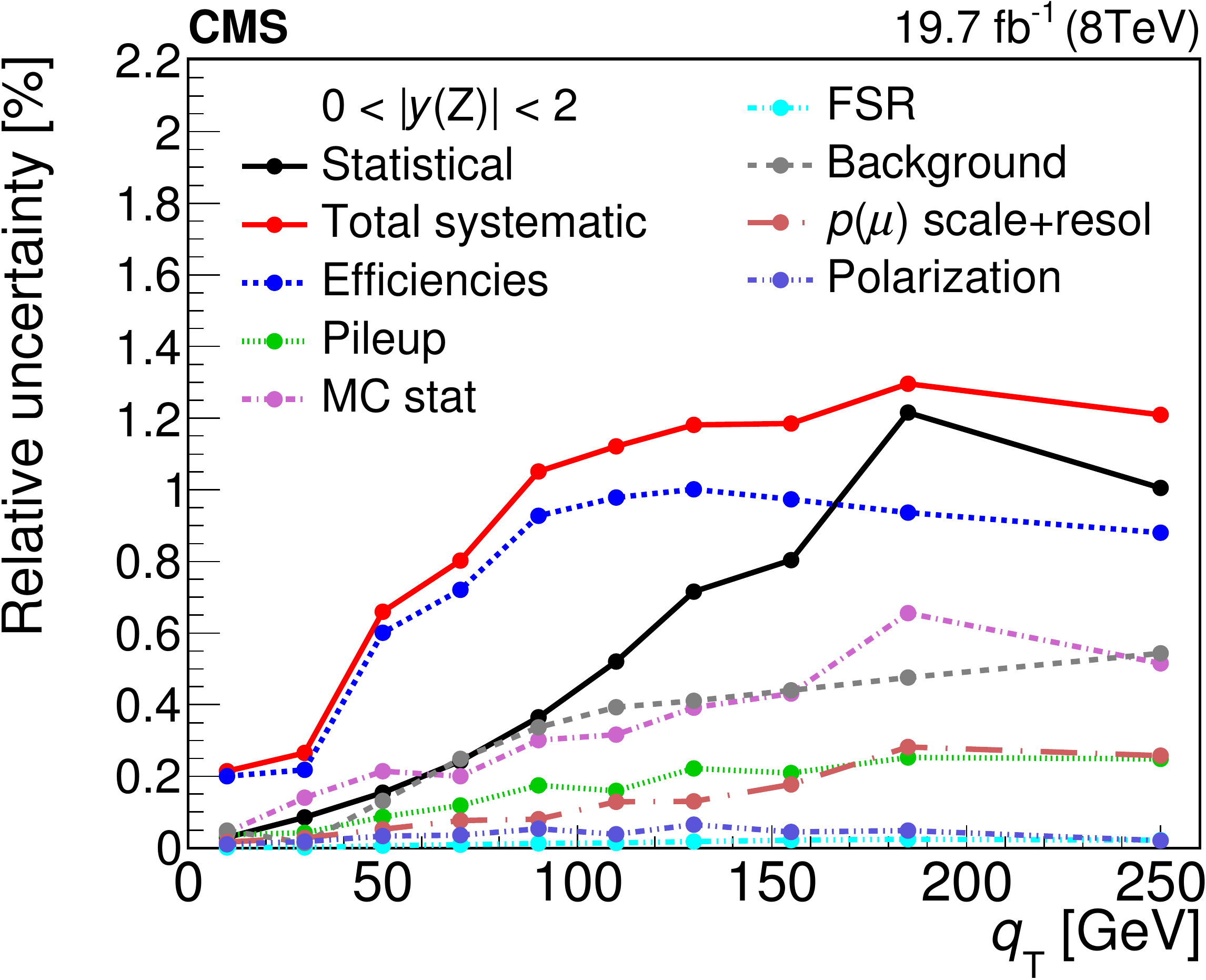}
   \caption{\small 
   Left plot: the decomposition of the various experimental statistical
   and systematic uncertainties on the normalized $p_T$ spectra
   of the $Z$ boson in dimuon channel from ATLAS at 8 TeV~\cite{Aad:2015auj} .
   Right plot: the corresponding experimental uncertainties
   for the normalized $p_T$ spectra
   of the $Z$ boson in dimuon channel from CMS at 8 TeV~\cite{Khachatryan:2015oaa} .
    \label{fig:ptz3}
  }
\end{center}
\end{figure}
%%%%%%%%%%%%%%%%%%%%%%%%%%%%%%%%%%%%%%%%%%%%%%%%%%%%%%%%%%%%%%%%%%%%%

The experimental precision for the measured $Z$ $p_T$ distribution
has reached the percent level 
for both ATLAS~\cite{Aad:2015auj}
and CMS~\cite{Khachatryan:2015oaa} at the LHC Run I,
due to both the clean dilepton
final state as well as the high statistics of the signal.
The ATLAS measurement extracts the cross sections at three different (`Born', `bare' and `dressed') particle
levels in terms of the effect of final--state photon radiation. The Born and bare
levels are defined from the lepton kinematics before and after final--state QED
radiation, while the dressed
level is defined by further combining the momentum of the lepton with photons
radiated within a certain cone.
These distributions can be presented with  the $Z$ boson rapidity integrated over, or separated
into different rapidity intervals, and can be on or off the $Z$ peak.
In addition, measurements of the distributions with respect to the angular variable
$\phi^*_{\eta}$~\cite{Aad:2015auj}, which is proportional to $p_{T,Z}$ at small transverse momentum, are also available.
As $\phi^*_{\eta}$ only depends on the direction of the lepton momenta, which are better measured than
the momenta themselves, this allows  the experimental systematics to be reduced.

 We summarise the available measurements on the $Z$ boson $p_T$
spectrum relevant to constraining the PDFs below:

\begin{itemize}

\item
The normalized $Z$ $p_T$ distribution in different rapidity
intervals by the ATLAS collaboration~\cite{Aad:2011gj,Aad:2014xaa} from LHC Run I (7 TeV). 

\item
The normalized and unnormalized distributions
of lepton pairs with respect to $p_T$ or $\phi^*_{\eta}$ in different rapidity intervals by the ATLAS
collaboration~\cite{Aad:2015auj} from LHC Run I (8 TeV). 

\item
The normalized $Z$ $p_T$ distribution integrated over rapidity
by the CMS collaboration~\cite{Chatrchyan:2011wt} from LHC Run I (7 TeV). 

\item
The normalized and unnormalized double--differential $Z$  distribution in $p_T$ and rapidity by the CMS
collaboration~\cite{Khachatryan:2015oaa} from LHC Run I (8 TeV). 

\item
The normalized 
 $Z$  $p_T$ distribution and the ratio to the  $W$ $p_T$ distribution by the CMS
collaboration~\cite{Khachatryan:2016nbe} from LHC Run I (8 TeV). 

\item
The  unnormalized $Z$ distribution in $\phi^*_{\eta}$ in the forward
region by the LHCb collaboration~\cite{Aaij:2012mda,Aaij:2015vua} from LHC Run I (7 and 8 TeV). 

\item
The normalized  $Z$ distribution in $p_T$ or 
$\phi^*_{\eta}$ in the forward region by the LHCb collaboration~\cite{Aaij:2015gna} from LHC Run I (7 TeV). 

\item
The  unnormalized distribution of the lepton pair with respect to $\phi^*_{\eta}$ in different rapidity intervals by the D0
collaboration~\cite{Abazov:2014mza} from Tevatron Run II (1.96 TeV). 

\end{itemize} 

A summary of the experimental uncertainties for
the ATLAS and CMS 8 TeV measurements~\cite{Aad:2015auj,Khachatryan:2015oaa} of the normalized $Z$ $p_T$ distribution is shown in Fig.~\ref{fig:ptz3}. The luminosity uncertainty and some of the systematic errors largely cancel
in the normalized distributions.
Both ATLAS and CMS have measured the $Z$ $p_T$ up to about 1 TeV, 
while ATLAS has a finer binning at small $p_T$, motivated by the usefulness
of low $p_T$ measurements for MC event generator tuning.

For ATLAS, the total systematic uncertainty is well within 1\% for $p_T$ smaller than 200 GeV and
is $O(10\%)$ at the higher $p_T$ tail.
The statistical
uncertainties are tiny, starting at $\sim$$0.2\%$ for $p_T$$\sim$10 GeV, and are within 1\%
in most of the relevant $p_T$ region for both ATLAS and CMS.
The PDF uncertainties from
individual PDF groups are about 2\%, which is in general already larger than the experimental
errors, even before considering  the  spread between different PDFs; such data can therefore provide valuable PDF constraints.
We emphasise that the high precision of these measurements is an unprecedented
challenge for QCD calculations, which need to achieve a permille level accuracy
to match the experimental precision.

\subsubsection*{Theoretical calculations and tools}\label{sec:datatheory.zpt.theory}

The NLO QCD corrections to the $Z$ $p_T$ distribution were calculated a long time
ago~\cite{Gonsalves:1989ar,Baer:1991qf,Arnold:1990yk}, while more recently the EW corrections have been studied extensively~\cite{Maina:2004rb,Kuhn:2004em,Kuhn:2005az,Becher:2013zua}.
The NLO QCD corrections are found to be sizeable at LHC energies, and large QCD scale variations are found in the predicted $p_T$ spectra, rendering such predictions unsuitable for
PDF determination.
However, the NNLO QCD corrections have recently
been calculated by two independent groups, in one case using the antenna subtraction method~\cite{Ridder:2015dxa,Ridder:2016nkl,Gehrmann-DeRidder:2016jns}
and in the other the $N$--jettiness subtraction method~\cite{Boughezal:2015ded,
Boughezal:2016isb}; these are found to be in good agreement.
While the original calculations are for $Z$+jet production, these
can readily be translated to the case of inclusive production of $Z$ boson
for larger values of $p_T$.

\begin{figure}[t]
\begin{center}
\includegraphics[scale=0.43]{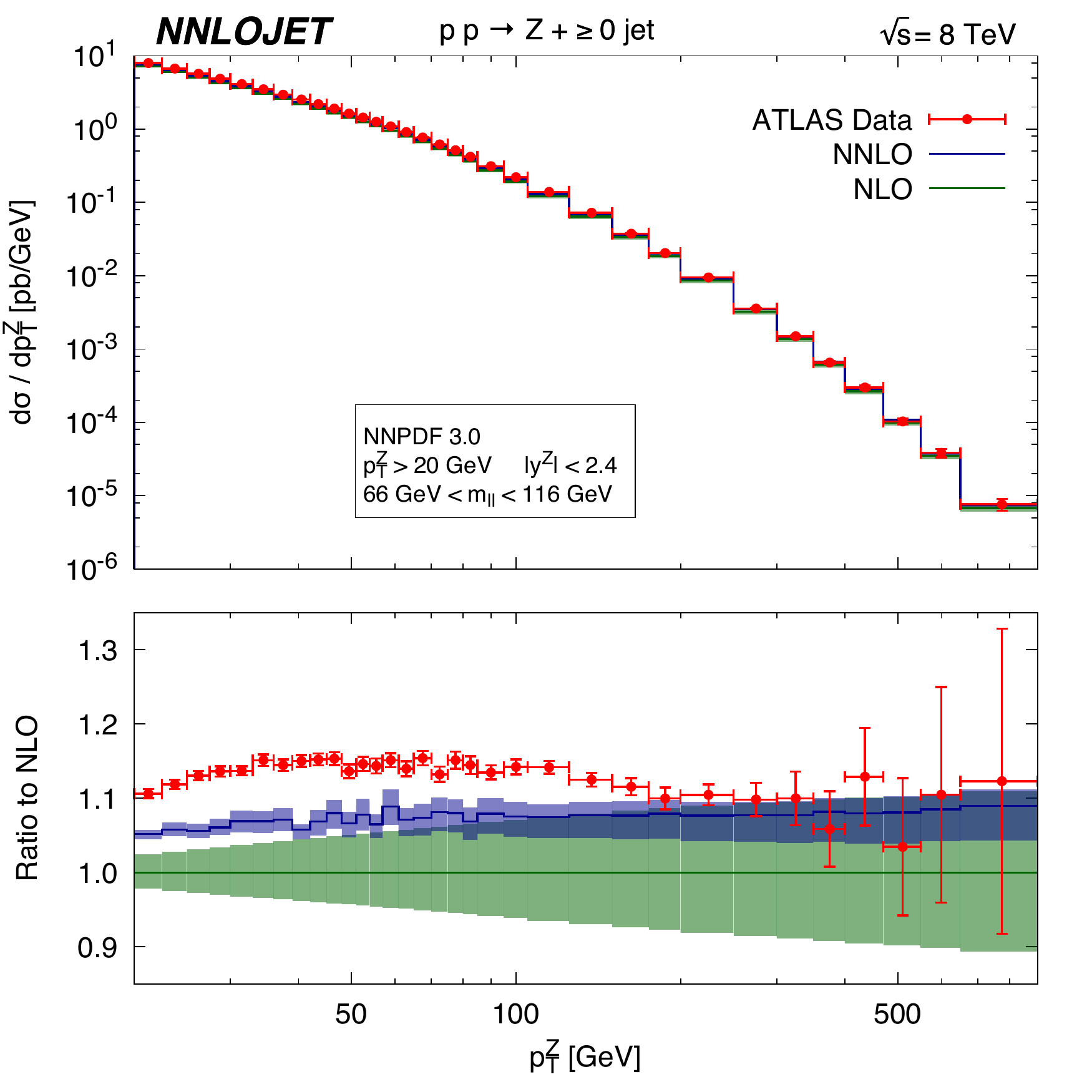}
\includegraphics[scale=0.43]{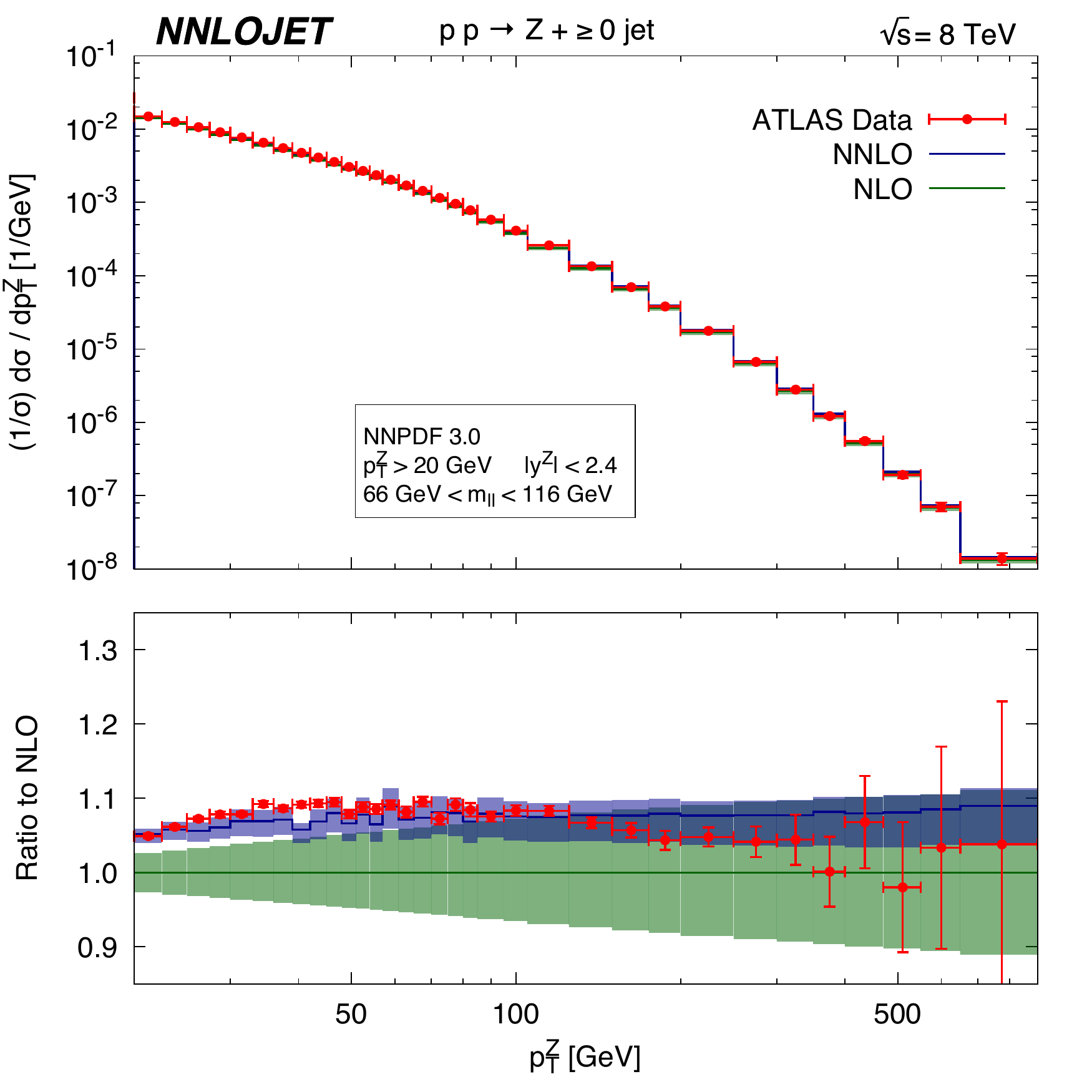}
 \caption{\small 
   Left (right): the absolute (normalized) transverse momentum distribution
   of the inclusive $Z$ boson production at LHC 8 TeV~\cite{Ridder:2016nkl}
   computed with NNPDF3.0, compared with the corresponding ATLAS measurements.
   The green and blue bands indicate the NLO and NNLO predictions with the
   corresponding theoretical uncertainties from scale variations.
    The fiducial cuts on the final state charged
   leptons are $p_{T,l}>20\,{\rm GeV}$ and $|\eta_l|<2.4$.  
    \label{fig:ptz2}
  }
\end{center}
\end{figure}

At NNLO, the theoretical uncertainty due to the QCD scale variation is found to be greatly reduced, allowing 
 the $Z$ boson $p_T$ distribution to be used for the first time for precision
PDF determination.
Moreover, these calculations include the leptonic decays of the $Z$
boson and thus parton--level selection cuts may be applied to the theoretical predictions, allowing a direct comparison
with the measured fiducial cross sections, without relying on any experimental phase space
extrapolation.
To illustrate this, Fig.~\ref{fig:ptz2} (taken
from~\cite{Ridder:2016nkl}) shows the NLO and NNLO predictions for the unnormalized and
normalized $Z$ boson $p_T$ spectra at the 8 TeV LHC. 
The central values of the renormalization and factorization scales are set to the
transverse mass of the $Z$ boson, with scale variations calculated by varying
these simultaneously by a factor of 2 up and down.
The NNLO corrections are moderate for the unnormalized distribution, being about
5\% at low $p_T$ and 9\% at high $p_T$. The remaining scale variations range
from 1\% to 6\% depending on the value of $p_T$. The EW corrections are small
at moderate transverse momentum but can be sizeable in the tail region,
reaching around $\sim 10$\% for $p_T$ values greater than 600 GeV.
However, the statistical errors in the tail region are currently quite large,
preventing a direct probe of these EW effects.
For the normalized distribution, the denominator corresponds to the inclusive
$Z$ production cross section at NNLO in the same fiducial region and
with independent scale variations. Namely, the uncertainties estimated for
the numerator and denominator are considered as uncorrelated.
This is justified by the fact that the inclusive cross section receives the dominant contribution
from the small $p_T$ region and is calculated upto one order lower in $\alpha_S$.
The size of the QCD corrections are found to be similar to the unnormalized case, as
the NNLO corrections to the inclusive $Z$ production cross section are very small.

%%%%%%%%%%%%%%%%%%%%%%%%%%%%%%%%%%%%%%%%%%%%%%%%%%%%%%%%%%%%%%%%%%%%%
%%%%%%%%%%%%%%%%%%%%%%%%%%%%%%%%%%%%%%%%%%%%%%%%%%%%%%%%%%%%%%%%%%%%%

A detailed phenomenological study and comparison of the 
NNLO calculation to the ATLAS and CMS 8 TeV measurements has been presented in~\cite{Ridder:2016nkl,Gehrmann-DeRidder:2016jns}.
Good agreement between the NNLO theory and data for the normalized distribution
in ranges from 20 GeV $<p_T< 900$ GeV, in all rapidity intervals, is observed.
However, there is some discrepancy in the comparison to the ATLAS unnormalized distributions
for using NNPDF3.0 PDFs, see Fig.~\ref{fig:ptz2}, with the data tending to overshoot the theory over a wide $p_T$ range.
The agreement in normalization is better if using NNPDF3.1 PDFs, of which the ATLAS data was
included in the fit, or using most recent PDFs from other major PDF groups.  
On the other hand, the NNLO prediction for the shape of the $p_T$ distribution is in good agreement with the data down to a $p_T$ value of 4 GeV, and is largely improved in comparison to the NLO predictions. 

\subsubsection*{Impact on PDFs}\label{sec:datatheory.zpt.impact}

The impact of the $Z$ boson $p_T$ data at LHC Run I
has recently been studied
 within a global analysis framework~\cite{Boughezal:2017nla,Ball:2017nwa}.
Here, an additional uncorrelated theoretical uncertainty of $\sim$$1\%$ has been
added to all $p_T$ bins, arguing that this required to account for the theoretical uncertainty due to
the residual MC integration error in the NNLO calculations.
Without including these errors, it was found that NNLO
predictions can not describe the data well, especially in the case of the
normalized distributions. Some tension is also found 
between the ATLAS 7 TeV normalized $p_T$ distribution~\cite{Aad:2014xaa} and the 8 TeV
$p_T$ distribution from both ATLAS and CMS~\cite{Aad:2015auj,Khachatryan:2015oaa}.
The ATLAS 7 TeV data also pulls the PDFs in a very different direction
with respect to the HERA inclusive DIS data~\cite{Boughezal:2017nla}.
In~\cite{Ball:2017nwa} it is concluded that the inclusion of the ATLAS 7 TeV
normalized data in the global analyses does not appear to be justified.

\begin{figure}[t]
\begin{center}
  \includegraphics[scale=0.4]{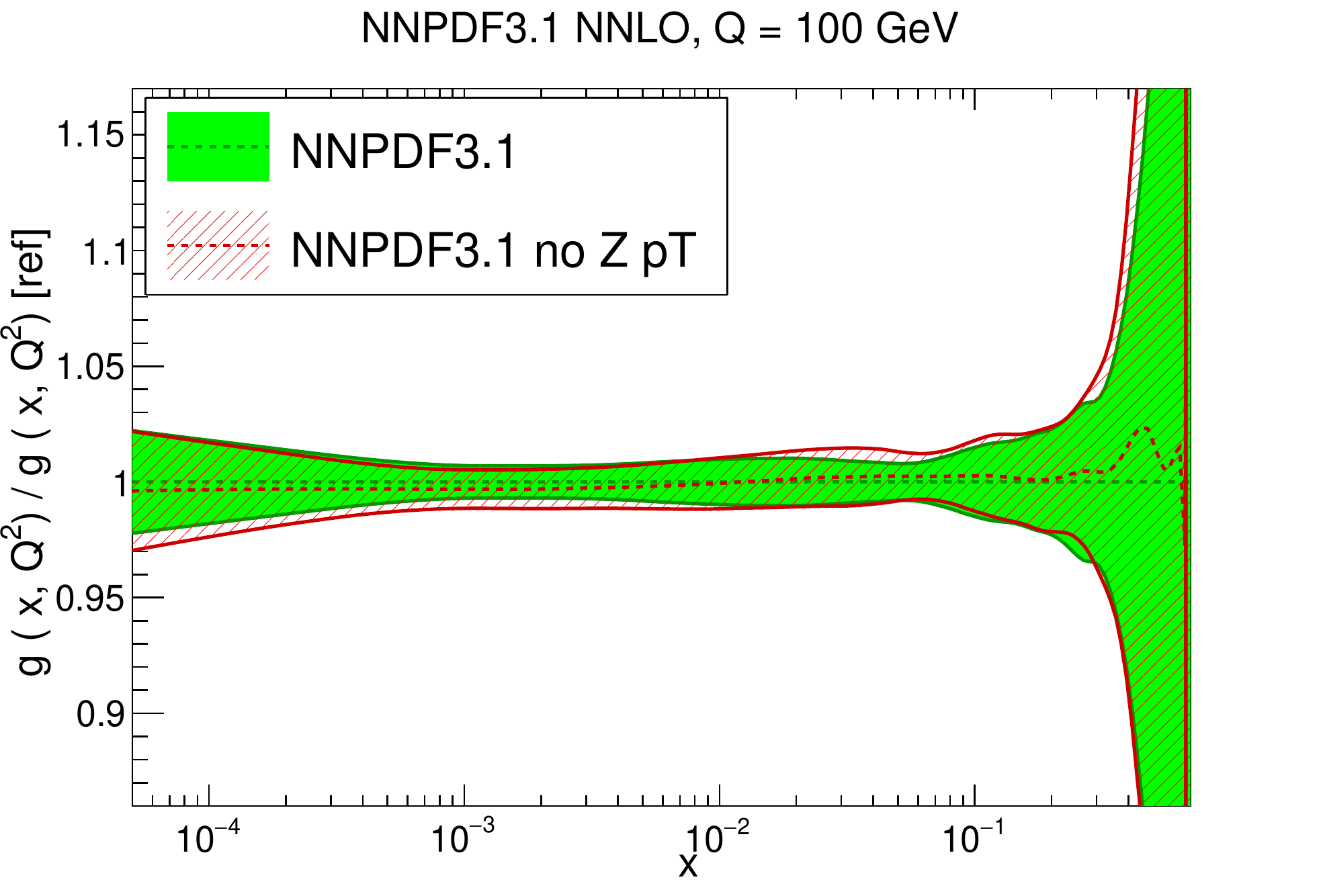}
  \includegraphics[scale=0.4]{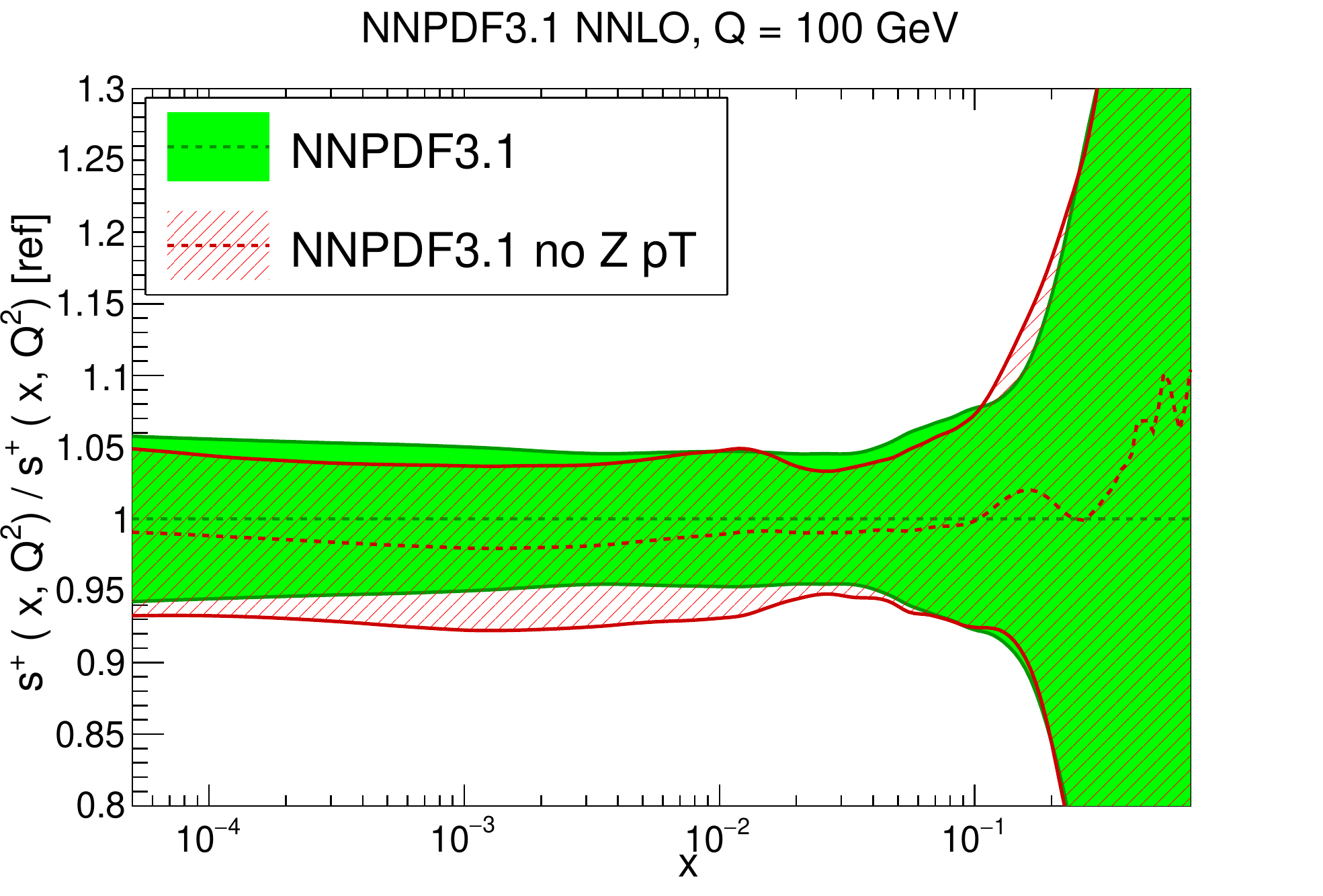}
   \caption{\small
   Impact of the $Z$ boson transverse momentum measurements from
   ATLAS and CMS 8 TeV on the gluon PDF and the total strangeness in the
   NNPDF3.1 global analyses~\cite{Ball:2017nwa}.  
    \label{fig:ptimp}
  }
\end{center}
\end{figure}

In Fig.~\ref{fig:ptimp} the impact of the ATLAS and CMS 8 TeV data on the NNPDF3.1
global analyses~\cite{Ball:2017nwa} is shown, by comparing the changes of the PDFs in the
analysis with and without the $Z$ $p_T$ data sets included. The uncertainty in the gluon
PDF is seen to be slightly reduced in the $x$ region of $10^{-2}\sim 10^{-1}$.
In the same region the gluon PDF receives constraints from precision
measurements on top quark pair production and HERA inclusive DIS, both of which are
present in the same analysis. It is also found that the 8 TeV data
lead to a moderate reduction in the PDF uncertainty on the total strangeness.
  
%%%%%%%%%%%%%%%%%%%%%%%%%%%%%%%%%%%%%%%%%%%%%%%%%%%%%%%%%%%%%%%%%%%%%
%%%%%%%%%%%%%%%%%%%%%%%%%%%%%%%%%%%%%%%%%%%%%%%%%%%%%%%%%%%%%%%%%%%%%

It is important to highlight that in Fig.~\ref{fig:ptimp} the impact of the
ATLAS and CMS $Z$ $p_T$ data is rather moderate due to the fact that many other datasets
in the fit already constrain the PDFs, and in particular the gluon, in the relevant $x$ region.
As shown in~\cite{Boughezal:2017nla}, the effects of these measurements
is rather more marked if the input dataset does not contain other data such
as jets of top quark pair production data that also pin down the gluon,
see Sect.~\ref{sec:structure.gluon} for further discussion.

\subsection{Direct photon production}\label{sec:datatheory.photon}
\label{subsec:photons}

In this section we discuss the PDF constraints that can be derived
from  isolated prompt photon production at hadron colliders.

\subsubsection*{PDF sensitivity}\label{sec:datatheory.photon.sensitivity}

The leading
parton--level processes for direct (also known as `prompt') photon production,
that is,  where the photon is produced by point--like emission from a quark,
are given by
\begin{align}
{\rm QCD~Compton}&:\qquad qg \to q \gamma \;,\\
{\rm Annihilation}&:\qquad q\overline{q}\to g\gamma\;.
\end{align}
The QCD Compton process gives the dominant contribution, in particular at the LHC,
and is directly sensitive to the gluon PDF.
For LO kinematics, the momentum fraction carried by the incoming gluon is directly proportional to the transverse energy $E_\perp^\gamma$ of the produced photon, and thus for higher $E_\perp^\gamma$ this process provides a direct probe of the gluon PDF at high $x$.
Moreover, this represents the highest rate electroweak process at the LHC, while the produced photon directly reflects the production kinematics, without for example requiring any additional hadronization corrections, as in the case of jet production. This can therefore provide a valuable tool with which to constrain the gluon.

However, the PDF interpretation of prompt photon
production is not without its complications.
In particular, the `direct' process described above is not the only way in which high $E_\perp$ photons can be produced in hadronic collisions.
We must also include the `fragmentation' contribution, whereby a standard $2\to 2$ QCD scatter involving a final--state quark (or anti--quark) produces a photon through a collinear $q\to q\gamma$ emission.
While the parton--level process for this fragmentation component carries an extra power of $\alpha_s$ compared to the direct production mechanism, the collinearly enhanced photon emission is effectively of order $\alpha/\alpha_s$, and thus fragmentation
enters at the same order.

Technically speaking, this fragmentation emerges from the higher order corrections to the direct process. These correspond to multiple collinear splittings of a high $p_\perp$ parton which end up with a photon, and that can be absorbed into universal `fragmentation functions'.
These cannot be calculated perturbatively, but rather must be fit to data, for example in $e^+e^-$ annihilation to hadrons. This introduces a potentially significant additional source of uncertainty, since our understanding of the non--perturbative quark--to--photon fragmentation
mechanism is poor.

\begin{figure}[t]
  \begin{center}
 \includegraphics[width=0.45\textwidth]{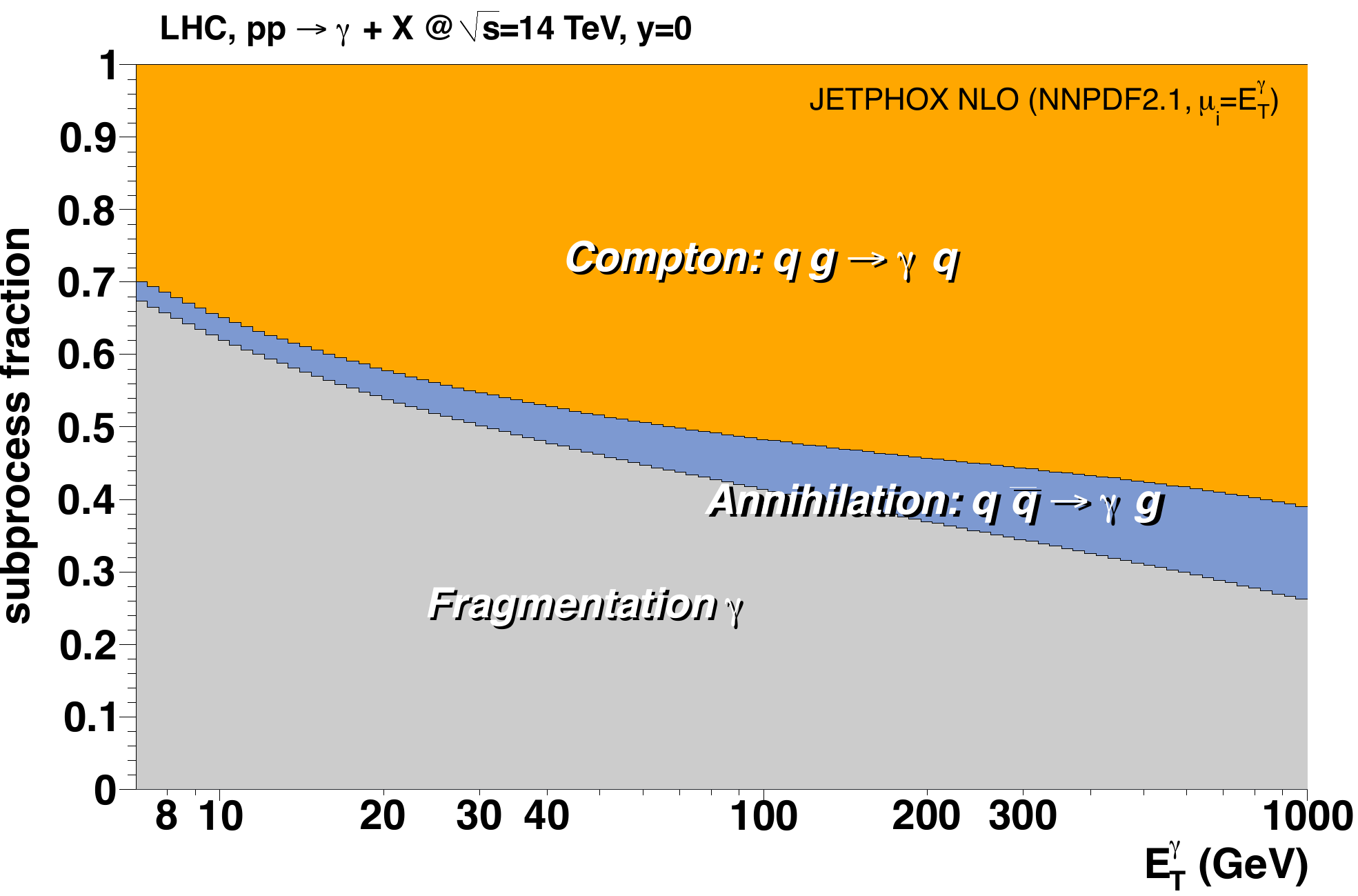}
\includegraphics[width=0.45\textwidth]{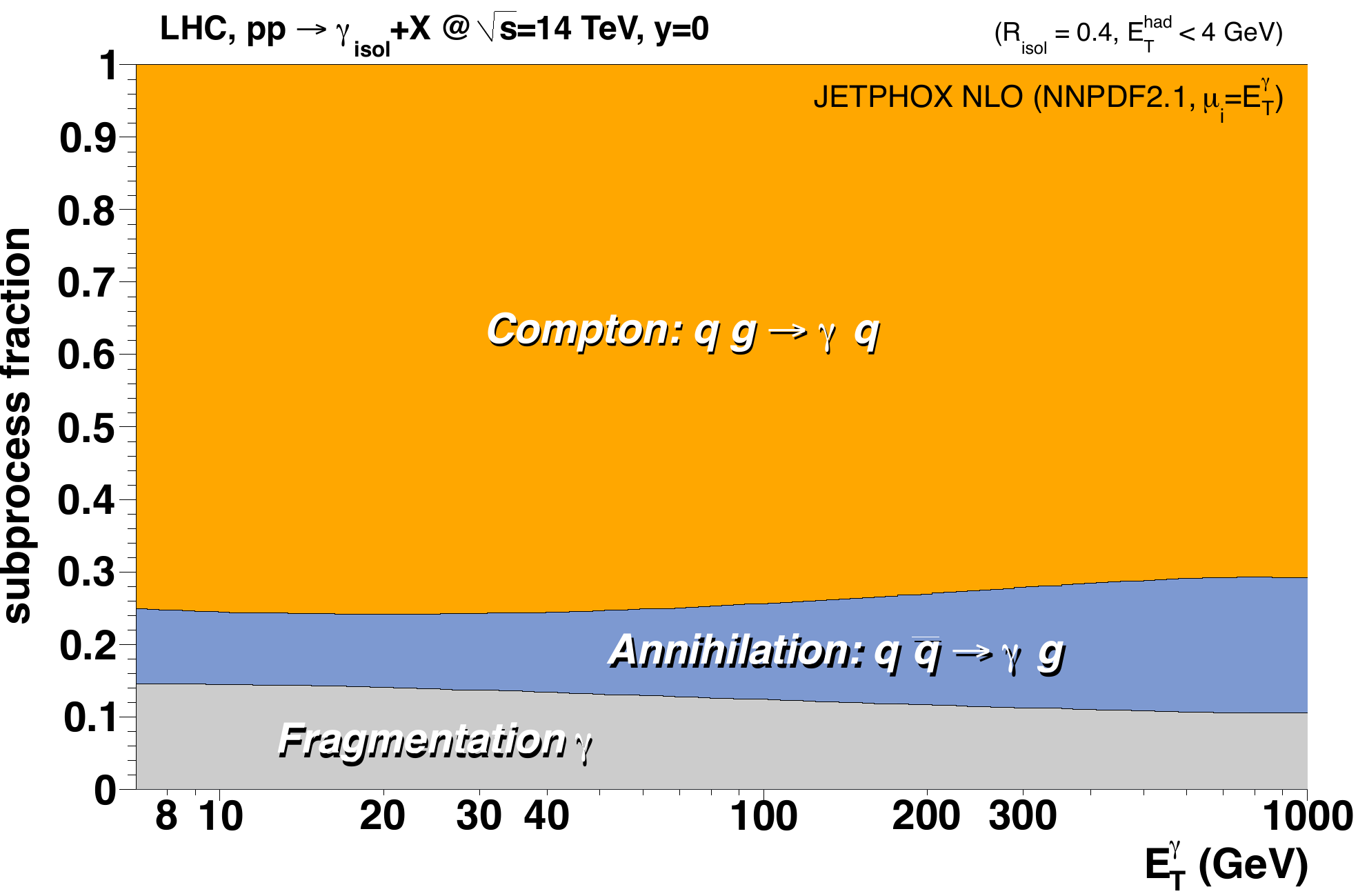}
  \end{center}
\caption{\small
Relative contributions from Compton ($qg$), annihilation ($q\overline{q}$) and fragmentation to prompt photon production at central rapidities at the 14 TeV LHC, before (left) and after (right) the application of isolation cuts. Figures taken from~\cite{d'Enterria:2012yj}.}\label{fig:isol}
\end{figure}

In fact, the situation is greatly improved by noting that physically this fragmentation process corresponds to the same multiple emission process that generates final--state jets, and indeed such fragmentation photons are typically accompanied by significant additional hadronic activity in the vicinity of the detected photon.
This is to be contrasted with direct emission, where at LO the produced photon and outgoing quark are produced completely back--to--back.
The direct mechanism may therefore be greatly enhanced by introducing isolation criteria whereby the total sum of the transverse energy of the hadrons present in some cone $R$ centred on the photon is less than a given value.
These also reduces the additional `non--prompt' background due to the electromagnetic decay of hadrons. The impact of such a cut is shown in Fig.~\ref{fig:isol}, where it seen that the contribution from the less well known fragmentation contribution is small.
This also demonstrates the dominance of the direct Compton production process for prompt photon production at the LHC.

%%%%%%%%%%%%%%%%%%%%%%%%%%%%%%%%%%%%%%%%%%%%%
%%%%%%%%%%%%%%%%%%%%%%%%%%%%%%%%

Historically, isolated photon production represented one of the first
non--DIS processes used to constrain PDFs in global analyses,
and was used in such early fits as 
Refs.~\cite{Aurenche:1988vi,Harriman:1990hi,Morfin:1990ck}.
However, the  difficulties in describing fixed--target photon
production measurements from
the E706 experiment~\cite{Apanasevich:1997hm,Apanasevich:2004dr}
raised questions about the reliability of this process for PDF fits within
the perturbative QCD framework, and the
potential sensitivities to non--perturbative effects.
Combined with the increasing availability of high precision jet data from the Tevatron, which also constrain the high $x$ gluon, this lead to the process falling out of favour in the PDF fitting community. 
The last PDF set to include any such data is the MRST99~\cite{Martin:1999ww} fit. 

However, the subsequent studies of~\cite{Ichou:2010wc,d'Enterria:2012yj} (see also~\cite{Carminati:2012mm} for a study of the related $\gamma$ + jet process) have shown that by increasing the
centre--of--mass energy $\sqrt{s}$ from fixed target to collider energies and, as discussed above, imposing a suitable isolation condition on the produced photon, the process may be brought under reasonable theoretical control.
Moreover, a comparison of the NLO perturbative QCD predictions to the ATLAS
Run I measurements~\cite{Aad:2016xcr,Aaboud:2017cbm}
exhibits
an adequate description of the data, albeit with fairly large $\sim 10-15\%$ scale variation uncertainties.
These largish theoretical uncertainties have been recently reduced down
to a few percent thanks to  the availability of the NNLO calculation, discussed below.
Therefore, isolated photon production may well provide a useful tool for PDF constraints
for the next generation of global fits.

\subsubsection*{Experimental data}\label{sec:datatheory.photon.data}

The available hadron collider data on isolated photon production is summarised below:

\begin{itemize}

\item The most recent data at 1.96 TeV from the CDF~\cite{Aaltonen:2017swx} and D0~\cite{Abazov:2005wc}
collaborations extends up to $E_\perp^\gamma<0.5$ TeV and 0.3 TeV, respectively, while the photon pseudorapidity is restricted to have $|\eta^\gamma|\lesssim 1$.
In the CDF case this corresponds to the full Run II $9.5\,{\rm fb}^{-1}$ data set, and so represents the final legacy measurement.

\item The ATLAS 7 TeV measurement~\cite{Aad:2013zba} covers up to $E_\perp^\gamma<1$ TeV,
while those at 8~\cite{Aad:2016xcr} and 13~\cite{Aaboud:2017cbm} TeV extend up to $E_\perp^\gamma<1.5$ TeV.
These datasets correspond to the full available integrated luminosities of $4.6\,{\rm fb}^{-1}$ and  $20.2\,{\rm fb}^{-1}$ at 7 and 8 TeV, respectively, while the 13 TeV measurement uses a $3.2\,{\rm fb}^{-1}$ data set.

\item The most precise CMS data at 7 TeV~\cite{Chatrchyan:2011ue}, corresponding to $36\,{\rm pb}^{-1}$ of integrated luminosity and extending to $E_\perp^\gamma<0.4$ TeV.

\item Data from a smaller sample at $2.76$ TeV have also been taken by ATLAS~\cite{Aad:2015lcb} and CMS~\cite{Chatrchyan:2012vq}, to be used as baseline for heavy ion collision studies.

\end{itemize}

 For the ATLAS and CMS measurements,
the photon pseudorapidity is restricted to satisfy $|\eta^\gamma|\lesssim 2.4$. In addition to these collider datasets, as discussed above there exist a number of older measurements taken from
fixed--target experiments (see~\cite{Ichou:2010wc,d'Enterria:2012yj} for a summary) but
these are characterised by large experimental uncertainties, as well as by an inadequate
treatment of photon isolation in some cases, and thus they should not be considered
for PDF studies.

\subsubsection*{Theoretical calculations and tools}\label{sec:datatheory.photon.theory}
\label{isolgam:theory}

For the past 15 years, the theoretical state--of--the--art was provided by the~\texttt{JETPHOX}~\cite{Catani:2002ny} MC generator, which implements both the direct and the fragmentation contributions consistently at NLO. The NLO EW corrections have also been calculated in~\cite{Kuhn:2005gv}.
Recently, the first NNLO calculation of direct photon production has been reported~\cite{Campbell:2016lzl}, although the (numerically small) fragmentation component still remains at NLO.
The NNLO prediction for the ATLAS 8 TeV data~\cite{Aad:2016xcr} is compared to the
corresponding NLO calculation and found to lie consistently within the NLO scale uncertainty band, with the central value being $\sim 5\%$ higher.
Moreover, the NNLO scale uncertainty is found to be greatly reduced, leading now
to scale variations at the 3\% level at most.

In order to achieve the best  description of the ATLAS measurement, in~\cite{Campbell:2016lzl}
the leading-log EW Sudakov corrections of~\cite{Becher:2013zua} are included, and the coupling $\alpha$ is evaluated at the scale $m_Z$, as recommended in~\cite{Becher:2013zua}.
In particular, the EW corrections are found to reduce the cross section by as much as $10\%$, that is, significantly outside the QCD scale variation band, at the highest $E_\perp$, improving the shape description.
The results of the \texttt{PeTeR} code~\cite{Schwartz:2016olw}, which combines the NLO calculation with ${\rm N}^3$LL threshold resummation in addition to these EW corrections, is found to lie close to the combined NNLO+EW prediction, but with a larger uncertainty band, indicating that the data may not be too sensitive to such additional resummation effects.
This is not completely unexpected, since direct photons are produced far from the
kinematic threshold where the effects of resummation are expected to be less important.

Therefore, while the NNLO calculation is a very encouraging step towards including isolated photon data in high precision PDF fits, there remain still some further theoretical issues to be investigated, related to the impact of EW corrections and, as discussed in~\cite{Campbell:2016lzl} the choice of photon isolation, which can also affect the comparison with the experimental data.

\subsubsection*{Impact on PDFs}\label{sec:datatheory.photon.impact}

%%%%%%%%%%%%%%%%%%%%%%%%%%%%%%%%%%%%%%%%%%%%%%%%%%%%%%%%%%%%%%%%%%%%%%%%%
\begin{figure}[t]
  \begin{center}
 \includegraphics[width=0.4\textwidth]{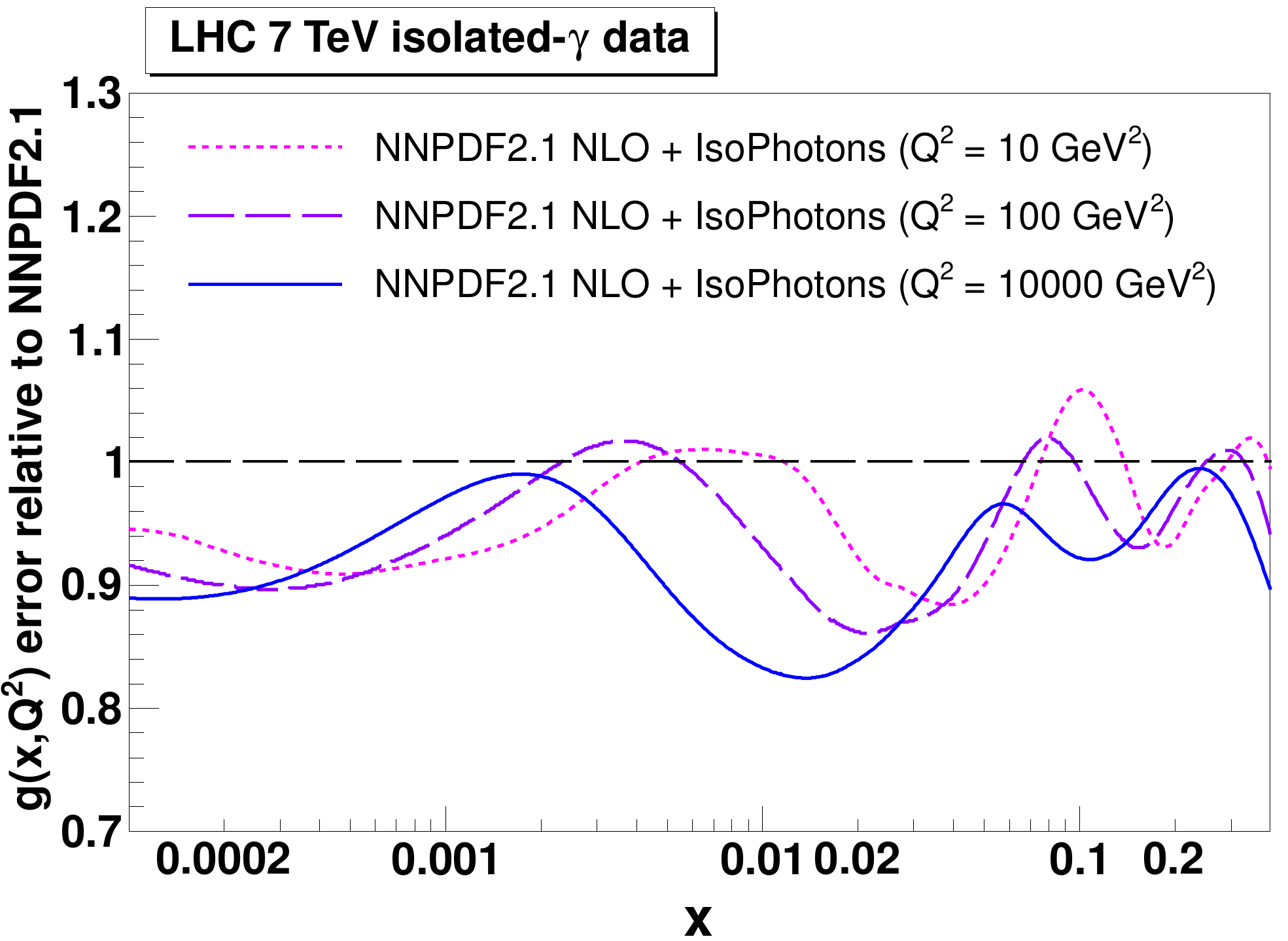}
  \end{center}
\caption{\small Relative reduction in the NNPDF2.1 NLO gluon PDF uncertainty when including a 36 ${\rm pb}^{-1}$ LHC 7 TeV isolated photon data set via Bayesian reweighting.
Taken from~\cite{d'Enterria:2012yj}.}\label{fig:isolPDF}
\end{figure}
%%%%%%%%%%%%%%%%%%%%%%%%%%%%%%%%%%%%%%%%%%%%%%%%%%%%%%%%%%%%%%%%%%%%%%%%%

Currently, no up to date studies of the impact of isolated photon data on the PDFs have been performed, in particular taking into account the new NNLO calculation and the high precision LHC data. However, in~\cite{d'Enterria:2012yj} (see also~\cite{Ichou:2010wc}) the impact of a range of data, including the earlier 36 ${\rm pb}^{-1}$ ATLAS and CMS measurements at 7 TeV, on the PDFs has been studied in detail through a reweighting of the NNPDF2.1 set.
In Fig.~\ref{fig:isolPDF} the impact of these LHC data on the gluon PDF is shown.
A moderate reduction in the uncertainty, of up to 20\%, is found in the intermediate $x$ region. Interestingly, this overlaps with the kinematic region relevant for Higgs boson production via gluon fusion at the LHC, and indeed a $\sim 20$\% reduction in the Higgs production cross section is found.
Given these results correspond to a reasonably limited LHC data set, it would be interesting to see the impact of the latest LHC data, as well as that of the NNLO corrections.

\subsection{Top quark production}\label{sec:datatheory.top}
In this section, we discuss the PDF information that can be obtained from top quark
pair production measurements, and we also briefly review the constraints that
can could potentially be derived from single top quark production.

\subsubsection*{PDF sensitivity}\label{sec:datatheory.top.sensitivity}

The production of top quark pairs at the LHC is dominated by 
gluon--gluon fusion, which represents around 85\% of the total cross section\footnote{Note
that this statement is not true at the lower values of $\sqrt{s}$ of the Tevatron,
where top quark pair production is dominated instead by quark anti--quark annihilation.}.
Therefore, provided that other sources of theoretical uncertainties
such as missing higher orders and the values of the top mass $m_t$ can
be kept under control, including top quark production data into the global
PDF fit has the potential to constrain the gluon in the large--$x$ region,
which is currently affected by large uncertainties.

\begin{figure}[t]
\begin{center}
  \includegraphics[scale=0.80]{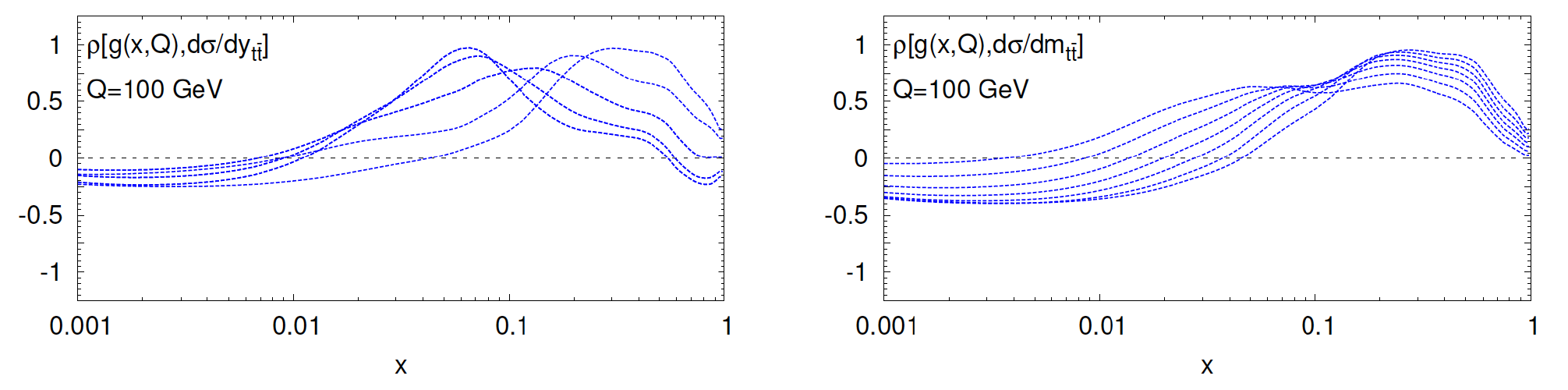}
   \caption{\small The correlation coefficient between the gluon
   PDF at $Q=100$ GeV and the theory predictions for
   the absolute differential distributions in $y_{t\bar{t}}$
   (left) and $m_{t\bar{t}}$ (right plot) at $\sqrt{s}=8$ TeV,
   as a function of $x$.
   Each curve corresponds to a specific measurement bin.
   The higher the absolute value of the correlation coefficient, the
   bigger the sensitivity to the gluon for those specific values of $x$.
    \label{fig:topdiffcorr}
  }
\end{center}
\end{figure}

To illustrate the kinematical sensitivity of
top quark pair production to the gluon,
in Fig.~ \ref{fig:topdiffcorr}
we show the correlation coefficient $\rho\lc g(x,Q), d\sigma\rc$
between the gluon
   PDF at $Q=100$ GeV and the theory predictions for
   the differential distributions in $y_{t\bar{t}}$
    and $m_{t\bar{t}}$ at $\sqrt{s}=8$ TeV,
   as a function of $x$.
   In this comparison, each curve corresponds to specific measurement bin.
   In the figure, the higher the absolute value of the correlation coefficient, the
   bigger the sensitivity to the gluon for those specific values of $x$.
   We observe that this sensitivity is high for values of $x$ up to $x\simeq 0.6-0.7$, beyond
   the reach of other processes sensitive to the gluon, such as inclusive jet production.
   Moreover, it is important
   to emphasise that the availability of differential distributions significantly extends
   the kinematical coverage beyond that provided by the total inclusive cross sections.

%%%%%%%%%%%%%%%%%%%%%%%%%%%%%%%%%%%%%%%%%%%%%%%%%%%%%%%%%%%%%%%%%%%%%
%%%%%%%%%%%%%%%%%%%%%%%%%%%%%%%%%%%%%%%%%%%%%%%%%%%%%%%%%%%%%%%%%%%%%

\subsubsection*{Experimental data}\label{sec:datatheory.top.data}

The available data on top quark pair production are summarised below:

\begin{itemize}
\item Earlier measurements, presented at the total cross section level, have been performed first at the Tevatron~\cite{Aaltonen:2013wca}
and then by ATLAS and CMS~\cite{ATLAS:2012aa,ATLAS:2011xha,TheATLAScollaboration:2013dja,
Chatrchyan:2013faa,Chatrchyan:2012bra,Chatrchyan:2012ria} at 7, 8, and 13 TeV.

\item Single--inclusive differential distributions of top quark pair production
have been presented by ATLAS~\cite{Aad:2015mbv} and CMS~\cite{Khachatryan:2015oqa}.
These include measurements both at the top quark level extrapolated
to the full phase space ($p_T^t$, $y_{t\bar{t}}$, $m_{t\bar{t}}$), and in terms of directly observable quantities
(charged lepton $p_T$ and rapidity, $b$--tagged jet kinematics etc).

\item Double--differential distributions for top quark pair production can also be performed as a function
of $p_{t\bar{t}}$ and $m_{t\bar{t}}$, as illustrated by the recent CMS measurement~\cite{Sirunyan:2017azo}
of normalized double--differential distributions.
\end{itemize}

The differential measurements are often presented both
as absolute distributions as well as normalized to the total fiducial cross section,
in order to benefit from a number of cancellations between experimental
systematic uncertainties.

\subsubsection*{Theoretical calculations and tools}\label{sec:datatheory.top.theory}

The NNLO QCD calculation of the total $t\bar{t}$ production cross section has been available
since 2013~\cite{Czakon:2013goa,Czakon:2012pz,Baernreuther:2012ws,Czakon:2011xx}, including
the resummation of logarithmically enhanced threshold corrections up to NNLL~\cite{Cacciari:2011hy}.
More recently, the full NNLO corrections to the single inclusive distributions
in top quark pair production have been computed~\cite{Czakon:2016dgf,Czakon:2015owf,Czakon:2014xsa}.
Differential NNLO results are available for the rapidity of the top quark and the top pair 
system, $y_t$ and $y_{t\bar{t}}$, the transverse 
momentum of the top quark, $p_T^t$, and the invariant mass of the top 
pair $m_{t\bar{t}}$, though not for other variables such as $p_T^{t\bar{t}}$ since
these vanish at leading order.

Moreover, when differential distributions probe the TeV region, electroweak corrections
(including taking into account photon--initiated processes,
see Sect.~\ref{sec:QED.photon}) also become relevant, and need to be included
in the theoretical calculations.
In~\cite{Czakon:2017wor} (see also~\cite{Pagani:2016caq}),
the NNLO QCD calculations were combined with the state--of--the--art NLO
EW corrections, in the latter case including not only the $\mathcal{O}\lp \alpha_s^2
\alpha\rp$ but also the $\mathcal{O}\lp \alpha_s\alpha^2\rp$
and $\mathcal{O}\lp \alpha^3\rp$ contributions.
This study showed that an accurate description of the tails of the kinematical
distributions, such as the high--$p_{T}^t$ and high--$m_{t\bar{t}}$ regions,
must include NLO EW corrections.

An important limitation of the calculations discussed above is that they are restricted to stable
top quarks.
However, when experimental measurements
are presented at the top quark level, they are  extrapolated from the fiducial cross sections
using some theoretical model, thus possibly biasing the result by an amount
which is difficult to quantify.
Ideally, one would like a fully differential calculation with NNLO corrections
included both for production and decay, in order to directly compare
with experimentally observable quantities.
An important milestone in this respect
was the recent calculation of top quark pair--production and decay~\cite{Gao:2017goi} which
provides predictions for observables constructed from top quark leptonic and b--tagged jet final states,
based on an approximation to the exact NNLO corrections for production and the exact NNLO corrections
to the decay.

Concerning the tools for the inclusion of top quark differential data into PDF fits,
there exist two basic approaches.
The first one is based on computing {\tt APPLgrids} for the NLO calculation using
either {\tt MCFM} or {\tt Sherpa} (see also Sect.~\ref{subsec:fast}), and then supplementing these with the
NNLO/NLO bin--by--bin $K$--factors from~\cite{Czakon:2016dgf,Czakon:2015owf}.
An improved strategy has been made feasible by the recent availability
of {\tt FastNLO} tables~\cite{Czakon:2017dip} that allow the efficient calculation
of NNLO top quark pair distributions for arbitrary PDF sets and input $\alpha_s(m_Z)$ values.
The latter option provides a more precise evaluation of the in principle PDF--dependent NNLO corrections, although as shown explicitly in~\cite{Czakon:2016olj}
the dependence of these $K$--factors on the PDF set is very small.

\subsubsection*{Impact on PDFs}\label{sec:datatheory.top.impact}

The availability of the NNLO calculation of the total cross sections
for top quark pair production 
has made it possible to include top quark data
from the Tevatron and the LHC consistently into a
NNLO PDF fit.
By applying Bayesian reweighting to NNPDF2.3, it was shown in Ref.~\cite{Czakon:2013tha}
that top quark data could reduce the PDF uncertainties in the large--$x$
gluon by up to 20\% for $x\simeq 0.2$ (see also previous related work in~\cite{Beneke:2012wb}).
Several other global fits, such as ABMP16 and MMHT14, also include total
$t\bar{t}$ cross sections in their baseline fits.
While these results provided an encouraging indication of the PDF constraining potential
of $t\bar{t}$ production, the full exploitation of this potential clearly required
the use of differential distributions.

The impact of the $\sqrt{s}=8$ TeV top quark
pair differential data from ATLAS and CMS at 8 TeV
on the NNPDF3.0 analysis has been quantified in~\cite{Czakon:2016olj}.
Here, it was shown that the constraints on the large--$x$ gluon were at this point competitive with
those provided by inclusive jet production, despite the
smaller number of experimental data points. See also~\cite{Guzzi:2014wia} for
related work based on approximate NNLO calculations.
An important result of the investigations of~\cite{Czakon:2016olj} was that the constraints from the
normalized distributions were in general superior to those from their absolute
counterparts, most likely because of the cancellation of systematic uncertainties
that takes place in this case.
In addition, top quark differential distributions at 8 TeV from the LHC have been included
in the recent NNPDF3.1 global analysis, and other groups have also studied the impact
of this data on their PDF fits in a preliminary form.

%%%%%%%%%%%%%%%%%%%%%%%%%%%%%%%%%%%%%%%%%%%%%%%%%%%%%%%%%%%%%%%%%%%%%

\begin{figure}[t]
\begin{center}
  \includegraphics[scale=0.26,angle=-90]{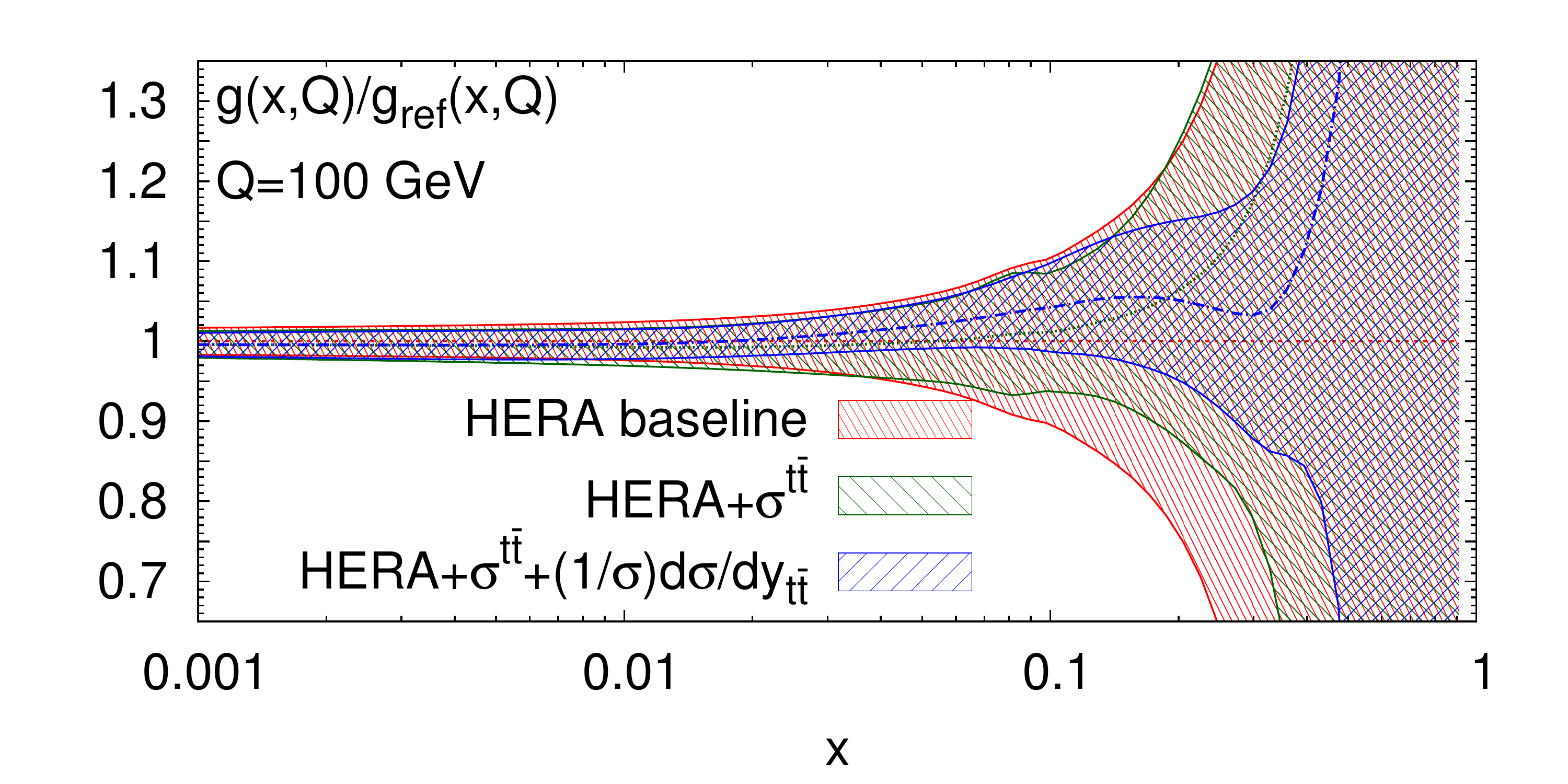}
  \includegraphics[scale=0.26,angle=-90]{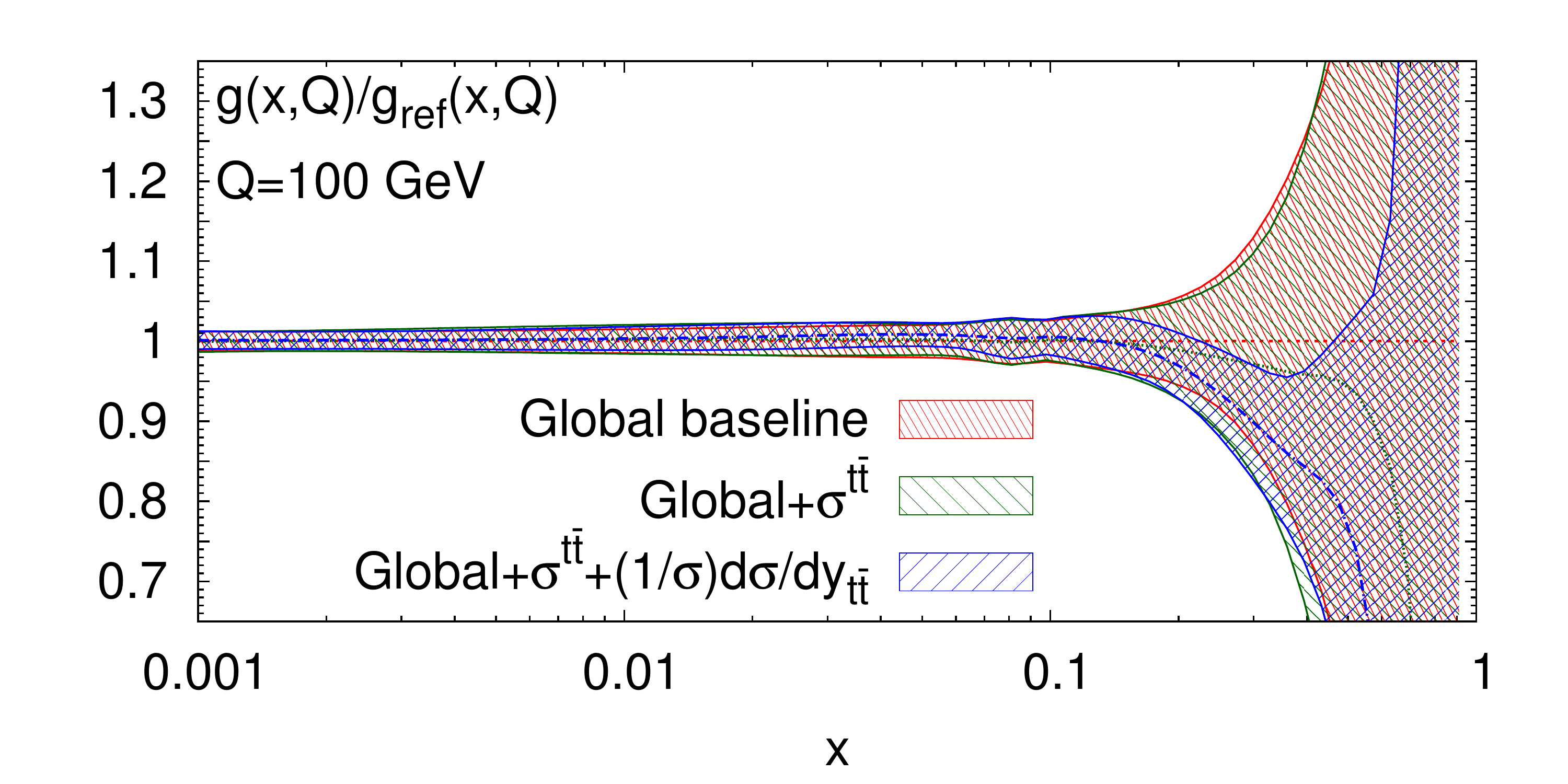}
   \caption{\small The impact of the LHC 8 TeV inclusive
   top quark pair data on the gluon PDF~\cite{Czakon:2016olj}. The results with the total cross section data only, and with the normalized $y_{t\overline{t}}$ distribution included 
   in addition are shown, relative to HERA--only (Left) and global fit (Right) baselines is shown.\label{fig:ttbar-impact}
  }
\end{center}
\end{figure}

%%%%%%%%%%%%%%%%%%%%%%%%%%%%%%%%%%%%%%%%%%%%%%%%%%%%%%%%%%%%%%%%%%%%%

A challenge in the study of~\cite{Czakon:2016olj} was the observed tension between some of the
ATLAS and CMS distributions, such as $m_{t\bar{t}}$, which prevented their simultaneous inclusion
in the global fit.
While the underlying cause of these discrepancies is still under investigation,
this issue was avoided by identifying pairs of distributions which could be fitted
with good quality at the same time and that exhibited comparable constraining power.
Further investigations of this issue, including 13 TeV data and comparisons between data
and theory in terms of lepton and jets observables, should be able to
shed more light on the origin of such tension.
   
   In order to illustrate the impact of the top quark pair production data on the large--$x$ gluon, in Fig.~\ref{fig:ttbar-impact} we should how the PDF uncertainties and central value of the gluon are affected by the inclusion of LHC 8 TeV data, taken from the above study. In particular, we show the results with only the total cross section data, and with the normalized $y_{t\overline{t}}$ distribution data included in addition, relative to HERA--only (left panel) and global (right panel) baseline fits. We can see that the impact of the total cross section data is moderate, in particular for the global fit. On the other hand, when the differential distribution data is included one clearly sees that the impact  is much more significant, highlighting the
   increase in constraining power of the differential distributions in comparison
   to the total cross section data, in particular in the large--$x$ region, where
   PDF uncertainties can be reduced by more than a factor of two.

\subsubsection*{Single top production}\label{sec:datatheory.top.single}

\begin{figure}[t]
\begin{center}
  \includegraphics[scale=0.40]{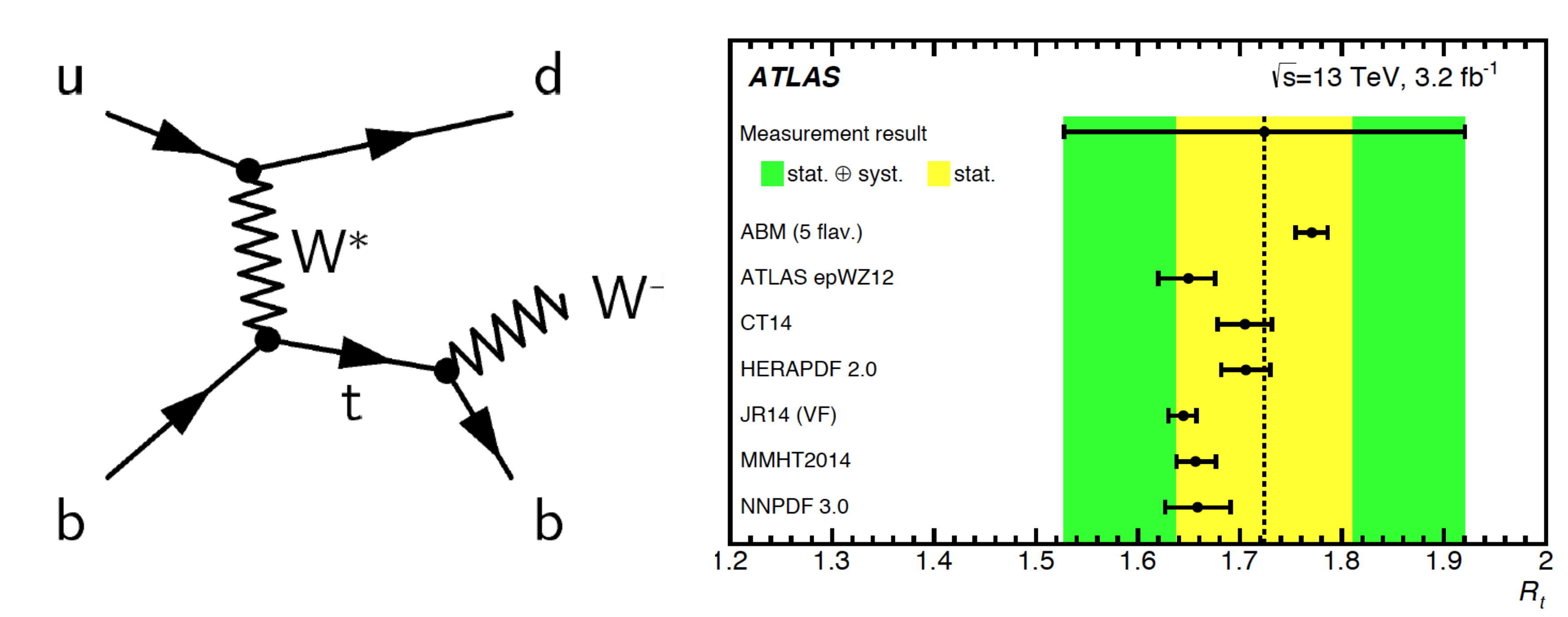}
   \caption{\small Left plot: one of the Feynman diagrams for single top production
   at leading order, illustrating its sensitivity to the $b$ quark
   PDF.
   Right plot: comparison of the theoretical predictions for the ratio $R_t=\sigma_t/\sigma_{\bar{t}}$
   from different PDF sets and the corresponding ATLAS measurements at $\sqrt{s}=13$ TeV
   from~\cite{Aaboud:2016ymp}.
    \label{fig:singletop}
  }
\end{center}
\end{figure}

In addition to top quark pair production, single top
production can also provide in principle useful PDF--sensitive information.
Such a process can proceed via the scattering of a bottom
quark with a light quark, see Fig.~\ref{fig:singletop} (left) for a typical diagram,
and will therefore provide information about the
$b$ quark PDF.
In addition, due to the presence of the $b$ quark in the initial state, it provides
an important testing ground for the different heavy quark flavour
schemes used in the calculation, analogously to those described in Sect.~\ref{sec:global.heavyq}
for the case of DIS structure functions.
That is, it is possible to use
a $n_f=4$ massive scheme, a $n_f=5$ massless scheme, or a matched scheme interpolating between
the two, see the discussion in Refs.~\cite{Maltoni:2012pa,Forte:2015hba}.

State of the art calculations of this process are based on NNLO QCD theory both
for the total cross sections and for differential
distributions~\cite{Brucherseifer:2014ama,Berger:2016oht}, and
LHC measurements at 8 TeV and 13 TeV of total cross sections (including ratios
of top to anti--top production) as well as
single inclusive distributions are already available~\cite{Aaboud:2017pdi,Aaboud:2016ymp}, although
some of them only in preliminary form.

%%%%%%%%%%%%%%%%%%%%%%%%%%%%%%%%%%%%%%%%%%%%%%%%%%%%%%%%%%%%%%%%%%%%%
%%%%%%%%%%%%%%%%%%%%%%%%%%%%%%%%%%%%%%%%%%%%%%%%%%%%%%%%%%%%%%%%%%%%%

Moreover, since the production of top and antitop quarks  is generated by different initial--state partons, cross section ratios such as $R_t\equiv \sigma_t/\sigma_{\bar{t}}$
can provide important information on the quark flavour separation of the proton,
specifically in the ratio $u/d$ between valence up and down
quarks at large-$x$.
To illustrate this point, we show in Fig.~\ref{fig:singletop} (right)
a comparison of the theoretical predictions for the $R_t$ ratio
   from different PDF sets and the corresponding ATLAS measurements at $\sqrt{s}=13$ TeV
   from~\cite{Aaboud:2016ymp}.
 While experimental uncertainties are still large, due to the limited statistics, we can see that
 the measurement may eventually become sensitive to differences between PDF sets.

In addition, similar comparisons could also be performed for differential distributions,
 either at the level of top kinematic variables or observable quantities
 constructed from leptons and $b$--jets.
 In the case of the ATLAS 8 TeV measurements~\cite{Aaboud:2016ymp}, these distributions
 are provided including the full experimental covariance matrix, and therefore
 all the ingredients are available in order to quantify for the first time
 the impact of the LHC
 single top production data on the PDFs.

\subsection{Charm production in $pp$ collisions}\label{sec:datatheory.charm}
In this section we discuss the impact of open $D$ meson production at hadron colliders on the gluon at small $x$.

\subsubsection*{PDF sensitivity}\label{sec:datatheory.charm.sensitivity}

The production of charmed mesons at hadron colliders
is dominated by the $gg\to c\overline{c}$ subprocess, and therefore
it provides a sensitive probe to the gluon PDF at small-$x$.
In particular, the forward measurements from the LHCb experiment
at 5, 7, and 13 TeV~\cite{Aaij:2013mga,Aaij:2016jht,Aaij:2015bpa}
provide information on the gluon at values of
$x$ as small as $x\simeq 10^{-6}$, well below the kinematic
reach of the HERA structure function data, and thus in a region
where PDF uncertainties are large due to the limited
amount of experimental information available.

This very small-$x$ region is not only important for the description of soft
and semi-hard QCD processes~\cite{Skands:2014pea}, it is also crucial for
the calculations of signal and
background processes~\cite{Gauld:2015kvh,
CooperSarkar:2011pa,Garzelli:2016xmx} for
ultra-high energy neutrino astrophysics.
In the latter case, the small-$x$ gluon is relevant both for the calculation
of signal event rates, via the interaction cross-section between UHE
neutrinos and target nucleons (ice or water), as well as for the calculation
of the rates for the dominant background process, the production of
charm quarks in cosmic ray collisions in the atmosphere which then
decay into so--called `prompt' neutrinos and which dominate the
atmospheric neutrino flux at high energies.
With this motivation, various groups have studied the impact
of the LHCb charm measurements on the gluon
PDF using different theoretical
frameworks~\cite{Gauld:2016kpd,Cacciari:2015fta,Zenaiev:2015rfa,Zenaiev:2016kfl}, as we discuss in more detail below.

\subsubsection*{Experimental data}\label{sec:datatheory.charm.data}

The unique forward coverage of the LHCb experiment allows the very small-$x$ region to be accessed, by means of the production of low-mass
final states such as $D$ mesons and $J/\Psi$ mesons.
LHCb has presented measurements
of $D$ meson production at three
center of mass energies, $\sqrt{s}=5,7$ and
13 TeV~\cite{Aaij:2013mga,Aaij:2016jht,Aaij:2015bpa}.
In particular, these datasets
are available double differentially in transverse momentum ($p_T^D$) and
rapidity ($y^D$) for a number of final states such as
$D^0, D^+, D_s^+$ and $D^{*+}$, which also
contain the contribution from charge-conjugate states.
The availability of the cross-sections for the same
processes at different values of $\sqrt{s}$ is particularly
appealing for PDF studies, since many experimental and theoretical
systematic uncertainties cancel in their
ratios~\cite{Mangano:2012mh}.

The original LHCb measurements of charm production were affected
by an experimental issue, whose effects were particularly
marked in the central rapidity region.
This also affected the $B$ meson production cross-section,
where QCD calculations were far off the LHCb data for
$\eta_B\simeq 2.0$~\cite{Gauld:2017omh}.
The issue has since been corrected, and the LHCb collaboration
has presented revised versions of their measurements, which
exhibit now an improved agreement with the theoretical
predictions.
Note that this issue did not affect the PDF interpretation of the
LHCb $D$ meson data, for which the dominant sensitivity comes
from the region with forward rapidities, which corresponds  to the small-$x$ PDF region.

\subsubsection*{Theoretical calculations and tools}\label{sec:datatheory.charm.theory}

The calculation of charm production at NLO was first performed almost three decades ago~\cite{Nason:1987xz,Nason:1989zy}.
This fixed-order calculation can be matched to NNLL
resummation~\cite{Cacciari:1998it} using a GM-VFN scheme, although
these corrections are numerically small in the kinematic region
accessible to the LHCb measurements.
The parton-level charm production cross sections then
need to be corrected for hadronization into the experimentally observed
$D$ mesons, and this can be done either analytically, using
the heavy quark fragmentation functions extracted from LEP
data~\cite{Cacciari:2012ny}, or using hadronization models as implemented
in Monte Carlo event generators such as {\tt Pythia8}~\cite{Sjostrand:2007gs}
or {\tt POWHEG}~\cite{Alioli:2010xd}.
The recent calculation of NNLO corrections to differential
distributions for $t\bar{t}$ production in hadron collisions,
see Sect.~\ref{sec:datatheory.top},
opens up the possibility in the near future of a NNLO calculation
for charm production as well.

One important challenge in order to include the LHCb $D$ meson
measurements into a PDF fit is the large scale uncertainties
from the NLO calculation, which can be up to $\mathcal{O}(100\%)$.
In order to tackle this limitation, it is convenient to define
 different cross sections ratios in order
to cancel most of these theoretical errors.
In particular, the following two families of observables
have been advocated:
\begin{align}
  \label{eq:Obs}
N_X^{ij} &=  \frac{d^2\sigma({\rm X~TeV})}{dy_i^D d (p_T^D)_j} \bigg{/} \frac{d^2\sigma({\rm X~TeV})}{dy_{\rm ref}^D d (p_T^D)_j} \; , \\[1mm]
R_{13/X}^{ij} &= \frac{d^2\sigma({\rm 13~TeV})}{dy_i^D d (p_T^D)_j} \bigg{/} \frac{d^2\sigma({\rm X~TeV})}{dy_{\rm i}^D d (p_T^D)_j} \; , \nonumber
\end{align}
which benefit from the partial cancellation of the
residual scale dependence from missing higher-orders, while
retaining sensitivity to the gluon, as different
regions of $x$ are probed in the numerator and denominator
of these observables.
In Eq.~(\ref{eq:Obs}), $y_{\rm ref}^D$ is a specific rapidity bin
chosen to normalize the absolute cross section measurements.
As explained in~\cite{Gauld:2016kpd,Gauld:2015yia,
Zenaiev:2015rfa,Zenaiev:2016kfl}, its value
should be such that the results of the PDF fit are stable
with respect to small variations of this choice.
Thanks to these normalizations, scale variations are greatly
reduced, and in particular it can be shown~\cite{Gauld:2016kpd}
that the entire set of LHCb $D$ meson measurements at
5, 7 and 13 TeV can be satisfactorily
described
using NLO QCD theory, with a $\chi^2$ per point of order 1.

\subsubsection*{Impact on PDFs}\label{sec:datatheory.charm.impact}

\begin{figure}[t]
\begin{center}
  \includegraphics[scale=0.46]{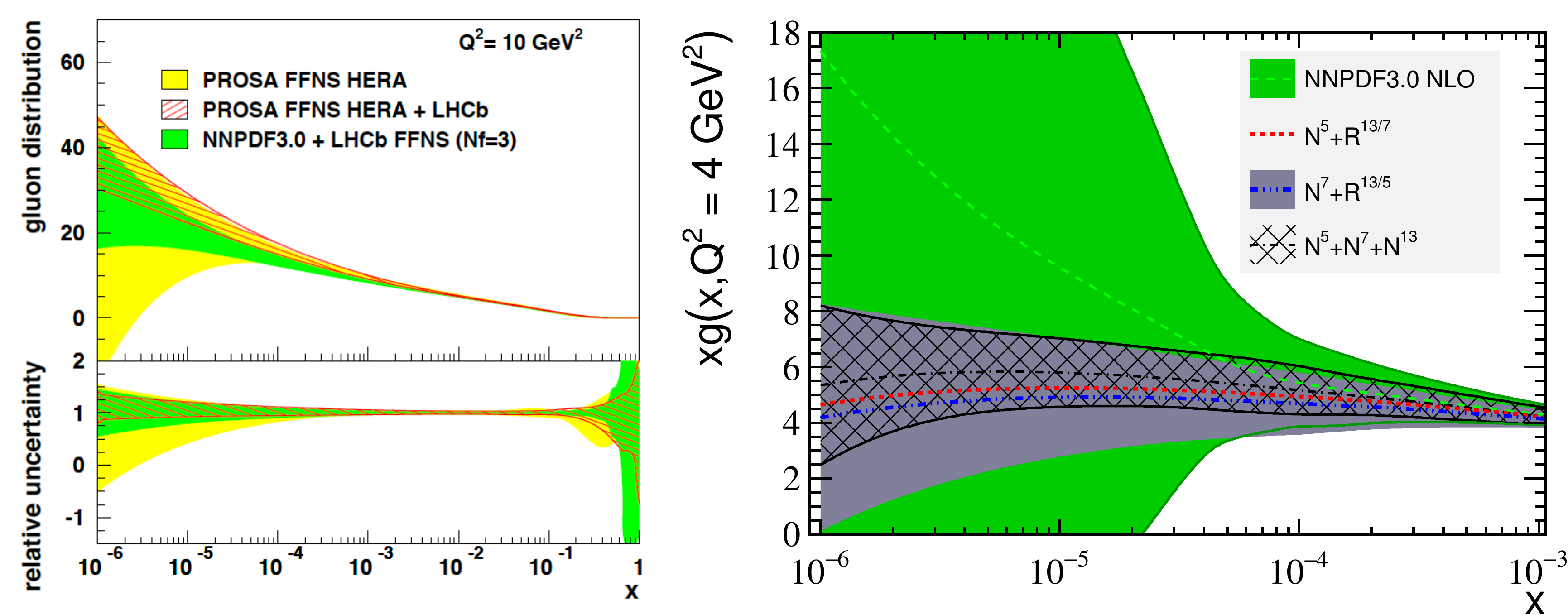}
   \caption{\small Left:
   the results of the gluon PDF from
   the PROSA fit of HERA structure functions and
   LHCb $D$ meson 7 TeV data in the FFNS~\cite{Zenaiev:2015rfa}, compared
   with the corresponding fit based on the
   NNPDF3.0 framework~\cite{Gauld:2015yia}.
Right: comparison between the small-$x$ gluon
   at $Q^2=4$ GeV$^2$
   in NNPDF3.0 with the corresponding result after different
   combinations of the charm production data at
   LHCb have been included in the fit, from~\cite{Gauld:2016kpd}.
   Specifically, we show here the central value and one-sigma PDF uncertainty
   bands for the $N^7+N^{13/5}$ and the $N^5+N^7+N^{13}$
   combinations, as well as the central value for the
   $N^5+N^{13/7}$ case, see text for more details.
    \label{fig:gPDF_rwgt}
  }
\end{center}
\end{figure}

In order to illustrate the impact of the LHCb open charm
production data on the small-$x$ gluon,
in Fig.~\ref{fig:gPDF_rwgt} we show
the results of the gluon PDF from
   the PROSA fit, based on the HERA structure functions and
   LHCb $D$ meson 7 TeV data in the FFNS~\cite{Zenaiev:2015rfa}, and the corresponding fit based on the
   NNPDF3.0 framework~\cite{Gauld:2015yia}.
   Comparing to the result of the HERA-only fit, we see that
   there is a marked reduction of the gluon PDF uncertainties
   down to $x\simeq 10^{-6}$.
   We also find that the PROSA and NNPDF results agree
   reasonably well with each other, both in terms
   of the central value and the size of the gluon PDF
   uncertainties at small $x$.

Next, in Fig.~\ref{fig:gPDF_rwgt} we show
a comparison between the small-$x$ gluon
   at $Q^2=4$ GeV$^2$
   for NNPDF3.0 with the corresponding result after different
   combinations of the LHCb charm production data at 5, 7 and 13 TeV
    have been included in the fit~\cite{Gauld:2016kpd}.
   We show the central value and one-sigma PDF uncertainty
   bands for the $N^7+R^{13/5}$ and the $N^5+N^7+N^{13}$
   combinations, as well as the central value for the
   $N^5+R^{13/5}$ case, see Eq.~(\ref{eq:Obs}) for the observable
   definitions.
   We can see that the charm data reduces the uncertainty on the small-$x$ gluon
   by up to almost an order of magnitude at $x\simeq 10^{-6}$.
   One can also observe that the central value of the
   resultant small-$x$ gluon is robust with respect to
   the specific choice of LHCb $D$ meson observables
   used in the fit.

%%%%%%%%%%%%%%%%%%%%%%%%%%%%%%%%%%%%%%%%%%%%%%%%%%%%%%%%%%%%%%%%%%%%%
%%%%%%%%%%%%%%%%%%%%%%%%%%%%%%%%%%%%%%%%%%%%%%%%%%%%%%%%%%%%%%%%%%%%%

The results of Fig.~\ref{fig:gPDF_rwgt} highlight that forward
$D$ meson production allows a precision determination
of the small-$x$ gluon PDF, with implications from UHE neutrino
astrophysics to future high-energy colliders (see also the
discussion of Sect.~\ref{sec:fcc}).
Furthermore, from the technical point of view, the inclusion
of these data into the next generation of global
PDF fits is facilitated by the availability of the
{\tt aMCfast} interface, which allows fast grids
for NLO calculations matched to parton showers to be generated, as required
for this process.

\subsection{$W$ production in association with
charm quarks}\label{sec:datatheory.Wcharm}
\begin{figure}[t]
\begin{center}
  \includegraphics[scale=0.45]{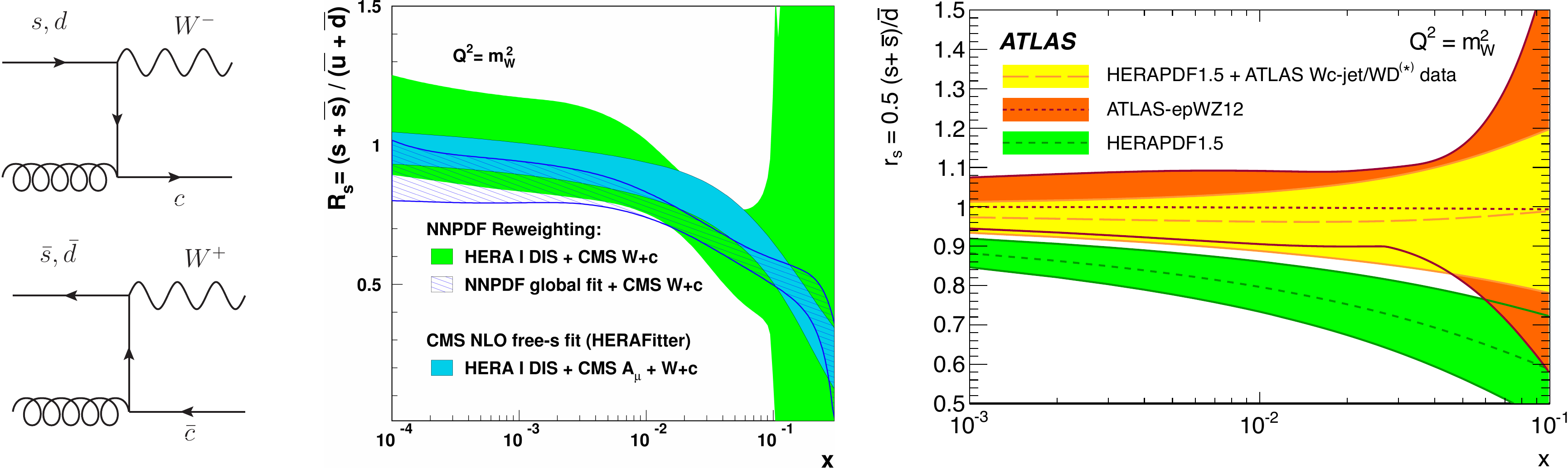}
   \caption{\small Left: the LO diagrams for $W$+charm production
   in $pp$ collisions.
   Center: the strangeness ratio $R_S=(s+\bar{s})/(\bar{u}+\bar{d})$
   at $Q=m_W$ comparing the results of the {\tt HERAfitter} analysis
   from~\cite{Chatrchyan:2013uja}, based on the CMS $W+c$ measurements,
   with the corresponding results based on the NNPDF reweighting
   method.
   Right: the strangeness ratio $r_S=0.5(s+\bar{s})/\bar{d}$,
   comparing the ATLAS-epWZ12 fit (which included the ATLAS 2010
   $W,Z$ inclusive rapidity distributions) with a new fit including
   the ATLAS $W+D$ measurements~\cite{Aad:2014xca}.
   See text for more details.
    \label{fig:wcharm}
  }
\end{center}
\end{figure}

The production of $W$ bosons in association with $D$ mesons
is a direct probe of the strange PDF~\cite{Stirling:2012vh}.
As shown in the left panel of Fig.~\ref{fig:wcharm}, at leading
order the $W$ + charm process is proportional to the proton
strangeness, and therefore for a long time measurements of this process have been advocated~\cite{Baur:1993zd} as a way to provide direct information
on the strange content of the proton.
Moreover, by taking ratios or asymmetries
of the $W^++\bar{c}$ and
$W^-+c$ differential distributions, it is also possible
to extract information on the strangeness asymmetry $s-\bar{s}$.

During Run I, the ATLAS, CMS, and LHCb collaborations have measured the production of $W$ bosons in association
with $D$ mesons using different final
states~\cite{Chatrchyan:2013uja,Aad:2014xca,Aaij:2015cha}.
In the CMS case~\cite{Chatrchyan:2013uja},
the experimentally
accessible $W+D$ cross sections were unfolded to the corresponding
$W+c$ parton--level cross sections, facilitating the comparisons
with theoretical calculations.
On the other hand, the ATLAS experiment has presented
their results~\cite{Aad:2014xca}
in terms of final--state
quantities such as $D$ mesons and $c$--tagged jets.
While the two types of measurement provide in principle
comparable information, the interpretation of the latter
is somewhat more delicate since PDF fits are typically based
on parton--level (as opposed to hadron--level) calculations.
However, this is no longer a fundamental limitation as using
the {\tt aMCfast}~\cite{amcfast}
interface it is possible to use NLO+PS calculations
directly into a PDF analysis, see Sect.~\ref{subsec:fast}.

%%%%%%%%%%%%%%%%%%%%%%%%%%%%%%%%%%%%%%%%%%%%%%%%%%%%%%%%%%%%%%%%%%%%%
%%%%%%%%%%%%%%%%%%%%%%%%%%%%%%%%%%%%%%%%%%%%%%%%%%%%%%%%%%%%%%%%%%%%%

In order to illustrate the impact of the LHC $W+D$ measurements
in the strange PDF, in the central panel
of Fig.~\ref{fig:wcharm} we show the
strangeness ratio $R_S=(s+\bar{s})/(\bar{u}+\bar{d})$
   at $Q=m_W$, comparing the results of the {\tt HERAfitter} analysis
   from~\cite{Chatrchyan:2013uja}, based on the CMS $W+c$ measurements
   added to the HERA--I inclusive DIS data,
   with the corresponding results based on the NNPDF reweighting
   method.
  We can see that the CMS $W+c$ data indeed allow
   $s(x,Q^2)$ to be pinned down with good precision, and that the central value
  is consistent with the global fit results, which favours
  a suppressed strangeness in comparison to the $\bar{u}$ and
  $\bar{s}$ light quark sea.
  For instance, one sees that the CMS data prefers $R_S\simeq 0.6$
  at $x\simeq 0.1$, though $R_s$ tends to 1 as we move towards
  smaller values of $x$.

In Fig.~\ref{fig:wcharm} we also show
the strangeness ratio $r_S=0.5(s+\bar{s})/\bar{d}$,
   comparing the ATLAS--epWZ12 fit, which included the ATLAS 2010
   $W,Z$ inclusive rapidity distributions~\cite{Aad:2011dm},
   with a new fit based on
   the ATLAS $W+D$ measurements~\cite{Aad:2014xca}.
   In contrast to the CMS measurements, the ATLAS $W+D$ data favour
   a symmetric strange sea, namely $r_s\simeq 1$, which is also
   consistent with the constraints from the ATLAS 2010
   $W,Z$ inclusive rapidity
   distributions~\cite{Aad:2011dm,Aaboud:2016btc}
   (as well as with the updated analysis from the 2011 dataset).
   While it is clear from Fig.~\ref{fig:wcharm} that the ATLAS
   and CMS $W$+charm measurements have opposite pulls
   on the strange PDF, the tension is only at the 1 or 2--$\sigma$ level.

   Further measurements of this process at 13 TeV should
   be able to shed more light on whether or not this tension
   between the ATLAS and CMS measurements
   persists.
   Here, a complete fit including hadron--level
   $W$+$D$ measurements (and thus avoiding unfolding
   to the parton level) would provide the cleanest possible
   interpretation of the PDF information from this process.
   A more detailed discussion of the strange PDF, and of
   the constraints from the different strange--sensitive
   processes, is presented in Sect.~\ref{sec:structure.strange}.

\subsection{Central exclusive production}\label{sec:datatheory.CEP}
\begin{figure}[t]
  \begin{center}
 \includegraphics[width=0.36\textwidth]{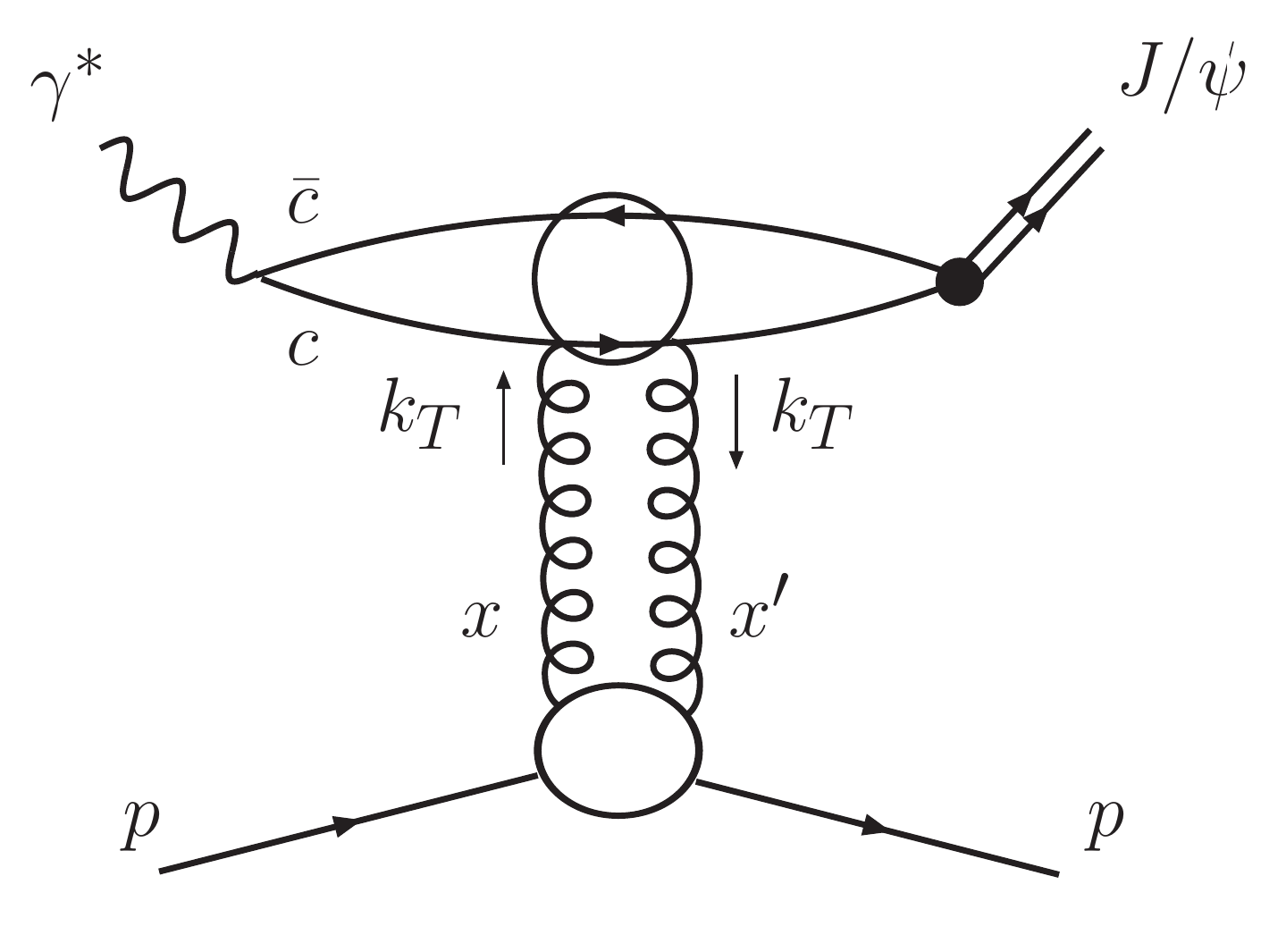}
  \end{center}
\caption{\small Photoproduction of heavy vector meson $J/\psi$~\cite{Jones:2013pga}.}\label{fig:cep}
\end{figure}

The Central Exclusive Production (CEP) process occurs when an object $X$ and nothing else is produced in a hadronic collision, while the hadrons themselves remain intact after the collision.
The photoproduction of heavy vector mesons, see Fig.~\ref{fig:cep}, is one example of such a process which has possible implications for PDF determination.
Here, one proton elastically emits a photon, while the other interacts via $t$--chansnel two gluon exchange.
This may therefore access the gluon PDF at a comparatively low scale $Q^2\sim M_V^2$ and $x\sim M_V/\sqrt{s}$, where it is so far quite poorly determined.
Indeed, the production kinematics are rather similar to those of
$D$ meson production described in the previous section.

%%%%%%%%%%%%%%%%%%%%%%%%%%%%%%
%%%%%%%%%%%%%%%%%%%%%%%%%%%%%%

LHCb has measured the exclusive production of $J/\psi$ and $\psi(2S)$ mesons at 7 TeV~\cite{Aaij:2014iea} and $\Upsilon$ production at $7$ and $8$ TeV~\cite{Aaij:2014iea}. Preliminary LHCb data on  $J/\psi$ and $\psi(2S)$ production at 13 TeV has also been reported in~\cite{LHCb:2016oce}. This uses the newly installed \texttt{HeRSCheL} shower counters, which greatly extend the rapidity coverage for vetoing additional particle production, reducing significantly the non--exclusive background.
In addition, the ALICE experiment has measured exclusive $J/\psi$ production in p--Pb collisions at $\sqrt{s}_{NN}=5.02$ TeV~\cite{TheALICE:2014dwa}.
Due to the large $Z^2$ enhancement for photon emission from the Pb ion, this is dominated by the process where the lead ion emits a photon.

While clearly the photoproduction process shown in Fig.~\ref{fig:cep} proceeds through an initial--state gluon interaction, this does not correspond to a standard inclusive process where collinear PDFs are generally introduced.
However, as discussed in~\cite{Jones:2013pga} under certain assumptions this process can be related to the standard gluon PDF and may therefore serve as a probe of it at low $x$ and $Q^2$. This analysis has subsequently been performed at NLO~\cite{Jones:2015nna} (see also~\cite{Ivanov:2004vd}) for the cases of $J/\psi$ and $\Upsilon$ production.
However, here it is found that the NLO correction is significantly larger than, and of the opposite sign to, the LO contribution, indicating a lack of perturbative stability and casting some doubt on the viability of this process as a PDF probe.

Nonetheless, work in the direction of at least partially solving this issue has been reported most recently in~\cite{Jones:2016ldq}, where the stability is shown to be improved through a judicious choice of factorization and renormalization scale, and by imposing a cut on the NLO contribution to avoid double counting. An alternative method, not investigated so far, might be to consider ratio observables similar to those described in the previous section for the case of inclusive $D$ meson production.
It is also worth noting that the perturbative stability is naturally improved somewhat by considering the production of the higher mass $\Upsilon$ meson. 
Future studies will determine whether or not central exclusive production can
then be reliably added to the toolbox of global PDF analyses.

\subsection{Fast interfaces to (N)NLO calculations}\label{sec:datatheory.fast}
\label{subsec:fast}

Given the highly CPU time consuming nature of global PDF fits, the direct evaluation of
the lengthy (N)NLO hadronic cross sections during  the fit itself is not feasible.
For this reason, until around 2008, fits included hadronic data using LO
hadronic cross sections supplemented by bin--by--bin $K$-factors, defined
as
\be
K_{\rm NLO}^i\equiv\frac{\sigma_i^{\rm NLO}}{\sigma_i^{\rm LO}} \; ,
\ee
using the same PDF set in the numerator and in the denominator.
To ensure consistency of the procedure, these $K$ factors can be computed iteratively
until convergence was achieved.
However, this approximation is known to have several deficiencies, the most
important one being the reduced sensitivity to those partonic initial states
that only enter the cross-section at NLO and beyond.

In order to improve upon this unsatisfactory situation, the method
of fast interfaces was proposed.
In these methods, the most CPU time consuming part of a (N)NLO calculation, namely the evaluation
of the partonic matrix elements by integrating over a very
large number of events, is precomputed {\it a priori} using
a complete interpolation basis for the input PDFs.
This way, the hadronic cross-sections can be reconstructed {\it a posteriori} by means
of a very efficient matrix multiplication of the PDFs evaluated in a grid of $(x,Q)$ points
and the precomputed partonic matrix elements at the same grid points.
These tools have become very popular within the community,
and are used in the majority of
modern PDF fits.

Following a common approach, two main tools have been developed, {\tt APPLgrid}~\cite{Carli:2010rw} and {\tt FastNLO}~\cite{Wobisch:2011ij}.
The former is interfaced to the {\tt MCFM}~\cite{Boughezal:2016wmq}
and {\tt NLOjet++}~\cite{Nagy:2001fj} programs.
The latter is interfaced to {\tt NLOjet++} as well as to
the NNLO $t\bar{t}$ calculation of Ref.~\cite{Czakon:2016dgf}.
Therefore, using these tools, fast interfaces to the most commonly used codes
in PDF fits can be constructed.
More recently, the {\tt aMCfast} interface~\cite{amcfast} to
{\tt MadGraph5\_aMC@NLO}~\cite{Alwall:2014hca} has also been developed.
Given the automated character of this code, {\tt aMCfast} allows the fast interpolation
of arbitrary NLO processes, defined by the user at run time.
It is also possible to produce fast grids for NLO calculations matched to parton shower MCs, which opens the way to include in the PDF fit hadron-level cross sections
such as $W$ boson in association with charmed mesons or forward $D$ meson production
at LHCb.

Currently, a significant amount of effort is being devoted to extending these fast
interfaces to NNLO calculations.
For instance, {\tt FastNLO} has been used to produce interpolating grids
for the NNLO calculation of top-quark pair differential distributions~\cite{Czakon:2017dip},
while both {\tt FastNLO} and {\tt APPLgrid} are being interfaced\footnote{
See for instance the talk by C.~Gwenlan at the Deep Inelastic Scattering Workshop 2017.}
to the {\tt NNLOJET} code~\cite{Currie:2016bfm,Currie:2017eqf}, which will make it possible to generate NNLO grids
for inclusive jets, dijets, and vector boson in association with jets.

To describe 
the basic strategy of these fast interpolation methods, we use the 
 {\tt APPLgrid} notation for concreteness, but the general method
is similar in the {\tt FastNLO} and {\tt aMCfast} cases.
All these methods
are based on representing the PDFs in $(x,Q^2)$ by means
of a suitable interpolation basis, computing a physical cross-section for a basis PDF set,
and then reconstructing the same observable {\it a posteriori} using
an arbitrary PDF set.
One expands an arbitrary PDF $f(x,Q^2)$ in terms of a suitable
basis of interpolating polynomials as follows
\be
\label{eq:interpolatingpolynomials}
f(x,Q^2)=\sum_{i=0}^n\sum_{j=0}^{m}f_{k+i,\kappa+j}I_i^{(n)}\lp \frac{y(x)}{\delta y}-k\rp
I_j^{(m)}\lp \frac{\tau(Q^2)}{\delta \tau}-\kappa\rp \; ,
\ee
where $n$ and $m$ are the interpolation orders in $x$ and $Q^2$ respectively, 
$y(x)=\ln 1/x+a(1-x)$ and $\tau(Q^2)=\ln\left(\ln Q^2/\Lambda^2\right)$, and $I_i^{(n)},I_j^{(m)}$
are interpolating functions, for instance Lagrange interpolating polynomials,
though this expression holds generically for other choices. Here, $a$ is a free parameter that controls the density of points in the large $x$ region.
In Eq.~(\ref{eq:interpolatingpolynomials}),
$k$ and $\kappa$ are the grid nodes associated with a given values
of $x$ and $Q^2$, and are defined as
\be
k(x)={\rm int}\left(\frac{y(x)}{\delta y}-\frac{n-1}{2}\right),\,\kappa(Q^2)=
{\rm int}\left(\frac{\tau(Q^2)}{\delta \tau}-\frac{m-1}{2}\right),
\ee
with ${\rm int}(u)$ being the largest integer that is smaller than $u$.

After the representation of the PDFs as in
Eq.~(\ref{eq:interpolatingpolynomials}) has been constructed, we need
to evaluate cross sections using the same interpolation basis.
Let us consider here
for simplicity a hypothetical DIS structure function $F$ that receives contributions
from a single flavour.
This NLO cross section may be computed by means of a MC program that generates a large
number $N$ of events, each one with weight $\omega_m$ and with associated values 
$x_m$ and $Q_m^2$.
If $p_m$ is the order of $\alpha_s$ for this specific event, the total cross-section
can be written as
\be
\label{eq:applgrid1}
F=\sum_{t=1}^N\omega_t \lp \frac{\alpha_s(Q^2_t)}{2\pi}\rp^{p_t} f(x_t,Q^2_t) \; .
\ee
The fast interpolation grid can then be constructed by, instead of computing $F$ as in
Eq.~(\ref{eq:applgrid1}), introducing a weight grid $W_{i_y,i_\tau}^{(p)}$, and for each event
only a fraction of the grid nodes is updated according to the expression
\be
W_{k+i,\kappa+j}^{(p_t)}\to W_{k+i,\kappa+j}^{(p_t)}+\omega_t
I_i^{(n)}\lp \frac{y(x_t)}{\delta y}-k\rp
I_j^{(m)}\lp \frac{\tau(Q^2_t)}{\delta \tau}-\kappa\rp \;.
\ee
Conceptually, the weight grid $W_{i_y,i_\tau}^{(p)}$ is the equivalent of computing
the structure function $F$ for a given combination of interpolating polynomials
rather than for the original parton distributions.

An important factor here is that the most CPU time intensive computation, namely the
calculation of the MC weights $\omega_m$, only needs to be done once to fill
the grid  $W_{i_y,i_\tau}^{(p)}$, and the PDF can be decided {\it a posteriori}
at virtually no extra computational cost.
Indeed, it can be shown that the structure function can be reconstructed
{\it a posteriori}
from the weight grid using the following expression:
\be
\label{eq:applgrid2}
F = \sum_p\sum_{i_y}\sum_{i_\tau}W_{i_y,i_\tau}^{(p)} \lp \frac{\alpha_s(Q^2_{i_\tau})}{2\pi}\rp^p
f(x_{i_y},Q^2_{i_\tau}) \; .
\ee
In other words, the only information which is needed is the value of the
PDFs and the strong coupling at the grid nodes $i_y,i_{\tau}$.
The method can be straightforwardly generalised to hadron-hadron collisions and to a generic composition
of the initial parton state.
In proton-proton collisions, the analog of Eq.~(\ref{eq:applgrid2}) is given by
\be
\label{eq:applgrid3}
\sigma = \sum_p\sum_{l=0}^{n_{\rm sub}}\sum_{i_{y_1}}\sum_{i_{y_2}}\sum_{i_\tau}W_{i_{y_1},i_{y_2},i_\tau}^{(p)(l)} \lp \frac{\alpha_s(Q^2_{i_\tau})}{2\pi}\rp^p
\mathcal{L}^{(l)} \lp x_{1,i_{y_1}},x_{2,i_{y_2}}, Q^2_{i_\tau} \rp \, ,
\ee
where we have indicated that there are $n_l$ contributing partonic subprocesses,
each with the corresponding luminosity $\mathcal{L}^{(l)}$, which depend on the cross-section
upon consideration.
The more complex the process is, the larger the number of independent PDF luminosities
that will be relevant for the calculation.
Moreover, the  extension of Eq.~(\ref{eq:applgrid3}) to NNLO calculations
is conceptually simple, although challenging in practice since NNLO codes are rather
more complex than NLO ones.

\begin{figure}[t]
\begin{center}
  \includegraphics[scale=0.46]{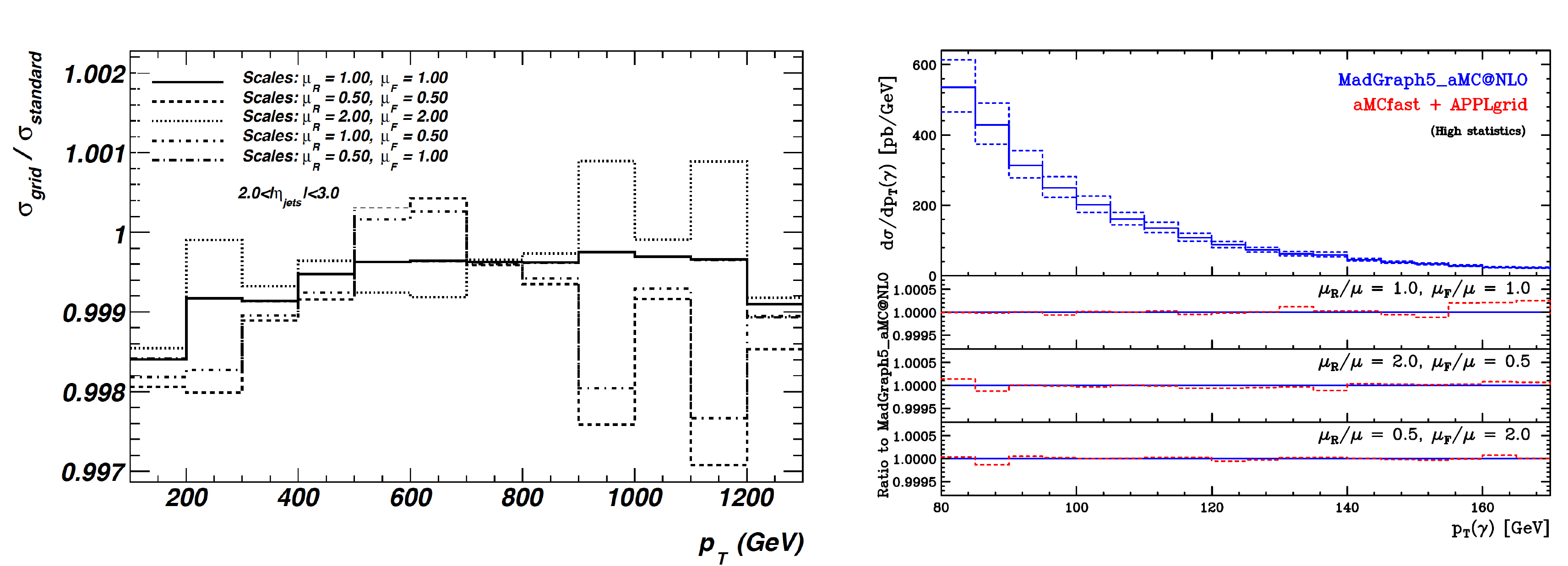}
   \caption{\small Left: the ratio between the {\tt NLOjet++} calculation
   of inclusive jet production at NLO with $\sqrt{s}=7$ TeV in the rapidity interval
   $2 \le y \le 3$ and the corresponding {\it a posteriori} calculation
   based on {\tt APPLgrid}, for different values of the factorization
   and renormalization scales.
   Right: the photon transverse momentum distribution in the $pp\to \gamma+{\rm jet}$
   process at 7 TeV, comparing the original {\tt MadGraph5\_aMC@NLO} calculation with the
   {\it a posteriori} result based on {\tt aMCfast} and {\tt APPLgrid}.
   The lower insets show the ratio between the two calculations for different
   choices of $\mu_R$ and $\mu_F$, highlighting the good agreement in all cases.
    \label{fig:fastinterfaces}
  }
\end{center}
\end{figure}

In order to illustrate the high precision that these fast interfaces can achieve,
we show two representative examples in Fig.~\ref{fig:fastinterfaces}.
First of all, we show the ratio between the {\tt NLOjet++} calculation
   of inclusive jet production at NLO for 7 TeV in the rapidity interval
   $2 \le y \le 3$ and the corresponding {\it a posteriori} calculation
   based on {\tt APPLgrid}, for different values of the factorization
   and renormalization scales.
   One sees that the differences between the original and the interpolated
   calculation are at the few permille level.
   Then we show the transverse momentum distribution of photons in the $pp\to \gamma+{\rm jet}$
   process at 7 TeV, comparing the original {\tt MadGraph5\_aMC@NLO} calculation with the
   {\it a posteriori} result based on {\tt aMCfast} and {\tt APPLgrid}.
   The lower insets show the ratio between the two calculations for different
   choices of $\mu_R$ and $\mu_F$.
   Here, we also find excellent agreement between the original and interpolated
   calculations, now at the sub-permille level.
   Note that in all these methods, the interpolation accuracy can be arbitrarily increased
   by using denser grids in $x$ and $Q^2$.

%%%%%%%%%%%%%%%%%%%%%%%%%%%%%%%%%%%%%%%%%%%%%%%%%%%%%%%%%%%%%%%%%%%%%
%%%%%%%%%%%%%%%%%%%%%%%%%%%%%%%%%%%%%%%%%%%%%%%%%%%%%%%%%%%%%%%%%%%%%

While the fast interfaces that we have just described
represent a very significant improvement in terms of CPU efficiency
in comparison to the original (N)NLO calculations, one limitation of this approach
can be seen from the master formula for proton--proton collisions, Eq.~(\ref{eq:applgrid1}):
each time that the PDF set is varied, one needs to recompute 
its values in the $(x,Q^2)$ nodes.
In a PDF fit, this means that each time the input parametrization is modified during the
iterative minimization, the  DGLAP evolution equations need to be solved again, before
the PDFs can be convoluted with the interpolated coefficient functions to obtain
the hadronic cross-section.
Therefore, even {\tt FastNLO} or {\tt APPLgrid} tables might not be fast enough
in the case of PDF fits with extremely intensive minimization algorithms, such
as the NNPDF approach, which involves training a neural network
within a $\mathcal{O}\lp 300\rp$ parameter space.

To improve upon this shortcoming, the {\tt APFELgrid} tool has recently been
developed~\cite{Bertone:2016lga}.
The goal of {\tt APFELgrid} is to combine the interpolated partonic cross-sections
provided by {\tt APPLgrid} with the DGLAP evolution factors provided by {\tt APFEL}, in a way
that hadronic cross-sections can be reconstructed from a matrix multiplication requiring
only as input the values of the PDFs at the $x$ grid nodes at the input evolution scale $Q_0$.
In this way, the need to solve the DGLAP PDF evolution
equations during the fit is bypassed.
The combination of these two ingredients
then leads to a very significant improvement in computation speed in comparison to
 Eq.~(\ref{eq:applgrid1}) without any loss of numerical accuracy, and therefore allows
 much faster PDF fits.
 Mathematically, in the {\tt APFELgrid} method  an arbitrary hadronic cross-section
 is expressed as 
 \be
 \label{eq:apfelgrid}
\sigma_{pp\to X}=\sum_{k,l}\sum_{\delta,\gamma}\widetilde{W}_{kl,\delta\gamma}f_k(x_\delta, Q_0^2)f_l(x_\gamma,Q_0)\; ,
\ee
in terms of the PDFs at the parametrization scale $Q_0$, where $k,l$ run over all active parton flavours
and $\delta,\gamma$ run over the nodes of the $x$ interpolating grid.

\begin{figure}[t]
\begin{center}
  \includegraphics[scale=0.65]{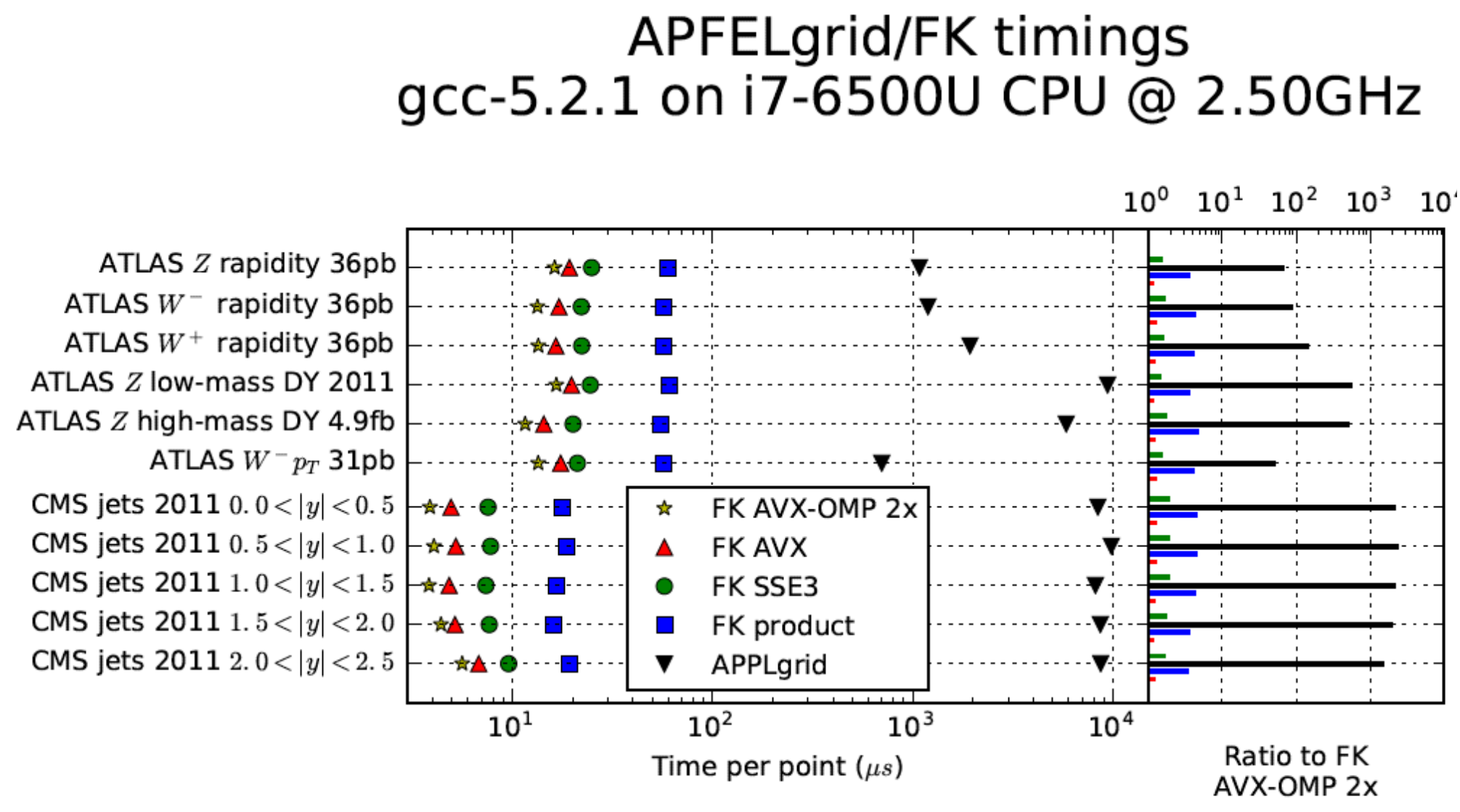}
   \caption{\small Comparison of the timings per data point between the
   original {\tt APPLgrid} computation of hadronic cross-sections,
   Eq.~(\ref{eq:applgrid2}), with the same calculation
   based on the {\tt APFELgrid} combination, Eq.~(\ref{eq:apfelgrid}), for a variety of
   LHC datasets~\cite{Bertone:2016lga}.
   We find that the improvement in computational speed is between a factor $10^2$ and a factor
   $10^3$
   depending on the specific process.
    \label{fig:apfelgrid}
  }
\end{center}
\end{figure}

To gauge the improvements in computational efficiency that can archived by this method,
in Fig.~\ref{fig:apfelgrid} we show a comparison of the timings per data point between the
   original {\tt APPLgrid} computation of hadronic cross-sections,
   Eq.~(\ref{eq:applgrid2}), with the same calculation
   based on the {\tt APFELgrid} combination, Eq.~(\ref{eq:apfelgrid}), for a variety of
   LHC datasets.
   The improvement in computational speed is between a factor 100 and a factor 1000
   depending on the specific dataset.
   %

%%%%%%%%%%%%%%%%%%%%%%%%%%%%%%%%%%%%%%%%%%%%%%%%%%%%%%%%%%%%%%%%%%%%%
%%%%%%%%%%%%%%%%%%%%%%%%%%%%%%%%%%%%%%%%%%%%%%%%%%%%%%%%%%%%%%%%%%%%%

%%%%%%%%%%
\vspace{0.6cm}
\section{Fitting methodology}
\label{sec:fitmeth}
In this section we present the main features of the global PDF analysis framework.
First, we discuss the PDF parameterization, and review the theoretical constraints
that are imposed on this, such as
the momentum and valence sum rules, as well as positivity.
Then we discuss how to quantify the agreement of data and
theory, and review various methods used in global analyses
for minimization of the figure of merit, $\chi^2$.
We  also review the different strategies available to estimate
and propagate PDF uncertainties, with emphasis on the three
most important ones, namely the Hessian, the Monte Carlo, and
the Lagrange multiplier methods.

Other related topics that are discussed in this section
include how to combine individual
PDF sets within a single set, the treatment
of theoretical parametric uncertainties,
the use of approximate methods
to estimate the impact of new experiments without redoing the PDF fit,
and the public delivery of PDFs by means of the
{\tt LHAPDF6} interface.

\subsection{PDF parametrization}\label{sec:fitmeth.PDFpara}
We start by discussing different aspects related to the parameterization
of the PDFs at the input scale $Q_0$, namely the choice of functional form, the theoretical
constraints from the momentum and valence sum rules and PDF positivity,
and the various quark flavour assumptions used in PDF fits.

\subsubsection{Choice of functional form}\label{sec:fitmeth.PDFpara.func}
In order to extract the PDFs, a particular choice for their parameterisation in $x$ at some
input scale $Q_0$ must be assumed, which can then be fit to the available data.
As described in Sect.~\ref{subsec:dglap}, given the PDFs at some reference scale $Q_0$, the
DGLAP evolution equations can be used to determine the PDFs at any other scale $Q$.
Thus the PDFs are typically parameterised at a low scale $Q_0^2 \sim 1-2\,{\rm GeV}^2$, which can then be evolved up to the scale relevant to e.g. LHC phenomenology.
These parametrizations usually adopt the generic form
\be\label{pdffunc}
xf(x,Q^2_0) = A_{f} x^{a_{f}}(1-x)^{b_{f}}I_{f}(x)\;.
\ee
The $(1-x)^{b_{f}}$ term, with $b_{f}>0$, ensures that the PDFs vanish in the elastic $x\to 1$ limit, as we would expect on basic physical grounds.
Such a form is also expected from the quark counting rules~\cite{Brodsky:1973kr}
(see also the discussion on~\cite{Ball:2016spl}).
There, in this elastic limit all the momentum is carried by the struck parton and the remaining $n_s$ quark become spectators.
An analysis of the scaling behaviour for elastic scattering then predicts $b_f=2n_s-1$, that is $b_f=3,5$ and 7 for the valence, sea and gluon PDFs, respectively.

The $x^{a_{f}}$ form dominates at low $x$; in this region, the PDFs may
be  related to the high energy parton--proton scattering amplitudes, which can be calculated using the tools of Regge theory.
This scenario predicts such a simple power--like form, with the precise value of the power $a_{f}$ being related to the leading Regge trajectory that is exchanged; for non--singlet distributions (e.g. the valence quarks) this predicts $a_f\sim 0.5$ and for singlet distributions (e.g. the gluon and the sea) this predicts $a_f\sim 0$.

We emphasise that the above discussion only corresponds to quite
general expectations (as opposed to direct
QCD predictions), which do not for example account for the scale dependence of the PDFs.
Thus while the high and low $x$ form of
Eq.~(\ref{pdffunc}) is usually adopted, in modern fits the values of the powers themselves are more generally left free where there is sufficient data to constrain them.

The $I_{f}(x)$ in Eq.~(\ref{pdffunc}) is the interpolating function, which determines the behaviour of the PDFs away from the $x\to 0$ and 1 limits, where it tends to a constant value.
This is assumed to be a smoothly varying function of $x$, for which a variety of choices have been made in the literature. The simplest ansatz, which has been very widely used, is to take a basic polynomial form in $x$ (or $\sqrt{x}$), such as
\be\label{Ipower}
I_{f}(x)=1+c_{f}\sqrt{x}+d_{f}x+...\;.
\ee
Forms of this type are for example taken by the CJ and HERAPDF groups as well as in the
MSTW08 analysis.
A similar approach, but where the polynomial enters as the exponent of a power of $x$ or a simple exponential function, are taken by ABM and earlier CT sets, respectively. 

Such a choice is appropriate for a relatively small number of parameters, say only two or three
in addition to $a_f$ and $b_f$.
However, as the precision and amount of the data included in the fit increases, it becomes essential to allow for an increasingly flexible parameterisation.
As discussed in~\cite{Pumplin:2009bb}, simply adding more parameters to (\ref{Ipower}) can quickly run into the issue that large coefficients appear, with large cancellations between the terms. This leads to an unstable $\chi^2$ minimisation and implausibly large variations in $x$ in certain regions.
This issue may be solved by instead expanding the interpolating function in terms of a basis of suitably chosen functions with the generic form
\be
I_{f}(x)=\sum_{i=1}^n \alpha_{f,i} P_i(y(x))\;,
\ee
where $y(x)$ is some simple function of $x$.
Two possible choices for the functions $P_i$ are Chebyshev and Bernstein polynomials, which are used in the MMHT14 and CT14 sets, respectively.
These are taken because each order of the polynomials is strongly peaked at different values of $y$, and hence $x$, significantly reducing the degree of correlation between the terms.
In addition, as the order is increased these tend to probe smaller scale variations in $x$, so that the smoothness requirement for $I(x)$ naturally leads to smaller coefficients $\alpha$ at higher $i$.
Thus, while formally equivalent to the simple polynomial expansion in Eq.~(\ref{Ipower}), these are much more convenient for fitting as the number of free parameters $n$ is increased. 

An alternative approach is taken by the NNPDF group. Here, the interpolating function is modelled with a multi--layer feed forward neural network (also known as
a {\it perceptron}), see Sect.~\ref{nnpdf} for more details.
In practice, this allows for a greatly increased number of free parameters, with the latest default fit having 37 per PDF, that is around an order of magnitude higher than other sets.
The form of Eq.~(\ref{pdffunc}) is still assumed, but
these are pre--processing factors that speed up the minimisation procedure and which do not in principle have to be explicitly included.
Nonetheless, the study of~\cite{Ball:2016spl} has shown that the NNPDF fit does exhibit high and low $x$ behaviour that is consistent with Eq.~(\ref{pdffunc}), providing further support for such an assumption in the choice of input PDF parametrization.

\subsubsection{Sum rules}\label{sec:fitmeth.PDFpara.sum}

The flavour quantum numbers of the proton, $uud$ with zero strangeness, are
expressed in the three number or valence sum rules,
\begin{align}
\int_0^1 {\rm d}x\,\left[u(x,Q^2)-\overline{u}(x,Q^2)\right]&=2\;,\\
\int_0^1 {\rm d}x\,\left[d(x,Q^2)-\overline{d}(x,Q^2)\right]&=1\;,\\ \label{strangesum}
\int_0^1 {\rm d}x\,\left[s(x,Q^2)-\overline{s}(x,Q^2)\right]&=0\;.
\end{align}
Thus for the valence distributions we must have $a_f>0$ for the exponents in Eq.~(\ref{pdffunc}) or these integrals will diverge.
In others words, this constraint is consistent
with the well known result that the $xf$ valence distributions vanish
as $x\to 0$.
Although not shown explicitly, a similar constraint applies to the heavy quark PDFs as to the strange PDF.
In the absence of any intrinsic heavy flavour, these are automatically satisfied.

The second moment of the sum of PDFs must also obey the momentum sum rule
\be
\label{eq:momsumrule}
\int_0^1 {\rm d}x\, x\left(\sum_{f=1}^{n_f}\left[
q_f(x,Q^2)+\overline{q}_f(x,Q^2)\right]+g(x,Q^2)+\cdots\right)=1\;,
\ee
which expresses the simple physical requirement, arising
from energy--momentum conservation,
that the total proton momentum must be equal to the sum of its constituents.
In Eq.~(\ref{eq:momsumrule}), the dots indicate possible additional
contributions from other partons in the proton, for instance from the
photon PDF $\gamma(x,Q^2)$
that should be accounted for in PDF sets with QED corrections
(see Sect.~\ref{sec:QED.photon}).
Eq.~(\ref{eq:momsumrule}) also implies that for non--valence distributions,
the exponent $a_f$ may be negative, but must be greater than -1 to avoid giving a divergent contribution to the momentum sum rule.

The above sum rules provide additional constraints on the input PDFs, and are typically applied to fix certain parameters, for example the overall normalization $A_f$ of the gluon or
of specific quark flavour combinations.
Provided these sum rules are satisfied at the input scale, it follows straightforwardly from
the structure
of the DGLAP evolution that they will be satisfied at any other scale $\mu$.
For instance, the $g\to q\overline{q}$ splitting can generate no net $q-\overline{q}$ component,
and while the DGLAP evolution reshuffles the momentum carried between the different partons, it does not generate any
violation of momentum conservation.

\subsubsection{Quark flavour assumptions}\label{sec:fitmeth.PDFpara.flavour}

Assuming that there are $n_f$ active quark flavours at the input
parametrization scale $Q_0$, there will in general be $2n_f+1$ PDFs to be parametrized and fitted
to data.
Assuming that the heavy quark PDFs are generated perturbatively,
in addition to the gluon, in many cases  the remaining 6 light quarks PDFs parametrized
are not those in the flavour basis, namely
\be
u,\quad \bar{u},\quad d,\quad \bar{d},\quad s,\quad \bar{s}\;,
\ee
but rather other  convenient linear combinations, e.g. the valence $u_V=u-\bar{u}$ and
$d_V=d-\bar{d}$ distributions, 
are often adopted.
To give one example, in addition to the gluon the MMHT14 analysis  takes as a
fitting basis the following
quark combinations:
\be
u_V,\quad d_V,\quad \overline{d}-\overline{u},\quad s+\overline{s},\quad s-\overline{s},\quad s+\overline{s}+2(u+\overline{u}+d+\overline{d})\;.
\ee
As another example, the NNPDF3.0 fit parametrizes the  PDFs at the input evolution scale
in the so--called {\it evolution basis}, defined as the eigenvectors of the DGLAP
evolution equations (see also Sect.~\ref{subsec:dglap}),
\bea
\Sigma&=& u+\bar{u}+d+\bar{d}+s+\bar{s} \, \nonumber \\
T_3&=& u+\bar{u} - d - \bar{d}  \, ,\nonumber \\
\label{eq:PDFevolutionbasis}
T_8&=& u+\bar{u} + d + \bar{d} -2s-2\bar{s}  \, , \\
V&=& u-\bar{u}+d-\bar{d}+s-\bar{s} \, \nonumber \\
V_3&=& u-\bar{u} - d + \bar{d}  \, ,\nonumber \\
V_8&=& u- \bar{u} + d - \bar{d} -2s + 2\bar{s}  \, ,\nonumber 
\eea
in addition to the gluon PDF.
However, as any particular basis can be trivially related to another by a linear transformation,
the resulting physics should not depend on these choices.
On the other hand, different flavour assumptions do often lead to different results in regions
with scarce experimental constraints, such as the large--$x$ region.

Historically, the strange quark has been rather
less well determined than the $u$ and $d$ quark PDFs,
and indeed in many earlier fits $s(x,Q^2)$ was not fitted from
data but rather was fixed according to 
\be
s=\overline{s}\propto  \lp \overline{u}+\overline{d} \rp \;.
\ee
Such a choice is still taken in the CJ15 and HERAPDF fits, due to the more restricted data set. With the increase in available data, the total strangeness $s+\overline{s}$ is now freely parameterised in all global fits.
While the sum rule Eq.~(\ref{strangesum}) requires there to be no overall strangeness in the proton, at a given $x$ value there is no requirement for the $s_V=s-\overline{s}$ distribution to vanish, and indeed non--perturbative approaches such as the `meson cloud model'~\cite{Signal:1987gz} predict a non--zero strange asymmetry.
However, the strange difference $s-\overline{s}$ is generally quite poorly determined and still broadly consistent with zero within current uncertainties. From the latest global fits, only  MMHT14 and the NNPDF3 family fit the strange difference $s_V$, while for all other sets it is still assumed that $s=\overline{s}$.
Note also that at NNLO, even if $s_V$ is set to zero at the initial evolution scale,
a non--zero strangeness (as well as charm and bottom) asymmetry will be generated dynamically by the DGLAP
evolution equations~\cite{Catani:2004nc}.

The above discussion assumes that the charm PDF $c(x,Q^2)$ is 
generated by perturbative $g\to c\overline{c}$ splittings, in which case it is
completely determined by the DGLAP evolution equations in
terms of the light quark and gluon PDFs above the charm mass $m_c$, while it vanishes for $Q<m_c$.
If the charm PDF is instead fitted, the input flavour assumptions need thus to be modified.
In the case of NNPDF3.1~\cite{Ball:2016neh}, the evolution basis of Eq.~(\ref{eq:PDFevolutionbasis})
is supplemented with $c^+=c+\bar{c}$, which is also freely parametrized with a 37--parameter
artificial neural
network, while it is assumed that $c^-=c-\bar{c}=0$.
This option is also adopted in other recent studies where the charm PDF is fitted,
such as in the CT14 IC analysis~\cite{Hou:2017khm}
(see also Sect.~\ref{sec:structure.charm}).
Note that in general different flavour assumptions concerning the parametrized
charm PDF are conceivable,
for instance Eq.~(\ref{eq:PDFevolutionbasis}) could be also generalised by adding
\be
T_{15}= u+\bar{u} + d + \bar{d} +s+\bar{s}-3(c+3\bar{c})  \, , 
\ee
though this option has the drawback that the connection with charm--sensitive
observables is less direct.

A further issue related to the PDF parametrization is that of the positivity.
While, beyond LO, PDFs are scheme--dependent quantities and thus in principle can become negative,
physical observables such as cross sections and structure functions should always be positive--definite.
This constraint is incorporated in the (N)NLO global fits in different ways.
For instance, in CT10 all PDFs are made by construction positive--definite, while MMHT14 allows
the small--$x$ gluon PDF to become negative.
In the case of the NNPDF family of fits, no positivity constraints are imposed at the
PDF level (except of course at the LO case),
but during the fit the strict positivity of a range of physical
cross sections is imposed by means of a Lagrange multiplier.
Specifically, in the NNPDF3 sets the positivity of the following cross sections
is imposed at $Q^2=5$ GeV$^2$: $F_2^u$,$F_2^d$, $F_2^s$, $F_L$,
$\sigma_{\rm DY}^{u\bar{u}}$, $\sigma_{\rm DY}^{d\bar{d}}$, and
$\sigma_{\rm DY}^{s\bar{s}}$.
Note that in general in this approach the positivity constraint applies to all conceivable cross sections,
including for instance those that involve hypothetical new particles, and is not
restricted to the actual cross sections that are accessible experimentally.

\subsection{Fit quality and minimization strategies}\label{sec:fitmeth.min}
In this section we discuss how the quality of the agreement between
experimental data and theoretical predictions can be quantified
within a PDF fit, and the associated issue of the minimization strategy
adopted to find the optimal set of PDF parameters starting from
a figure of merit, $\chi^2$.

\subsubsection{Fit quality and $\chi^2$ definition}\label{sec:fitmeth.min.chi2def}
The quality of the agreement between experimental measurements
and the corresponding theoretical
predictions within a global fit is usually expressed in terms of the log--likelihood function, or $\chi^2$.
When the correlations between the experimental systematic errors are not available,
the $\chi^2$ as a function of the PDF parameters is given by
\begin{equation}
\chi^2(\{a\})=\sum_{k=1}^{N_{pt}}{1\over \sigma_k^2}\left(D_k-T_k(\{a\})\right)^2\; ,
\end{equation}
where $N_{pt}$ is number of data points, and $\sigma_k$
are the total experimental errors, given by adding the statistical and systematic
errors in quadrature.
In this expression,
$T_k(\{a\})$ are theoretical predictions,
expressed in terms of the PDF parameters $\{a\}$, and
$D_k$ are the central values of the experimental measurement.

Modern experiments provide correlated sources of the
various systematic uncertainties, in addition to the 
statistical and uncorrelated systematics.
The simplest example is the luminosity error in collider experiments, which is fully correlated among all
the bins from the same dataset.
Typically, there are many other sources that are introduced in the process of any given analysis.
In such cases, the $\chi^2$ has the following form~\cite{Pumplin:2002vw}
\begin{equation}\label{eq:chi2}
\chi^2(\{a\},\{\lambda\})=\sum_{k=1}^{N_{pt}}{1\over s_k^2}\left(D_k-T_k-
\sum_{\alpha=1}^{N_{\lambda}}\beta_{k,\alpha}\lambda_{\alpha}\right)^2
+\sum_{\alpha=1}^{N_{\lambda}}\lambda_{\alpha}^2\; ,
\end{equation}
for $N_{\lambda}$ sources of correlated error. Here, $s_k$ represents the total uncorrelated error, which is constructed
by adding the statistical and uncorrelated systematic errors in quadrature.  
Each source of correlated systematic error is described by a nuisance
parameter $\lambda_{\alpha}$, with the error $\beta_{i,\alpha}$ correlated
among all data points.
Thus the induced systematic shift
to the experimental measurement is $\sum_{\alpha}\beta_{k,\alpha}\lambda_{\alpha}$.
The second sum on the right hand side of Eq.~(\ref{eq:chi2}) includes the penalty terms
to the $\chi^2$, assuming standard Gaussian
distributions for the nuisance parameters.

In global PDF analyses we are more interested in the PDF parameters than
the specific values that these nuisance parameters take.
Therefore, for any given set $\{a\}$ we can first
minimise the $\chi^2$ with respect to the nuisance parameters
$\lambda_\alpha$ to give the profiled log--likelihood
function $\chi^2(\{a\})\equiv\chi^2(\{a\},\{\hat \lambda\})$. 
While na\"{i}vely we might worry that this would be a computationally intensive exercise, the simple quadratic dependence of the $\chi^2$  on the $\lambda_{\alpha}$ allows the profiled nuisance parameter $\hat \lambda_{\alpha}$ to be solved for analytically, assuming purely Gaussian errors.  Explicitly, we have
\begin{equation}
\hat \lambda_{\alpha}=\sum_{i=1}^{N_{pt}}\frac{\left(D_i-T_i\right)}{s_i}
\sum_{\delta=1}^{N_{\lambda}}A^{-1}_{\alpha \delta}
\frac{\beta_{i,\delta}}{s_i}\; ,
\end{equation}
with
\begin{equation}
A_{\alpha\beta}=\delta_{\alpha\beta}+\sum_{k=1}^{N_{pt}}
\frac{\beta_{k,\alpha}\beta_{k,\beta}}{s_k^2}\; .
\end{equation}
By substituting $\hat \lambda_{\alpha}$ into Eq.~(\ref{eq:chi2}) we
obtain the profiled $\chi^2$ as a function of the PDF parameters,
\begin{equation}
\chi^2(\{a\})=\sum_{i,j=1}^{N_{pt}}(T_i-D_i)({\rm cov}^{-1})_{ij}(T_j-D_j)\; ,
\end{equation}
with the experimental covariance matrix and its inverse given by
\begin{equation}
({\rm cov})_{ij}\equiv s_i^2\delta_{ij}+\sum_{\alpha=1}^{N_{\lambda}}\beta_{i,\alpha}
\beta_{j,\alpha}\, ,\,\,\quad ({\rm cov}^{-1})_{ij}=\frac{\delta_{ij}}{s_i^2}
-\sum_{\alpha,\beta=1}^{N_{\lambda}}
\frac{\beta_{i,\alpha}}{s_i^2}A^{-1}_{\alpha\beta}\frac{\beta_{j,\beta}}{s_j^2}\; .
\end{equation}
Thus, the profiled $\chi^2$ is fully determined by the covariance matrix,
which is  itself
constructed analytically in terms of the experimental statistical and systematic errors.
In certain circumstances, for example in the case of some
of the CMS and LHCb measurements,
the experiments publish the covariance matrix directly, instead of the full breakdown of the experimental systematics. This is however less advantageous
from the PDF fitting point of view.

One final subtlety concerning the construction of the covariance matrix
arises due to the fact that experimental systematic errors are usually presented
as relative errors $\sigma_{i,\alpha}$ with respect to the data, that is
\begin{equation}
({\rm cov})_{ij}=s_i^2\delta_{ij}+\left(\sum_{\alpha=1}^{N_c}\sigma_{i,\alpha}^{(a)}
\sigma_{j,\alpha}^{(a)}+\sum_{\beta=1}^{N_{L}}\sigma_{i,\beta}^{(m)}\sigma_{j,\beta}^{(m)}
\right)D_iD_j\; .
\end{equation}
Here, we have further separated these sources into $N_c$ additive 
and $N_L$ multiplicative errors; in the former case this counts those errors that are absolute in size, while in the latter those (such as the luminosity) which genuinely correspond to a relative uncertainty on the data. 
These have quite different statistical interpretations, and indeed it is known that the above experimental
definition of the covariance matrix will result in a D'Agostini bias of the
multiplicative errors~\cite{Ball:2009qv} when used in a PDF fit.

To obtain unbiased results,
one should use the `$t_0$' definition of the covariance matrix~\cite{Ball:2009qv},  given by   
\begin{equation}
({\rm cov})_{ij}=s_i^2\delta_{ij}+\left(\sum_{\alpha=1}^{N_c}\sigma_{i,\alpha}^{(a)}
\sigma_{j,\alpha}^{(a)}D_iD_j+\sum_{\alpha=1}^{N_{L}}\sigma_{i,\beta}^{(n)}
\sigma_{j,\beta}^{(m)}T^{(0)}_iT^{(0)}_j\right)\; .
\end{equation}
That is, one rescales the multiplicative errors not by the data but by the theory prediction $T^0_i$,
from a previous iteration of the $\chi^2$ minimization.
An alternative prescription is the `{\it $t$}' definition~\cite{Ball:2012wy}, where the multiplicative errors are rescaled by the same theoretical prediction
as in the comparison to the data,
\begin{equation}
({\rm cov})_{ij}=s_i^2\delta_{ij}+\left(\sum_{\alpha=1}^{N_c}\sigma_{i,\alpha}^{(a)}
\sigma_{j,\alpha}^{(a)}D_iD_j+\sum_{\alpha=1}^{N_{L}}\sigma_{i,\beta}^{(m)}
\sigma_{j,\beta}^{(m)}T_iT_j\right)\; ,
\end{equation}
or the {\it extended-$t_0$} and {\it extended-$t$}
definitions, where  both the additive and multiplicative errors are rescaled by the
corresponding theory.
A detailed discussion of the various possible prescriptions
can be found in~\cite{Ball:2012wy,Butterworth:2014efa}.    

\subsubsection{Minimization strategies}
\label{sec:fitmeth.min.chi2min}
The central PDF fits are determined by finding the global minimum of the log--likelihood function $\chi^2(\{a\})$. For PDF sets with a moderate number of free parameters, say less than $N_{\rm par}\simeq 40$,
numerical gradient--based algorithms are typically used. As a simple example, in Newton's method, the trial solution for the global minimum is given by
\be
a^{\rm trial}_i=a^0_i-\sum_{j=1}^{n_{\rm par}}H^{-1}_{ij}d_j\;,
\ee
for the $i$th PDF parameter. Here $a^0$ is an arbitrary starting point in the PDF parameter space, $d$ is the gradient and $H^{-1}$ is the inverse of the Hessian matrix (defined in Sect.~\ref{sec:pdfuncertainties.hess}) at the same point.
This solution is exact assuming a purely quadratic shape for the $\chi^2$, although in practice it can deviate significantly from this when it is far away from the global minimum. The above solution is therefore typically applied iteratively until the desired degree of convergence is achieved. However, the method will fail if the Hessian matrix $H$ is not positive--definite, and can suffer from numerical instabilities.
Various quasi--Newton
methods have been proposed to overcome these complexities in real applications, such as the Levenberg--Marquardt method used in MSTW/MMHT analyses~\cite{Martin:2009iq},
which is based on
a dynamically determined combination of Newton's method and the steepest decent method. 

In gradient--based methods, the gradient and Hessian matrix must be calculated numerically by
means of finite differences.
Another class of widely used gradient based algorithms are the variable metric methods (VMM), where it is not necessary to calculate the
Hessian matrix numerically.
Instead, the matrix $H$
is updated iteratively based only on information of the gradients. VMM
is the default algorithm in the MINUIT package~\cite{James:1975dr} and is used
in CTEQ--TEA analyses~\cite{Pumplin:2000vx}.   

As the number of free parameters is increased, the above methods will begin to suffer from numerical instabilities and issues with local minima.
For the NNPDF analysis, where the typical number of parameters is an order of magnitude higher than in other sets, a genetic algorithm~\cite{tau,GonzalezGarcia:2006ay,quevedo}
is the appropriate choice, as demonstrated in~\cite{Ball:2010de}.
The basic idea is to start from an ensemble of arbitrarily chosen samples 
of the PDF parameters. Random mutations with possible crossing--overs
are then applied to generate a larger group of new samples. Those
candidates predicting a lower $\chi^2$ are then selected to form a
new ensemble with the same size.
This procedure is then
iterated until a suitable convergence criterion is met, while care is taken to
prevent over--fitting.

\subsection{PDF uncertainties}
\label{sec:pdfuncertainties}
It is
essential to determine not only the best--fit PDFs, but also the associated uncertainties   in a
systematic way,
arising from instance from the data errors or 
methodological fit choices.
Subsequently, it should be possible to propagate
these uncertainties to the cross section predictions.
The most widely--used methods
to estimate PDF uncertainties fall into three main categories, known as the
Hessian, Monte Carlo, and Lagrange multiplier methods.
Each of these methods
will be explained in turn in the following sections.

\subsubsection{The Hessian method}
\label{sec:pdfuncertainties.hess}
The Hessian method to quantify PDF uncertainties was first developed
in~\cite{Pumplin:2001ct}.
Here, we describe the basic ingredients  of this method
and of their subsequent refinements, mostly following the
discussion of~\cite{Martin:2009iq}.
Given the $\chi^2$ estimator, the best--fit values correspond to those
for which this estimator has a global minimum,
$\chi^2_{\rm min}$.
In the vicinity of this minimum, the $\chi^2$ can
be approximated in terms of a quadratic expansion of the form
\be
\label{eq:hessianexpansion}
\Delta\chi^2 \equiv \chi^2- \chi^2_{\rm min}
=\sum_{i,j=1}^{n_{\rm par}}H_{ij}\lp a_i-a_i^0\rp
\lp a_j-a_j^0\rp \, ,
\ee
where the $n_{\rm par}$ fit parameters are denoted by $\{a_1,\ldots,
a_{n_{\rm par}}\}$, and where the best--fit values that minimize the
$\chi^2$ are indicated by
$\{a_1^0,\ldots,
a^0_{n_{\rm par}}\}$.
In the quadratic expansion Eq.~(\ref{eq:hessianexpansion})
we have introduced the Hessian matrix, defined as the matrix
of second derivatives of the $\chi^2$ with respect to the
fit parameters, namely
\be
H_{ij}(\{\vec{a}^0\})\equiv \frac{1}{2} \frac{\partial^2\chi^2}{\partial a_i
\partial a_j}\Bigg|_{\{\vec{a}\}=
\{\vec{a}^0\}} \, .
\ee
This Hessian matrix contains all the information necessary
to quantify the PDF uncertainties.
Indeed, for a generic function $ \mathcal{F} [ \{  a_i \}]$
that depends on the PDFs and thus indirectly on the fit
parameters, the associated uncertainty can be computed
by means of linear error propagation
\be
\label{eq:hessianmaster1}
\sigma_{\mathcal{F}}=T\lp \sum_{i,j}^{n_{\rm par}}
\frac{\partial\mathcal{F}}{\partial a_i} \lp H \rp^{-1}
\frac{\partial\mathcal{F}}{\partial a_j} \rp^{1/2} \, ,
\ee
where $T=\sqrt{\Delta \chi^2}$ is the
`tolerance factor' that determines the matching between
the allowed range of parameter variations around the best--fit
values and the associated confidence interval of the PDF
uncertainties.
While textbook statistics suggest that $T=1$ corresponds
to a 68\% confidence interval, in the context of a global
fit there is ample evidence that somewhat larger values
for the tolerance are required in the Hessian method, in particular
to account for inconsistent experiments, theoretical uncertainties, and for methodological
uncertainties such as the specific choice of functional form~\cite{Pumplin:2009bb}.

The main limitation of Eq.~(\ref{eq:hessianmaster1}) is that
in general the derivatives $\partial\mathcal{F}/\partial a_i$
are unknown.
This problem can be bypassed by diagonalizing the Hessian matrix
and then representing PDF uncertainties in terms of orthogonal
eigenvectors.
After this diagonalization procedure, Eq.~(\ref{eq:hessianmaster1})
has the much simpler form
\be
\label{eq:hessianmaster2}
\sigma_{\mathcal{F}}=\frac{1}{2}\lp \sum_{i,j}^{n_{\rm par}}
\lc \mathcal{F}(S_i^+)-\mathcal{F}(S_i^-) \rc^2 \rp^{1/2} \, ,
\ee
where $S_i^{\pm}$ corresponds to the $i$--th eigenvector
associated to positive and negative variations with respect
to the best--fit value.
Using the eigenvectors $\{ S_i^{\pm} \}$ it is also
possible to compute asymmetric PDF uncertainties using
the prescription of Ref.~\cite{Nadolsky:2001yg}.

\begin{figure}[t]
\begin{center}
  \includegraphics[scale=0.70]{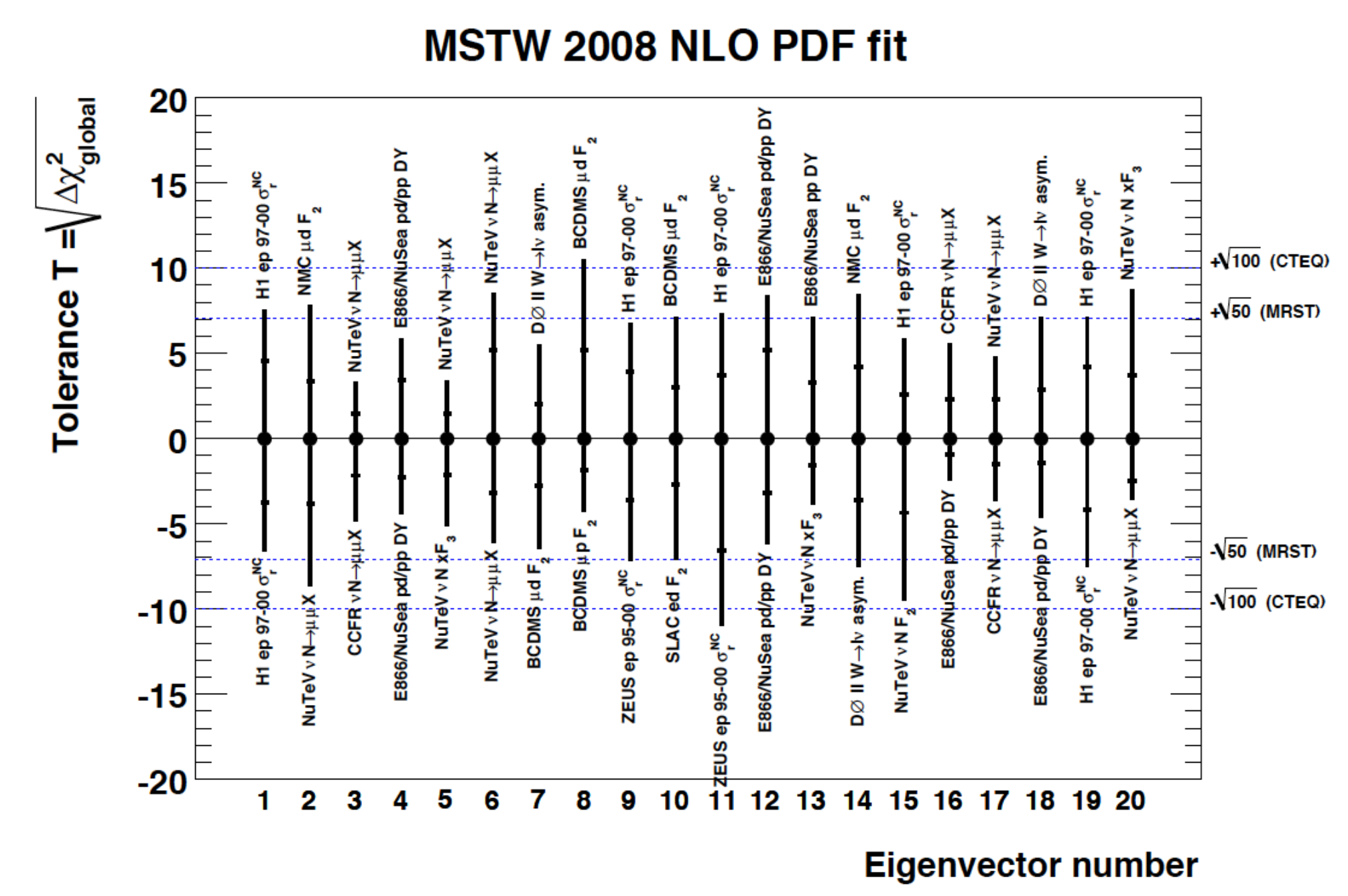}
   \caption{\small 
  The individual tolerance for each eigenvector, determined
  by the criterion that each separate experiment should be described
  within its 68\% C.L. (inner error bars) or 90\% C.L. (outer error bars) limit.
  In each case the figure indicates the name of the experiment
  that determines the 90\% C.L.  tolerance for the various eigenvector directions.
    \label{fig:mstw08-tolerance}
  }
\end{center}
\end{figure}

For the tolerance factor $T=\sqrt{\Delta\chi^2}$,
the original studies by the CTEQ and MRST group used values of $T=10$
and $T=\sqrt{50}$ respectively.
In more recent releases, the determination of this tolerance has been refined.
In the case of the MSTW08 analyses for example (as well as the subsequent MMHT14 set), the tolerance is
determined dynamically for each eigenvector by demanding that all
data sets are included within a given confidence level variation.
To illustrate this, in Fig.~\ref{fig:mstw08-tolerance}
we show the individual tolerance for each eigenvector
of the MSTW08 global analysis, determined
  by the criterion that each separate experiment should be described
  within the 68\% C.L. (inner error band) and 90\% C.L (outer error band) limit. The name of the experiment that determines the 90\% C.L.  tolerance for the various eigenvector directions is also shown.
  For the various eigenvector directions,
  the figure indicates the name of the experiment
  that determines the tolerance -- in other words, the experiment
  which is most sensitive to variations along this specific eigenvector
  direction.
  The fact that many different experiments are responsible
  for determining this tolerances emphasizing the crucial
  importance of using a very wide dataset in global
  PDF analyses.

%%%%%%%%%%%%%%%%%%%%%%%%%%%%%%%%%%%%%%%%%%%%%%%%%%%%%%%%%%%%%%%%%%%%%
%%%%%%%%%%%%%%%%%%%%%%%%%%%%%%%%%%%%%%%%%%%%%%%%%%%%%%%%%%%%%%%%%%%%%

In addition to the PDF uncertainties, it is
also possible to compute other statistical estimators such as
the correlation coefficients between two PDF flavours or between a PDF set
and a specific cross section.
If we denote as above by $n_{\rm par}$ the number of eigenvectors (which
have
both positive and negative variations) and by $\mathcal{F}$ and
$\mathcal{G}$ two physical quantities (such as a cross section
or the PDFs themselves), their correlation coefficient will be given
by
\be
\label{eq:correlationHessian}
\rho\lp \mathcal{F},\mathcal{G} \rp \equiv \cos \phi
= \frac{1}{4 \sigma_{\mathcal{F}}\sigma_{\mathcal{G}}}
\sum_{i=1}^{n_{\rm par}}\lp \mathcal{F}(S_i^+)-\mathcal{F}(S_i^-) \rp
\lp \mathcal{G}(S_i^+)-\mathcal{G}(S_i^-) \rp \, ,
\ee
where $\sigma_{\mathcal{F}}$ and $\sigma_{\mathcal{G}}$
are computed using Eq.~(\ref{eq:hessianmaster2}).
The interpretation of the correlation coefficient $\cos \phi$
is straightforward, with values close to zero implying that the
two quantities $\mathcal{F}$ and $\mathcal{G}$ are uncorrelated,
and values close to 1 (-1) implying that they are strongly
correlated (anti-correlated).
This therefore provides a useful tool to
investigate the PDF sensitivity of a given measurement; it is only
in the kinematic region where the absolute value of the
correlation coefficient is large that such a measurement is expected
to have significant impact on a PDF analysis.

\subsubsection{The Monte Carlo method}\label{sec:pdfuncertainties.MC}
In this method, the propagation of the experimental data uncertainties
to the PDFs is achieved by constructing a MC
representation of the probability distribution associated to the original
measurements.
This requires generating a large number $N_{\rm rep}$ of artificial MC replicas
of the original data, the so--called `pseudo--data',
which encode the same information on
central values, variances and correlations as that provided by the experiment.
In particular, given an experimental measurement of a hard--scattering
cross section denoted generically by $F_{I}^{\rm (exp)}$ with
total uncorrelated uncertainty $\sigma_{I}^{\rm (stat)}$, $N_{\rm sys}$ fully
correlated systematic uncertainties $\sigma^{\rm (c)}_{I,\alpha}$ and
$N_a$ ($N_r$) absolute (relative) normalization uncertainties
$\sigma^{\rm (norm)}_{I,n}$, the artificial
MC replicas are constructed using the following expression
\be
\label{eq:replicas}
F_{I}^{(\art)(k)}=S_{I,N}^{(k)} F_{I}^{\rm (\mrexp)}\lp 1+
 \sum_{\alpha=1}^{N_{\rm sys}}r_{I,\alpha}^{(k)}\sigma^{\rm (c)}_{I,c}+r_{I}^{(k)}\sigma_{I}^{\rm (stat)}\rp
 \ , \quad k=1,\ldots,N_{\rep} \ ,
\ee
where the normalization prefactor is given by
\be
\label{eq:totalnorm}
S_{I,N}^{(k)}=\prod_{n=1}^{N_a}\lp1+r_{I,n}^{(k)}\sigma^{\rm (norm)}_{I,n}\rp
\prod_{n'=1}^{N_r}\sqrt{1+r_{I,n'}^{(k)}\sigma^{\rm (norm)}_{I,n'}}.
\ee 
Here the variables $r_{I,c}^{(k)},r_{I}^{(k)},r_{I,n}^{(k)}$ are
 univariate Gaussian random numbers.
 Eq.~(\ref{eq:replicas}) represents the fluctuations of the pseudo--data
 replicas around the measured central values by the amount allowed by the
 experimental uncertainties.
 Note that for each replica the random fluctuations
 associated to a given fully correlated systematic
 uncertainty will be the same
 for all data points, that is, $r^{(k)}_{I,\alpha}=r^{(k)}_{I',\alpha}$.
 The same condition holds for the normalization uncertainties.
 As discussed in~\cite{Forte:2002fg},
 the second term in Eq.~(\ref{eq:totalnorm})
 (with the square root) correspond to the contribution of
 the $N_r$ relative normalization uncertainties, which
 needs to be treated differently from the one associated
 to the $N_a$ absolute normalization uncertainties.

\begin{figure}[t]
\begin{center}
  \includegraphics[scale=0.79]{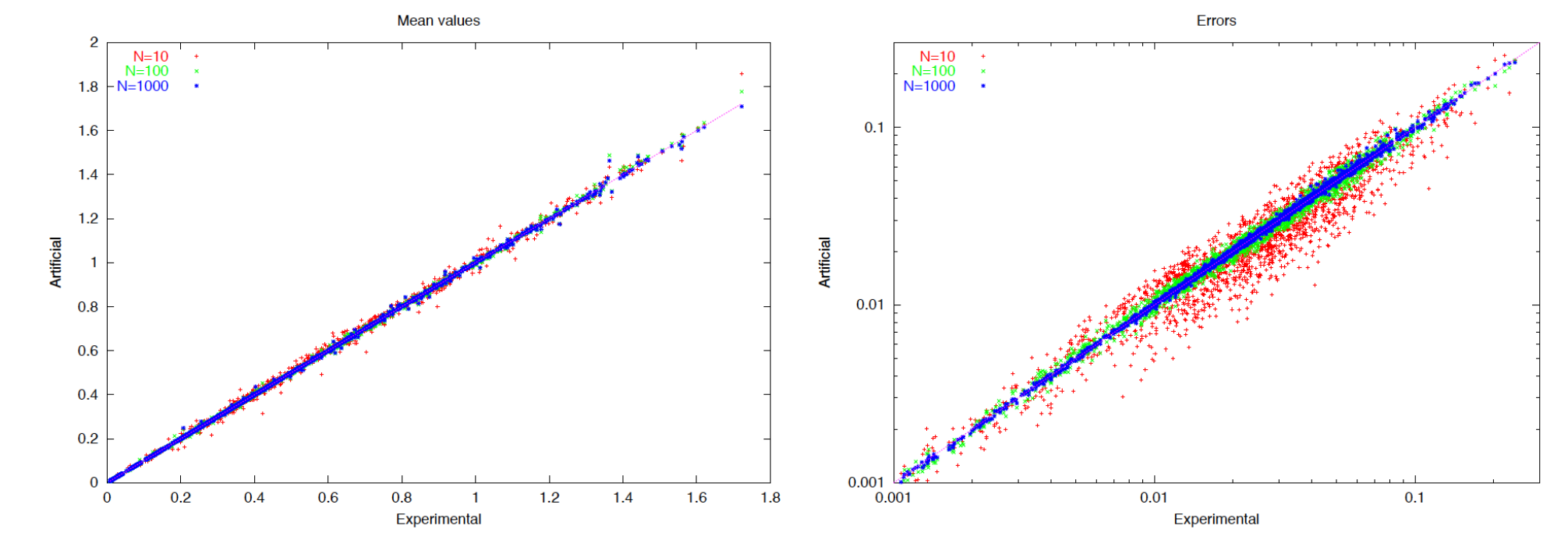}
   \caption{\small 
  The scatter between the mean values (left) and variances (right plot)
  of all the data points included in the analysis of~\cite{DelDebbio:2004qj}, comparing
  the original experimental values with the results obtained from the MC representation
  for different number $N_{\rm rep}$ of replicas.
    \label{fig:f2p-scatter}
  }
\end{center}
\end{figure}

An important question in the MC method is how many replicas
$N_{\rm rep}$ need to be generated in order to achieve a faithful representation
of the underlying probability density in the space of data.
To this purpose, a number of statistical estimators were constructed in
Ref.~\cite{DelDebbio:2007ee}.
It was found that $N_{\rm rep}=10$ replicas are enough to reproduce
central values, $N_{\rm rep}=100$ for the variances and that
$N_{\rm rep}=1000$ is required to satisfactorily reproduce the
data correlations.
Subsequent analysis have shown that this statement holds for
a generic input dataset.
To illustrate this point, in Fig.~\ref{fig:f2p-scatter} we show
the scatter between the mean values  and variances
  of all the data points included in the analysis of~\cite{DelDebbio:2004qj}, comparing
  the original experimental values with the results obtained from the MC representation
  for different number $N_{\rm rep}$ of replicas.
  We find that indeed for central values, the averages computed using
  only $N_{\rm rep}=10$ replicas agree with the original data, but that
  for variances this is not the case, where $N_{\rm rep}=100$ replicas
  are required.
  A similar study shows that $N_{\rm rep}=1000$ replicas are required
  to faithfully reproduce the experimental data correlations.

%%%%%%%%%%%%%%%%%%%%%%%%%%%%%%%%%%%%%%%%%%%%%%%%%%%%%%%%%%%%%%%%%%%%%
%%%%%%%%%%%%%%%%%%%%%%%%%%%%%%%%%%%%%%%%%%%%%%%%%%%%%%%%%%%%%%%%%%%%%

Once the MC sampling of the experimental data has been achieved,
a separate PDF fit is performed for each replica.
This can be done using either traditional polynomial functional forms
or other interpolators such as artificial neural networks -- the MC method
works in both cases.
The resulting sample of $N_{\rep}$ PDF replicas realises the concept of
the probability density in the space of parton distributions.
The calculation of the resulting PDF uncertainties and their propagation to generic
cross sections can then be performed using textbook methods.
Note that in this approach the PDF uncertainty propagation is fully general,
and in particular is not restricted to the Gaussian approximation.

For instance, in the MC method the expectation function of a generic
cross section $ \mathcal{F} [ \{  q \}]$
is evaluated as an average over the replica sample,
\be
\label{masterave}
\la \mathcal{F} [ \{  q \}] \ra
= \frac{1}{N_{\rm rep}} \sum_{k=1}^{N_{\rm rep}}
\mathcal{F} [ \{  q^{(k)} \}] \, ,
\ee
and the corresponding uncertainty is then determined as the variance of the
MC sample,
\be
\sigma_{\mathcal{F}} =
\left( \frac{1}{N_{\rm rep}-1}
\sum_{k=1}^{N_{\rm rep}}   
\lp \mathcal{F} [ \{  q^{(k)} \}] 
-   \la \mathcal{F} [ \{  q \}] \ra\rp^2 
 \right)^{1/2}.
\label{mastersig}
\ee
These formulae may also be used for the determination of central values and
uncertainties of the parton distribution themselves, in which case the
functional $\mathcal{F}$ is trivially
identified with the parton distribution $q$, that is,  
$\mathcal{F}[ \{ q\}]\equiv q$.

As in the Hessian case, see Eq.~(\ref{eq:correlationHessian}), the
correlation coefficient between two quantities $\mathcal{F}$
and $\mathcal{G}$ can be straightforwardly computed, using the standard expression
\be
\label{eq:correlationMC}
\rho\lp \mathcal{F},\mathcal{G} \rp 
= \frac{1}{\sigma_{\mathcal{F}}\sigma_{\mathcal{G}}}\lc
\frac{1}{N_{\rm rep}}\sum_{k=1}^{N_{\rm rep}} \lp \mathcal{F} [ \{  q^{(k)} \}]
\mathcal{G} [ \{  q^{(k)} \}] \rp
-
\lp \frac{1}{N_{\rm rep}}\sum_{k=1}^{N_{\rm rep}}\mathcal{F} [ \{  q^{(k)} \}] \rp
\lp \frac{1}{N_{\rm rep}}\sum_{l=1}^{N_{\rm rep}}\mathcal{G} [ \{  q^{(l)} \}] \rp
\rc \; ,
\ee
where we have used Eq.~(\ref{mastersig}) to evaluate the standard
deviation of $\mathcal{F}$ and $\mathcal{G}$ over the MC replica sample.
The statistical interpretation of Eq.~(\ref{eq:correlationMC}) is the same as before, with
values of $\rho\lp \mathcal{F},\mathcal{G} \rp $
close to 1 (-1) indicating that there exists a strong correlation
(anti-correlation) between $\mathcal{F}$ and $\mathcal{G}$.

\begin{figure}[t]
\begin{center}
  \includegraphics[scale=0.79]{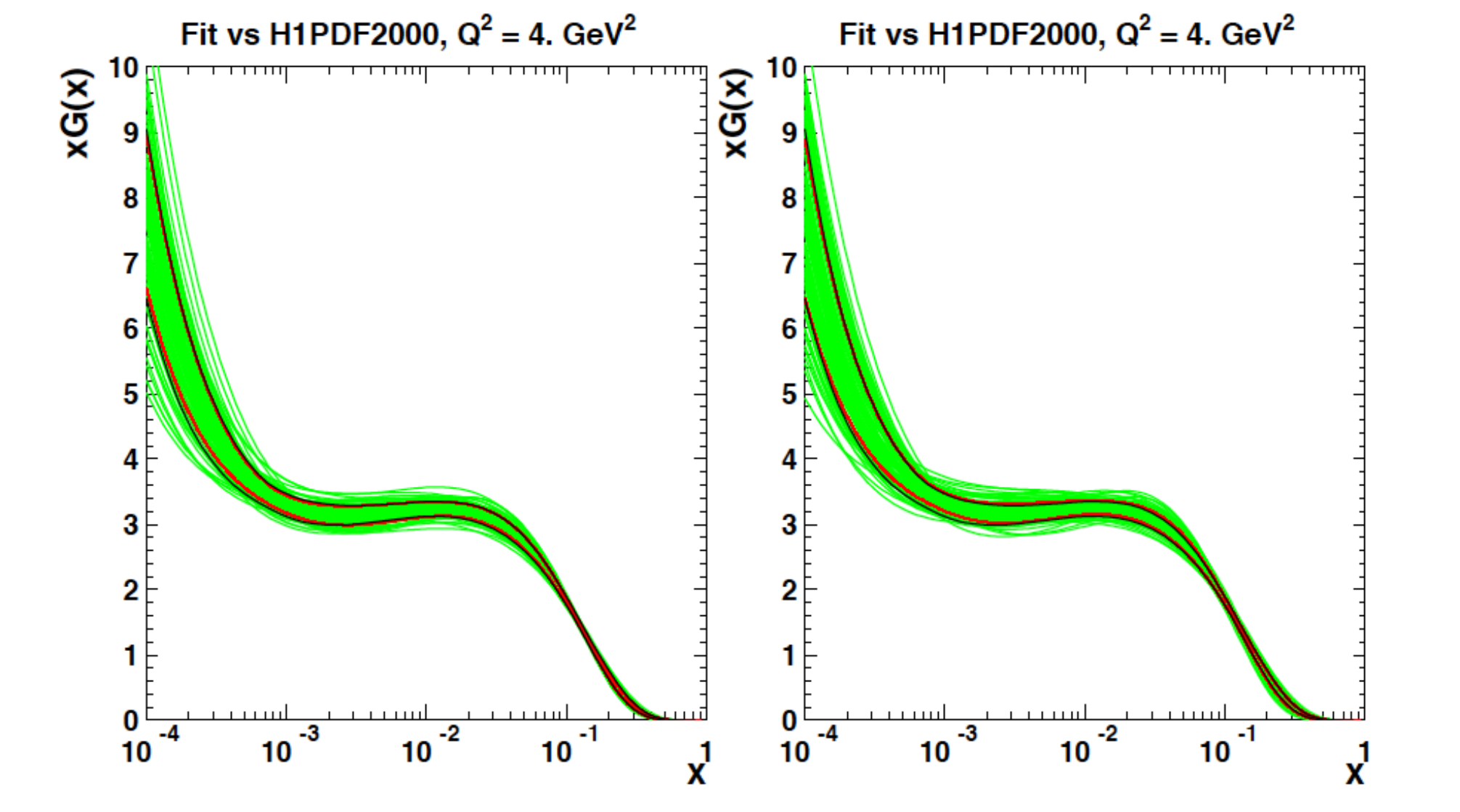}
   \caption{\small 
  The gluon PDF at $Q=2$ GeV in the HERA--LHC benchmark fit of
  Ref.~\cite{Dittmar:2009ii}, where the one--sigma PDF uncertainties
  computed with the Hessian method (black lines) are compared to
  those of the Monte Carlo method (red lines), finding good agreement.
  Each of the green curves corresponds to an individual MC replica.
  In the left fit, the normalization
  and systematic
  uncertainties in the MC replicas from Eq.~(\ref{eq:replicas}) fluctuate
  according to a multi--Gaussian distribution, while in the right
  fit they fluctuate instead according to a log--normal distribution.
    \label{fig:heralhc}
  }
\end{center}
\end{figure}

In the case of a fully consistent dataset, the MC method
to estimate the PDF uncertainties is expected to coincide with the
Hessian method described in Sect.~\ref{sec:pdfuncertainties.hess} for a standard
tolerance $\Delta \chi^2=1$.
This equivalence was explicitly demonstrated in the HERA--LHC workshop
proceedings~\cite{Dittmar:2009ii}.
In Fig.~\ref{fig:heralhc} we show the gluon PDF at $Q=2$ GeV in this benchmark fit,
based on HERA inclusive structure function data and a HERAPDF--like
parametrization. The one--sigma PDF uncertainties are
  computed with the Hessian method (black lines) and compared to
  those of the Monte Carlo method (red lines), finding good agreement.
  In this figure, each of the green curves corresponds to an individual MC replica.
  In the left fit, the normalization
  and systematic
  uncertainties in the MC replicas from Eq.~(\ref{eq:replicas}) fluctuate
  according to a multi--Gaussian distribution, while in the right
  fit they fluctuate instead according to a log--normal distribution, showing
  that the resulting PDFs depend only weakly on the specific assumptions about the
   probability distribution of the experimental systematic
  uncertainties.

%%%%%%%%%%%%%%%%%%%%%%%%%%%%%%%%%%%%%%%%%%%%%%%%%%%%%%%%%%%%%%%%%%%%%
%%%%%%%%%%%%%%%%%%%%%%%%%%%%%%%%%%%%%%%%%%%%%%%%%%%%%%%%%%%%%%%%%%%%%

Finally, we note that a Monte Carlo
representation of a Hessian PDF set can be constructed
following the strategy of Ref.~\cite{Watt:2012tq}, and
conversely, that a
Hessian representation of a Monte Carlo
PDF set can be accurately constructed using the {\tt mc2h} algorithm
developed in Refs.~\cite{Carrazza:2015aoa,Carrazza:2016htc}.
These techniques are discussed in more detail below in Sect.~\ref{sec:fitmeth.combined}.

\subsubsection{The Lagrange multiplier method}\label{sec:pdfuncertainties.Lagrange}

This method was originally developed in Refs.~\cite{Stump:2001gu,Pumplin:2000vx}
as a generalization of the $\chi^2$ minimization procedure.
As in the Hessian case, the first step is to find the PDF parameters $\{a_i^0\}$
that minimize the global $\chi^2(\{a_i\})$.
Then one has to select a specific physical quantity that depends on the PDFs,
such a DIS structure function or a hadron collider cross section,
that we denote generically by
$\mathcal{F}(\{a_i\})$, and which takes the value $\mathcal{F}_0=\mathcal{F}(\{a_i^0\})$
at the global fit minimum.
The goal of the Lagrange multiplier method is then to determine the PDF uncertainty
associated to $\mathcal{F}_0$ without making any assumption about the specific behaviour
of the $\chi^2$ around the global minimum, and in particular avoiding the Gaussian
assumption which is at the core of the Hessian method.

In order to achieve this, the global fit $\chi^2$ is modified by introducing the
physical quantity $\mathcal{F}$ as a Lagrange multiplier, so that the new
function that needs to minimized is now given by the sum of two contributions,
\be
\label{eq:lagrange1}
\Psi(\lambda, \{a_i\} )\equiv \chi^2(\{a_i\}) + \lambda \mathcal{F}(\{a_i\}) \, .
\ee
Now for each specific value of $\lambda$, denoted by $\lambda_{\alpha}$, the minimization
of Eq.~(\ref{eq:lagrange1}) will lead to a different set of best--fit PDF
parameters, which we indicate by $\{a_i^{\rm (min)}(\lambda_\alpha)\}$.
Mathematically, these parameters are the result of a constrained PDF fit
where the value of the physical observable has been fixed to $\mathcal{F}_\alpha
=\mathcal{F}(\{a_i^{\rm (min)}(\lambda_\alpha)\} )$.
The resulting PDF set of this constrained fit is now indicated by $S_{\alpha}$.

The main result of this procedure is to establish a parametric relation between the value
of the physical quantity $\mathcal{F}$ and the global fit $\chi^2$ by means of
the Lagrange multiplier $\lambda$.
This means that we can determine the PDF uncertainty associated to $\mathcal{F}$
imposing that the $\chi^2$ satisfies $\chi^2=\chi^2_{\rm min}+\Delta\chi^2$
with $\Delta\chi^2=T^2$ representing the tolerance introduced
in Sect.~\ref{sec:pdfuncertainties.hess}.
It is clear that the main advantage of the Lagrange multiplier method in comparison 
to the Hessian method is that one does not need to restrict to the quadratic
expansion or linear error propagation, since the PDF uncertainties
in this method are determined only by the values of the $\chi^2$ and not by its
specific shape.
On the other hand, an important restriction of the method is that the PDF error
analysis for each specific physical quantity $\mathcal{F}$ requires 
a large number of new PDF fits.
This is an important limitation
not only because this is a CPU time intensive operation,
but also because this cannot be carried out
outside the PDF fitting collaborations themselves.

\begin{figure}[t]
\begin{center}
  \includegraphics[scale=0.79]{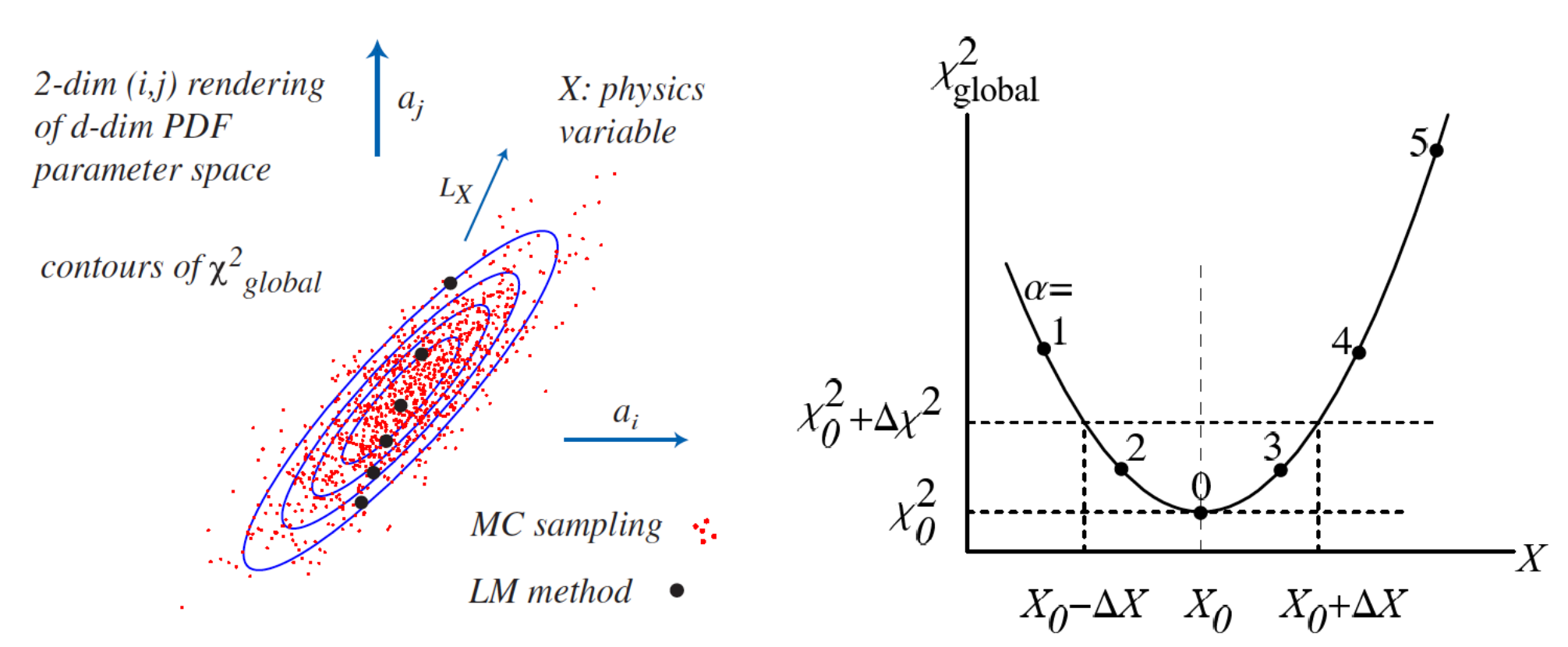}
   \caption{\small Schematic representation of the Lagrange Multiplier method,
   from Refs.~\cite{Stump:2001gu,Pumplin:2000vx}.
   In the left plot we show a two--dimensional projection of the PDF parameter space,
   indicating the contours in $\chi^2$ for fixed values of
   the physical quantity $\mathcal{F}$.
   In the right plot we show how the PDF uncertainty associated to $\mathcal{F}$
   for a given confidence interval is determined by the condition that the
   global $\chi^2$ should not grow beyond the tolerance $\Delta\chi^2$.
    \label{fig:LM}
  }
\end{center}
\end{figure}

The Lagrange multiplier method is schematically illustrated in
Fig.~\ref{fig:LM}.
In the left plot we show a two--dimensional projection of the PDF parameter space,
   indicating the contours in $\chi^2$ for fixed values of
   the physical quantity $\mathcal{F}$.
   The parametric relation is provided by the value of the Lagrange multiplier $\lambda$.
   In the right plot we show how the PDF uncertainty associated to $\mathcal{F}$
   for a given confidence interval is determined by the condition that the
   global $\chi^2$ should not grow beyond the tolerance $\Delta\chi^2$.
   As in the case of the Hessian method, the specific value of
   the tolerance $T=\sqrt{\Delta\chi^2}$ is an input to the method
   and must be determined independently.
   
%%%%%%%%%%%%%%%%%%%%%%%%%%%%%%%%%%%%%%%%%%%%%%%%%%%%%%%%%%%%%%%%%%%%%
%%%%%%%%%%%%%%%%%%%%%%%%%%%%%%%%%%%%%%%%%%%%%%%%%%%%%%%%%%%%%%%%%%%%%

\subsection{Combined and reduced PDF sets}
\label{sec:fitmeth.combined}
Individual PDF sets from different groups are widely used when comparing precision
theoretical predictions with LHC measurements and in the assessment of the accuracy
of PDF sets themselves.
However, for several LHC applications an assessment
of the {\it total} PDF uncertainty for certain observables, by taking into account 
predictions from all applicable PDF sets, is advantageous.
This will for example be the case in the
extraction of the couplings of the Higgs boson from the
experimental cross sections, or the calculation of signal and
background rates in searches for BSM physics.

For such purposes, a statistical procedure is needed with which to combine the
results from different PDF sets.
However, a statistical combination of PDF sets is rather more complicated than
in the case of for example the world average of the strong coupling constant or heavy--quark masses,
since it combines functions which have in principle an infinite number of degrees of freedom.
In addition, a suitable prescription must also
accommodate the fact that the individual PDF sets
are not identical either in their central values or in their uncertainties, and it
should account for possible correlations between PDF sets from different
groups.
Moreover, it would be desirable to achieve a final representation of this
combined PDF set in terms of a relatively small number of Hessian eigenvectors
or MC replicas.

The PDF4LHC Working Group 2010
recommendation proposed the use of a simple envelope prescription~\cite{Alekhin:2011sk,
Ball:2012wy,Botje:2011sn}.
That is, the PDF determinations from different groups are treated as
instances of a probability distribution affected by unknown sources of systematics,
rather than 
statistically distributed instances of an underlying probability distribution.
This envelope prescription can also only be applied at the level of individual observables,
thereby losing the information on PDF--induced correlations.
Given the better understanding of current PDF determinations, the
relatively good agreement between the current releases of the
three main global PDF sets,
and the high precision demands for LHC Run II studies, such a
prescription is therefore certainly inadequate. 

With the above considerations in mind, the updated 2015
PDF4LHC recommendation was proposed as a replacement~\cite{Butterworth:2015oua}.
There, a number of
criteria for the individual PDF sets to be considered for the combination were
adopted.
First of all,
the individual PDF sets should be based on a global determination with a large number
of datasets from a variety of experiments, that is,
DIS and hadron--hadron scattering in
fixed--target and collider experiments.
Second, the hard cross sections for DIS and hadron--hadron
scattering processes used in the extraction should be evaluated up to two loops in
QCD in a GM--VFN scheme, with a maximum number of $n_f=5$ active quark flavours.
Third, all known
experimental and procedural sources of uncertainties should be properly accounted
for, including the experimental uncertainties propagated from data, uncertainties
due to the incompatibility of different data sets, and uncertainties due to the functional
form of PDFs.

It was also decided that the combination should be carried out with
a central value of $\alpha_s(m_Z)=0.118$ at both NNLO and NLO,  with the corresponding
total uncertainty taken to be $\delta \alpha_s(m_Z)= 0.0015$, consistent with the then current PDG world--average~\cite{Beringer:1900zz}.
The heavy quark masses used in individual PDF sets
are not currently required to be the same\footnote{It would be desirable in the
future for all PDF groups provide error sets with common choices of heavy--quark masses and
furthermore to include the uncertainties due to the mass inputs, similarly to the case of $\alpha_s$.}
but should be compatible with their
world--average values.
The existing PDF sets satisfying all of the above requirements at
present have been identified as CT14~\cite{Dulat:2015mca}, MMHT2014~\cite{Harland-Lang:2014zoa},
and NNPDF3.0~\cite{Ball:2014uwa}.
The PDF4LHC 2015 PDF
sets are therefore statistical combination of these three sets, although the method
can be straightforwardly generalised to include additional PDF sets satisfying the
same quality conditions.

An important point here is that this
statistical combination can only be carried out efficiently using the MC method,
as different PDF determinations adopt different forms for the PDF parametrizations.
In the first step, the CT14 and MMHT2014 PDFs, which are originally in their Hessian form, are
converted into their MC representations by applying the Watt--Thorne method
with symmetric formula~\cite{Watt:2012tq}.
It has been validated that a MC ensemble with
$N_{\rm rep}=300$ replicas is sufficient to reproduce the central value and uncertainties of the original
Hessian PDFs to high precision.
The NNPDF3.0 PDF set is already in a MC
form with 1000 replicas. Following the idea of individual PDF determinations
as equally likely representations of an underlying probability distribution, a
combined PDF set is then built by taking 300 MC replicas from each input PDF sets and
merging them equally.

The resulting combined PDF set, an ensemble of 900 MC replicas,
is referred to as the MC900 or PDF4LHC15\_prior, and represents the combined 
probability distribution of the PDFs.
However, such a set of 900 PDFs would be unmanageably large for most applications, in particular 
given the time and storage cost required for complicated NNLO calculations and
experimental simulations.
Therefore, various methods have been developed to reduce the size
of the combined sets, while minimizing the information loss according to various statistical measures.

The first method applies the META--PDFs framework~\cite{Gao:2013bia}. Here, a flexible
functional form with Bernstein polynomials is chosen to parametrize the PDFs at an
initial scale. Each replica in the MC900 ensemble is then represented by a group
of PDF parameters through a fit to the chosen parametrizations, by minimizing
a metric function.
The prior probability distribution of PDFs is thus transformed
into the probability distributions in the PDF parameter space.
The covariance matrix of
the PDF parameters is calculated as usual,
\begin{equation}
{\rm cov}(a_l,a_m)=\frac{1}{N_{\rm rep}-1}\sum_{k=1}^{N_{\rm rep}}(a_l^{(k)}-a_l^{(0)})(a_m^{(k)}-a_m^{(0)}) \; ,
\end{equation}
where $a_l^{(0)}$ and $a_l^{(k)}$ denote the fitted PDF parameters from the central set and
the $k$-th MC replica respectively,
and $N_{\rm rep}$ is the total number of MC replicas.
The covariance matrix can be diagonalized by an orthogonal transformation.
Eigenvectors are then calculated and ordered according to their impact on the PDF uncertainties
with a designed error metric; the eigenvectors with smaller contributions can be dropped according to the
accuracy required.

In the final step of the META--PDF procedure,
a central PDF set and a group of orthogonal error PDF sets
are generated under the assumption of a multi--Gaussian distribution, which
 can then be used in a
similar way to the conventional Hessian PDF sets.
For example, the 68\% CL uncertainty or
$1\sigma$ error on a generic cross section $\mathcal{F}$ is given by
\begin{equation}
\label{eq:syhe}
\delta^{\rm PDF} \mathcal{F}=\sqrt{\sum_{i=1}^{N_{\rm eig}}(\mathcal{F}_{i}-\mathcal{F}_0)^2}\; ,
\end{equation}
where $\mathcal{F}_0$ is the prediction on observable $X$ given by the central set and
$\mathcal{F}_i$ is the prediction given by the $i$-th error set.
Note there is only one
error set along each eigenvector/orthogonal direction since
symmetric Gaussian distributions are assumed by construction. 

The second method that can be used to compress the MC900 prior set
is the {\tt mc2h} algorithm with Singular Value Decomposition,
followed by the Principle Component Analysis~\cite{Carrazza:2015aoa}.
The idea underlying this method is to first
discretise the PDFs at $N_xN_{\rm pdf}$ grid nodes, where $N_x$ denotes the total number of grid points
in the momentum fraction $x$ and $N_{\rm pdf}$ is the number of total independent
flavours.
Then, a $N_xN_{\rm pdf}\times N_xN_{\rm pdf}$ covariance matrix on all
those PDF values can be constructed from the MC replicas,
\begin{equation}
{\rm cov}_{ll'}=\frac{1}{N_{\rm rep}-1}\sum_{k=1}^{N_{\rm rep}}X_{lk}X^T_{kl'}=\frac{1}{N_{\rm rep}-1}XX^T \; ,
\end{equation}
where $X_{lk}$ is the PDF value on $l$-th grid point given by $k$-th MC replica subtracted
with the corresponding value from central PDF set,
$N_{\rm rep}$ is the total number of MC replicas.
The above covariance matrix can be rewritten in its singular value decomposition
form     as follows,                               
\begin{equation}
{\rm cov}_{ll'}=\frac{1}{N_{\rm rep}-1}(USV^T)(USV^T)^T\; ,
\end{equation}
where $S$ is a diagonal matrix constructed out from singular values of $X$, $V$ is
an orthogonal $N_{\rm rep}\times N_{\rm rep}$ matrix of coefficients, and $U$ is
a $N_{x}N_{\rm pdf}\times N_{\rm rep}$ matrix containing orthogonal eigenvectors
of the covariance matrix with nonzero eigenvalues.
Indeed, the matrix $V$ gives
a Hessian basis built upon linear combinations of original MC replicas, which
reproduces fully the covariance matrix given by the original MC replicas.
This basis
can be further truncated using Principle Component Analysis with a certain error
metric, resulting in a smaller Hessian PDF set,
an approach similar to the case of the META--PDF method.

A third option is provided by the compressed MC
(CMC) method~\cite{Carrazza:2015hva}. In this case,
an ensemble of pseudo--MC replica PDFs (CMC--PDFs) are generated. The CMC--PDFs
have a different statistical interpretation in comparison to the native MC PDFs. However, certain
statistical measures, such as the mean, covariance matrix, skewness, kurtosis and the
Kolmogorov distance  can be reconstructed in a similar way to the native MC PDFs.
The CMC--PDFs aim
to preserve some of the non--Gaussian features in the prior given by MC900 in addition
to the Gaussian features, for which the Hessian form is more adequate. Note that in the
current prescription some of the non--Gaussian behaviors from individual PDF sets have been
smoothed out due to the symmetric formula used in converting Hessian PDFs to MC replicas.
The compression starts with a figure of merit or error function ERF,
\begin{equation}
{\rm ERF}=\sum_k \frac{1}{N_k}\sum_i\left(\frac{C_i^{(k)}-O_i^{(k)}}{O_i^{(k)}}\right) \; ,
\end{equation}
where $k$ runs over the number of chosen statistical estimators, $N_k$ is a
normalization factor, $O_i^{(k)}$ is the value of the $k$-th estimator calculated
at the generic point $(x_i,Q_i)$ from the
prior and $C_i^{(k)}$ is the corresponding value of the same estimator in the
compressed set. The compressed set is simply a subset of the MC900 ensemble. For
any given number of total MC replicas, the compressed set is chosen by minimization
of the above error function using a genetic algorithm.   

%%%%%%%%%%%%%%%%%%%%%%%%%%%%%%%%%%%%%%%%%%%%%%%%%%%%%%%%%%%%%%%%%%%%%
\begin{figure}[t]
\begin{center}
  \includegraphics[scale=0.4]{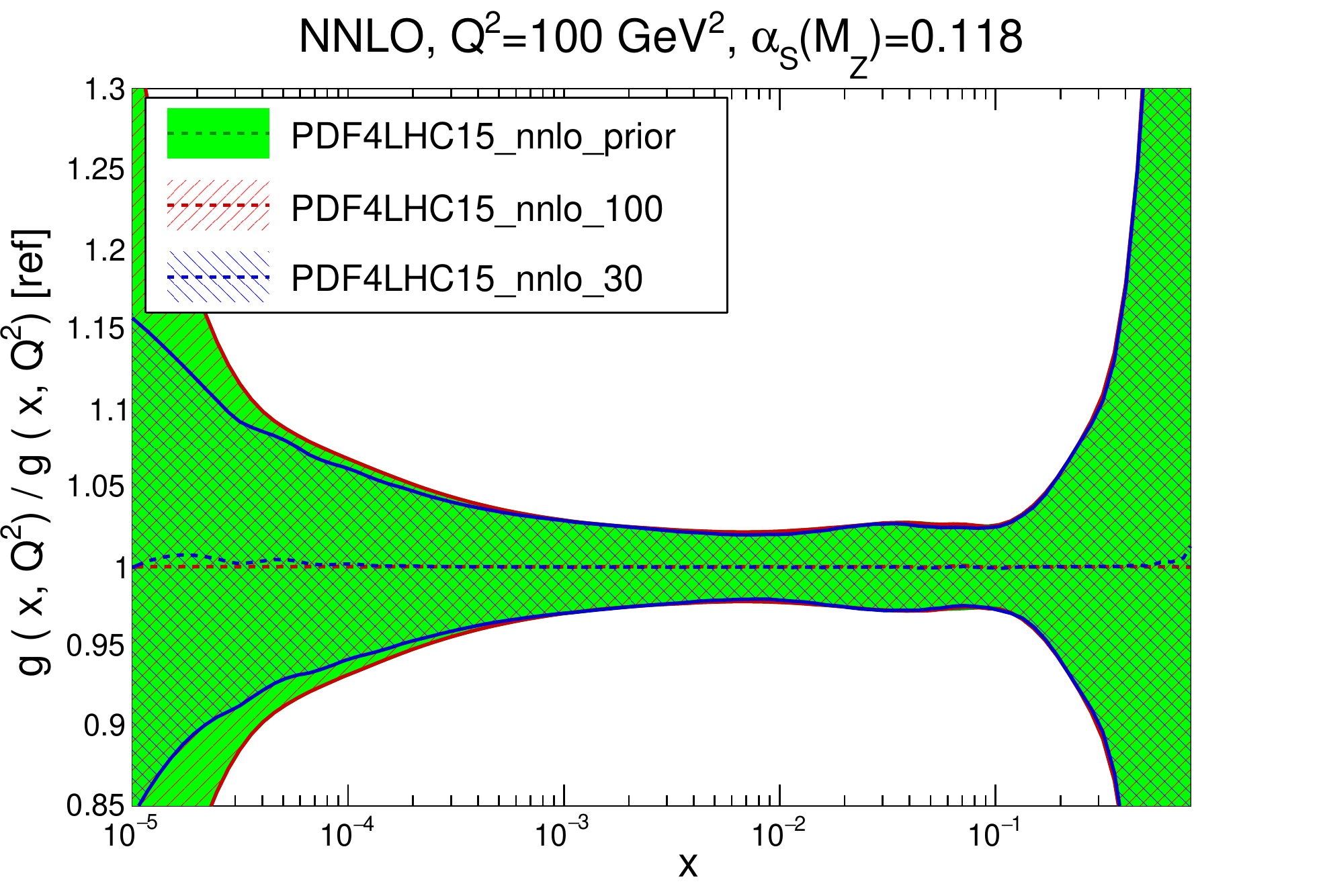}
  \includegraphics[scale=0.4]{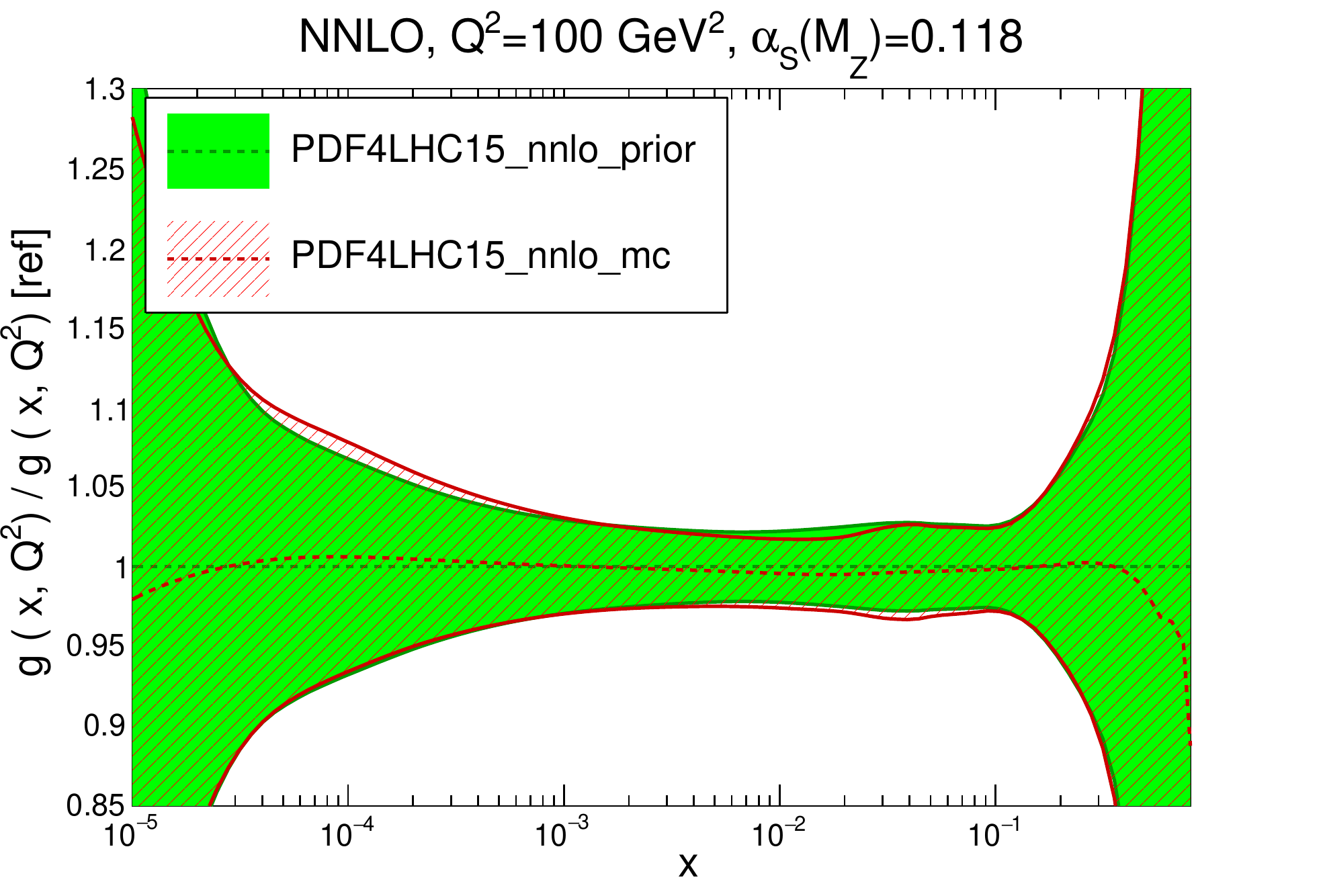}
  \includegraphics[scale=0.4]{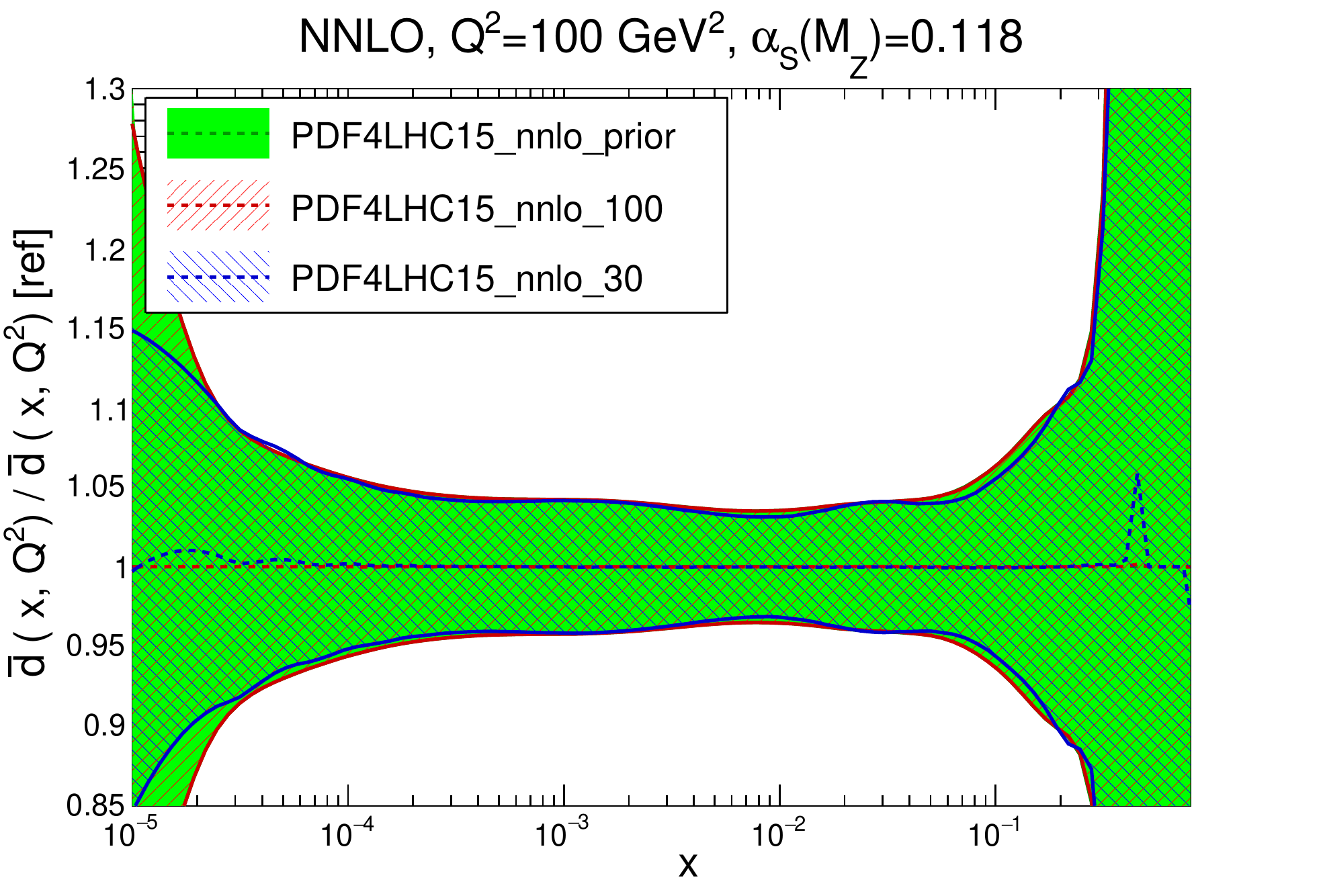}
  \includegraphics[scale=0.4]{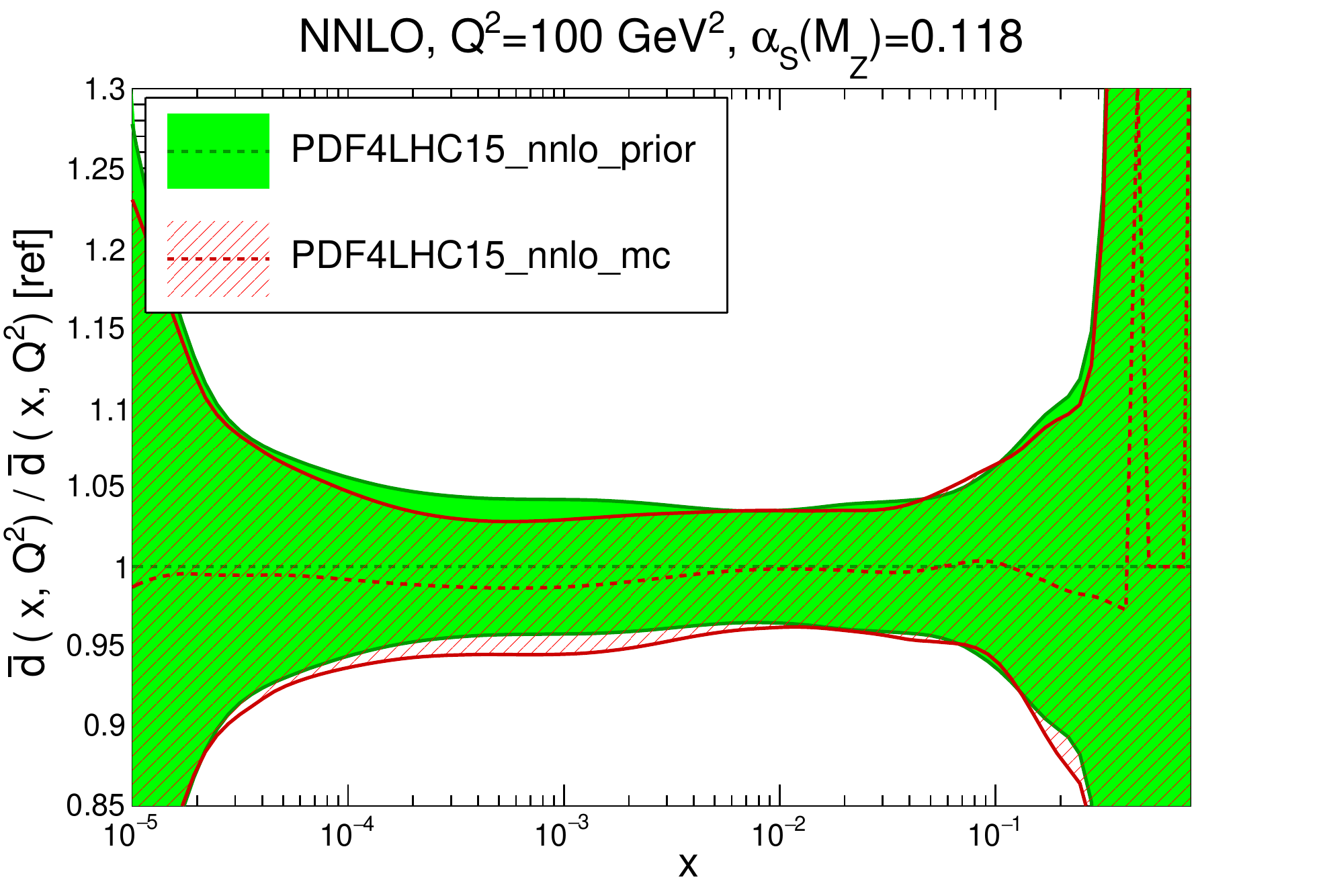}
   \caption{\small
   Comparison of the gluon and $\bar d$ quark PDFs at a scale of $Q^2=100\,{\rm GeV}^2$
   between the PDF4LHC15 prior and the two reduced Hessian sets, and between the prior and
   the compressed MC set, normalized to the central value of the prior~\cite{Butterworth:2015oua}. 
    \label{fig:comb1}
  }
\end{center}
\end{figure}
%%%%%%%%%%%%%%%%%%%%%%%%%%%%%%%%%%%%%%%%%%%%%%%%%%%%%%%%%%%%%%%%%%%%%

These three compression and reduction techniques were used to construct the
three PDF sets that are the main output of the
2015 PDF4LHC recommendation: a Hessian set with $N_{\rm eig}=30$ eigenvectors
({\tt PDF4LHC15\_30}), a
Hessian set with  $N_{\rm eig}=100$ eigenvectors ({\tt PDF4LHC15\_100}), and a compressed MC set with
 $N_{\rm rep}=100$ replicas ({\tt PDF4LHC15\_mc}). All of these are constructed from the same prior
(MC900), but have a slightly different focus in each case.
The symmetric PDF uncertainties of any observables
can be calculated using Eq.~(\ref{eq:syhe}) for Hessian sets and the usual
master formula for MC PDFs, Eq.~(\ref{mastersig}).

Fig.~\ref{fig:comb1} shows the comparison of the
central value and the uncertainties of the gluon and $\bar d$ quark PDFs for
the PDF4LHC15 prior and the three reduced sets.
The agreement between the
Hessian set with 100 eigenvectors and the prior is good for all PDF combinations in the
complete range of $x$.
The Hessian set with 30 eigenvectors also shows good 
agreement with the prior in the $x$ range related to precision physics measurements,
but gives a slightly smaller uncertainty in the extrapolation regions at small
and large $x$ as a tradeoff for the smaller number of PDF eigenvectors.
The compressed MC set also
agrees well with the prior in most
of the region for the mean and uncertainty except for small overall fluctuations.

The PDF4LHC Working Group recommendation for the usage of these different
PDF4LHC15 sets depends on the particular case under consideration:
\begin{itemize}
\item
Use individual PDF sets, and, in particular, as many of the modern PDF sets as possible,
for comparisons between data and theory for Standard Model measurements.

\item
Use the {\tt PDF4LHC15\_mc} sets for searches for BSM phenomena where
non--Gaussian behaviour could be important.

\item
Use the {\tt PDF4LHC15\_30} sets for calculation of PDF uncertainties in situations when
computational speed is needed, or a more limited number of error PDFs may be desirable.

\item
Use the {\tt PDF4LHC15\_100}
sets for calculation of PDF uncertainties in precision observables.
 
\end{itemize}
The cases listed above are not exclusive, and one or the other will be more
appropriate depending on the theoretical interpretation of a given experimental measurement.

In addition,  two further combined PDF4LHC15 sets are provided
with $\alpha_s(m_Z)=0.1165$ and $0.1195$,
in order to be able to estimate  the uncertainty due to $\delta \alpha_s$.
The corresponding
uncertainty at 68\% CL for a generic cross section $\mathcal{F}$ is given by
\begin{equation}
\delta^{\alpha_s}\mathcal{F}=\frac{\mathcal{F}(\alpha_s=0.1195)-\mathcal{F}(\alpha_s=0.1165)}{2}\;  ,
\end{equation}
where $\mathcal{F}(\alpha_s)$ is the value calculated using the PDF together with the 
hard matrix elements evaluated at that $\alpha_s$ value.
The combined PDF$+\alpha_s$
error is then computed by adding in quadrature the two sources of
uncertainty,
\begin{equation}
\delta^{{\rm PDF}+\alpha_s}\mathcal{F}=\sqrt{(\delta^{\rm PDF}\mathcal{F})^2
+(r\cdot\delta^{\alpha_s}\mathcal{F})^2}\; ,
\end{equation}
where the rescaling factor $r=1$ is recommended but can be varied according to
user's choice of uncertainty on $\alpha_s(m_Z)$, as compared to the range
$\delta\alpha_s=0.0015$ adopted as default.

It is also noted that the PDF4LHC15 PDF sets can be further reduced to compact
sets with around ten eigenvectors or less if the applications are restricted to a certain
group of observables, e.g. the cross section and distributions for Higgs boson
production at the LHC.
This can be achieved either through the data set diagonalization
method~\cite{Pumplin:2009nk,Gao:2014fma} or the singular value decomposition
method~\cite{Carrazza:2016wte}, the latter leading to the so--called
Specialised Minimal PDF (SM--PDF) sets.

\subsection{Treatment of theoretical parametric uncertainties}\label{sec:fitmeth.theoryunc}
The theoretical calculations that enter PDF determinations depend on a number
of external parameters, such as the value of the strong coupling constant
$\alpha_s(m_Z)$ and that of the heavy quark masses.
The choice of these input parameters therefore represents an additional component
of the total PDF uncertainty, known as the theoretical parametric
uncertainty.
As there can be strong correlations between the fitted PDFs and the values of these
inputs, a robust
evaluation of the PDF errors requires a consistent
combination of these with the PDF parametric uncertainties.
The global analysis can also provide an independent
determination of those QCD parameters, which can in turn contribute to the
world average values.

\subsubsection{The strong coupling constant $\alpha_S$}
\label{sec:fitmeth.theoryunc.alphas}
The current world average value for the
strong coupling constant is $\alpha_s(m_Z)=0.1181\pm 0.0011$.
This average is
extracted from six families of determinations, namely $\tau$ decay, lattice
results, $e^+e^-$ jets and event shapes, structure functions, EW precision fits,
and $t\bar t$ cross sections at LHC.
These are obtained with theoretical predictions at
NNLO or higher and combined using the $\chi^2$ averaging method~\cite{Olive:2016xmw}.
The 2015 PDF4LHC combined PDF sets are based on a slightly different value of
$\alpha_s(m_Z)=0.1180\pm 0.0015$~\cite{Butterworth:2015oua}, that is,
rounded to a value that is often
used in global fits and with a
 somewhat more conservative uncertainty band. 
Individual PDF groups also extract values of $\alpha_s(m_Z)$ including 
uncertainties solely from their global analyses.

The choice of strong coupling constant affects a global PDF analysis in two principle ways, through
 the DGLAP evolution of the PDFs themselves, and through the perturbative QCD predictions for the processes that enter the fit.
To study these effects, a scan over different values of $\alpha_s(m_Z)$ is typically performed.
For each choice
of $\alpha_s$, the best--fit PDFs are found and the $\chi^2$ profile
is constructed. 
The best--fit value of $\alpha_s(m_Z)$ is then identified and the uncertainty on this can be determined in a similar way to the standard PDF uncertainties, using either a `$\Delta\chi^2=1$' or a tolerance criteria.

\begin{figure}[t]
\begin{center}
  \includegraphics[scale=0.57]{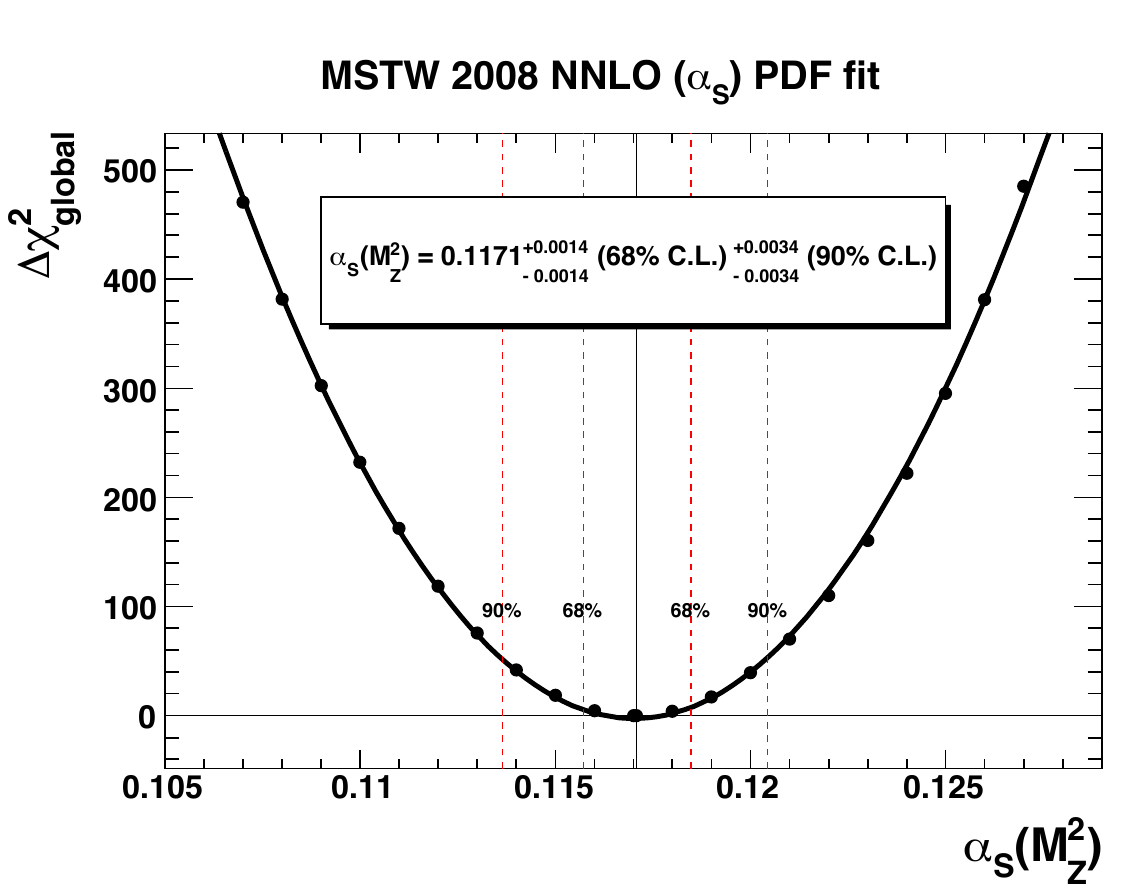}
  \includegraphics[scale=0.6]{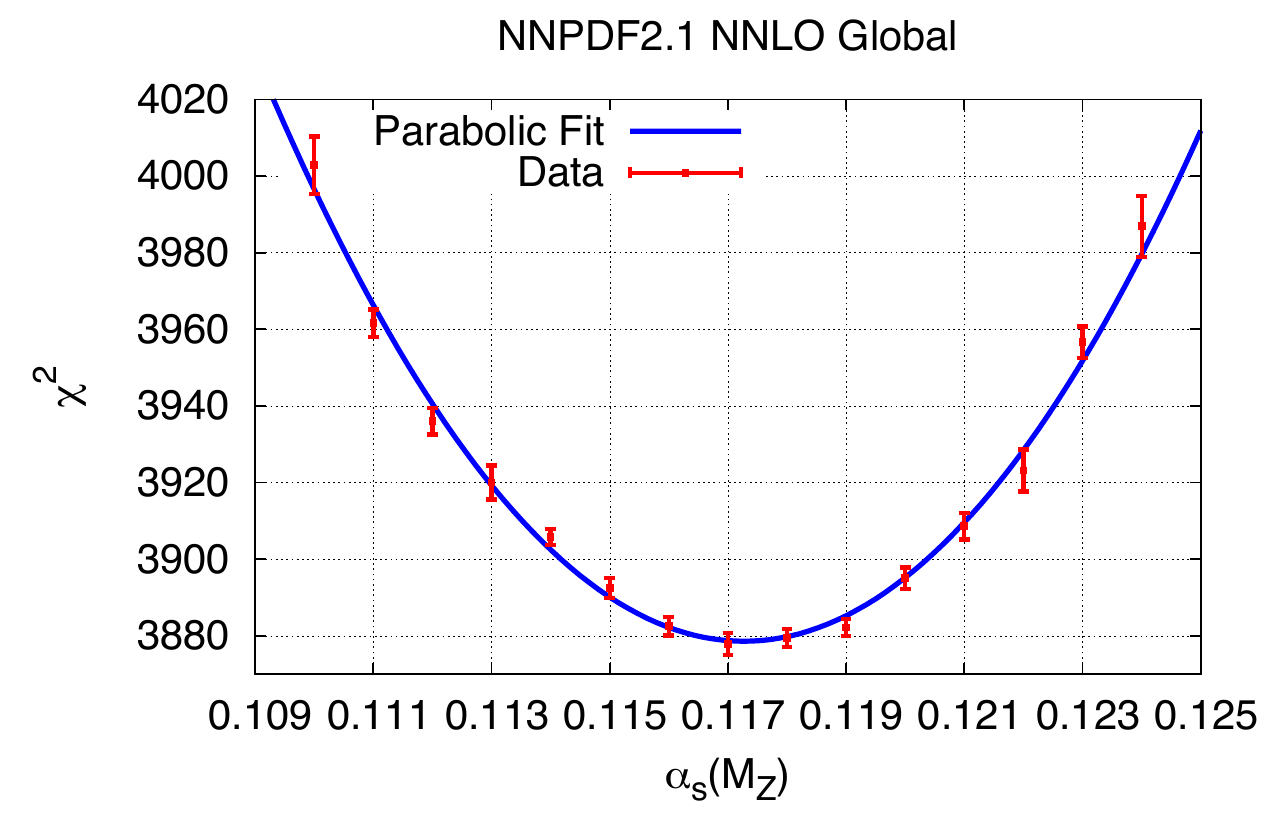}
   \caption{\small
  Left plot: the profile of global $\chi^2$ as in a scan of $\alpha_s(m_Z)$
  in MSTW2008 NNLO analysis~\cite{Martin:2009bu}.
  Right plot: the profile of global $\chi^2$ as in a scan of $\alpha_s(m_Z)$
  in NNPDF2.1 NNLO analysis~\cite{Ball:2011us}, where the error bars indicate fluctuations of
  the $\chi^2$ due to finite number of MC replicas used for each $\alpha_s$ value. 
    \label{fig:as1}
  }
\end{center}
\end{figure}

As an illustrative example, Fig.~\ref{fig:as1} shows the $\chi^2$ profiles from the MSTW and NNPDF NNLO global analyses. 
The extracted $\alpha_s(m_Z)$ values at NNLO are $0.1171\pm 0.0014$~\cite{Martin:2009bu}
and $0.1173\pm 0.0007$~\cite{Ball:2011us,Lionetti:2011pw} respectively. 
The CT and ABMP groups have also extracted values of the strong coupling,
finding at NNLO the values $0.115\pm 0.003$~\cite{Dulat:2015mca}
and $0.1147\pm 0.0008$~\cite{Alekhin:2017kpj}, that is, with rather lower central values than the MSTW and NNPDF
determinations.
The error reported by the CT group is much larger than the other groups due to 
the weaker tolerance condition used, in a way that is consistent within uncertainties
with the results from the other groups.
There is therefore a large
spread in the best--fit values from the different PDF groups, and so
the combined $0.1156 \pm 0.0021$ which enters the world average has a much larger error than
those reported by the individual groups.
At NLO, the global analyses prefer a $\alpha_s(m_Z)$
value that is about $0.002\sim 0.003$ higher than at NNLO,
compensating for the missing higher--order corrections.

%%%%%%%%%%%%%%%%%%%%%%%%%%%%%%%%%%%%%%%%%%%%%%%%%%%%%%%%%%%%%%%%%%%%%
%%%%%%%%%%%%%%%%%%%%%%%%%%%%%%%%%%%%%%%%%%%%%%%%%%%%%%%%%%%%%%%%%%%%%

The choice of a value for the strong coupling constant $\alpha_s(m_Z)$ obviously
has a significant impact on the predictions for various important processes at
hadron colliders, such as Higgs boson production via gluon fusion and
top quark pair production, both of which are proportional to
$\alpha_s^2$ at LO.
As mentioned above, here it is crucial to account for the correlations
between $\alpha_s$ and PDFs when evaluating the full uncertainties
of observables.
For example, it is well known that the gluon PDF is anti--correlated
with $\alpha_s$ in the small and intermediate $x$ regions due to the
constraints from scaling violations of inclusive structure functions,
which can partly compensate the change of cross sections due to change of
$\alpha_s$ in the matrix elements.

In principle, it is possible within a global PDF analysis to treat $\alpha_s(m_Z)$ in
exactly the same way as other PDF parameters, {\it e.g.}, in the Hessian method,
by calculating the full Hessian matrix and then determining the eigenvector directions
and the uncertainties along each direction. 
Then, the PDF+$\alpha_s$ uncertainty on any observable can
be evaluated using the Hessian error PDFs, in exactly the same way as for the standard case
where only PDF uncertainties are included. 
The ABM and later ABMP group follows
precisely this procedure. 
The downside of this approach is that it is
not possible to separate the PDF and $\alpha_s$
uncertainties, and each error PDF will be associated with a different
value of $\alpha_s$.

\begin{figure}[t]
\begin{center}
  \includegraphics[scale=0.38]{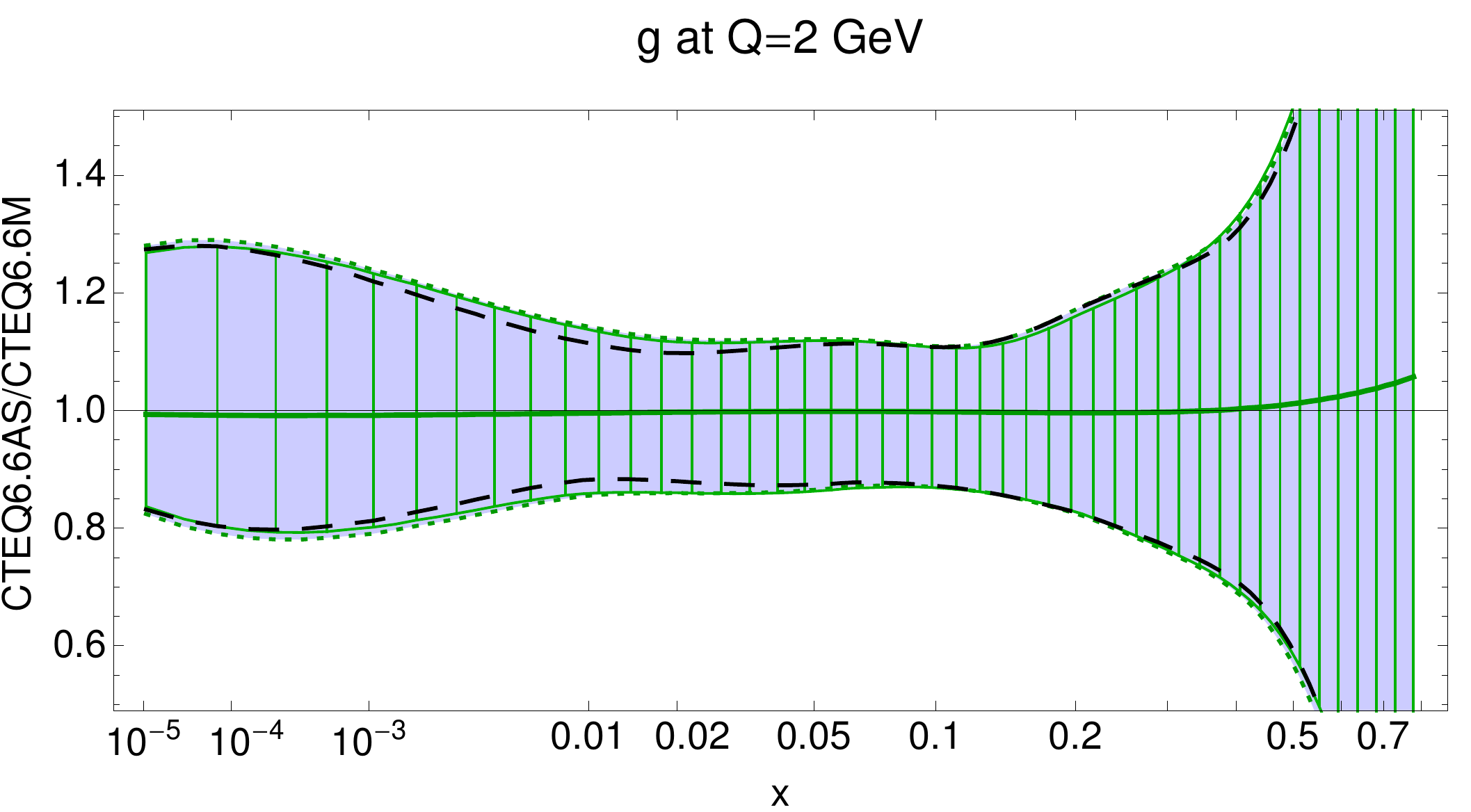}
  \includegraphics[scale=0.38]{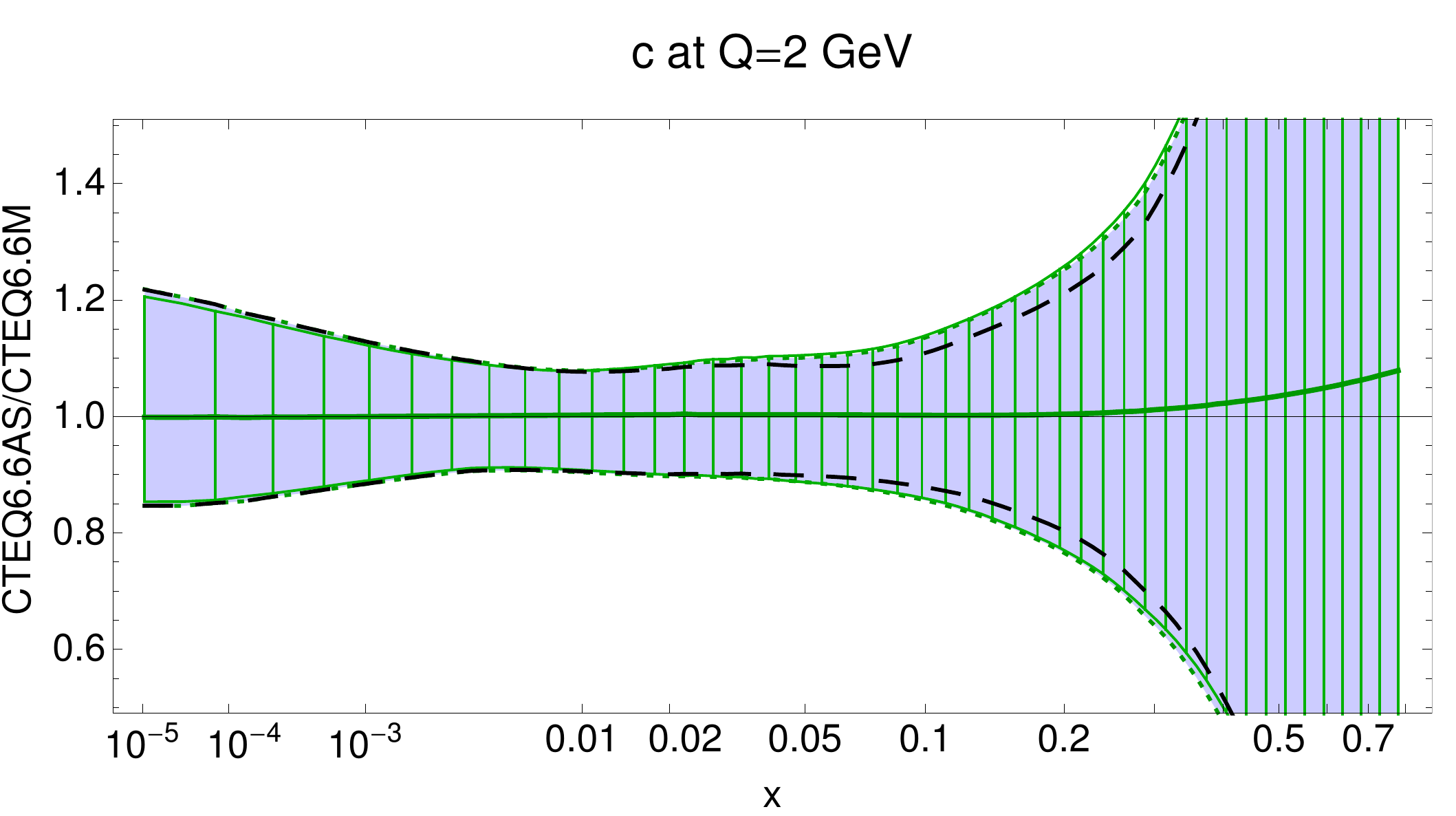}
   \caption{\small
   Comparison of the combined PDF+$\alpha_s$ uncertainties for the gluon and charm quark
   PDFs using the full eigenvectors including $\alpha_s(m_Z)$ in the Hessian
   matrix (filled error band with dotted borders) and with the separate PDF and
   $\alpha_s$ uncertainties added in quadrature
   (hatched band with solid borders) from~\cite{Lai:2010nw}.
Here the dashed lines
   indicate the PDF uncertainties only.
    \label{fig:as2}
  }
\end{center}
\end{figure}

A more convenient but completely equivalent method has been proposed in~\cite{Lai:2010nw}.
There, it has been shown that, under the quadratic approximation for the
$\chi^2$, the full PDF+$\alpha_s$ uncertainty can be calculated by
simply adding the usual PDF uncertainty and the $\alpha_s$ uncertainty in
quadrature, with the eigenvectors for the PDF uncertainties 
constructed with $\alpha_s$ fixed to its best--fit value.
The $\alpha_s$ uncertainty is then calculated through one additional
eigenvector (with two directions) constructed by fixing $\alpha_s(m_Z)$ to
its upper and lower limits and then fitting the remaining PDF parameters in the usual way.
The equivalence of the above two approaches is shown in Fig.~\ref{fig:as2} for the gluon and charm quark PDFs. 
This latter approach is now adopted by CT, MMHT, and NNPDF collaborations
due to its simple form and ease of use.   
Note that the upper and lower
limits on $\alpha_s(m_Z)$ can come either from the fit itself, as in
the case of MMHT 2014, or can be chosen according to the world average, as 
in CT14 and NNPDF3.1. 
Modified ranges for the $\alpha_s$ uncertainty 
can easily be obtained via linear rescaling~\cite{Butterworth:2015oua}.  

%%%%%%%%%%%%%%%%%%%%%%%%%%%%%%%%%%%%%%%%%%%%%%%%%%%%%%%%%%%%%%%%%%%%%
%%%%%%%%%%%%%%%%%%%%%%%%%%%%%%%%%%%%%%%%%%%%%%%%%%%%%%%%%%%%%%%%%%%%%

\subsubsection{Heavy quark masses}
\label{sec:fitmeth.theoryunc.heavyq}
Global PDF fits also rely on the input values of the heavy quark (charm, bottom and top) masses.
For instance, in an analysis based on the GM--VFNS (see Sect.~\ref{sec:global.heavyq}), the
charm and bottom quark masses enter through the running of $\alpha_s$,
the boundary conditions for the switching of active flavours, as well as the predictions for the
inclusive DIS structure functions and for open charm/bottom production
in DIS or hadron--hadron collisions.
On the other hand, the dependence on the top quark mass is less pronounced unless top quark
production data itself are included in the analysis.
In this case,  since processes such as  $gg\to t\overline{t}$ are strongly dependent on both
$\alpha_s$, the top quark mass, and the gluon PDF, it is important
to account for the non--trivial correlation
between the values of $m_t$ and $\alpha_S$ used in the fit.

For the charm and bottom masses,
the world averages and the corresponding
uncertainties for $m_c$ and $m_b$ in $\overline {\rm MS}$ scheme are~\cite{Olive:2016xmw}
\begin{equation}
m_c(m_c)=1.27\pm 0.03\,{\rm GeV}\,\,{\rm and}\,\,m_b(m_b)=4.18\pm 0.035\,{\rm GeV} \,,
\end{equation}
which can be translated into the following pole masses as described in~\cite{Harland-Lang:2015qea}      
\begin{equation}
m_c^{\rm pole}=1.5\pm 0.2\,{\rm GeV}\,\,{\rm and}\,\,m_b^{\rm pole}=4.9\pm 0.2\,{\rm GeV} \,,
\end{equation}
by using the 3--loop conversion between the pole and  $\overline {\rm MS}$
values for the bottom quark mass, together with the known relation between
the bottom and charm masses~\cite{Hoang:2005zw}.
The larger uncertainties arise here
due to the fact that the pole mass is  not well completely well defined, due to the diverging series, i.e. there is a renormalon ambiguity of order $\sim 0.1-0.2$ GeV.
On the other hand, this effect largely cancels in the difference of the two masses, and therefore the above uncertainties are highly correlated.

The majority of PDF groups use the pole mass as input, as
the relevant coefficient functions and matrix elements are calculated in
the on--shell scheme.
In particular, CT14 takes a default value of $m_{c(b)}=1.3~(4.75)$ GeV,
MMHT14 takes $1.4~(4.75)$ GeV and NNPDF3.1 takes $1.51~(4.92)$ GeV.
Both the CT14 and MMHT 2014 NNLO analyses prefer a lower charm quark mass of
about 1.3 GeV~\cite{Dulat:2015mca,Harland-Lang:2015qea} if it is treated as
a free parameter in the fit, which is
consistent with the converted value from the $\overline {\rm MS}$ world average value.
In ABMP16 the $\overline {\rm MS}$ masses are extracted directly from the
fit, giving $m_c(m_c)=1.252\pm 0.018$ GeV and $m_b(m_b)=3.84\pm 0.12$ GeV~\cite{Alekhin:2017kpj}.   

As in the case of the strong coupling constant,
the uncertainty due to the heavy quark masses
can be calculated by constructing an additional eigenvector from fits with
alternative mass values.
The full uncertainty  can then be obtained by adding it in quadrature to the
PDF uncertainty obtained with the default choice of heavy quark masses.
For example, the CT14, MMHT 2014 and NNPDF3.1 analyses all provide a series of best--fit
PDFs with $m_c$ or $m_b$ fixed to alternative values around their default choices.
However, there has so far not been an agreement on a common choice of the
heavy quark masses and their errors in global PDF analyses, although this is foreseen for the next PDF4LHC recommendation.

\begin{figure}[t]
\begin{center}
  \includegraphics[scale=0.3]{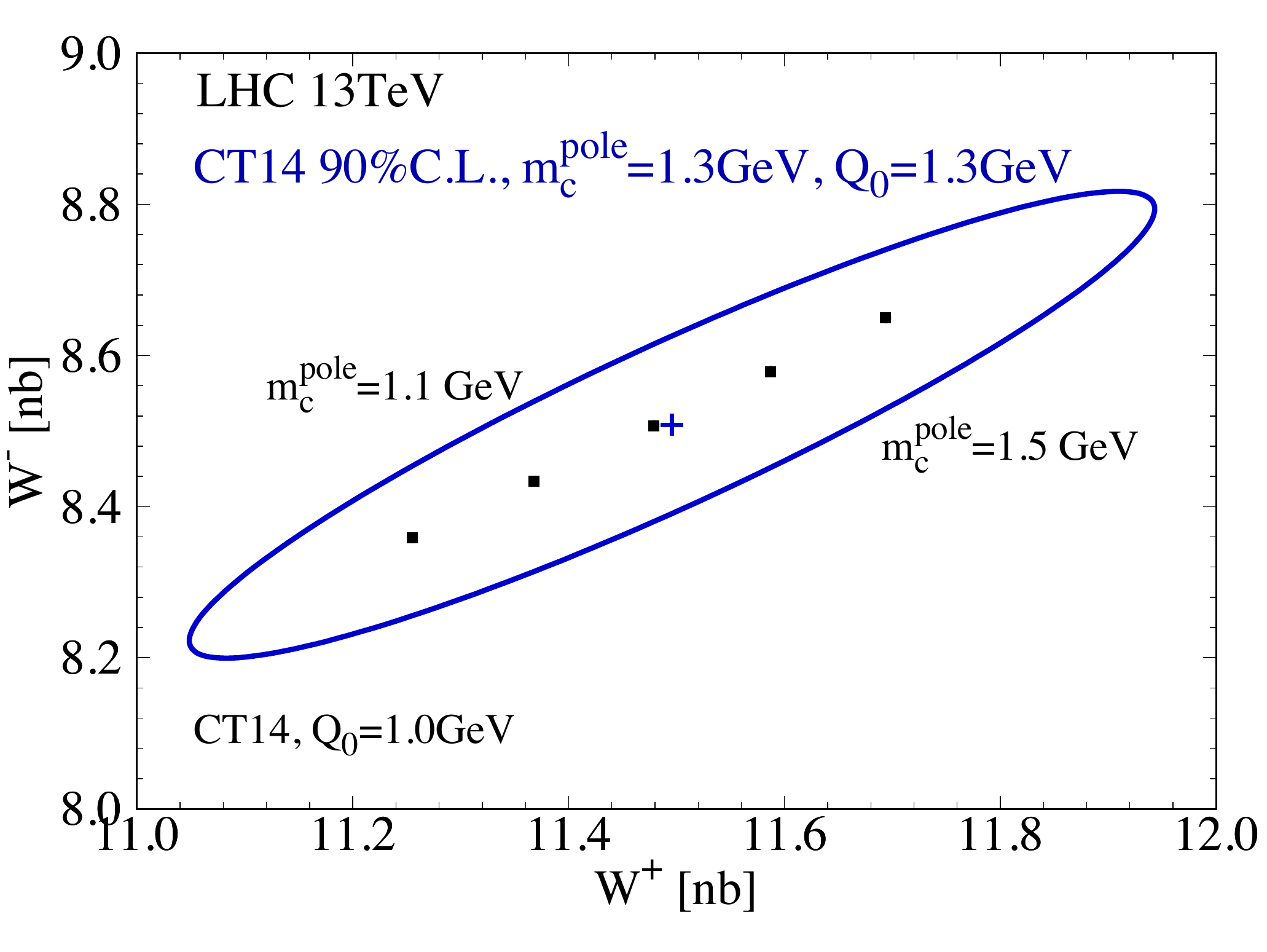}
  \hspace{0.3in}
  \includegraphics[scale=0.3]{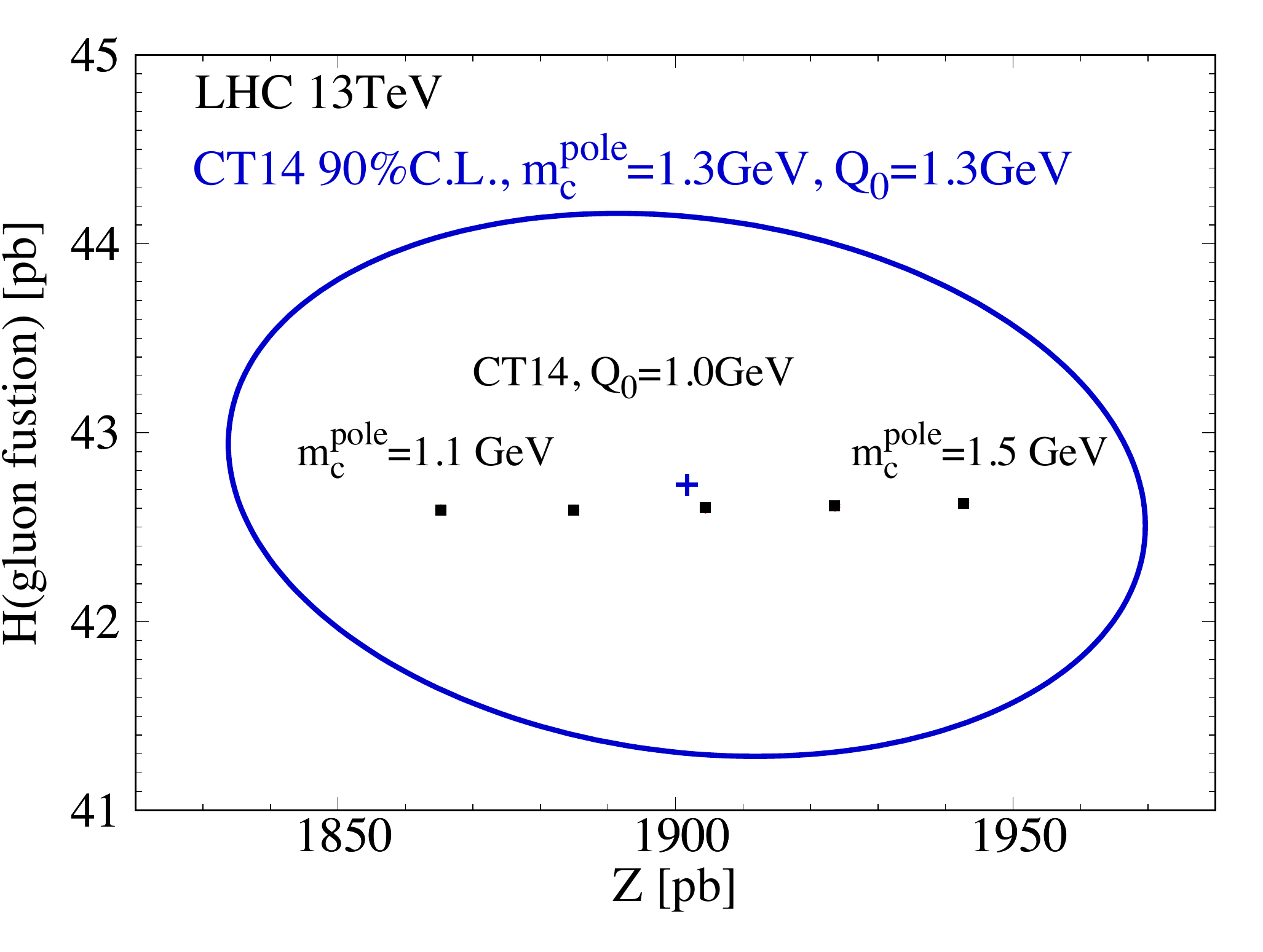}
   \caption{\small
   Dependence of the total cross sections for weak boson and Higgs boson production
   at the 13 TeV LHC on the choice of charm quark mass in the CT14 NNLO analysis~\cite{Hou:2017khm}.
   The dots correspond to CT14 NNLO fits with charm quark mass of 1.1-1.5 GeV and
   an initial parametrization scale of $Q_0=1.0$ GeV.
   The ellipse indicates the PDF uncertainties at 90\% C.L.
    \label{fig:as3}
  }
\end{center}
\end{figure}

The dependence of the predicted total cross sections for electroweak gauge
boson
and Higgs boson production at the 13 TeV LHC on the choice of charm quark pole mass
used in the CT14 NNLO analysis~\cite{Hou:2017khm} is shown in Fig.~\ref{fig:as3}.
It is found that varying $m_c$ by 0.2 GeV has a negligible effect on the Higgs boson cross section
and induces at most a 2\% change in the weak boson cross sections. This is well
within the PDF uncertainties.
In the MMHT 2014 analysis~\cite{Harland-Lang:2015qea} larger uncertainties are found, but these are still in general small compared to the PDF uncertainties.
In addition, in both the CT14 and NNPDF3.1 analyses, it is observed that the
effect of varying $m_c$ can be partly canceled by changes in the
non--perturbative component of the charm PDFs, since it is re--absorbed into
the fitted charm boundary condition.
 
%%%%%%%%%%%%%%%%%%%%%%%%%%%%%%%%%%%%%%%%%%%%%%%%%%%%%%%%%%%%%%%%%%%%%
%%%%%%%%%%%%%%%%%%%%%%%%%%%%%%%%%%%%%%%%%%%%%%%%%%%%%%%%%%%%%%%%%%%%%

\subsection{Approximate methods}\label{sec:fitmeth.approx}
In this section we discuss two approximate methods that can, under certain
circumstances, be used instead of a full-fledged fit to estimate the
impact on the PDFs of new experimental measurements.
These two methods go under the name of
the Bayesian reweighting of MC
replicas~\cite{Ball:2011gg,Ball:2010gb} and the profiling
of Hessian sets~\cite{Camarda:2015zba}.
The main advantage of these two techniques is that they can be used  to quantify
the impact of new experiments on a pre--existing fit based only on publicly available
information, in particular the {\tt LHAPDF} PDF grids, and thus
such studies can be carried out outside the PDF fitting collaborations.

On the other hand, these methods have a number of limitations. In particular,
they can fail if the impact of the new data is too large, and moreover
are not able to account for the effect of methodological changes, such as the input
PDF parametrization, or of modifications in the theoretical calculations.
Therefore, they represent an important addition to the PDF fitting toolbox,
but some care should be taken when using them and when interpreting
their results.

\subsubsection{Bayesian Monte Carlo reweighting}

The Bayesian reweighting
method developed in Refs.~\cite{Ball:2011gg,Ball:2010gb} can be applied
to a native Monte Carlo PDF set to quantify the impact at the PDF level
of a new experimental measurement.
The basic idea is that, starting from a sample of
$N_{\rm rep}$ MC replicas, each of equal weight,
the compatibility of each replica with the new
experimental dataset can be quantified by computing  a series of new weights
for each replica.
In this approach, the weight of the $k$-th replica is given by
\be
\label{eq:RWweights}
\omega_k = \frac{\lp \chi_k^2\rp^{(n-1)/2}e^{-\chi^2_k/2}/N_{\rm rep}}{\frac{1}{N_{\rm rep}}
\sum_{k=1}^{N_{\rm rep}}\lp \chi_k^2\rp^{(n-1)/2}e^{-\chi^2_k/2}} \, , \quad
k=1,\ldots,N_{\rm rep} \; ,
\ee
where $\chi^2_k$ is the goodness--of--fit estimator between the replica
$k$ and the new experimental measurement.
Thus, if for a given replica the agreement with the new
experiment is very poor, its $\chi^2_k$ will be large, and thus
the weight of this specific replica will be exponentially suppressed.
Note that by definition these new weights $\omega_k$
are appropriately normalized, and from the statistical point of view, they
can be interpreted as the probability of the replicas $f_k$, given
the $\chi^2_k$ for the new experimental measurement.
The validity of the Bayesian MC method
 has been explicitly demonstrated by comparing
the reweighted results with those of a direct refit, finding
good agreement in all cases.

One of the limitations of the Bayesian reweighting method is that
it entails a loss of information compared to the initial prior,
because some of the original $N_{\rm rep}$ MC replicas will carry
a very small weight, meaning that they have been effectively discarded.
One suitable estimator to quantify this efficiency loss is the so--called
Shannon entropy, which allows  the effective number
of replicas left out after the reweighting to be evaluated.
Following standard information theory, the Shannon entropy of the reweighted
set, or more precisely, the effective number of replicas, is given by
\be
N_{\rm eff}\equiv \exp \lc \frac{1}{N_{\rm rep}}\sum_{k=1}^{N_{\rm rep}}
\omega_k \ln \lp N_{\rm rep}/\omega_k\rp\rc \; ,
\ee
where by construction, $0\le N_{\rm eff}\le N_{\rm rep}$.
The interpretation of this effective number of replicas is that
a reweighted PDF set carries the same amount of information
as a direct refit based on $N_{\rm eff}$ replicas.
Clearly, the smaller $N_{\rm eff}$ is, the more the new dataset constrains the
PDFs, but on the other hand if $N_{\rm eff}$ becomes
small enough, the reweighting method loses validity and a full
refit becomes necessary.
The Bayesian reweighting approach has been applied in a variety of 
PDF studies, both
for proton (unpolarized and polarized) PDFs,
see for instance
Refs.~\cite{d'Enterria:2012yj,Gauld:2016kpd,Czakon:2013tha,
Watt:2013oha,Nocera:2014gqa},
and for nuclear PDFs~\cite{Armesto:2013kqa}. 

An advantage of this Bayesian reweighting method is that
it provides a way to estimate if the experimental uncertainties
have been either underestimated or overestimated, assuming that theoretical
uncertainties are under control.
To achieve this, we can rescale the total experimental
uncertainties of the data by a factor $\alpha$, and then
use inverse probability in order to evaluate the probability
density associated to the rescaling parameter $\alpha$, namely
\be
\label{eq:palpha}
\mathcal{P}\lp \alpha\rp \propto \frac{1}{\alpha}\sum_{k=1}^{N_{\rm rep}}
\omega_k(\alpha) \; ,
\ee
where the weights $\omega_k(\alpha)$ are computed using
Eq.~(\ref{eq:RWweights}) but replacing $\chi^2_k$ by $\chi^2_k/\alpha^2$,
and therefore represent the probability of $f_k$ given the new data
with rescaling error.
If this probability density Eq.~(\ref{eq:palpha}) peaks far above (below)
one, then this suggest that the uncertainties in the data have been
under (over) estimated, providing a useful handle to assess the
compatibility of a new measurement with a prior PDF analysis.

It is important to point out here that for
native Hessian PDF sets, as well as for Monte Carlo sets converted from
Hessian sets using the procedure of~\cite{Watt:2012tq},
the weighting Eq.~(\ref{eq:RWweights}) might not be the optimal
one.
Indeed, a different
functional form for the weights has been advocated, see for instance~\cite{Giele:1998gw,Giele:2001mr,Sato:2013ika}
and the discussion in~\cite{Paukkunen:2014zia,Kusina:2016fxy}.
For example, in the nCTEQ analysis of nuclear parton distributions~\cite{Kusina:2016fxy},
it has been shown that using a weight of the form,
\be
\omega_k = \frac{e^{-\chi^2_k/(2T)}}{ \sum_k e^{-\chi^2_k/(2T)}/N_{\rm rep}}
\, , \quad
k=1,\ldots,N_{\rm rep} \; ,
\ee
namely a modified Giele-Keller expression, leads to a reasonable agreement
with the corresponding fit results,
where $T$ is the tolerance criterion used when defining the Hessian
error PDFs.
Note also that the application of the naive GK reweighting method
would fail in this case, since they do not
account for the tolerance $T\ne 1$ adopted in the original Hessian fit.

\begin{figure}[t]
\begin{center}
  \includegraphics[scale=0.48]{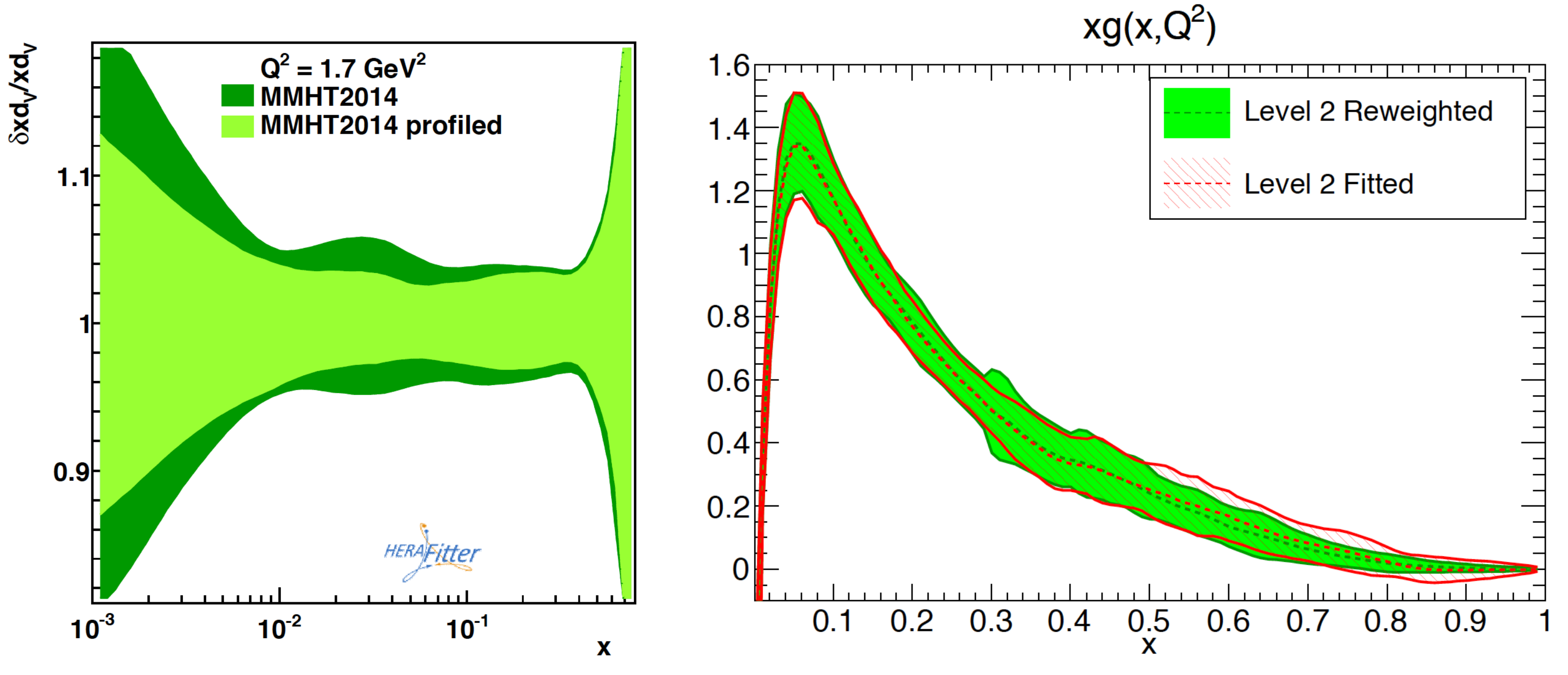}
   \caption{\small 
     Left plot: the impact of the Tevatron $W$ and $Z$ data
     on the MMHT2014 NLO fit, estimated by the Hessian
     profiling method of Ref.~\cite{Camarda:2015zba}.
     Right plot: the gluon PDF in the NNPDF3.0 closure tests,
     quantifying the impact of the collider inclusive jet data,
     and comparing the results of the Bayesian reweighting
     with those of a direct refit. 
    \label{fig:rw}
  }
\end{center}
\end{figure}

In order to provide an illustrative example of the
Bayesian reweighting method applied to native Monte Carlo sets,
in Fig.~\ref{fig:rw} we show the gluon PDF in the NNPDF3.0
closure tests~\cite{Ball:2014uwa},
     estimating the impact of the collider inclusive jet data
     and
     comparing the results of the Bayesian reweighting
     with those of a direct refit.
     In this study, the prior was a set of $N_{\rm rep}=1000$ replicas
     obtained with NNPDF2.3--like dataset but without any
     collider inclusive jet production data included.
     The pseudo--data were generated using the MSTW08 NLO set,
     though similar results were obtained with
     other priors.
     We observe that there is good agreement between the
     approximate Bayesian reweighting method and the exact refit
     results.

%%%%%%%%%%%%%%%%%%%%%%%%%%%%%%%%%%%%%%%%%%%%%%%%%%%%%%%%%%%%%%%%%%%%%
%%%%%%%%%%%%%%%%%%%%%%%%%%%%%%%%%%%%%%%%%%%%%%%%%%%%%%%%%%%%%%%%%%%%%

\subsubsection{Hessian profiling}

For a Hessian PDF set, the so--called
profiling technique provides a method to approximately
quantity the impact of a new experimental measurement.
This method is based on the minimization of a
 $\chi^2$ estimator that compares
the theoretical predictions obtained with a given input Hessian
PDF set with the new experimental measurements.
This estimator takes into account both the experimental
uncertainties and the effects from the PDF variations (as encoded
by the Hessian
eigenvectors) and is defined as follows:
\be
\label{eq:hessianchi2}
\chi^2\lp {\rm \beta_{exp}},{\rm \beta_{th}}\rp
=\frac{1}{\Delta_i^2}\sum_{i=1}^{N_{\rm dat}}\lp \sigma_i^{\rm exp}
+\sum_j\Gamma_{ij}^{\rm exp}\beta_{j,\rm exp}
 -\sigma_i^{\rm th}
 +\sum_k\Gamma_{ik}^{\rm th}\,\beta_{k,\rm th} \rp^2
 +\sum_j \beta_{j,\rm exp}^2+\sum_k \beta_{k,\rm th}^2 \; ,
 \ee
 where $\beta_{j,\rm exp}$ are the nuisance parameters corresponding
 to the set of fully correlated experimental systematic
 uncertainties, and $\beta_{k,\rm th}$ are the nuisance parameters
 corresponding to the PDF Hessian eigenvectors.
 $\Delta_i$ is the total experimental
 uncorrelated uncertainty, and $N_{\rm dat}$ is the number of data
 points of the measurement which is being added into the
 PDF fit.
 Finally, the matrices $\Gamma_{ij}^{\rm exp}$ and
 $\Gamma_{ik}^{\rm th}$ encode the effects of the corresponding
 nuisance parameters on the experimental data and on the
 theory predictions, respectively.
 
 Upon minimization of the $\chi^2$ estimator Eq.~(\ref{eq:hessianchi2}),
 the corresponding values of the theoretical nuisance parameters,
 denoted by 
 $\beta_{k,\rm th}^{\rm min}$, can be interpreted as leading
 to PDFs that have been optimized (hence the name ``profiled'')
 to describe this new specific measurement.
 Note also that in general the profiling will modify both
 the central value and the total PDF uncertainty.
 For example, the new measurement might reduce the allowed
 range of variation of a given eigenvector, if the original
 variation leads to large values of the figure of merit
 Eq.~(\ref{eq:hessianchi2}).

As in the case of the Bayesian reweighting method described
in the previous section,
there are a number of limitations of the Hessian profiling method
that limit the cases where it can be used to replace
a complete refit.
 First, it assumes that the optimal PDF parametrization will
 not be modified by the addition of the new data.
 It is well known that this condition does not necessarily holds,
 as new experiments
 might require the use
 of more flexible input PDF parametrizations in order
 to achieve an optimal description, and this effect cannot
 be accounted for with the profiling method.
 Secondly, the standard version of the Hessian profiling method
 assumes that the PDF uncertainties are defined by the
 $\Delta\chi^2=1$ criterion, which is generally not the case for global
 Hessian PDF fits, see Sect.~\ref{sec:pdfuncertainties}.
 For this reason, the impact of the data as estimated
 by Hessian profiling will in general differ in comparison
to the result of a full refit.
 However, this limitation can be eliminated by using a tolerance criterion
 that mimics the one used in the prior Hessian PDF set,
 see for example Ref.~\cite{Kusina:2016fxy}.

As an example of the applications of the profiling
method, in Fig.~\ref{fig:rw} we show 
the impact of the Tevatron $W$ and $Z$ data
     on the MMHT2014 NLO set from Ref.~\cite{Camarda:2015zba},
     estimated by Hessian
     profiling.
This comparison shows
that these measurements lead
     to a reduction of the PDF uncertainties in the down valence PDF
     $d_V(x,Q)$.
     An important point to emphasise here is that this exercise was performed
     using completely public tools, in this case the
     MMHT2014 {\tt LHAPDF} grids, and the experimental
     information of the Tevatron  $W$ and $Z$ measurements,
     without any additional input from the authors
     of the original MMHT2014 analysis.
    This is the main advantage of both the Hessian profiling  and  Bayesian reweighting techniques. That is, they allow for the possibility of carrying
    out PDF studies without the need of first having to produce a complete
    baseline global PDF fit.

%%%%%%%%%%%%%%%%%%%%%%%%%%%%%%%%%%%%%%%%%%%%%%%5

\subsection{Public delivery: {\tt LHAPDF}}\label{subsec:fitmeth.delivery}

The last step of any PDF analysis is of course to make it publicly available.
In the earlier days of PDF fitting, this was achieved by means of $(x,Q)$ interpolation tables
and corresponding driver codes.
Back then, the specific format of these tables and of the driver codes
were specific to each PDF group.
This was however far from optimal, since standardization was rather difficult,
with programs requiring PDFs as input having to be adapted each time a new
PDF set was released.
A first step towards PDF access standardization was achieved with the
release in 1993 of {\tt PDFLIB}~\cite{PlothowBesch:1992qj} as part of the CERN
Program Library software.
This allowed a unique interface for calling PDFs to be used without the need
to add external files on a case by case basis.
In addition to the PDFs, the value of $\alpha_s(Q)$ used in each specific fit
could also be accessed.

The next step in this standardisation process came with the release in 2005 of {\tt LHAPDF},
the {\it Les Houches Accord on PDFs}~\cite{Whalley:2005nh,Bourilkov:2006cj}, which was developed
as a functional replacement for {\tt PDFLIB}.
In order to ensure backwards compatibility, {\tt LHAPDF} included {\tt LHAPDF}Glue, a PDFLIB-like
interface.
One of the main motivations to release {\tt LHAPDF} was the realisation that dealing with
the large number of error PDF sets that had then recently became available was
extremely cumbersome with {\tt PDFLIB}.
In particular, {\tt LHAPDF} was organised around the concept of {\it PDF set}, which
was constituted by the central (average) member as well as the corresponding error PDF sets.
As in the case of {\tt PDFLIB}, {\tt LHAPDF} was written in Fortran 77, although later a {C/C++} interface was
also developed.

While the Fortran incarnation of {\tt LHAPDF} was very popular and widely used, at some point
its further development became very challenging in particular due to the intrinsic
limitations of Fortran 77 as its native language.
In particular, since Fortran 77 required space for all available
PDFs to be allocated at compilation time, the memory footprint eventually become impossible to handle
and {\tt LHAPDF} v5.9.1 was the last release.
To overcome these limitations, a complete rewriting of {\tt LHAPDF} from scratch in
{\tt C++} was completed in 2014, dubbed {\tt LHAPDF6}~\cite{Buckley:2014ana}.
In addition to reducing static memory requirements by orders of magnitude,
this {\tt C++} incarnation of {\tt LHAPDF} offered improved CPU
performance and improved interpolation and extrapolation
functionalities.
Moreover, its cascading meta-data system ensured that software releases
are completely decoupled from the availability of novel PDF sets.
To ensure backwards compatibility, Fortran 77 interfaces, still very
popular within many MC programs, were also
provided.

\begin{figure}[t]
\begin{center}
  \includegraphics[scale=0.45]{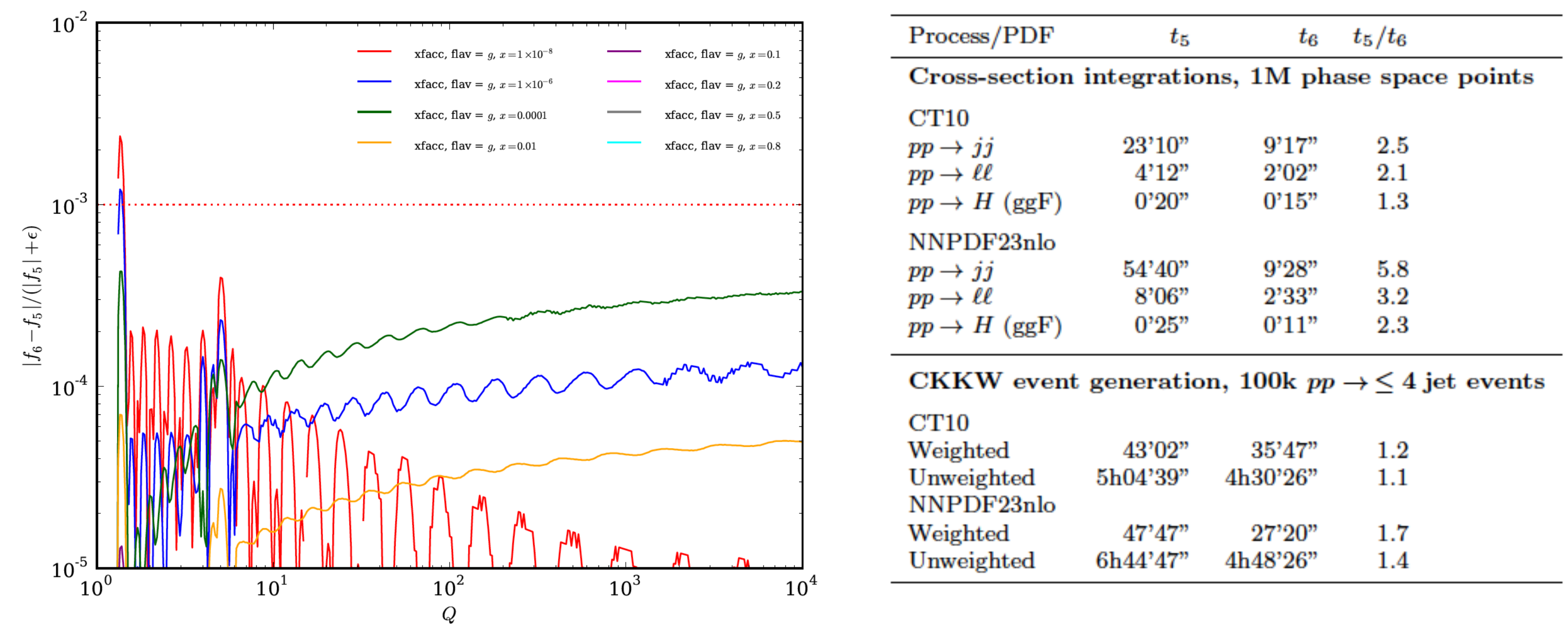}
   \caption{\small 
  Left plot: the relative difference between {\tt LHAPDF} v5 and v6
  for $g(x,Q)$ for different values of $x$ as a function of $Q$, using
  CT10 as input PDF.
  Right plot: the timing improvement in v6 as compared to v5, $t_5/t_6$,
  for a cross-section integration of 1M phase space points
  with {\tt Sherpa} and for CKKW event generation of
  100k $pp \to 4$ jet events.
    \label{fig:lhapdf6}
  }
\end{center}
\end{figure}

In terms of interpolation accuracy, {\tt LHAPDF6} reproduces the v5
results down to residual differences of at most 0.1\%.
This is illustrated in Fig.~\ref{fig:lhapdf6}, where we show
the relative difference between {\tt LHAPDF} v5 and v6
  for the gluon PDF at different values of $x$ as a function of $Q$, using
  CT10 as the input PDF.
  In Ref.~\cite{Buckley:2014ana} it was also shown that
  {\tt LHAPDF}6 improves the CPU time required by a factor of between
  2 and 6. This is seen in the right table in Fig.~\ref{fig:lhapdf6},
  which shows the timing improvements in v6 compared to v5, $t_5/t_6$,
  for a cross-section integration of 1M phase space points
  with {\tt Sherpa}~\cite{Gleisberg:2008ta} and for CKKW event generation of
  100k $pp \to 4$ jet events.
  The reason for this improvement is the adoption of a more efficient $(x,Q^2)$
  interpolation algorithm.
  Indeed, while in v5  each group had to provide their own
  interpolation code, in v6 the interpolation settings are universal.

%%%%%%%%%%%%%%%%%%%%%%%%%%%%%%%%%%%%%%%%%%%%%%%%%%%%%%%%%%%%%%%%%%%%%
%%%%%%%%%%%%%%%%%%%%%%%%%%%%%%%%%%%%%%%%%%%%%%%%%%%%%%%%%%%%%%%%%%%%%

Currently {\tt LHAPDF}6 has established itself as the almost universal
software with which to access PDFs.
Its current version is 6.2 and more than 700 PDF sets can be accessed.
In addition to unpolarized parton distributions, the flexibility
of the {\tt LHAPDF}6 framework makes it suitable to release other types
of non-perturbative QCD objects, and indeed polarized PDFs~\cite{Nocera:2014gqa},
nuclear PDFs~\cite{Eskola:2016oht}, and even
fragmentation functions~\cite{Bertone:2017tyb},
are also available.

%%%%%%%%%%%%%%%%%%%%%%%%%%%%%%%%%%%%%%%%%%%%%%%%%%%%%%%%%%%%%%%

%%%%%%%%%%
\vspace{0.6cm}
\section{PDF analyses: state of the art}\label{sec:pdfgroups}
In this section
we review the latest developments from the main PDF fitting groups, as well as the PDF efforts carried out
within the LHC collaborations.

\subsection{CT}\label{sec:pdfgroups.CT}

The CTEQ--TEA global analysis was established by Wu--Ki Tung and collaborators
in the early 1990s with
the CTEQ1 PDFs~\cite{Morfin:1990ck}.
The most recent release of general purpose PDFs from this
 collaboration are the CT14 PDF sets~\cite{Dulat:2015mca}, which include the nominal sets as well as
alternative sets with different choices of $\alpha_s$ and the maximum number of active flavours $n_f^{\rm max}$.
In the CT approach, the PDFs are parameterised at the starting scale $Q_0=1.3\, {\rm GeV}$ using the
functional form
(\ref{pdffunc}) described in Sect.~\ref{sec:fitmeth.PDFpara.func}.
In pre--CT14 analyses, the interpolating function $I_f(x)$ was
chosen as an exponential of a polynomial in $x$ or $\sqrt x$, such that positivity
conditions on the PDFs at the initial scale were enforced.
In the recent CT14 analysis, on the other hand, an improved  parametrization choice
was introduced, with for example for the $u$--valence quark
\begin{equation}
I_{u_v}(y)=d_0p_0(y)+d_1p_1(y)+d_2p_2(y)+d_3p_3(y)+d_4p_4(y)\; ,
\end{equation}   
where $y=\sqrt x$ and $p_n$ are the fourth order Bernstein polynomials, given by
\begin{equation}
p_0(y)=(1-y)^4, \,\,p_1(y)=4y(1-y)^3,\,\,p_2(y)=6y^2(1-y)^2,\,\,p_3(y)=4y^3(1-y),\,\,p_4(y)=y^4 \; .
\end{equation}
Namely, the interpolating function is chosen as a fourth--order polynomial in $y$
with an expansion in the basis of Bernstein polynomials. 
As discussed in Sect.~\ref{sec:fitmeth.PDFpara.func}, this greatly increases the stability of the fit within the Hessian approach.
Moreover, in the CT14
case the positivity of PDFs at $Q_0=1.3\,{\rm GeV}$ is not imposed {\it a priori} but rather
emerges automatically as a
consequence of the fit to data.
The CT14 PDFs have a total number of 28 free parameters; using a 
more flexible parametrization, by adding higher--order
polynomials, is found to have a small effect on both the best--fit and the estimated
PDF uncertainties in the region that is well constrained by the experimental data.

The CT14 global analysis includes a wide variety of experimental data.
The majority comes from the inclusive DIS and semi--inclusive DIS measurements of the structure functions
and the reduced cross section measurements from fixed--target experiments (BCDMS~\cite{Benvenuti:1989rh,Benvenuti:1989fm},
NMC~\cite{Arneodo:1996qe}, CCFR~\cite{Yang:2000ju,Seligman:1997mc,Goncharov:2001qe}, NuTeV~\cite{MasonPhD},
CDHSW~\cite{Berge:1989hr}) or
HERA experiments~\cite{Aktas:2004az,Abramowicz:1900rp,Aaron:2009aa,Collaboration:2010ry}.
A $Q$ cut of 2 GeV and $W$ cut of 3.5 GeV are
adopted in the selection of DIS data to minimize non--perturbative effects
from either nuclear corrections or higher--twists corrections.
Thus no further nuclear
or higher twists corrections are included in theory predictions in CT14 except for those
already applied in the unfolding of experimental data. For the NC DIS process, the CT14 analysis uses a
treatment of heavy quark mass effects up to NNLO, through a variant of the GM--VFNS
known as S--ACOT--$\chi$~\cite{Guzzi:2011ew}. 
For CC DIS, the theoretical calculations are
only implemented at NLO, which is judged to be sufficient given the relatively small number of data points and their large experimental errors.
Drell--Yan production data from
fixed--target experiments (E605~\cite{Moreno:1990sf}, E866~\cite{Towell:2001nh})
and $W/Z$ boson production data from Tevatron~\cite{Abe:1996us,Acosta:2005ud,Abazov:2007pm,
Abazov:2006gs,Aaltonen:2010zza} including
the new D0 electron charge asymmetry data~\cite{D0:2014kma}, are also fit.

\begin{figure}[t]
\begin{center}
  \includegraphics[scale=0.6]{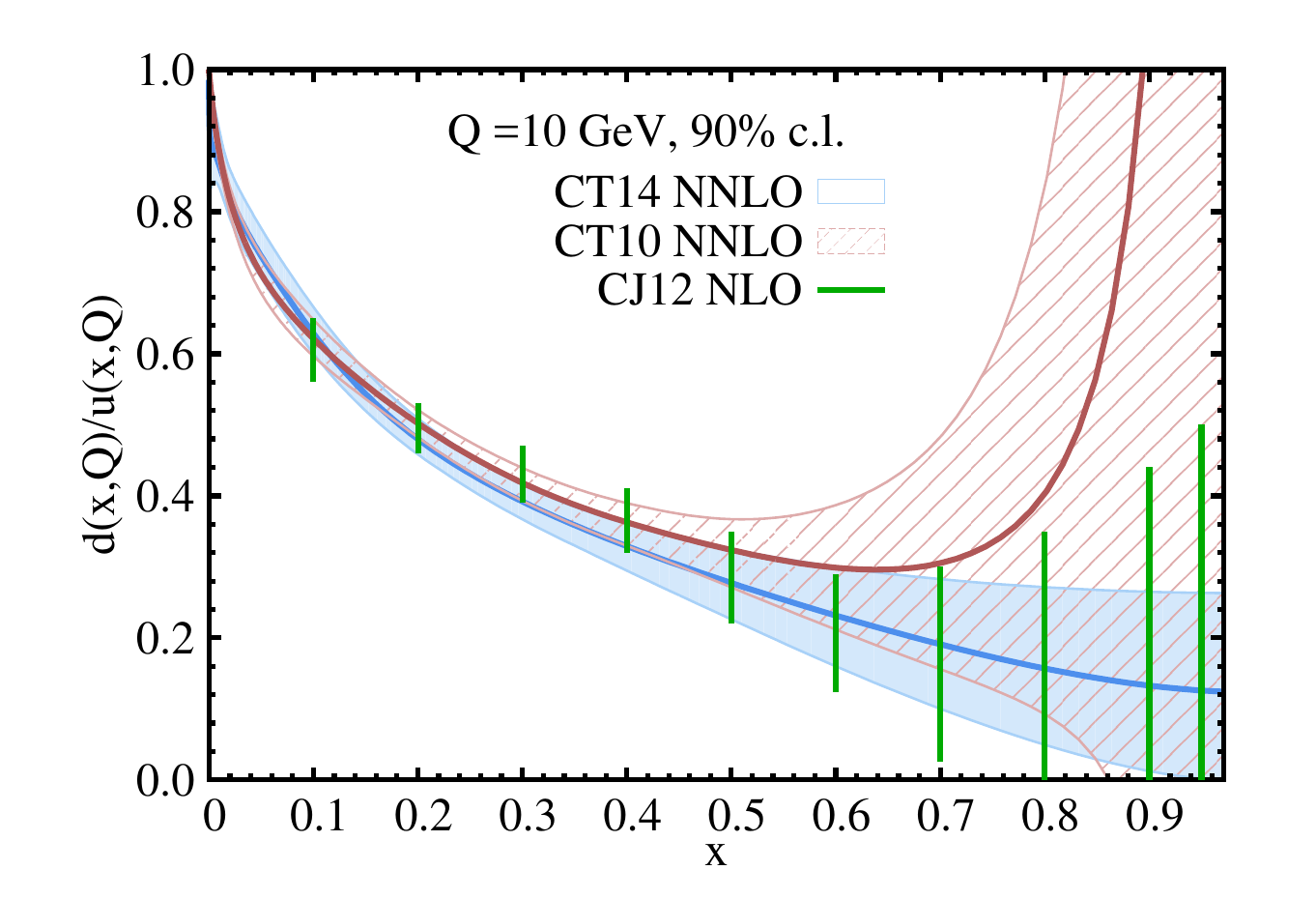}
   \caption{\small
   A comparison of the  $d/u$ ratio at $Q=10$ GeV for
   the CT14 NNLO (solid blue), CT10 NNLO (dashed red), and
    CJ12 NLO (green lines) analyses, together with the
    corresponding
90\% CL PDF uncertainties. From Ref.~\cite{Dulat:2015mca}. 
    \label{fig:ct1}
  }
\end{center}
\end{figure}

The Tevatron $W,Z$ data provide further discriminations on
quark flavours in large--$x$ region, with the $W$ asymmetry data
probing the average slope of $d/u$ ratio at $x\gtrsim 0.1$, see Sect.~\ref{sec:datatheory.gauge.sensitivity}
for a more detailed discussion.
Predictions from ResBos~\cite{Balazs:1995nz,
Balazs:1997xd,Landry:2002ix,Guzzi:2013aja} are used for the $W/Z$ boson
production data, with a $p_T$ cut imposed on the charged leptons, and incorporating soft gluon
resummation effects at small $p_T$ of the vector boson.
These resummed predictions provide
a better description of the $p_T$ spectrum of the charged leptons. The updated
D0 electron charge asymmetry data~\cite{D0:2014kma} shows a large impact on the $d/u$ PDF ratio at
large $x$ compared to CT10 and CT10W~\cite{Lai:2010vv,Gao:2013xoa}.
In the CT10 fits
the D0 lepton charge asymmetry data resulted in larger asymptotic value of $d/u$ though
tensions were found between different subsets of the data or the D0 data and other
DIS experiments.
As shown in Fig.~\ref{fig:ct1}, for CT14 the updated D0 electron charge asymmetry data shows better
agreements with other data sets in the global analysis and drives the $d/u$ ratio
to a lower value, close to CTEQ6.6~\cite{Nadolsky:2008zw} at large $x$.
The $d/u$ ratio in CT14 also shows good agreement with the extraction from CJ12~\cite{Owens:2012bv}, which is
based on independent large--$x$ and low--$W$ DIS data, including power corrections and deuteron corrections.    

Similar data on $W$ and $Z$ boson production from LHC Run I are also included from the ATLAS~\cite{Aad:2011dm},
CMS~\cite{Chatrchyan:2013mza,Chatrchyan:2012xt} and LHCb~\cite{Aaij:2012vn}
experiments, which further extend the coverage to the intermediate and small--$x$ region.
In addition, single inclusive jet production data from the Tevatron~\cite{Aaltonen:2008eq,Abazov:2008ae}
and the LHC~\cite{Aad:2011fc,Chatrchyan:2012bja} were fitted, providing the dominant constraint
on the gluon PDF at large $x$, with the latter extending the coverage to the intermediate
$x$ region.
For inclusive jet production at hadron colliders only the NLO predictions were available
at the time of the CT14 fit, and therefore these were used in the NNLO fit. This will be updated with the recent NNLO
calculations~\cite{Currie:2016bfm} in future instalments of the CTEQ--TEA global analysis.

%%%%%%%%%%%%%%%%%%%%%%%%%%%%%%%%%%%%%%%%%%%%%%%%%%%%%%%%%%%%%%%%%%%%%
%%%%%%%%%%%%%%%%%%%%%%%%%%%%%%%%%%%%%%%%%%%%%%%%%%%%%%%%%%%%%%%%%%%%%

\begin{figure}[t]
\begin{center}
  \includegraphics[scale=0.60]{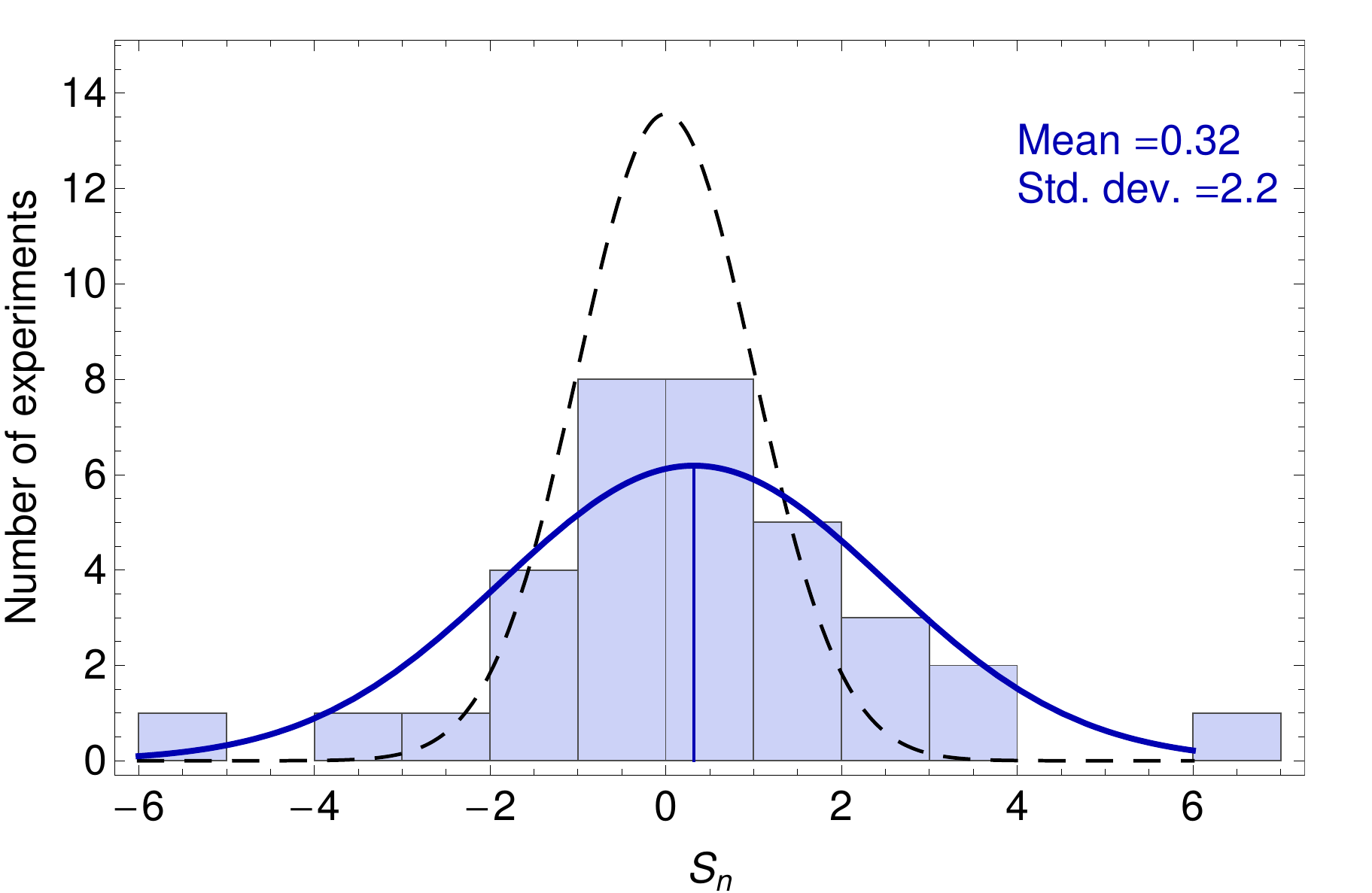}
   \caption{\small
   Best--fit values of the equivalent Gaussian variable $S_n$, Eq.~(\ref{sec:equivalent}),
   for all 33 experiments in the CT14 NNLO global analysis~\cite{Dulat:2015mca}. 
    \label{fig:ct2}
  }
\end{center}
\end{figure}

The CTEQ--TEA group uses the Hessian method with certain tolerance conditions for the
nominal fits in the determination of PDF uncertainties at 90\% CL, see Sect.~\ref{sec:pdfuncertainties}.
In addition, the Lagrange Multiplier (LM) scan method discussed in Sect.~\ref{sec:pdfuncertainties.Lagrange} is also adopted 
 for specific important observables or for PDFs in regions poorly constrained
by data. 
In the pre--CT10 analyses, it
was found that within a global $\chi^2$ tolerance of $\Delta \chi^2=100$ (for more than 2000 data
points) the fits agree with all experiments at 90\% CL.
In later CTEQ--TEA analyses, a more
efficient dynamic tolerance criteria is adopted.
It is constructed from an equivalent Gaussian variable, e.g.,
\begin{equation}
\label{sec:equivalent}
S_n=\sqrt{2\chi^2(N_n)}-\sqrt{2N_n-1}\; ,
\end{equation}
where $N_n$ is the total number of data points in data set $n$ and $\chi^2(N_n)$
represents the $\chi^2$ of the fit to that data set. $S_n$ follows a normal distribution
if the number of data points is large enough.
Thus a value of $S_n$ greater than
1.3 will be excluded at 90\% CL.
Subsequently,
a second layer of penalty is added to the global
$\chi^2$ when determining the boundaries of confidence intervals, called a Tier--2 penalty,
 defined as
\begin{equation}
P=\sum_{n=1}^{N_{\rm exp}}(S_n/S_{n,\rm best})^{16},
\end{equation}
where the sum runs over all data sets included and we normalize $S_n$ to its value in
the best--fit $S_{n,\rm best}$ to account for the poor fit to certain experiments.
The power of 16 is
introduced so that the penalty will reach the tolerance of 100 as soon as any data set
shows disagreement at 90\% CL.
The tolerance criteria then changes to $\Delta\chi^2$+P=100.
Fig.~\ref{fig:ct2} shows the distribution of $S_{n,\rm best}$ for all 33 experiments included
in CT14 analysis.
The
distribution is wider than a normal distribution, indicating the presence of disagreement,
or tension, between some of the included experiments, a well--established fact in global
PDF fits.

%%%%%%%%%%%%%%%%%%%%%%%%%%%%%%%%%%%%%%%%%%%%%%%%%%%%%%%%%%%%%%%%%%%%%
%%%%%%%%%%%%%%%%%%%%%%%%%%%%%%%%%%%%%%%%%%%%%%%%%%%%%%%%%%%%%%%%%%%%%

With the best--fit and $2N_{\rm eig}$ uncertainty eigenvectors sets, the asymmetric errors for any QCD
observable $\mathcal{F}$ can be calculated through the master formula:
\bea
(\delta \mathcal{F})_+&=&\sqrt{\sum_{i=1}^{N_{\rm eig}}[{\rm max}(\mathcal{F}_{+i}-\mathcal{F}_0,\mathcal{F}_{-i}-\mathcal{F}_0,0)]^2}\; ,\,\,\qquad \nonumber \\
(\delta \mathcal{F})_-&=&-\sqrt{\sum_{i=1}^{N_{\rm eig}}[{\rm max}(\mathcal{F}_0-\mathcal{F}_{+i},\mathcal{F}_0-\mathcal{F}_{-i},0)]^2}\; ,
\eea
where $\mathcal{F}_0$ is the prediction from central set, and
$\mathcal{F}_{+i}$ and $\mathcal{F}_{-i}$ are from
two error sets in the direction of $i$-th eigenvector.
These PDF errors can be scaled
down to 68\% CL with a factor of 1.64 assuming Gaussian distributions.

\begin{figure}[t]
\begin{center}
  \includegraphics[scale=0.55]{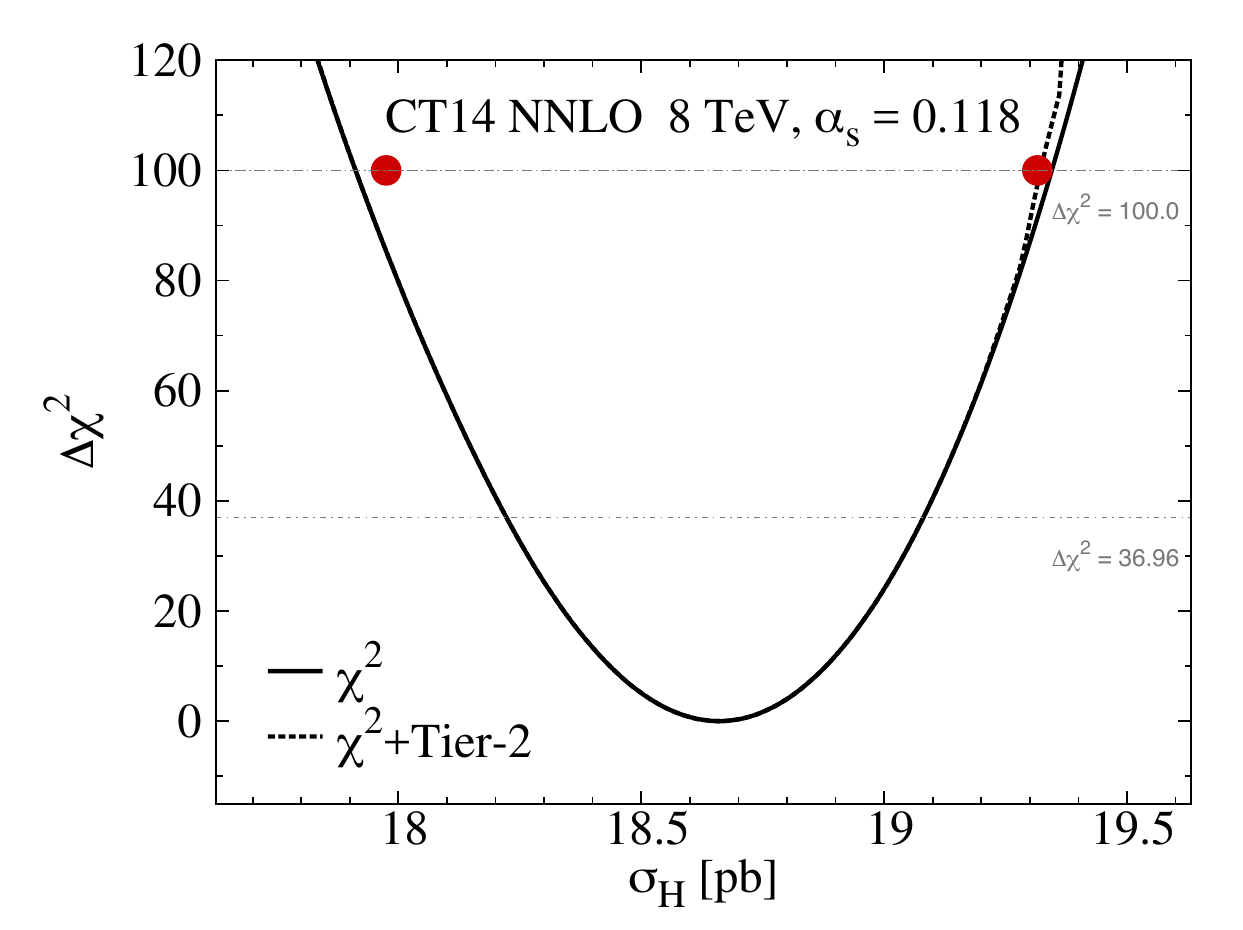}
  \hspace{0.3in}
  \includegraphics[scale=0.55]{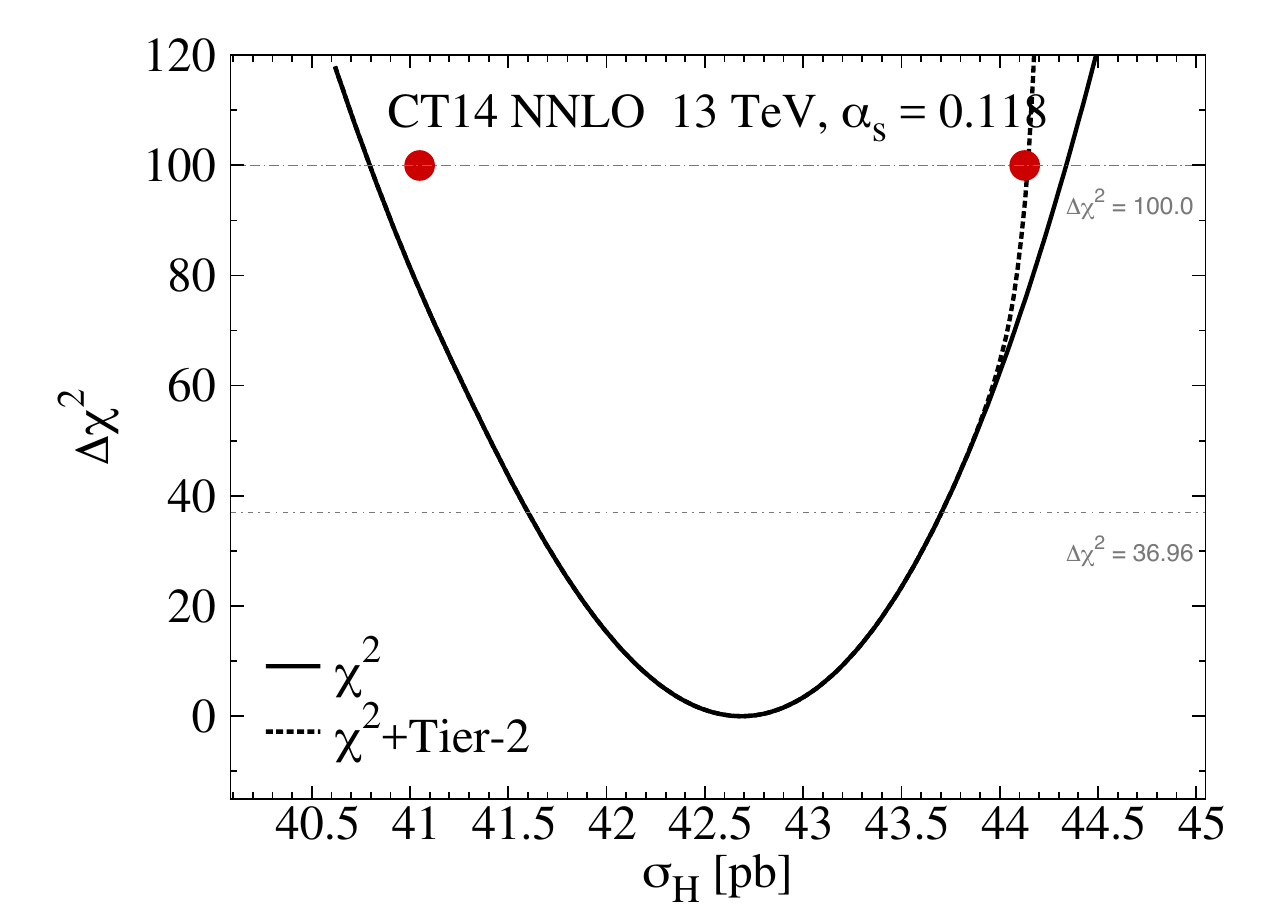}
   \caption{\small
   Dependence of the increase in $\chi^2$ with respect to the global
   minimum $\chi^2_{\rm min}$, defined as
   $\Delta\chi^2 =\chi^2- \chi^2_{\rm min}$,
   for the constrained CT14 fit on the expected
   cross section $\sigma_H$ at the LHC 8 and 13 TeV~\cite{Dulat:2015mca}, for $\alpha_s(m_Z)$=0.118.
The solid
   and dashed curves are for the constrained fits without and with the Tier--2 penalties,
   respectively. The red dots correspond to the upper and lower 90\% CL limits calculated
   by means of the Hessian method.  
    \label{fig:cthiggs}
  }
\end{center}
\end{figure}

As mentioned above, the CTEQ--TEA analysis also uses the Lagrange multiplier method~\cite{Stump:2001gu} to
cross--check the error estimation from nominal Hessian sets. In the CT14 analysis,
Lagrange multiplier scans have been performed for the cross sections of Higgs boson production
via gluon fusion the top quark pair production at the LHC. In such scans the
best--fits and the associated $\chi^2$ are found for each fixed value of the observable studied.
Then the PDF uncertainties on the observable are determined from the $\chi^2$ profile obtained
using the same tolerance criteria as in the Hessian method.
Fig.~\ref{fig:cthiggs} shows the
good agreement of the 90\% CL uncertainties for the Higgs cross sections from the CT14
Hessian PDFs and the CT14 LM scans.
The latter can be read off from the intersection of
the horizontal line $\Delta \chi^2=100$ and the various curves. As the LM method does not rely on the linear approximation, it serves as a robust check of the Hessian results.

%%%%%%%%%%%%%%%%%%%%%%%%%%%%%%%%%%%%%%%%%%%%%%%%%%%%%%%%%%%%%%%%%%%%%
%%%%%%%%%%%%%%%%%%%%%%%%%%%%%%%%%%%%%%%%%%%%%%%%%%%%%%%%%%%%%%%%%%%%%

Other dedicated applications of the CTEQ--TEA global analysis have
been presented recently.
The CTEQ--TEA analyses use the
world average of the strong coupling constant $\alpha_s(m_Z)$ as an input. Usually the nominal fit
is performed with $\alpha_s(m_Z)=0.118$ at both NLO and NNLO, but additional fits with alternative
$\alpha_s$ choices are also provided, which can be used to compute the combined
PDF+$\alpha_s$ uncertainties.
The fit itself  provides a much weaker constraint on $\alpha_s$ than the world average,
see also the discussion in Sect.~\ref{sec:fitmeth.theoryunc.alphas}.
Similarly, the pole mass of the charm and bottom quarks are chosen to be close to the
world average values, with $m_c=1.3\,{\rm GeV}$ and $m_b=4.75\,{\rm GeV}$.

The CTEQ--TEA group also
provides specialized fits with non--perturbative charm quark PDFs~\cite{Hou:2017khm,Dulat:2013hea}.
For these studies, a sea--like or
valence--like charm distribution is added to the nominal parametrization and then
fitted to the data.
Limits on the momentum fraction carried by the fitted charm at
the initial scale $Q_0=1.3$ GeV are then derived. In the most recent CT14 analysis, the limits
are 1.6\% for the SEA model and 2.1\% for the BHPS model both at 90\% CL~\cite{CT14IC}.
Finally, there
are also CT14 QED PDFs~\cite{Schmidt:2015zda} based on a radiative ansatz for the inelastic
component of the photon PDFs.
The 90\% CL limit on the momentum fraction of the proton
carried by the photon is found to be 0.11\% at $Q_0=1.3$ GeV,
as derived from a fit to the ZEUS measurement
on isolated photon production~\cite{Chekanov:2009dq}.
Recent progress in PDF fits with QED corrections and the role of the photon
PDF is discussed in Sect.~\ref{sec:QED}.

\subsection{MMHT}\label{sec:pdfgroups.MMHT}

The MMHT14 PDFs~\cite{Harland-Lang:2014zoa} are the successor to the MSTW08~\cite{Martin:2009iq} set, which
in turn derives from the earlier MRST and MRS studies.
The first NLO fit~\cite{Martin:1987vw} to DIS data was performed in the late 80s, while in the mid 90s the MRS(A)~\cite{Martin:1994kn} fit was released, including data from HERA and the Tevatron for the first time. This corresponded to a truly global analysis, fitting to fixed target, DIS and hadroproduction data to constrain the PDFs as precisely as possible. Subsequent releases have all built on this approach, but with significant advances achieved over the years due to improvements in both theory and experiment, as
well as with significant methodological developments.
The MRST98 release~\cite{Martin:1998sq} was the first set to include a full treatment of heavy flavours within the GM--VFNS developed in~\cite{Thorne:1997ga}, and discussed further in Sect.~\ref{sec:global.heavyq}. This was motivated by the new HERA measurements of the charm structure function, which demonstrated the importance of a consistent treatment of charm production at low and high scales;  indeed, the introduction of this flavour scheme resulted in an improved description of such data.
The MRST02 release~\cite{Martin:2002aw} included a full treatment of PDF errors for the first time, described further below
(see also Sect.~\ref{sec:pdfuncertainties}). The MRST04~\cite{Martin:2004ir} set went to NNLO for the first time, although earlier fits based on approximate NNLO treatments in fact gave quite similar results. A proper treatment of the PDF discontinuities which occur at NNLO across heavy flavour thresholds was presented later in~\cite{Martin:2007bv}.

These elements were all incorporated in the major MSTW08~\cite{Martin:2009iq} release. This presented a global fit to a range of DIS data from HERA and fixed proton and nuclear targets, fixed target Drell--Yan and dimuon production and $W$, $Z$ and jet production at the Tevatron, with $O(2500)$ data points in total. Fits were performed up to NNLO in the strong coupling, with an improved dynamical error treatment, and with an up to date heavy flavour scheme applied.
This release aimed to provide a general--purpose PDF set for use at the LHC, which
began operation soon after the release, and was subsequently very widely used in LHC phenomenological studies and experimental analyses.
This fit was updated in the latest MMHT14~\cite{Harland-Lang:2014zoa} set, which includes a number of theoretical and experimental updates.
In particular for the first time LHC data on $W$, $Z$, $t\overline{t}$ and jet production are included, as well as updated HERA data on the charged, neutral, charm and longitudinal structure functions, and updated Tevatron $W$ and $Z$ measurements. As in earlier fits, for DIS data a $Q^2$ cut of 2 ${\rm GeV}^2$ and $W^2$ cut of 15 ${\rm GeV}^2$ are imposed to avoid sensitivity to higher twist corrections.

In the case of Tevatron jet production, in the absence of a full NNLO calculation at the time, an approximation to the NNLO corrections based on the threshold corrections of~\cite{Kidonakis:2000gi} was applied in the NNLO MSTW08 fit, with the judgment being made that the difference between this and the full NNLO result would be expected to be smaller than the systematic uncertainties on the data, which itself provided the only direct constraint on the gluon at high $x$.
At the LHC much of the jet data are quite far from threshold, while those that are not probe a kinematically similar region to the Tevatron data, and so at NNLO these are not included in the MMHT14 fit. For the $t\overline{t}$ data the top mass is allowed to be determined from the fit, with the pole mass value of $m_t=172.5\pm 1$ GeV taken as an input.
This gives a value at NNLO that is consistent with the world average, while at NLO it is somewhat lower.

The MSTW PDFs were parameterised in terms of simple polynomials in $x$, with 29 free parameters. However, in~\cite{Martin:2012da} it was shown that this parameterisation was not sufficiently adaptive to describe the Tevatron and LHC $W$ asymmetry data. In particular, it was necessary to introduce a more flexible basis for the interpolating function described in Sect.~\ref{sec:fitmeth.PDFpara.func}, with
\be
\label{eq:chebmmht}
I_f(x)=\sum_i^n \alpha_{f,i}T_i(y(x))\;,
\ee 
where $T_i$ is a Chebyshev polynomial of order $i$ and $y(x)=1-2\sqrt{x}$ is chosen so as to sample a wide range of $x$, and has the additional advantage that this provides a half--integer separation in powers of $x$, as expected on Regge theory grounds.

In order to determine how many parameters $n$ were needed in Eq.~(\ref{eq:chebmmht}),
in~\cite{Martin:2012da} pseudo--data points with a constant percentage error were generated for the required distributions, in terms of a very large order polynomial with additional smoothness constraints applied.
The fractional deviation from the true PDF, as well as the decrease in $\chi^2$, were then determined as the number of parameters were increased, until no further significant improvement was observed and the level of agreement was well below the PDF uncertainty for the set.
In this way $n=4$ was arrived at as a good choice with which to parameterise the $u_V$, $d_V$, $s+\overline{s}$ and light quark sea $S$ distributions.
Fitting to the MSTW08 data set, these resulted in some improvement in the fit quality, but with the only significant change in the PDF being in the $u_V$ at lower $x$.
This was found to lie outside the previous PDF uncertainty band, and the additional flexibility provided a greatly improved description of $W$ asymmetry data from the LHC Run I. 

In the MMHT14 set, this Chebyshev parameterisation is used at $Q_0^2=1\,{\rm GeV}^2$  for the $u_V$, $d_V$, $s+\overline{s}$ and light quark sea $S$ distributions, while for the gluon a term with $n=2$ Chebyshev polynomial is included, but with a second term still present, as in MSTW08, which has a different low $x$ power and provides the additional flexibility at low $x$ that is required by the HERA data; this has the effect that the gluon at NLO and NNLO can become negative at low $x$ and $Q^2$. Standard polynomial parameterisations are taken for the less constrained $s-\overline{s}$ and $\overline{d}-\overline{u}$ distributions, although as the data become more precise we can expect this to change. 

%%%%%%%%%%%%%%%%%%%%%%%%%%%
\begin{figure}[t]
\begin{center}
  \includegraphics[scale=0.70]{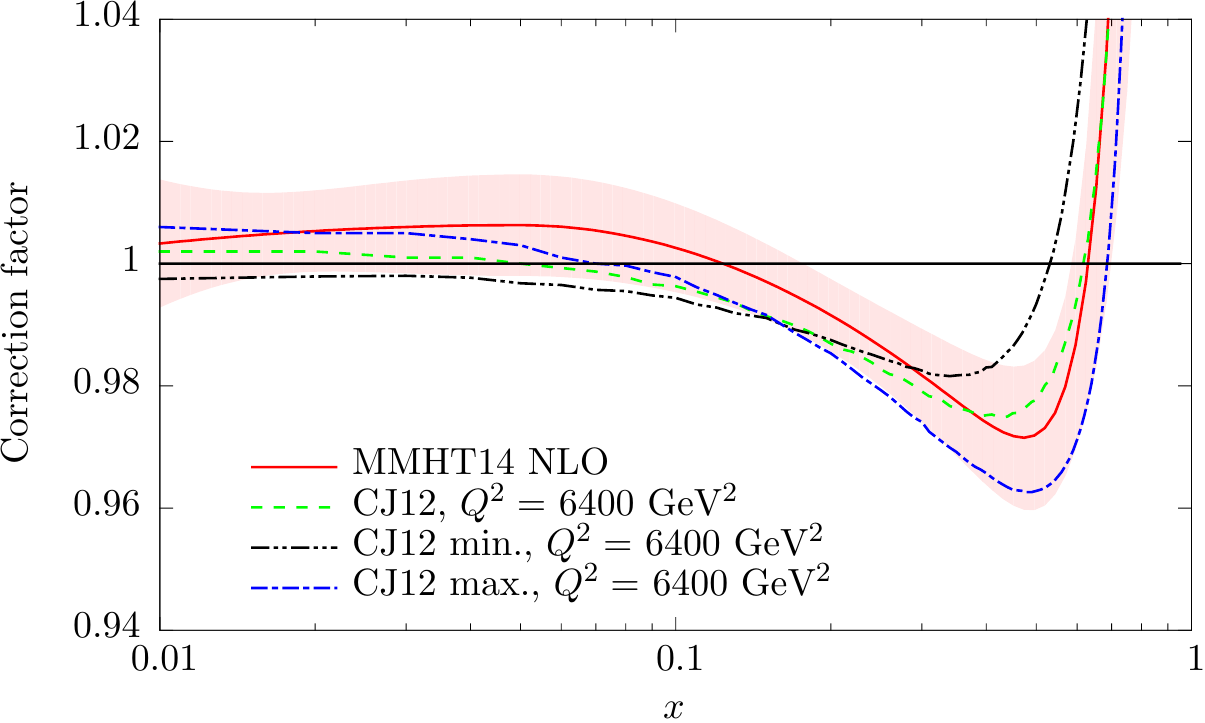}
   \caption{\small Fitted MMHT14 deuteron correction factors with uncertainty, compared to the CJ12~\cite{Owens:2012bv} predictions. Taken from~\cite{Harland-Lang:2014zoa}.
    \label{fig:deutcorr}
  }
\end{center}
\end{figure}
%%%%%%%%%%%%%%%%%%%%%%%%%%%

A further improvement described in~\cite{Martin:2012da} that is included in the MMHT14 set is in the treatment of the non--perturbative corrections that should in general be applied when considering DIS data on deuteron targets, to account for the binding of the proton and neutron within the deuteron. While in MSTW08 and earlier fits, a fixed shadowing correction at small $x$ was applied, a more flexible approach is now taken. In particular the deuteron corrections are freely parameterised in terms of a function $c(x)$, which is determined along with its corresponding uncertainties from the PDF fit. This results in a significantly improved description of the BCDMS deuteron structure function data, the E866 Drell--Yan asymmetry and the Tevatron lepton asymmetry data, with some significant changes in $d_V$. The result of the MMHT14 fit is shown in Fig.~\ref{fig:deutcorr} and compared against different model predictions used in the CJ12~\cite{Owens:2012bv} analysis. Interestingly, very good agreement is found with the CJ12mid prediction, demonstrating the power of global PDF fits to extract additional physical information beyond the PDFs themselves.

In MMHT14, the Hessian approach is applied to calculate the PDF errors, with the `dynamical' tolerance criteria described in Sect.~\ref{sec:pdfuncertainties.hess} taken. 
For MMHT14 the 68\% uncertainties are calculated using this procedure. In the fit there are 37 free PDF parameters in total, however in the error determination certain parameter directions are found to be largely degenerate, leading to departures from quadratic $\chi^2$ behaviour. This is corrected by fixing some parameters when calculating the error eigenvectors, reducing the number to 25, that is 50 directions.

Other theoretical updates in MMHT14 include the treatment of the $D\to \mu$ branching ratio, which is required in the fit to dimuon production in DIS. This is now determined from the fit but with the measurement of~\cite{Bolton:1997pq}, which is not determined from dimuon production data, included as a data point. The result is somewhat lower than that taken in MSTW08, corresponding to a larger strangeness, but the most dramatic effect is that the $\sim 10\%$ uncertainty on the branching ratio allows for a much larger strangeness uncertainty when fitting to the same data. Other smaller improvements include an updated treatment of nuclear corrections and a multiplicative, rather than additive, treatment of systematic uncertainties where appropriate. 

\begin{figure}[t]
\begin{center}
  \includegraphics[scale=0.63,trim={0 -2cm 0 0},clip]{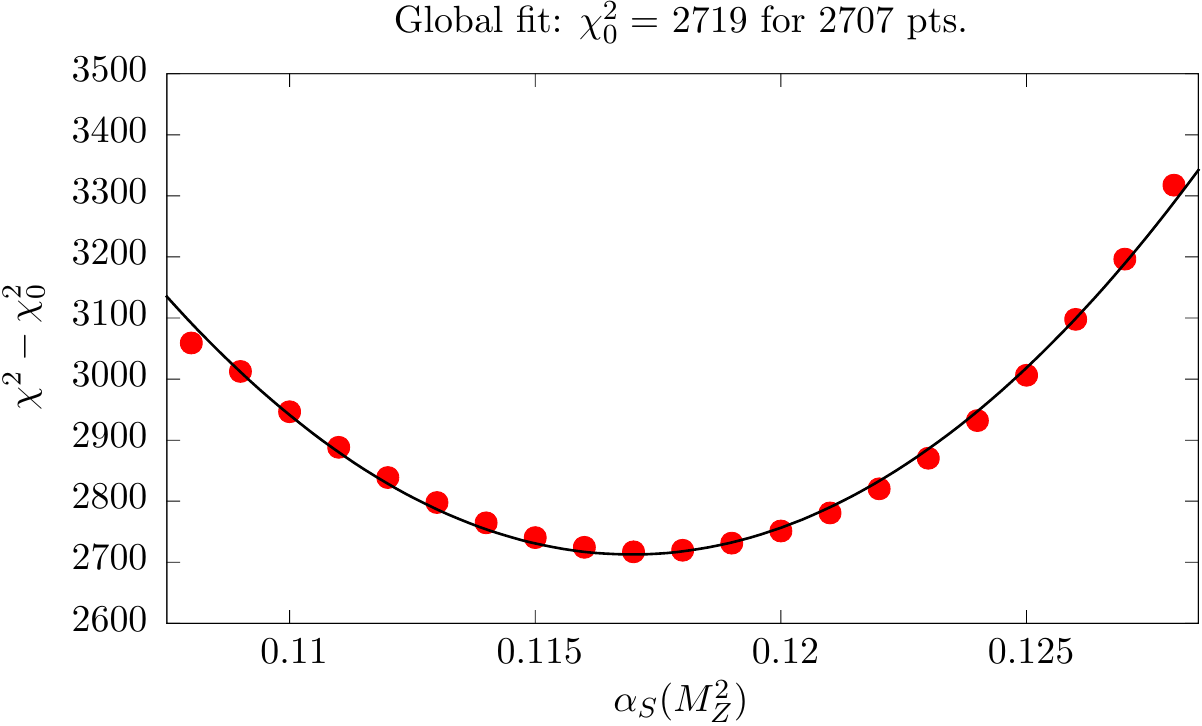}\quad
   \includegraphics[scale=0.83]{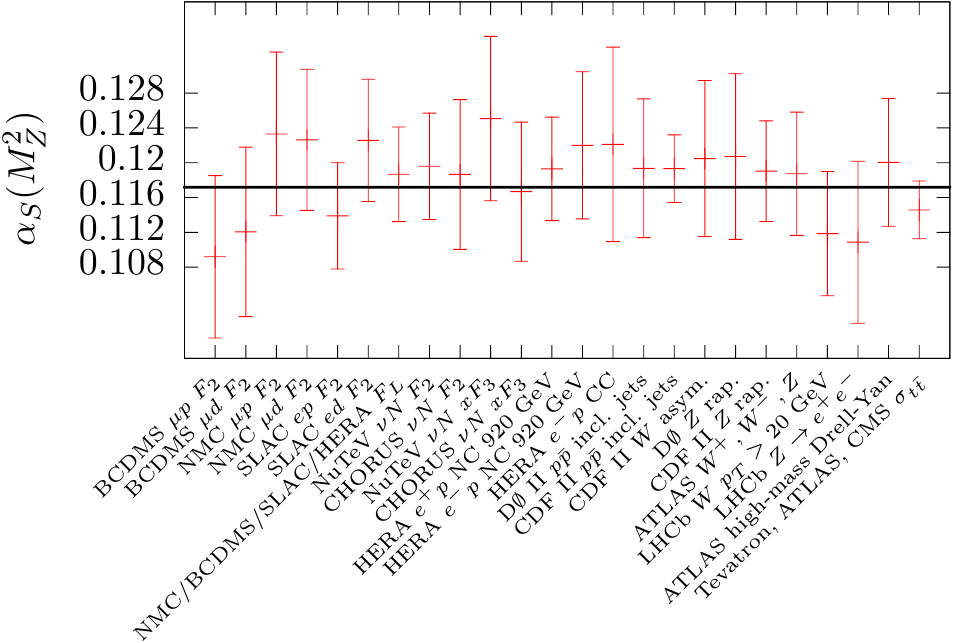}
   \caption{\small Left: global $\chi^2$ values as a function of $\alpha_S(m_Z^2)$. Right:
   best fit $\alpha_S(m_Z^2)$ together with the upper and lower 1$\sigma$ constraints from the most constraining data sets. Both figures correspond to NNLO fits and are taken from~\cite{Harland-Lang:2015nxa}.
    \label{fig:MMHTalphas}
  }
\end{center}
\end{figure}

In  MMHT the value of the strong coupling is allowed to be determined by the fit, it being argued that valuable information can be provided from global PDF fits about this object. This in addition serves as a consistency test on the overall fit; if the extracted value is in strong tension with the world average then this would indicate that further work is needed. In the fit the preferred values at NLO and NNLO are indeed found to be consistent with the world average, and including this as an additional data point is not found to affect the fit significantly.
In particular, the detailed study of~\cite{Harland-Lang:2015nxa} found best fit values of $\alpha_S(m_Z^2)=0.1201\pm 0.0015$ at NLO and $\alpha_S(m_Z^2)=0.1172\pm 0.0013$, to be compared with the world average value of $\alpha_S(m_Z^2)=0.1181\pm 0.0013$. The NNLO $\chi^2$ profile for $\alpha_S(m_Z^2)$ and the corresponding individual constraints from the most constraining data sets are shown in Fig.~\ref{fig:MMHTalphas}. The PDF sets for a range of $\alpha_S$ values, from 0.108 to 0.128 in steps of 0.001, are publicly available.

%%%%%%%%%%%%%%%%%%%%%%%%%%%%%%%%%%%%%%%%%%5
%%%%%%%%%%%%%%%%%%%%%%%%%%%%%%%%%%%%%%%%%%5

Concerning the treatment of the heavy quarks, the GM--VFNS is taken, applying the `optimal'
TR scheme of~\cite{Thorne:2012az} which improves the smoothness of the transition region where the number of active flavours is increased by one. The charm and bottom pole masses are fixed to
$m_c~(m_b)=1.4~(4.75)$ GeV, but a detailed study of the heavy quark mass dependence was
performed in~\cite{Harland-Lang:2015qea}.
In particular, when $m_c$ and $m_b$ were taken as free parameters in the fit, the preferred values were found to be somewhat lower than these defaults, possibly suggesting a $\overline{\rm MS}$ scheme may be more appropriate, although the impact of such a choice is expected to be fairly small. Sets with a range of charm and bottom masses, as well as with a maximum number of  active flavours $n_f^{\rm max}=3$ or 4, are also made available.

A subsequent study within the MMHT framework on the PDF impact of the final HERA I + II combined data~\cite{Abramowicz:2015mha}, which was released following MMHT14, was performed in~\cite{Thorne:2015caa}. This was found to lead to some reduction in the PDF uncertainties, principally in the gluon, with little change in the central values.
It was therefore decided not to release an updated set but rather to wait until more precise and varied LHC data became available, as well as theoretical calculations such as NNLO jet production. Subsequent work towards a new PDF set has been presented in for example~\cite{Harland-Lang:2017dzr}, where a first fit at NNLO to jet data is presented, while the impact of new LHC data in the fit is seen to be significant. This was then studied in more detail in~\cite{Harland-Lang:2017ytb}. In addition, the first steps towards the inclusion of the photon PDF within the MMHT framework are presented, see Sect.~\ref{sec:QED} for more discussion. A further public release is therefore anticipated in the near future.

\subsection{NNPDF}\label{nnpdf}\label{sec:pdfgroups.NNPDF}

As discussed in Sect.~\ref{sec:fitmeth}, the NNPDF fitting methodology is based on  the
combination of three main components: i) the use of artificial neural networks
as universal unbiased interpolants, ii) the Monte Carlo method to estimate
and propagate PDF uncertainties, and iii) Genetic Algorithms (GA)
minimization for the training of
the neural networks combined with a look--back cross--validation stopping
to avoid over--fitting.
Here we review the historical development of the NNPDF family of PDF fits,
highlighting recent progress.

The NNPDF methodology was first presented in~\cite{Forte:2002fg}, where it was used
to produce a neural network based determination of the proton, deuteron
and non--singlet DIS structure functions from the NMC and BCDMS fixed target data.
As a first phenomenological application, this determination was used
to extract the strong coupling constant $\alpha_S(m_Z)$ from scaling violations
of truncated moments of structure functions~\cite{Forte:2002us}.
This analysis was subsequently extended~\cite{DelDebbio:2004qj} to include as well
the $F_2^p$ measurements from the H1 and ZEUS experiments at the HERA collider.
Note that a determination of structure functions is purely data--driven, with no theoretical
input required at this point.

When moving from fitting structure functions to PDFs, there are a number of simplifications,
for instance one needs to fit only a 1D function $q_i(x,Q_0)$ as opposed to a 2D function $F_2^p(x,Q^2)$,
but also technical complications, the most important one being the need to compute
DIS structure functions starting from a neural--network based parametrization of $q_i(x,Q_0)$.
First of all, the usual ANN training algorithm of back--propagation cannot be used in this
case, due to the convolution of the PDFs with the DGLAP evolution kernels and the
DIS coefficients functions.
To overcome this limitation, it was demonstrated how Genetic Algorithms can be efficiently
used for ANN training under a non--trivial mapping between the latter and the experimental data,
and used to extract the QCD vacuum condensates from hadronic tau decay data~\cite{tau}.
An efficient method to solve the DGLAP evolution equations in $N$-space was also developed,
called the {\tt Fast Kernel} method.
With these ingredients at hand, it became possible for the first time to apply the
NNPDF methodology to a determination of the PDFs, starting from
a fit of the non--singlet combination $q_{NS}(x,Q_0)$~\cite{DelDebbio:2007ee} and them moving to a first
full--fledged NLO PDF fit based on NC DIS structure function data~\cite{Ball:2008by} in 
the NNPDF1.0 analysis.

Subsequently, the global NNPDF fits were improved both by adding new experimental data,
updating the theoretical calculations and/or refining the fitting methodology.
The NNPDF1.2 analysis~\cite{Ball:2009mk}
relaxed the previous assumption that
the strange sea was proportional to the light quark sea, $s=\bar{s}=\kappa \lp \bar{u}
+\bar{d}\rp$, and parametrized both $s^+$ and $s^-$ using neural networks, exploiting
the constraints from the NuTeV dimuon CC neutrino scattering data.
Two important phenomenological consequences of this analysis were, first of all, the demonstration
that the PDF uncertainties associated with $s^-$ were enough to remove the
NuTeV anomaly~\cite{Davidson:2001ji}
in the determination of the weak mixing angle $\sin^2\theta_W$; and second,
a direct extraction of the CKM matrix element $V_{cs}$ with a precision compatible
with that of the PDG average.

In 2010, the NNPDF2.0 set was released~\cite{Ball:2010de},
which constituted the first truly global PDF fit from the NNPDF collaboration.
In addition to the NC and CC DIS structure function data included in previous releases, this fit in addition fit to fixed--target Drell--Yan cross sections from the Fermilab E605
and E866 experiments, inclusive jet production measurements from CDF and D0 at the Tevatron,
as well as the differential rapidity distributions of the $Z$ boson also from
the Tevatron.
From the theoretical point of view, NNPDF2.0 was still based on the ZM--VFNS, and thus charm and bottom structure function data from HERA were not included.
A good overall description of all experiments in the global fit was found.
NNPDF2.0 was also the first global PDF set to include the recently release HERA combination
of H1 and ZEUS structure function data for the Run I data period~\cite{Aaron:2009aa}.

While NNPDF2.0 demonstrated that the NNPDF methodology could be successfully applied
to a global determination of parton distributions, there were still a number of important
limitations from the theoretical point of view.
First, the use of a ZM--VFNS neglected heavy quark mass effects in the DIS
structure functions, which were known to be important for the
description of the low--$x$, low--$Q^2$ HERA data.
Second, all NNPDF fits so far were based on NLO theory, and NNLO accuracy was essential
to match the corresponding precision of important partonic hard--scattering cross sections
such as Higgs production in gluon fusion.
The first of these theory limitations was removed with the release of NNPDF2.1~\cite{Ball:2011mu},
which was based on the FONLL GM--VFNS for the calculation of DIS structure functions, which
allowed the HERA charm and bottom structure function data to be fit.
This analysis also showed that the impact of heavy quark mass effects was less drastic
then previously reported, with the cross section predictions between NNPDF2.0 and 2.1 typically
agreeing at the one--sigma level.
The NNPDF2.1 fit was also used to produce a determination of the strong coupling $\alpha_s(m_Z)$
from the global dataset~\cite{Lionetti:2011pw}.

%%%%%%%%%%%%%%%%%%%%%%%%%%%%%%%%%%%%%%%%%%%%%%%%%%%%%%%%%%%%%%%%%%%%%
\begin{figure}[t]
\begin{center}
  \includegraphics[scale=0.45]{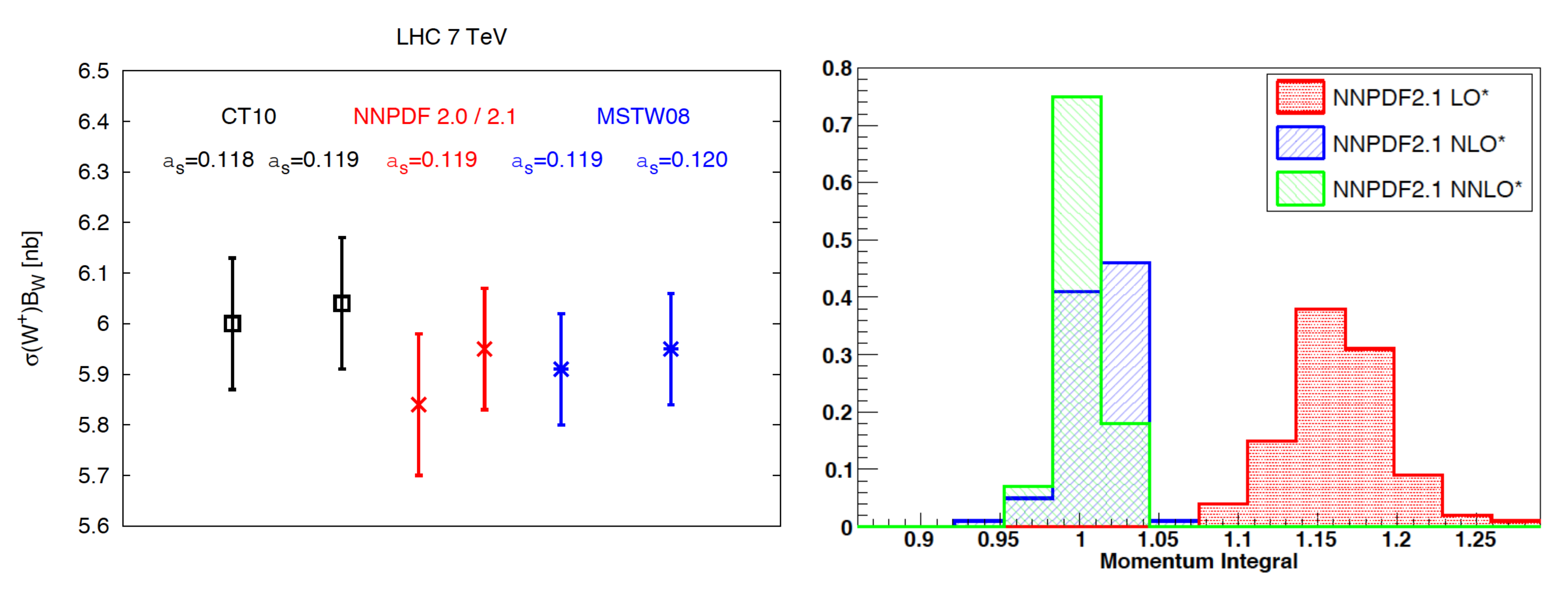}
   \caption{\small 
   Left plot: comparison between the NNPDF2.0 and 2.1 predictions
   for the inclusive $W^+$ production cross section at the LHC 7 TeV,
   which illustrates the phenomenological impact of heavy quark mass effects.
   Right plot: the distribution of the momentum integral Eq.~(\ref{eq:momentumintegral})
   among the MC replicas for the variants of the NNPDF2.1 LO, NLO, and NNLO fits that
   do not impose the momentum sum rule.
    \label{fig:20momsumrule}
  }
\end{center}
\end{figure}
%%%%%%%%%%%%%%%%%%%%%%%%%%%%%%%%%%%%%%%%%%%%%%%%%%%%%%%%%%%%%%%%%%%%%

The second of these theoretical limitations was removed shortly after, with the
release of a NNLO version of NNPDF2.1~\cite{Ball:2011uy}.
This PDF set was based on the same dataset as its NLO counterpart, but with the DIS and hadronic cross sections computed at NNLO, in the former
case using the FONLL--C GM--VFNS.
In the same publication, the first NNPDF LO sets were also presented.
The availability of NNPDF2.1 fits at LO, NLO and NNLO allowed a systematic
study of the perturbative convergence of the global fit, finding in particular reasonable
agreement at the one--sigma level between the NLO and NNLO versions.
The  consistency of the global QCD analysis framework was also tested by performing
fits without imposing the momentum sum rule and then verifying a posteriori that
the global fit result was consistent with QCD expectations, finding indeed that
at NNLO
\be
\label{eq:momentumintegral}
\lc M \rc \equiv \int_0^1 dx\,\lp g(x,Q^2)+\Sigma(x,Q^2)\rp=1.002\pm 0.014 \, .
\ee
The NNPDF2.1 NNLO analysis was also used to perform a determination of
the strong coupling constant~\cite{Ball:2011us}, finding a value
$\alpha_s(m_Z)=0.1173\pm 0.0007^{\rm stat}\pm 0.0009^{\rm pert}$, a result which is 
included in the current PDG global average of $\alpha_s$~\cite{Olive:2016xmw}.

Similarly to the early 90s, when the availability of the HERA structure function measurements
became a game--changer for global fits, from 2010 onwards the LHC experiments have started
producing a wealth of PDF--sensitive information for global fits.
With this motivation, in 2012 the NNPDF2.3 set was released~\cite{Ball:2012cx},
and was the first
PDF fit to include LHC data, in particular electroweak gauge boson production
from ATLAS, CMS and LHCb as well as jet production from ATLAS.
As with all subsequent releases, NNPDF2.3 was available at LO, NLO and NNLO
and was based on the FONLL general mass scheme.
The NNPDF2.3 set became the baseline PDF set in several popular MC
event generators, such as {\tt Pythia8}~\cite{Sjostrand:2014zea}
and {\tt aMC@NLO}~\cite{Alwall:2014hca}.
Based on NNPDF2.3, an study of theoretical uncertainties affecting the
PDF from sources such as deuteron corrections and higher twists was
presented in~\cite{Ball:2013gsa}. 

Following the release of NNPDF2.3, it was realised that the increase in complexity required
to include the many new experiments that were either available or about to be released
could not be satisfactory tackled with the current, {\tt FORTRAN77}--based code.
With this motivation, a complete rewriting of the NNPDF global analysis framework
into {\tt C++} and {\tt Python} was undertaken, a two--year long effort that
culminated with the release of the NNPDF3.0 set~\cite{Ball:2014uwa}.
In addition to including many new LHC experiments on jets, vector boson production,
$W$+charm and top production, the main result of NNPDF3.0 was the systematic validation
of the complete fitting methodology based on statistically robust closure tests.

\begin{figure}[t]
\begin{center}
  \includegraphics[scale=0.47]{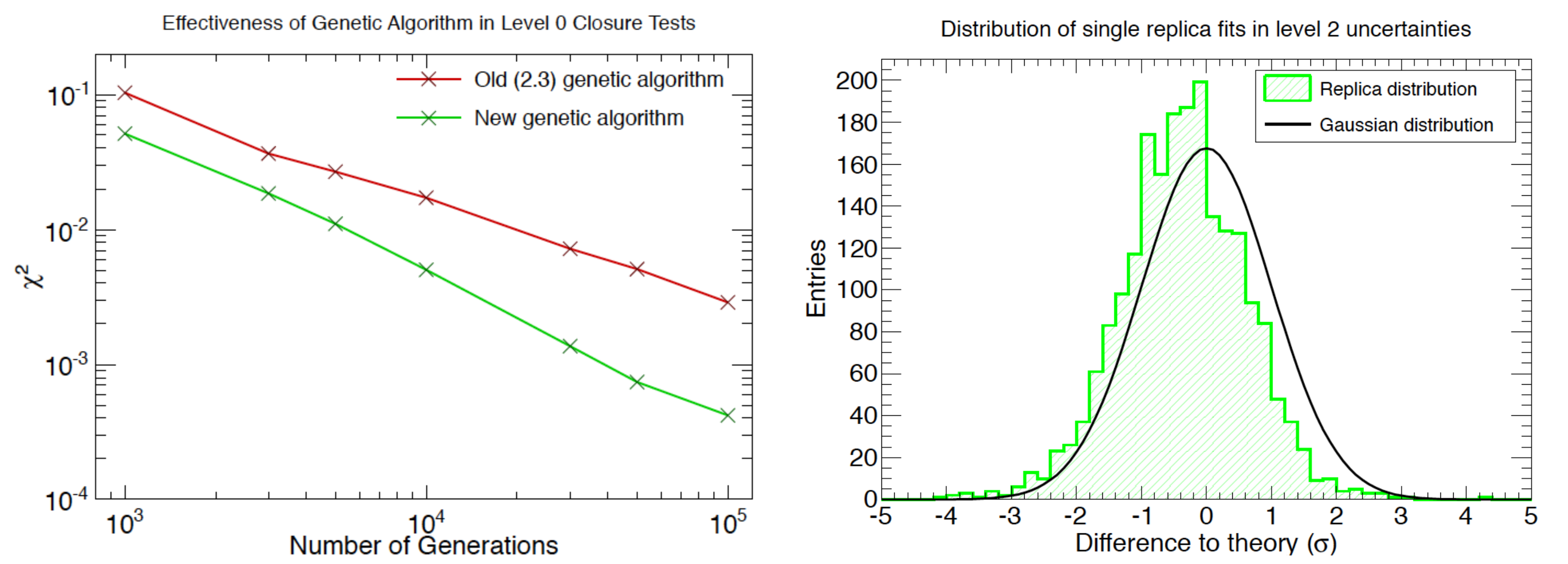}
   \caption{\small Representative results of the closure tests presented
   in the NNPDF3.0 analysis.
   Left plot: in a level 0 closure tests, where the pseudo--data is generated
   without any statistical fluctuations, the $\chi^2$ should decrease monotonically
   as a function of the number of GA iterations, down to arbitrarily small values.
   Right plot: the distribution of the difference between theory and data
   in units of the error of the latter among each of the Monte Carlo replicas.
   This distribution is consistent with the Gaussian expectations.
    \label{fig:nnpdf30closuretests}
  }
\end{center}
\end{figure}

In these closure tests, pseudo--data was generated based on some ``truth'' PDFs, and then
a PDF fit was performed: if the resulting PDF central values and uncertainties
were consistent with the (known) input PDFs, then the closure test can be considered successful.
In Fig.~\ref{fig:nnpdf30closuretests} we show 
representative results of the closure tests presented
   in the NNPDF3.0 analysis.
   In the left plot we show the
   results of a level 0 closure test, where the pseudo--data is generated
   without any statistical fluctuations, and as expected the $\chi^2$ is found decrease monotonically
   as a function of the number of GA iterations, down to arbitrarily small values.
   And in the right plot we show
   the distribution of the difference between theory and data,
   in units of the error of the latter among each of the MC replicas.
   This is consistent with the expectations from an underlying
   Gaussian distribution.

%%%%%%%%%%%%%%%%%%%%%%%%%%%%%%%%%%%%%%%%%%%%%%%%%%%%%%%%%%%%%%%%%%%%%
%%%%%%%%%%%%%%%%%%%%%%%%%%%%%%%%%%%%%%%%%%%%%%%%%%%%%%%%%%%%%%%%%%%%%

A recent development in the NNPDF family of global analyses concerns the treatment of
the charm PDF.
In previous PDF sets, it was assumed that the charm
PDF $c(x,Q)$ was generated dynamically from the gluons and light quarks, as dictated by the DGLAP
evolution starting from the charm mass threshold $\mu_c\simeq m_c$.
However, a possible non--perturbative component of the charm PDF would invalidate this assumption,
a hypothesis which can ultimately be tested against experimental data.
In addition, treating the charm PDF on an equal footing with the gluon and light quark
PDFs offers other potential advantages, such as a reduced dependence on the value of
$m_c$ and an improved  agreement between data and theory from the more flexible input PDF
parametrization.

With this motivation, a variant of the NNPDF3.0 analysis with a fitted charm PDF was studied
in~\cite{Ball:2016neh}.
By parametrising the charm PDF with an artificial neural network with 37 free parameters, it
was found that
fitting charm leads to an improved $\chi^2$ for several experiments and stabilises the dependence
of the fit with respect to the value of $m_c$.
It was also observed that fitting charm allows 
a reasonable description of the EMC charm structure function data, something which is not
possible if perturbative charm is used.
This result was shown to be robust
with respect to variations of the treatment of the EMC data, such as increasing some
of its systematic errors
or introducing additional kinematical cuts~\cite{Rottoli:2016lsg}.

The resulting charm PDF can then be compared with non--perturbative models~\cite{Brodsky:1980pb},
and some intriguing evidence
for a large--$x$ non--perturbative charm component in the proton was found.
Predictions for a number of LHC process such as $Z+$charm and large--$p_T$ $D$ meson production
were also provided, showing the potential of Run II data to disentangle the charm
content of the proton.
See Sect.~\ref{sec:structure.charm} for more details about the charm
content of the proton.
As discussed there, the main qualitative features of the
comparison between fitted and perturbative charm~\cite{Ball:2016neh} were
confirmed by the subsequent NNPDF3.1 set~\cite{Ball:2017nwa} and shown
to be independent of whether or not the EMC data is included in the fit.

The most recent incarnation of the NNPDF global analysis is the  NNPDF3.1 set.
The main motivation for this release was the availability of a large number of high--precision
collider measurements (and the corresponding
NNLO calculations) providing PDF--sensitive information, including some 
that for the first time could be used in a PDF fit, such as differential
distributions in top quark pair production and the $p_T$ of $Z$ bosons.
The second main motivation was to provide a state--of--the--art PDF set without assuming
that charm is generated perturbatively, that is, providing baseline global
PDF fits with fitted charm.
Some of the new experiments included in NNPDF3.1 were the $t\bar{t}$ distributions
from ATLAS and CMS, the legacy LHCb inclusive $W$ and $Z$ measurements from Run II,
the D0 $W$ asymmetries in the muon and electron channel, the $p_T$ of $Z$ bosons
from ATLAS and CMS at $\sqrt{s}=8$ TeV, as well as several other inclusive gauge boson
and jet production measurements from ATLAS and CMS.

\subsection{ABM}\label{sec:pdfgroups.ABM}

%%%%%%%%%%%%%%%%%%%%%%%%%%%%%%%%%%%%%%%%%%%%%%%%%%%%%%%%%%%%%%%
\begin{figure}[t]
\begin{center}
  \includegraphics[scale=0.3]{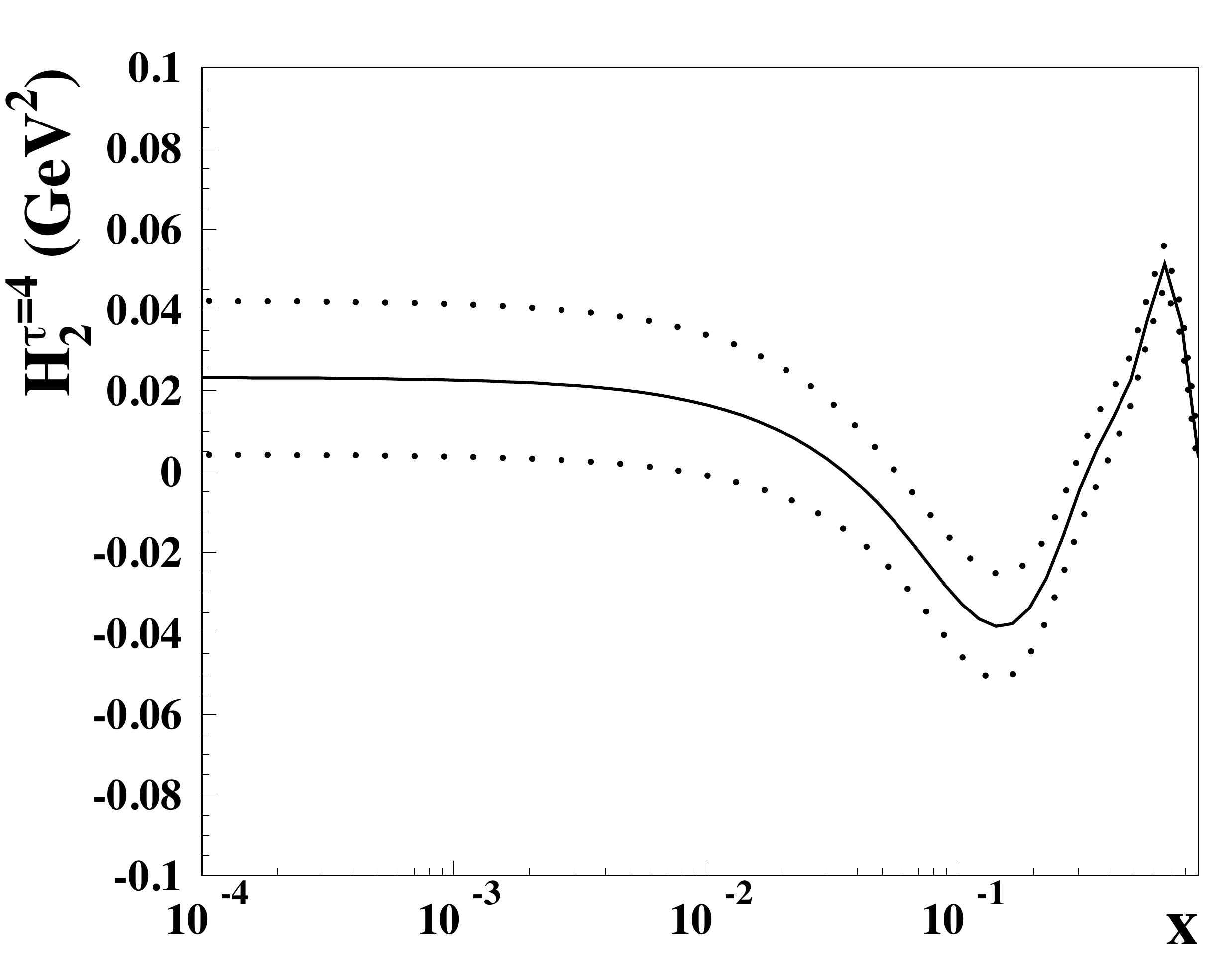}\qquad
   \includegraphics[scale=0.3]{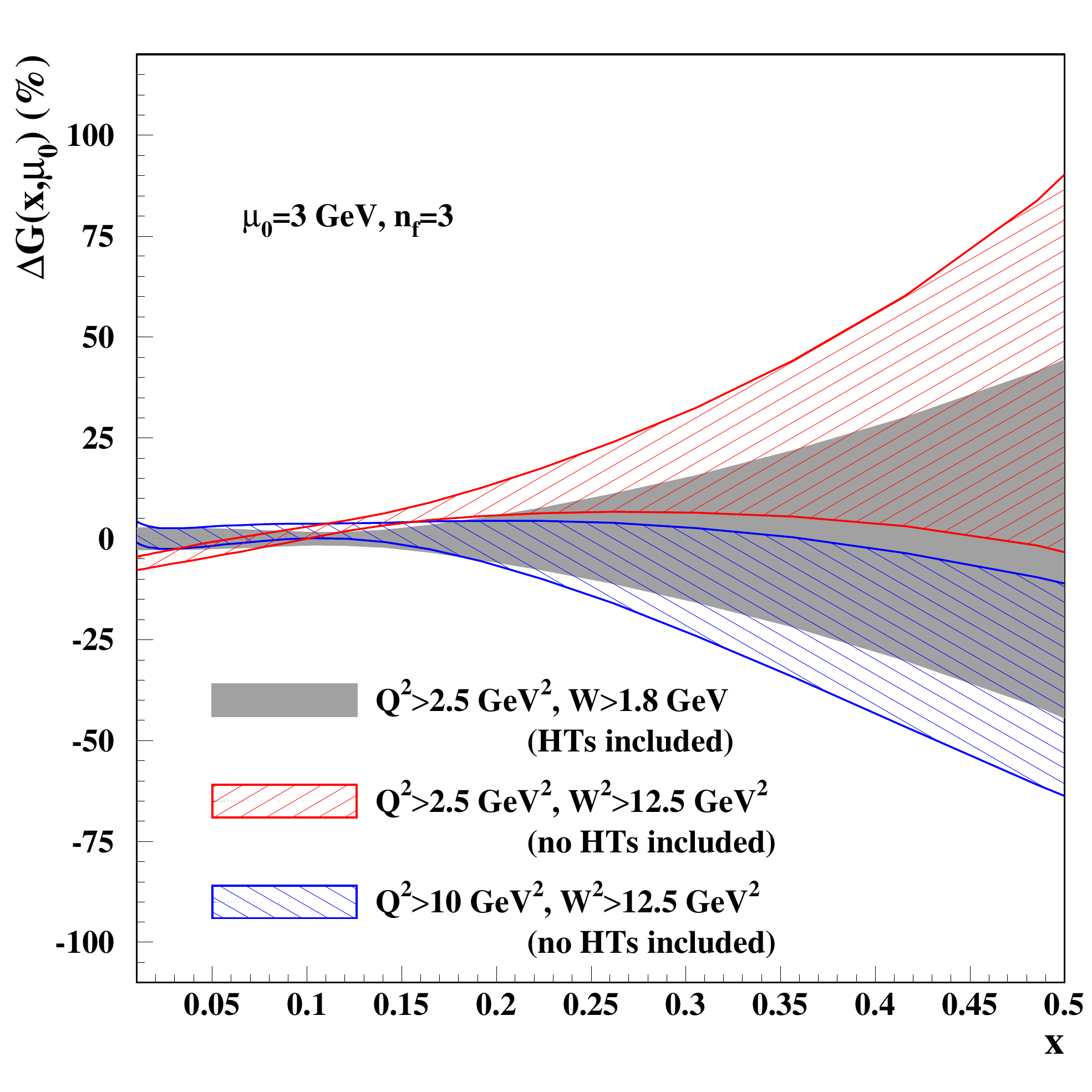}
   \caption{\small Left: the higher twist coefficient for the $F_2$ structure function from the ABMP16 fit, including the associated
     1--$\sigma$ uncertainty.
     Right: percentage difference in ABMP $n_f=3$ gluon distribution between the default result and fits performed with higher $W^2$ cuts but
     without the  higher twist corrections.
The 1--$\sigma$ uncertainty bands are also shown.
     Taken from~\cite{Alekhin:2017kpj}.
    \label{fig:ABMhtwist}
  }
\end{center}
\end{figure}
%%%%%%%%%%%%%%%%%%%%%%%%%%%%%%%%%%%%%%%%%%%%%%%%%%%%%%%%%%%%%%%

The ABMP16~\cite{Alekhin:2017kpj} set is the latest PDF fit from the ABM collaboration,
following on from the previous
ABM11~\cite{Alekhin:2012ig}, ABM12~\cite{Alekhin:2013nda}, and ABKM09~\cite{Alekhin:2009ni} sets.
These PDF sets were in turn based on the earlier fits of
Refs.~\cite{Alekhin:1996za,Alekhin:2000ch,Alekhin:2005gq} to HERA and fixed proton and deuteron target DIS data, with the ABKM09 fit~\cite{Alekhin:2009ni} and those that follow it including in addition fixed target Drell--Yan and dimuon production data from neutrino DIS on fixed nuclear targets.
The ABM sets of parton distributions are parameterised in terms of polynomials in $x$,
with the latest fits including 25 free parameters.
In the context of this fit to a somewhat reduced dataset,
the use of the classical `$\Delta\chi^2=1$' criteria for determination of the PDF errors
is applied. All sets from ABKM09 onwards go to NNLO in the strong coupling.

Two notable features of these fits are the use of a purely FFNS for the charm and bottom quark contributions to DIS structure functions using the $\overline{\rm MS}$
running masses~\cite{Alekhin:2010sv},
 and the treatment of higher--twist effects.
 In the latter case, no attempt is made to impose a cut to remove the region of sensitivity to such effects. Rather, a lower cut ($W>1.8$ GeV) is imposed for the DIS data than
 is typically applied in other PDF fits.
 The structure functions are then given by
\be\label{eq:abmf}
F_i(x,Q^2)=F_i^{\rm TMC}(x,Q^2)+\frac{H_i^{\tau=4}}{Q^2}\;,
\ee
where $i=2,T$. Thus $x$ dependent and $Q^2$ independent twist--4 corrections $H^4_i$ are included. While the effect of these dies off with increasing $Q^2$, at lower scales they can have a significant effect. These are then parameterised in terms of cubic splines defined at $x_{k}$ ($k=1...7$) points roughly linearly spaced between $x=$ 0 and 1, which are then determined from the fit. The result for the $F_2$ correction is shown in Fig.~\ref{fig:ABMhtwist} (left), and is found to be inconsistent with zero, in particular at higher $x$.
The effect of these corrections, and of conversely omitting them and including a more stringent $W^2$ cut on the DIS data is shown in Fig.~\ref{fig:ABMhtwist} (right) for the extracted gluon PDF.
The fit with the cut of $W^2>12.5$ GeV$^2$
and no higher twist corrections is found to prefer a somewhat larger gluon at higher $x$, and in some regions lies outside the 1--$\sigma$ uncertainty band of the default fit.

In addition to the higher twist corrections included in Eq.~(\ref{eq:abmf}),
the structure function functions also include target mass corrections, that is the impact of terms $\sim M_N^2/Q^2$, where $M_N$ is the nucleon mass. These are taken into account according the Georgi--Politzer prescription~\cite{Georgi:1976ve} (see also~\cite{Alekhin:2012ig}), with
\be
F_2^{\rm TMC}(x,Q^2)=\frac{x^2}{\xi^2\gamma^3}F_2(x,Q^2)+6\frac{x^3M_N^2}{Q^2\gamma^4}\int_\xi^1\frac{{\rm d}\xi'}{\xi'}F_2(\xi',Q^2)\;,
\ee
where $\xi=2x/(1+\gamma)$ and $\gamma=(1+4x^2M_N^2/Q^2)^{1/2}$, and a similar result holds for $F_T$. Thus, as $Q^2 \to \infty$ the corrected $F_i^{\rm TMC}$ reduce to the regular $F_i$.
The same TMC are also included in some other PDF analyses, such as in the NNPDF family~\cite{Ball:2008by}.

As mentioned above, the ABMP fit  uses a purely fixed flavour scheme to describe the DIS
structure function data.
That is, the fit is performed with $n_f=3$ light quark PDFs and with the heavy $c,b$ treated as massive final--state partons which can be produced at NLO and higher,
regardless of the value of $Q^2$.
It is then
argued that the bulk of the DIS data can be described within this scheme.
The Tevatron and LHC collider as well as fixed--target DY data, on the other hand, for which $\mu_F^2\gg m_{c,b}^2$, is treated using a $n_f=5$
flavour set evolved from the same input by means of the NNLO matching conditions~\cite{Alekhin:2009ni}.
The use of a FFNS as opposed to a GM--VFNS for the DIS
structure functions has been shown to be one of the
dominant reasons for the differences between the ABM family and the
global fits~\cite{Ball:2013gsa,Thorne:2014toa}.
PDF sets for $n_f=3,4$ and 5 active flavours are made publicly available, with the caveat that
each set should be used within its specified validity regime, for instance the ABMP16
$n_f=5$ set can only be used in applications for which $Q \ge m_b$.

A further feature of note in the ABM family of PDF sets
is that the strong coupling $\alpha_S$  is always
determined from the fit together with the PDFs.
In the ABM11 analysis, this was found to be $\alpha_S(m_Z^2)=0.1134\pm 0.0011$ at NNLO, that is in some tension with the PDG world average
value~\cite{Olive:2016xmw} of $\alpha_S(m_Z^2)=0.1181\pm 0.0013$ (the dominant uncertainty in which is determined by the lattice QCD input)
used by the CT and NNPDF collaborations, and the value extracted by MMHT.
While in~\cite{Alekhin:2012ig,Alekhin:2017kpj} the omission of higher twist corrections is found to lead to a sizeable increase in $\alpha_s$, in contrast in~\cite{Thorne:2014toa} the use of the FFNS within the MSTW framework is found to lead to a smaller extracted value, consistent with that seen by ABM(P), while higher twist effects are found to have less of an impact.

%%%%%%%%%%%%%%%%%%%%%%%%%%%%%%%%%%%%%%%%%%%%%%%%%%%%%%%%%%%%%%%%%%%
\begin{figure}
\begin{center}
  \includegraphics[scale=0.36]{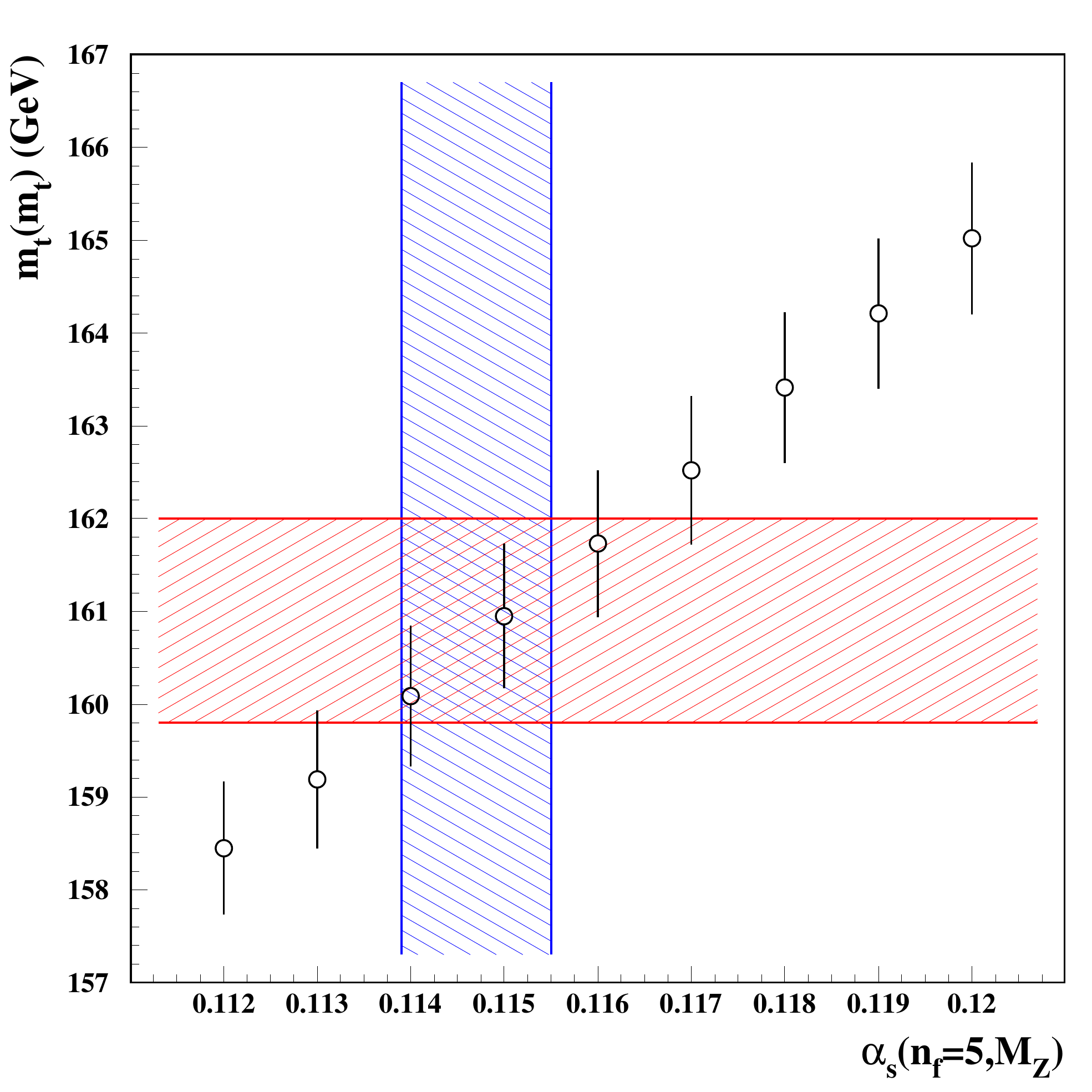}
  \caption{\small The $\overline{{\rm MS}}$ value of the top quark mass $m_t(m_t)$ obtained in the ABMP16 fit for variants of $\alpha_S(m_Z^2)$ (data points),
    compared to the corresponding best--fit values (hatched bands).
    Taken from~\cite{Alekhin:2017kpj}.
    \label{fig:ABMmt}
  }
\end{center}
\end{figure}
%%%%%%%%%%%%%%%%%%%%%%%%%%%%%%%%%%%%%%%%%%%%%%%%%%%%%%%%%%%%%%%%%%%

In the ABM11 fit~\cite{Alekhin:2012ig}, the heavy quark masses $m_{c,b}$ were included in the $\overline{\rm MS}$ scheme for the first time, in contrast to other PDF fits which adopt instead the pole mass scheme.
This allows the heavy quark mass values to be constrained directly from the PDG results without relying on the perturbative transformation between the $\overline{\rm MS}$ and pole masses, which is known to be poorly convergent.
Thus, the quark masses are left free in the fit but with the PDG values added in as pseudo--data points.
For the charm mass $m_c(m_c)$, the DIS data included in this fit are then found to give a comparable error to the PDG average.

The ABM12 fit~\cite{Alekhin:2013nda} included hadron collider data for the first time,  with a range of LHC $W$ and $Z$ boson measurements fit, while a comparison to top quark pair production data from the LHC and the
Tevatron was also presented.
The latest ABMP16 fit~\cite{Alekhin:2017kpj} includes an increased LHC set, with single top for the first time, as well as Tevatron lepton asymmetry data.
In addition, the HERA I+II combined data set and updated NOMAD and CHORUS data on dimuon production are fit.
For the $t\overline{t}$ data, the top quark mass $m_t$ is treated in the $\overline{\rm
  MS}$ scheme and is determined from the fit, giving $m_t(m_t)=160.9\pm 1.1$ GeV. This is consistent with the PDG value of $160.0^{+4.8}_{-4.3}$ GeV, although this clearly has quite a large uncertainty, being based on a single Tevatron measurement. Converting to the pole mass, they find $m_t^{\rm pole}=(170.4\pm 1.2)$ GeV, which is in some tension with the PDG world average value of $(173.1\pm 0.6)$ GeV~\cite{Olive:2016xmw}.
The result is shown in Fig.~\ref{fig:ABMmt}, with the masses extracted at different $\alpha_S(m_Z^2)$ also given. The correlation between $m_t$ and $\alpha_s$ is clear; as discussed in~\cite{Alekhin:2017kpj},
further information on this correlation
can be provided by considering single top production data. 

Interestingly, the extracted value of the strong coupling from the ABMP16
analysis, $\alpha_S(m_Z^2)=0.1147\pm 0.0008$, is somewhat larger in this fit than in the previous sets, due dominantly to the HERA I+II combined data,
as well as by the increased number of LHC experiments,
although this is still lower than the world average value.
The ABMP16 PDFs are available at NNLO for 3,4 and 5 fixed flavours, and for a range of $\alpha_S(m_Z^2)$ values, from 0.112--0.120 in steps of 0.001, in the $n_f=5$ flavour scheme,
including the corresponding Hessian eigenvectors.
Note that, as explained, in Sect.~\ref{sec:fitmeth.theoryunc}, in the
ABM fits the uncertainties due to the value of $\alpha_S(m_Z^2)$
and the heavy quark masses are also included when determining the Hessian
eigenvectors.

\subsection{CTEQ-JLab (CJ)}\label{sec:pdfgroups.CJ}

The CTEQ-Jefferson Lab (CJ) collaboration has performed a series of global
PDF analyses~\cite{Accardi:2009br,Accardi:2011fa} with
the latest PDF set being CJ15~\cite{Accardi:2016qay}, following the
previous CJ12 set~\cite{Owens:2012bv}. The analyses are carried out at NLO in QCD only, and
focus on utilising DIS data at the highest--$x$ values applicable to a perturbative QCD
treatment. The kinematic selection cuts are chosen to be $Q^2>1.69\,{\rm GeV}^2$
and $W^2>3\,{\rm GeV}^2$ so as to keep data points at low $Q$ and high $x$,
unlike the more restrictive cuts used by most other analyses.
This results in about 1300 more data points from proton and
deuteron targets, roughly a 50\% increase in comparison to standard cuts.
These
additional data points provide useful information on the PDFs at large $x$, into the
$x\gtrsim 0.7$ region where the constraints for most global analyses are indirect
or purely from extrapolation. In particular, the deuterium data can improve on the
determination of $d$ quark at large $x$, for which the proton DIS data are less
sensitive.     

For the treatment of heavy quark mass effects in DIS structure functions, CJ12 uses
a ZM--VFNS with heavy quark masses implemented as the flavour thresholds.
CJ15 uses a more adequate S--ACOT~\cite{Kramer:2000hn} GM--VFNS to better describe
data over a wide kinematic range, including the threshold regions.
It is found that
the implementation of the GM--VFNS leads to large changes in the gluon PDF at large $x$.  
Going to low $Q$ and large $x$ involves further complications to the theoretical predictions
for the DIS structure functions, as finite $Q^2$ corrections to the leading--twist
calculation, i.e. power corrections of $\mathcal O(1/Q^2)$, must be taken into account.
In particular, the
CJ analyses adopt the standard Operator Product Expansion (OPE) expression for the target mass corrections (TMCs)
which allows structure functions at finite $Q^2$ be expressed in terms of their
massless ($M^2/Q^2\sim 0$) values through the scaling variable
$\rho^2=1+4x^2M^2/Q^2$~\cite{Georgi:1976ve,Brady:2011uy}.
Other subleading $1/Q^2$ corrections, including
higher twists, are parametrized by a phenomenological function form~\cite{Accardi:2016qay}.

Another important aspect of the CJ analyses concerns the nuclear corrections for processes
with deuteron targets, which become significant in the intermediate and large--$x$ region
and are equally important for low and high $Q$ values.
The nuclear corrections
account for Fermi motion, binding, and nucleon off--shell effects, and can be implemented
as convolutions with nuclear smearing functions.
In the CJ12 analysis, three PDF fits are
provided with different models of deuteron corrections, CJ12min, CJ12mid and CJ12max,
corresponding to mild, medium, and strong corrections, respectively.
These corrections are only applied at the level of
structure functions.

The CJ15 analysis employs a phenomenological parametrization
for part of the deuteron corrections with free parameters fitted to data, reducing
the model dependence and increasing the flexibility of the fit.
Moreover, these deuteron corrections
are formulated at the parton level and can therefore also be applied to non--DIS processes.    
For example, the total quark PDF in deuteron can be written as $q^d=q^{d({\rm on})}
+q^{d({\rm off})}$, with the on--shell and off--shell components given by~\cite{Melnitchouk:1993nk,Kulagin:1994fz},
\begin{align}\nonumber
q^{d({\rm on})}(x,Q^2)&=\int{dz \over z}f^{({\rm on})}(z)q^N(x/z,Q^2)\;,\\ \label{eq:cjq}
q^{d({\rm off})}(x,Q^2)&=\int{dz \over z}f^{({\rm off})}(z)\delta f^N(x/z,Q^2)q^N(x/z,Q^2)\; ,
\end{align}
where $q^N(x,Q^2)$ corresponds to the quark PDF in a free nucleon.
The on--shell and off--shell smearing
functions $f^{({\rm on})}$ and $f^{({\rm off})}$ can be calculated systematically within
the weak binding approximation, using the deuteron wave functions~\cite{Kulagin:2004ie,Kahn:2008nq}.
The nominal CJ15 PDF fit is based on AV18 wave functions~\cite{Wiringa:1994wb}, but alternative fits with
CD--Bonn~\cite{Machleidt:2000ge}, WJC--1 and WJC--2~\cite{Gross:2008ps} wave functions are also provided. 
The off--shell correction $\delta f^N(x)$ in Eq.~(\ref{eq:cjq}) is parametrized as~\cite{Kulagin:2004ie}
\begin{equation}
\delta f^N(x)=C(x-x_0)(x-x_1)(1+x_0-x)\; .
\end{equation}
The two zeros $x_0$, $x_1$ and the normalization $C$ are free parameters fitted to data
with the constraint of maintaining the total number of valence quarks in the nucleon.
It was found that different wave function models give similar quality
of fits to the global data set and result in changes in the PDFs that are well within the uncertainties,
as their differences can be largely compensated by the parametrization of
the off--shell corrections.  

%%%%%%%%%%%%%%%%%%%%%%%%%%%%%%%%%%%%%%%%%%%%%%%%%%%%%%%%%%%%%%%%%%%%%
\begin{figure}[t]
\begin{center}
  \includegraphics[scale=0.49]{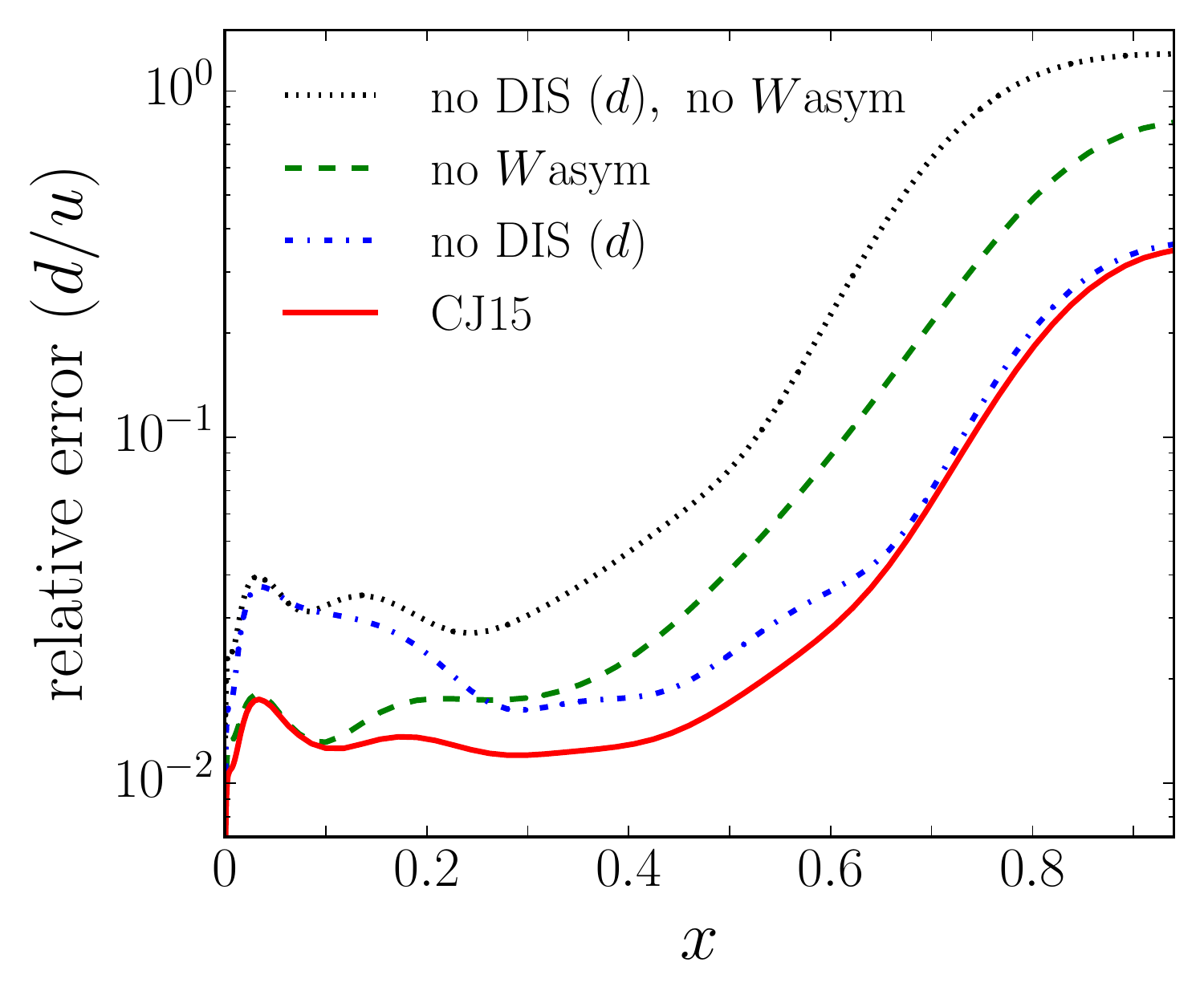}
  \includegraphics[scale=0.49]{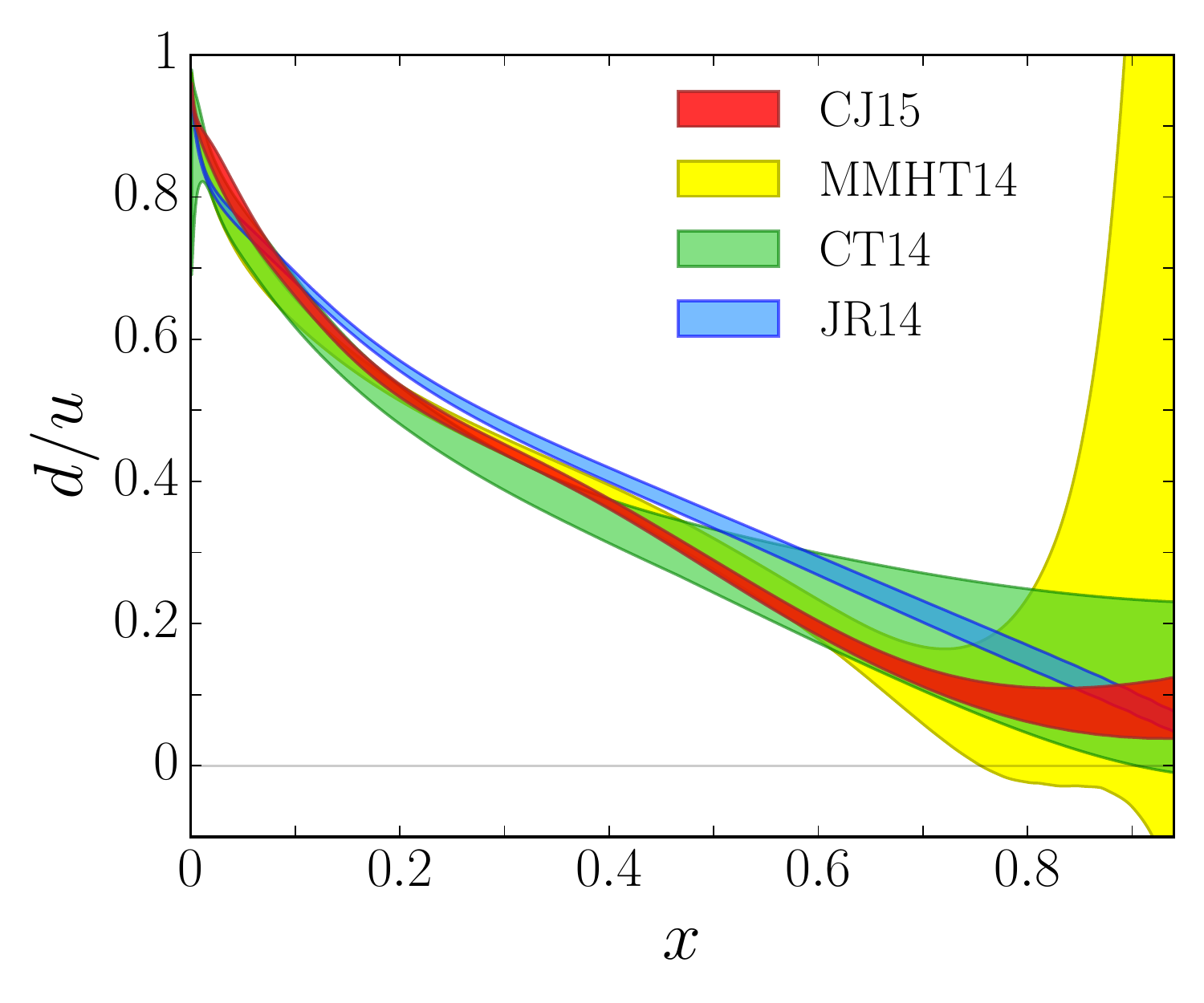}
   \caption{\small 
   Left plot: relative error (90\% C.L.) on the $d/u$ PDF ratio as a function of $x$ at
   $Q^2=10\,{\rm GeV}^2$ from the CJ15 fit, compared with uncertainties obtained in fits excluding
   certain data sets~\cite{Accardi:2016qay}.
   Right plot: comparison of the $d/u$ ratio at $Q^2=10\,{\rm GeV}^2$ for different PDF
   sets with PDF uncertainties shown fat 90\%
   C.L. except for MMHT14, which corresponds to 68\% C.L.~\cite{Accardi:2016qay}
    \label{fig:cj2}
  }
\end{center}
\end{figure}
%%%%%%%%%%%%%%%%%%%%%%%%%%%%%%%%%%%%%%%%%%%%%%%%%%%%%%%%%%%%%%%%%%%%%

In the CJ15 analysis, in addition to the deuteron data, two new data sets are found to place significant constraints on the $d$ quark PDF at large $x$.
These are the
measurement of the $F_2$ structure function of a nearly free neutron inside a deuteron nucleus
from the BONuS experiment~\cite{Baillie:2011za,Tkachenko:2014byy} at Jefferson Lab using a
spectator tagging technique, and
the lepton and reconstructed $W$ boson charge asymmetry measurements from
D0 Run 2 with full luminosities~\cite{Abazov:2013rja,D0:2014kma}.
Fig.~\ref{fig:cj2} (Left) shows the
impact of different data sets on the 90\% C.L. PDF uncertainty of the $d/u$ ratio in various CJ15 fits.
It was found that at $x\lesssim 0.3$ the DIS data from deuteron target can reduce
the PDF uncertainty on $d/u$ by almost 50\%.
For $x\gtrsim 0.3$ the $W$ asymmetry data provides the
dominant constraint. The constraint from deuteron DIS data
dies out for $x\gtrsim 0.6$, as in this region the fit is mostly sensitive to the deuteron off--shell corrections.
In Fig.~\ref{fig:cj2} (Right) the comparison of the $d/u$ ratios for the
CJ15, MMHT14, CT14 and JR14 PDFs is shown.
These are in good agreements within PDF uncertainties as $x$ goes to 1.
The CJ15 PDF set has smaller PDF errors on $d/u$ in general, with
an extrapolated value
\begin{equation}
d/u \longrightarrow_{x\rightarrow 1} 0.09\pm 0.03\; ,
\end{equation}
at 90\% C.L., due to the new data sets that constrain the $d$ quark at large $x$.
Note that the fact that the $d/u$ ratio has a finite $x\to 1$ limit occurs by construction in the CJ15
and CT14 parametrizations~\cite{Ball:2016spl}.
MMHT14 and NNPDF3.0 do not impose the exact isospin symmetry
when $x$ goes to 1, namely $b_{u_V}=b_{d_V}$ (as defined in Sect.~\ref{sec:fitmeth.PDFpara.func}),
and the predicted $d/u$ can range from zero to infinity~\cite{Ball:2016spl}.    

\begin{figure}[t]
\begin{center}
  \includegraphics[scale=0.49]{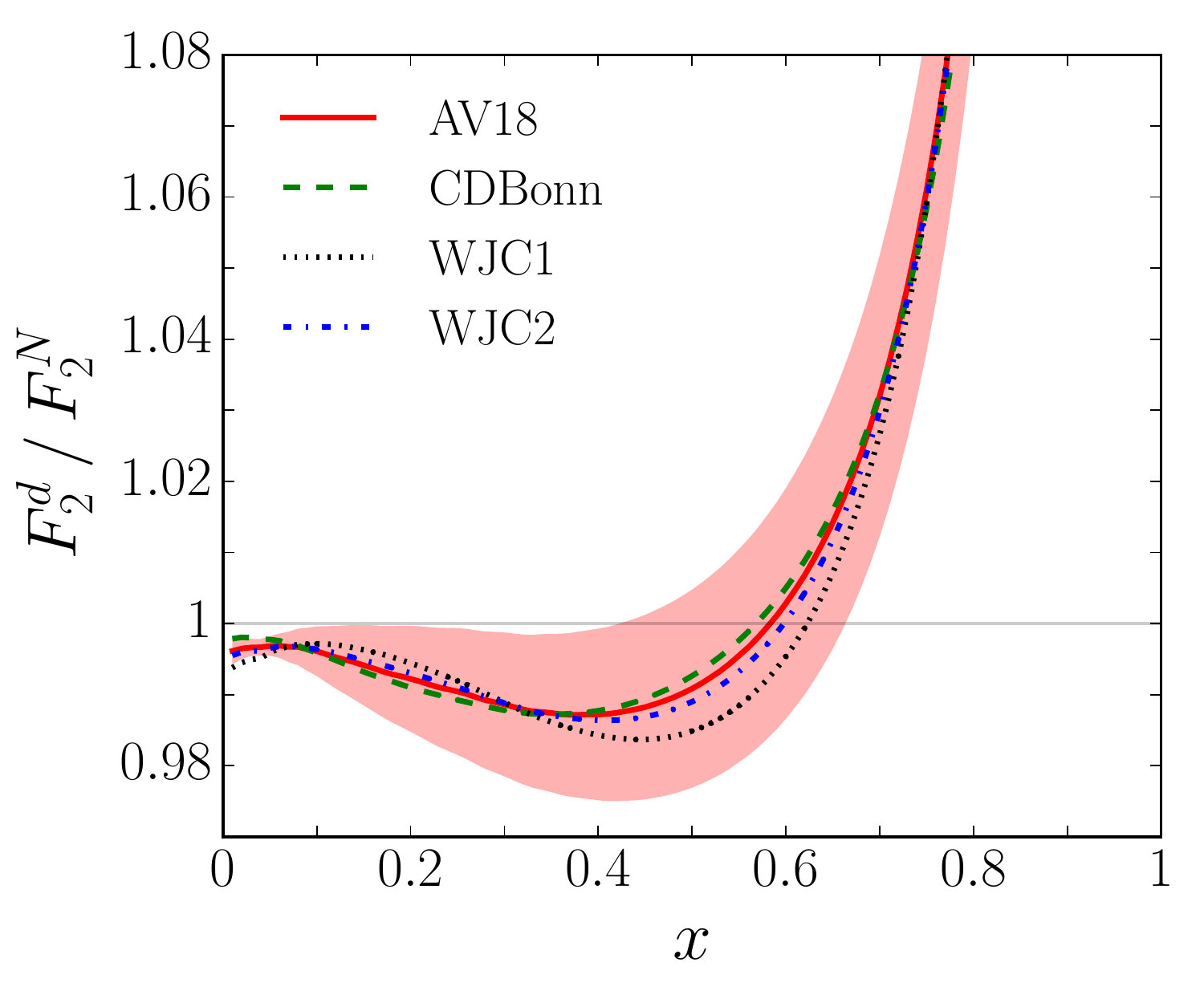}
   \caption{\small 
   Ratio of deuteron to isoscalar nucleon structure functions $F_2^d/F_2^N$ at
   $Q^2=10\,{\rm GeV}^2$ for the CJ15 fits with different models of wave functions~\cite{Accardi:2016qay};
   the colored band is the 90\% C.L. error for CJ15 nominal fit, obtained
   using the AV18 nuclear wave functions. 
    \label{fig:cj3}
  }
\end{center}
\end{figure}

The CJ15 analysis is able to
pin down the deuteron corrections through the interplay of 
additional data sets that are less sensitive to the nuclear corrections,
i.e., the D0 $W$ asymmetry data and BONuS data, 
and the deuteron DIS
structure function data.
This is illustrated in Fig.~\ref{fig:cj3}, which gives the deuteron to isoscalar nucleon ratio $F_2^d/F_2^N$ from
the CJ15 fits with different input wave functions together with the 90\% C.L. uncertainty
of the CJ15 nominal fit shown in the coloured bands. Significant corrections are found for
$x\gtrsim 0.7$.
The ratio turns out to be insensitive to the choice of wave functions, since it
is only the combination of the wave function and the off--shell corrections that are
constrained by current data.               

%%%%%%%%%%%%%%%%%%%%%%%%%%%%%%%%%%%%%%%%%%%%%%%%%%%%%%%%%%%%%%%%%%%%%
%%%%%%%%%%%%%%%%%%%%%%%%%%%%%%%%%%%%%%%%%%%%%%%%%%%%%%%%%%%%%%%%%%%%%

\subsection{HERAfitter/xFitter}\label{sec:pdfgroups.xfitter}

\begin{figure}[t]
\begin{center}
  \includegraphics[scale=0.35]{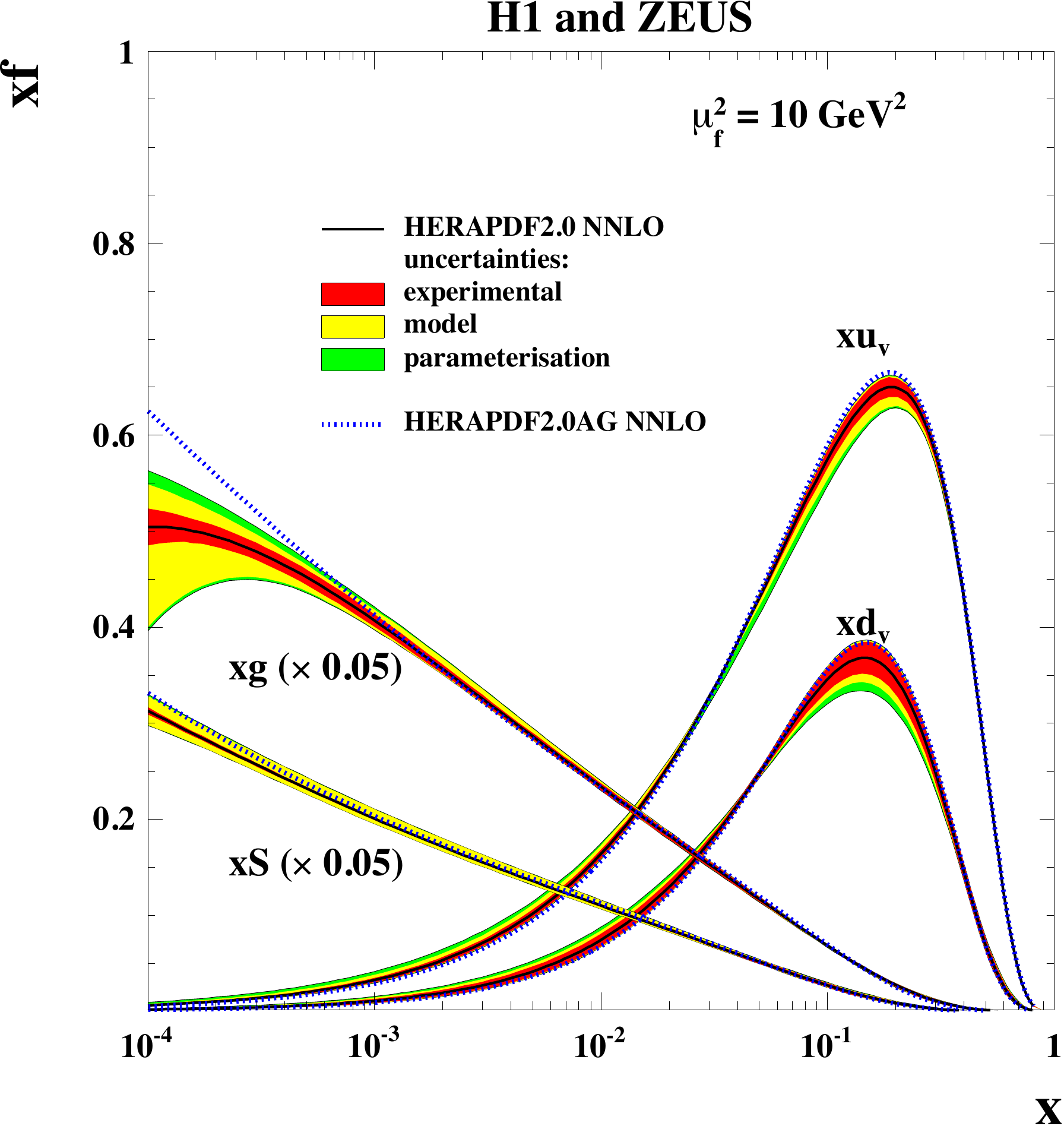}\qquad
  \includegraphics[scale=0.50]{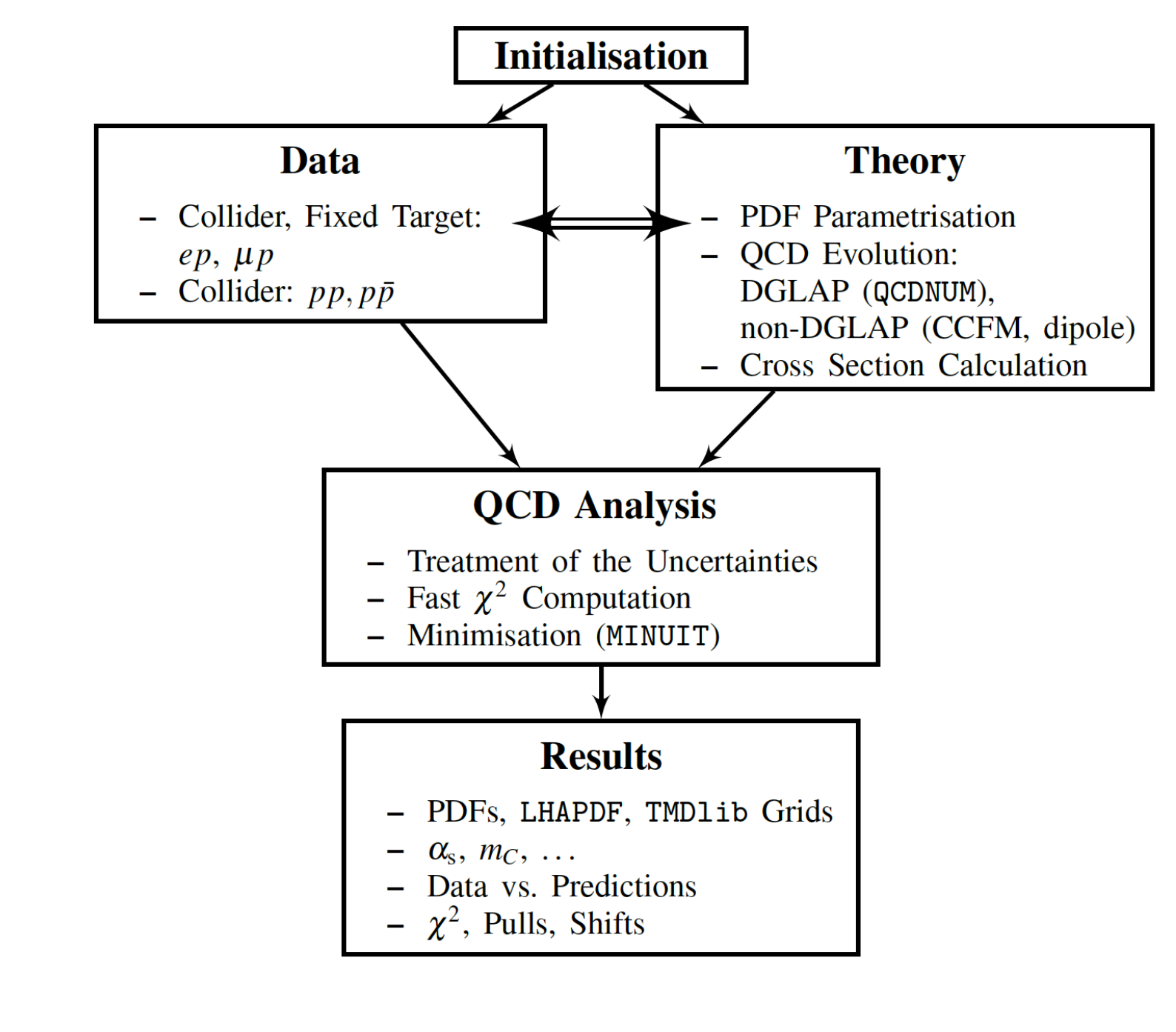}
   \caption{\small Left plot: the HERAPDF1.0 determination of parton distributions,
   based on the analysis of the combined HERA structure functions from
   Run I.
   Right plot: schematic representation of the {\tt xFitter} code structure, see text
   for more details.
    \label{fig:xfitter}
  }
\end{center}
\end{figure}

For many years, the H1 and ZEUS collaborations have performed QCD
 analyses of their structure function data, first separately
and then together based on the H1+ZEUS combined datasets.
The backbone of these analyses was the neutral-- and charged--current inclusive structure
function measurements, in some cases supplemented by the charm production
structure functions and DIS jet cross sections.
The main results from these studies were the HERAPDF family of PDF fits, which include
HERAPDF1.0~\cite{Aaron:2009aa}, based on the Run I data, and
HERAPDF2.0~\cite{Abramowicz:2015mha}, based on the final combination
of inclusive measurements from Runs I+II.
In Fig.~\ref{fig:xfitter} (Left) we show the results of the HERAPDF2.0 analysis for the
$u_V$, $d_V$ and $S$ quarks and the gluon. The central value of the HERAPDF2.0A.G set, corresponding to
an alternative gluon parameterisation, is also shown.

In the HERAPDF methodology, the total PDF uncertainty is divided into three types: experimental uncertainties, propagated from the statistical
and systematic uncertainties in the fitted data, model uncertainties, for instance
due to $\alpha_s$ and $Q_0$ variations, and parametrization uncertainties, reflecting
the spread from different comparable choices of input functional form for the PDFs.
Note that in principle, for a sufficiently flexible input parametrization,
it should be possible to neglect
the model uncertainties related to the choice of $Q_0$ as well as the parametrization
uncertainties.

The expertise developed by  these QCD analyses
of the HERA structure function data lead to the development and release of
{\tt HERAfitter}~\cite{Alekhin:2014irh}, a publicly available open--source PDF fitting toolbox.
This was developed as an extension of the H1 and ZEUS internal PDF fitting
codes, that were extensively tested and applied in various QCD analyses
of HERA inclusive and charm data, including the HERAPDF sets.
Despite its name, {\tt HERAfitter}  was not restricted
to the analysis of HERA data, and could also be used for the PDF interpretation
of measurements from fixed--target DIS and proton--proton collisions.
The flexibility of this open--source software tool also allows QCD
analyses beyond unpolarized fixed--order PDF fits to be performed, such as fits of transverse momentum
dependent (TMD) parton distributions and fragmentation functions.
Recently, {\tt HERAfitter} was renamed {\tt xFitter}, to emphasise that this code is a general fitting toolbox,
not  necessarily related to or involving the analysis of HERA inclusive structure function data.

The {\tt xFitter} framework includes modules that allow for various theoretical and
methodological options, and is capable of fitting to a large number of data sets from
HERA, Tevatron and LHC.
For example, polarized and unpolarized PDF evolution can be performed
using either {\tt APFEL} or {\tt QCDNUM}, and a number of fixed and variable
flavour number schemes are implemented, such as the FFN scheme from {\tt OpenQCDrad}
and the S--ACOT, TR and FONLL general--mass schemes.
In addition to PDF fitting, a large number of other functionalities are available,
such as the approximate inclusion of new datasets in existing PDF sets
by means of either Bayesian reweighting and Hessian profiling (see Sect.~\ref{sec:fitmeth.approx}),
and a wide variety
of PDF plotting options.

In Fig.~\ref{fig:xfitter} (Right) we show a schematic representation of the 
{\tt xFitter} code structure.
The first part is the initialization, where the fit settings are specified
in the steering file.
This involves a number of choices, in particular selecting the fitted data sets
and the theory and methodology settings such as the specific PDF parametrisation
or the scheme for heavy quark structure functions.
Then the PDF fit is performed, where the fit parameters are determined
by means of {\tt MINUIT}--based minimization including the propagation
of experimental results.
The final result is the {\tt LHAPDF6} grid file, together with various
PDF and data/theory comparison plots.

%%%%%%%%%%%%%%%%%%%%%%%%%%%%%%%%%%%%%%%%%%%%%%%%%%%%%%%%%%%%%%%%%%%%%%%%%%%%%%%%%%%%%%
%%%%%%%%%%%%%%%%%%%%%%%%%%%%%%%%%%%%%%%%%%%%%%%%%%%%%%%%%%%%%%%%%%%%%%%%%%%%%%%%%%%%%%%%%%

The {\tt HERAfitter/xFitter} framework has been used in many ATLAS and CMS
PDF interpretation studies, discussed in Sect.~\ref{sec:pdfgroups.exp}.
In addition, the {\tt HERAfitter/xFitter} developer's team has released a number
of dedicated PDF studies, including:
\begin{itemize}
\item A QCD analysis of the legacy $W$ and $Z$ boson production measurements at the
Tevatron~\cite{Camarda:2015zba}, based on the precise $W$ asymmetries in the electron
and muon channels by the D0 collaboration, together with the HERA structure function data.
This analysis demonstrated that these measurements, which are now included in most global
PDF fits, provide useful information on quark flavour separation at medium and
large $x$.

\item A determination of the running charm quark mass $m_c(m_c)$
from HERA structure function data within the framework of the FONLL
GM--VFNS~\cite{Bertone:2016ywq}.
This study demonstrated that the best--fit value of $m_c(m_c)$
was consistent with the result of a fit using the FFNS, as expected from general theoretical considerations. This value was also consistent with previous determinations of the running mass from HERA data
and with the global PDG average.

\item A determination of the photon PDF $x\gamma(x,Q^2)$~\cite{Giuli:2017oii} from the
measurement of high mass Drell--Yan cross sections at 8 TeV from
the ATLAS collaboration.
This was the first analysis where LHC data was included in a QED fit of the photon
PDF directly, rather than using reweighting methods, by means of an extension
of the {\tt aMCfast} interface to  account for the photon--initiated contributions
in {\tt MadGraph5\_aMC@NLO}.
The results of this analysis showed that the high--mass DY data indeed allowed important
constraints on the photon PDF at intermediate $x$, although the resulting
PDF uncertainties are
not competitive with those from the more recent determinations of $\gamma(x,Q^2)$,
discussed in Sect.~\ref{sec:QED.photon}.

\end{itemize}

\begin{figure}[t]
\begin{center}
  \includegraphics[scale=0.85]{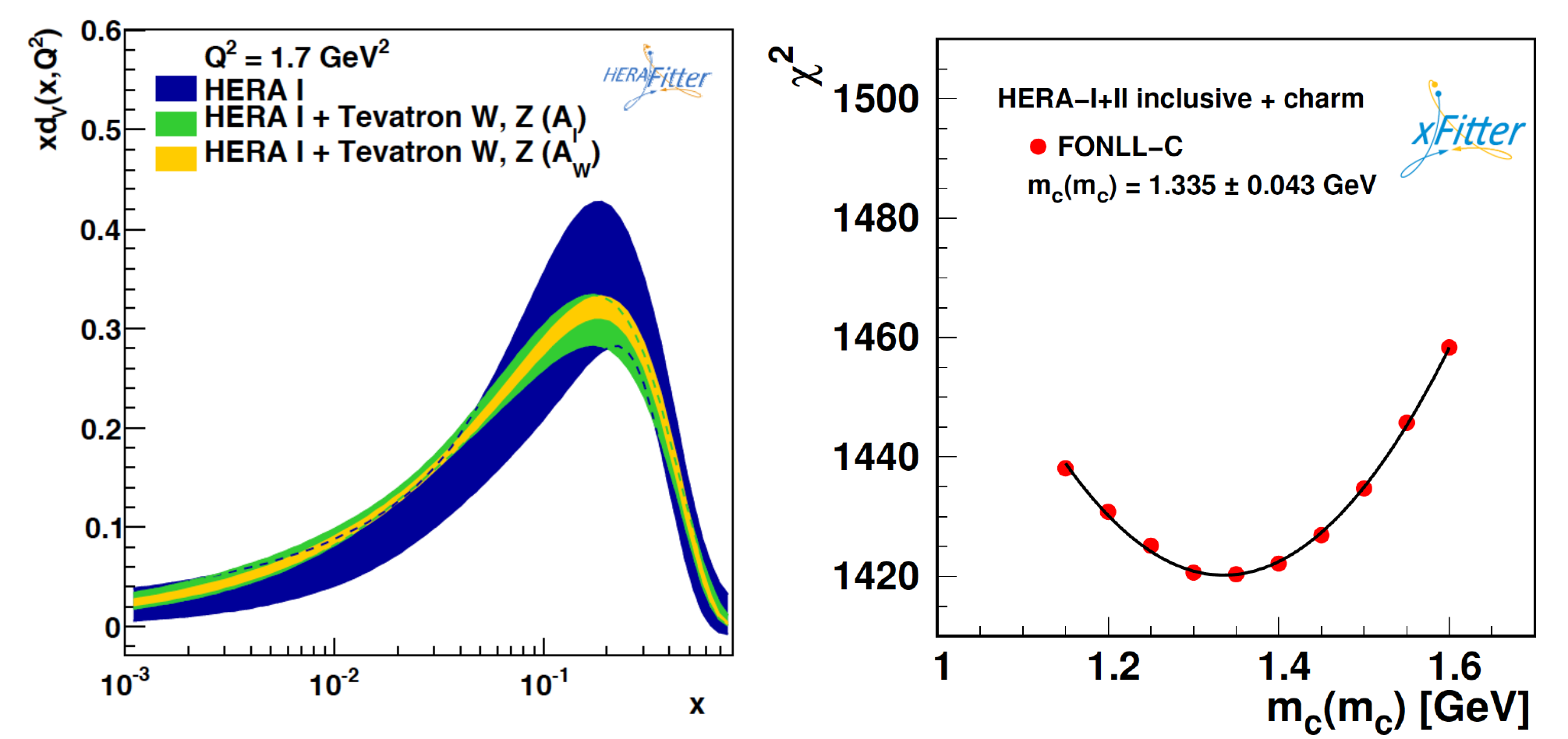}
   \caption{\small Two representative analyses of PDF--related studies
   performed by the {\tt xFitter} Developer's Team.
   Left plot: the impact on $d_V$ of the Tevatron $W$ and $Z$ data on a HERA--only
   fit, comparing the impact of the lepton--level measurements with that of the boson--level
   measurements, from Ref.~\cite{Camarda:2015zba}.
   Right plot: the $\chi^2$ profile of a fit based on the inclusive
   HERA and charm data, as a function of the running mass $m_c(m_c)$ from
   Ref.~\cite{Bertone:2016ywq}.
   In this analysis, the charm structure functions were computed with {\tt APFEL}
   in the FONLL--C general mass scheme.
    \label{fig:xfitter2}
  }
\end{center}
\end{figure}

Additional work based on {\tt xFitter} include studies of PDFs
with correlated uncertainties between different
perturbative orders~\cite{Belov:2014xwo},
non--DGLAP evolution equations~\cite{Hautmann:2017xtx},
a study of the impact of the heavy quark matching scales in PDF
fits~\cite{Bertone:2017ehk},
and
the determination of transverse momentum
dependent PDFs~\cite{Angeles-Martinez:2015sea}.
In Fig.~\ref{fig:xfitter2} we show two representative PDF--related analyses
   performed by the {\tt xFitter} Developer's Team.
   First, we show the impact on the $d_V$ PDF of the Tevatron $W$ and $Z$ data when
   added to an HERA--only
   fit, comparing the impact of the lepton--level measurements with that of the boson--level
   measurements, from the {\tt xFitter} analysis of Ref.~\cite{Camarda:2015zba}.
  We also show the $\chi^2$ profile of a {\tt xFitter} fit based on the inclusive
   HERA and charm data, as a function of the running mass $m_c(m_c)$ from
   Ref.~\cite{Bertone:2016ywq}.
   In this analysis charm structure functions were computed with {\tt APFEL}
   in the FONLL--C general mass scheme.
   As discussed above, this analysis finds a value of the running charm
   mass $m_c(m_c)=1.335\pm 0.043$, which is consistent with the PDG average
   as well as with previous determinations based on HERA data, such
   as those in~\cite{Alekhin:2012vu,Gao:2013wwa,Alekhin:2017kpj}.
   
%%%%%%%%%%%%%%%%%%%%%%%%%%%%%%%%%%%%%%%%%%%%%%%%%%%%%%%%%%%%%%%%%%%%%%%%%%%%%%%%%%%%%%
%%%%%%%%%%%%%%%%%%%%%%%%%%%%%%%%%%%%%%%%%%%%%%%%%%%%%%%%%%%%%%%%%%%%%%%%%%%%%%%%%%%%%%%%%%

Concerning future developments, the {\tt xFitter} code is now being rewritten
from Fortran to C++, to ensure modularity and to facilitate its maintenance and
the addition of novel theoretical ingredients.
Several new external codes and additional features are being implemented, such
as the possibility of new parametrization options like Chebyshev polynomials,
the fast convolution option for hadronic cross sections as realised
in {\tt APFELgrid}~\cite{Bertone:2016lga}, more flexible PDF parametrizations
including the charm and the photon PDF, and improvements in the QED evolution
interface.

\subsection{PDF efforts by the LHC collaborations}\label{sec:pdfgroups.LHC}

As discussed in Sect.~\ref{sec:datatheory}, the LHC experiments have provided
a large number of experimental measurements with important
PDF sensitivity, most of which are now part of the toolbox
of global PDF fits.
In addition to presenting the results of such measurements, the ATLAS 
and CMS collaborations have developed an active
program of PDF interpretation studies, aimed at quantifying the constraints of their
data on proton structure.

In most cases, these studies have been performed using the {\tt HERAfitter/xFitter} frameworks,
described in Sect~\ref{sec:pdfgroups.xfitter}.
Thus the PDFs are parameterised at $Q_0^2=1.9\,{\rm GeV}^2$ in terms of relatively
simple polynomials in $x$.
Fits are then performed with an increasing number of free parameters introduced up to the point when no further improvement in $\chi^2$ is observed. This generally leads to $\sim 15$ free parameters in the fit, with the precise number depending on the particular analysis. Experimental uncertainties are calculated using the standard `$\Delta\chi^2=1$' criteria, and
additional model and parameterisation uncertainties are included, as in the HERAPDF approach.
In all cases, either the HERA I DIS~\cite{Aaron:2009aa}, or in later studies the I + II combination~\cite{Abramowicz:2015mha} are included in a baseline fit,
before assessing the impact of the corresponding LHC measurements, which
are then fit in addition.

However, it should be emphasised that such studies can only give an indication of the effect of the considered data on the PDFs. In particular, as they compare to a restricted DIS--only baseline set, the impact will inevitably be exaggerated in comparison to its inclusion within a global fit, and indeed if some dominant constraint from a global analysis is omitted, then the presented results may be misleading. Moreover, due to the relatively restricted parameterisation that is used, the PDF impact in $x$ regions, for example at high $x$, that are not directly probed by the data can have non--negligible parameterisation dependence.

Nonetheless, such studies represent an important contribution to the PDF fitting community,
not only because they give an indication the PDF impact of specific measurements,
but also because they provide an internal cross-checks that the information
required for PDF fits, in particular the full experimental covariance matrix,
is ready to be used.
In the following, we describe the individual efforts from ATLAS and CMS,
highlighting a representative sample of their PDF studies.

\subsubsection{ATLAS}\label{sec:pdfgroups.LHC.ATLAS}

A representative selection of ATLAS PDF interpretation studies is given below:

\begin{itemize}

\item The ATLAS measurements of $W^+$, $W^-$ and $Z$ rapidity distributions
at 7 TeV from the 2010 dataset were used in Ref.~\cite{Aad:2012sb}, to determine
the strange content of the proton.
The full cross-correlations between the three
rapidity distributions were accounted for, and while $W^+$ and $W^-$ measurements constrained
the up and down quarks and antiquarks, the $Z$ measurement constrained the strangeness.
This analysis found that the strange sea was not suppressed relative to the up and down
quark sea.

\item The recent study~\cite{Aaboud:2016btc},
based on the updated $W^+$, $W^-$ and $Z$ rapidity distributions
at 7 TeV  from the 2011 dataset, corresponding to a greatly improved precision in comparison to 2010.
An analysis of this data was found to prefer a strange quark sea that is symmetric with respect
to the light quark sea, consistent with the previous findings
of~\cite{Aad:2012sb}.
The PDF uncertainties in the strangeness
determination were significantly reduced in comparison to the PDFs determined from the
analysis of the 2010 data.
The issue of the strange content of the proton will be
discussed in more detail in Sect.~\ref{sec:structure.strange}.

\item PDF fits based on jet production measurements have also been performed
by ATLAS.
For instance, in~\cite{Aad:2013lpa} an analysis of the inclusive jet cross sections at
$\sqrt{s}=2.76$ TeV and $\sqrt{s}=7$ TeV (from the 2010 run) was performed.
It was shown that an improved constraint could be achieved by considering
the ratio $R_{7/2.76}$ of jet cross-sections, given that many
experimental and theoretical uncertainties cancel when taking
such ratios between different centre-of-mass energies~\cite{Mangano:2012mh}.

\end{itemize}

\begin{figure}[t]
\begin{center}
  \includegraphics[scale=0.45]{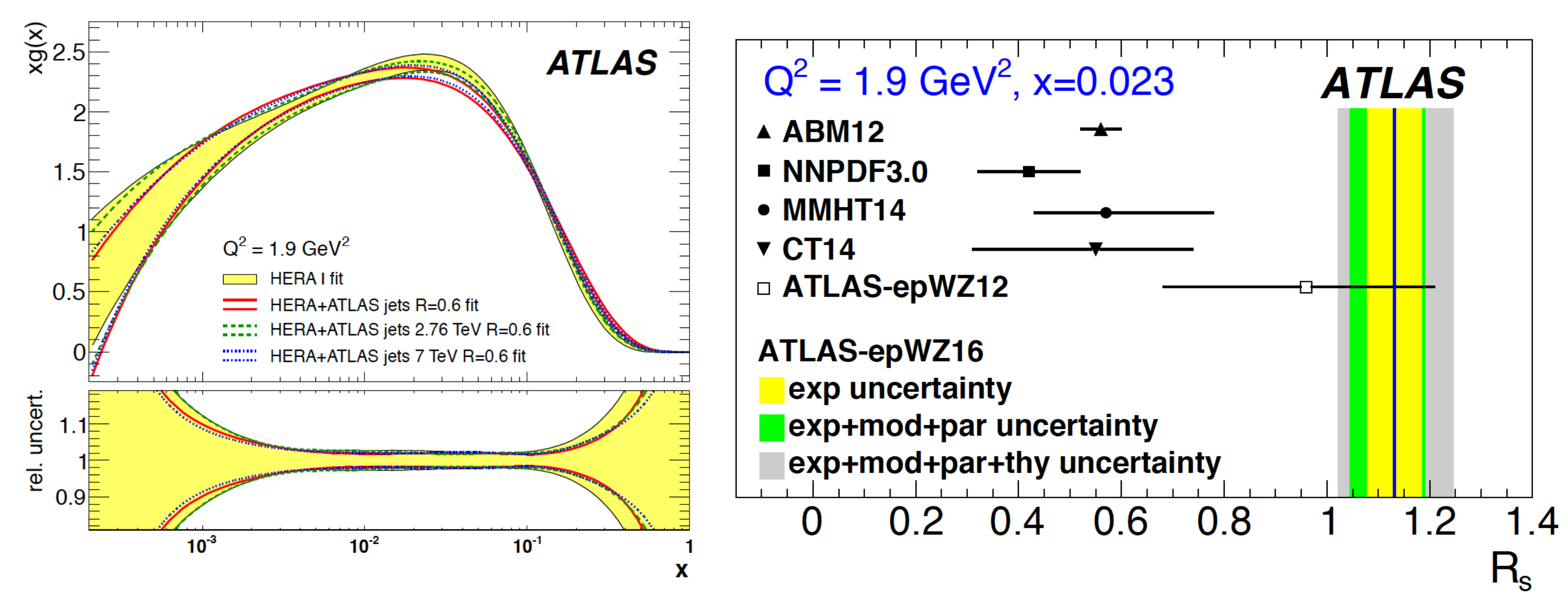}
   \caption{\small Two representative results of the PDF fitting efforts
   performed within the ATLAS collaboration.
   Left plot: a PDF fit quantifying the effect on the gluon from the HERA-only fit
   of the ATLAS inclusive jet measurements at $\sqrt{s}=2.76$ TeV
   and 7 TeV, from Ref.~\cite{Aad:2013lpa}.
   Right plot: the determination of the strangeness ratio $R_s(x=0.023,Q^2=1.9\,{\rm GeV}^2)$
   for a fit to HERA data and the 2011 ATLAS measurements of the $W^{\pm}$ and $Z$
   rapidity distributions at 7 TeV, where the results of the {\tt xFitter} analysis,
   denoted by ATLAS-epWZ16, are compared with the predictions from various PDF fits. Taken from~\cite{Aaboud:2016btc}.
    \label{fig:atlaspdffits}
  }
\end{center}
\end{figure}

In Fig.~\ref{fig:atlaspdffits} we show some representative results
of PDF interpretation studies performed within the ATLAS collaboration.
In the left plot, we show the results of a
PDF fit quantifying the effect on the gluon 
   of the ATLAS inclusive jet measurements at $\sqrt{s}=2.76$ TeV
   and 7 TeV, in comparison to a HERA--only fit.
   In the right plot we show the
   determination of the strangeness ratio $R_s(x=0.023,Q^2=1.9\,{\rm GeV}^2)$
   for a HERA--only fit and including the 2011 ATLAS measurements of the $W^{\pm}$ and $Z$
   rapidity distributions at 7 TeV. The results of the ATLAS analysis,
   denoted by ATLAS-epWZ16, are compared with the predictions from various PDF fits.

%%%%%%%%%%%%%%%%%%%%%%%%%%%%%%%%%%%%%%%%%%%%%%%%%%%%%%%%%%%%%%%%%%%%%%%%%%%%%%%%%%%%%%
%%%%%%%%%%%%%%%%%%%%%%%%%%%%%%%%%%%%%%%%%%%%%%%%%%%%%%%%%%%%%%%%%%%%%%%%%%%%%%%%%%%%%%%%%%

\subsubsection{CMS}\label{sec:pdfgroups.LHC.CMS}

%%%%%%%%%%%%%%%%%%%%%%%%%%%%%%%%%%%%%%%%%%%%%%%%%%%%%%%%%%%%%%%%%%%%%%%%%%%%%%%%%%%%%%
\begin{figure}[t]
\begin{center}
  \includegraphics[scale=0.65]{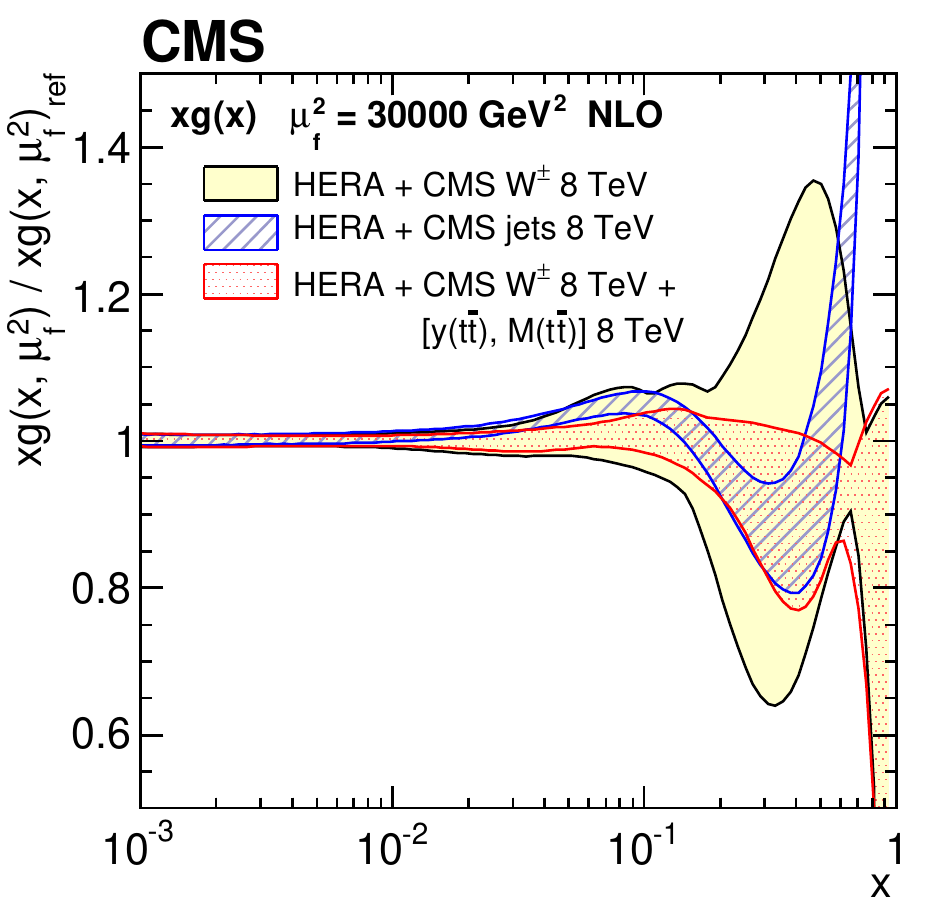}\qquad
  \includegraphics[scale=0.25]{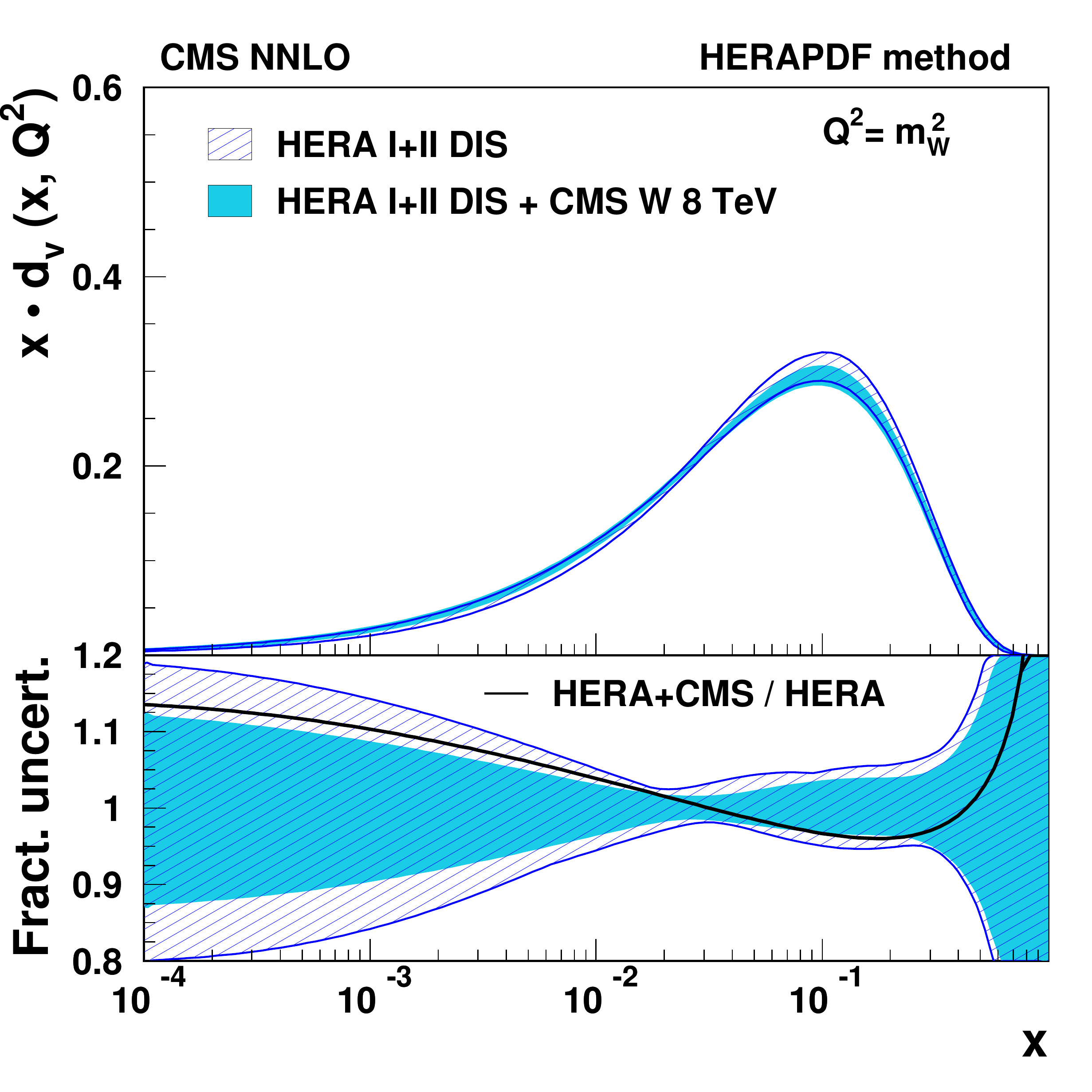}
   \caption{\small 
    \label{fig:cmspdffits}
    Left: the gluon distribution at $\mu^2= 30000\,{\rm GeV}^2$, as obtained from a PDF fit to HERA DIS data and CMS $W^\pm$ boson charge asymmetry measurements~\cite{Khachatryan:2016pev}, the CMS inclusive jet production cross sections~\cite{Khachatryan:2016mlc}, and the $W^{\pm}$ boson charge asymmetry plus the double-differential $t\overline{t}$ cross section~\cite{Sirunyan:2017azo}, in all cases at 8 TeV. All presented PDFs are normalized to the results from the fit using the DIS and $W^{\pm}$ boson charge asymmetry measurements, and the total uncertainty band in each fit is shown. Taken from~\cite{Sirunyan:2017azo}.
    Right: the down valence distribution at $\mu^2=M_W^2$ as obtained from a PDF fit to the HERA DIS data and CMS $W^\pm$ boson charge asymmetry measurement at 8 TeV, with the total PDF uncertainties shown. In the lower panel the distributions are normalized to 1. Taken from~\cite{Khachatryan:2016pev}.
  }
\end{center}
\end{figure}
%%%%%%%%%%%%%%%%%%%%%%%%%%%%%%%%%%%%%%%%%%%%%%%%%%%%%%%%%%%%%%%%%%%%

A representative selection of PDF interpretation studies from CMS is given below:

\begin{itemize}

\item In~\cite{Chatrchyan:2013mza} the 7 TeV measurement of the $W$ charge asymmetry, as well as $W+c$ production~\cite{Chatrchyan:2013uja}, is fit at NLO, and improvements in the determination of the up and down valence quark PDFs due to the $W$ asymmetry, and the strange quark PDFs due to the $W+c$ data, are demonstrated. 

\item In~\cite{Khachatryan:2016pev} a fit to the 8 TeV differential $W$ boson production data~\cite{Khachatryan:2016pev} is performed at NNLO, again showing improvements in the 
determination of the up and down valence quark PDFs.

\item In~\cite{Khachatryan:2014waa} the 7 TeV inclusive jet measurement is fit at NLO, and the significant impact of these data on the gluon PDF in particular is demonstrated. A study is also performed here using the MC method for PDF determination, allowing for a more flexible PDF parameterisation, and consistent results are found but with larger PDF uncertainties.

\item A NLO fit to the 8 TeV jet data is performed in~\cite{Khachatryan:2016mlc}, and a direct comparison to the 7 TeV case is shown, with the impact found to be very similar.

\item In~\cite{Sirunyan:2017azo} a NLO fit to the 8 TeV double--differential top pair production data is compared to a baseline fit that includes the 8 TeV $W$ boson production data~\cite{Khachatryan:2016pev}. The impact of including the data differential in different kinematic variables is assessed, and a sizeable reduction in the uncertainty on the gluon PDF in particular is found for $x>0.01$, with the largest constraint coming from the rapidity, $y_{t\overline{t}}$, and invariant mass, $M_{t\overline{t}}$ of the top pair.

\item More recent preliminary results including PDF fits to triple--differential dijet production at 8 TeV~\cite{CMS:2016wpr} and the top pair production cross section at 5.02 TeV~\cite{CMS:2017olb} have been presented.

\end{itemize}

To illustrate these CMS PDF studies,
in Fig.~\ref{fig:cmspdffits}  we show the impact on the gluon PDF of the CMS $W^\pm$ data~\cite{Khachatryan:2016pev}, the double--differential top pair production data~\cite{Sirunyan:2017azo} and the inclusive jet production data~\cite{Khachatryan:2016mlc}, in all cases at 8 TeV. This is seen to lead to a sizeable reduction in the uncertainty at higher $x$, in a way that is consistent between the data sets in the probed $x$ region.
The $t\overline{t}$ differential data are competitive with the jet measurement. In Fig.~\ref{fig:cmspdffits} (Right) we show the impact of the  CMS $W^\pm$ boson charge asymmetry measurement at 8 TeV on the down valence distribution, in comparison to a HERA--only fit. The impact on the shape, and reduction in uncertainties, achieved by the asymmetry data is clear.

\label{sec:pdfgroups.exp}

%%%%%%%%%%
\vspace{0.6cm}
\section{The proton structure}\label{sec:structure}
In this section we  compare the results of a number of 
state--of--the--art PDF analysis.
This comparison is organised in terms of specific
PDF flavour combinations relevant for phenomenology.
We begin by discussing the gluon PDF,
before considering the quark flavour
separation, 
and subsequently studying the strange and charm content
of the proton.
In the following, we compare  the ABMP16, CT14, MMHT14,
     and NNPDF3.1 NNLO sets, all with $\alpha_{S}(m_Z)=0.118$.
     We will present results both at low ($Q=1.7$ GeV)
     and high ($Q=100$ GeV) scales.
     While we will
     only show a representative selection of PDF comparisons,
other results, including with PDF sets not shown
here, can be straightforwardly produced using the {\tt APFEL-WEB}
online PDF plotting interface~\cite{Carrazza:2014gfa}.

We emphasize that in order to ensure a consistent comparison between PDFs,
we use the same value of $\alpha_{S}(m_Z)=0.118$ for all the sets.
This is however not the best-fit value for ABMP16, which is instead
$\alpha_{S}(m_Z)=0.1147$.
We will briefly comment on the main differences in ABMP16 with respect to
the two different values of $\alpha_s(m_Z)$ in the following sections.
Note also that  in the comparisons at low scales $Q$ the ABMP16 
cannot be included, since their set
with $\alpha_{S}(m_Z)=0.118$ is only available in the $n_f=5$ scheme,
and therefore can only be used for $Q > m_b$.

\subsection{The gluon PDF}\label{sec:structure.gluon}
\begin{figure}[t]
\begin{center}
  \includegraphics[scale=0.40]{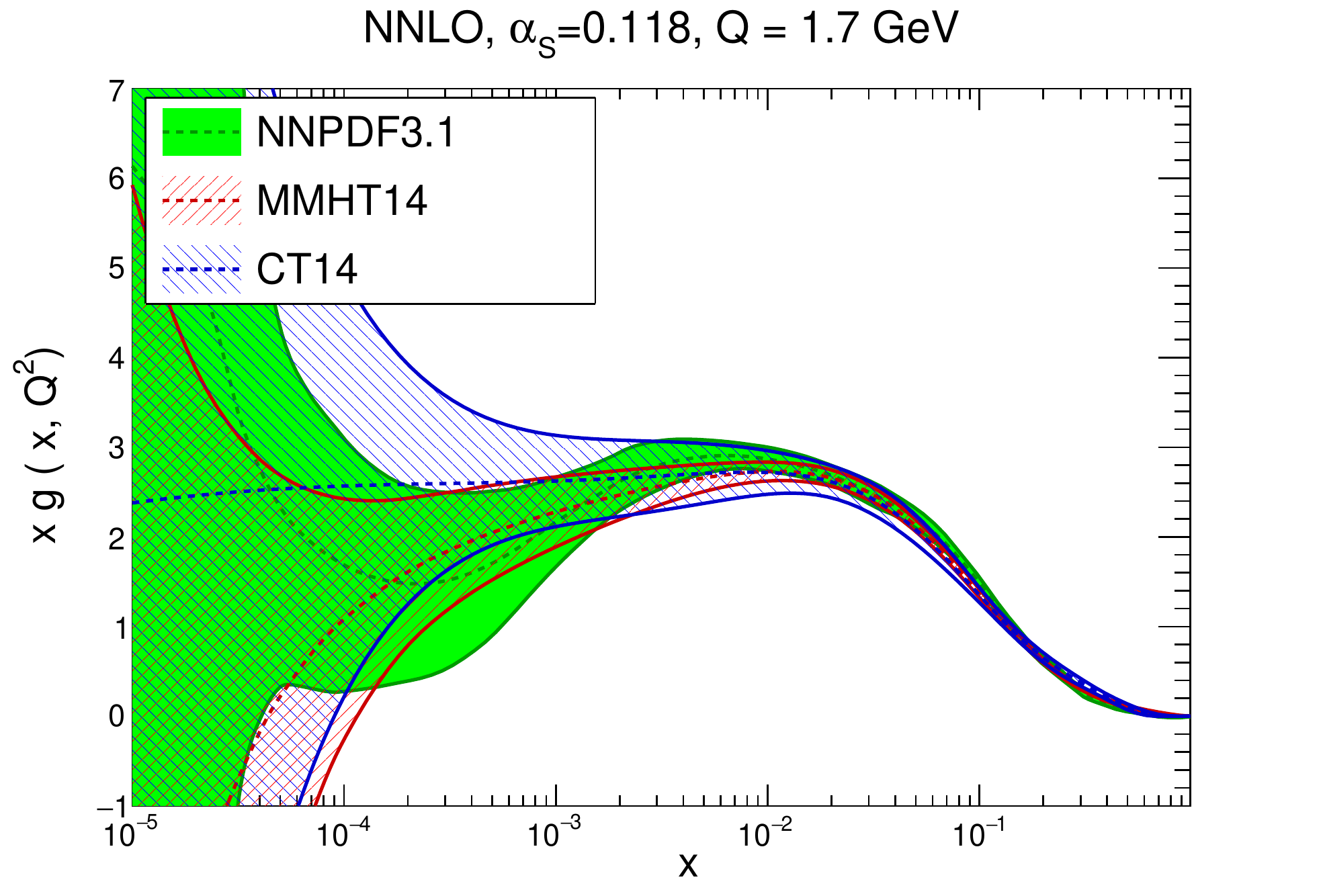}
  \includegraphics[scale=0.40]{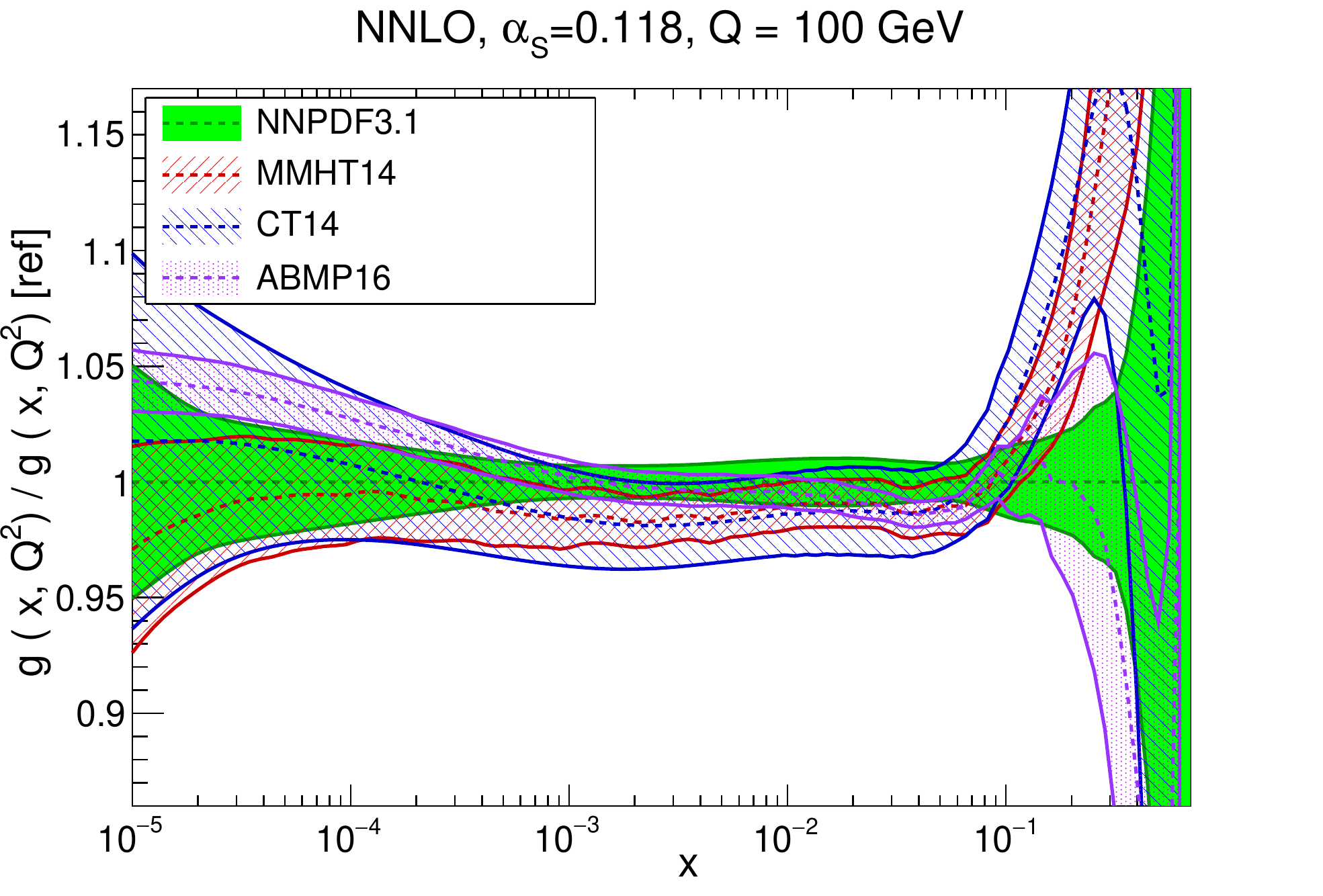}
   \caption{\small 
     The gluon PDF $xg(x,Q^2)$ at $Q=1.7$ GeV (left)
     and $Q=100$ GeV (right) comparing the ABMP16, CT14, MMHT14,
     and NNPDF3.1 NNLO sets with $\alpha_{S}(m_Z)=0.118$.
     In the right plots, results are normalized to the central value
     of NNPDF3.1.
    \label{fig:pdfcomp-xg}
  }
\end{center}
\end{figure}

In Fig.~\ref{fig:pdfcomp-xg} we show the gluon PDF, $g(x,Q^2)$, at
a low scale $Q=1.7$ GeV (left)
     and at a typical LHC scale of $Q=100$ GeV (right).
     We find that
in general there is reasonable agreement between the four
considered sets within PDF uncertainties.
This remains true at small $x$ and small $Q$, where the PDF uncertainties become rather
large, due to the limited  experimental constraints.
In this case, while the central value of the MMHT14 gluon becomes negative for $x\lsim 5\times 10^{-5}$,
the CT14 result exhibits a flat behaviour, and the NNPDF3.1 gluon increases rapidly.

%%%%%%%%%%%%%%%%%%%%%%%%%%%%%%%%%%%%%%%%%%%%%%%%%%%%%%%%%%%%%%%%%%%%%
%%%%%%%%%%%%%%%%%%%%%%%%%%%%%%%%%%%%%%%%%%%%%%%%%%%%%%%%%%%%%%%%%%%%%

Perhaps the most important discrepancy between the gluon PDFs from the four
groups arises in the large-$x$ region, where the NNPDF3.1 result, and even more markedly ABMP16, is rather softer in comparison to CT14 and MMHT14.
For example, at $x\simeq 0.2$ the differences between the NNPDF3.1 and
CT14 central values are at the 2-$\sigma$ level.
This may be related to the fact that
different datasets are used to constrain the large-$x$ gluon, with NNPDF3.1
in particular fitting to top-quark differential distributions, in contrast to the other sets. 
This has been shown to lead to a softer large-$x$ gluon in comparison to the
same fit without any $t\bar{t}$ data included~\cite{Czakon:2016olj},
see also Sect.~\ref{sec:datatheory.top}.
Indeed,  the CT14 and the previous NNPDF3.0 sets (which do not fit such data)
 are in better agreement within uncertainties.
It will therefore be interesting to compare the large-$x$
gluon PDF once other groups also include the $t\bar{t}$
differential measurements.
We also note that in the large-$x$ region PDF uncertainties
are quite large, leading to siginificant theoretical
uncertainties for the production of new BSM heavy particles,
as will be discussed in Sect.~\ref{sec:LHCpheno.BSM}.

\begin{figure}[t]
\begin{center}
  \includegraphics[scale=0.40]{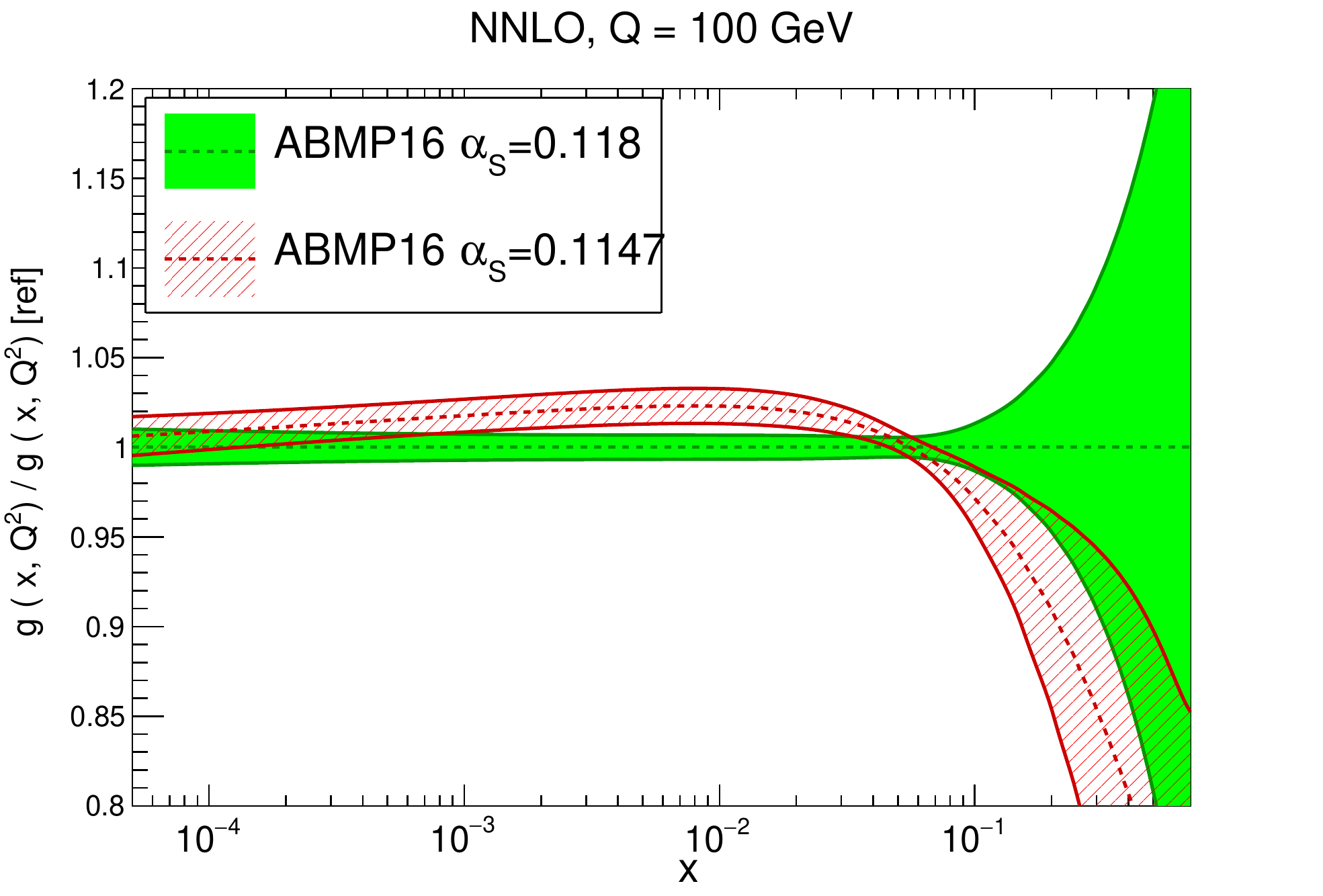}
  \includegraphics[scale=0.40]{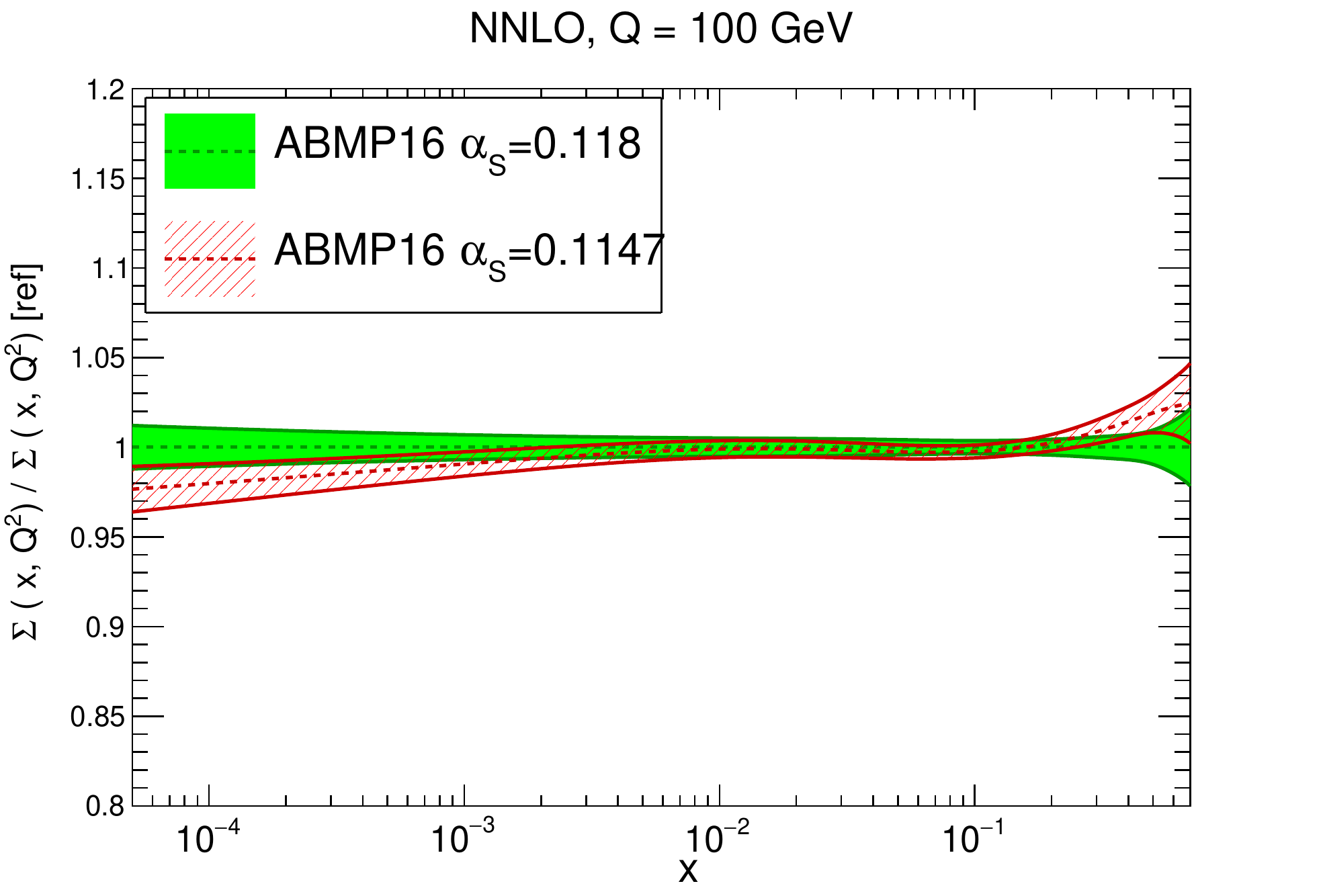}
   \caption{\small 
     The gluon (left) and quark singlet (right) PDFs in ABMP16
     at $Q=100$ GeV, comparing the results obtained
     with their best-fit $\alpha_s(m_Z)=0.1147$ with those
     with $\alpha_s(m_Z)=0.118$ used to compare with the other
     PDF sets.
    \label{fig:pdfcomp-xg-abmp16}
  }
\end{center}
\end{figure}

Another interesting feature of the comparison
in Fig.~\ref{fig:pdfcomp-xg} is the fact that the gluon PDF
uncertainties in ABMP16 are much smaller than
those of the other three groups.
As discussed in Sect.~\ref{sec:pdfgroups.ABM}, the underlying reason
for this is the use of a $\Delta\chi^2=1$ criterion to define
the Hessian PDF uncertainties, instead of using a tolerance
  $\Delta\chi^2 > 1$, as is done in the cases of CT14 and MMHT14. This point will be relevant
  for all further comparisons in this section.
We also note that the agreement between ABMP16 and the other three sets
becomes  worse
if the PDF set corresponding to their best-fit $\alpha_s(m_Z)=0.1147$
value is used.
To illustrate this, in Fig.~\ref{fig:pdfcomp-xg-abmp16} we show
the gluon PDF in ABMP16
     at $Q=100$ GeV, comparing the results obtained
     with their best-fit value $\alpha_s(m_Z)=0.1147$ with those
     with $\alpha_s(m_Z)=0.118$ used to compare with the other
     PDF sets.
When the former is used, the gluon is suppressed at large $x$,
resulting in an increased disagreement with the results
of the other three fits shown in Fig.~\ref{fig:pdfcomp-xg}
in the large-$x$ region.

%%%%%%%%%%%%%%%%%%%%%%%%%%%%%%%%%%%%%%%%%%%%%%%%%%%%%%%%%%%%%%%%%%%%%
%%%%%%%%%%%%%%%%%%%%%%%%%%%%%%%%%%%%%%%%%%%%%%%%%%%%%%%%%%%%%%%%%%%%%

\begin{figure}[t]
\begin{center}
  \includegraphics[scale=0.44]{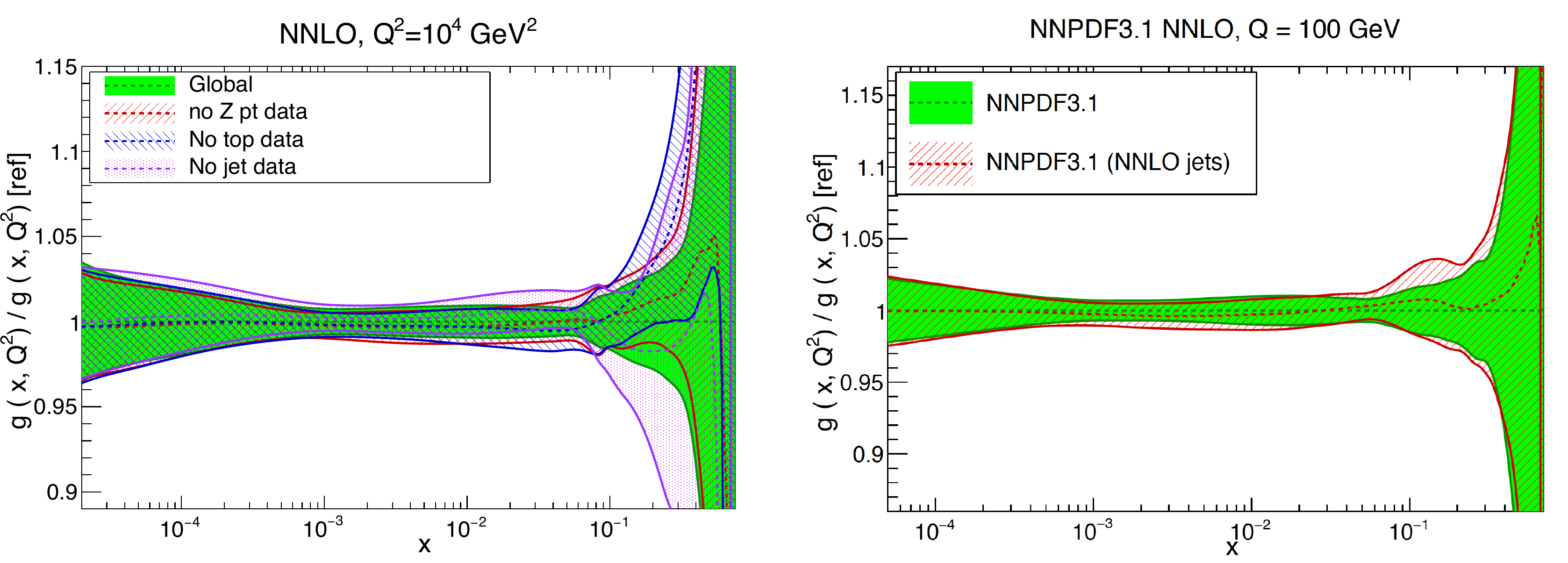}
   \caption{\small 
     Left: comparison of the NNPDF3.1 NNLO global fit at $Q=100$ GeV with
     the corresponding fits where the $Z$ $p_T$, top quark, or inclusive
     jet data have been removed.
     Right: same, now comparing with the NNPDF3.1 NNLO fit
     where the ATLAS and CMS 7 TeV inclusive jet data have been treated
     using exact NNLO theory, from~\cite{Ball:2017nwa}.
    \label{fig:largexgluon}
  }
\end{center}
\end{figure}

It is worth emphasising that until recently, the gluon at large-$x$ was only constrained in the
PDF fit by inclusive jet production data, and to a lesser extent by
DIS data via scaling violations.
However, there are now at least three datasets available with which constrain
the large-$x$ gluon, namely inclusive jets, the $p_T$ distribution
of $Z$ bosons, and top quark differential distributions. In all cases, NNLO calculations are now available.
To illustrate the robustness of the resulting gluon, in 
Fig.~\ref{fig:largexgluon} (Left) we show a comparison of the NNPDF3.1 NNLO global
fit at $Q=100$ GeV with
     the corresponding fits where the $Z$ $p_T$, top quark, or inclusive
     jet data have been removed.
     We observe that the four fits agree within
     PDF uncertainties,
    highlighting that these three families of 
     processes have statistically consistent pulls on the large-$x$ gluon.

%%%%%%%%%%%%%%%%%%%%%%%%%%%%%%%%%%%%%%%%%%%%%%%%%%%%%%%%%%%%%%%%%%%%%
%%%%%%%%%%%%%%%%%%%%%%%%%%%%%%%%%%%%%%%%%%%%%%%%%%%%%%%%%%%%%%%%%%%%%

Another consideration that is relevant for the determination of the large-$x$ gluon in
a PDF analysis
are the settings for the theoretical calculations used for the inclusive jet
cross sections.
Until 2016, only the NLO calculation was available, and different groups
treated jet data in different ways, either adding the NLO scale errors as additional systematic uncertainties
as in CT14 and NNPDF3.1, using the threshold approximation
to
the full NNLO result as in MMHT14, or excluding jet data altogether as advocated
by ABMP16 (see Sect.~\ref{sec:pdfgroups}).
The availability of the complete NNLO calculation, discussed
in Sect.~\ref{sec:datatheory.jets}, means that
future  PDF analyses will be able to fit to inclusive jet
data using the exact NNLO theory.

There are some indications that for specific settings of the NLO
calculation, the use of the exact NNLO theory may have moderate
phenomenological impact.
In particular, if the jet $p_T^{\rm jet}$ is adopted as the central renormalization
and factorization scale, and a larger value of $R$ is used, the
NNLO/NLO $K$-factor is $O(5\%)$ or less~\cite{Currie:2016bfm,Currie:2017ctp}.
To illustrate this, in Fig.~\ref{fig:largexgluon} we compare the baseline
NNPDF3.1 NNLO fit, where the jet production theory is treated at NLO using $p_T^{\rm jet}$
as central scale, with scale variations included
as an additional systematic error, to the same fit where the exact NNLO theory
has been used, for the ATLAS and CMS 7 TeV
data~\cite{Aad:2014vwa,Chatrchyan:2012bja}.
We see that the resulting differences are small at the PDF level, and at the
$\chi^2$ level one finds~\cite{Ball:2017nwa} a small but non-negligible improvement
once NNLO theory is used. On the other hand, for lower choices of jet radius and/or other scale choices, in particular the jet $p_T^{\rm max}$ per event, the NNLO/NLO K-factor can be larger and the impact of the exact NNLO theory may be more significant. A full comparison of these choices within the context of a NNLO global fit will therefore clearly be desirable in the future, although  a first study in this direction, presented in~\cite{Harland-Lang:2017ytb}, indicates that the impact of the scale choice on the extracted PDFs is not significant.

\subsection{Quark flavour separation}\label{sec:structure.flavoursep}
\begin{figure}[t]
\begin{center}
  \includegraphics[scale=0.39]{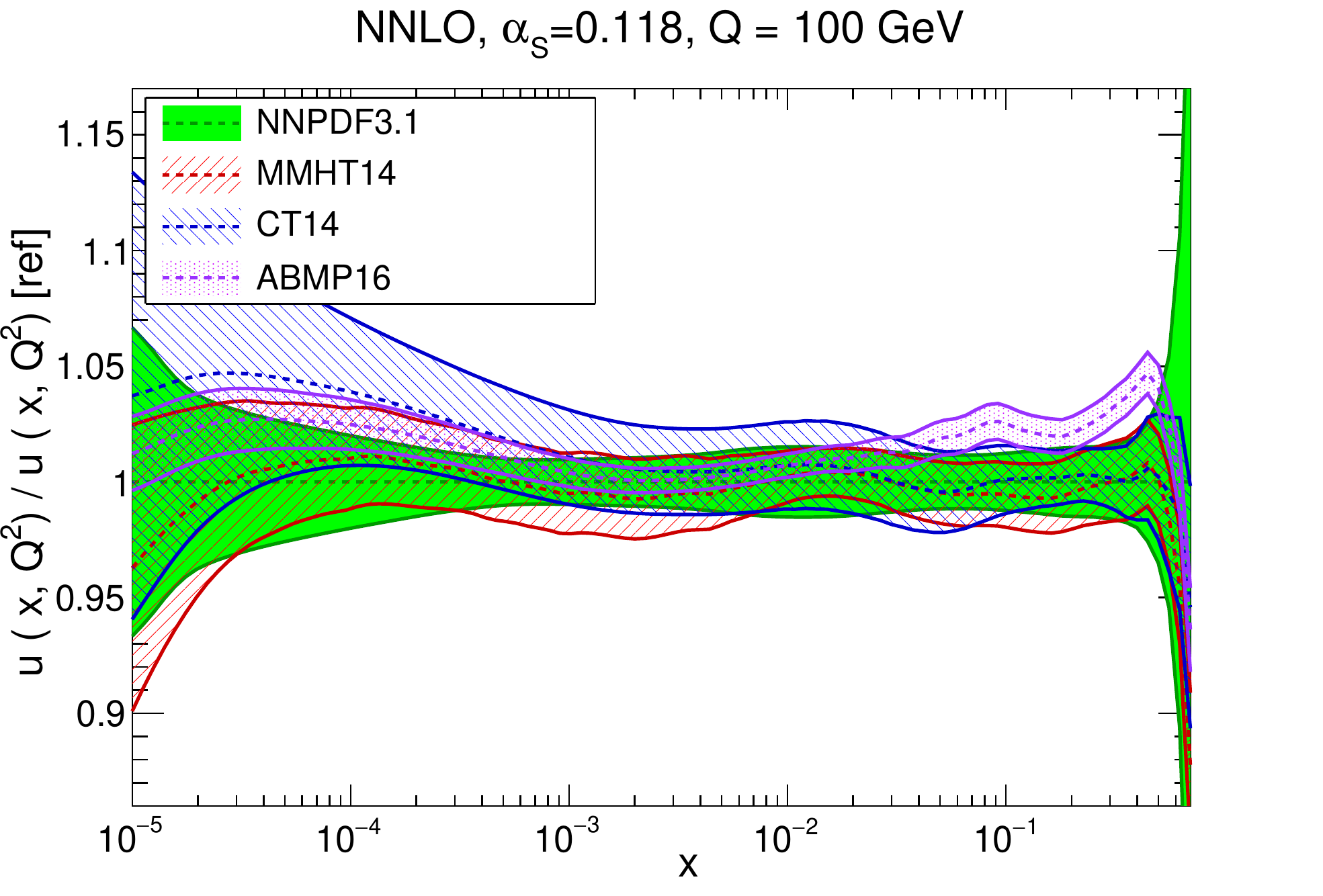}
  \includegraphics[scale=0.39]{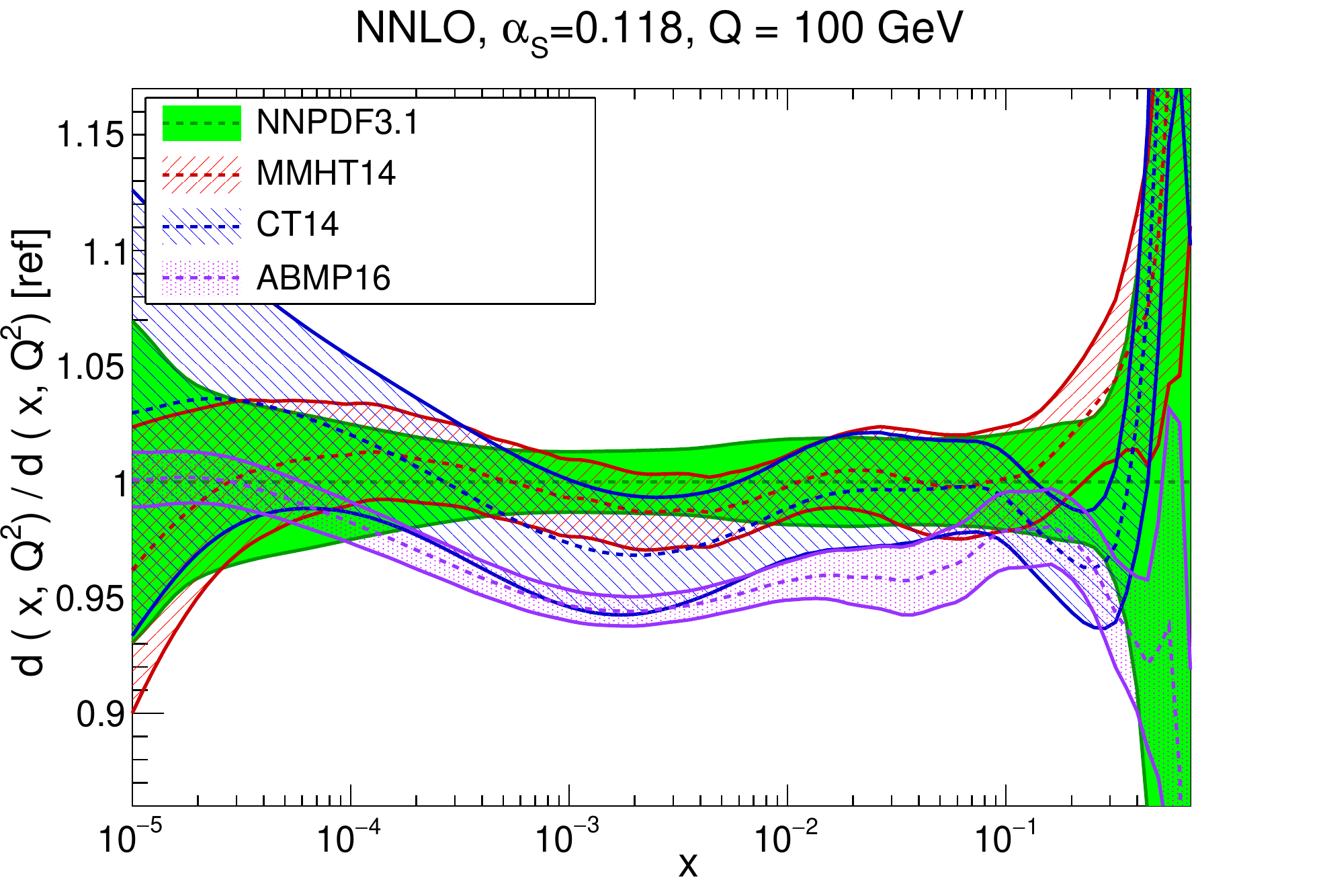}
  \includegraphics[scale=0.39]{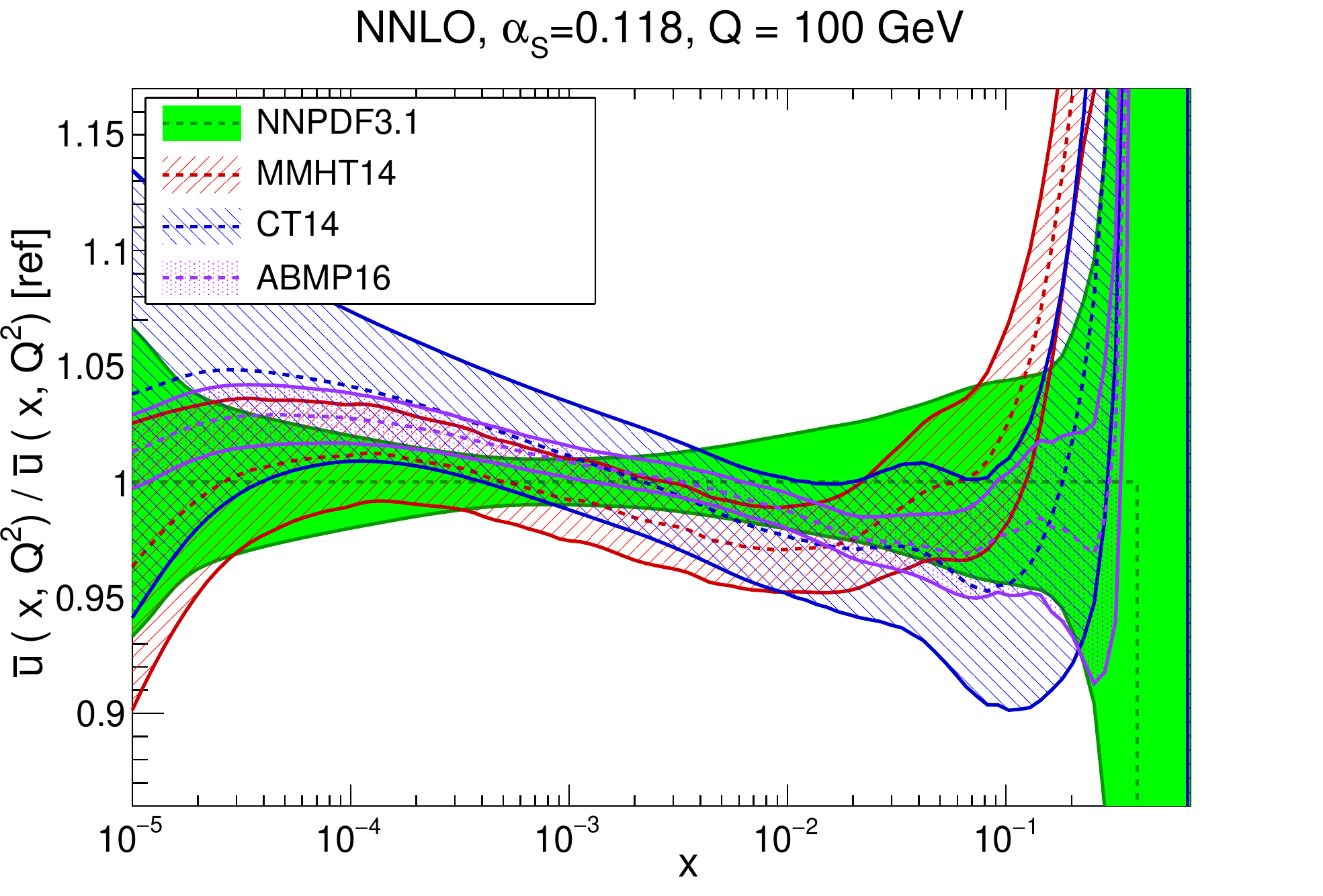}
  \includegraphics[scale=0.39]{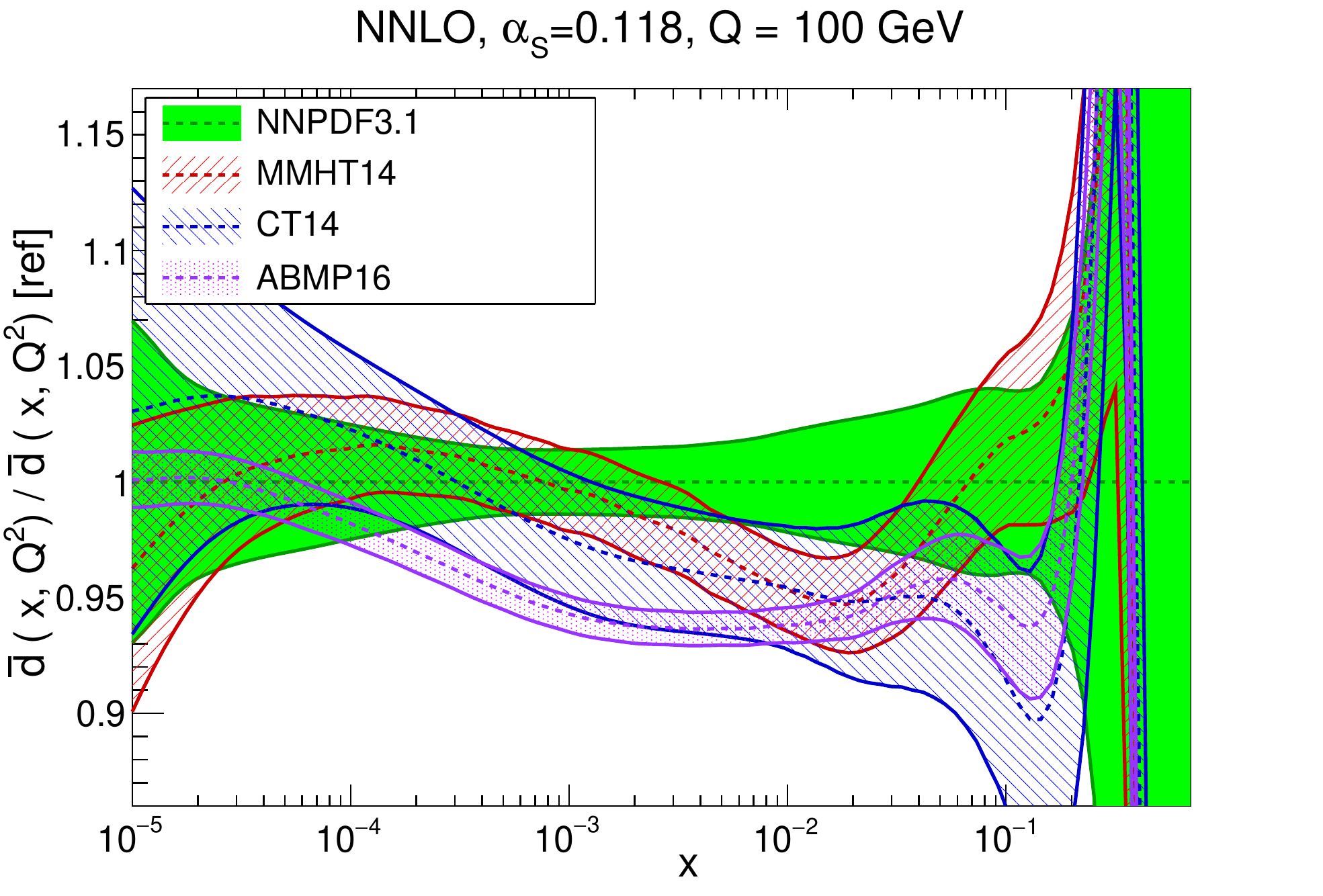}
   \caption{\small 
     Same as Fig.~\ref{fig:pdfcomp-xg} (right), now comparing the
     up, down, anti--up, and anti--down quark PDFs.
    \label{fig:pdfcomp-quarkseparation}
  }
\end{center}
\end{figure}

In this section we discuss the light quark content of the proton,
and in particular quantify the degree with which current fits
achieve quark flavour separation, i.e. the ability
to separate $u$ from $d$ quarks, and quarks from
anti--quarks.
First, in Fig.~\ref{fig:pdfcomp-quarkseparation} we show
the up, down, anti--up, and anti--down quark
PDFs at $Q=100$ GeV.
The up quark, $u(x,Q)$, is one of the better constrained
PDFs, in particular at large $x$, due to fixed--target DIS data.
For this PDF, we find good agreement within uncertainties in the entire range of $x$,
with the only exception being ABMP16, which overshoots the other
three sets in the large--$x$ region.
%%%%%%%%%%%%%%%%%%%%%%%%%%%%%%%%%%%%%%%%%%%%%%%%%%%%%%%%%%%%%%%%%%%%%
%%%%%%%%%%%%%%%%%%%%%%%%%%%%%%%%%%%%%%%%%%%%%%%%%%%%%%%%%%%%%%%%%%%%%

For the down quark, $d(x,Q)$, the spread between the central values
is larger, and the PDF uncertainties are also
comparatively increased.
Here we find good agreement between CT14, MMHT14, and NNPDF3.1 within uncertainties
for the entire range of $x$, while ABMP16 is around 5\% lower than the central
NNPDF3.1 value at intermediate values of $x$.
The PDF uncertainties are the largest at high $x$, with CT14, MMHT14 
and ABMP16 sets pointing in different directions, and
the NNPDF3.1 central value lying somewhere in the middle.
One possible source of difference between the groups
is the treatment of deuteron nuclear
corrections in the fitting of the
deuteron
structure functions~\cite{Harland-Lang:2014zoa},
though this effect is known to be localised in the region around $x\simeq 0.1$~\cite{Ball:2013gsa}.

For the light antiquark PDFs, $\bar{u}$ and $\bar{d}$, 
there is reasonable agreement between the various sets within uncertainties
for $\bar{u}$, while this agreement is only marginal for $\bar{d}$.
In the latter case, the ABMP16 result is again around $\simeq 5\%$ smaller
than the NNPDF3.1 central value.
As in the case of the quark PDFs, we see significant differences
at large $x$; in this region there are limited experimental constraints, and thus
the methodological differences in each PDF fit
can have a rather more marked impact.
Similarly to the gluon, these large PDF uncertainties 
at high $x$ have phenomenological consequences,
for instance for the production
of a heavy $W'$ or $Z'$ boson, or the production of
a squark--antisquark
pair $\tilde{q}\tilde{q}^*$, both processes being driven
by the quark--antiquark
luminosity, see Sect.~\ref{sec:LHCpheno.BSM}.

\begin{figure}[t]
\begin{center}
  \includegraphics[scale=0.39]{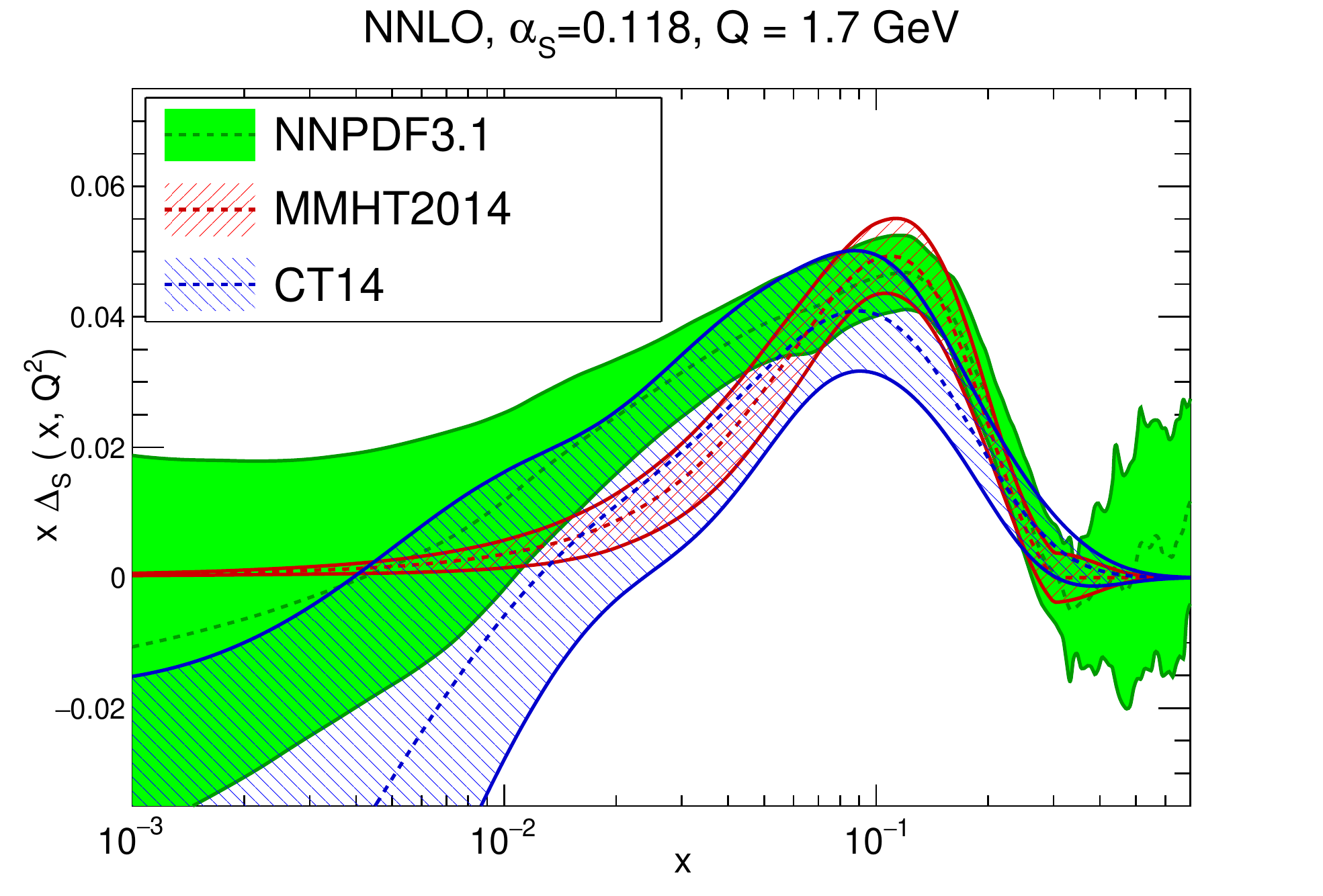}
  \includegraphics[scale=0.39]{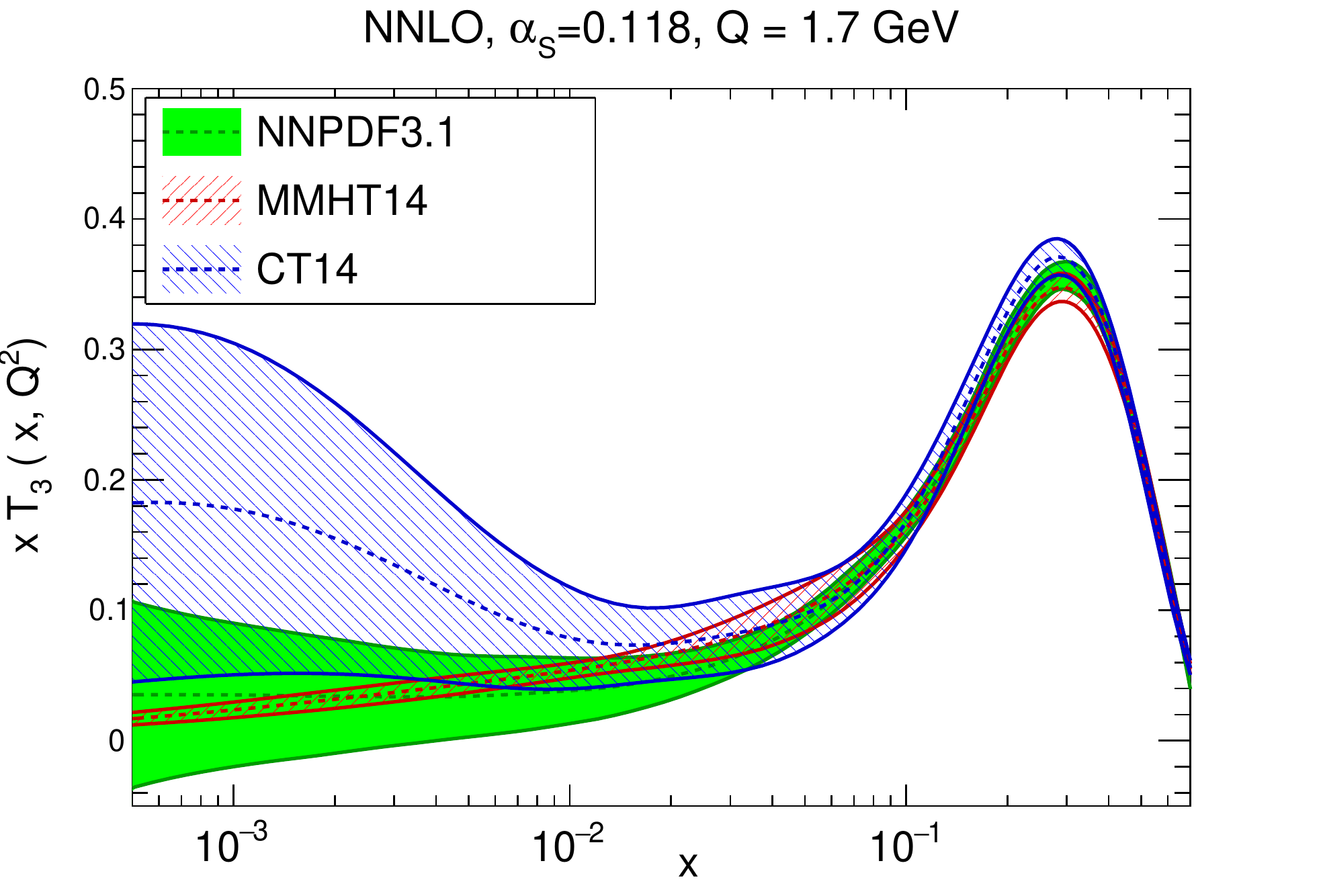}
    \caption{\small 
     Same as Fig.~\ref{fig:pdfcomp-xg} (left), now comparing the
     sea quark asymmetry $\Delta_S=\bar{d}-\bar{u}$ (left) and the quark
     isotriplet $T_3=u+\bar{u}-d-\bar{d}$ (right plot).
    \label{fig:pdfcomp-quarkseparation-2}
  }
\end{center}
\end{figure}

Another useful way to compare the quark flavour separation between the various
PDF groups is to plot flavour combinations that can be more
directly related to physical cross sections.
In Fig.~\ref{fig:pdfcomp-quarkseparation-2} we compare the
sea quark asymmetry $\Delta_S=\bar{d}-\bar{u}$ and
the quark
isotriplet $T_3=u+\bar{u}-d-\bar{d}$ at $Q=1.7$ GeV for
CT14, MMHT14, and NNPDF3.1.
The former quark
flavour combination is closely related to the $W$ asymmetries in
collider Drell--Yan production (see Sect.~\ref{sec:datatheory.gauge}),
while the latter is directly
sensitive to the difference between the proton and deuteron
DIS structure functions, the non--singlet structure
function
$F_2^{\rm NS}\equiv F_2^p-F_2^d$.
From this comparison, we see that for the sea quark
asymmetry $\Delta_S$
the general shape is similar between the three groups, although there are large
differences in the estimate of the PDF uncertainties, both at small
and large $x$, which in some cases can be traced
back to the PDF parametrization assumptions.
The agreement is also
reasonably good both in terms of central values
and of uncertainties for the quark
non--singlet combination $T_3$, although here again the
small--$x$ behaviour does differ among the three groups, due to the different parametrization choices.

%%%%%%%%%%%%%%%%%%%%%%%%%%%%%%%%%%%%%%%%%%%%%%%%%%%%%%%%%%%%%%%%%%%%%
%%%%%%%%%%%%%%%%%%%%%%%%%%%%%%%%%%%%%%%%%%%%%%%%%%%%%%%%%%%%%%%%%%%%%

\begin{figure}[t]
\begin{center}
  \includegraphics[scale=0.39]{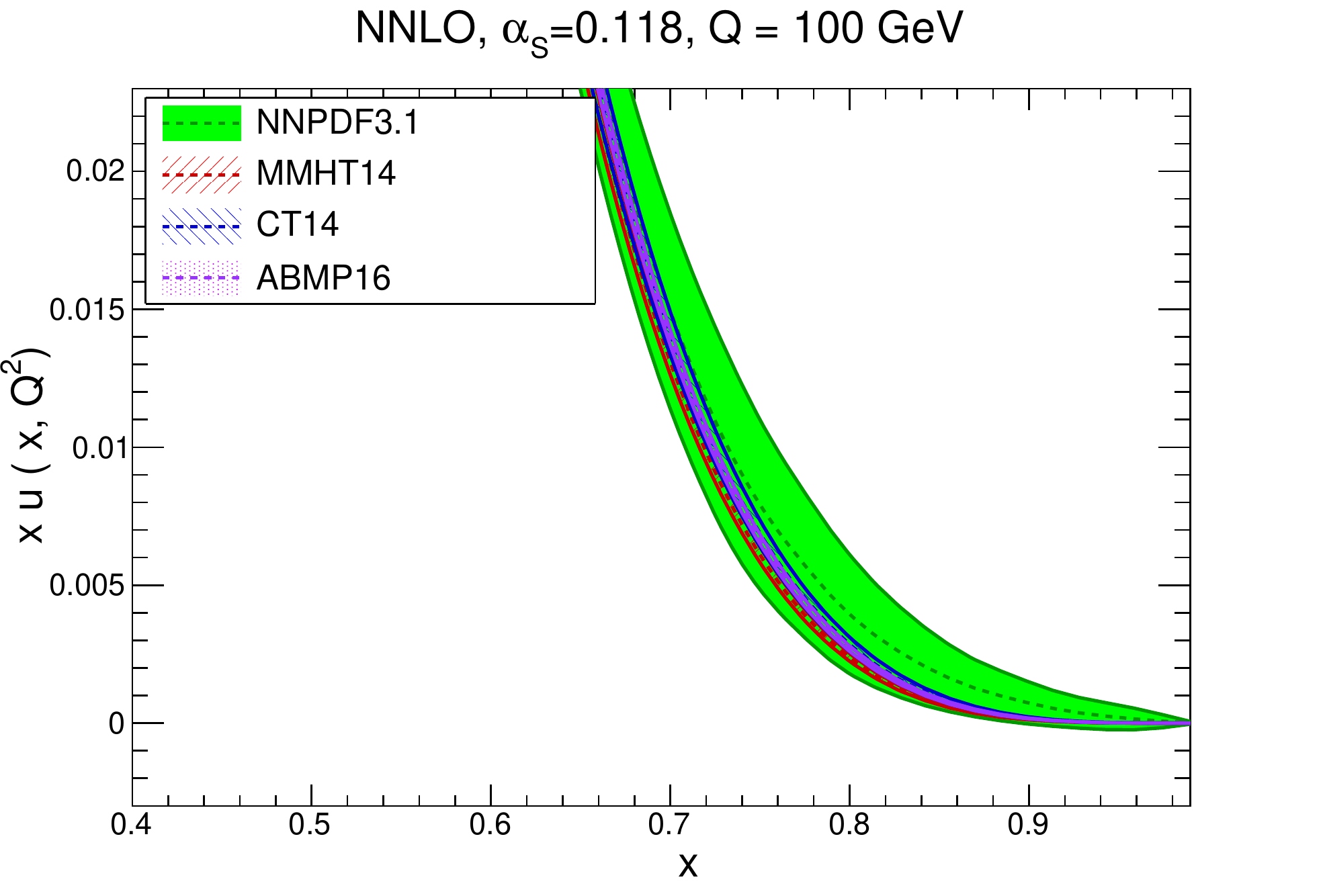}
  \includegraphics[scale=0.39]{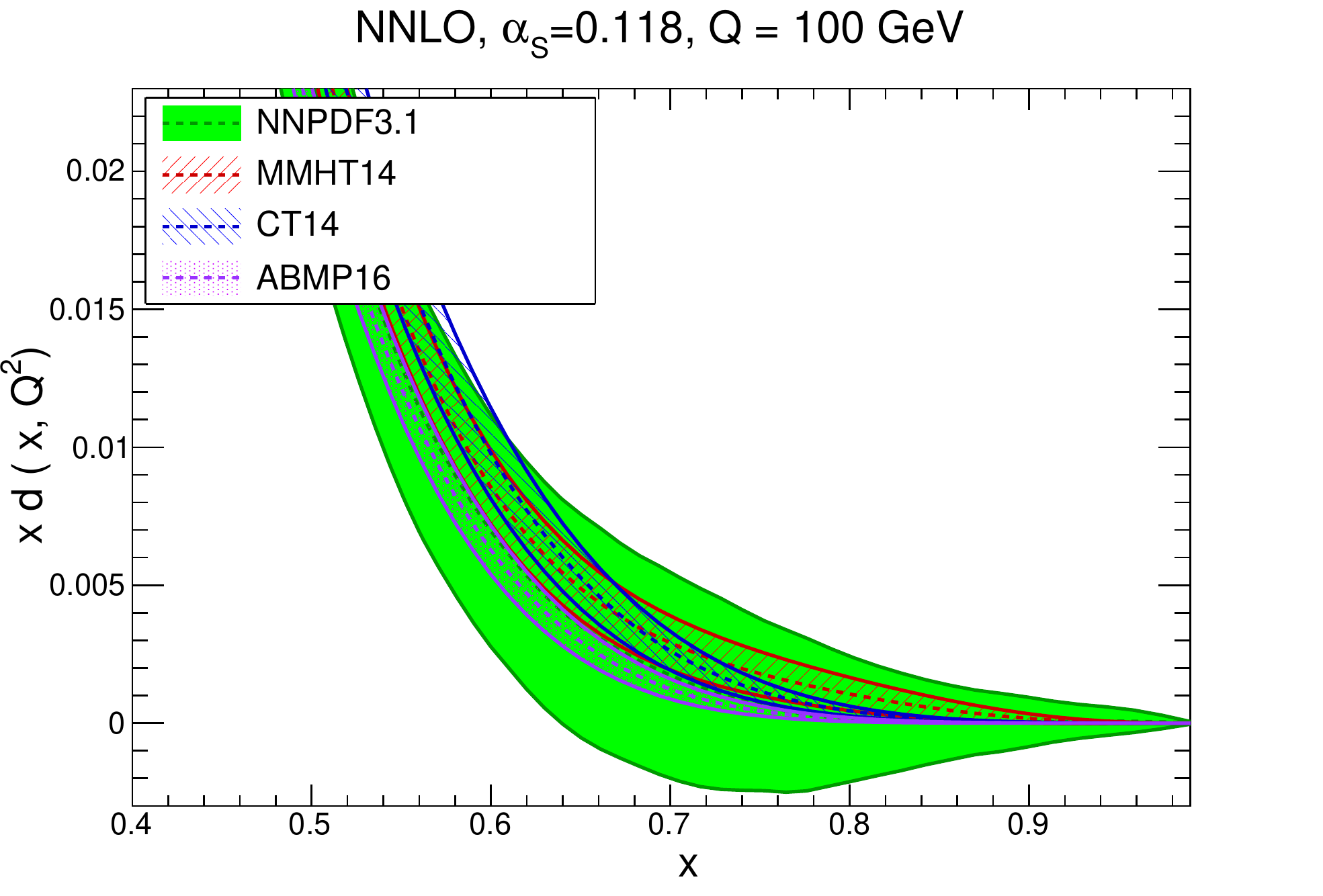}
    \caption{\small 
      Same as Fig.~\ref{fig:pdfcomp-quarkseparation}, now focusing
      on the large--$x$ region of the up quark (left) and down quark
      (right plot) PDFs.
    \label{fig:pdfcomp-quarkseparation-largex}
  }
\end{center}
\end{figure}

From these comparisons, we can see that the differences in the quark
flavour separation between the various groups are mostly localised in the
large--$x$ region.
With this in mind, in Fig.~\ref{fig:pdfcomp-quarkseparation-largex} we again show
the up and down quark PDFs,
focusing now on the large--$x$
region and adopting a linear scale in the $x$ axis.
From this comparison we can see that PDF uncertainties are the largest
in NNPDF3.1.
In terms of central values, there is reasonable agreement for the
up quark, less
so for the down quark.
Note that as discussed in Sect.~\ref{sec:fitmeth.PDFpara},
in the NNPDF framework,
the PDFs themselves are not forced to be positive
(although the physical cross sections are indeed positive--definite)
and therefore the down PDF can become negative at large $x$, although
its central value always remains positive.
An alternative approach that can be used to compare the behaviour
of PDF sets at large $x$, and in doing so compare with
non--perturbative models such as the quark counting rules,
is the effective exponent method discussed in Ref.~\cite{Ball:2016spl}.

%%%%%%%%%%%%%%%%%%%%%%%%%%%%%%%%%%%%%%%%%%%%%%%%%%%%%%%%%%%%%%%%%%%%%
%%%%%%%%%%%%%%%%%%%%%%%%%%%%%%%%%%%%%%%%%%%%%%%%%%%%%%%%%%%%%%%%%%%%%

\subsection{Strangeness}\label{sec:structure.strange}
The size and shape of the strange PDF has recently attracted a lot
of debate, see~\cite{Alekhin:2017olj,Alekhin:2014sya} for recent summaries.
Most PDF analyses find a suppressed strangeness relative to
the non--strange light quark sea, a pull driven mostly
by the deep--inelastic neutrino inclusive and  charm production
(`dimuon') data (see Sect.~\ref{sec:datatheory.DIS}).
However, high--precision collider data from the LHC
 instead exhibit the opposite trend, with a recent
QCD analysis from ATLAS based on the $W,Z$ 7 TeV rapidity distributions
from the 2011 dataset finding 
a strange sea that is in fact larger than the non--strange sea.
Given the importance of strangeness for many phenomenological
applications, such as the measurement of the $W$ mass,
it will be important to resolve this issue in the future.
Alternative probes of $s(x,Q)$ are therefore particularly useful,
such as the recent suggestion of using semi--inclusive DIS measurements with kaons
in the final state~\cite{Borsa:2017vwy}.

\begin{figure}[t]
\begin{center}
\includegraphics[scale=0.40]{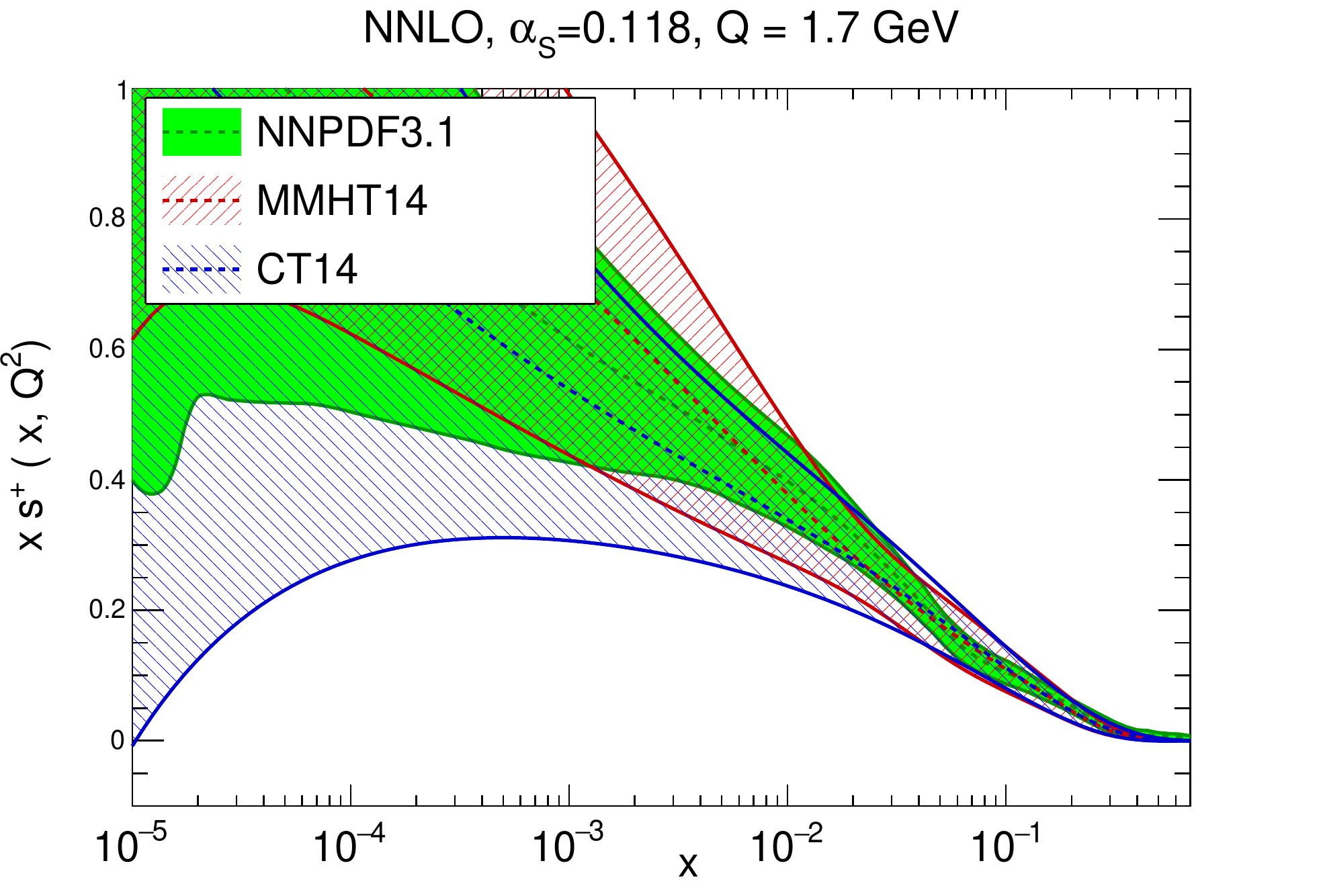}
\includegraphics[scale=0.40]{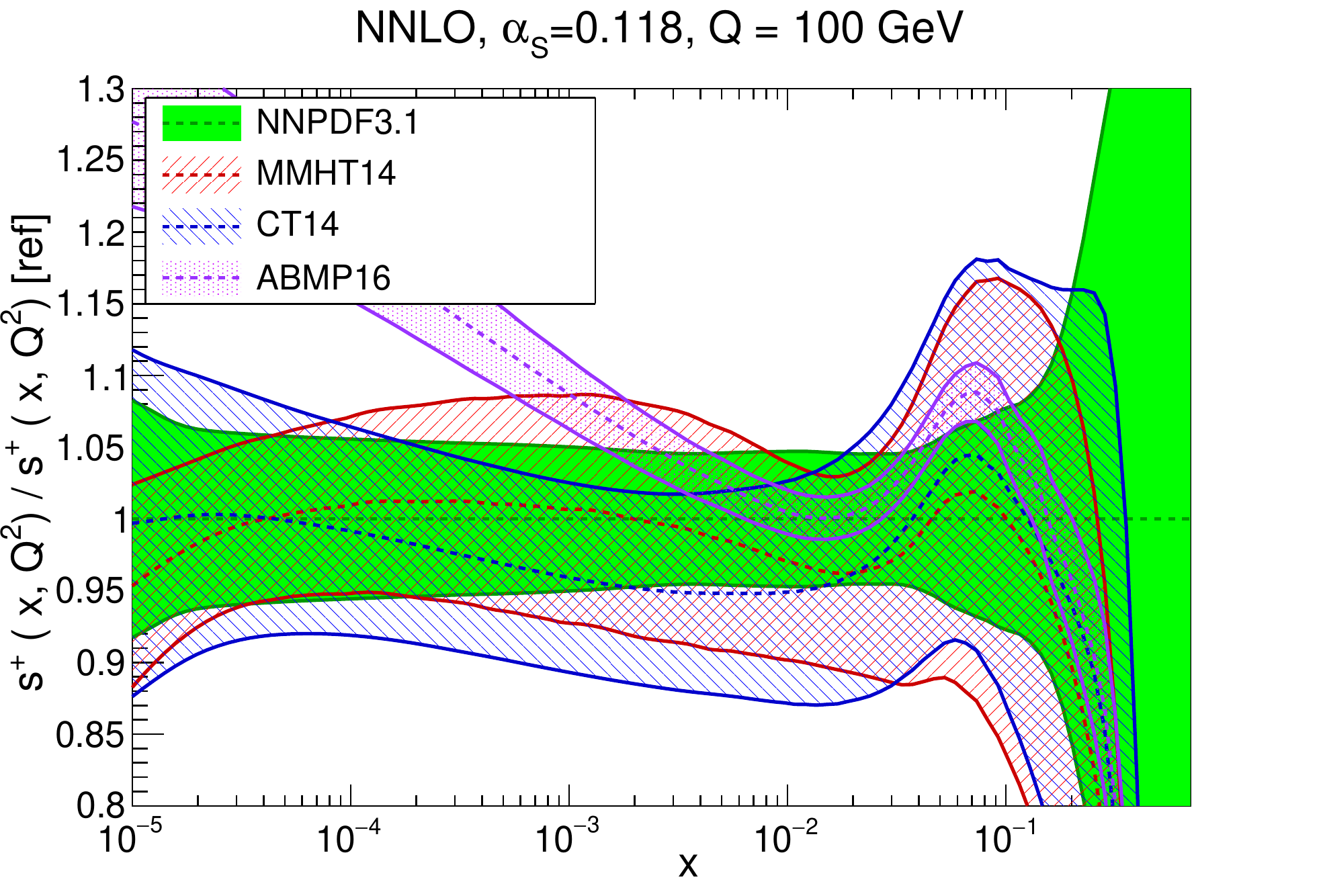}
     \caption{\small 
      Same as Fig.~\ref{fig:pdfcomp-xg} for the total strangeness
      $xs^+(x,Q^2)$.
      \label{fig:pdfcomp-xsp}
  }
\end{center}
\end{figure}

In Fig.~\ref{fig:pdfcomp-xsp} we show the
total strange PDF, $xs^+(x,Q^2)$, at $Q=100$ GeV, 
     in the same format as that of Fig.~\ref{fig:pdfcomp-xg}.
     The strangeness--sensitive datasets included in
     the four analysis are rather different,
     both
     in terms of the neutrino fixed--target data and 
      the LHC collider data.
     For example, only the ABMP16 fit includes the NOMAD dimuon
     data~\cite{Samoylov:2013xoa}, while only
     NNPDF3.1 includes the ATLAS $W,Z$ 2011 rapidity distributions.
     We can see that there is reasonable
     agreement within uncertainties between the four groups
     except for ABMP16 for $x\lsim 10^{-3}$, which has a much harder
     strangeness than the other fits.
     We also note that the differences in the size of the
     strange PDF uncertainty can vary by up to a factor $\sim 5$,
     with ABMP16 exhibiting the smallest uncertainties. 

%%%%%%%%%%%%%%%%%%%%%%%%%%%%%%%%%%%%%%%%%%%%%%%%%%%%%%%%%%%%%%%%%%%%%
%%%%%%%%%%%%%%%%%%%%%%%%%%%%%%%%%%%%%%%%%%%%%%%%%%%%%%%%%%%%%%%%%%%%%

A more physically transparent method to quantify the strange
content of the proton is given by the ratio of the strange to the
non--strange sea quark PDFs, defined as
\be
\label{eq:rs}
R_s (x,Q^2) \equiv
\frac{ s(x,Q^2)+\bar{s}(x,Q^2) }{ \bar{u}(x,Q^2)+\bar{d}(x,Q^2) } \, .
\ee
In this ratio, a symmetric strange sea would correspond to $R_s\simeq 1$,
while a suppressed strangeness instead leads to $R_s\ll 1$.
Note that this strangeness ratio $R_s (x,Q^2)$ depends on the
specific scale $Q^2$ and thus when providing predictions
(or measurements) for $R_s$ it is necessary to specify the value
of $Q^2$ which is being assumed, since results can be quite
different depending on the value of $Q^2$.
In addition, from general considerations one expects that
$R_s\to 1$ for small-$x$ and large $Q^2$, since in this region
all sea quarks become very similar to each other due
to perturbative evolution, which becomes dominant in this region.

\begin{figure}[t]
\begin{center}
\includegraphics[scale=0.45]{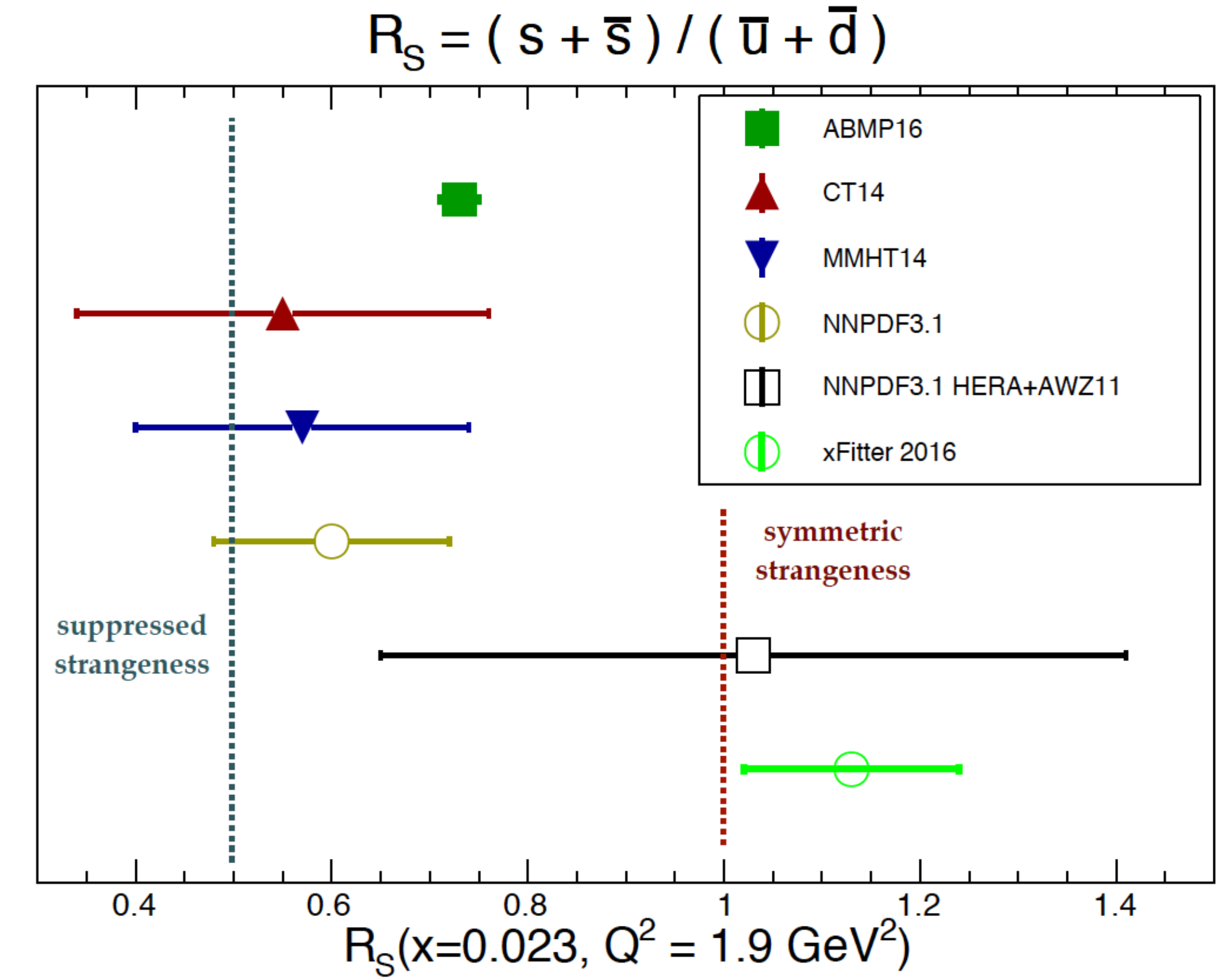}
     \caption{\small 
     The ratio of strange to non--strange sea quarks
     $R_s (x,Q^2)$, Eq.~(\ref{eq:rs}) for $x=0.023$
     and $Q^2=1.9$ GeV$^2$.
     We compare the results of various global PDF fits with those
     of the ATLAS/{\tt xFitter} interpretation study as well as with
     those of a NNPDF3.1 fit based on the same dataset as the ATLAS
     study.
     The vertical lines indicate two possible scenarios for the strange
     PDFs, namely a suppression of size $R_{s}\simeq 0.5$ and 
     a strange sea which is symmetric with the non--strange light quark sea,
     $R_{S}\simeq 1$.
      \label{fig:Rsx}
  }
\end{center}
\end{figure}

Traditionally, the constraints from neutrino--induced dimuon production
have preferred a value $R_s\sim 0.5$ in most global fits.
In Fig.~\ref{fig:Rsx} we show the ratio of strange to non--strange sea quarks
     $R_s (x,Q^2)$, Eq.~(\ref{eq:rs}),
     for $Q^2=1.9$ GeV$^2$ and $x=0.023$.
     We compare the results of the various global PDF fits with those
     of the ATLAS/{\tt xFitter} study~\cite{Aaboud:2016btc}, which includes the recent ATLAS $W,Z$ high precision data, as well as with
     those of a NNPDF3.1 fit based on the same dataset as the ATLAS
     study.
     The vertical lines indicate two possible scenarios for the strange
     PDFs, namely a suppression of size $R_{s}\simeq 0.5$ and 
     a strange sea which is symmetric with the non--strange one,
     $R_{S}\simeq 1$. We observe that the four PDF sets considered here exhibit a preference for
     a suppressed strangeness.
     On the other hand, this comparison also shows that
     if only the HERA and ATLASWZ11 data are considered, the
     NNPDF3.1 analysis yields an unsuppressed strangeness,
     although with PDF uncertainties rather larger than
     those from the {\tt xFitter} analysis.
     This comparison demonstrates that the opposite pull between the low--energy
     neutrino data and the high--energy collider data is genuine,
     although the tension is only at the 1-- or 2--$\sigma$ level, as indicated
     by the fact that the NNPDF3.1 global and HERA+ATLASWZ11 results
     agree within PDF uncertainties.

%%%%%%%%%%%%%%%%%%%%%%%%%%%%%%%%%%%%%%%%%%%%%%%%%%%%%%%%%%%%%%%%%%%%%
%%%%%%%%%%%%%%%%%%%%%%%%%%%%%%%%%%%%%%%%%%%%%%%%%%%%%%%%%%%%%%%%%%%%%

\begin{figure}[t]
\begin{center}
\includegraphics[scale=0.40]{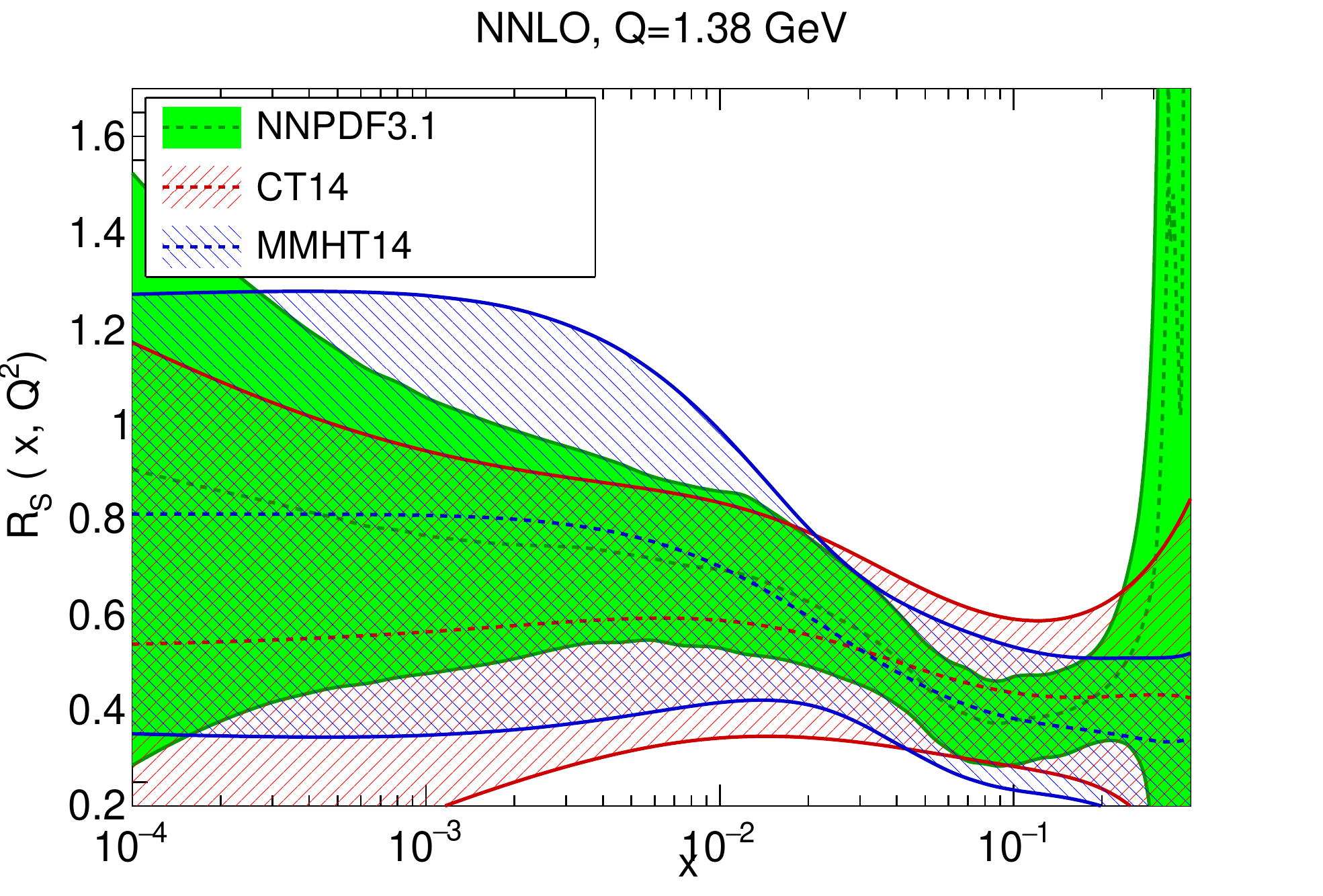}
\includegraphics[scale=0.40]{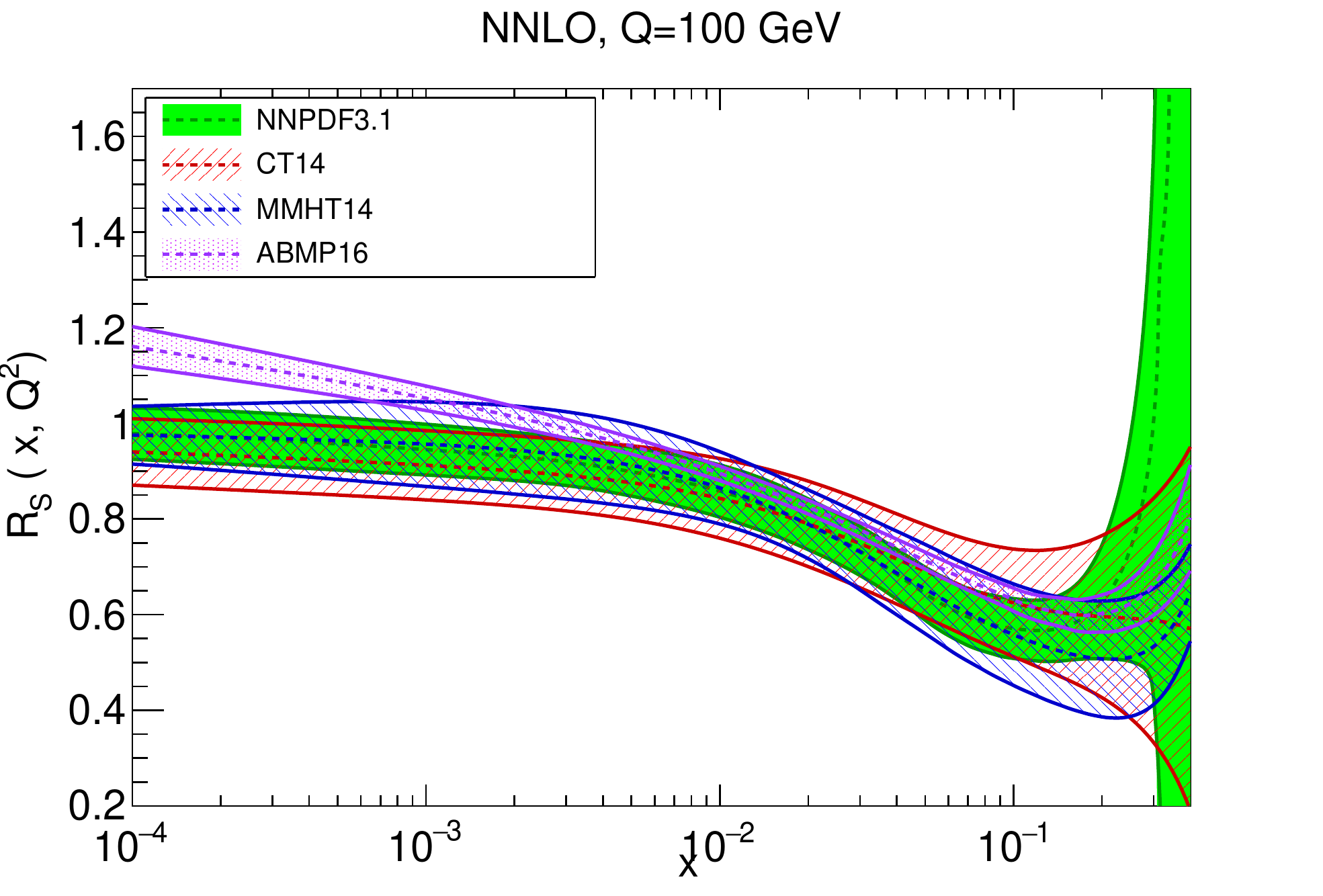}
     \caption{\small 
     The ratio of strange to non--strange sea quarks
     $R_s (x,Q^2)$, Eq.~(\ref{eq:rs}), as a function of $x$
     for $Q=1.38$ GeV (left plot) and for $Q=100$ GeV (right plot).
               \label{fig:Rsxplot}
  }
\end{center}
\end{figure}

One restriction of the comparison summarised
in Fig.~\ref{fig:Rsx}
is that it  corresponds to a single  point $x\simeq 0.023$.
To bypass this limitation, in
Fig.~\ref{fig:Rsxplot} we therefore show the $R_s (x,Q^2)$ ratio
as a function of $x$ both at low and at high scales.
There are a number of interesting features that can be observed from
this comparison.
First, we can see that DGLAP evolution automatically increases
the value of $R_s$ as we go to higher scales, in particular at medium and
small $x$; the reason being that, as explained above,
in this region all sea quarks become similar to each other.
Second, we find  a consistent strangeness content for the four groups in most of the range
of $x$, although the size of the corresponding uncertainties in each case
can vary by a significant amount.
Another important
point from this comparison is that
clearly any statement about whether or not strangeness is
suppressed depends on the region of $x$ that is being considered.
For instance, in the MMHT14 case for $Q=1.9$ ${\rm GeV}^2$ the value
of $R_s$ changes from around 0.4 at $x\simeq 0.1$ to around 0.8
for $x\simeq 0.007$.
So different $x$ regions exhibit different amounts of suppression
with respect to the light sea quarks, and therefore the question
of the suppression (or lack thereof) of the strange PDF
is a more nuanced issue than is sometimes stated.
In any case, it is clear from the comparison of
Fig.~\ref{fig:Rsxplot} that a symmetric strange sea in the entire
range of $x$ is not favoured by any of the four fits shown here,
in particular in the region around $x\simeq 0.1$ and above.
We can expect future data from the LHC and elsewhere to help
shed some light on this issue.

%%%%%%%%%%%%%%%%%%%%%%%%%%%%%%%%%%%%%%%%%%%%%%%%%%%%%%%%%%%%%%%%%%%%%
%%%%%%%%%%%%%%%%%%%%%%%%%%%%%%%%%%%%%%%%%%%%%%%%%%%%%%%%%%%%%%%%%%%%%

\subsection{The charm content of the proton}\label{sec:structure.charm}
The charm content of the proton (see~\cite{Brodsky:2015fna} for a recent review)
is a topic that has recently received quite a lot of attention,
in particular with the development of more sophisticated
theoretical and methodological treatments of the charm quark PDF,
relaxing the assumption that charm is entirely
generated by perturbative DGLAP evolution~\cite{Hou:2017khm,Ball:2016neh}.
In this respect, the NNPDF collaboration advocates that the charm
PDF should be treated on an equal
footing to the light quarks in global fits~\cite{Ball:2016neh}, and this
leads to a reduced $m_c$ dependence of important collider cross-sections
as well as improving the agreement with high--precision data.
On the other hand, the CT group~\cite{Hou:2017khm} argues instead that a
non-zero charm PDF at the initial parametrization scale $Q_0$,
under certain assumptions, may represent twist-4 contributions
that are not suppressed by the usual
$Q^2$ and $W^2$ kinematic cuts, but could also arise due to missing radiative corrections
which are however not universal but rather process-dependent.
From the phenomenological point of view, these
and related approaches, reviewed in
this section, allow for improved comparisons
with non--perturbative models of the charm
content of the proton~\cite{Brodsky:2015fna}.

As discussed in Sect.~\ref{sec:fitmeth.PDFpara}, there are two different
approaches to treating the charm PDF within a PDF analysis.
One can assume that the charm PDF is generated
entirely
from perturbative evolution, and thus compute the charm
PDF from the gluon and light quark PDFs starting from
the charm threshold $\mu_c\simeq m_c$ by means of
the DGLAP evolution equations.
Alternatively, it is possible to release this assumption
and treat the charm PDF on an equal footing to the light quark
PDFs, namely introducing a functional form for $c(x,Q_0)$
with parameters to be determined by
experimental data.

Until recently, in most global fits the charm PDF was generated
using perturbative evolution, and then separately,
in dedicated intrinsic charm studies, variants of these global fits
were performed with specific 
models for the charm PDF.
In these studies, the parameters
of the model charm PDF, typically
its overall normalization, were constrained by experimental data, see
for instance Refs.~\cite{Pumplin:2007wg,
  CT14IC,Hou:2017khm,Jimenez-Delgado:2014zga}.
An alternative approach is taken
by the NNPDF3.1 global analysis, which fits the charm PDF
 using  the same parametrization as for
the light quarks.
When the charm PDF is fitted, the dominant constraints
arise from processes sensitive to initial--state
charm, such as the charm structure
functions of the EMC experiment~\cite{Aubert:1982tt},
other fixed--target DIS datasets, and collider electroweak
gauge boson production.

For the first approach, the
 CT14IC analysis~\cite{Hou:2017khm} provides a recent and representative example.
Here, the charm PDF
is parametrized according to two theoretical
scenarios.
The first scenario assumes either the exact
or the approximate BHPS model~\cite{Brodsky:1980pb},
which predicts a valence--like charm PDF at the input scale.
In the case of the approximate solution of the model,
we have
\be
c(x,Q_0)=\frac{1}{2}Ax^2 \lc \frac{1}{3}(1-x)(1+10x+x^2)
-2x(1+x)\ln(1/x)\rc \, ,
\ee
while  a more complicated, non--analytic expression is used
for the exact BHPS solution.
The other scenario explored in the CT14IC study
is the SEA model, where the charm
PDF is parametrized by a `sea--like' function taken
to be proportional to the light quark sea PDFs, namely
\be
c(x,Q_0)=A\lp \bar{d}(x,Q_0)+\bar{u}(x,Q_0)\rp \, .
\ee
In these three scenarios, the overall normalization $A$ of
the fitted charm PDF is a free parameter to be determined
from the experimental data.

\begin{figure}[t]
\begin{center}
\includegraphics[scale=0.44]{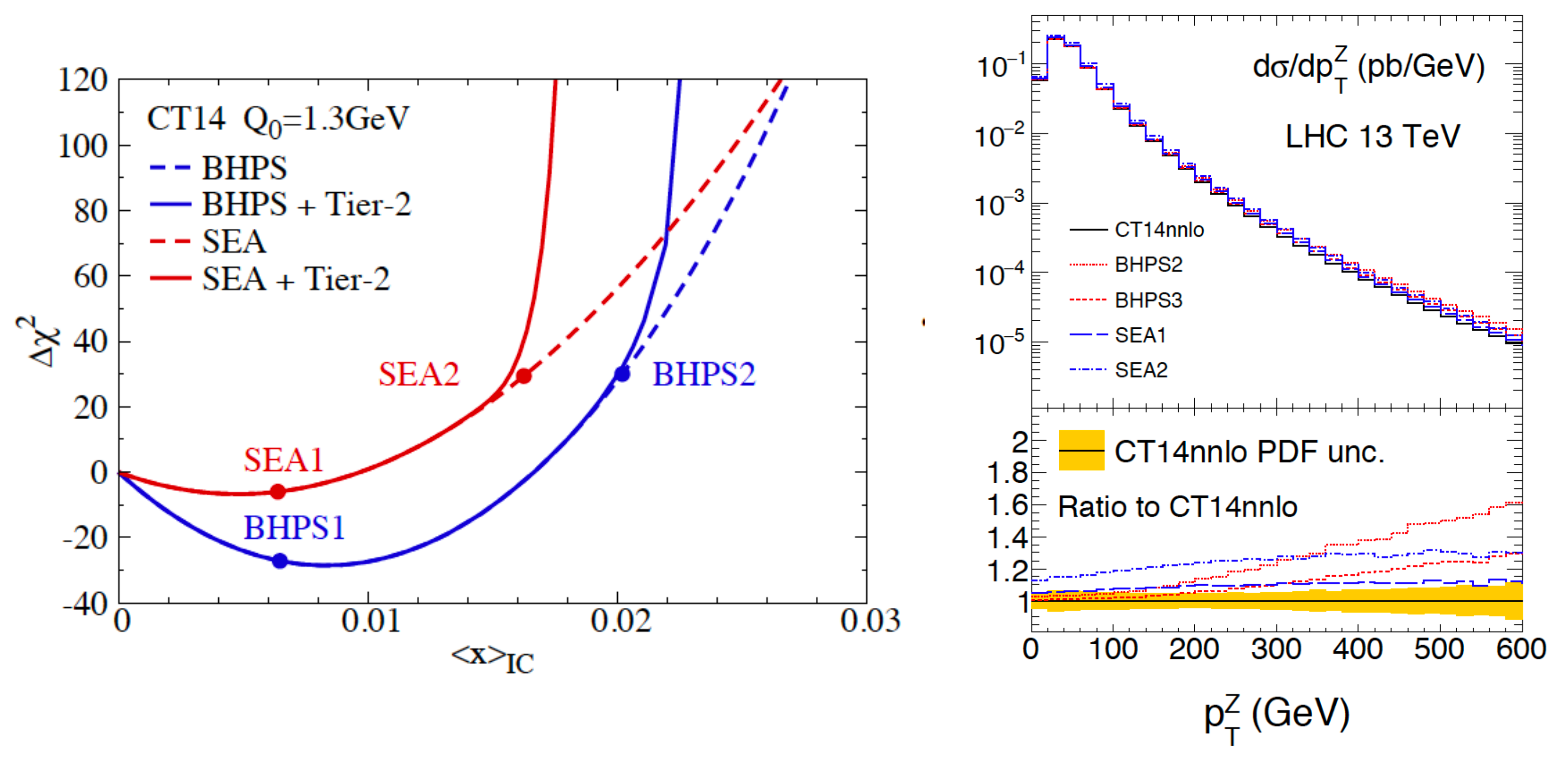}
     \caption{\small 
     Left: the deviation of the $\chi^2$ in the CT14IC fits,
with respect to the
     best--fit value of the CT14 fit with perturbative charm,
     as a function of $\la x\ra_{\rm IC}$
     Results are shown for the BHPS and SEA models, with the
    ``1'' points labeling the preferred value of  $\la x\ra_{\rm IC}$,
    with those labelled with ``2'' indicate the largest
    values of the charm momentum fraction allowed by the
    fit tolerance criteria.
    Right: the transverse momentum distribution of $Z$ bosons
    in the $pp \to Z+c$ process at 13 TeV, comparing the CT14 NNLO
    result with various of the CT14 IC models, as a function
    of $p_T^Z$.
      \label{fig:ct14ic}
  }
\end{center}
\end{figure}

To illustrate the main results of this study,
In Fig.~\ref{fig:ct14ic} we plot
the deviation of the $\chi^2$ in the CT14IC fits,
computed with respect to the
     best--fit value of the CT14 fit with perturbative charm,
     as a function of the charm
     momentum fraction $\la x\ra_{\rm IC}=C(Q_0)$, where
     we have defined
     \be
     \label{eq:charmmomfrac}
C(Q^2) \equiv \int_0^1dx~x~\lp c(c,Q^2)+\bar{c}(x,Q^2)\rp \, .
\ee
     Results are shown for the BHPS and SEA models, with the
    `1' points labelling the preferred value of  $\la x\ra_{\rm IC}$,
    and those labelled with `2' indicating the largest
    values of the charm momentum fraction allowed by the
    fit tolerance criteria.
    We observe that the BHPS model is preferred, leading
    to a best--fit value of $\la x\ra_{\rm IC}\simeq 0.6\% $.

%%%%%%%%%%%%%%%%%%%%%%%%%%%%%%%%%%%%%%%%%%%%%%%%%%%%%%%%%%%%%%%%%%%%%
%%%%%%%%%%%%%%%%%%%%%%%%%%%%%%%%%%%%%%%%%%%%%%%%%%%%%%%%%%%%%%%%%%%%%

As mentioned above, a different approach to fitted charm is adopted by the
NNPDF collaboration~\cite{Ball:2016neh,Ball:2017nwa}.
In this case, the charm PDF is treated on an equal footing
to the  light quark PDFs, and therefore it is parametrized
with a 37--parameter artificial neural network with  2-5-3-1
architecture,
\be
c^{+}(x,Q_0) = c(x,Q_0)+\bar{c}(x,Q_0)=
x^{a_{c^+}}(1-x)^{b_{c^+}}\,{\rm NN}_{c^+}(x) \, ,
\ee
where $a_{c^+},b_{c^+}$ are the corresponding preprocessing
exponents, whose range is determined by an iterative
procedure.
The charm asymmetry, on the other hand, is assumed to vanish
$c^{-}(x,Q_0) = 0$.
The charm PDF is then determined at the input evolution
scale $Q_0=1.64$ GeV from the experimental data,
with the recent LHC high--precision electroweak
gauge boson production measurements providing the
best constraints~\cite{Ball:2017nwa}.
 DIS structure functions are treated
      with the FONLL general--mass VFN scheme, modified
      to account for initial--state massive
      contributions~\cite{Ball:2015tna,Ball:2015dpa}.

\begin{figure}[t]
\begin{center}
\includegraphics[scale=0.45]{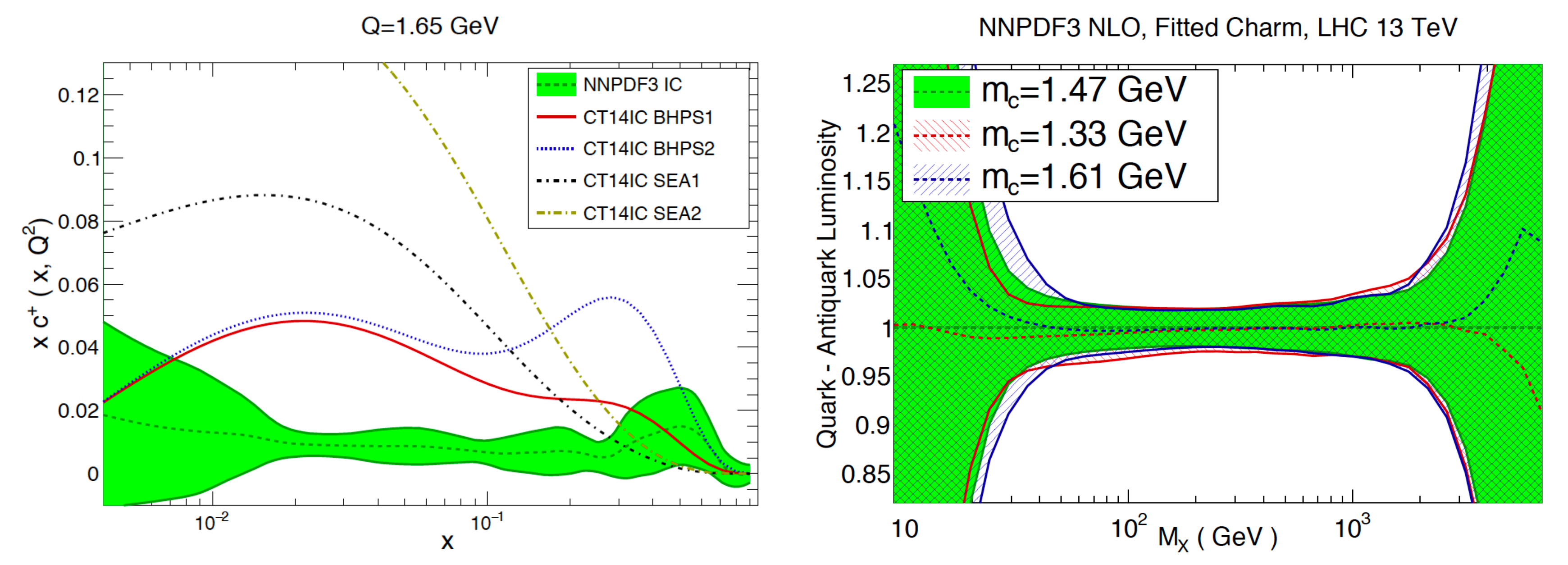}
     \caption{\small 
      Left: comparison of the fitted charm PDF at $Q=1.65$ GeV
      between the NNPDF3IC set and the different models
      of the CT14IC analysis.
      Right: the dependence of the quark--anti--quark
      PDF luminosity at the LHC 13 TeV in the NNPDF3IC fits
      with the value of the charm mass $m_c$ used in the fit.
      \label{fig:nnpdf31ic}
  }
\end{center}
\end{figure}

One of the consequences of this model--independent approach is
that it improves the stability of the
fitted PDFs with respect to the value of the charm mass $m_c$,
since its variations can be re--absorbed into the fitted charm
boundary condition.
To illustrate this point, in Fig.~\ref{fig:nnpdf31ic}
we show the dependence of the quark--anti--quark
      PDF luminosity at the LHC 13 TeV in the NNPDF3IC fits
      with the value of the charm mass $m_c$ used in the fit.
      We find that even for a relatively large variation
      of $\delta m_c =\pm 0.14$ GeV, the $q\bar{q}$ luminosity
      is very stable in most of the $M_X$ range.
      A similar stability is found at the level of relevant
      LHC cross-sections such as the
      differential distributions
      for Higgs and $Z$ production, as discussed in~\cite{Ball:2016neh}. 
       
\begin{figure}[t]
\begin{center}
\includegraphics[scale=0.40]{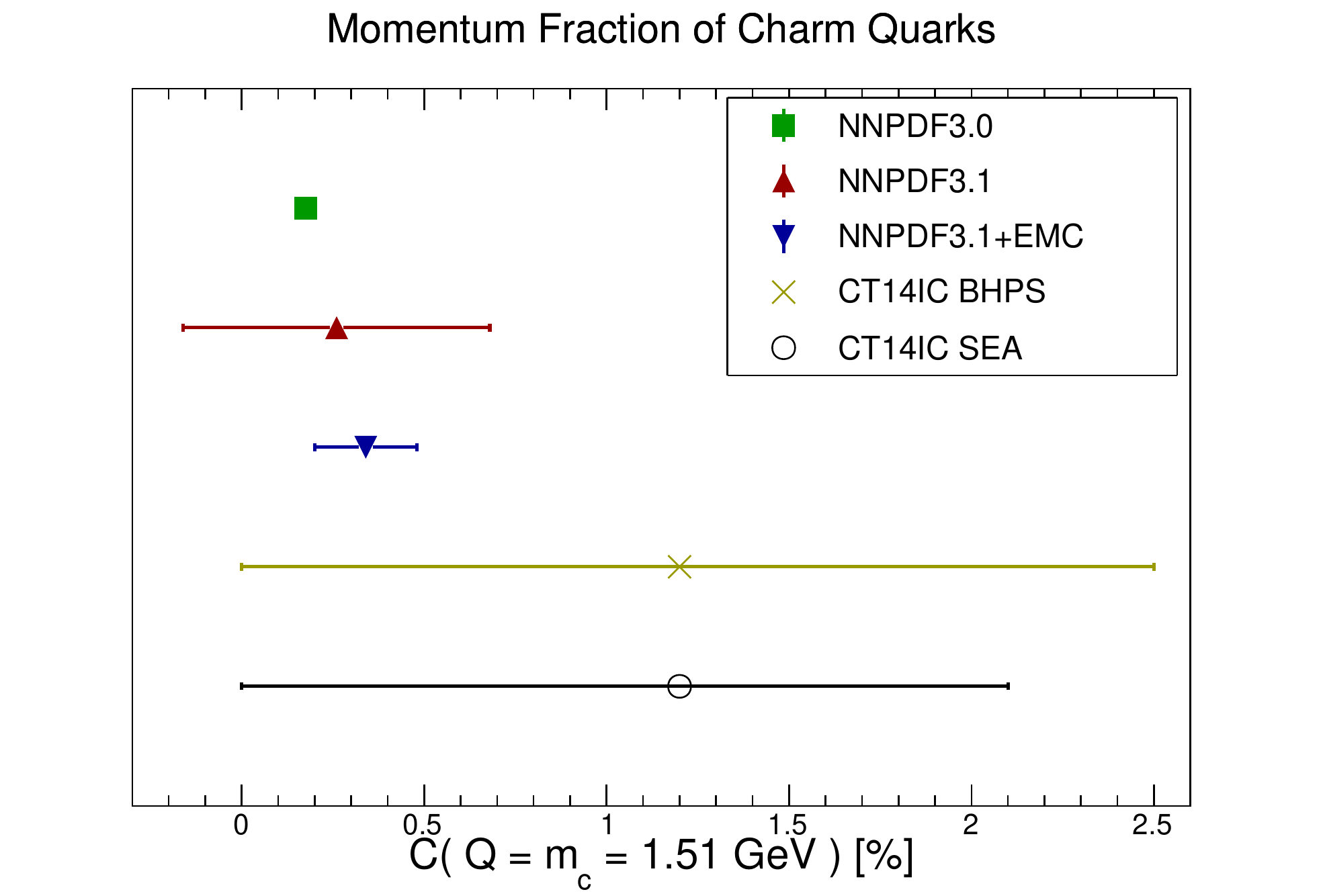}
\includegraphics[scale=0.40]{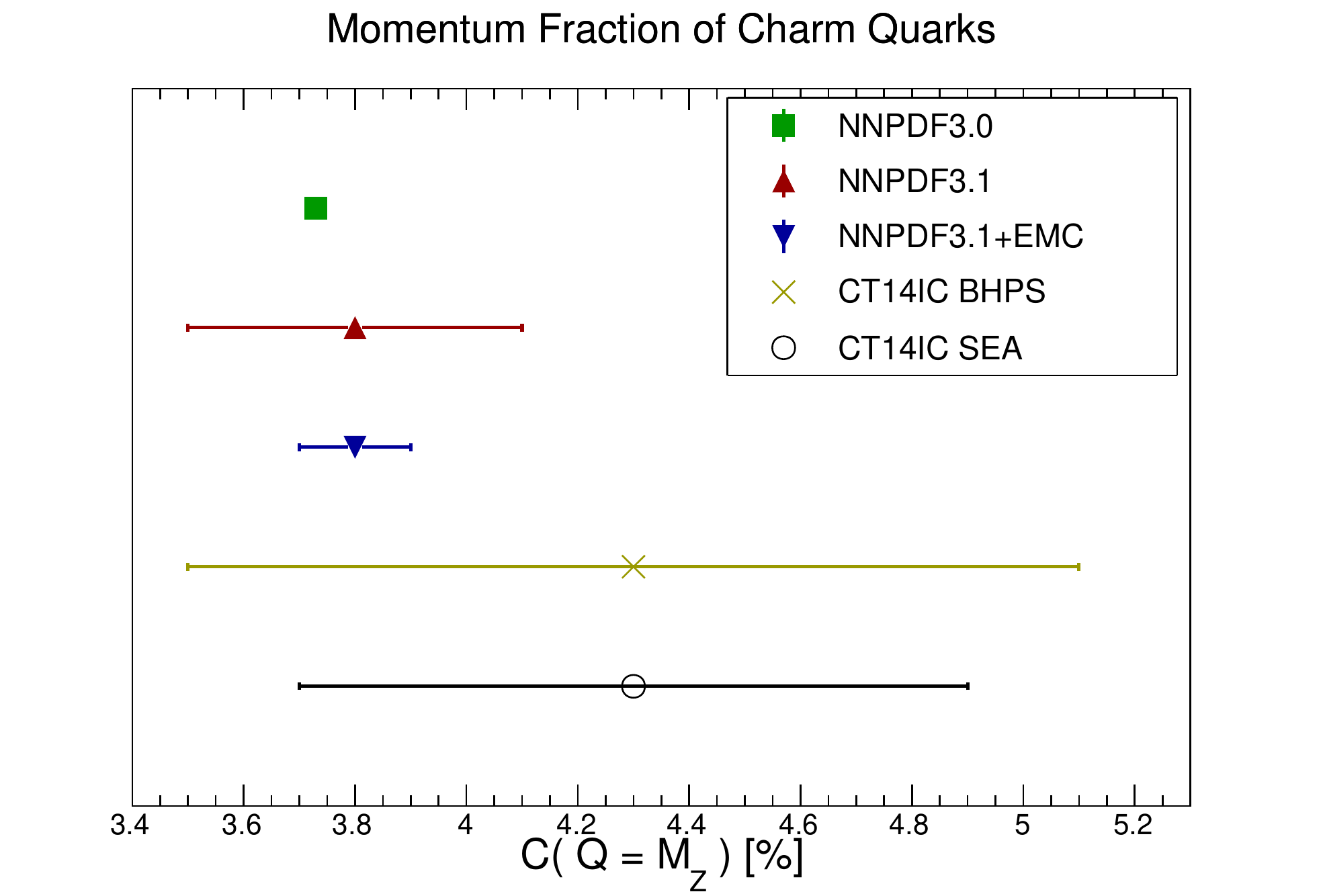}
     \caption{\small 
       The momentum fraction carried by charm quarks, $C(Q)$
       Eq.~(\ref{eq:charmmomfrac}),
       both at a low scale $Q=1.51$ GeV (left) and at a high scale
       $Q=M_Z$ (right plot).
       We compare NNPDF3.0 (perturbative charm) with NNPDF3.1 (based on
       fitted charm) with and without the inclusion of the EMC charm
       data, as well as with the BHPS and SEA scenarios of the
       CT14IC fits.
       See text for more details.
      \label{fig:charmfrac}
  }
\end{center}
\end{figure}

The amount of charm present inside the proton is most
usefully quantified by its
momentum fraction, defined in
Eq.~(\ref{eq:charmmomfrac}).
In the case of perturbative charm, by construction we have $C(Q^2< \mu_c^2)=0$,
while if there is a non--perturbative charm component in the proton
in general we have $C(Q^2)\ne 0$ at all values of the scale $Q^2$.
In Fig.~\ref{fig:charmfrac} we show the momentum fraction carried by charm quarks
both at a low scale $Q=1.51$ GeV and at a high scale
$Q=M_Z$, comparing the results from  NNPDF3.0, based on
perturbative charm,  with those from NNPDF3.1, based on
       fitted charm, with and without the inclusion of the EMC charm
       data, as well as with the BHPS and SEA scenarios of the
       CT14IC fits. The uncertainty bands of the CT14IC results
   indicate the
       range between no intrinsic charm
       and the maximum amount of IC allowed within the CT14 tolerance,
       with the central
       value corresponding to the best fit.
Note that the standard CT14IC fits do not include the EMC data, with the rationale
that  the experimental
systematics are not well documented.
It was found in alternative fits that
neither of the BHPS or SEA models can describe
 the EMC data well. This is also true for the fit with purely perturbative charm.
Thus, within the CT framework, the EMC charm
structure function data is found to provide
no definitive discrimination between the purely
perturbative and intrinsic charm models.
 
%%%%%%%%%%%%%%%%%%%%%%%%%%%%%%%%%%%%%%%%%%%%%%%%%%%%%%%%%%%%%%%%%%%%%
%%%%%%%%%%%%%%%%%%%%%%%%%%%%%%%%%%%%%%%%%%%%%%%%%%%%%%%%%%%%%%%%%%%%%

%%%%%%%%%%%%%%%%%%%%%%%%%%%%%%%%%%%%%%%%%%%%%%%%%%%%%%%%%%%%%%%%%%%%%
%%%%%%%%%%%%%%%%%%%%%%%%%%%%%%%%%%%%%%%%%%%%%%%%%%%%%%%%%%%%%%%%%%%%%

The comparisons of Fig.~\ref{fig:charmfrac} highlight, first of all, that 
when the charm PDF is generated perturbatively
its uncertainties are very small, but that
this is not necessarily the case once charm is fitted.
Reassuringly, once charm is fitted (NNPDF3.1),
the results with perturbative charm (NNPDF3.0)
are consistent within uncertainties.
The NNPDF3.1 analysis also finds that adding the EMC charm
data reduces the PDF uncertainties on $\la x\ra_{\rm IC}$
by around a factor 3, but that even without it one can achieve a quite
competitive determination of $c(x,Q)$, due to the precision
collider electroweak data.
The CT14IC results are consistent within PDF errors
with NNPDF3.1IC, although the highest 
CT14IC values are disfavoured by the NNPDF fit with the latest LHC data.
The rapid growth of $C(Q)$
from $Q=1.51$ GeV to $Q=M_Z$, driven by the perturbative
component, is also apparent from this comparison.

In Fig.~\ref{fig:nnpdf31ic} we show
a comparison of the fitted charm PDF at $Q=1.65$ GeV
      between the NNPDF3IC set and the different model scenarios
      considered in 
      the CT14IC analysis.
      We see that NNPDF3IC prefers a valence--like shape,
      along the lines of the BHPS model, though uncertainties
      are still large.
      The CT14IC BHPS results tend to have a maximum
      at slightly lower values of $x$; note also that
      they develop a perturbative tail since the plot
      is performed at a value $Q > Q_0$.
      The CT14IC SEA models predict that the enhancement of
      the charm PDF is localised at medium and small $x$,
      while in the valence region the fitted charm agrees
      with the perturbative calculation.

\begin{figure}[t]
\begin{center}
\includegraphics[scale=0.40]{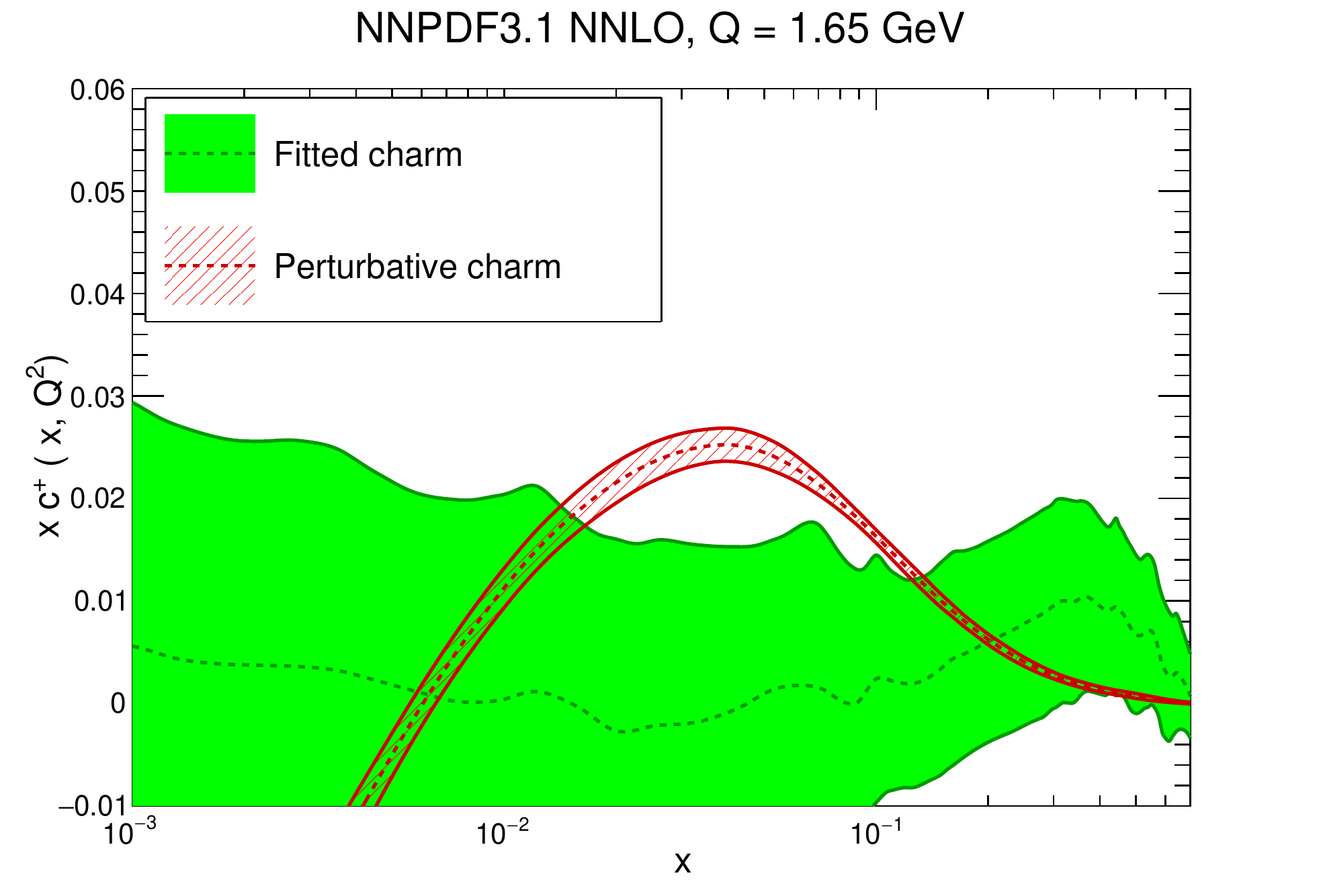}
\includegraphics[scale=0.40]{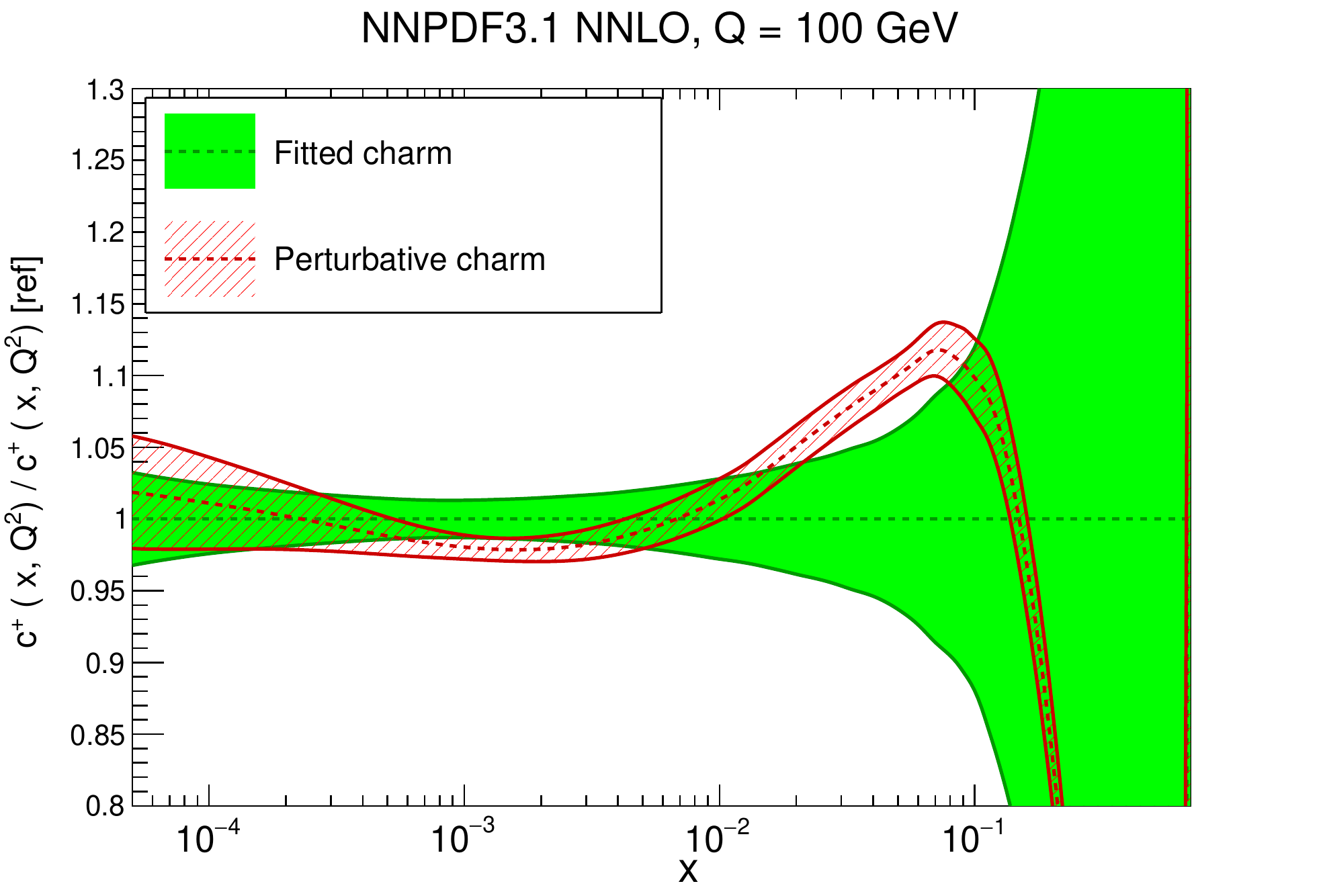}
     \caption{\small 
     Comparison between the charm PDF $xc(x,Q^2)$ in the NNPDF3.1 NNLO analysis with fitted and with
     perturbative charm, both at $Q=1.65$ GeV (left)
     and at $Q=100$ GeV (right plot), in the latter case
     normalized to the fitted charm results.
     Note that these fits do not include the EMC charm structure function data.
     \label{fig:nnpdf31icComp}
  }
\end{center}
\end{figure}

In order to better disentangle the consequences of parametrizing the
charm PDF in the NNPDF case,
in Fig.~\ref{fig:nnpdf31icComp} we show the comparison between the charm
PDF $xc(x,Q^2)$ in NNPDF3.1 NNLO with fitted and with
     perturbative charm, both at $Q=1.65$ GeV 
     and at $Q=100$ GeV, in the latter case
     normalized to the fitted charm results.
     One finds that both approaches are consistent within errors,
     with differences at the two-sigma level at most in the region around
     $x\simeq 0.05$, where the fitted charm PDF undershoots the perturbative
     ansatz, both at low and at high scales.
     We emphasize that the latter behaviour (which is absent for instance
     in the CT14IC fits) is unrelated to the EMC
     charm structure function measurements, which are not included
     in the NNPDF3.1 baseline dataset and thus in the fits shown in
     Fig.~\ref{fig:nnpdf31icComp}

%%%%%%%%%%%%%%%%%%%%%%%%%%%%%%%%%%%%%%%%%%%%%%%%%%%%%%%%%%%%%%%%%%%%%
%%%%%%%%%%%%%%%%%%%%%%%%%%%%%%%%%%%%%%%%%%%%%%%%%%%%%%%%%%%%%%%%%%%%%

If the charm content of the nucleon is indeed different
from the one generated purely from
the perturbative evolution, there
are a number of phenomenological consequences that
could be studied at the LHC.
To begin with, it would modify the kinematic distributions
of the $Z$ bosons
    in the $pp \to Z+c$
    process~\cite{Boettcher:2015sqn,Bailas:2015jlc,Beauchemin:2014rya}, leading to an enhancement
    of the cross section which grows with the value
    of $p_T^Z$.
    To illustrate this point, in Fig.~\ref{fig:ct14ic}
we show the transverse momentum distribution of $Z$ bosons
    in the $pp \to Z+c$ process at $\sqrt{s}=13$ TeV, comparing the CT14 NNLO
    result with the CT14 IC models, as a function
    of $p_T^Z$.
    Depending on the specific model considered, enhancements
    of up to 50\% are predicted, those this may be diluted by fragmentation contributions.

    A closely related process is photon production
    in association with charm
    mesons~\cite{Stavreva:2009vi,Bednyakov:2013zta},
    which is however theoretically less clean as it
    is affected by the poorly understood parton--to--photon
    fragmentation component.
    Another important process where non--perturbative charm
    could play a role is open $D$ meson production~\cite{Vogt:1994zf,Kniehl:2012ti},
    in particular at large $p_T$ and at forward rapidities,
    which enhance the sensitivity to the large--$x$ region.
    By exploiting the information from these various LHC processes using
    data from the present and future runs,
    we can hope to shed more light
    on the charm content of the proton in the future.

%%%%%%%%%%
\vspace{0.6cm}
\section{Electroweak corrections and the photon PDF}\label{sec:QED}
In this section we explore a topic that has recently received a lot
of attention, namely
the role of QED and weak corrections, and in particular
photon-initiated processes.
We first discuss the role of QED corrections
and the photon PDF, before reviewing the case of pure weak corrections
to the hard scattering matrix elements
arising from virtual massive weak boson exchange.

\subsection{Photon--induced processes}\label{sec:QED.photon}

It has been over a decade since the calculation of the splitting functions at NNLO in $\alpha_s$~\cite{mvvns,Vogt:2004mw} provided the necessary tools to
be able to carry out NNLO PDF fits.
Moreover, we have seen in Sect.~\ref{sec:datatheory} that for the majority of PDF--sensitive observables, the perturbative calculation is available at this NNLO order. Given that data from the LHC are available at increasing precision, to below the percent level, NNLO PDF fits are essential to match
this and have naturally become the standard. However, a simple counting of powers of $\alpha_S$ indicates that
\be
\alpha^2_s(M_Z)\sim \frac{1}{70}\;,\qquad \alpha_{\rm EM}(M_Z^2)\sim \frac{1}{130}\;.
\ee
That is, we may roughly expect the NNLO QCD  and NLO EW corrections to be of the same order of magnitude. While such an argument clearly neglects the non--trivial differences in the structure of the QCD and EW corrections, this nonetheless serves as a warning that we must at least consider the impact of going to NLO EW if we are to claim percent--level precision to LHC cross sections.

\begin{figure}
  \begin{center}
    \includegraphics[width=0.28\textwidth]{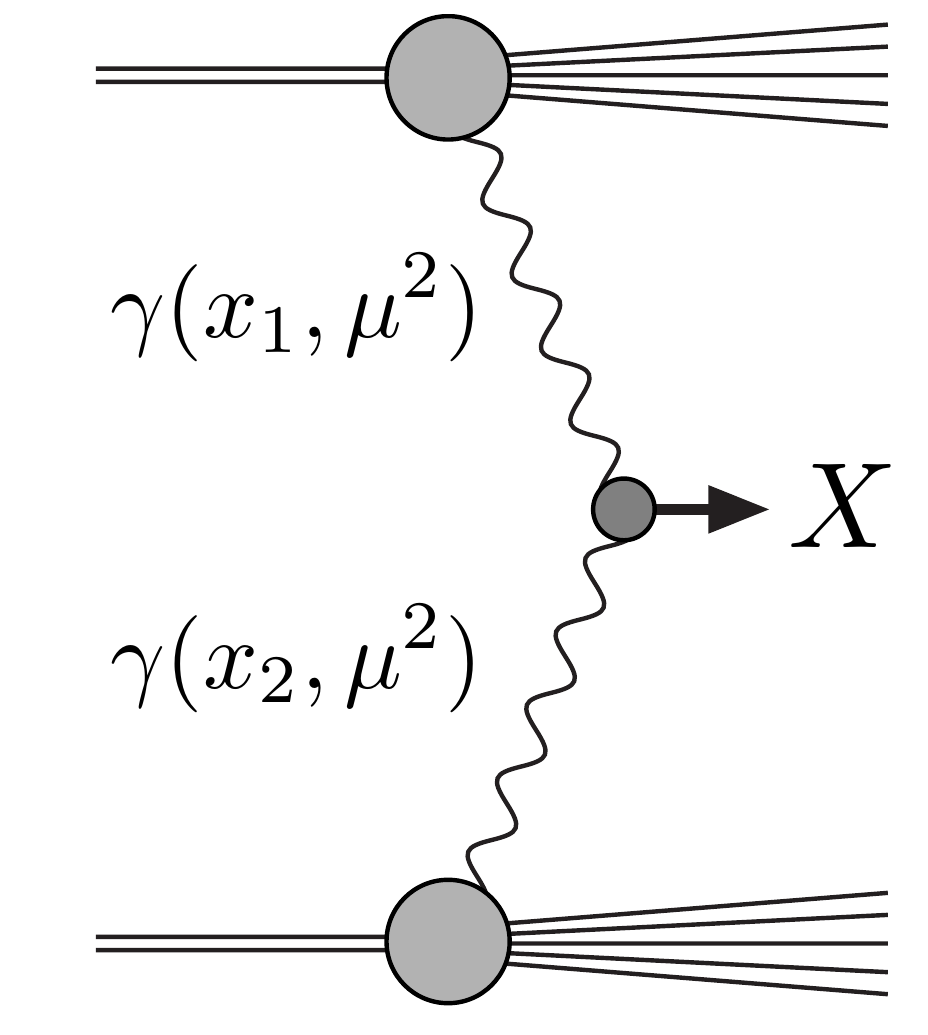}\qquad
    \includegraphics[width=0.45\textwidth]{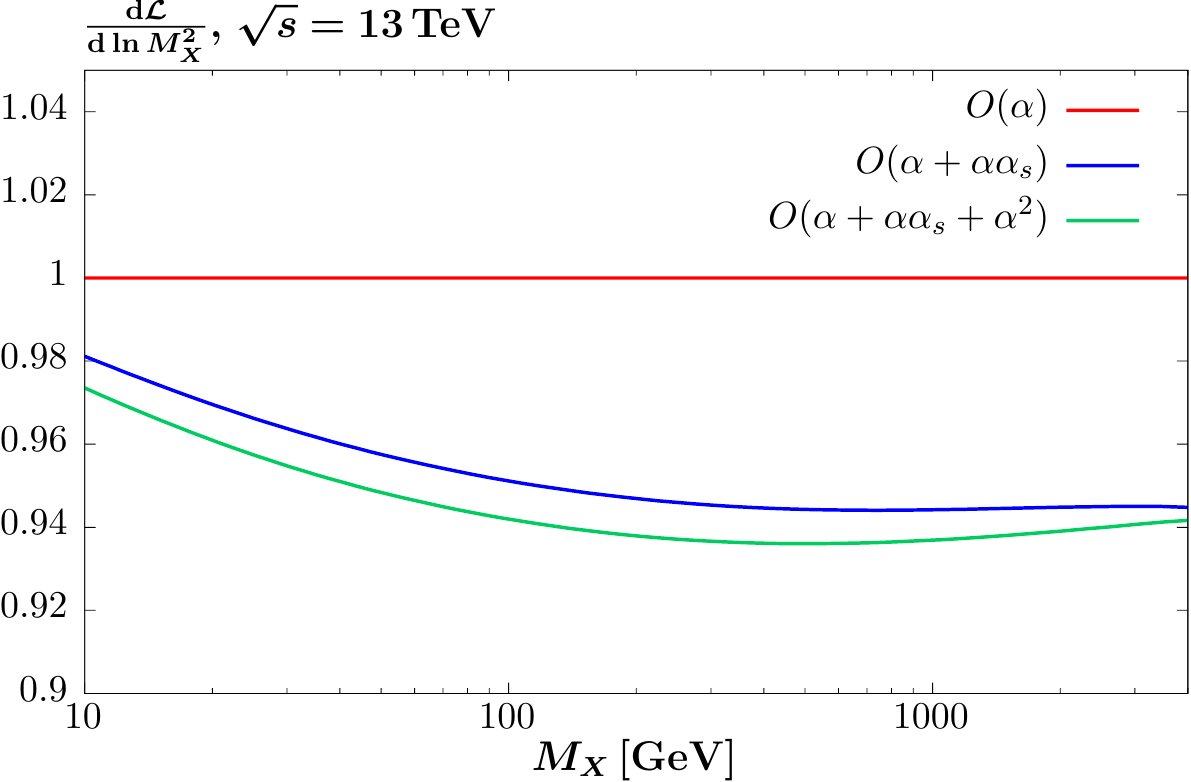}
  \end{center}
  \caption{\small Left: schematic diagram for the photon--initiated production of a system $X$, and the corresponding photon PDFs.
    Right: the $\gamma\gamma$ luminosity as a function of the invariant mass, $M_X$, of the produced final state.
The ratio of results with $O(\alpha\alpha_S)$ and $O(\alpha^2)$ to the leading $O(\alpha)$ DGLAP evolution shown. Calculated using the approach described in~\cite{Harland-Lang:2017dzr}.
  }\label{fig:photfeyn} \label{fig:qedevol}
\end{figure}

A specific type of EW correction of particular relevance to PDF studies is the contribution from photon--initiated processes, such as those
shown schematically in Fig.~\ref{fig:photfeyn}.
As this involves a photon in the initial state, this requires the introduction of a PDF for the photon in the proton\footnote{For brevity, we will refer to this throughout as the photon PDF, but this should not be confused with the partonic content of the photon itself, which often receives a similar label, see e.g.~\cite{Gluck:1999ub}.}. This is included in complete analogy to the QCD partons, and moreover as it involves a massless boson in the initial state, higher--order QED $q\to q\gamma$ and $\gamma\to q\overline{q}$ splitting will generate collinear singularities that must be absorbed into the corresponding PDFs.
In other words, this will produce QED corrections to the DGLAP evolution of the PDFs.
Another important type of EW corrections relevant for PDF fits, namely
those associated to virtual massive weak boson exchange, are
discussed in Sect.~\ref{sec:QED.photon.EW}.

%%%%%%%%%%%%%%%%%%%%%%%%%%%%%%%%%%%%%%%%%%%%%%%%%%%
%%%%%%%%%%%%%%%%%%%%%%%%%%%%%%%%%%%%%%%%%%%%%%%%%%%%

\subsubsection*{QED corrections to DGLAP evolution}\label{sec:QED.photon.dglap}

The introduction of the photon PDF requires the following straightforward extension of the DGLAP evolution equations,
\begin{align}\nonumber
 Q^2\frac{\partial}{\partial Q^2}g(x,Q^2) &=\sum_{q,\overline{q},g}P_{g a}(x,\as(Q^2))\otimes f_a(x,Q^2)+P_{g\gamma}(x,\as(Q^2))\otimes \gamma(x,Q^2),\,\\ \nonumber
 Q^2\frac{\partial}{\partial Q^2}q(x,Q^2) &=\sum_{q,\overline{q},g}P_{q a}(x,\as(Q^2))\otimes f_a(x,Q^2)+P_{q\gamma}(x,\as(Q^2))\otimes \gamma(x,Q^2),\,\\ \label{eq:photdglap}
   Q^2\frac{\partial}{\partial Q^2}\gamma(x,Q^2) &=P_{\gamma\gamma}\otimes \gamma(x,Q^2)+\sum_{q,\overline{q},g}P_{\gamma a}(x,\as(Q^2))\otimes f_a(x,Q^2).\,
\end{align}
The splitting functions can then be expanded in powers of both the QCD and QED coupling
\be
P_{ij}=\sum_{m,n}\left(\frac{\alpha_S}{2\pi}\right)^m\left(\frac{\alpha}{2\pi}\right)^n P_{ij}^{(m,n)}\;.
\ee
The lowest order QED splitting function $P_{\gamma q}^{(0,1)}$ is due to the same type of Feynman diagram as in the LO QCD case for $P_{g q}^{(1,0)}$, with the gluon simply replaced by a photon, and similarly for $P_{qq}$ and $P_{q\gamma}$. Thus these are trivially related by suitable adjustments of the colour factors and inclusion of the electric charges $e_q$ of the quark, with
\be
P_{qq}^{(0,1)}=\frac{e_q^2}{C_F}P_{qq}^{(1,0)}\;,\qquad
P_{q\gamma}^{(0,1)}=\frac{e_q^2}{T_R}P_{q g}^{(1,0)}\;,\qquad
P_{\gamma q}^{(0,1)}=\frac{e_q^2}{C_F}P_{g q}^{(1,0)}\;,\qquad
P_{\gamma\gamma}^{(0,1)}=-\frac{2}{3}\sum_f e_f^2 \,\delta(1-x)\;,
\ee
where for the $P_{\gamma\gamma}$ case only the Abelian contribution is present and the sum is over all fermions in the loop, that is quarks and leptons\footnote{To be consistent, and in particular to preserve momentum fully, this requires the introduction of lepton PDFs in the proton. However as discussed in~\cite{Bertone:2015lqa} the contribution from these is generally of limited phenomenological relevance, and can be safely neglected. Note in any case that including leptons
in the running of the QED coupling $\alpha(Q)$ is still required.}. The calculation of the $O(\alpha_s \alpha)$ ($m,n=1$) terms, where the $P_{g\gamma}$ and $P_{\gamma g}$ splittings enter for the first time, is given in~\cite{deFlorian:2015ujt}, while the $O(\alpha^2)$ ($m=0$, $n=2$) terms are given in~\cite{deFlorian:2016gvk}. Publicly available implementations of the DGLAP evolution including these QED corrections are provided by the \texttt{APFEL}~\cite{Bertone:2013vaa} and \texttt{QEDEVOL}~\cite{Sadykov:2014aua} packages.

The impact of the $O(\alpha \alpha_S)$ and $O(\alpha^2)$ corrections on the $\gamma\gamma$ luminosity is shown in Fig.~\ref{fig:qedevol}. The $O(\alpha\alpha_S)$ corrections have a fairly small but clearly non--negligible impact on the luminosity, giving up to a $\sim 5\%$ negative correction. As we would expect, the $O(\alpha^2)$ corrections are significantly smaller, but can reach the percent level. The results of Fig.~\ref{fig:qedevol} have been computed
using the MMHT framework~\cite{Harland-Lang:2017dzr}, which is closely based on the LUXqed formalism described below.  

 As discussed in detail in~\cite{Harland-Lang:2016lhw}, the $P_{\gamma\gamma}$ self--energy contribution to the DGLAP evolution of the photon PDF is intimately connected to the choice of renormalization scale for the initial--state photon coupling to the hard process. 
It is well known in QED that
 for on--shell external photons the coupling receives no renormalization, and is completely determined to be $\alpha(0)$.
However, the $P_{\gamma\gamma}$ term breaks this simple picture, and we should instead use $\alpha(\mu_F)$ in the calculation.
 Physically, the photon substructure is being resolved by the introduction of a photon PDF and the contribution from $\gamma\to q\overline{q}$ splittings in the evolution, such that a purely on--shell renormalization scheme is no longer appropriate. This has been confirmed at NLO EW order in~\cite{Kallweit:2017khh}, where it is shown that using the on--shell scheme leads to uncanceled fermion--mass singularities in the hard cross section. 
 
\subsubsection*{The photon PDF}\label{sec:QED.photon.PDF}

The first attempts at describing the photon PDF can be divided into two distinct categories, either being phenomenological approaches that model the photon PDF, as in the MRST2004QED~\cite{Martin:2004dh} and more recent CT14QED~\cite{Schmidt:2015zda} sets, or the data--driven
approach of the NNPDF2.3/3.0QED~\cite{Ball:2013hta,Bertone:2016ume} sets.
The first attempt to include the photon in a PDF set was provided by MRST2004QED. Here, a simple model for the photon PDF at input scale $Q_0$ due to one--photon emission off the valence quarks was
assumed.
In other words, the quark valence distributions are frozen at $Q_0$ and the LO QED DGLAP evolution for the photon is integrated between the light quark mass $m_q$ and $Q_0$, so that
\be\label{eq:photinel}
\gamma(x,Q_0^2)=\frac{\alpha}{2\pi}\left[\frac{4}{9}\log\left(\frac{Q_0^2}{m_u^2}\right) u(x,Q_0)+\frac{1}{9}\log\left(\frac{Q_0^2}{m_d^2}\right)d(x,Q_0)\right]\otimes\frac{1+(1-x)^2}{x}\;.
\ee
The CT14QED set generalised this approach, allowing additional freedom in the normalization of the photon, which was fit to ZEUS data~\cite{Chekanov:2009dq} on isolated photon production\footnote{In fact, as we will discuss below, this has subsequently been supplemented with the elastic component to give the inclusive set CT14QEDinc.}. Thus, within such approaches the photon PDF is completely predicted within the specific model, up to any freedom in the model parameters, such as the choice of quark masses for MRST2004QED or the overall normalization for CT14QED. 

On the other hand, the NNPDF2.3QED~\cite{Ball:2013hta} set (subsequently updated to NNPDF3.0QED~\cite{Bertone:2016ume}), freely parameterises the photon PDF at input. In other words, the photon is treated on exactly the same footing as the QCD partons. This is then extracted from a fit (or more precisely, a Bayesian reweighting) to DIS and LHC $W,Z$ data; in the former case no explicit photon--initiated contribution was included, and so the constraint came purely from the effect on the PDF evolution.
However, in general the contribution of photon--initiated process are small, leading to significant
uncertainties on the extracted photon PDF. 

More recently, there has been a great deal of progress in our understanding of the photon PDF. One crucial point that was missed in the above approaches is the long range nature of QED.
That is, the proton is itself an electrically charged object which can coherently emit a photon, with the proton remaining intact afterwards. The possibility for such elastic photon emission is of course very well established. Elastic $ep$ scattering is an extremely well measured process, providing for example the first measurement of the proton charge radius~\cite{Hofstadter:1955ae,Mcallister:1956ng} in the 1950s, with further precise measurements of this process~\cite{Bernauer:2013tpr} continuing to this day.
Theoretically, the well known equivalent photon approximation (EPA)~\cite{Budnev:1974de} provides a precise foundation for describing the elastic scattering process in terms of a flux of coherently emitted photons from the proton.

The connection of this fact to the photon PDF was discussed in~\cite{Gluck:2002fi} and more recently in~\cite{Martin:2014nqa,Harland-Lang:2016qjy}. Following the equivalent photon approximation, it is straightforward to show that elastic photon emission leads to a contribution to the photon PDF at a scale $Q_0\sim $1 GeV given by
\be \label{eq:photcoh}
\gamma_{\rm el}(x,Q_0^2)=\frac{1}{x}\frac{\alpha}{\pi}\int_{\frac{x^2 m_p^2}{1-x}}^{Q_0^2}\!\!\frac{{\rm d}Q^2 }{Q^2}\left[\left(1-x-\frac{x^2 m_p^2}{Q^2}\right)F_E(Q^2)+\frac{x^2}{2}F_M(Q^2)\right]\;,
\ee
where $F_E$ and $F_M$ are the elastic and magnetic form factors of proton, which are related to the electric and magnetic charge distributions in the proton. These are steeply falling functions of $Q^2$ that are probed very precisely in a range of elastic $ep$ scattering experiments, see e.g.~\cite{Bernauer:2013tpr}.

To demonstrate the connection of this elastic
component to the inclusive photon PDF, 
if we for simplicity omit the small backreaction
that the photon has on the quark and gluon PDFs via the evolution
equations, then we can solve Eq.~(\ref{eq:photdglap}) to get~\cite{Harland-Lang:2016apc}
\begin{align}\label{eq:photsol}
\gamma(x,\mu^2)&=\frac{1}{\alpha(\mu^2)}\left(\alpha(Q_0^2)\,\gamma(x,Q_0^2)+\int_{Q_0^2}^{\mu^2}\frac{{\rm d}Q^2}{Q^2}\alpha(Q^2)\sum_{q,\overline{q},g}P_{\gamma a}(x,\as(Q^2))\otimes f_a(x,Q^2)\right)\;,\\ \label{eq:photsol1}
&\equiv\gamma_{\rm input}(x,\mu^2)+\gamma_{\rm evol}(x,\mu^2)\;.
\end{align}
Thus the photon is given separately in terms of an input at low scale $Q_0$ and an evolution component due to the usual DGLAP $q\to q\gamma$ emission for $Q^2>Q_0$.
The latter is completely determined in terms of the quark and gluon PDFs, leaving the input photon at $Q_0$, which is dominantly due to elastic emission.  Thus this already provides quite a strong constraint on the photon PDF; as we will see below, the impact in comparison to the model--independent
NNPDF approach can be dramatic.

However, even for relatively low photon virtualities, $Q^2<Q_0^2$, the emission may also be inelastic, such that the proton no longer remains intact afterwards. In other words we have
\be
\gamma(x,Q_0^2)=\gamma_{\rm el}(x,Q_0^2)+\gamma_{\rm inel}(x,Q_0^2)\;,
\ee
In~\cite{Martin:2014nqa,Harland-Lang:2016qjy} fairly simple phenomenological models for this inelastic component, given by suitable generalizations of (\ref{eq:photinel}), were taken, while the CT14QED set allowed an additionally free normalization to be fitted to ZEUS data on isolated photon production, as described above. 

Given that the elastic component is directly determined from the form factors $F_E$ and $F_M$, that are themselves measured from elastic $ep$ scattering, it is natural to ask whether the inelastic component can be similarly determined. In other words, rather than relying on a phenomenological model, can $\gamma_{\rm inel}$ instead be calculated directly from the form factors for inelastic $ep$ scattering, that is, from the proton structure functions?
In the analysis of~\cite{Manohar:2016nzj} it was shown that this is indeed the case, with the corresponding {\tt LUXqed} PDF set made publicly available. In particular, they find that the photon PDF can be expressed as\footnote{Following the publication of~\cite{Manohar:2016nzj} it was discovered that this expression had been derived in the earlier papers of~\cite{Anlauf:1991wr,Blumlein:1993ef,Mukherjee:2003yh}, but without the correct limits on the $Q^2$ integral.}
   \begin{align}
\nonumber
  x\gamma(x,\mu^2) &=  \frac{1}{2\pi \alpha(\mu^2)} \! \int_x^1 \frac{dz}{z} \Bigg\{ \int^{\frac{\mu^2}{1-z}}_{\frac{x^2 m_p^2}{1-z}}   \frac{dQ^2}{Q^2} \alpha^2(Q^2) \Bigg[\! \left(  zp_{\gamma q}(z)+ \frac{2 x ^2 m_p^2}{Q^2}\right)\! F_2\left(\frac{x}{z},Q^2\right)-z ^2 F_L\!\left(\frac{x}{z},Q^2\right)\Bigg] \\  \label{eq:luxqed}
 & - \alpha^2(\mu^2) z^2F_2\left(\frac{x}{z},\mu^2\right)\Bigg\}\,.
  \end{align}
  Thus the photon PDF is in fact a completely derived quantity, which can be written purely in terms of the proton structure functions, which are measured precisely. More recently, the detailed study of~\cite{Manohar:2017eqh} has demonstrated how this expression may be derived in a process--independent way by using the operator definition of the photon PDF, as well as generalising this expression to the case of the polarized and transverse momentum dependent PDFs. This approach is also shown to provide a simple derivation of the known $O(\alpha\alpha_s)$ and $O(\alpha^2)$ splitting functions, $P_{\gamma i}$.

\begin{figure}[t]
  \begin{center}
    \includegraphics[width=0.99\textwidth]{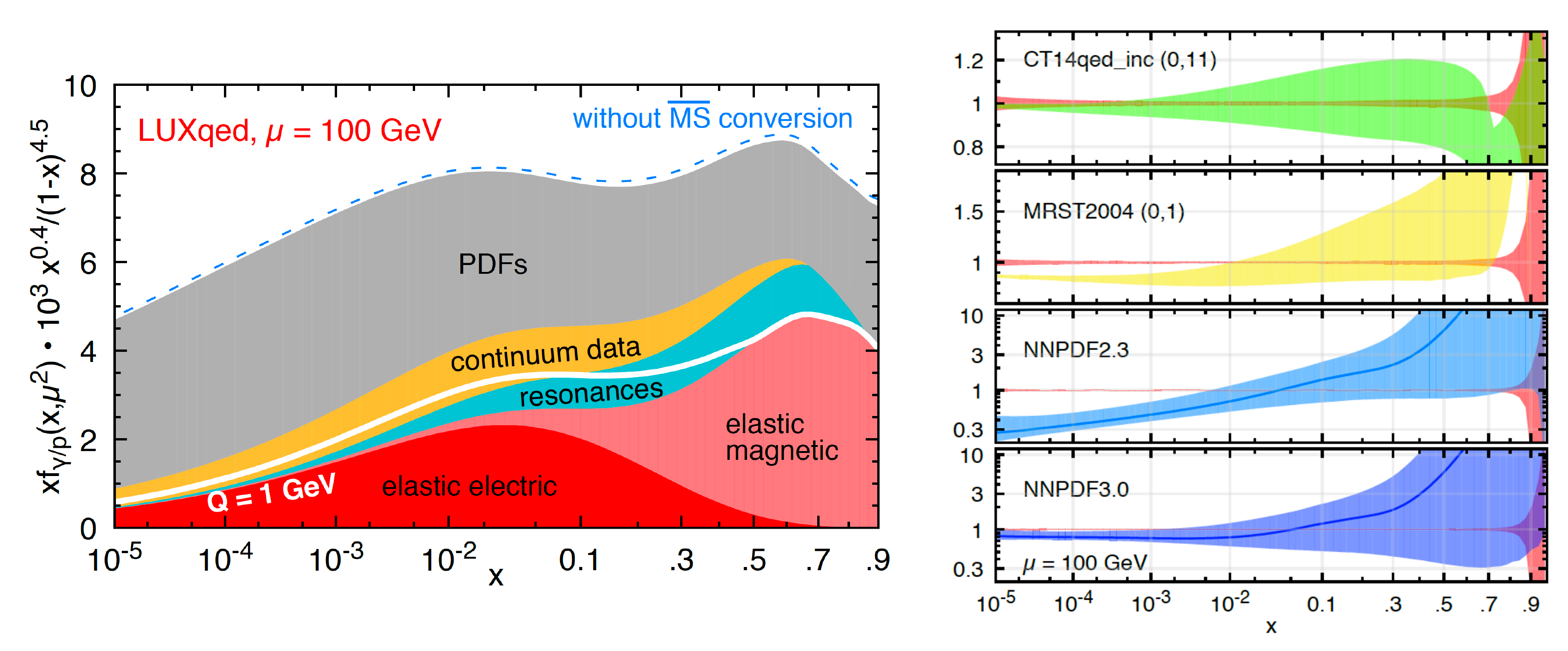}
  \end{center}
  \vspace{-0.6cm}
  \caption{\small Left: overview of the various contributions to the photon
    PDF $\gamma(x,Q^2)$ in the LUXqed approach as a function of $x$
    at $Q=100$ GeV.
    Right: comparison of the photon PDFs from CT14qed\_inc, MRST2004, NNPDF2.3/3.0
    and LUXqed, normalized to the central value of the latter.
  }\label{fig:LUXqed}
\end{figure}

  While the connection of Eq.~(\ref{eq:luxqed}) to the considerations above is not immediately clear, some similarity in the overall form with Eq.~(\ref{eq:photsol}) is apparent.
  Indeed, by substituting for $F_{2,L}$ in terms of the quark and gluon PDFs, at high $Q^2$  this readily reduces to  $\gamma_{\rm evol}$ in (\ref{eq:photsol}); indeed this is how the \texttt{LUXqed}
  photon PDF is calculated in this region.
  In addition, using the known expression for the elastic contributions to $F_{2,L}$ reproduces Eq~(\ref{eq:photcoh}) when combined with Eq.~(\ref{eq:photsol}); this elastic contribution is also included. By using fits to the experimentally determined inelastic structure functions at low $Q^2$, including in the resonance region, it is shown in~\cite{Manohar:2016nzj} that the remaining inelastic contribution, and therefore the photon PDF in its entirety, is very precisely determined.
  In Fig.~\ref{fig:LUXqed} we show an overview of the various contributions to the photon
    PDF $\gamma(x,Q^2)$ in the LUXqed approach as a function of $x$
    at $Q=100$ GeV.
    We see that at small $x$ it is dominated by the PDF contribution, while
    at large $x$ the elastic contribution accounts for up to half
    of the size of $\gamma(x,Q)$.
  
    It is worth emphasizing that the expression Eq.~(\ref{eq:luxqed}) does not rely on any explicit separation between an input and evolution component to the photon as in (\ref{eq:photsol}), and corresponds to the exact result for the photon within the quoted accuracy of~\cite{Manohar:2016nzj,Manohar:2017eqh}, valid to all orders in QED and QCD, and including non--perturbative corrections.
    Indeed, applying standard DGLAP above the starting scale $Q_0$ the power--like $\sim m_p^2/Q^2$ correction would be missed, while for $Q^2>Q_0^2$ the contribution from the elastic component would be omitted and the inelastic resonance component, which contributes at higher $x$ in this region, would not be correctly modelled.
  
 %%%%%%%%%%%%%%%%%%%%%%%%%%%%%%%%%%%%%%%%%%%%%%%%%%%
%%%%%%%%%%%%%%%%%%%%%%%%%%%%%%%%%%%%%%%%%%%%%%%%%%%%

However, from the point of view of a global PDF set it may be preferable to use Eq.~(\ref{eq:luxqed}) in a form that can be more directly implemented within the standard fitting framework. That is, applying this approach after suitable modification to calculate the input photon, which can then be included as part of the default input parton set for any future fits and studies, see ~\cite{Harland-Lang:2017dzr} for initial results in the
context of the MMHT framework.
An alternative iterative approach is proposed in~\cite{Manohar:2017eqh}. This strategy has been implemented to construct the
recent NNPDF3.1luxQED global analysis~\cite{Bertone:2017bme}.

To illustrate the differences and similarities between these various determinations
of $\gamma(x,Q)$, in Fig.~\ref{fig:LUXqed} we show the comparison of the photon PDFs from CT14qed\_inc, MRST2004, NNPDF2.3/3.0
    and LUXqed, normalized to the central value of the latter.
    It is clear from this comparison that the theoretical uncertainties associated
    with the LUXqed determination are much smaller than in any other of the
    previous approaches.

\subsection*{Phenomenology}\label{sec:QED.photon.pheno}

\begin{figure}[t]
  \begin{center}
 \includegraphics[width=0.45\textwidth]{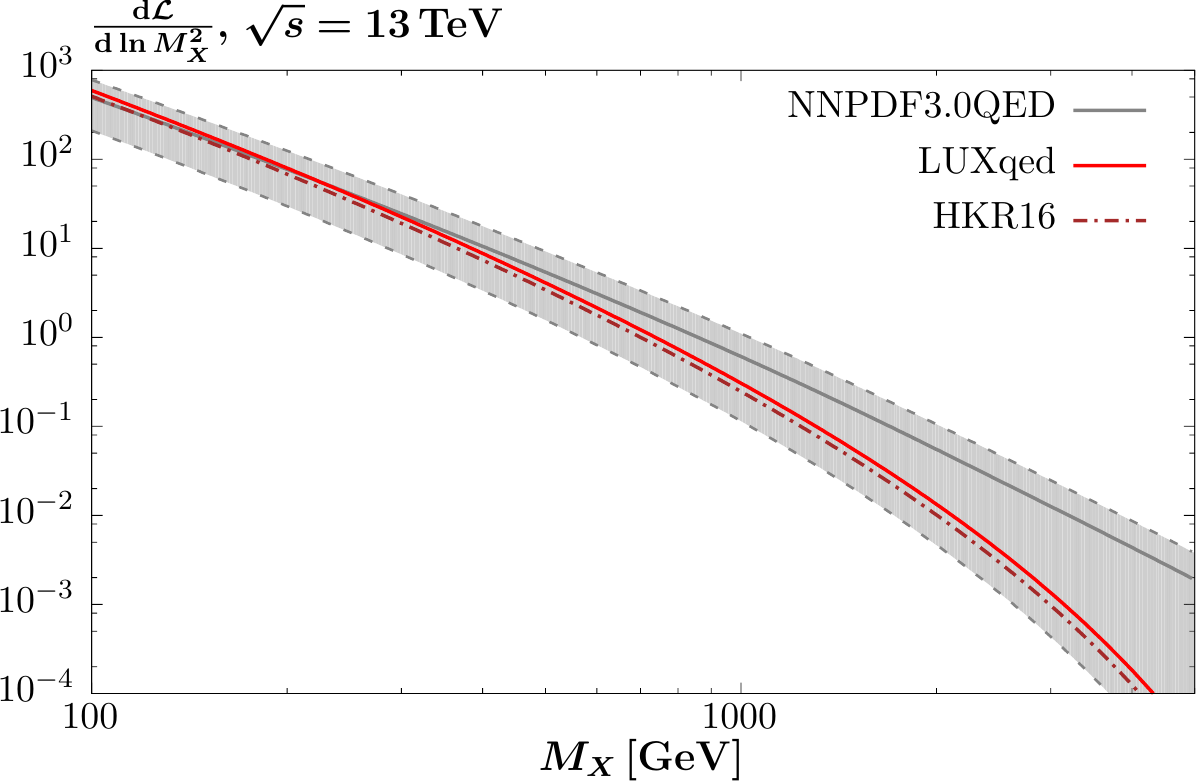}
  \includegraphics[width=0.45\textwidth]{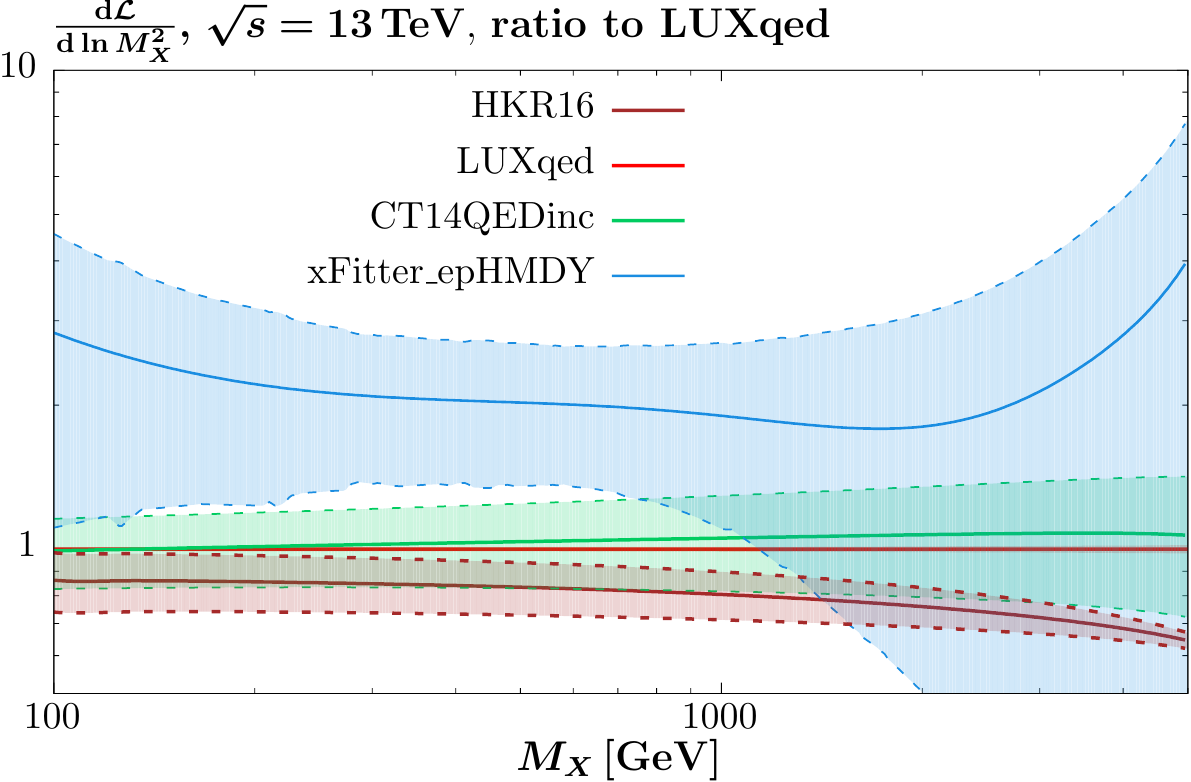}
  \end{center}
\caption{$\gamma\gamma$ luminosities at $\sqrt{s}=13$ TeV. (Left) Absolute values for the HKR16,  NNPDF3.0QED and LUXqed sets. (Right) Ratios of the CT14QED, HKR16 and {\tt xFitter\_HMDYep} sets to the LUXqed prediction. 68\% uncertainty bands are shown, with the exception of the HKR16 set, where the error is due to model variation in the inelastic input (lower edge corresponds to elastic only).}\label{fig:photlumi}
\end{figure}

In Fig.~\ref{fig:photlumi} (left) we show the NNPDF3.0QED $\gamma\gamma$ luminosity at $\sqrt{s}=13$ TeV, including the 68\% C.L. error bands. A very large PDF uncertainty is evident, in particular at higher system mass $M_X$. As discussed above, the input component in Eq.~(\ref{eq:photsol1}) is poorly determined within this approach, due to limited constraints placed by the available experimental
data.
It is therefore unsurprising that the PDF errors should be most significant at higher mass,  as here the relative contribution from this input component is larger, due the reduced phase space for PDF evolution. In addition, the central value of the luminosity is seen to lie towards the upper end of the uncertainty band. As discussed in~\cite{Mangano:2016jyj,Harland-Lang:2016qjy}, this exhibits a much gentler decrease with $M_X$ in comparison to the QCD parton luminosities. However, also plotted is the LUXqed result, and the difference is dramatic. The central value lies towards the lower end of the NNPDF band at higher mass, with a PDF uncertainty that is smaller than the line width of the plot. We also show the result of~\cite{Harland-Lang:2016qjy}, labelled HKR16, which includes the elastic contribution to the input photon and a simple model for the remaining inelastic component. This demonstrates a similar trend. Thus, simply applying basic physical conditions on the form of the photon PDF, and including the dominant coherent input
Eq.~(\ref{eq:photcoh}) gives a qualitatively similar result.

%%%%%%%%%%%%%%%%%%%%%%%%%%%%%%%%
%%%%%%%%%%%%%%%%%%%%%%%%%%%%%%%%%5

Taking a closer look, in Fig.~\ref{fig:photlumi} (right) we show the ratio of the HKR16, CT14QED and xFitter\_HMDYep results to the LUXqed luminosity. The xFitter\_HMDYep set is extracted
in~\cite{Giuli:2017oii} by applying a similar agnostic methodology to NNPDF, but including the more constraining ATLAS high mass Drell--Yan data~\cite{Aad:2016zzw} in the fit; this therefore represents the most up to date set within such an approach. Again, the LUXqed uncertainty band is barely visible on the curve, varying from $1-2$\% over the considered mass interval.  The CT14QEDinc prediction, which includes an elastic component, is consistent, but with larger $\sim 20-40\%$ uncertainties, due to the more limited constraints placed by the ZEUS isolated photon production data on the inelastic input\footnote{In addition, the ZEUS data are selected by requiring that at least one track associated with the proton side is reconstructed within the detector acceptance. While this will remove the elastic component entirely, which will produce no extra tracks, it is also possible for the proton dissociation products in the inelastic case to fall outside the acceptance and therefore not pass such a cut. Thus while CT14QED use these data to  extract the total inelastic component, at least part of this will also be removed by this cut.}. The HKR prediction lies somewhat below the LUXqed result, outside of the quoted model variation band, in particular at larger $M_X$. This is due in large part to the lack of any explicit resonant contribution in the inelastic input, which becomes more important at higher $x$ and hence $M_X$. Finally, the xFitter\_HMDYep set is seen to have a sizeable uncertainty band (albeit smaller than the NNPDF3.0 set~\cite{Giuli:2017oii}), and interestingly appears to lie somewhat above the LUXqed result. From this it is clear that any attempt to extract the photon PDF within such an approach will almost certainly not be competitive. More generally, we can see that the LUXqed set exhibits by far the smallest PDF uncertainties. 

\begin{figure}[t]
  \begin{center}
 \includegraphics[width=0.45\textwidth]{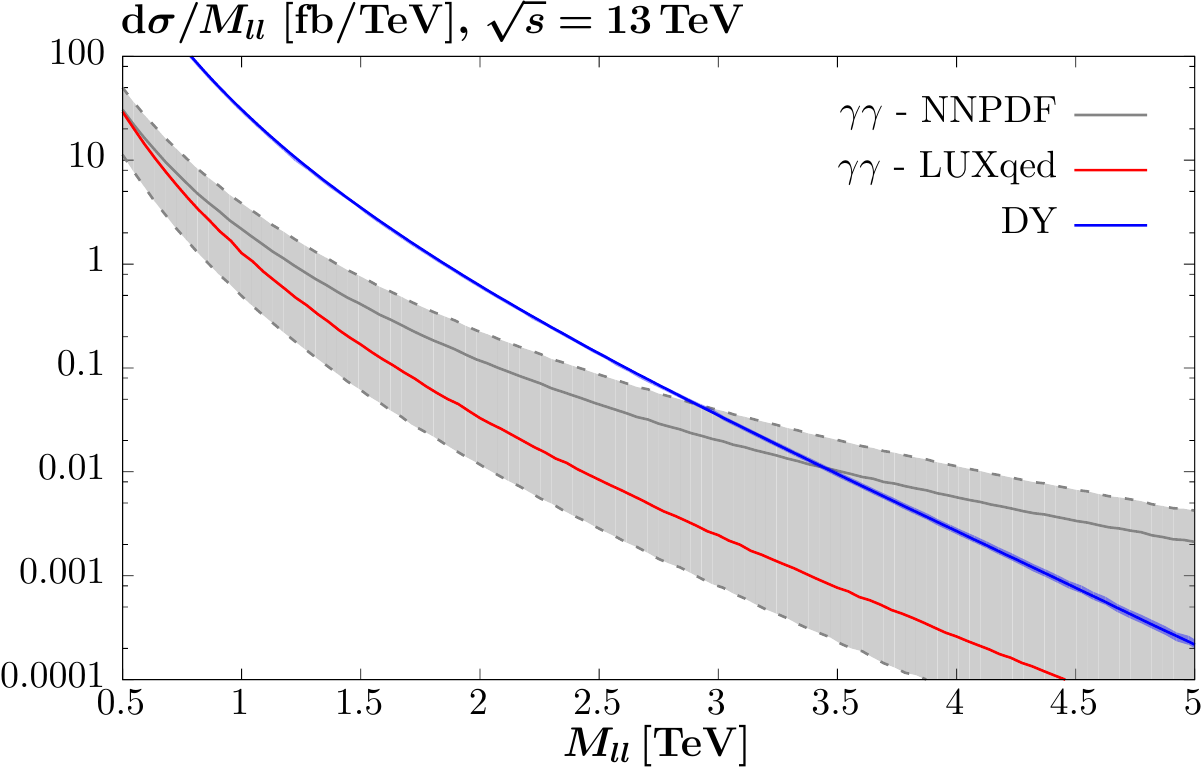}
  \includegraphics[width=0.45\textwidth]{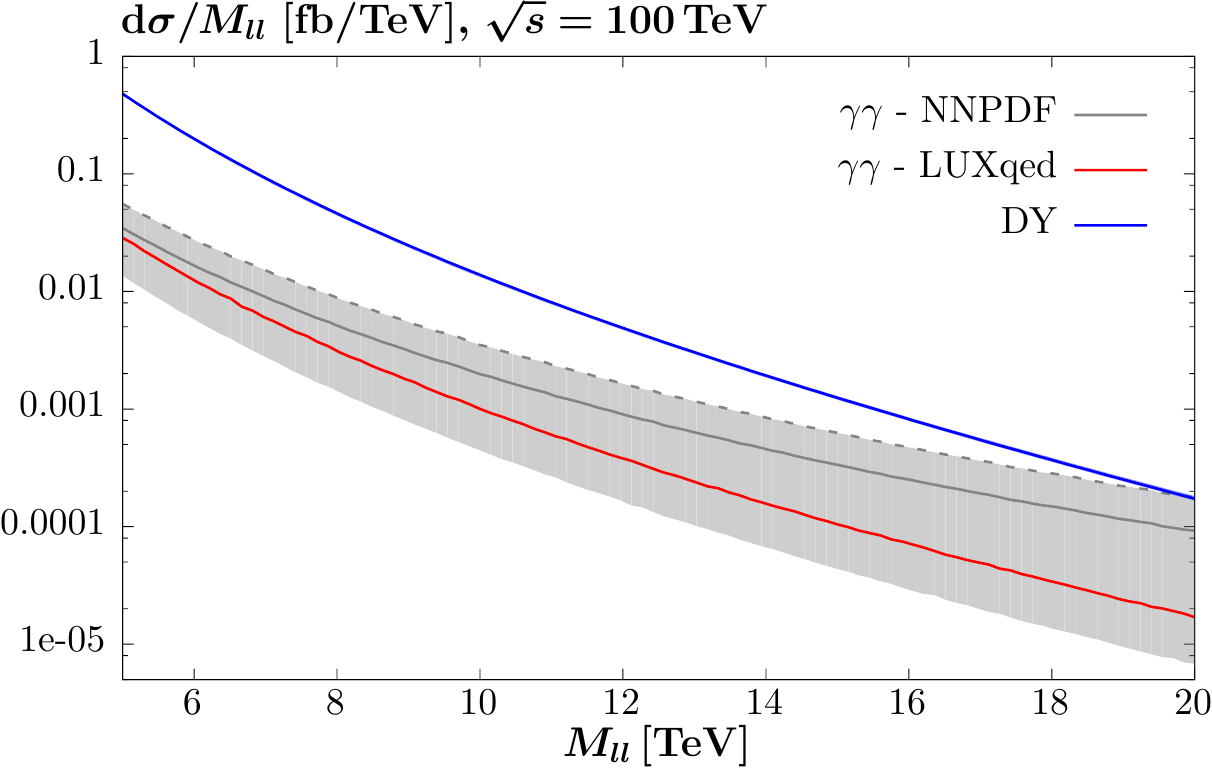}
  \end{center}
\caption{The differential lepton pair production cross sections at $\sqrt{s}=$ 13 TeV and 100 TeV with respect to the invariant mass of the pair $M_{ll}$, for lepton $|\eta| <2.5$ and $p_\perp  > 20$ GeV. The photon--initiated contributions predicted using the LUXqed and NNPDF3.0QED sets, including the 68\% C.L. uncertainty bands. The NLO Drell--Yan cross section, calculated with \texttt{MCFM}~\cite{Campbell:2011bn}, is also shown.}\label{fig:lepphot}
\end{figure}

Prior to these most recent developments, a range of phenomenological studies pointed out similar trends in the NNPDFQED predictions for the photon--initiated contributions to lepton and $W$ pair~\cite{Mangano:2016jyj,Bourilkov:2016qum,Accomando:2016tah} and $t\overline{t}$~\cite{Pagani:2016caq} production. At high system invariant mass these could be significant, and even dominant over the standard channels, with a very large PDF uncertainty. From Fig.~\ref{fig:photlumi} the reason for this is clear, being driven by the large PDF uncertainty in the $\gamma\gamma$ luminosity at high mass, and the relatively gentle decrease with mass in the central value. However, from the discussion above we know that using the NNPDF set will dramatically overestimate the uncertainty on the photon--initiated contribution. In Fig.~\ref{fig:lepphot} we show the lepton pair production cross section at high mass, at the $\sqrt{s}=13$ TeV LHC and a $\sqrt{s}=100$ TeV FCC. We can see that indeed at the LHC, the NNPDF prediction for the photon--initiated contribution can even be larger than the standard Drell--Yan contribution. However, the up--to--date LUXqed prediction exhibits no such behaviour. The prediction is under good theoretical control, and gives a small, though not negligible, contribution.

Thus, by considering the physics that generates the photon PDF, and recognising the dominance of the elastic emission process, we already achieve a significant reduction in PDF uncertainty in comparison to the model--independent approach, even when accounting for the most sensitive
data available in the latter case.
Moreover, the additional information provided by Eq.~(\ref{eq:luxqed}) in combination with the high precision data on the inelastic (and elastic) proton structure functions give extremely tight constraints on the photon PDF, resulting in a $\sim 1\%$ level PDF uncertainty. It is worth emphasising that while consistency tests are of course to be encouraged, this is not the result of a particular theoretical model, to be treated on the same footing as older PDF sets. The LUXqed set is a fundamentally experimental determination of the photon PDF; it is simply that by doing this directly in terms of the measured structure functions the tightest constraints can be achieved. Such information must be included in any future photon PDF set, and we have therefore moved beyond the era of large photon PDF uncertainties.
Indeed, the photon PDF, which used to be the poorest known of all the proton PDFs, now has the smallest uncertainty.

% Subsection on electroweak corrections
\subsection{Electroweak corrections}\label{sec:QED.photon.EW}

In addition to the QED photon--initiated corrections discussed above,
it can also be important to include other EW contributions, in particular those
arising from virtual EW bosons, in a PDF fit.
These corrections are most important at larger invariant masses
of the produced system, $Q\gg M_W$,
where virtual EW contributions receive logarithmic enhancements,
see Ref.~\cite{Mishra:2013una} for a review.
In particular, the virtual exchange of soft or collinear weak
bosons leads to Sudakov logarithms of the form $\alpha_W \ln^2 Q^2/M_W^2$,
where $\alpha_W=\alpha/\sin^2\theta_W$, which can lead to
large (negative) corrections for large values of $Q$.
Given that many of the LHC datasets that enter into the global
PDF are sensitive to the TeV region, from high--mass Drell--Yan
production
to the large $p_T$ tail of $Z$ production and inclusive jets
and dijets, the inclusion of such EW corrections is in general
required to achieve the best possible description of experimental
data in this region.

The state--of--the--art for EW corrections is NLO, that is
$\mathcal{O}\lp \alpha_W\rp $ relative to the Born--level process, including in addition in some
cases mixed terms of the form $\mathcal{O}\lp \alpha_W\alpha_s\rp$
and related terms.
These corrections are available for most of the hadron
collider processes
that enter a typical global fit, including inclusive jet and dijet
production~\cite{Dittmaier:2012kx},
inclusive 
electroweak gauge boson production at high
$p_T$~\cite{Becher:2013zua,Campbell:2016lzl}
and high invariant mass $M_{ll(\nu)}$~\cite{Li:2012wna,Gavin:2012sy}
and differential top quark pair
production~\cite{Pagani:2016caq,Czakon:2017wor}.
Most of these calculations are implemented
in publicly available programs.
For instance, EW corrections to inclusive gauge boson
production  are available
in programs such as {\tt FEWZ}~\cite{Li:2012wna,Gavin:2012sy} and
{\tt HORACE}~\cite{CarloniCalame:2007cd}.
The latest version of {\tt MCFM}~\cite{Campbell:2016dks}
also includes the calculation of weak corrections to Drell--Yan, top quark pair,
           and dijet production  at hadron colliders.
Recently, there has also been progress in the
automation of the calculation of these corrections,
both in the framework of {\tt MadGraph5\_aMC@NLO}~\cite{Frixione:2015zaa}
and of {\tt Sherpa}/{\tt OpenLoops}~\cite{Kallweit:2014xda}.

\begin{figure}[t]
  \begin{center}
 \includegraphics[width=0.99\textwidth]{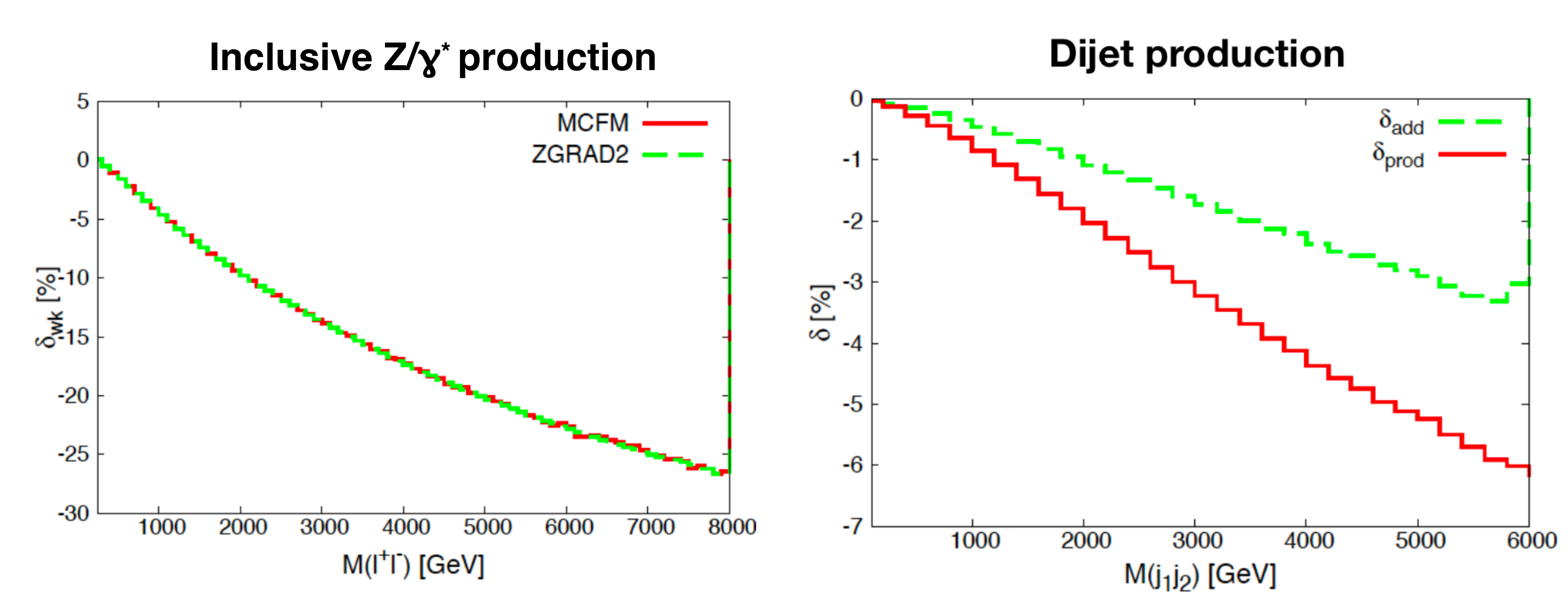}
  \end{center}
  \vspace{-0.6cm}
  \caption{\small Two representative examples of NLO EW corrections
    for processes relevant for PDF determinations, computed with {\tt MCFM}
    at $\sqrt{s}=13$ TeV~\cite{Campbell:2016dks}.
    In the left plot, we show the percentage NLO EW correction for high--mass dilepton
    production as a function of $M_{ll}$, comparing also with the
    corresponding {\tt ZGRAD} calculation.
    In the right plot, we show the same quantity, now for dijet production
    as a function of the invariant mass of the dijet $M_{jj}$.
    The two curves correspond to two possible ways to combine NLO
    QCD and EW corrections, known as additive ($\delta_{\rm add}$)
    and multiplicative ($\delta_{\rm prod}$).
    \label{fig:Sect7-EWcorr}}
\end{figure}

In Fig.~\ref{fig:Sect7-EWcorr} we show two
representative examples of NLO EW corrections
    for processes relevant for PDF determinations, computed with {\tt MCFM}
    at $\sqrt{s}=13$ TeV~\cite{Campbell:2016dks}.
    In the left plot, we show the percentage NLO EW correction for
    high--mass dilepton
    production as a function of $M_{ll}$. The {\tt ZGRAD} calculation~\cite{Baur:2001ze} is also shown.
    We see that these corrections are negligible for $M_{ll}\lsim 500$
    GeV, but that they can become significant as we increase $M_{ll}$,
    reaching $\delta_{\rm wk}\sim -20\%$ at 5 TeV.
    In the right plot, we show the same quantity, now for dijet production
    as a function of the invariant mass of the dijet, $M_{jj}$.
    The two curves correspond to two possible ways to combine NLO
    QCD and EW corrections, known as additive ($\delta_{\rm add}$)
    and multiplicative ($\delta_{\rm prod}$).
    Here the corrections are more moderate (since the Born
    is a pure QCD process) but they can still reach a few
    percent in the region accessible at the LHC.
    The results of Fig.~\ref{fig:Sect7-EWcorr} illustrate
    how a careful inclusion of NLO EW corrections is important
    for the description of the LHC data in the TeV region
    used for PDF determinations.

%%%%%%%%%%%%%%%%%%%%%%%%%%%%%%%%%%%%%%%%%%%%%%%%%%%%%%%%%%%%%%
%%%%%%%%%%%%%%%%%%%%%%%%%%%%%%%%%%%%%%%%%%%%%%%%%%%%%%%%%%%%%%%%%%

%%%%%%%%%%
\vspace{0.6cm}
\section{Implications for LHC phenomenology}\label{sec:LHCpheno}
In this section we present an overview of some of the
most representative implications of PDFs and their
uncertainties for LHC phenomenology.
First we discuss the role of PDFs for the
predictions of the SM Higgs boson production cross-sections
at the LHC.
Then we assess the role of PDF uncertainties
in searches for new heavy resonances predicted by various
BSM scenarios.
Finally, we highlight the importance
of PDFs for the precision measurements of SM parameters such
as the $W$ mass and the strong coupling constant.

\subsection{Higgs production cross-sections}\label{sec:LHCpheno.Higgs}

In the Standard Model, once the Higgs mass $m_h$ is measured,
all other parameters of the Higgs sector, such as the strength
of its coupling to fermions and vector bosons and its branching ratios, are uniquely
determined~\cite{deFlorian:2016spz,Dittmaier:2011ti,Dittmaier:2012vm}.
However, deviations of these Higgs couplings with respect
to the SM predictions are generically expected in BSM scenarios~\cite{Dawson:2013bba}.
Therefore, the precision measurements of the Higgs couplings represents a unique opportunity for BSM
searches, since any deviation with respect to the the
tightly fixed SM predictions would represent
a smoking gun for New Physics.
Crucially, realising this program requires not only
high precision experimental measurements of Higgs boson
production and its decay in various channels, but also the calculation
of the SM cross sections and decay rates with matching theoretical precision.
Moreover, PDFs are one of the largest sources
of theoretical uncertainty affecting the predictions for Higgs boson production~\cite{deFlorian:2016spz}.

%%%%%%%%%%%%%%%%%%%%%%%%%%%%%%%%%%%%%%%%%%%%%%%%%%%%%%%%%%%%%%%%%%%%%
\begin{figure}[t]
\begin{center}
  \includegraphics[scale=0.61]{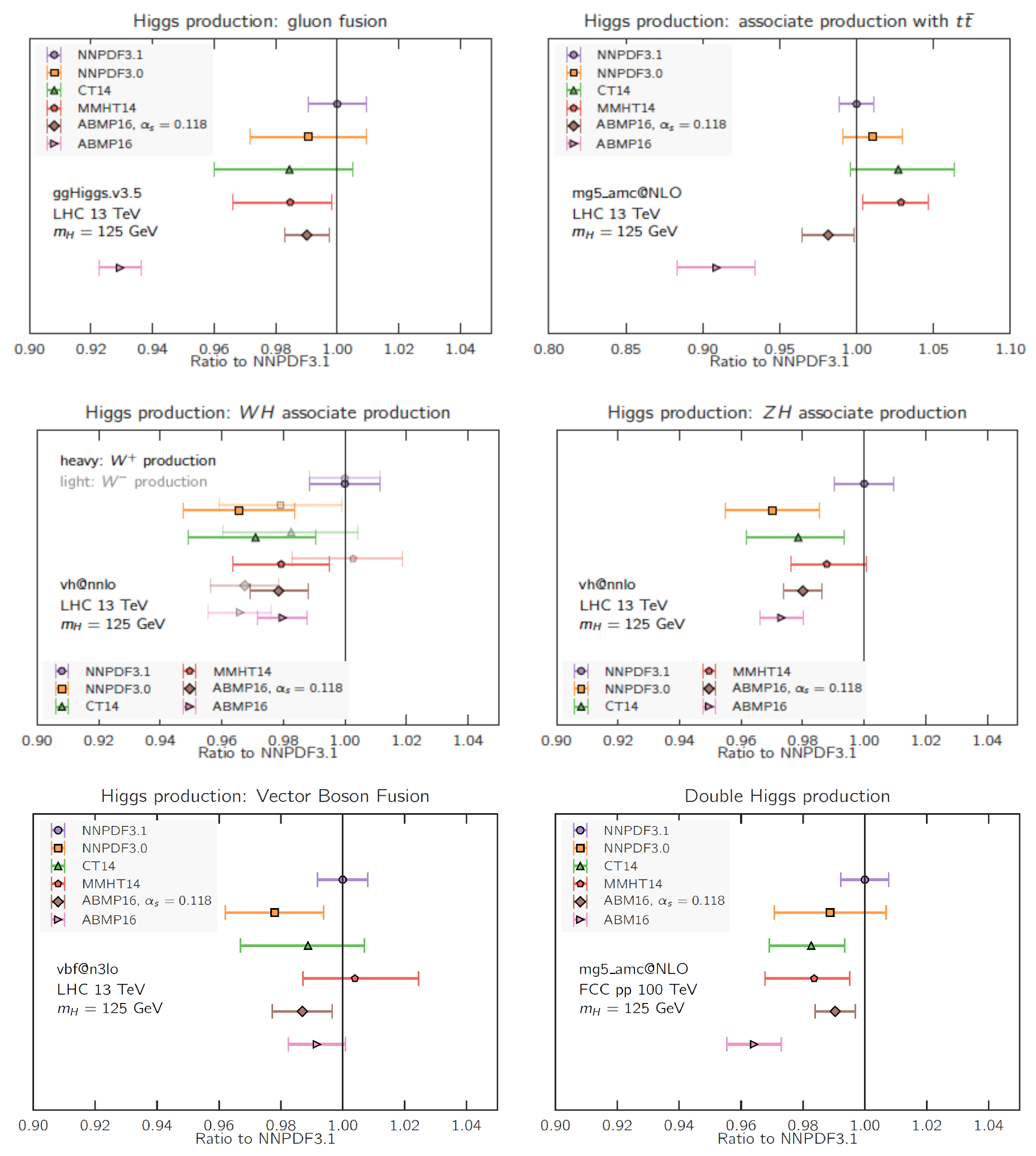}
  \caption{\small The PDF dependence of the
  most important Higgs production inclusive
    cross sections at the LHC 13 TeV.
    The results are normalized to the central value of the NNPDF3.1 prediction,
    and only PDF uncertainties are shown.
    In the case of ABMP16, we show results for the nominal $\alpha_s$ value
    as well as for the common $\alpha_s(m_Z)=0.118$ value.
    See~\cite{Ball:2017nwa} for more details of the theoretical
    calculations.
    \label{fig:higgs}
  }
\end{center}
\end{figure}
%%%%%%%%%%%%%%%%%%%%%%%%%%%%%%%%%%%%%%%%%%%%%%%%%%%%%%%%%%%%%%%%%%%%%

We therefore present a comparison of inclusive
Higgs production cross sections at 13 TeV
with the latest releases of all PDF groups.
The settings of this comparison, and the codes
used for each process, are described in Ref.~\cite{Ball:2017nwa}.
Specifically, 
we show the dominant Higgs boson production modes at hadron colliders:
gluon fusion, associated production with a $t\bar{t}$ pair,
$VH$ associated production, and Higgs production in vector--boson
fusion.
In addition, we also show the results for double Higgs production
in the dominant gluon--fusion channel, which should become accessible
at the High Luminosity LHC (HL--LHC)~\cite{Behr:2015oqq,Papaefstathiou:2012qe,Dolan:2012ac,Bishara:2016kjn,Frederix:2014hta}.
Results are provided for NNPDF3.0 and 3.1, CT14, MMHT14 and for the
ABMP16 NNLO sets for $\alpha_s(m_Z)=0.118$, while in the latter
case we also show the result corresponding to
their best--fit value of $\alpha_s(m_Z)=0.1149$.
The theoretical settings for each cross section calculation
are based on state--of--the--art
matrix element calculations, for instance the gluon--fusion
and VBF results are computed at N3LO using the
{\tt ggHiggs}~\cite{Ball:2013bra}
and {\tt vbf@n3lo}~\cite{Dreyer:2016oyx} codes respectively.
We only show the PDF uncertainties, while
other sources of theoretical
errors
affecting these cross sections
are listed in e.g. the latest Higgs Cross Section Working
Group report~\cite{deFlorian:2016spz}.
We note that the uncertainty associated to the input value
of $\alpha_s(m_Z)$ can be comparable to the
PDF uncertainties in some channels.

There are a number of noteworthy features in the comparison
of  Fig.~\ref{fig:higgs}.
First, it shows that
in general there is
good agreement between the three
global fits, NNPDF3.1, CT14 and MMHT14 for all the
Higgs boson production modes.
The comparison between NNPDF3.1 and its predecessor
NNPDF3.0 highlights good agreement for the gluon
initiated channels, with a reduction of the PDF
uncertainties in the former case, while for quark--initiated
cross sections such as $VH$ and VBF there is an upper shift
by around one sigma.
Another significant feature of this comparison is
that the recent ABMP16 set is also in reasonable agreement
with the other groups, provided that
the same common value of the strong
coupling constant $\alpha_s(m_Z)=0.118$ is used.
On the other hand, if their best--fit value
$\alpha_s(m_Z)=0.1149$ is used in the calculation,
significant differences in the predicted cross sections arise for the gluon--initiated channels.
In particular, the ABMP16 prediction is around 5--7\% (10--12\%) lower
than the other predictions for the gluon--fusion ($t\bar{t}$ associated
production) cross section.

\begin{figure}[t]
\begin{center}
  \includegraphics[scale=0.45]{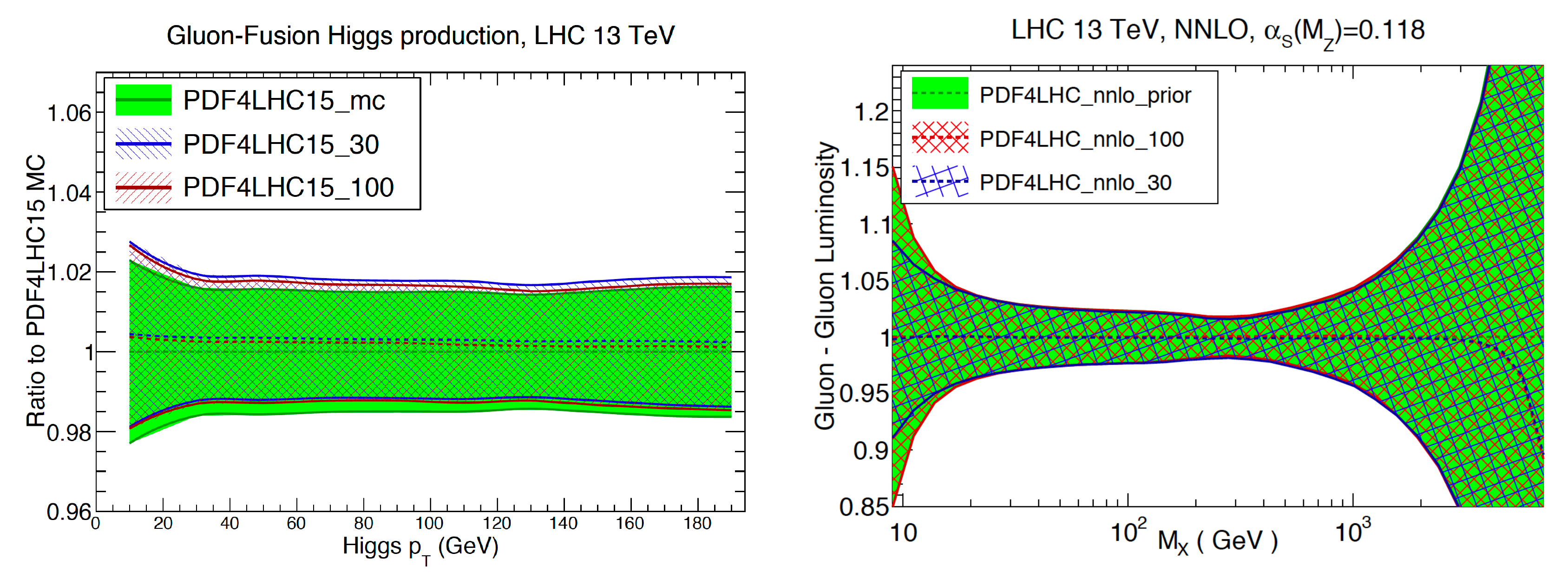}
  \caption{\small Left: the PDF uncertainties in the
  $p_T^h$ distribution of Higgs bosons produced in the gluon--fusion
  mode at the LHC 13 TeV for $0 \le p_T^h \le 200$ GeV,
  computed using the PDF4LHC15 NNLO sets.
  Right: the gluon--gluon PDF luminosity with the same
  set now extending up to higher values of the invariant mass
  of the final state $M_X$.
    \label{fig:higgs_differential}
  }
\end{center}
\end{figure}

It is also worth mentioning here that PDF uncertainties
are relevant not only for the extraction
of Higgs couplings from inclusive cross sections, but
also for the differential measurements~\cite{Aad:2016lvc,Khachatryan:2015yvw}
that will become
available due to the large statistics that will be accumulated
by the end of Run II as well as for the HL--LHC.
To illustrate this point, in
Fig.~\ref{fig:higgs_differential} we show the
PDF uncertainties in the Higgs boson
  $p_T^h$ distribution in the gluon--fusion
  mode at the LHC 13 TeV for $0 \le p_T^h \le 200$ GeV,
  computed using the PDF4LHC15 NNLO sets.
  In this case we find that PDF uncertainties are at around the
  $\sim 2\%$ level.
  However, these uncertainties will increase as
  the LHC becomes sensitive to higher $p_T$ values:
  as shown in the right plot of Fig.~\ref{fig:higgs_differential},
  at high invariant masses (high $p_T$ values) the
  PDF uncertainties in the gluon--gluon luminosity become
  rather larger, affecting the extraction of Higgs properties in that region.

%%%%%%%%%%%%%%%%%%%%%%%%%%%%%%%%%%%%%%%%%%%%%%%%%%%%%%%%%%%%%%%%%%%%%
%%%%%%%%%%%%%%%%%%%%%%%%%%%%%%%%%%%%%%%%%%%%%%%%%%%%%%%%%%%%%%%%%%%%%

\subsection{PDF uncertainties and searches for new massive particles}\label{sec:LHCpheno.BSM}

Many BSM physics scenarios predict
the existence of new heavy particles with masses around the
TeV scale.
For example, supersymmetry, composite Higgs,
and extra dimensions, are all classes of models, among many other others, that
motivate the search for new heavy resonances at the LHC
in the high--mass tail of various kinematic distributions.
Here, PDF uncertainties play an important
role in setting robust exclusion limits based on available
null results, and would become even more important
in the case of a discovery.
In particular, PDFs represent the dominant
theoretical uncertainty for the production of new
heavy particles in the TeV region, as such processes
are sensitive to the  large--$x$ behaviour of quarks and gluons.
As discussed in Sect.~\ref{sec:structure}, PDF
uncertainties are large in this region due to the
limited  experimental constraints.

\begin{figure}[t]
\begin{center}
  \includegraphics[scale=0.40]{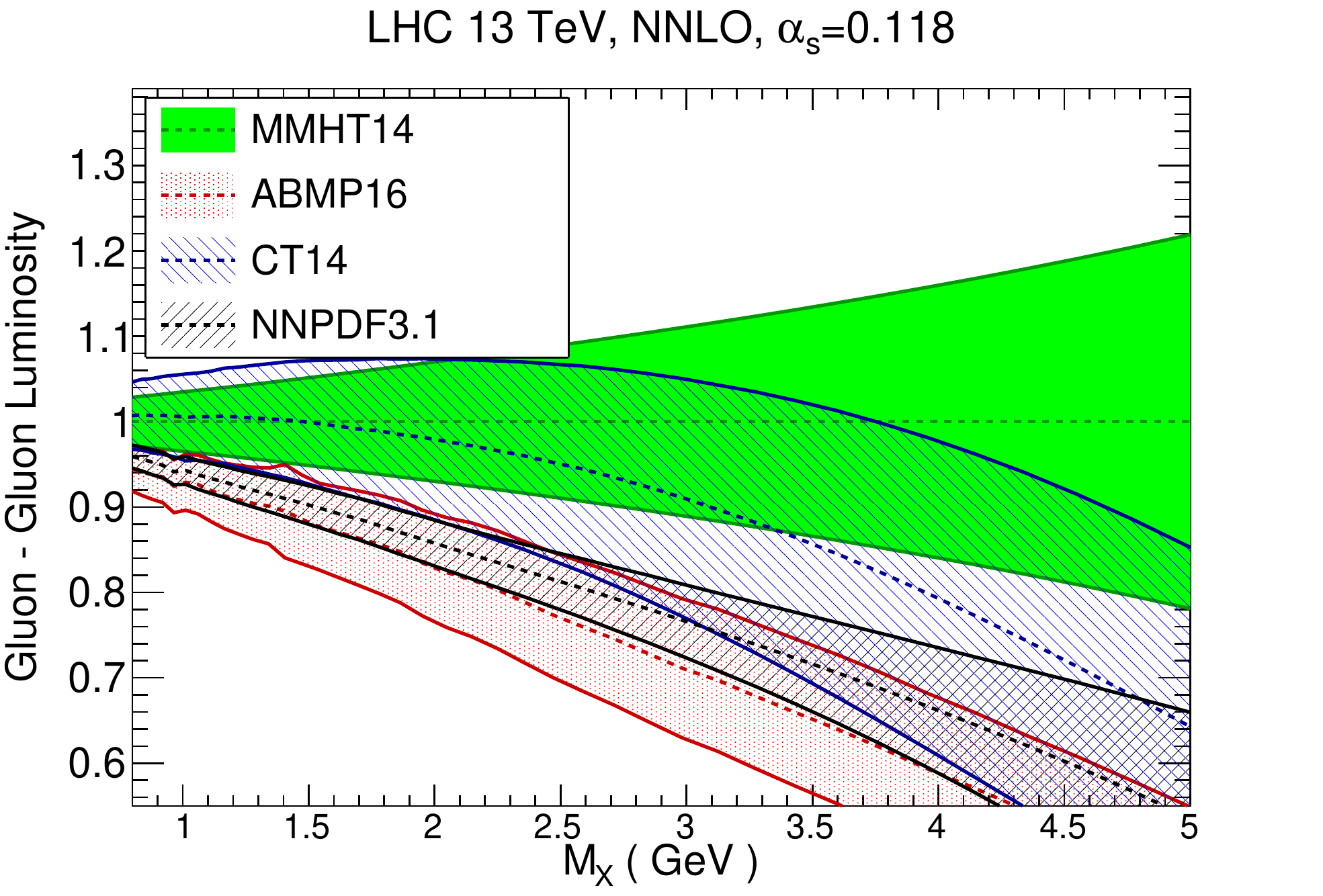}
  \includegraphics[scale=0.40]{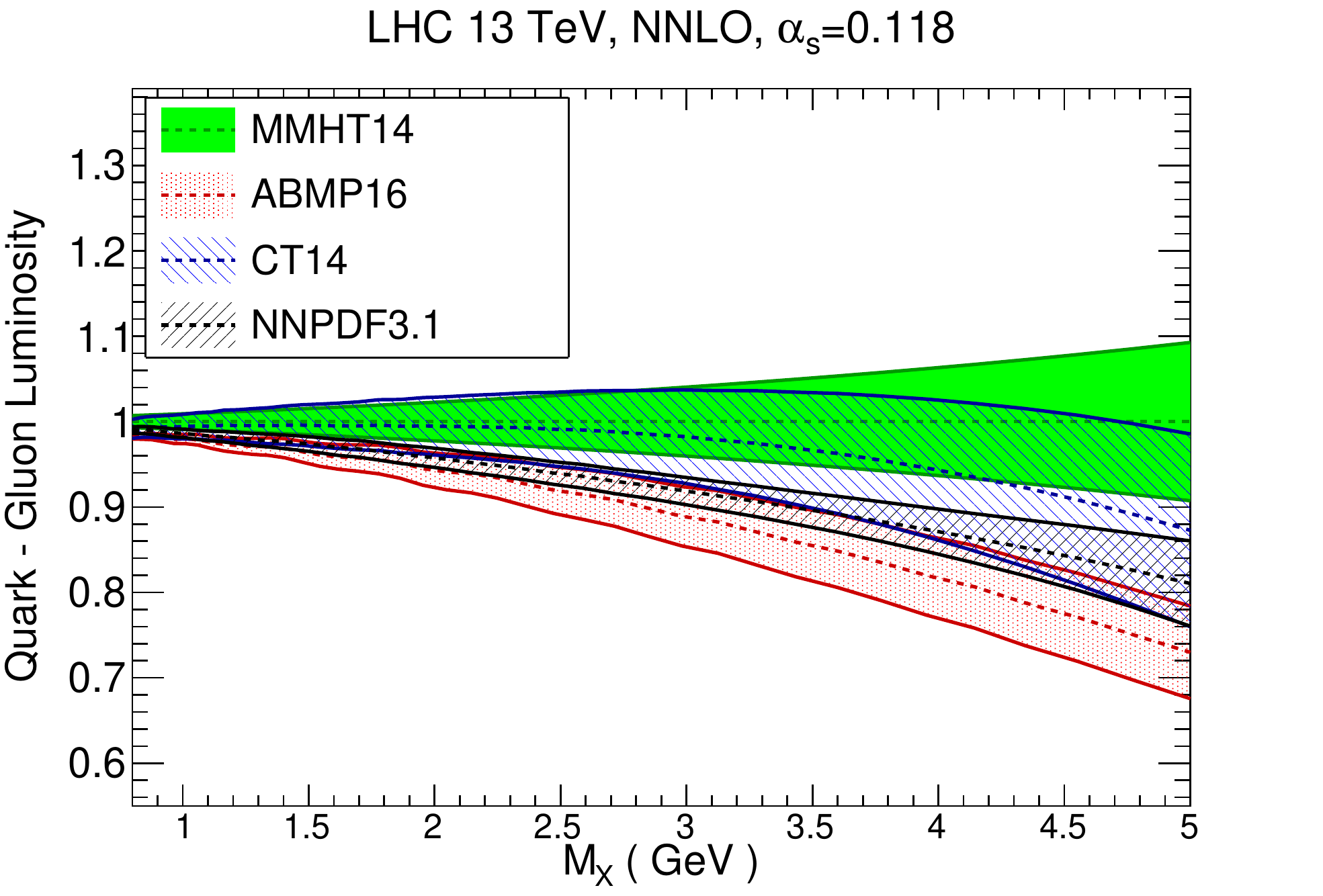}
  \includegraphics[scale=0.40]{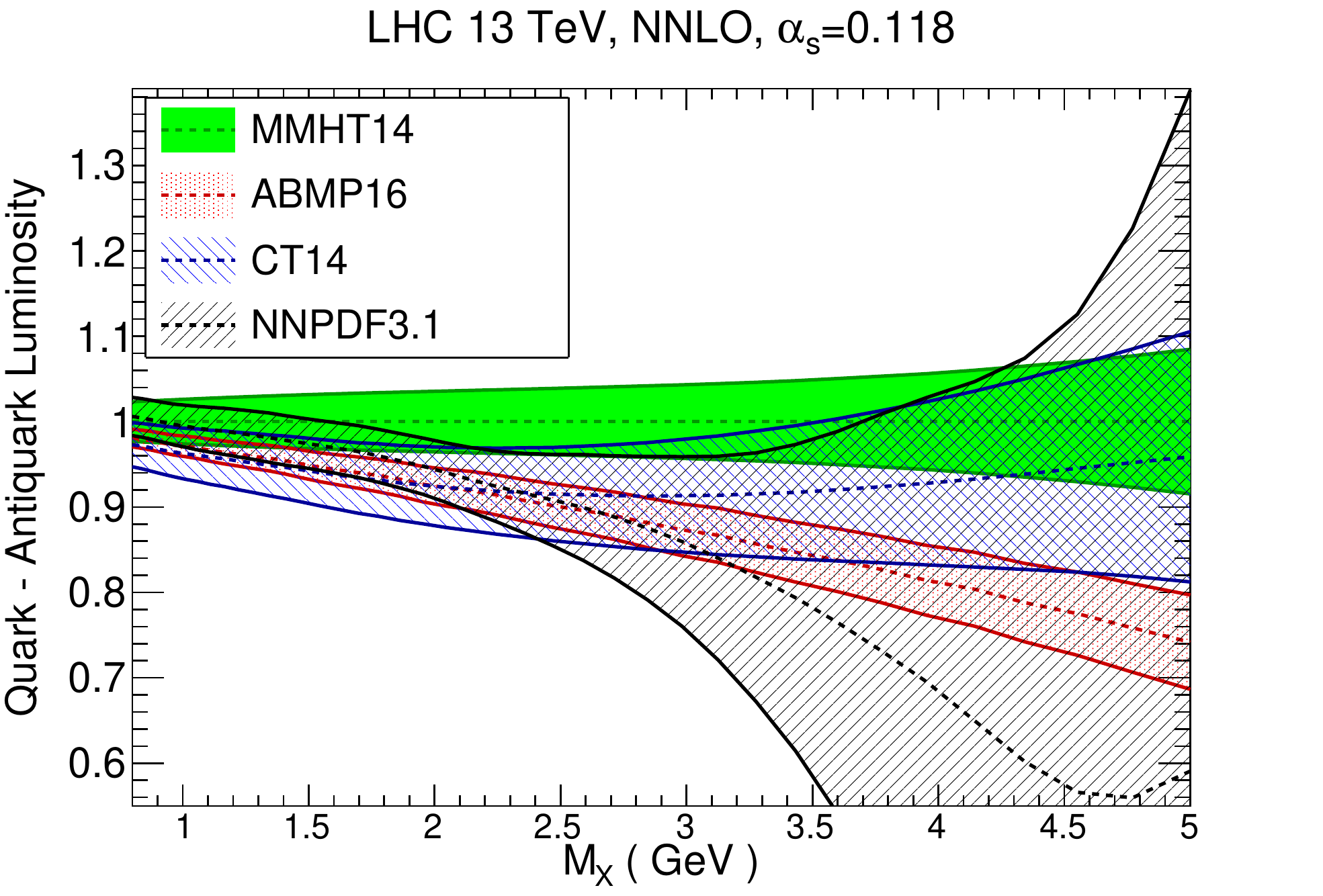}
  \includegraphics[scale=0.40]{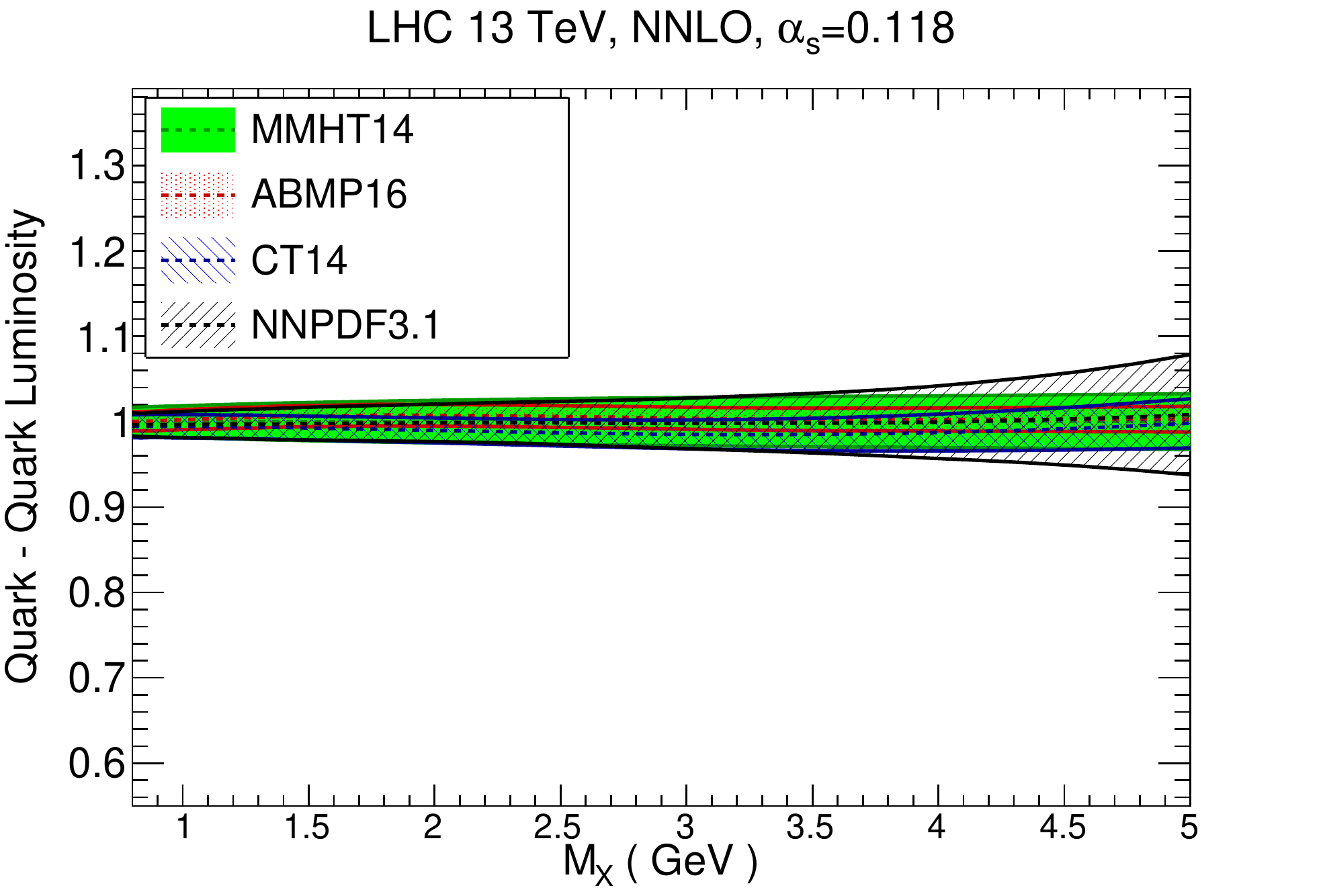}
  \caption{\small Comparison of PDF luminosities in the large
    invariant mass $M_X$ region between MMHT14, ABMP16, CT14 and
    NNPDF3.1.
    From left to right and from top to bottom we show the results
    of the gluon--gluon, gluon--quark, quark--anti--quark
    and quark--quark luminosities, normalized
    to the central value of MMHT14.
    In this comparison, NNLO PDFs with $\alpha_s(m_Z)=0.118$
    sets are used.
    \label{fig:BSMlumi}
  }
\end{center}
\end{figure}

In order to quantify the size
of the PDF uncertainties in the large invariant mass
region,
as well as the relative agreement between the PDF groups, it is useful
to compare the PDF luminosities for $M_X\ge 1$ TeV.
We will restrict ourselves to ABMP16, CT14, MMHT14
and NNPDF3.1, in all cases using $\alpha_s(m_Z)=0.118$.
Results are shown in Fig.~\ref{fig:BSMlumi} for
$\sqrt{s}=13$ TeV normalized to the central value
of the MMHT14 calculation.
From the comparison in Fig.~\ref{fig:BSMlumi}, we find that
PDF uncertainties are small, at the few--percent level, up to
$M_X\simeq 5$
TeV for the quark--quark luminosities.
This is due to the fact that $\mathcal{L}_{qq}$ is dominated
by the rather accurately known up and down quark valence PDFs,
which are constrained by measurements of e.g. fixed--target
DIS structure functions.

%%%%%%%%%%%%%%%%%%%%%%%%%%%%%%%%%%%%%%%%%%%%%%%%%%%%%%%%%%%%%%%%%%%%%
%%%%%%%%%%%%%%%%%%%%%%%%%%%%%%%%%%%%%%%%%%%%%%%%%%%%%%%%%%%%%%%%%%%%%

For the gluon--gluon luminosity, $\mathcal{L}_{gg}$, we find a rather large spread
in the predictions between the different groups, with MMHT14 (ABMP16)
leading to the largest (smallest) central values.
For instance, at $M_X\sim 5$ TeV, which is close to the upper
limit of the kinematic coverage of the LHC, the envelope of the PDF uncertainty bands
spans $\sim 100\%$.
Even for more moderate invariant masses the spread is quite large, with the values of $\mathcal{L}_{gg}$
at $M_X\sim 2.5$ TeV varying between $\sim$ $+10\%$ and $-30\%$ in comparison to the central MMHT14 result.
It is thus clear that these uncertainties would represent
one of the limiting factors for BSM characterization
in the case of a discovery.

For the quark--gluon luminosity, $\mathcal{L}_{qg}$, we
observe a similar trend to the gluon--gluon case, but
with reduced PDF uncertainties due to the contribution
of the well--constrained large--$x$ quark PDFs.
Interestingly, in the above two cases the results from all sets do not necessarily agree within the quoted 68\% C.L. uncertainty bands.
As discussed further in Sect.~\ref{sec:structure.gluon}, it will be informative to see how this comparison changes when data sensitive to the high $x$ gluon, such as the $t\overline{t}$ differential distributions, currently only in the NNPDF3.1 set, are included in all the fits shown in this comparison.

PDF uncertainties, as well as the differences between groups,
are also large for the quark--anti--quark PDF luminosity
$\mathcal{L}_{q\bar{q}}$, also shown in Fig.~\ref{fig:BSMlumi}.
The reason for this behaviour is two--fold.
First, the large--$x$ anti--quarks
are notoriously difficult to pin down, although recent
high--precision measurements from the Tevatron
and the LHC are improving the situation.
Second, various groups parameterize the quark
sea content of the proton with rather different
assumptions~\cite{Ball:2016spl} (see Sect.~\ref{sec:pdfgroups}),
and this can have important
implications for the quark--anti--quark luminosities.
We find that the spread of the different results
ranges between $+5\%$ and $-30\%$ for $M_X=3$ TeV,
with PDF uncertainties becoming $\mathcal{O}(100\%)$
for higher invariant masses.
Note here that the PDF uncertainties are the largest for
NNPDF3.1, despite this being the global analysis which
includes the largest amount of LHC electroweak data sensitive
to anti--quarks.
This highlights the fact that
methodological differences in the flavour assumptions
and parametrization of anti--quarks are one of the dominant
factors in explaining the difference between the various
groups for $\mathcal{L}_{q\bar{q}}$ at large invariant masses $M_X$.

\begin{figure}[t]
\begin{center}
  \includegraphics[scale=0.45]{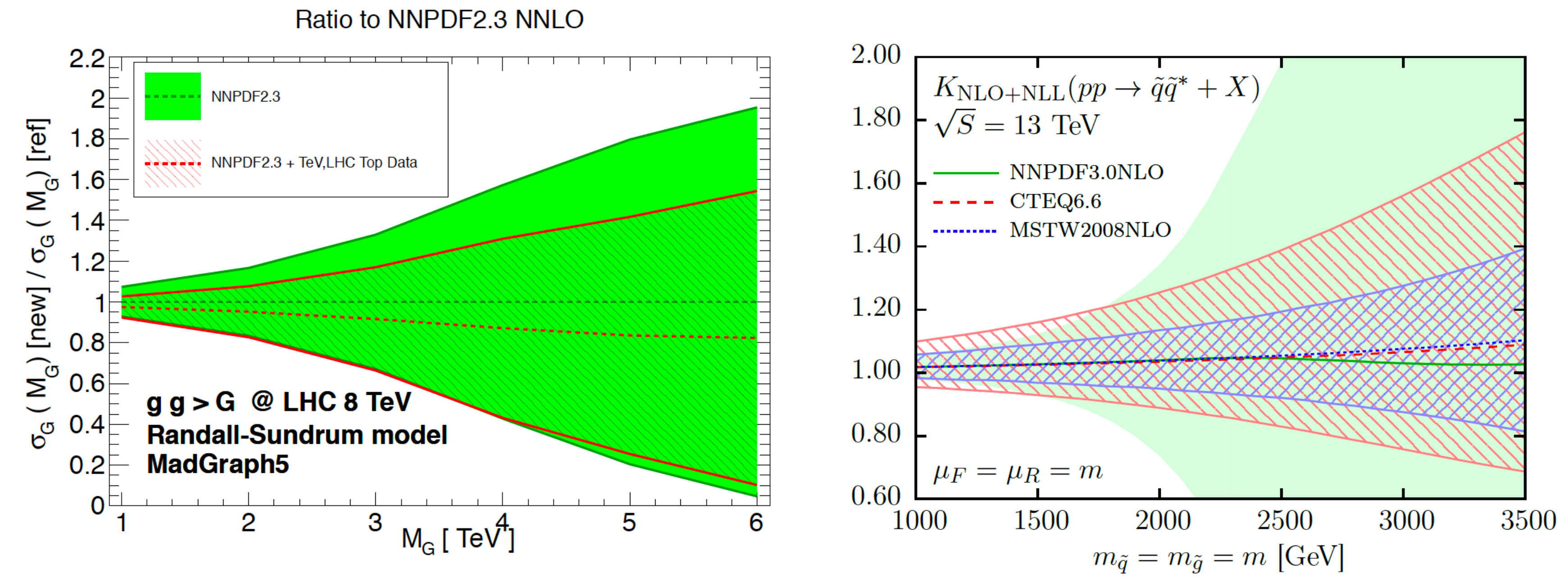}
  \caption{\small Left: the PDF uncertainties for high-mass
  graviton production in the Randall-Sundrum scenario
  induced by gluon fusion at the LHC 8 TeV, computed at LO
  with {\tt MadGraph5}.
  We compare the results of the NNPDF2.3 fit with that
  of the same fit including the constraints
  from top quark production cross sections~\cite{Czakon:2013tha}.
  Right: the $K$-factor for the NLO+NLL cross section,
  including PDF uncertainties, for the production
  of a squark-anti-squark pair $\tilde{q}\tilde{q}^*$
  at the 13 TeV LHC with various PDF sets, from~\cite{Beenakker:2015rna}.
    \label{fig:BSMhighmass}
  }
\end{center}
\end{figure}

In order to illustrate the phenomenological consequences of
these large PDF uncertainties at high $M_X$, in Fig.~\ref{fig:BSMhighmass}
we show the  PDF uncertainties for high--mass
  graviton production in the Randall--Sundrum
  scenario~\cite{Randall:1999vf,Randall:1999ee},
  induced by gluon fusion at the LHC 8 TeV, computed at LO
  with {\tt MadGraph5}~\cite{Alwall:2007st}.
   We compare the results of the NNPDF2.3 fit with those
  of the same fit including the constraints
  from top quark production cross sections~\cite{Czakon:2013tha}.
  We observe that PDF uncertainties become $\mathcal{O}(100\%)$
  at large values of the graviton mass, consistent
  with the estimates from the gluon-gluon luminosity
  shown  in Fig.~\ref{fig:BSMlumi}.
  We also see how these PDF uncertainties can be reduced
  by the inclusion of top quark pair production
  total cross sections,
  highlighting the cross-talk between precision SM measurements
  and improving BSM searches.
  Further constraints on the large-$x$ gluon can be obtained
  {\it e.g.} fitting top quark differential distributions,
  see Sect.~\ref{sec:datatheory.top}.

%%%%%%%%%%%%%%%%%%%%%%%%%%%%%%%%%%%%%%%%%%%%%%%%%%%%%%%%%%%%%%%%%%%%%
%%%%%%%%%%%%%%%%%%%%%%%%%%%%%%%%%%%%%%%%%%%%%%%%%%%%%%%%%%%%%%%%%%%%%

In Fig.~\ref{fig:BSMhighmass}
we also show the $K$-factor for the NLO+NLL cross sections
for the production
  of a squark-anti-squark pair $\tilde{q}\tilde{q}^*$
  at the LHC 13 TeV.
The calculation has been performed using  various PDF sets,
  including PDF uncertainties, with each set normalized to the corresponding NLO
  result~\cite{Beenakker:2015rna}.
  This production channel is dominated by the quark--anti--quark
  luminosity.
  We compare the predictions of NNPDF3.0,
  CTEQ6.6, and MSTW2008, all at NLO.
   Here we also see that PDF uncertainties become very large
  at high mass, in particular in the case of NNPDF3.0,
  reflecting the underlying behaviour of the quark-anti--quark
  luminosities.
 
  Note that by construction the central values of
  the three predictions shown in Fig.~\ref{fig:BSMhighmass}
  are rather close, since differences mostly
  cancel in this $K$-factor ratio.
   The usefulness
  of this comparison is thus related to estimating the PDF uncertainties
  in each case.
  The corresponding differences between central values
  can be read from the quark-antiquark
  PDF luminosity of Fig.~\ref{fig:BSMlumi}, which
  is the relevant initial-state for the  SUSY cross-sections
  shown in Fig.~\ref{fig:BSMhighmass}.
  From that comparison we see that differences in the
  central values of up to 40\% are present at $M_X\simeq 4$ TeV,
  well within the LHC coverage, highlighting our poor
  understanding of the proton structure in the large-$x$ region.

  The results shown in Fig.~\ref{fig:BSMhighmass}
  (together with those of Fig.~\ref{fig:BSMlumi}) highlight that,
  in the case of an eventual discovery of novel high-mass
  particles at the LHC, it will be crucial to improve
  our knowledge of the large-$x$ PDFs in order to be able
  to characterize the underlying BSM scenario.
To achieve this it will be essential to exploit as much as possible high-precision collider data, mostly from the LHC, in fits,
in order to pin down the large-$x$ gluons and
anti-quarks and thus reduce the PDF uncertainties
associated to high-mass BSM particle production.
For example, in Ref.~\cite{Czakon:2016olj} it was shown that
by including the $y_t$ and $y_{t\bar{t}}$ differential
distributions for top quark pair production
in a global PDF fit, it is possible to reduce the PDF
uncertainties that affect the high-mass tail of the $m_{t\bar{t}}$
distribution by up to a factor of two.
This distribution is widely use for searches, for instance
of new resonances that couple strongly to the top quark~\cite{Aaboud:2017hnm}.
In the future, it might also be feasible
to provide indirect constraints on BSM models,
for instance on the
coefficients of the SM-EFT~\cite{Brivio:2017vri} higher--dimension operators,
by including these in the global PDF fit, along the lines
of early studies aiming to constrain colored sparticles
from Tevatron jet production~\cite{Berger:2010rj}.

\subsection{Precision measurements of SM parameters}\label{sec:LHCpheno.SM}

The precision measurement of SM parameters, such as the
mass of the $W$ boson $M_W$ or the running of the strong coupling
constant $\alpha_S(Q)$, represent powerful ways of constraining
BSM dynamics at the LHC.
For instance, following the discovery of the Higgs boson,
in the absence of new physics the SM is an over--constrained
theory: one can use a set of input parameters, such as the
Higgs mass $m_h$ and the top quark mass $m_t$, in the
context of the global electroweak precision fit~\cite{Baak:2014ora},
to predict other parameters, such as $M_W$.
By comparing these indirect predictions of the $W$ mass with
direct experimental measurements, one can provide a stress test
of the SM, where any tension might indicate hints for BSM dynamics
at scales higher than those that are currently directly accessible.
Tests of this type have already resulted in the famous $g_{\mu}-2$ anomaly,
where a persistent $3$ to $4$-sigma discrepancy is found
between the theoretical predictions of the muon anomalous magnetic
moment~\cite{Benayoun:2015gxa}
and the corresponding experimental measurements.

\begin{figure}[t]
\begin{center}
  \includegraphics[scale=0.48]{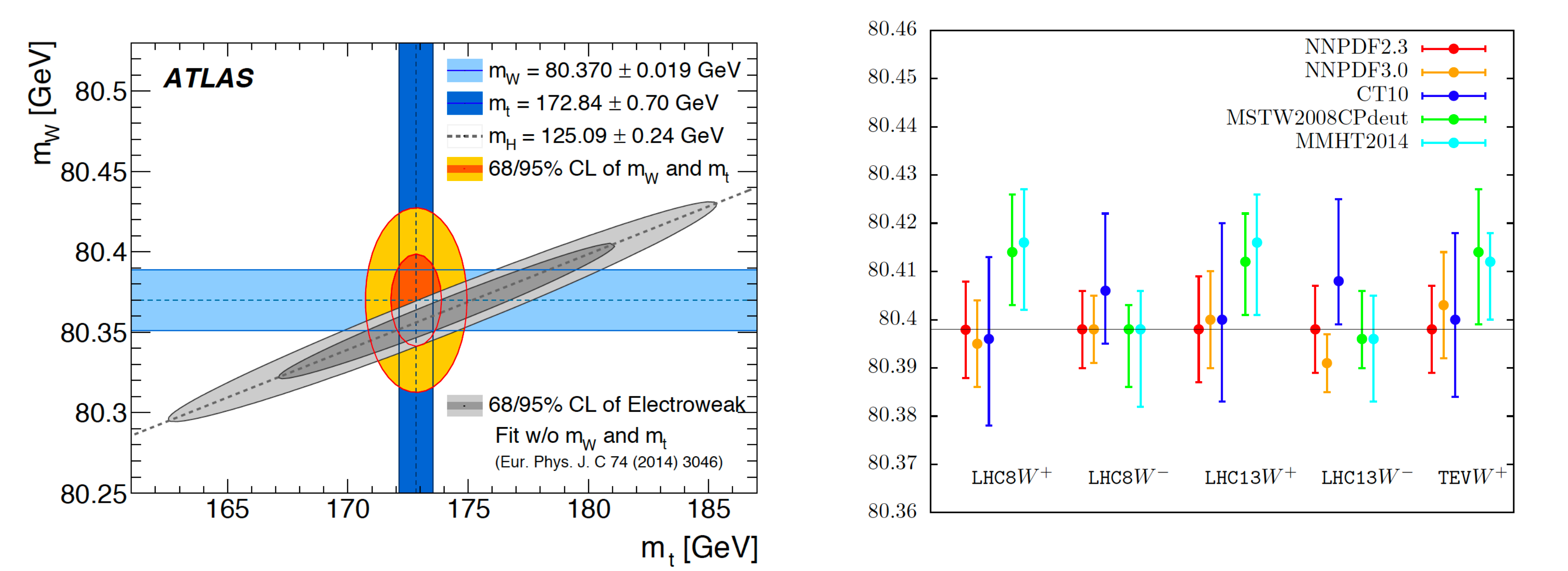}
  \caption{\small Left: comparison between the direct
  measurements of $m_W$, $m_t$, and $m_h$ from ATLAS
  with the predictions from the global electroweak fit,
  from~\cite{Aaboud:2017svj}.
  Right: estimate of the PDF uncertainties in the $m_W$
  determination using different PDF sets and collider scenarios,
  from~\cite{Bozzi:2015hha}. 
  This estimate has been obtained from template fits to the
  $p_T^l$ distribution, imposing the constraint that $p_T^W\le 15$ GeV.
    \label{fig:mw}
  }
\end{center}
\end{figure}

In order to make these comparisons between indirect predictions
and direct measurements as stringent as possible, it is
important to improve the precision of the latter.
For many SM parameters, PDF uncertainties represent one of the
dominant theoretical
uncertainties in their determination, providing
another motivation for the need of improved PDFs.
Focusing on the case of the $W$ mass measurements, the role
of PDF uncertainties has been quantified in detail
in a number of studies, both from the
phenomenological~\cite{Bozzi:2015hha,Bozzi:2015zja,Bozzi:2011ww}
and from the experimental point of view.
In the latter case, the first direct measurement of $M_W$
at the LHC has recently been presented by the
ATLAS collaboration~\cite{Aaboud:2017svj}, yielding
a total uncertainty of only 19 MeV, of which around half of
it (9 MeV) is estimated to come from PDF uncertainties.
In Fig.~\ref{fig:mw} we show a comparison
 between the direct
  measurements of $m_W$, $m_t$, and $m_h$ from ATLAS
  with the predictions from the global electroweak fit,
  from~\cite{Aaboud:2017svj}.
  We observe that there is good agreement between the direct measurements
and the indirect predictions, providing a highly non--trivial
validation test of the SM.
Future measurements of $m_W$ and $m_t$, as well as their
combination with other experiments, should be able
to reduce the uncertainties in this comparison.

%%%%%%%%%%%%%%%%%%%%%%%%%%%%%%%%%%%%%%%%%%%%%%%%%%%%%%%%%%%%%%%%%%%%%
%%%%%%%%%%%%%%%%%%%%%%%%%%%%%%%%%%%%%%%%%%%%%%%%%%%%%%%%%%%%%%%%%%%%%

In Fig.~\ref{fig:mw} we also show a phenomenological estimate
of the PDF uncertainties associated with the $m_W$
  measurements using different PDF sets and collider scenarios,
  from~\cite{Bozzi:2015hha}. 
  This estimate has been obtained from template fits to the
  $p_T^l$ distribution, imposing the additional
  constraint that $p_T^W\le 15$ GeV.
  A number of NNLO PDF sets are used in this comparison,
  in order to achieve a robust estimate of the
  PDF uncertainties.
  In general one finds that there is good agreement within
  PDF uncertainties, although in some cases this
  agreement is only marginal, as in the case of NNPDF3.0
  and CT10 at the LHC 13 TeV for the $W^-$ case.
  From this study, one estimates that at the LHC 7 TeV
  PDF uncertainties using state--of--the--art
  sets are around 6 MeV, a number close to the one estimated
  in the ATLAS measurement~\cite{Aaboud:2017svj}.

Another SM parameter that could potentially provide
indirect information
on BSM dynamics is the QCD coupling $\alpha_s(Q)$, and
specifically its running at the TeV scale.
It is well known that the presence of new colored heavy degrees of freedom
will modify the QCD $\beta$--function and therefore lead to a different
running with $Q$ compared to the corresponding SM prediction.
This fact forms for example the basis of the improved
agreement at high scales between the strong, weak and electromagnetic
couplings in the case of low--scale supersymmetry,
which suggest the unification of the three forces
at a GUT scale of around $\Lambda_{\rm GUT}\sim 10^{16}$
GeV~\cite{Dimopoulos:1981yj}.

\begin{figure}[t]
\begin{center}
  \includegraphics[scale=0.48]{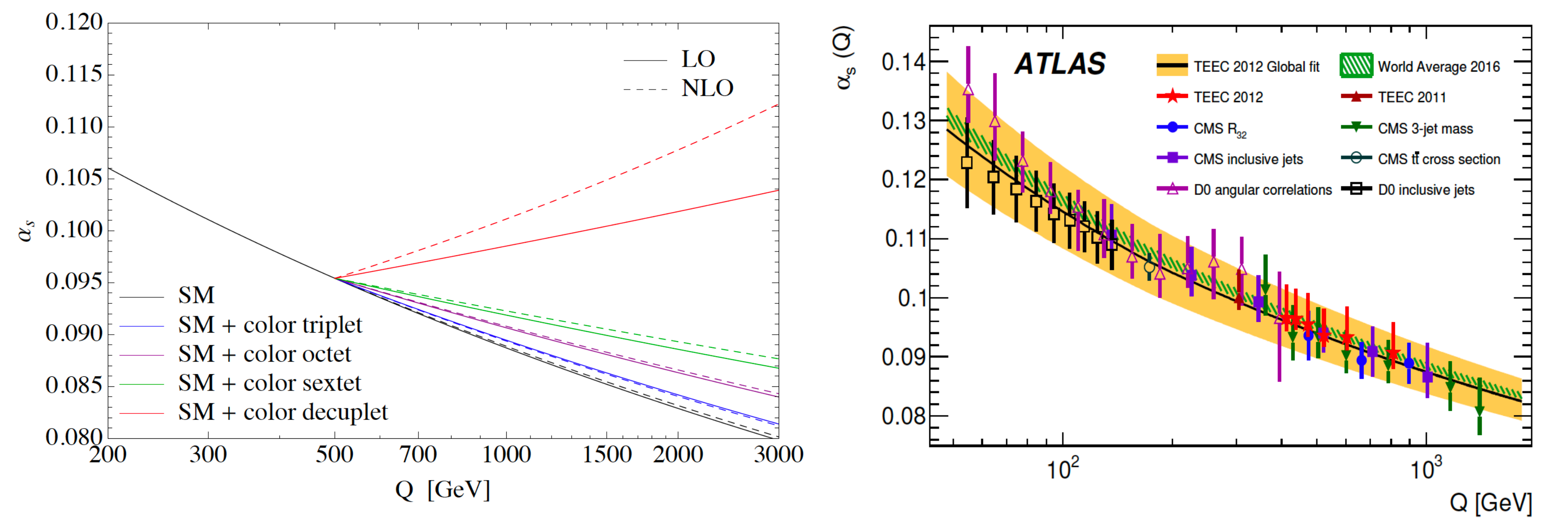}
  \caption{\small Left: the modifications
  in the running of $\alpha_s(Q)$ induced
  by a new heavy colored fermion of mass $m=0.5$ TeV as compared to the SM
  expectation, for various representations of its color gauge group,
  from~\cite{Becciolini:2014lya}.
  Right: comparison of recent direct determinations of $\alpha_s(Q)$ at the Tevatron
  and the LHC as a function of $Q$, together with the PDG 2016 world
  average and with the results of the
  ATLAS TEEC 2012 extraction~\cite{Aaboud:2017fml}.
    \label{fig:alphasrunning}
  }
\end{center}
\end{figure}

If  new BSM heavy particles are present at the TeV scale,
the difference induced in the QCD coupling running
could be accessible at the LHC, see e.g.~\cite{Becciolini:2014lya}
and Fig.~\ref{fig:alphasrunning}, where we show the
modifications in the running of $\alpha_s(Q)$ induced
  by a new heavy colored fermion of mass $m=0.5$ TeV for
  various representations of its color gauge group.
With this motivation,
as well as to compare with other measurements
of $\alpha_s$ at lower energies, 
the ATLAS and CMS
collaborations have presented a number of measurements
of both $\alpha_s(m_Z)$ and of $\alpha_s(Q)$ for restricted
$Q$ ranges, mostly from jet
production~\cite{Khachatryan:2014waa,Aaboud:2017fml,
Chatrchyan:2013txa} but also
from top quark pair production~\cite{Chatrchyan:2013haa}
(see also~\cite{Rojo:2014kta} for a review, and Fig.~\ref{fig:alphasrunning}
for a graphical overview).

%%%%%%%%%%%%%%%%%%%%%%%%%%%%%%%%%%%%%%%%%%%%%%%%%%%%%%%%%%%%%%%%%%%%%
%%%%%%%%%%%%%%%%%%%%%%%%%%%%%%%%%%%%%%%%%%%%%%%%%%%%%%%%%%%%%%%%%%%%%

In these collider--based
determinations of the strong coupling, PDF uncertainties, which are
significant
at the TeV scale (see Sect.~\ref{sec:LHCpheno.BSM}),
represent an important source of theoretical uncertainties.
For instance, in the recent ATLAS determination of $\alpha_s(m_Z)$ from
transverse energy--energy correlations (TEEC) at 8 TeV~\cite{Aaboud:2017fml}, the PDF uncertainty is
$\delta_{\rm pdf}=0.0018$, almost a factor 2 larger than the experimental
uncertainty of $\delta_{\rm exp}=0.0011$.
While in this analysis PDF uncertainties are sub--dominant with respect
to the scale uncertainties, $\delta_{\rm scale}\simeq 0.006$, the latter were
computed using NLO theory and can be reduced significantly by exploiting
the NNLO calculation.
Likewise, in the CMS measurement at 7 TeV based on a QCD
analysis of the inclusive jet cross sections~\cite{Khachatryan:2014waa}, one finds
that the PDF uncertainties $\delta_{\rm pdf}=0.0028$ are larger than
the experimental uncertainties $\delta_{\rm exp}=0.0019$, though still
sub--dominant with respect to the large scale variations of the
NLO calculation $\delta_{\rm scale}=^{+0053}_{-0.0024}$.
In both cases, it is clear that if one is able to reduce PDF uncertainties,
and exploit the reduction of scale errors arising from the NNLO calculation,
it should be possible to achieve a rather
competitive determination of $\alpha_s(m_Z)$ and 
also perform stringent tests of its running in the TeV region.

As a related topic, we would like to mention
that there are also proposals to measure the running
of the electroweak running couplings at the LHC~\cite{Alves:2014cda}
and use these to impose model--independent constraints
on new particles with electroweak quantum numbers without any assumptions
about their decay properties.
Again, here PDFs are one of the dominant sources of theoretical
uncertainty, for instance in the high--mass tail of $W$ and $Z$
production at the LHC due to the quark--antiquark luminosity.
Improving our knowledge of large--$x$ antiquarks will therefore
be helpful in providing such indirect constraints
of new heavy electroweak sectors.

%%%%%%%%%%%%%%%%%%
\vspace{0.6cm}
\section{The future of PDF determinations}
\label{sec:PDFfuture}

In the final section of this Report we discuss
three topics that could play an important role in shaping
global PDF analyses in the coming years.
First we discuss the problem of theoretical uncertainties
in PDF fits, whose estimate is becoming increasingly necessary given the size of PDF uncertainties in recent
global analysis.
Second, we summarise recent progress in lattice QCD computations
of PDFs, including the first efforts towards a determination
of their $x$--dependence, and discuss how in the near future lattice inputs
could contribute to global PDF fits.
Third, we briefly review the status and plans for future high-energy
colliders, such as the
Large Hadron electron Collider (LHeC) or
the Future Circular Collider (FCC),
and the role that PDFs would play in these.

\subsection{PDFs with theoretical uncertainties}
\label{sec:theoreticalunc}

The development of sophisticated
methodologies for PDF fits, as well as the
availability of a wealth of high--precision data,
have reduced the PDF uncertainties in global
analyses, arising mostly from experimental data and procedural
choices, to the few percent--level in the most constrained regions.
At this level of accuracy, various
theoretical uncertainties become more and more important, representing
a major limitation for present and (even more) for future studies.
Therefore, a robustly account of these theoretical uncertainties
is one of the main goals of PDF fitters for the near future.

In this section, we focus specifically on
the role of the theoretical uncertainties due to missing higher orders (MHOU)
in the
QCD coupling constant, namely those arising
from the truncation of the asymptotic perturbative
expansion.
There have been a number studies recently
on how to estimate MHOU, although we are still
far from a conclusive answer. In the following we first review progress
on MHOU of calculations of non--hadronic and hadronic processes and then
several recent studies related to PDF determination.

We emphasise that these theoretical uncertainties from MHO
should not be confused with the parametric theoretical
uncertainties, that is
those arising from the choice of the values of
input parameters such as $\alpha_s$ and $m_c$.
These have been reviewed in Sect.~\ref{sec:fitmeth.theoryunc},
and the procedure for estimating the impact on the
PDF fit and propagating these parametric
uncertainties to collider cross sections is well established.

\subsubsection{MHOU in matrix element calculations}
The most frequently used and probably also the simplest method of estimating
 MHO corrections is the variation of the
renormalization
and factorization scales in a given fixed--order
calculation.
In the case of the total inclusive cross sections in $e^+e^-$ collisions or decay rates
with a single hard scale $Q$, usually one
varies the QCD renormalization scale $\mu_R$ within the interval $[Q/r,\,rQ]$.
The induced changes in the physical observable,
either from three--point evaluations or from a scan over the entire range,
are taken as the {\it uncertainty} of the MHO, assuming then
typically either a Gaussian (or
two half--Gaussians) or a flat distribution.
The conventional choice is
$r=2$, which is found to work well in most cases, but that underestimates
the true higher order corrections in certain cases, especially if
the fixed--order calculation is carried out at leading order.

At hadron
colliders, there exists in
addition the factorization scale $\mu_F$, arising from the 
factorization of collider QCD divergences
due to initial--state hadrons.
The two scales $\mu_R$
and $\mu_F$ can be varied either simultaneously or independently within the
above range, with the latter case usually further restricted to
$1/r\leq \mu_F /\mu_R\leq r$.
However, even for a single scale problem, there
can still be different choices of the central or nominal scale,
e.g., $Q/2$ or $2Q$, motivated by either QCD resummation or speed
of convergence of the series~\cite{Czakon:2016dgf}, which leads to further
ambiguities in the estimation of
theoretical uncertainties from scale variations.
There are also some studies on utilizing the so--called principal of maximum
conformality to determine the QCD renormalization scale at different
orders, which claims a much smaller MHOU~\cite{Mojaza:2012mf} than traditional
scale variations. 

Determining a suitable prescription for scale variations becomes more
complicated
when moving to differential observables, since here
more hard scales, including
those related to the kinematics, are involved.
This therefore usually
requires a dynamical choice for the central scale, which often also depends
on the specific distribution considered.
For example, in a recent study on the hadronic
production of top quark pairs~\cite{Czakon:2016dgf} it was shown that the
preferable scale is half of the transverse mass of the top quark when
studying the transverse momentum distribution of the top quark, and
one fourth of the sum of transverse mass of top quark and anti--quark
when studying rapidity distribution of the top quark.
Starting from a given choice of the nominal scale,
the scale variations
can then be evaluated in a similar way to the inclusive case and serve as
an estimate of MHOU.
One further complexity arises concerning the correlations
of the MHOU or scale variations in different regions of the
distribution.
Typically, they are assumed to be fully (anti--)correlated
in the entire region which leads to very small theoretical uncertainties
in the case of a normalized distribution.
There have however been attempts to
decorrelate these scale variations based on consideration of the
kinematic dependence of the QCD corrections~\cite{Olness:2009qd}.      

Alternative proposals for estimating MHOU based on
results at known orders exist, such a the so--called
Cacciari--Houdeau (CH) approach~\cite{Cacciari:2011ze}.
The basic idea is to express the full perturbation series in terms of
the expansion parameter $\alpha_s(Q)$ and assume that
all the expansion coefficients
follow the same uniform bounded probability density distribution, in the Bayesian
sense.
Bayesian inference can be used to calculate the probability density of
the unknown higher--order coefficients given those known coefficients
at lower orders. Thus the MHOU, including its probability density distribution, which will be non--Gaussian in general,
can be constructed.
The original CH method was developed
for the study of non--hadronic
processes.
Subsequently,
in the modified CH ($\overline{\rm CH}$)~\cite{Bagnaschi:2014wea}
approach, it was generalised to include hadronic processes.
There the
expansion parameter has been adjusted to $\alpha_s(Q)/\lambda$, with the
parameter $\lambda$ determined from a global survey of selected
processes with known higher--order corrections.
To be specific, the best value
of $\lambda$ computed from
the predicted probability density of higher orders is required
to match the distribution from a frequency count in the survey.
For hadronic
processes, the optimal value of $\lambda$ is found to be about 0.6,
meaning that the {\it true} perturbative expansion parameter
is actually around $1.7\alpha_s(Q)$ rather than $\alpha_s(Q)$.

To give another example,
the series acceleration method~\cite{David:2013gaa} can also be applied to
approximate the
full result for physical
observables based on the available
information from a finite number
terms of the asymptotic series, {\it e.g.},
using Levin--Weniger sequence transforms.
In the `Passarino--David' method of Ref.~\cite{David:2013gaa} a uniform
distribution (in the Bayesian sense) for the theory prediction is assumed between the last known partial sum of the perturbation series
and its approximated value from a Weniger $\delta$--transform.

\begin{figure}[h]
\begin{center}
  \includegraphics[scale=0.45]{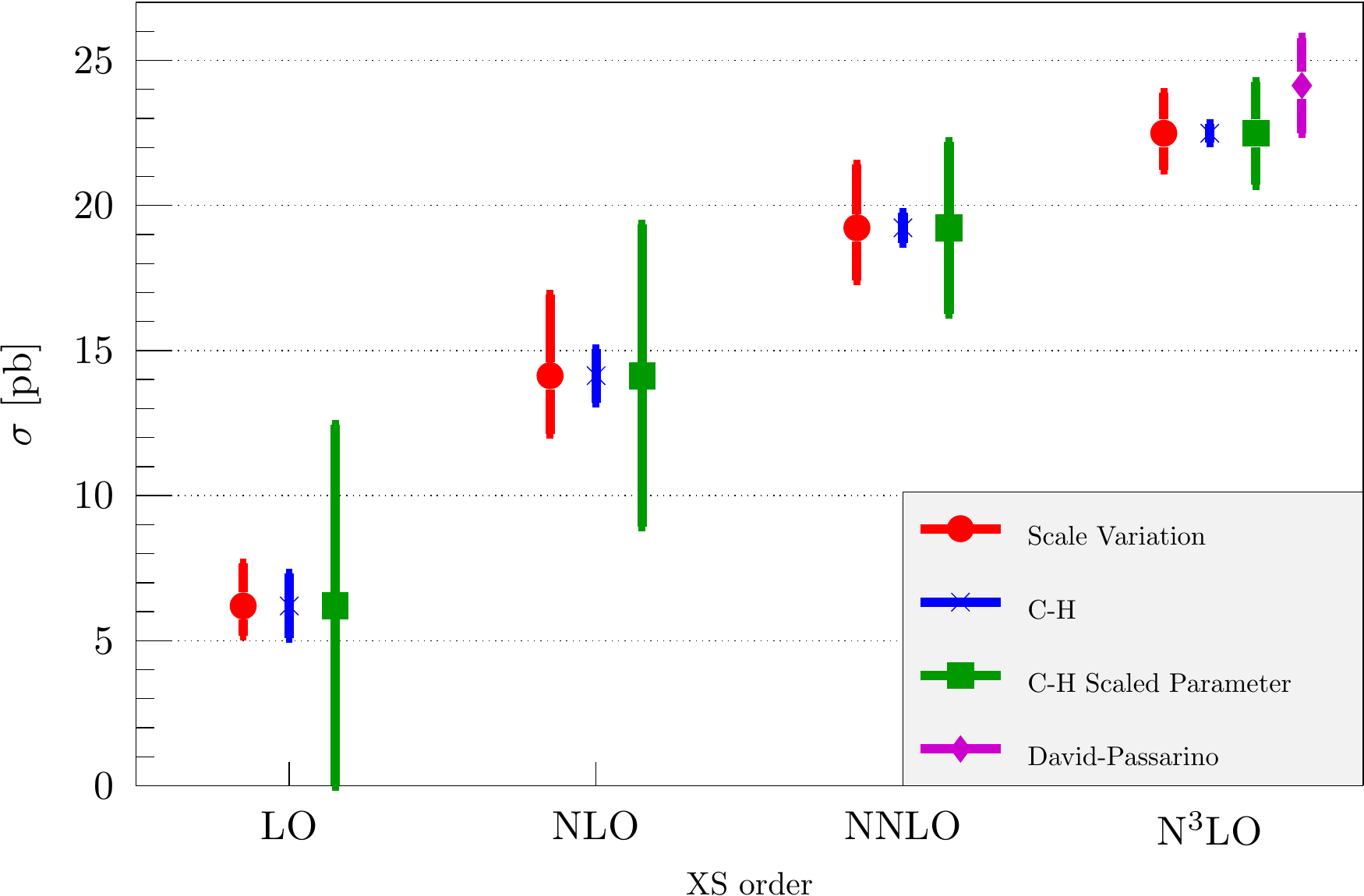}
   \caption{\small
   The cross section for Higgs boson production in gluon fusion calculated
   at increasing perturbative orders~\cite{Forte:2013mda}.
   At each order the theoretical uncertainty is shown for 
   using scale variation (red circles), the CH method (blue crosses),
   and the $\overline {\rm CH}$ method (green squares); at N$^3$LO the
   Passarino--David uncertainty based on series acceleration method
   is also shown (purple diamonds).
    \label{fig:thunc1}
  }
\end{center}
\end{figure}

To illustrate how these various methods compare
to each other, Fig.~\ref{fig:thunc1}
shows the predictions for the
production cross sections of the SM Higgs boson via gluon
fusion at the 8 TeV LHC, calculated at LO, NLO, NNLO and approximate N$^3$LO
with a nominal scale of $\mu_R=\mu_F=m_H$~\cite{Forte:2013mda}. The MHOU as estimated
from different approaches are shown, including scale variations, CH, $\overline {\rm CH}$
and the series acceleration method of~\cite{David:2013gaa} at various
perturbative orders.
Note that different approaches may have different
interpretations on the uncertainties.
In the case of the $\overline{\rm CH}$ method,
the $\lambda$ value has been adjusted to
give almost equal expansion coefficients for the known orders~\cite{Forte:2013mda}.
The $\overline{\rm CH}$ method predicts a larger uncertainty, while the scale variation uncertainty is larger than the CH result.
Encouragingly, the difference in the MHOU bands is observed to decrease with increasing perturbative order.
Indeed, the different
approaches turn out to give similar sizes
for the MHOU associated to the N$^3$LO calculation, with
the exception of the original CH.
However, for the case of the Higgs boson production, none of the methods work very well,
i.e., the N$^3$LO cross section is not captured by uncertainty bands from lower orders. 
In addition, we note that
the series acceleration method also induces a shift in the
central value of the prediction.

%%%%%%%%%%%%%%%%%%%%%%%%%%%%%%%%%%%%%%%%%%%%%%%%%%%%%%%%%%%%%%%%%%%%%
%%%%%%%%%%%%%%%%%%%%%%%%%%%%%%%%%%%%%%%%%%%%%%%%%%%%%%%%%%%%%%%%%%%%%
              
\subsubsection{MHOU in PDF determination}

Global determinations of PDFs are based on
perturbative calculations of matrix elements and DGLAP splitting kernels,
suitably combined to predict a variety of physical cross sections.
In these perturbative calculations, in principle one should
account for their associated  MHOU, which then  propagates into the
resulting PDFs via the fitting of the
theoretical predictions to the experimental
data.
Therefore,
sophisticated treatments of the MHOU from different sources
are required in order
to study the impact on the PDFs, not unlike the treatment
of the experimental
systematic uncertainties.
Crucially, the correlations between theoretical predictions of
different experimental bins of one process and between different
processes must be accounted for.
Furthermore, when making any theoretical prediction,
one should also take care of the correlations between the
MHOU of the PDFs and
of those coming from
the MHOU of the process studied, since they may rely on the same
perturbative expansion for the relevant matrix elements.

\begin{figure}[t]
\begin{center}
  \includegraphics[scale=0.335]{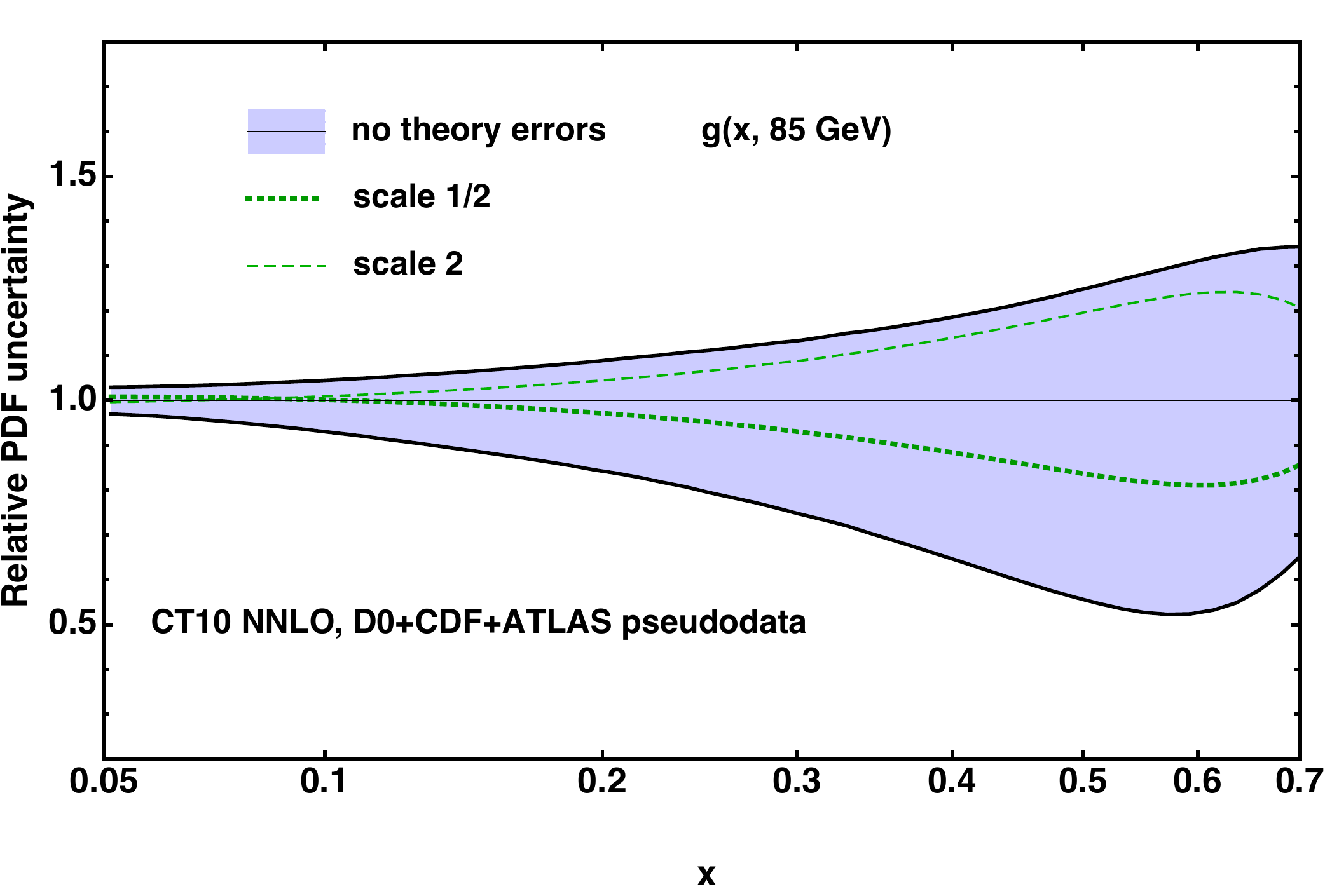}
  \hspace{0.1in}
  \includegraphics[scale=0.38]{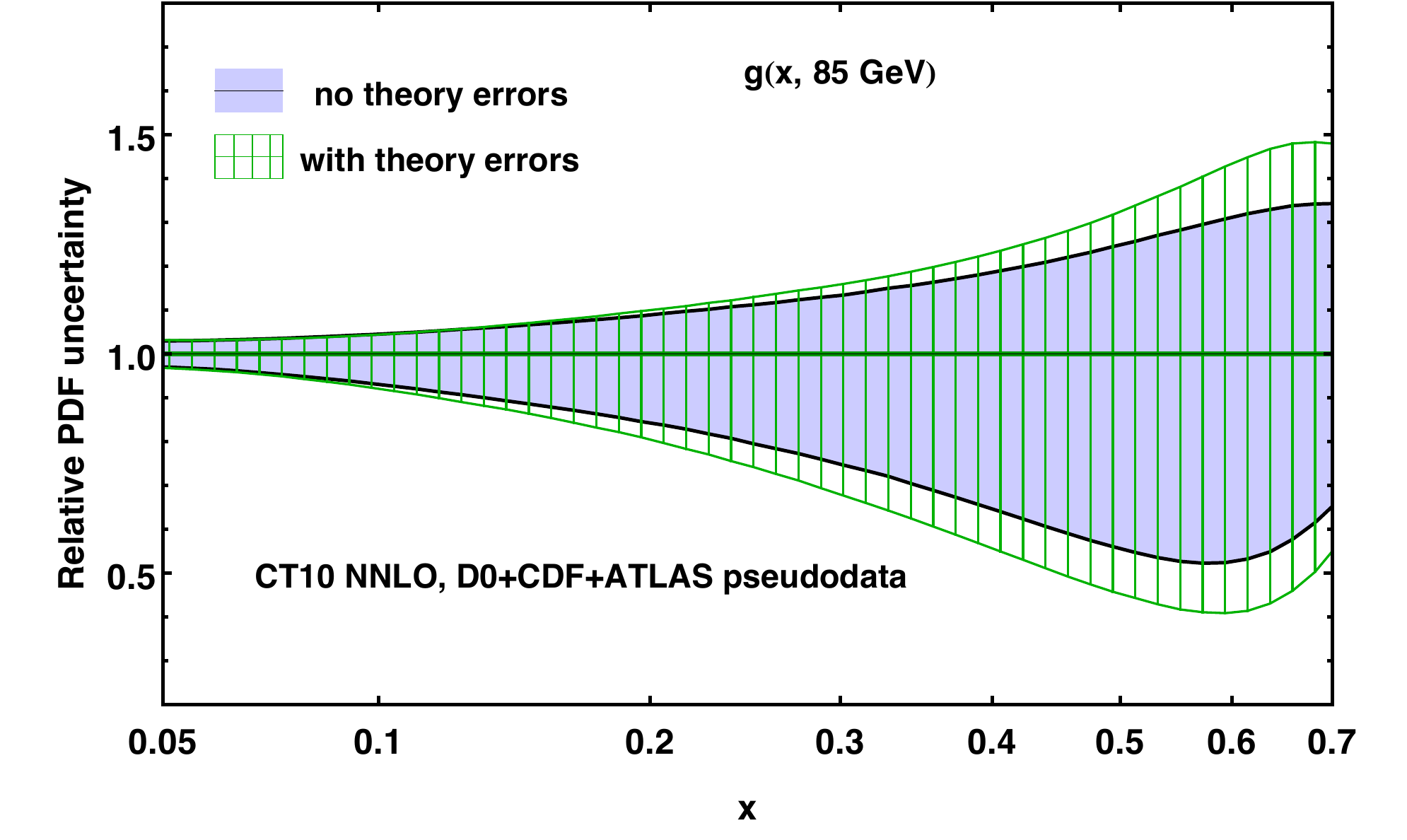}
   \caption{\small 
   Left plot: dependence of the gluon PDF on the choice of QCD scales
   used in the calculation of inclusive jet cross sections in
   CT10 NNLO fits~\cite{Gao:2013xoa} by changing scales to half and twice of
   the nominal value.
   Right plot: impact of the theoretical uncertainties from
   the inclusive jet
   cross sections included in the CT10 NNLO fits~\cite{Rojo:2015acz}
   in the resulting gluon PDF uncertainties.
    \label{fig:thunc2}
  }
\end{center}
\end{figure}

Due to the significant complexity of this problem,
there is still no satisfactory solution, 
and the MHOU have not been included in any of the public PDFs from
global determinations. 
However, it is possible
to restrict ourselves to a region where a single process is most likely
dominant in the MHOU, such as inclusive
jet cross sections and the gluon PDF at large $x$. In such a case, it may still be possible to construct a
simple prescriptions such as the use of scale variations.
To illustrate this, in Fig.~\ref{fig:thunc2} (Left)
we show the impact of the choice of the QCD scales in calculations of 
the inclusive jet cross sections on the gluon PDFs at $Q=85$ GeV for
alternative CT10 NNLO fits~\cite{Gao:2013xoa}.
Note that the theoretical predictions
here are only at NLO, though the PDFs are determined
at NNLO.

%%%%%%%%%%%%%%%%%%%%%%%%%%%%%%%%%%%%%%%%%%%%%%%%%%%%%%%%%%%%%%%%%%%%%
%%%%%%%%%%%%%%%%%%%%%%%%%%%%%%%%%%%%%%%%%%%%%%%%%%%%%%%%%%%%%%%%%%%%%

From Fig.~\ref{fig:thunc2} we can see that in
the higher $x>0.2$ region, the spread of the gluon PDFs
by using scales of $0.5$, 1 and $2$ times the central
scale (in this case the jet $p_T$) is not negligible in comparison to
the nominal PDF uncertainties. This illustrates the potential
significance of the MHOU in PDF determination.
In a related study, also based on alternative CT10 NNLO fits, scale variations
of the NLO inclusive jet cross sections are further decomposed into
several correlated systematics described by five nuisance parameters~\cite{Rojo:2015acz}.
By treating those systematics in a similar way to the experimental
correlated systematic errors, it is possible to include the MHOU
in the standard PDF uncertainty on the same footing as the experimental
systematic uncertainties.
As shown in 
Fig.~\ref{fig:thunc2} (Right) the inclusion of theory errors from the
jet cross sections in the CT10 NNLO fit results in an increase of
the gluon PDF uncertainty
at large $x$, consistent with  Fig.~\ref{fig:thunc2} (Left).   

\begin{figure}[h]
\begin{center}
  \includegraphics[scale=0.37]{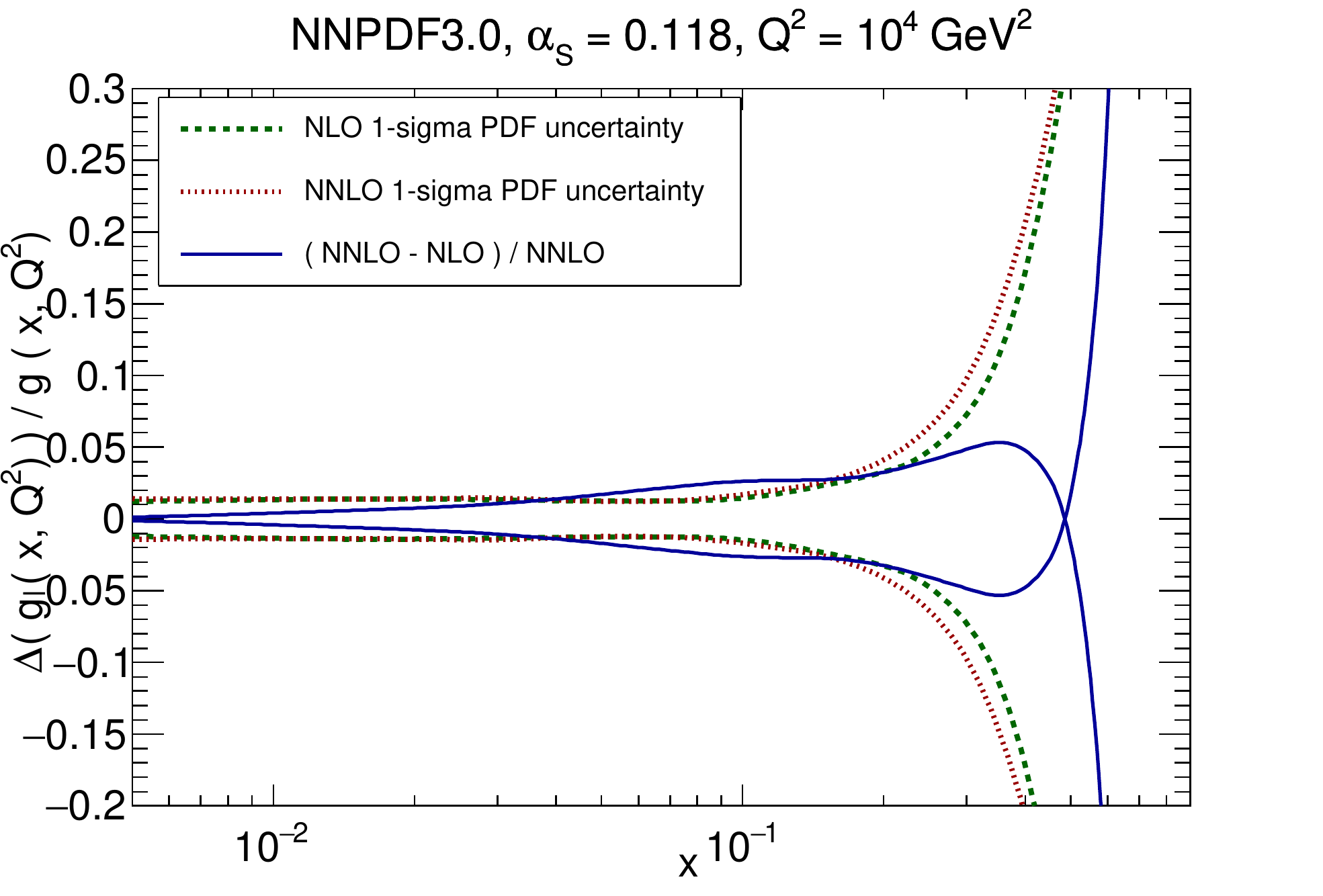}
  \hspace{0.1in}
  \includegraphics[scale=0.37]{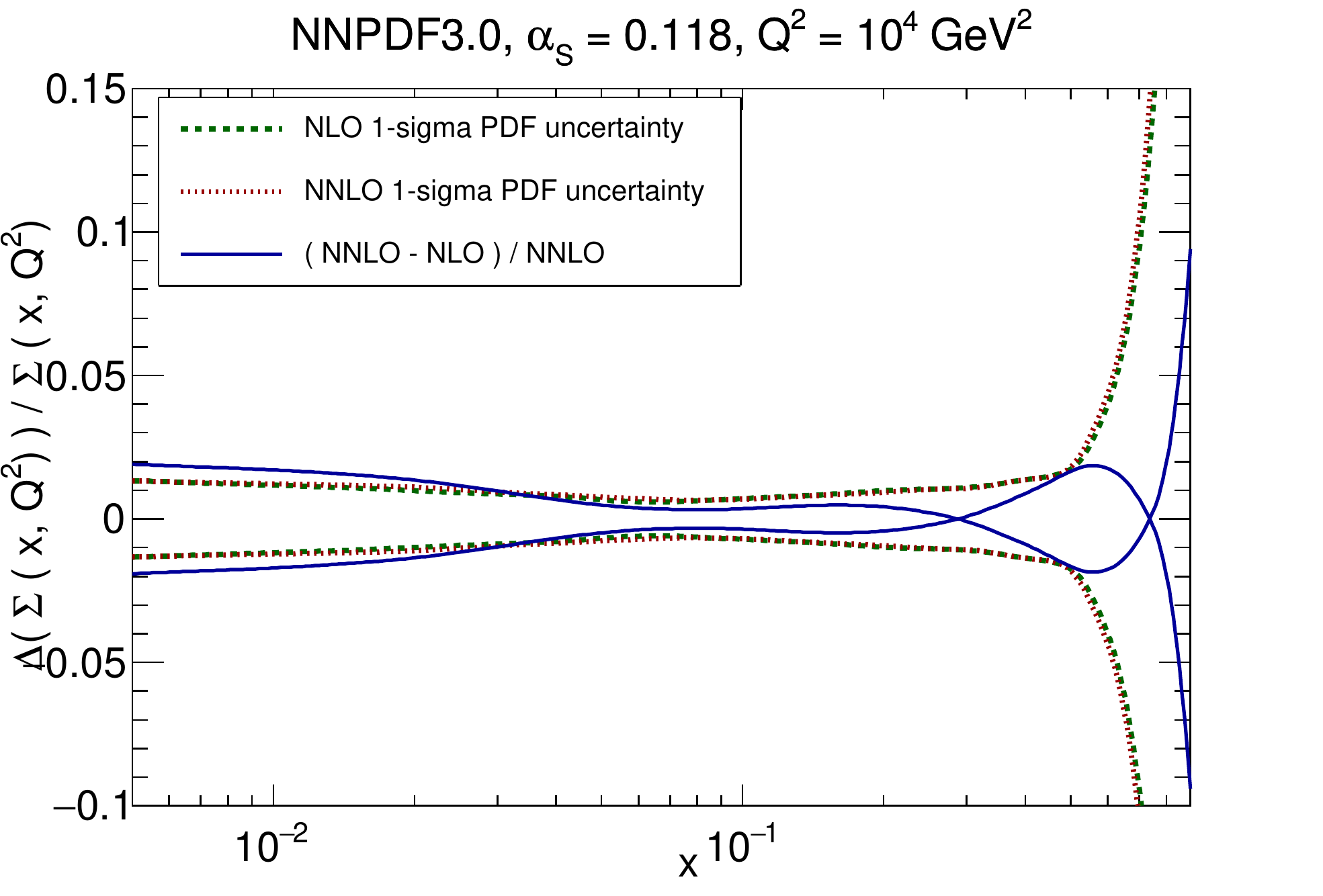}
   \caption{\small
   Left  plot: comparisons between the nominal PDF uncertainties with
   the difference of the central PDFs determined at NLO and NNLO for the
   gluon  PDF at $Q=100$ GeV in the NNPDF3.0 fit.
   Right plot: same for the total quark singlet, $\Sigma(x,Q^2)$.
    \label{fig:thunc3}
  }
\end{center}
\end{figure}

Another possibility to provide a rough estimate of the MHOU
consists in checking the convergence of the fitted PDFs with
increasing orders.
Fig.~\ref{fig:thunc3} shows the comparison of the nominal
PDF uncertainties
with the difference of the central PDFs fitted at NLO and NNLO for
gluon and total singlet PDFs at $Q=100$ GeV
in the NNPDF3.0 fits~\cite{Ball:2014uwa}.
This difference between the PDF central values at NLO and NNLO
provides a conservative upper bound 
of the MHOU associated to the NNLO PDFs.
From this comparison, we see that
there are regions where the shifts of
NLO to NNLO PDFs are comparable to or even larger than the conventional
PDF uncertainties.
%
%Here, one could also apply the CH or $\overline{\rm CH}$
%approach based on the fitted PDFs at LO, NLO and NNLO.
%
Based on similar approaches Ref.~\cite{Forte:2013mda} found that
the MHOU of the NNLO PDFs have a
negligible impact on the Higgs production cross section through gluon
fusion, but these could be relevant for top quark pair production.

%%%%%%%%%%%%%%%%%%%%%%%%%%%%%%%%%%%%%%%%%%%%%%%%%%%%%%%%%%%%%%%%%%%%%
%%%%%%%%%%%%%%%%%%%%%%%%%%%%%%%%%%%%%%%%%%%%%%%%%%%%%%%%%%%%%%%%%%%%%

\subsection{Lattice QCD calculations of the proton structure}
\label{sec:lattice}

As discussed in Sect.~\ref{sec:pdffitting.DIS}, PDFs arise from
 non-perturbative QCD dynamics.
 Therefore,
since we are currently not able
 to  analytically solve strongly--coupled non--Abelian gauge theories,
 it is very challenging to compute PDFs from first principles.
 Perhaps the only possibility,
 beyond model calculations, consists in
 exploiting recent progress in lattice QCD~\cite{Gupta:1997nd}.
 This method is based on discretizing the QCD Lagrangian in
 a finite--volume Euclidean lattice, which naturally introduces 
 an ultraviolet cutoff, and then computing non--perturbative QCD quantities directly on this lattice, before taking the continuum limit.
 Such lattice QCD calculations require minimal external input,
 in particular only the hadronic mass scale $\Lambda_\text{QCD}$ and the values of the quark masses, or
 alternatively, the physical pion and kaon masses.

Here we briefly review some of these recent developments in lattice QCD calculations of PDFs.
For a more detailed overview of this progress,
together with the study of their interplay with state--of--the--art global analysis, see Ref.~\cite{Lin:2017snn}, a White Paper that 
has been produced as the outcome of the dedicated workshop
{\it ``Parton Distributions and Lattice QCD calculation in the LHC era''},\footnote{\url{http://www.physics.ox.ac.uk/confs/PDFlattice2017}} which took place in
Oxford in March 2017.
This workshop
 brought together experts from the two communities (lattice QCD and global PDF fitting)
 to explore the synergies
and complementary aspects between the two approaches.
The discussion and results shown in this section represent
a brief excerpt of the material contained
in that White Paper (see also~\cite{Nocera:2017war}).

Given that PDFs have a formal definition in terms of the nucleon matrix elements
of certain non--local operators
(see the discussion in
Sect.~\ref{sec:pdffitting.DIS}), it is in principle 
possible to compute PDFs using lattice QCD.
From the practical point of view, however, given the extremely
CPU intensive nature of these calculations,
most lattice QCD results on PDFs have been limited for a long time to the first two moments of non--singlet flavour
combinations for large (unphysical) quark masses.
These restrictions have been overcome in the recent years, with several groups providing now
results for PDF moments with physical pion and kaon masses.

Moreover, the development of novel
strategies to overcome the limitations in computing the first few 
moments~\cite{Lin:2012ev,Constantinou:2014tga,Syritsyn:2014saa} has allowed the determination of more challenging quantities, 
such as gluon and flavour--singlet matrix elements as well as higher PDF moments.
More recently, both conceptual and numerical breakthroughs in lattice QCD computations have allowed
even greater progress, providing the first attempts to evaluate PDFs
and related quantities directly in Bjorken--$x$ space~\cite{Lin:2014zya,Alexandrou:2015rja,Chen:2016utp,Alexandrou:2016jqi}.
These developments have pushed lattice QCD calculations
to the point where, for the first time, it is possible to provide information on the PDF shape
of specific flavour combinations, both for quarks and for antiquarks, and meaningful comparisons with 
global fits can start to be made.

Recent progress in lattice QCD calculations of PDFs and related
quantities has also been partly driven
by a greatly improved control of the systematic uncertainties
that enter the calculation
of relatively simple quantities such as nucleon matrix elements, which correspond to
low moments of the PDFs.
These  include, among others, using
physical pion masses, reducing the excited--state contamination, and using large lattices
to remove finite--size effects.
Moreover, to make contact with the physical world and experimental
data, the numerical results are extrapolated to the continuum 
and infinite--volume limits.
In addition, the past decade has seen significant progress in
the development of efficient algorithms for the generation of
ensembles of gauge field configurations, which represent the QCD
vacuum, and tools for extracting the relevant information from lattice QCD
correlation functions.

In order to be reliably used in phenomenological
applications, lattice QCD calculations must demonstrate control over all relevant
sources of
systematic uncertainty introduced by the discretisation of QCD on the
lattice, such as those discussed above.
For a coherent assessment of the present state of lattice QCD
calculations of various quantities, the degree to which each
systematic uncertainty has been controlled in a given calculation should be an important
consideration.
The quality of individual lattice calculations can be quantitatively
assessed based on criteria such as those from the
FLAG analysis of flavour physics on the lattice~\cite{Aoki:2016frl}.

The traditional approach for lattice QCD calculations of
PDFs
has been to determine the matrix elements of local twist--two operators, that can be related to the Mellin moments of PDFs. In principle, given a sufficient number of Mellin moments, PDFs can be reconstructed from the inverse Mellin transform.
In practice, however, the calculation is limited to the lowest three moments, because power--divergent mixing occurs between twist--two operators on the lattice.
Three moments are unfortunately insufficient to reconstruct the full
$x$--dependence of the PDFs without significant model dependence~\cite{Detmold:2003rq}. The lowest three moments do provide, however, useful information, both as benchmarks of lattice QCD calculations and as constraints in global extractions of PDFs.
For instance, provided systematic uncertainties are kept under control, one can envisaging adding
these lattice QCD calculations of PDF moments as an additional theoretical
constraint to the global fit, on the same footing as the momentum
and valence sum rules.

Now we briefly summarize the state--of--the--art of lattice QCD calculations of the first moment
of unpolarized PDFs, which are those for which systematic uncertainties
are under better control. See Ref.~\cite{PDFlattice2017} for a more exhaustive set of
comparisons, including those of lattice QCD calculations with global fits.
The observables that are discussed here are defined as follows:
\begin{enumerate}

\item The second moment of the flavour triplet combination, $T_3=u^+-d^+$,
\begin{equation}
\left.\langle x\rangle_{u^+-d^+}(\mu^2)\right|_{\mu^2=Q^2}
\equiv
\int_0^1 dx\, x\left\{u(x,Q^2)+\bar{u}(x,Q^2)-d(x,Q^2)-\bar{d}(x,Q^2)\right\} \, .
\label{eq:unpfmumdtot}
\end{equation}

\item The second moment of the individual quark $q^+=q+\bar{q}$ PDFs,
\begin{equation}
\left.\langle x\rangle_{q^+=u^+,d^+,s^+,c^+}(\mu^2)\right|_{\mu^2=Q^2}
\equiv
\int_0^1 dx\, x\left\{q(x,Q^2)+\bar{q}(x,Q^2)\right\} \, .
\label{eq:unpfmiqtot}
\end{equation}

\item The second moment of the gluon PDF,
\begin{equation}
\left.\langle x \rangle_g(\mu^2)\right|_{\mu^2=Q^2}
\equiv
\int_0^1 dx\, x\, g(x,Q^2) \, .
\label{eq:unpfmg}
\end{equation}

\end{enumerate}
In Table~\ref{tab:unpolLQCDstatus1} we show a selection of recent results
for the  moments defined in Eqns.~(\ref{eq:unpfmumdtot}-\ref{eq:unpfmg}).
As can be seen, for $\langle x\rangle_{u^+-d^+}$ the lattice QCD uncertainties
vary between 5\% and 15\%, with the quoted results not agreeing among themselves within
errors.
For the first moment of the gluon, $\langle x \rangle_g$, the uncertainties
are around 10\%, and for the individual total quark combinations they vary between
 10\% and 20\%.
 Therefore, while current determinations of the first moments are unlikely to provide
 constraints on global PDF fits (where uncertainties are the few--percent level), future
 calculations with improved systematic errors might be able to make a difference.
 On the other hand,  existing calculations can already be used to provide meaningful
 constraints on polarized PDFs, where uncertainties are rather larger than in
 the unpolarized case due to the scarcer dataset.
 
\begin{figure}[t]
\centering
\includegraphics[width=0.80\textwidth]{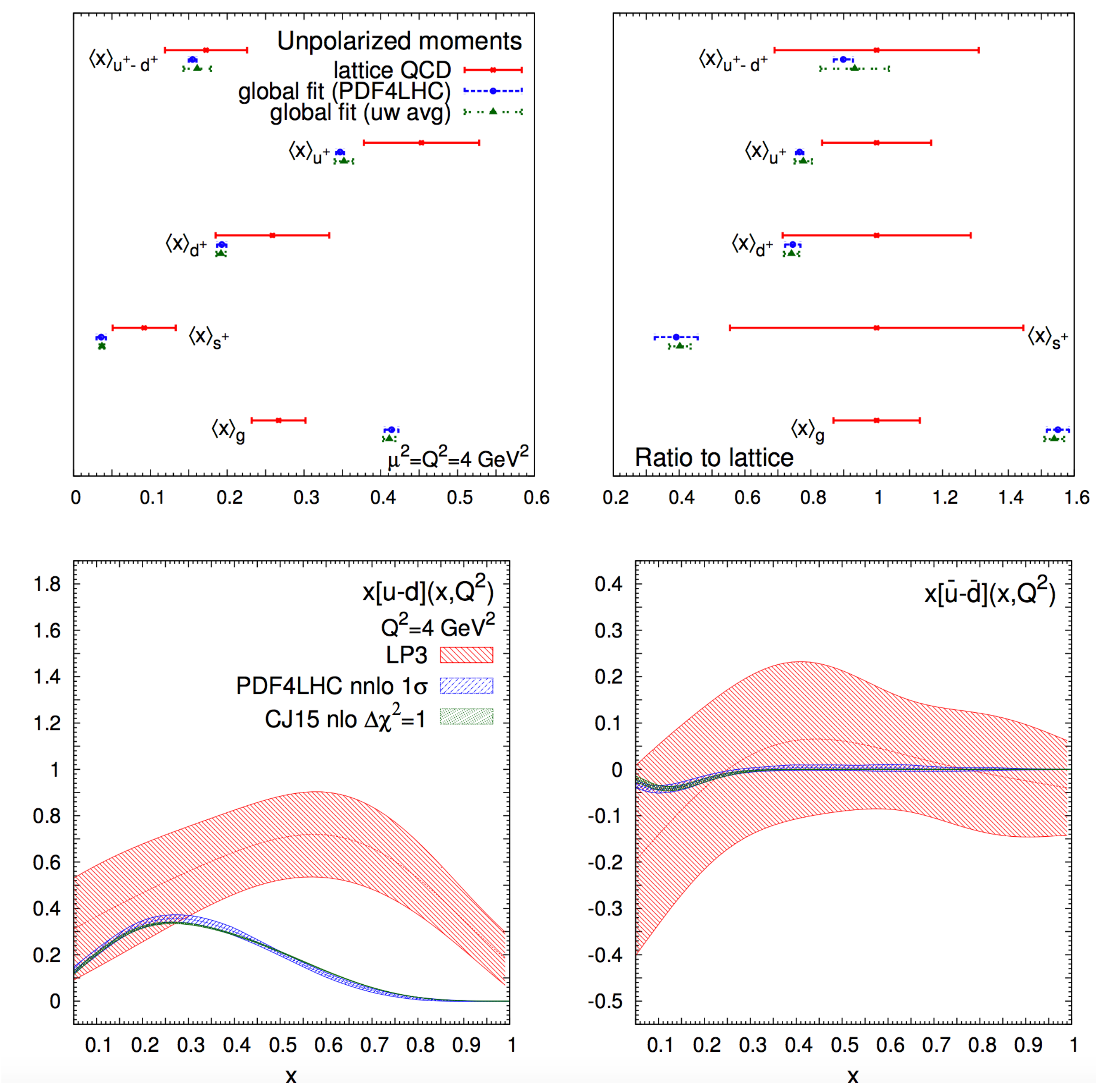}
\caption{\small Upper plots: comparison
between lattice QCD calculation of different moments
of unpolarized PDFs at $Q=2$~GeV with the corresponding
results from global PDF fits, both in absolute scale
(left) and normalized to the lattice result (right plot).
Bottom plots: the same comparison
now for the unpolarized isovector
   quark distributions, $x(u-d)$ (left)
   and $x(\bar{u}-\bar{d})$ (right plot) at $Q=2$~GeV.
See the discussion in the White Paper~\cite{Lin:2017snn} for more details.
}
\label{fig:xspaceLatticePDFs}
\end{figure}

To quantify this,
in Fig.~\ref{fig:xspaceLatticePDFs}  we show the comparison
between lattice QCD calculation of different moments
of unpolarized PDFs at $Q=2$~GeV with the corresponding
results from global PDF fits, both in absolute scale
(left) and normalized to the lattice result (right).
We find that there is agreement within uncertainties for most cases,
except for the gluon, but that lattice QCD calculations are still
far from being competitive with the global PDF fit results.
However, it is also clear that future lattice calculations with
reduced systematic uncertainties could be able
to produce useful constraints in the global fit framework.

%-------------------------------------------------------------------------------
\begin{table}[t] 
\renewcommand{\arraystretch}{1.2} 
\centering 
\begin{threeparttable}
\begin{tabular}{llclll}
Mom. & Collab. & Ref. & $N_f$ & Value\\
\toprule
$\langle x\rangle_{u^+-d^+}$ 
& LHPC\,14  
  & \cite{Green:2012ud} 
  & 2+1 
  & 0.140(21)\\
& ETMC 17  
  & \cite{Alexandrou:2017oeh} 
  & 2   
  & 0.194(9)(11)\\
& RQCD 14  
  & \cite{Bali:2014gha} 
  & 2   
  & 0.217(9)\\
\midrule
$\langle x\rangle_{u^+}$
&  ETMC 17  
  & \cite{Alexandrou:2017oeh} 
  & 2 
  & $0.453(57)(48)$\\
\midrule
$\langle x\rangle_{d^+}$
& ETMC 17  
  & \cite{Alexandrou:2017oeh} 
  & 2 
  & $0.259(57)(47)$\\
\midrule
$\langle x\rangle_{s^+}$
& ETMC 17  
  & \cite{Alexandrou:2017oeh} 
  & 2 
& $0.092(41)(0)$\\
\midrule
$\langle x\rangle_{g}$
& ETMC 17  
  & \cite{Alexandrou:2017oeh} 
  & 2 
  & 0.267(22)(27)\\
\bottomrule
\end{tabular}
\end{threeparttable}
\caption{\small Selected
recent lattice QCD calculations of the
first moments of unpolarized PDFs, defined in
Eqns.~(\ref{eq:unpfmumdtot}-\ref{eq:unpfmg}), evaluated at
 $\mu^2=4$ GeV$^2$.
See Ref.~\cite{PDFlattice2017} for more details.
}
\label{tab:unpolLQCDstatus1}
\end{table}
%------------------------------------------------------------------------------

While the lowest moments of the PDFs provide crucial benchmarks to
assess the reliability of lattice QCD calculations
of the nucleon structure, as well as potentially
useful information for global PDF fits,
they do not allow the complete $x$--dependence of the PDFs to be reconstructed.
In particular, the calculation of PDF moments is mostly insensitive to the small--$x$
region.
To bypass these limitations, recently a number of approaches have been
developed, aiming  to determine the $x$--dependence of PDFs directly from lattice QCD.
One of the most important approaches goes under the name of
`quasi--PDFs', first formulated in Refs.~\cite{Ji:2013dva,Ji:2014gla}.
For simplicity, we focus in the following on the flavour nonsinglet case,
so that we can neglect the mixing with gluons.

In this approach, the unpolarized quark quasi--PDF $\widetilde{q}(x,\Lambda,p_z)$ is defined as a momentum--dependent
nonlocal static matrix element:
\begin{align}
\label{eq:qPDF}
\widetilde{q}(x,\Lambda,p_z)  \equiv \int \frac{dz}{4\pi} e^{-i x z p_z} 
\frac{1}{2}\sum_{s=1}^2\left\langle p,s\right\vert \bar{\psi}(z)\gamma_z e^{ig\int_0^z
A_z(z^\prime) dz^\prime} \psi(0) \left\vert p,s\right\rangle \, ,
\end{align}
where $\Lambda$ is an UV cut--off scale,
typically chosen to be the inverse lattice spacing $1/a$.
As we see from Eq.~(\ref{eq:qPDF}), these quasi--PDFs are defined for
nucleon states at finite momentum (as $p_z$ is finite, the momentum fraction $x$ can be larger than unity).
Therefore, in order to make contact with the standard collinear PDFs and thus with
phenomenology, they
must be related to the corresponding light--front PDF, for which the nucleon
momentum is taken to infinity.

In the  large--momentum  effective field theory (LaMET) approach, the
quasi--PDF $\widetilde{q}(x,\Lambda,p_z)$ can be related to the $p_z$--independent
light--front PDF $q(x,Q^2)$ through the following relation~\cite{Ji:2013dva,Ji:2014gla}
\begin{equation} \label{eq:qPDFmatching}
\widetilde{q}(x,\Lambda ,p_z) = 
  \int_{-1}^1 \frac{dy}{\left\vert y\right\vert} 
    Z\left( \frac{x}{y}, \frac{\mu}{p_z}, \frac{\Lambda}{p_z}\right)_{\mu^2 = Q^2} q(y,Q^2) +
  \mathcal{O}\left( \frac{\Lambda_\text{QCD}^2}{p_z^2},\frac{m^2}{p_z^2}\right) \, ,
\end{equation}
where $\mu$ is the renormalisation scale;
$Z$ is a matching kernel; and $m$ is the nucleon mass.
Here the $\mathcal{O}\left(m^2/p_z^2\right)$ terms are target--mass corrections and the $ \mathcal{O}\left(\Lambda_\text{QCD}^2/p_z^2\right)$ terms are higher twist effects, both of which are suppressed at large nucleon momentum. A complementary approach to the LaMET methods instead views the quasi--PDF as a `lattice cross section' from which the light--front PDF can be factorized~\cite{Ma:2014jla, Ma:2014jga}. An alternative construction, using so--called `pseudo--PDFs', has been explored in~\cite{Radyushkin:2017cyf,Orginos:2017kos}.

Preliminary results from lattice calculations of quasi--PDFs have been rather encouraging~\cite{Lin:2014zya,Alexandrou:2015rja,Chen:2016utp,Alexandrou:2016jqi}, although a number of important limitations still need to be overcome.
To illustrate this progress in lattice calculations of $x$--space
PDFs, in Fig.~\ref{fig:xspaceLatticePDFs} 
we also display
the comparison between lattice QCD and global PDF fit results
for the unpolarized isovector  quark distributions, $x(u-d)$ (left)
   and $x(\bar{u}-\bar{d})$ (right plot) at $Q=2$~GeV as
   a function of Bjorken-$x$ (see~\cite{Lin:2017snn} for more details).
From this comparison, we see that the lattice QCD results are still far from the global
fits (and thus also from the experimental data), but given
recent progress we may expect sizeable improvements
in the coming years.
In addition, the same methodology can be also applied to other nucleon matrix elements,
including the polarized PDFs.

%%%%%%%%%%%%%%%%%%%%%%%%%%%%%%%%%%%%%%%%%%%%%%%%%%
%%%%%%%%%%%%%%%%%%%%%%%%%%%%%%%%%%%%%%%%%%%%%%%%%

Despite these encouraging developments, there still remain a number of important challenges that must be overcome
before one can achieve a complete  determination of the $x$--dependence of PDFs directly from lattice QCD
that is competitive with global PDF fits.
In particular, excellent control over the various sources of systematic uncertainties that affect
the calculation must be reached.
Some of these are common to the calculations of PDF moments, as discussed above,
but there are also a number of additional systematic errors specific to quasi--PDFs, such as those
associated with the finite nucleon momentum of the lattice calculations and with the
renormalisation of quasi--PDFs.
Given the recent fast progress, it is quite possible that this may be achieved and that  in the future
lattice QCD calculations of $x$--space PDFs may be used to constrain global analyses.

\subsection{Parton distributions at future high-energy colliders}
\label{sec:fcc}

We now discuss the role that PDFs would have in some of the recent
proposals for future colliders involving hadrons in the initial state.
There are three main families of potential future high--energy colliders currently under
active discussion.
Electron--positron colliders, such as the ILC~\cite{Baer:2013cma},
CLIC~\cite{Linssen:2012hp},
TLEP/FCC--ee~\cite{Gomez-Ceballos:2013zzn}, or CEPC~\cite{CEPC-SPPCStudyGroup:2015csa}, offer the potential
for ultra--high precision measurements of the Higgs, electroweak
and top quark sectors.
However, hadron colliders with energy much greater than the LHC would allow
 the continued exploration of the high--energy frontier and significantly
extend the coverage of 
searches for new BSM particles, including Dark Matter candidates,
while at the same time providing  unprecedented opportunities for the study of the Higgs sector, for example the Higgs self--interactions.
With this motivation, there is ongoing work towards a circular
collider hosted at the CERN site which would accelerate protons
up to the extreme energies of $\sqrt{s}=100$
TeV~\cite{Contino:2016spe,Mangano:2016jyj}, dubbed FCC--hh,
while a similar machine is under study by the Chinese HEP community~\cite{CEPC-SPPCStudyGroup:2015csa}.

Another avenue for future high--energy collisions would be new machines based
on electron--proton collisions, exploiting the successful strategy
adopted at HERA.
One of the open proposals is the Large Hadron electron Collider (LHeC)~\cite{AbelleiraFernandez:2012cc}, where the $E_{p}=7$ TeV proton beam from the LHC would collide with an electron/positron
beam with $E_e=60$ GeV coming from a new LinAc, which would be able
to reach down to $x_{\rm min}\simeq 2\cdot 10^{-6}$ at
$Q^2=2$ GeV$^2$.
A more extreme incarnation of the same idea corresponds to colliding these
$E_e=60$ GeV electrons
with the $E_p=50$ TeV beam of the FCC--hh.
The resulting collider, dubbed FCC--eh, would be able to reach down to
 $x_{\rm min}\simeq 2\cdot 10^{-7}$ at
$Q^2=2$ GeV$^2$.
These two machines would therefore extend
the PDFs coverage at small $x$ by more than two orders of magnitude
in comparison to HERA.

The Electron Ion Collider (EIC)~\cite{Boer:2011fh}, which
might start construction soon either at the BNL or the
JLAB site, falls under the same category.
This would offer
the possibility to polarize both leptons and protons and
to accelerate heavy nuclei, although its $\sqrt{s}$ would
be smaller than that of HERA.
It would therefore not be able to provide new information
on the unpolarized proton PDFs, though it would be a unique machine
to explore the spin and 3D content of the proton, as well as
map in detail the nuclear modifications of the free--proton PDFs.

In this section, we review the role that PDFs would play
first at a future higher energy lepton--proton collider
such as the LHeC/FCC--eh, and then at a future
hadron collider with a centre--of--mass energy
of $\sqrt{s}=100$ TeV.

\subsubsection{PDFs at high--energy lepton--hadron colliders}

As mentioned above, one of the possibilities for a future high--energy collider
now under active discussion
would be to exploit the LHC/FCC proton beam and collide it with a high energy lepton
beam, which would be delivered by a new LinAc to be built at the CERN site.
In the case of using the LHC beams, the
LHeC~\cite{AbelleiraFernandez:2012cc,AbelleiraFernandez:2012ty,Armesto:2013dia} 
would then represent
a scaled--up version of HERA, and as such would offer significant opportunities for
improved determinations of the proton structure down to very low $x$
and up to high $Q^2$, as well as providing a wealth of information on nuclear PDFs
in a kinematic region where they are currently essentially unconstrained.
Several options are now being considered, with some preference now for synchronous
operation during the final years of the HL--LHC upgrade, as then the LHeC
program can be extended to include measurements of the Higgs sector.

\begin{figure}[t]
\centering
\includegraphics[width=0.93\textwidth]{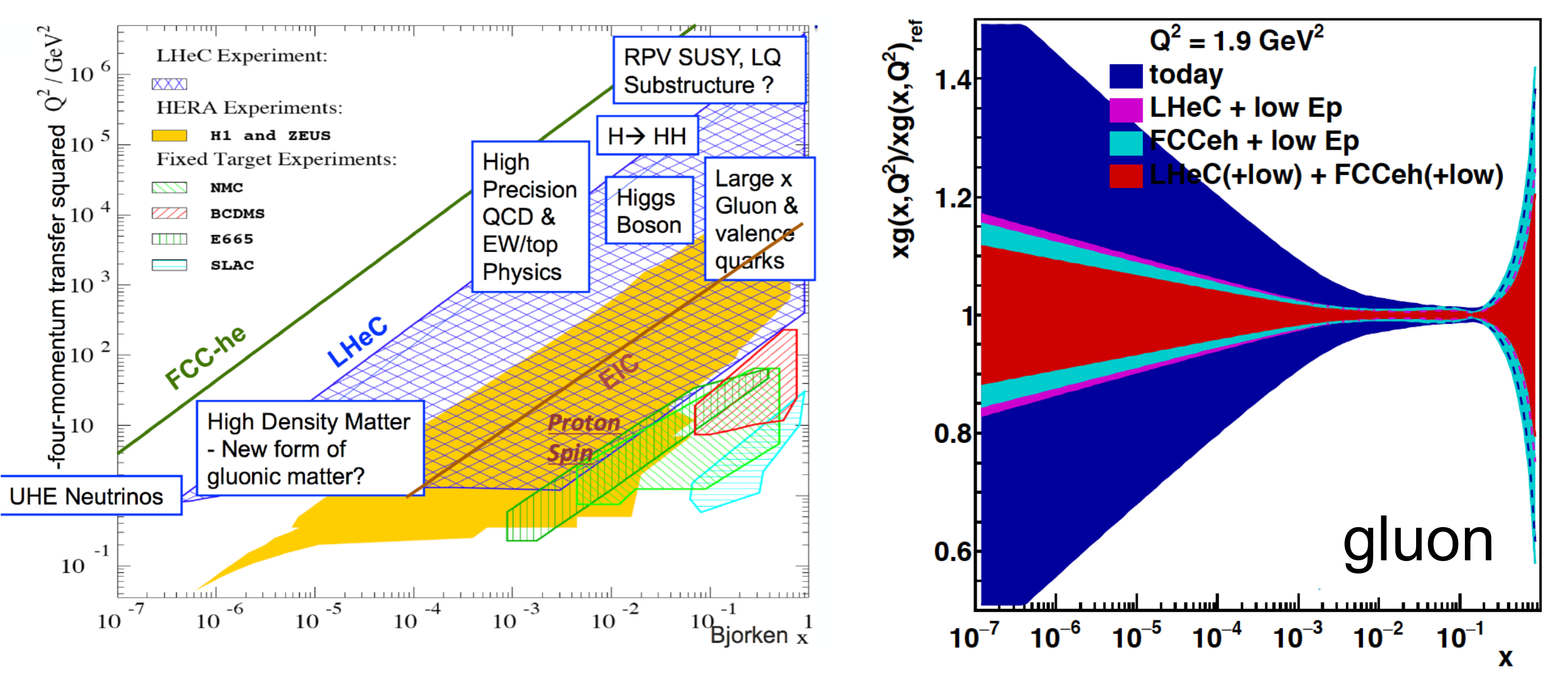}
\caption{\small Left: kinematic coverage in the $(x,Q^2)$
of several existing and proposed DIS experiments.
Starting from the fixed--target experiments and then moving to HERA, the
LHeC and finally the FCC--eh, as the centre of mass energy increases,
the kinematic reach extends both towards higher $Q^2$ and smaller $x$ values.
Right: results of an {\tt xFitter} PDF feasibility study that
compares the impact on the gluon PDF of adding either LHeC or FCC--eh (or both)
pseudo--data in addition to the HERA inclusive structure function dataset.
}
\label{fig:kinplotLHeC}
\end{figure}

In Fig.~\ref{fig:kinplotLHeC} we show
the kinematic coverage in  $(x,Q^2)$
of several existing and future DIS experiments,
including the EIC, the LHeC, and the FCC--eh.
We observe that by starting from the fixed--target experiments and then moving to HERA, the
LHeC and finally the FCC--eh, as the centre of mass energy increases,
the kinematic reach extends both towards higher $Q^2$ and smaller $x$ values.
At the FCC--eh in particular, it should be able to cover the region down to
$x\simeq 10^{-7}$ without leaving the perturbative region $Q \gsim 1$ GeV.
It is important to emphasise that the same coverage would be achieved
for nuclear PDFs, extending the coverage by four or five orders of magnitude in
$x$ as compared to existing measurements.

One of the most important aspects of the LHeC/FCC--eh scientific case
is the capability to probe the proton/nuclear PDFs with an unprecedented precision,
not only by means of inclusive structure functions but also with measurements
of the strange, charm, and bottom structure functions, that provide
a direct handle on the heavy flavour PDFs.
The LHeC/FCC--eh would also allow a measurement of the strong
coupling constant $\alpha_s(M_Z)$ with per--mile
experimental uncertainties,
for instance using jet production~\cite{Currie:2016ytq}, and high--precision
measurements of the electroweak sector parameters.
Providing theoretical
predictions with matching uncertainties,  in particular
for jet production, will be a challenge.
To illustrate these possibilities,
in Fig.~\ref{fig:kinplotLHeC} we also show the
results of an {\tt xFitter} PDF feasibility study that
compares the impact on the gluon PDF of adding either LHeC or FCC--eh (or both)
pseudo--data in addition to the HERA inclusive structure function dataset.
The reduction of the PDF uncertainties down to very small--$x$ values reflects
the extended kinematic reach of these future high energy
lepton proton colliders.
A similar reduction of the PDF uncertainty is expected for the quark PDFs.

%%%%%%%%%%%%%%%%%%%%%%%%%%%%%%%%%%%%%%%%%%%%%%%%%%
%%%%%%%%%%%%%%%%%%%%%%%%%%%%%%%%%%%%%%%%%%%%%%%%%

Another important aspect of the interplay between PDFs and the LHeC/FCC--eh
is related to the small--$x$ resummation
framework~\cite{Ciafaloni:2007gf,White:2006yh,Altarelli:2008aj,Bonvini:2016wki,Bonvini:2017ogt}.
This framework is based on extending the collinear DGLAP formalism to account for
the all--order resummation of terms of the form $\alpha_s^k\ln^m(1/x)$ in the
splitting functions and the partonic coefficient functions,
as implemented in the BFKL equation~\cite{Lipatov:1976zz,Fadin:1975cb,Kuraev:1976ge}.
Using this approach, the reliability of theoretical predictions
for DIS structure functions and collider cross sections can be extended
down to much smaller values of $x$ in comparison to the calculations
based on the collinear DGLAP framework.
Recently, a version of the NNPDF3.1 global analysis, called
NNPDF3.1sx, based on
NLO+NLL$x$ and NNLO+NLL$x$ theory, has been
presented~\cite{Ball:2017otu},
which suggest that accounting for small--$x$ effects leads to an
improved description of the inclusive HERA structure
function data (see also the discussion in~\cite{Caola:2010cy,Caola:2009iy}).
One may expect such effects to become even more relevant for higher--energy
lepton--proton colliders (see Fig.~\ref{fig:kinplotLHeC}).

\begin{figure}[t]
\centering
  \includegraphics[width=0.495\textwidth]{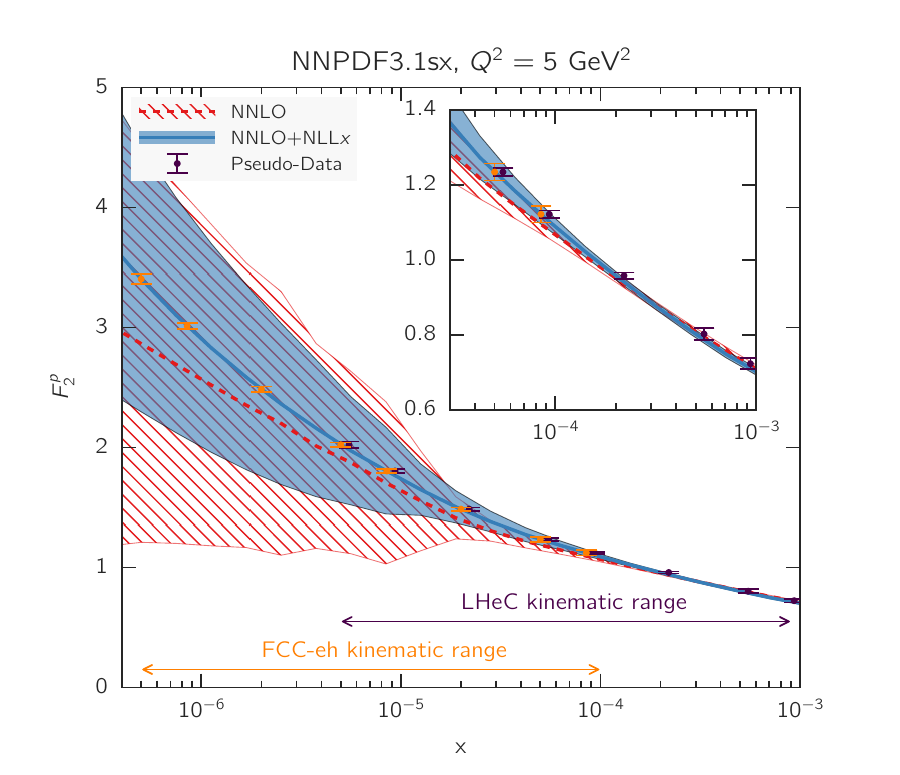}
  \includegraphics[width=0.495\textwidth]{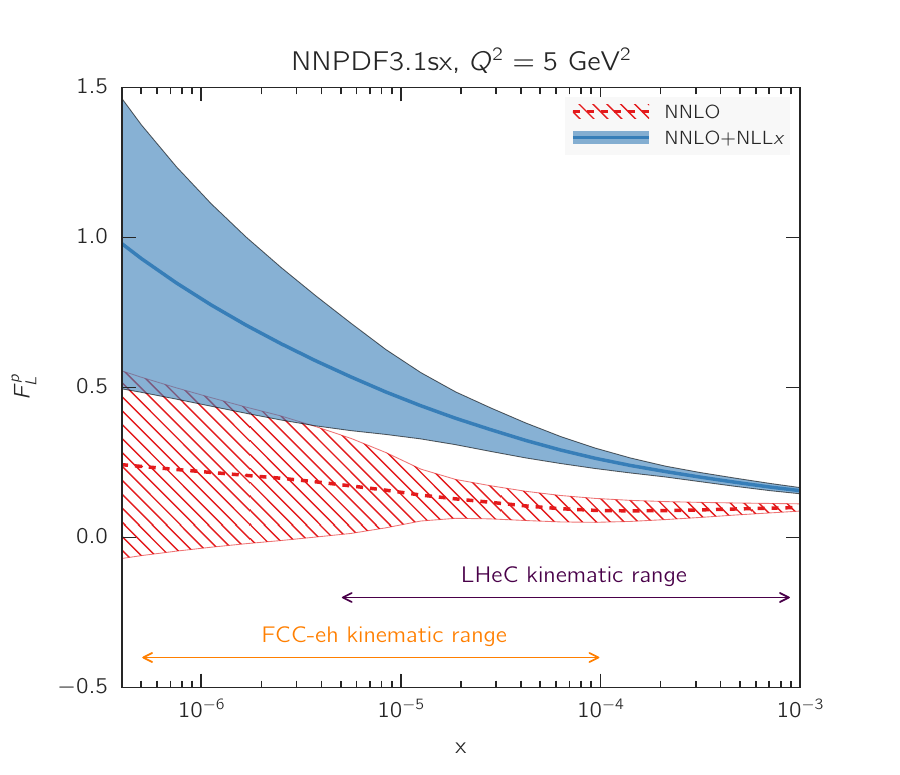}
 \caption{\small Theoretical predictions for the $F_2(x,Q^2)$
 and $F_L(x,Q^2)$ structure functions
 evaluated with the NNPDF3.1sx NNLO
   and NNLO+NLL$x$ sets.
These structure functions are showed at $Q^2=5$ GeV$^2$
for the range in $x$ that would be accessible at the LHeC
and FCC-eh.
   In the case of $F_2$, we show in
   addition the expected total experimental uncertainties
   with the NNLO+NLL$x$ values as central prediction. 
   The inset in the $F_2$ plot provides
   a magnified view of the region $x>3\times 10^{-5}$ which
   is also covered by the HERA measurements.}
 \label{fig:LHeCpredictions}
\end{figure}

With the motivation of providing a first estimate of the relevance of small--$x$ resummation
for the LHeC/FCC--eh, in
Fig.~\ref{fig:LHeCpredictions} we show
predictions for the $F_2$ and $F_L$ structure functions using the NNPDF3.1sx NNLO
   and NNLO+NLL$x$ fits at $Q^2=5$ GeV$^2$ for the kinematics of the LHeC
   and the FCC--eh.
   For these calculations,
   we have used {\tt APFEL} to produce  NNLO(+NLL$x$) predictions,
each using as input the corresponding NNPDF3.1sx fits,
for the most updated version of the simulated LHeC/FCC--eh pseudo--data kinematics.
    In the case of $F_2$, we also show the expected total experimental uncertainties
    based on the simulated pseudo--data, assuming the NNLO+NLL$x$ curve as central prediction.
    The total uncertainties of the simulated pseudo--data are at the few percent
    level, and therefore they are rather smaller than the PDF uncertainties in the
    entire kinematic range.
    Moreover,  the inset in the $F_2$ plot provides
   a magnified view of the region $x>3\times 10^{-5}$ which
   is also covered by the HERA measurements.

%%%%%%%%%%%%%%%%%%%%%%%%%%%%%%%%%%%%%%%%%%%%%%%%%%%%%%%%%%%%%%%%%%%%
%%%%%%%%%%%%%%%%%%%%%%%%%%%%%%%%%%%%%%%%%%%%%%%%%%%%%%%%%%%%%%%%%%%%%%

From the comparisons in Fig.~\ref{fig:LHeCpredictions}
we see how the FCC--eh would allow the small--$x$ region to be probed
by about an order of magnitude further than the LHeC (which in turn
extends HERA by about the same amount).
The differences between NNLO and NNLO+NLL$x$ theory are moderate for
$F_2$, in particular taking into account the large PDF uncertainties.
This implies in turn that
refitting the pseudo--data is required to first reduce the PDF errors
and then discriminate
between the two theoretical scenarios.
This has been demonstrated
explicitly in the study of~\cite{Ball:2017otu}, which highlights
how inclusive $F_2^p$ measurements would
represent a powerful probe of small--$x$ dynamics.
We also see that differences are more marked for $F_L$, with the central values
differing by several times the PDF uncertainty in most
of the accessible kinematic range.
This illustrates the unique sensitivity of $F_L^p$ measurements to probe
small--$x$ QCD dynamics.

We emphasize that
both structure functions $F_2^p$ and $F_L^p$
provide complementary information on small-$x$ QCD: the former,
which can be measured with smaller uncertainties, via scaling
violations, and the latter, which is affected by larger uncertainties,
directly via the different overall normalization.
We also note that small--$x$ predictions based on non--linear
effects~\cite{Albacete:2012rx,Albacete:2010sy}
(`saturation') have typically the opposite trend to small--$x$ resummation,
suppressing the structure functions in comparison to the NNLO fixed--order
calculation.
Therefore, the measurements in Fig.~\ref{fig:LHeCpredictions} and related
ones would open a unique window to the novel dynamical regime of QCD at very
small $x$.

\subsubsection{PDFs at a 100 TeV hadron collider}

\begin{figure}[t]
\centering
\includegraphics[width=0.85\textwidth]{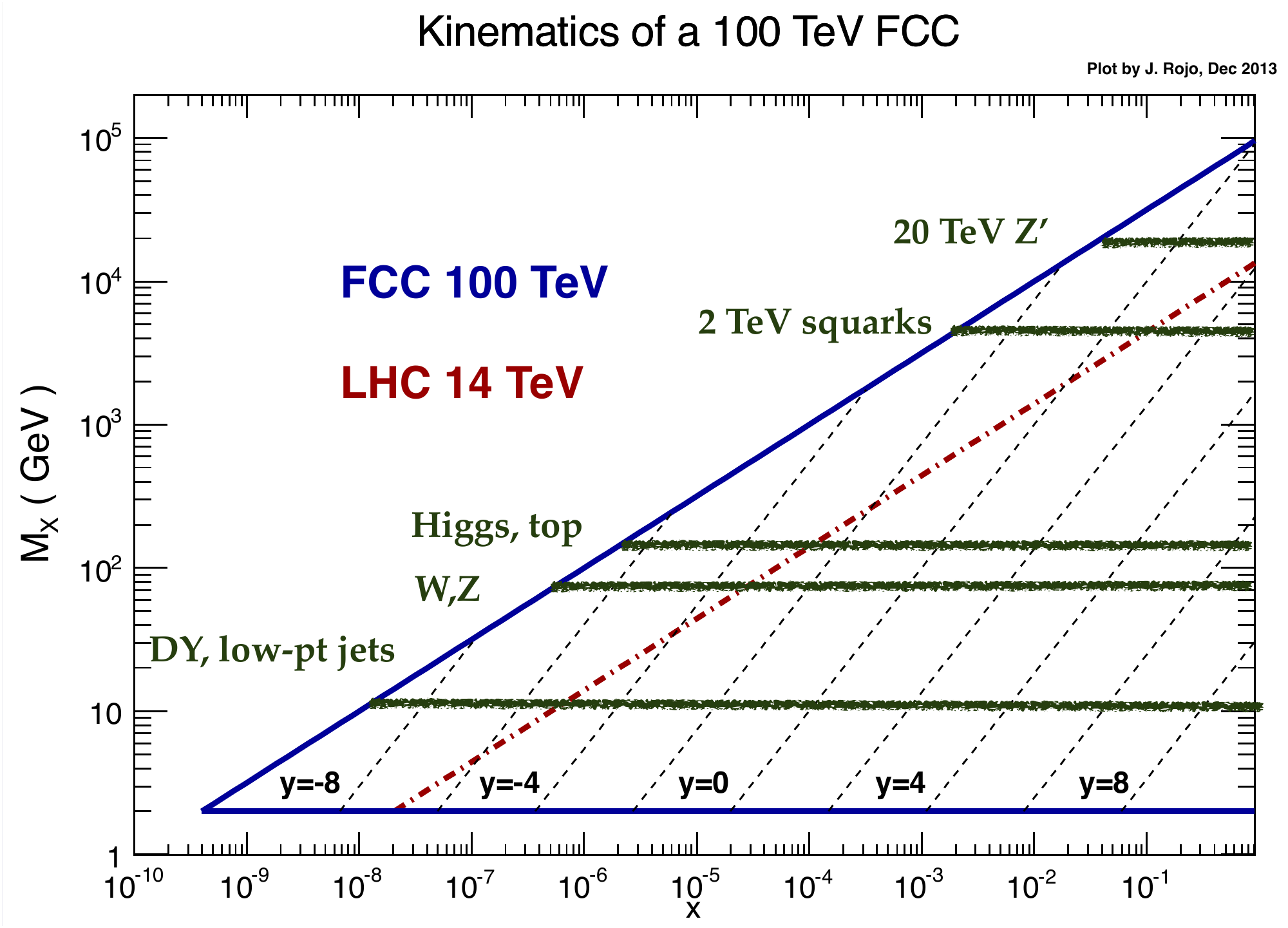}
\caption{\small Kinematic coverage in the $(x,M_X)$ plane of a
  $\sqrt{s}=100$ TeV 
  hadron collider (solid blue line),
  compared with the corresponding coverage of the LHC at
  $\sqrt{s}=14$ TeV (dot--dashed red line), from~\cite{Rojo:2016kwu}.
}
\label{fig:kinplot}
\end{figure}

Next we discuss parton distributions at the FCC--hh,
a proposal for a future hadron collider with a centre of mass energy of
$\sqrt{s}=100$ TeV.
In order to illustrate the extended kinematic coverage that
would be achieved at a 100 TeV proton--proton collider
in comparison to the one at the LHC, in Fig.~\ref{fig:kinplot}, taken from~\cite{Rojo:2016kwu}, 
we compare the $(x,M_X)$ coverage at 100 TeV and 14 TeV, where
the dotted lines indicate the regions of constant rapidity at the
FCC--hh.
In addition,  we also indicate the relevant $M_X$ regions for some representative
  processes, from low masses (Drell--Yan, low $p_T$ jets), to
  electroweak scale processes (Higgs, $W,\,Z$, top) and new
  high--mass particles (squarks, $Z'$).
  A significant increase in the kinematic
  coverage relative to the LHC is clear.
  A particularly interesting aspect is that
 at the FCC--hh even
 electroweak--scale processes
  such as $W,Z$, or $h$ production become sensitive to the small--$x$
  region.

%%%%%%%%%%%%%%%%%%%%%%%%%%%%%%%%%%%%%%%%%%%%%%%%%%
%%%%%%%%%%%%%%%%%%%%%%%%%%%%%%%%%%%%%%%%%%%%%%%%%

As discussed in the FCC--hh Yellow Report~\cite{Mangano:2016jyj}, there are two main aspects
of PDF phenomenology relevant for a 100 TeV collider.
As for the LHC, at $\sqrt{s}=100$ TeV parton distributions
are one of the dominant theoretical uncertainties for
several cross sections.
In particular, electroweak scale cross sections, such as $W$ or $h$ production,
become sensitive to the small--$x$ region where PDF uncertainties
are currently large.

\begin{figure}[t]
\begin{center}
  \includegraphics[scale=0.60]{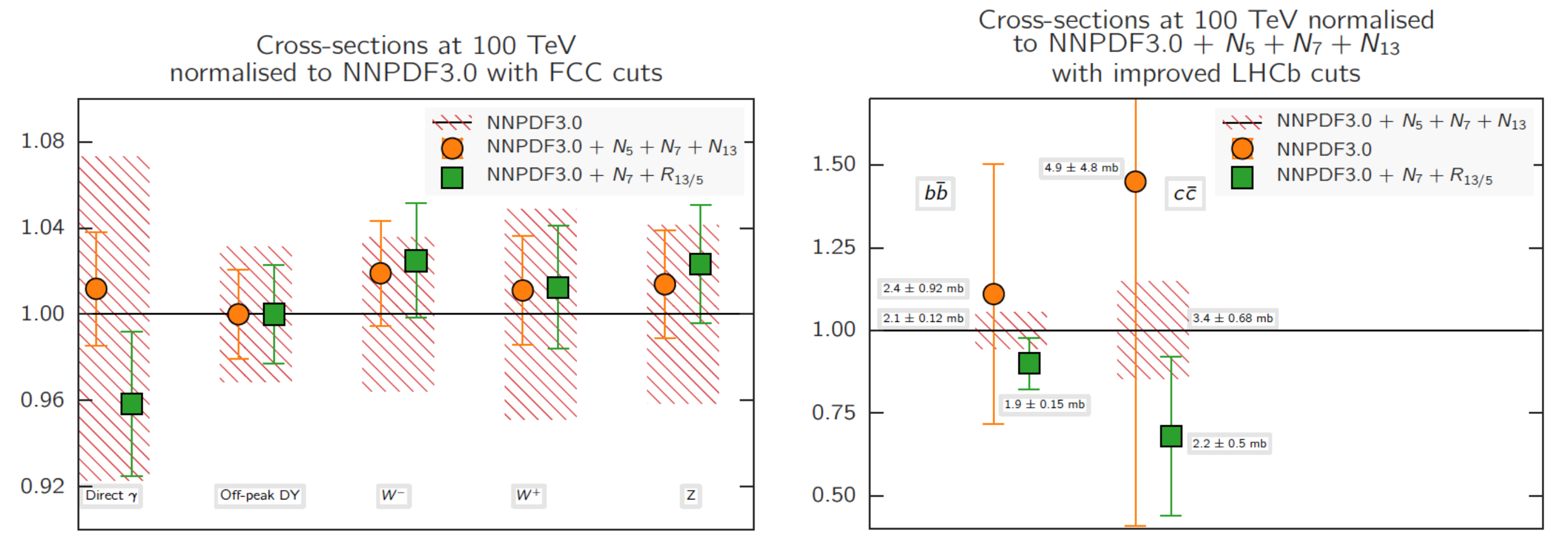}
   \caption{\small 
     Comparison of cross sections for different representative
     processes at the FCC with $\sqrt{s}=100$ TeV,
     between the NNPDF3.0 predictions and those of the NNPDF3.0+LHCb
     sets.
     Note that the acceptance cuts are different in each process.
     In the left plot we show the results for direct photon
     production, off--peak Drell--Yan cross sections, and inclusive
     weak boson production.
     In the right plot we show the fiducial cross sections for
     $c\bar{c}$ and $b\bar{b}$ production.
     Note that the range of the $y$ axis is different in the two plots, reflecting the larger
     uncertainties in heavy quark production.
    \label{fig:fcc-xsecs}
  }
\end{center}
\end{figure}

To illustrate this point,
in Fig.~\ref{fig:fcc-xsecs} we show the
comparison of cross sections for different representative
     processes at the FCC with $\sqrt{s}=100$ TeV,
     between the NNPDF3.0 predictions and those of the NNPDF3.0+LHCb
     sets~\cite{Gauld:2016kpd}.
     Specifically, the NNPDF3.0+LHCb sets used here include $D$ meson production
     data from LHCb at $\sqrt{s}=5,7$ and 13 TeV, as discussed in Sect.~\ref{sec:datatheory.charm}.
     The acceptance cuts are different in each process, see~\cite{Gauld:2017rbf} for more details.
     In the left plot we show the results for direct photon
     production, off--peak Drell--Yan, and inclusive
     weak boson production.
     In the right plot we show the fiducial cross sections for
     $c\bar{c}$ and $b\bar{b}$ production.
     In all cases, and in particular for heavy quark pair production,
     we can observe the reduction of PDF uncertainties
     that is derived once the NNPDF3.0+LHCb sets are used,  highlighting the
     sensitivity to the small--$x$ region.

%%%%%%%%%%%%%%%%%%%%%%%%%%%%%%%%%%%%%%%%%%%%%%%%%%%%%%%%%%%%%%%%%%%%%
%%%%%%%%%%%%%%%%%%%%%%%%%%%%%%%%%%%%%%%%%%%%%%%%%%%%%%%%%%%%%%%%%%%%%

Another new aspect of PDFs at the FCC--hh is the onset of novel phenomena that are absent
at the lower energies of the LHC.
These include the possibility of treating the top quark as a
massless parton~\cite{Han:2014nja,Dawson:2014pea},
the need for resummation of ``collinear'' weak gauge boson radiation
and the consequent introduction of
electroweak PDFs~\cite{Bauer:2017isx,Ciafaloni:2001mu,Chen:2016wkt}, as well as the increased role
of photon--induced processes~\cite{Bertone:2015lqa,Mangano:2016jyj,Harland-Lang:2016kog}.
Moreover, just as in the case of the LHeC/FCC--eh, the role of small--$x$ resummation is expected
to become more important at the FCC--hh than at the LHC, given the sensitivity of even standard
candles such as $W$, $Z$ and Higgs production to the small--$x$ region.

\begin{figure}[t]
\begin{center}
  \includegraphics[scale=0.44]{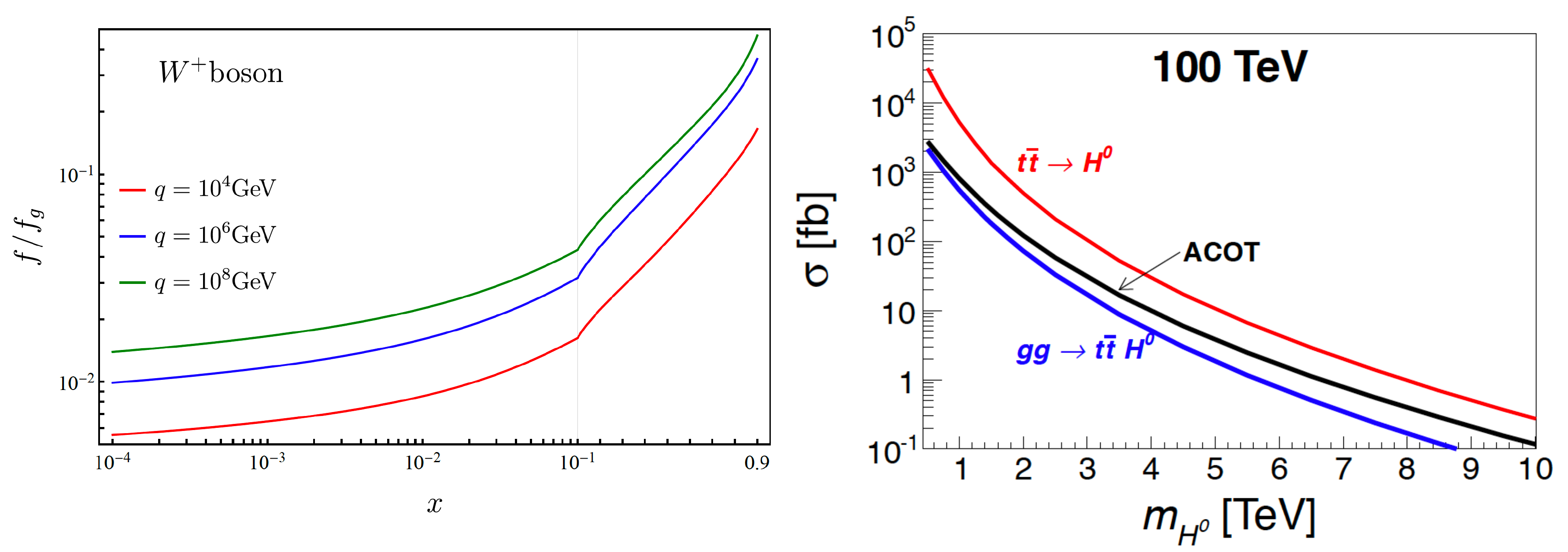}
   \caption{\small 
    Left: the PDF of the $W^+$ boson normalized to that of the gluon,
    as a function of $x$ for different scales: $q=10^4$ GeV,
    $10^6$ GeV, $10^8$ GeV, from Ref.~\cite{Bauer:2017isx}.
    Right: the cross section of inclusive Higgs production by $t\bar{t}$ associated
    production, comparing the results of the $n_f=5$ scheme ($gg\to h t\bar{t}$),
    the $n_f=6$ scheme ($t\bar{t}\to h$), and of their interpolation by means
    of the ACOT general--mass scheme, from~\cite{Han:2014nja}.
\label{fig:FCCtopEW}
  }
\end{center}
\end{figure}

Here we provide two representative illustrations of these new PDF--related phenomena at the FCC--hh.
First, as mentioned above, at 100 TeV the electroweak gauge bosons become effectively massless,
and thus it is possible to construct electroweak PDFs and corresponding evolution
equations.
In Fig.~\ref{fig:FCCtopEW} we show the PDF of the $W^+$ boson normalized to that of the gluon,
    as a function of $x$ for different scales: $q=10^4$ GeV,
    $10^6$ GeV, $10^8$ GeV, computed using the framework of~\cite{Bauer:2017isx}.
    We observe that the dependence of the $W^+$ PDF with the energy $q$ is rather mild.
    For most of the range of $x$, the $W^+$ PDF is at most a few percent of the gluon
    PDF, while for $x\ge 0.1$ it becomes larger, up to 40\% of the gluon PDF.
    This does not necessarily mean that the effects of the $W$ PDF will be phenomenologically
    relevant: this can be assessed only at the cross section level, comparing calculations
    with massive gauge bosons and those where these are treated as massless (and thus resummed
    into the electroweak PDFs).
    Interestingly, the chiral nature of the weak interactions implies that
    once electroweak effects are accounted for in the DGLAP evolution, the proton
    can become spontaneously polarized, with implications for BSM searches~\cite{Fuks:2014uka}.

%%%%%%%%%%%%%%%%%%%%%%%%%%%%%%%%%%%%%%%%%%%%%%%%%%%%%%%%%%%%%%%%%%%%%
%%%%%%%%%%%%%%%%%%%%%%%%%%%%%%%%%%%%%%%%%%%%%%%%%%%%%%%%%%%%%%%%%%%%%

Following a similar line of thought, at 100 TeV it is tempting to treat the top quark
as a massless parton, in the same way as at the LHC the bottom quark is treated
as massless in most calculations.
In Fig.~\ref{fig:FCCtopEW} we show the
cross section for inclusive heavy scalar (``Higgs'') production by $t\bar{t}$ associated
    production, comparing the results of the $n_f=5$ scheme ($gg\to H^0 t\bar{t}$),
    the $n_f=6$ scheme ($t\bar{t}\to H^0$), and of their interpolation by means
    of the ACOT general--mass scheme~\cite{Han:2014nja}
    (see also Sect.~\ref{sec:pdffitting.heavyq}).
    The comparison is performed as a function of the heavy Higgs boson mass $m_{H^0}$.
    We find that the $n_f=6$ calculation, where the top quark is treated as massless and resummed
    into a top quark PDF $t(x,Q)$, is rather far from the matched calculation up to at least $m_{H^0}=10$ TeV.
    This suggest that the massless top approximation is not suitable even for the extreme
    FCC energies.
    However, the concept of top PDF is still useful in order to improve fixed
    order calculations, using general--mass schemes such as ACOT or FONLL, but this seems to be inadvisable without accounting for the top quark mass effects.

%%%%%%%%%%%%%%%%%%%%%%%%%%%%%%

%%%%%%%%%%
\vspace{0.6cm}
\section{Conclusions}
\label{sec:conclusions}

The wealth of experimental data collected by the LHC experiments
so far represents only a small fraction of the complete
dataset that it will provide over the coming two decades of planned operation.
This will allow the SM to be studied and stress--tested to an even greater, completely unprecedented, level.
Together with recent progress in theoretical calculations, we are therefore entering the era of
the LHC precision physics, which will aim to compare data and theory
at the percent level or even less.
Given the null results of BSM searches so far, a systematic high--precision
analysis of the SM predictions and the LHC data might be one of most promising
approaches to look for BSM dynamics at the LHC, which may only be
manifest as subtle differences with the SM predictions.
The detailed mapping of the quark and gluon structure of the
proton represents a crucial component of this LHC precision physics
program.

In this Report we have presented an overview of the most important recent
developments in PDF determination, with emphasis on their implications for
LHC phenomenology.
After a brief review of the theoretical foundations of the global QCD analysis framework,
we have reviewed recent progress both from the theoretical and the experimental
point of view for those hard--scattering cross sections used in PDF fits. We have
compared the similarities and differences between the methodologies used
by the various PDF fitting collaborations, and then presented the state--of--the--art
fits from each group and assessed what we can learn about the internal structure of the
nucleons from various points of view.
We have then discussed the role of QED corrections in PDF fits, in particular concerning
photon--initiated processes, and presented some of the most representative
examples of the applications of PDFs for LHC phenomenology, from the measurement
of the Higgs boson couplings to the determination of the mass of the $W$ boson.

In the last section of this Report we have discussed some topics
that will very likely play an important role in the near future of PDF determinations.
One of these topics is the role of theoretical uncertainties, in particular those arising from
missing higher--order terms in the perturbative expansion.
Given the size of PDF uncertainties in the current generation of global analyses, it is likely that these theory
errors will increasingly become comparable (if not larger) than the nominal PDF uncertainties, and thus
finding a statistically sound method to account for these is of utmost importance.
Another topic that might affect the PDF fitting paradigm is that of the interplay with lattice
QCD calculations, where recent progress both in computing Mellin moments of various flavour
combinations as well as in direct $x$--space calculation of PDFs suggest that in the future the constraints
from non--perturbative
lattice calculations might be able to provide information for global PDF fits.
Finally, we have summarized the importance of PDFs for future higher--energy colliders whose
physics case is being discussed at present, such as a Large Hadron electron Collider or a new
proton--proton collider with a centre--of--mass energy of up to 100 TeV.

We hope that this Report has managed to convey to the reader that the topic of PDF determinations
is an interesting and lively one, with many implications, from our understanding of non--perturbative
dynamics of the strong interactions to searches for new BSM physics and ultra--high energy astrophysics.
Moreover, PDF fits represent a unique internal
consistency check of the SM and of the collinear
QCD factorization framework.
The latter appears to be in remarkably
good shape, given that we are now
able to simultaneously describe a few tens of individual experiments spanning a wide
kinematical domain, some of them with extremely
small uncertainties at the per--mile level.

PDF fits thrive at the crossroads of advanced data analysis, state--of--the--art perturbative
calculations, and modern robust statistical methodology, and thus provide guidance for
other similar global analysis efforts, such as fits of the Wilson coefficients within
the SM Effective Field Theory framework.
Moreover, progress in unpolarized PDF fits is also one of the main drivers of recent
improvements in other related aspects of the proton structure, from
polarized PDFs to nuclear PDFs, as well as in other aspects of non--perturbative QCD
such as light and heavy hadron fragmentation.

As we enter in the LHC precision era, ever--improving PDF determinations will keep providing a unique
contribution to
this exciting exploration of the high--energy frontier. 
We fully expect that the requirements of
this LHC precision program will further drive improvements in global PDF determinations,
leading to an ever more detailed picture of the inner life of the proton.

%%%%%%%%%%
\vspace{0.6cm}
\section*{Acknowledgments}
\noindent
We are grateful to Giulia Zanderighi for giving us the opportunity to
write this Report.
We are also indebted to many colleagues for extensive discussions and fruitful collaborations of the topic
of PDF determinations in the last years.
A necessarily incomplete list of colleagues that we would like to thank for this includes
Alberto Accardi,
Sergei Alekhin,
Richard D. Ball,
Josh Bendavid,
Valerio Bertone,
Marco Bonvini,
Ulla Blumenschein,
Stefano Carrazza,
Mandy Cooper-Sarkar,
David D'Enterria,
Sayipjamal Dulat,
Stephen Farry,
Stefano Forte,
Rikkert Frederix,
Stefano Frixione,
Marco Guzzi,
Nathan P. Hartland,
Tie-Jiun Hou,
Joey Huston,
Zahari Kassabov,
Valery Khoze,
Jan Kretzschmar,
Rhorry Gauld,
Sasha Glazov,
Joey Huston,
Katerina Lipka,
Jon Pumplin,
Michelangelo Mangano,
Alan Martin,
Simone Marzani,
Ronan McNulty,
Alexander Mitov,
Pavel Nadolsky,
Emanuele R. Nocera,
Luca Perrozzi,
Ringaile Placakyte,
Klaus Rabbertz,
Voica Radescu,
Misha Ryskin,
Albert de Roeck,
Luca Rottoli,
Robert Thorne,
Gavin Salam,
Carl Schmidt,
Peter Skands,
Daniel Stump,
Maria Ubiali, and
C.~P. Yuan,
among many other colleagues that have helped
pushing the boundaries of global PDF fits to a level that would have
seemed impossible just a few years ago.\\[0.3cm]

\noindent
J.~R. is supported by the European Research Council (ERC)
Starting Grant ``PDF4BSM'' and
by the Dutch Organization for Scientific Research (NWO).
The work of J. G. is
sponsored by Shanghai Pujiang Program.
L.~H.~L thanks the Science and Technology Facilities Council (STFC) for support via grant award ST/L000377/1.

%%%%%%%%%%
\vspace{0.6cm}
%\bibliographystyle{elsarticle-num}
%\bibliography{PDFreview}

\end{document}